\documentclass[11pt,a4paper,footinclude=true,headinclude=true]{book} 

\usepackage[dvipsnames]{xcolor}
\definecolor{color2}{rgb}{0.42, 0.46, 0.75}
\definecolor{color1}{rgb}{0.0, 0.4, 0.75}
\usepackage{hyperref}
\hypersetup{colorlinks=true
,urlcolor=color1
,anchorcolor=color1
,citecolor=color1
,filecolor=color1
,linkcolor=color1
,menucolor=color1
,linktocpage=true
,pdfproducer=medialab
,pdfa=true
}
\usepackage{lipsum}
\usepackage[linedheaders,parts,pdfspacing]{classicthesis} 
\usepackage{amsthm}
\usepackage[T1]{fontenc}  
\usepackage{pdfpages}
\usepackage[utf8]{inputenc}
\usepackage{ragged2e}
\usepackage{soul}
\usepackage{slashed}
\usepackage[makeroom]{cancel}
\usepackage[tableposition=top]{caption}
\usepackage{booktabs}
\usepackage{titletoc}
\usepackage{rotating}
\usepackage{appendix}
\usepackage{graphicx}
\usepackage{amssymb,amsmath}
\usepackage{multirow}
\usepackage{framed,fancybox,xcolor}
\usepackage{xfrac} 
\usepackage{arydshln}
\usepackage{tikz-cd}
\usepackage{float}
\usepackage{enumitem}
\usepackage{hhline}
\usepackage{bbold} 
\usepackage{caption, subcaption}
\usepackage{arabtex}
\usepackage{array}
\usepackage{tabu}

\usepackage[nottoc,notlot,notlof]{tocbibind}
\usepackage{pdfpages}
\usepackage{hyphenat}

\allowdisplaybreaks

\definecolor{col1}{rgb}{0.631373, 0.752941, 0.215686}
\definecolor{col2}{rgb}{0.58,0.035,0.074}
\definecolor{col3}{rgb}{0.9,0.7,0.1}

\DeclareOldFontCommand{\rm}{\normalfont\rmfamily}{\mathrm}
\DeclareOldFontCommand{\it}{\normalfont\itshape}{\mathit}
\DeclareOldFontCommand{\bf}{\normalfont\bfseries}{\mathbf}

\usepackage{titlesec}
\usepackage{charter}

    \titleformat
{\chapter} 
[display] 
{\fontsize{21pt}{21pt}\scshape\bfseries} 
{} 
{-6.0ex} 
{
    {\color{col1}\rule{\textwidth}{0.8mm}}
    \raggedleft
    {{\fontsize{30pt}{30pt}\selectfont\textcolor{col1}{}}}
} 
[
\vspace{-1.6ex}%
{\color{col1}\rule{\textwidth}{0.8mm}}
] 

    \titleformat
{\section} 
[display] 
{\fontsize{15pt}{15pt}\scshape\bfseries} 
{} 
{-3.0ex} 
{
    {{\fontsize{15pt}{15pt}\selectfont\textcolor{col1}{\thesection}}}
} 
[
] 

    \titleformat
{\subsection} 
[display] 
{\fontsize{11pt}{11pt}\scshape\bfseries} 
{} 
{-3.0ex} 
{
    {{\fontsize{11pt}{11pt}\selectfont\textcolor{col1}{\thesubsection}}}
} 
[
] 

\usepackage{tocvsec2}
\usepackage{titletoc}
\usepackage{tocstyle}
\usetocstyle[deactivatetocstyle]{classic}
\settocfeature{entryvskip}{10pt}
\settocstylefeature[-1]{leaders}{\hfill}
\settocstylefeature[-1]{pagenumberhook}{\nullfont}

\usepackage[square,numbers,compress,sort]{natbib}
\def\Ugold{{\rm U}}
\def\Vchi{{\rm V}}
\def\Tchi{{\rm T}}

\newcommand{\gsim}{\lower.7ex\hbox{$\;\stackrel{\textstyle>}{\sim}\;$}}
\newcommand{\lsim}{\lower.7ex\hbox{$\;\stackrel{\textstyle<}{\sim}\;$}}

\newcommand{\bea}{\begin{eqnarray}}
\newcommand{\eea}{\end{eqnarray}}

\def\be{\begin{equation}}
\def\ee{\end{equation}}
\def\ba{\begin{align}}
\def\ea{\end{align}}

\renewcommand\eqref[1]{Eq.~(\ref{#1})}
\newcommand\eqrefs[2]{Eqs.~(\ref{#1})-(\ref{#2})}
\newcommand\figref[1]{Fig.~\ref{#1}}

\newcommand\tabref[1]{Table~\ref{#1}}

\def\significance{\sigma^{\rm stat}}
\def\cosw{c_{\rm w}}
\def\sinw{s_{\rm w}}

\newcommand{\nn}{\nonumber}
\newcommand{\bear}{\begin{eqnarray}}
\newcommand{\eear}{\end{eqnarray}}
\newcommand{\mL}{\mathcal{L}}
\newcommand{\mO}{\mathcal{O}}
\newcommand{\mX}{\mathcal{X}}
\newcommand{\Frac}[2]{\frac{\displaystyle #1}{\displaystyle #2}}

\newcommand{\bbbbjj}{$pp\to b\bar{b}b\bar{b}jj$ }

\def\VSM{V^{\begin{minipage}{1cm}\scalebox{0.6}{\rm SM}\end{minipage}}}
\def\VEChL{V^{\begin{minipage}{1cm}\scalebox{0.6}{\rm EChL}\end{minipage}}}
\def\VIAMMC{V^{\begin{minipage}{1cm}\scalebox{0.6}{\rm IAM-MC}\end{minipage}}}

\usepackage{geometry}

\evensidemargin = \oddsidemargin

\begin{document}
	\pagestyle{scrheadings}

\begingroup
\thispagestyle{empty} 
\begin{tikzpicture}[remember picture,overlay]
\node[inner sep=0pt] (background) at (current page.center) {\includegraphics[width=\paperwidth]{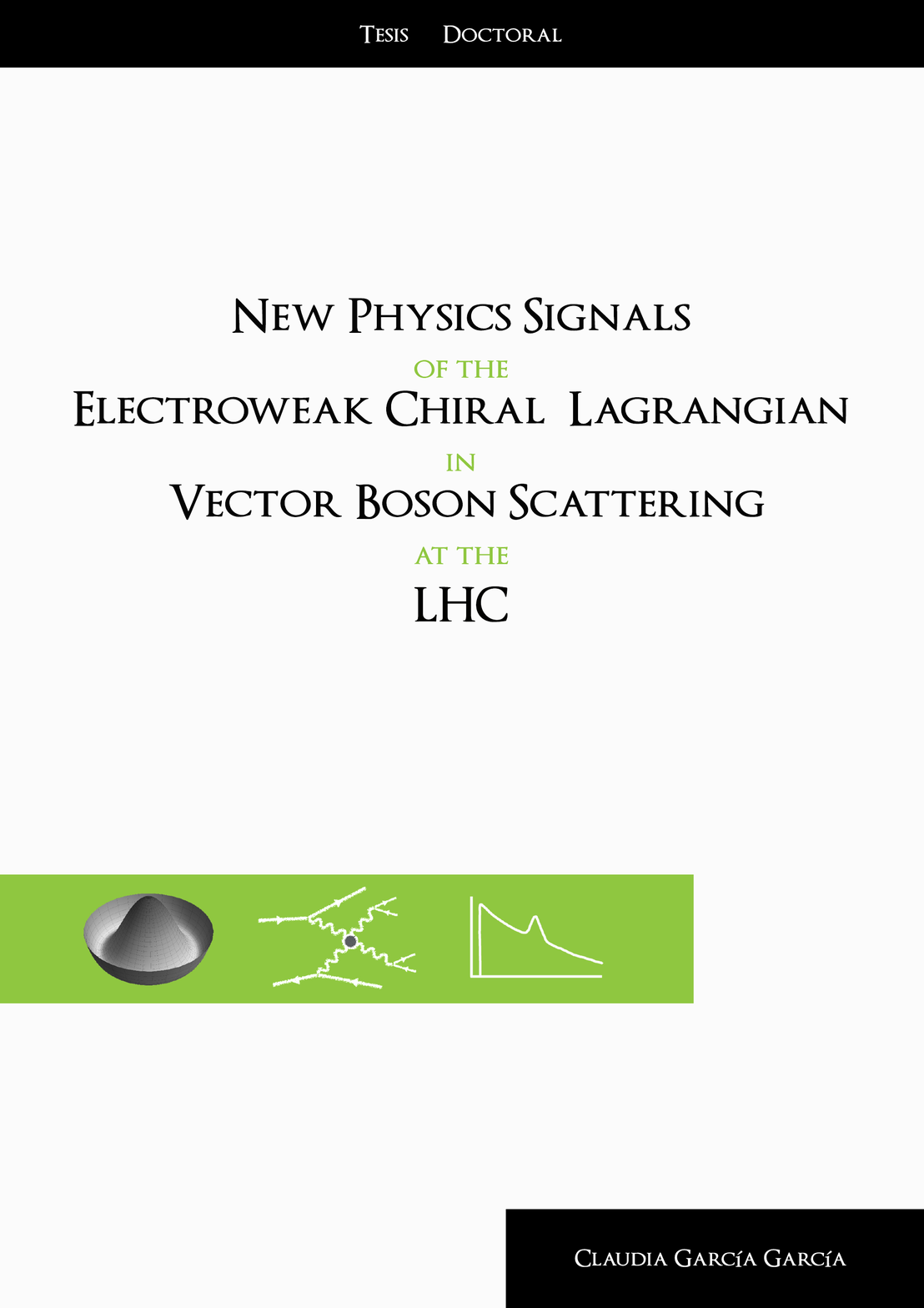}};
\end{tikzpicture}
\vfill
\endgroup

\newpage
\thispagestyle{empty}
~
\newpage
\thispagestyle{empty}
~
\newpage
~\vfill
\thispagestyle{empty}

\noindent \textsc{Universidad Aut\'onoma de Madrid}\\

\noindent The research contained in this thesis has been developed by Claudia Garc\'ia Garc\'ia under the supervision of Dr. Mar\'ia Jos\'e Herrero Solans\\ 

\noindent Copyright \copyright\ 2019 Claudia Garc\'ia Garc\'ia\\ 

\noindent \textit{September 2019} 

~
\newpage
\thispagestyle{empty}
~
\vspace{0.5cm}
\begin{center}
{\Large Memoria de Tesis Doctoral }
\end{center}
\vspace{1.5cm}

\begin{center}
 {\color{col1}\rule{\textwidth}{0.8mm}}\\[5pt]

{ {\huge\textbf{N\scalebox{0.8}{EW}  P\scalebox{0.8}{HYSICS}  S\scalebox{0.8}{IGNALS} \scalebox{0.8}{OF} \scalebox{0.8}{THE} \\[8pt] E\scalebox{0.8}{LECTROWEAK}  C\scalebox{0.8}{HIRAL}  L\scalebox{0.8}{AGRANGIAN}  \scalebox{0.8}{IN} \\[18pt] V\scalebox{0.8}{ECTOR}  B\scalebox{0.8}{OSON}  S\scalebox{0.8}{CATTERING}  \scalebox{0.8}{AT} \scalebox{0.8}{THE} LHC
}}}

\vspace{0.3cm}
 {\color{col1}\rule{\textwidth}{0.8mm}}
		
\end{center}
 
\vspace*{2.85cm}

{ \Large\textbf{C\scalebox{0.8}{LAUDIA} G\scalebox{0.8}{ARC\'IA} G\scalebox{0.8}{ARC\'IA} }}

\vspace*{0.5cm}

		{\large Dirigida por la Dra. Mar\'ia Jos\'e Herrero Solans,\\
		\indent profesora Catedr\'atica del Departamento de F\'isica Te\'orica de la\\
\indent Universidad Aut\'onoma de Madrid}

\vspace*{0.5cm}

		{Madrid, 4 de septiembre de 2019}

\vspace*{3.9cm}
\begin{center}
\begin{figure}[ht]
\begin{tabular}{lr}
\includegraphics[scale=0.15]{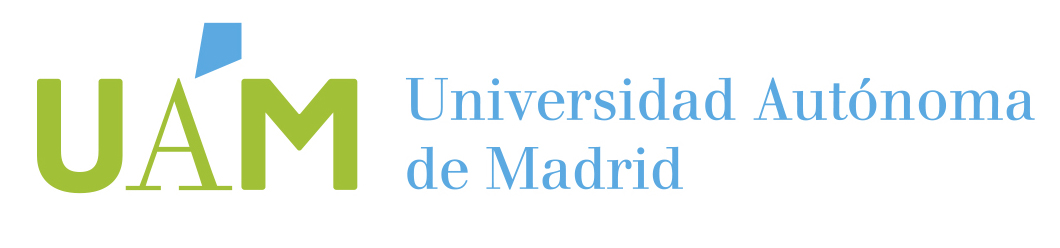}
\qquad\qquad\qquad\qquad\qquad
&
\qquad\includegraphics[scale=0.3]{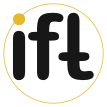}\\
Departamento de Física Teórica&~~~~~~~~~~ Instituto de Física Teórica\\ 
Universidad Autónoma de Madrid &  UAM-CSIC
\end{tabular}
\end{figure}
\end{center}

\newpage
\thispagestyle{empty}
~

\newpage
\thispagestyle{empty}

\vspace*{5cm}
\begin{flushright}
\textit{
A mi madre, por mis ra\'ices,\\
y a mi padre, por mis alas...\\
...y viceversa.
}
\end{flushright}

\newpage
\thispagestyle{empty}
\phantom{asdfaf}

\frontmatter

\thispagestyle{empty}
\begin{center}
	\Large {\bfseries\scshape List of Publications}\\[-8pt]
	 {\color{col1}\rule{6cm}{0.8mm}}
	\vspace{0.5cm}
\end{center}

\justifying


The research contained in this Thesis has lead to the following scientific articles:

\begin{itemize}

 \item[\cite{Arganda:2018ftn}]{\bfseries Probing the Higgs self-coupling in double Higgs production through vector boson scattering at the LHC  } \\
  E. Arganda, C. Garcia-Garcia, M. J. Herrero  \\ 
   \href{https://www.sciencedirect.com/science/article/pii/S0550321319301737?via\%3Dihub}{\textbf{Nucl.Phys. B945 (2019) 114687} }
   [\href{https://arxiv.org/pdf/1807.09736}{\textbf{arXiv:1807.09736}}].

\vspace*{0.2cm}

 \item[\cite{Garcia-Garcia:2019oig}]{ \bfseries Unitarization effects in EFT predictions of WZ scattering at the LHC} \\
  C. Garcia-Garcia, M. J. Herrero, R. Morales \\
   $[$\href{https://arxiv.org/abs/1907.06668}{\textbf{arXiv:1907.06668}}$]$.

\vspace*{0.2cm}

 \item[\cite{Delgado:2017cls}]{\bfseries Production of vector resonances at the LHC via $\boldsymbol{WZ}$-scattering: a unitarized EChL analysis} \\
  R. L. Delgado, A. Dobado, D. Espriu, C. Garcia-Garcia, M. J. Herrero, X. Marcano, J. J. Sanz-Cillero \\ 
   \href{https://link.springer.com/article/10.1007\%2FJHEP11\%282017\%29098}{\textbf{JHEP 1711 (2017) 098 } }
   [\href{https://arxiv.org/pdf/1707.04580}{\textbf{arXiv:1707.04580}}].

\vspace*{0.2cm}

 \item[\cite{Delgado:2019ucx}]{\bfseries Dynamical vector resonances from the EChL in VBS at the LHC: the WW case} \\
  R. L. Delgado, C. Garcia-Garcia, M. J. Herrero \\ 
   $[$\href{https://arxiv.org/pdf/1907.11957}{\textbf{arXiv:1907.11957}}$]$.
   
\end{itemize}
   
 and conference proceedings:
   
\begin{itemize}

 \item[\cite{Delgado:2018lvo}]{\bfseries Collider phenomenology of vector resonances in WZ scattering processes} \\
  R. L. Delgado, A. Dobado, D. Espriu, C. Garcia-Garcia, M. J. Herrero, X. Marcano, J. J. Sanz-Cillero \\ 
   Conference proceeding, $[$\href{https://arxiv.org/abs/1811.08720}{\textbf{arXiv:1811.08720}}$]$.

\vspace*{0.2cm}

 \item[\cite{Delgado:2018nnh}]{\bfseries Resonant $\boldsymbol{pp \to W^+Zjj}$ events at the LHC from a unitarized study of the EChL} \\
  R. L. Delgado, A. Dobado, D. Espriu, C. Garcia-Garcia, M. J. Herrero, X. Marcano, J. J. Sanz-Cillero\\ 
   Conference proceeding, $[$\href{https://arxiv.org/abs/1805.04870}{\textbf{arXiv:1805.04870}}$]$.

\end{itemize}

\newpage
\thispagestyle{empty}
\phantom{asdfaf}


\thispagestyle{empty}
\phantom{asdfaf}
\begin{center}
\textbf{\large Abstract}\\[-5pt]
 {\color{col1}\rule{2.5cm}{0.8mm}}
\end{center}

 The Standard Model of fundamental interactions, albeit an incredibly elegant and successful theory, lacks explanations for some experimental and theoretical open questions. Interestingly, many of these problems seem to be related to the electroweak symmetry breaking sector of the theory, whose dynamical generation is still unknown. Important questions such as what is the true nature of the Higgs boson, why is its mass so light and so close to that of the electroweak gauge bosons or whether the properties of this particle are the ones predicted in the Standard Model remain unanswered. The LHC is our tool to unveil these mysteries and vector boson scattering processes are the perfect window to access them, since they are considered as the most sensitive observables to new physics in the electroweak symmetry breaking sector. In this Thesis we employ the effective electroweak chiral Lagrangian with a light Higgs, which assumes a strongly interacting electroweak symmetry breaking sector, to perform a model independent analysis of the phenomenology of vector boson scattering processes at the LHC as well as to present quantitative predictions for the sensitivity to possible beyond the Standard Model physics scenarios.

\newpage
\thispagestyle{empty}
\phantom{asdfaf}


\thispagestyle{empty}
\phantom{asdfaf}

\begin{center}
\textbf{\large Resumen}\\[-5pt]
 {\color{col1}\rule{2.5cm}{0.8mm}}
\end{center}

A pesar de que el Modelo Est\'andar de las interacciones fundamentales es una de las teor\'ias m\'as elegantes y exitosas en el contexto de la f\'isica de part\'iculas, existen ciertas cuestiones tanto experimentales como te\'oricas que no es capaz de explicar. Muchas de ellas parecen estar relacionadas con el sector de ruptura espont\'anea de la simetr\'ia electrod\'ebil, cuyo origen din\'amico nos es a\'un desconocido. Preguntas tales como cu\'al es la verdadera naturaleza del bos\'on de Higgs, por qu\'e es \'este tan ligero, estando su masa tan pr\'oxima a la de los bosones gauge electrod\'ebiles o si sus acoplamientos al resto de part\'iculas del Modelo Est\'andar son como los predichos por esta teor\'ia todav\'ia deben ser contestadas. El LHC es nuestra herramienta para desvelar estos misterios y los procesos de {\it scattering} de bosones vectoriales son la ventana perfecta para acceder a ellos, ya que dichos procesos se consideran como los m\'as sensibles a la posible nueva f\'isica relacionada con el sector de ruptura espont\'anea de la simetr\'ia electrod\'ebil. En esta Tesis empleamos el Lagrangiano quiral electrod\'ebil efectivo con un Higgs ligero, que asume un sector de ruptura espont\'anea de la simetr\'ia electrod\'ebil que interacciona fuertemente, para llevar a cabo un an\'alisis exhaustivo de la fenomenolog\'ia de los procesos de {\it scattering} de bosones vectoriales en el LHC, as\'i como para presentar predicciones cuantitativas de la sensibilidad de este experimento a algunos escenarios de f\'isica m\'as all\'a del Modelo Est\'andar.

\newpage
\thispagestyle{empty}
\phantom{asdfaf}

\tableofcontents 
\chaptermark{}
\thispagestyle{empty}
%
%
\mainmatter
\renewcommand{\sectionmark}[1]{\markright{\spacedlowsmallcaps{#1}}{}}

    \titleformat
{\chapter} 
[display] 
{\fontsize{21pt}{21pt}\scshape\bfseries} 
{} 
{-6.0ex} 
{
    {\color{col1}\rule{\textwidth}{0.8mm}}
    \raggedleft
    {{\fontsize{30pt}{30pt}\selectfont\textcolor{col1}{}}}
} 
[
\vspace{-0.5ex}%
{\color{col1}\rule{\textwidth}{0.8mm}}
] 

%
\setcounter{page}{1}
%
%

\chapter*{Introducci\'on}\label{Introduccion}
\addcontentsline{toc}{chapter}{\scshape\bfseries Introducci\'on}
\chaptermark{Introducci\'on}

Existe un gran parecido entre intentar entender c\'omo funciona realmente la naturaleza y buscar un tesoro. Lo primero que uno necesita en ambos casos, es, sin duda, un mapa. Una gu\'ia fiable que apunte hacia las gemas enterradas, una base s\'olida desde la que empezar.

En el caso de la f\'isica de altas energ\'ias, nuestro mapa es el Modelo Est\'andar de las interacciones fundamentales~\cite{Glashow:1961tr,GellMann:1964nj,Weinberg:1967tq,Salam:1968rm} (SM de sus siglas en ingl\'es), quiz\'a una de las teor\'ias m\'as exitosas de la historia de la ciencia. Este marco te\'orico, a pesar de ser en principio simple, es capaz de clasificar todas las part\'iculas elementales conocidas y de describir tres de las cuatro fuerzas fundamentales a trav\'es de las que interaccionan: la electromagn\'etica, la d\'ebil y la fuerte.

El SM es una teor\'ia cu\'antica de campos invariante bajo transformaciones locales o gauge del grupo $SU(3)_C\times SU(2)_L\times U(1)_Y$, as\'i como bajo transformaciones del grupo de Poincar\'e (invariante Lorentz). Tambi\'en es, por construcci\'on, renormalizable. Estos principios de simetr\'ia dan lugar a la descripci\'on de las interacciones del SM, mediadas por sus correspondientes bosones gauge de spin 1. Las interacciones fuerte y electromagn\'etica poseen mediadores no masivos y neutros: el glu\'on y el fot\'on, respectivamente. Las interacciones d\'ebiles son mediadas, no obstante, por bosones vectoriales masivos: el W${}^+$, el W${}^-$ y el Z, cuyas cargas son, en unidades de la carga del electr\'on, uno, menos uno y cero respectivamente.

El contenido de materia del SM se organiza en tres familias que contienen cuatro fermiones cada una, siendo la \'unica diferencia existente entre ellas la masa de las part\'iculas involucradas. Sin embargo, los fermiones de las tres familias poseen las mismas propiedades (salvo la masa ya mencionada) y n\'umeros cu\'anticos lo que hace que  esta teor\'ia posea una estructura basada en tres copias id\'enticas cada vez m\'as masivas. Cada una de estas familias alberga dos quarks, uno con carga positiva y otro con carga negativa, y dos leptones, uno con carga negativa y otro neutro. Este patr\'on de cuatro fermiones por familia est\'a fijado por las simetr\'ias gauge ya que sin \'el no podr\'ia asegurarse la cancelaci\'on de anomal\'ias. 

La \'ultima pieza del modelo (y tambi\'en la \'ultima en ser observaba experimentalmente) corresponde a un bos\'on escalar:  el bos\'on de Higgs, remanente del proceso de generaci\'on de masa incluido en el SM a trav\'es del mecanismo de Brougt-Englert-Higgs~\cite{Higgs:1964ia,Higgs:1964pj,Englert:1964et,Guralnik:1964eu, Higgs:1966ev} (BEH, o simplemente Higgs, como es usual encontrarlo en la literatura). Este mecanismo ejemplifica c\'omo la ruptura espont\'anea de la simetr\'ia electrod\'ebil puede dar lugar a las masas observadas de los Ws, el Z y los fermiones a excepci\'on de los neutrinos, cuya masa en el SM es nula.

Con este contenido de part\'iculas y estos principios de simetr\'ia puede construirse el Lagrangiano del SM introduciendo ciertos par\'ametros cuyo valor ha de ser fijado emp\'iricamente. Dichos par\'ametros corresponden a los tres acoplamientos gauge, que controlan la intensidad de las interacciones; las matrices de Yukawa, que albergan las masas y los mixings de los fermiones, y los dos par\'ametros asociados al potencial del Higgs: su masa y su autoacoplamiento. Una vez que los valores de estos par\'ametros son conocidos, el SM es capaz de predecir la mayor\'ia de fen\'omenos que tienen lugar en la f\'isica de part\'iculas con extrema precisi\'on.

No obstante, en este momento nuestro empe\~no por encontrar el preciado tesoro se ve interrumpido: el mapa est\'a incompleto.

A pesar de la impresionante precisi\'on de las predicciones del modelo y del acuerdo entre \'estas y las medidas experimentales correspondientes, hoy en d\'ia sabemos que el SM no puede ser la teor\'ia que describa por completo la naturaleza, ya que no posee explicaciones satisfactorias para varios fen\'omenos f\'isicos que han sido observados en las pasadas d\'ecadas ni para determinados prejuicios te\'oricos.

Los dos ejemplos m\'as importantes de estos fen\'omenos aun por explicar son las oscilaciones de neutrinos y la existencia de la materia oscura, entre otros. Aunque los neutrinos se consideren no masivos por construcci\'on en el SM, se ha determinado experimentalmente que en realidad han de poseer una masa no nula debido al hecho de que oscilan en sabor~\cite{Fukuda:1998mi,Fukuda:1998ah,Ahmad:2002jz,Eguchi:2002dm}. Esto irremediablemente supone que f\'isica m\'as all\'a del SM (BSM) ha de ser considerada. Una de las formas m\'as comunes de dotar a los neutrinos de masa es extender el contenido de part\'iculas del SM a\~nadiendo neutrinos dextr\'ogiros, ausentes a priori en la teor\'ia. Con estos nuevos campos es posible construir un t\'ermino de masa de Dirac para los neutrinos a trav\'es de su interacci\'on con el bos\'on de Higgs, al igual que sucede con el resto de fermiones. Estas nuevas part\'iculas tendr\'ian adem\'as la peculiaridad de que podr\'ian ser estados de Majorana, debido a sus n\'umeros cu\'anticos, de forma que podr\'ian recibir parte de su masa de manera independiente al campo de Higgs.

Por otro lado, evidencias astrof\'isicas y cosmol\'ogicas apuntan hacia el hecho de que cierta materia {\it invisible}, que conformar\'ia aproximadamente el 25\% del contenido energ\'etico total del universo, ha de existir~\cite{Bessel:1844ck,Zwicky:1933gu,Zwicky:1937zza,Rubin:1970zza,Begeman:1991iy,Taylor:1998uk,Clowe:2003tk,Ade:2015xua}. Esta materia oscura interacciona gravitacionalmente, pero las caracter\'isticas de sus interacciones a trav\'es de las otras fuerzas aun est\'an por descubrir. Por esta raz\'on, no existe un candidato apropiado en el SM a part\'icula de materia oscura, y es necesario construir extensiones de la teor\'ia que s\'i lo alberguen.

En cuanto a los prejuicios te\'oricos que sufre el SM, es importante mencionar que no se trata m\'as que de eso: de prejuicios, llamados, normalmente, problemas de {\it fine tuning} o de naturalidad~\cite{tHooft:1979rat}. Estos problemas se refieren a  las medidas experimentales de determinados par\'ametros del SM que han resultado ser m\'as extremas de lo que se ten\'ia pensado sin ning\'un principio de simetr\'ia que ayude a explicarlas. En cualquier caso, estas medidas podr\'ian ser meras coincidencias; valores que la naturaleza ha elegido de entre otros muchos posibles. Sin embargo, tambi\'en podr\'ian estar indicando que existe f\'isica m\'as all\'a del SM que aun no conocemos.
 
Uno de estos dilemas te\'oricos es el llamado {\it strong CP-problem}, o problema de la violaci\'on de conjugaci\'on de carga y paridad (CP) en las interacciones fuertes. A pesar de que en el sector de la interacci\'on fuerte del SM puede a\~nadirse un t\'ermino  que viola CP en el Lagrangiano preservando el resto de simetr\'ias, el momento dipolar magn\'etico del neutr\'on que se deriva de dicho t\'ermino no ha sido observado aun. Este hecho impone una cota al par\'ametro del Lagrangiano que controla la violaci\'on de CP en el sector fuerte, el par\'ametro $\theta$, que ha de ser m\'as peque\~no que $10^{-10}$~\cite{Baluni:1978rf,Crewther:1979pi,Kanaya:1981se,Schnitzer:1983pb,Cea:1984qv,Musakhanov:1984qy,Morgan:1986yy}. Ya que no hay ninguna explicaci\'on para un valor tan sumamente peque\~no, \'este puede interpretarse como una pista hacia nueva f\'isica. Comunmente, se utiliza la conocida simetr\'ia de Peccei-Quinn~\cite{Peccei:1977hh}, cuya ruptura espont\'anea da lugar al axi\'on (aun no observado), como posible soluci\'on a este problema.

Otra de estas cuestiones es la que concierne al peculiar patr\'on de masas y mezclas de sabor fermi\'onicos, es decir, a la estructura de las matrices de Yukawa. Las masas de los fermiones del SM van desde la escala del MeV hasta cientos de GeV, y, aunque los {\it mixings} (mezclas de sabor entre fermiones) en el sector de los quarks resultan ser bastante jer\'arquicos, esto no sucede en el sector lept\'onico. Este misterio, as\'i como la existencia de tres y (por ahora) s\'olo tres familias se conoce como el puzle del sabor.

La \'ultima propiedad no deseada del SM que mencionaremos aqu\'i es la del llamado problema de las jerarqu\'ias. \'Este se relaciona con el hecho de que en el SM el bos\'on de Higgs es una part\'icula fundamental escalar cuya masa, desprotegida por las simetr\'ias, se introduce {\it ad hoc} en el Lagrangiano. Si apareciera nueva f\'isica por encima de la escala electrod\'ebil, el bos\'on de Higgs podr\'ia acoplarse a ella, y las correcciones radiativas a la masa del Higgs resultar\'ian ser cuadr\'aticamente dependientes de la escala de la nueva f\'isica. Si se asume que no hay f\'isica m\'as all\'a del SM hasta la escala de Planck, del order de $10^{19}$ GeV, a la cual se cree que la gravedad participa de forma relevante en las interacciones entre part\'iculas, el valor de la masa del Higgs se acercar\'ia mucho a dicha escala. No obstante, el valor observado difiere por muchos \'ordenes de magnitud de esta cantidad: $m_H=125.09\pm0.21\,(stat.)\pm0.11\,(\rm syst.)$ GeV~\cite{Aad:2015zhl}. De hecho, la masa del bos\'on de Higgs est\'a muy pr\'oxima a la escala electrod\'ebil, $v=246$ GeV y por tanto a las masas de los bosones gauge electrod\'ebiles. Es por esto que una extraordinaria cancelaci\'on tiene que darse entre la masa desnuda y las correcciones cu\'anticas para poder obtener una predicci\'on compatible con el dato experimental.

Debido a la existencia de los problemas descritos en los p\'arrafos previos parece estar claro que a  nuestro mapa le falta la parte donde se encuentra enterrado el tesoro, y que debemos explorar {\it terra incognita} lo mejor que podamos para conseguir encontrarlo.

Cuando se analizan los problemas del SM es interesante darse cuenta de que muchos de ellos est\'an o pueden estar relacionados con el sector de ruptura espont\'anea de la simetr\'ia electrod\'ebil de la teor\'ia. En el SM, el campo de Higgs se interpreta como un doblete complejo que desencadena la ruptura espont\'anea de la simetr\'ia electrod\'ebil al adquirir un valor esperado en el vac\'io invariante bajo interacciones electromagn\'eticas. En este proceso, descrito por el mecanismo BEH, tres grados de libertad contenidos en el doblete escalar, correspondientes a los bosones de Goldstone electrod\'ebiles, se manifiestan como las componentes longitudinales de los bosones gauge electrod\'ebiles, que adquieren as\'i una masa. El otro grado de libertad aparece como una nueva part\'icula presente en el espectro de la teor\'ia: el bos\'on de Higgs.

Este mecanismo resolv\'ia varias de las limitaciones del SM primitivo: adem\'as de proveer una explicaci\'on para las masas de los bosones electrod\'ebiles, que se sab\'ian no nulas debido al corto alcance de la interacci\'on correspondiente, tambi\'en aportaba una forma de generar masas para los fermiones. Adem\'as, a trav\'es de \'el pod\'ia curarse la violaci\'on de unitariedad presente en las amplitudes de probabilidad de {\it scattering} de bosones electrod\'ebiles masivos. Sin embargo, como ya ha sido comentado en los p\'arrafos anteriores, aun parece haber algo m\'as all\'a del mecanismo de ruptura de la simetr\'ia electrod\'ebil del SM. Asimismo, tras el descubrimiento del bos\'on de Higgs~\cite{Aad:2012tfa,Chatrchyan:2012xdj}, numerosas b\'usquedas experimentales se est\'an centrando en recabar tantos datos como sea posible para arrojar algo de luz sobre la verdadera naturaleza del sector escalar del SM~\cite{Chatrchyan:2012jja,Aad:2013xqa,Aad:2015zhl,Khachatryan:2014jba,Khachatryan:2014iha,Aad:2015xua,Aad:2015pla,Khachatryan:2014qaa,Aad:2015gra,ATLAS:2013rma,Khachatryan:2015lga,Aad:2014xva,Khachatryan:2014aep,Heinemeyer:2013tqa,LHCHiggsCrossSectionWorkingGroup:2012nn}, y, por lo tanto, es un excelente momento para investigar en esta direcci\'on.

Pero no sabemos hacia qu\'e punto cardinal debemos extender nuestro mapa, as\'i que nuestro objetivo ser\'a hacerlo en la manera m\'as general posible. Con este criterio haciendo de br\'ujula, la opci\'on m\'as clara ser\'ia utilizar teor\'ias de campos efectivas. Estas teor\'ias describen la din\'amica a bajas energ\'ias de una teor\'ia ultravioleta completa, siendo esta \'ultima, en el caso que nos ata\~ne, aquella que explicara los atributos desconocidos del la ruptura de la simetr\'ia electrod\'ebil. El hecho de que las teor\'ias efectivas sean independientes de modelos concretos reside en que la f\'isica a altas energ\'ias es codificada en una serie de par\'ametros libres (a priori) de baja energ\'ia. Si estos par\'ametros se midieran experimentalmente tendr\'iamos acceso a pistas sobre la teor\'ia ultravioleta completa.

Existen, sin embargo, varias maneras de emplear las teor\'ias efectivas para describir la din\'amica de la ruptura de la simetr\'ia electrod\'ebil. Lo primero que se viene a la mente es sin duda utilizar el SM como punto de partida. Sin necesidad de extender su contenido de part\'iculas pero renunciando al criterio de renormalizabilidad, pueden incluirse nuevos operadores en el Lagrangiano que respeten las simetr\'ias requeridas. Estos operadores, que poseen una dimensi\'on energ\'etica mayor que cuatro, y que, por tanto, est\'an suprimidos por potencias correspondientes de la escala de nueva f\'isica, se organizan en una expansi\'on basada en su dimensi\'on can\'onica. En este sentido entran en juego nuevas estructuras motivadas por la presencia de f\'isica m\'as all\'a del SM a energ\'ias altas y que pueden modificar las interacciones relacionadas con el sector de ruptura de la simetr\'ia electrod\'ebil. Este marco te\'orico corresponde a la realizaci\'on lineal de la ruptura de la simetr\'ia electrod\'ebil, en el cual el campo de Higgs y los bosones de Goldstone forman un doblete de $SU(2)_L$ que interacciona d\'ebilmente. Los t\'erminos de orden m\'as bajo en la expansi\'on coinciden con los t\'erminos del SM, y, por esta raz\'on, esta teor\'ia efectiva suele llamarse SMEFT, de SM effective field theory, en ingl\'es. La fenomenolog\'ia asociada al SMEFT ha sido (y est\'a siendo), de hecho, ampliamente estudiada. Algunos ejemplos de recientes an\'alisis en este contexto pueden encontrarse en~\cite{Eboli:2006wa,Corbett:2013pja,Ellis:2014dva,Ellis:2014jta,Corbett:2014ora,Degrande:2016dqg,Ellis:2018gqa,Barzinji:2018xvu,Grazzini:2018eyk,Perez:2018kav,Dawson:2018liq,Dawson:2018jlg,Dawson:2019xfp,Baglio:2018bkm,Almeida:2018cld,Hartland:2019bjb,Buchalla:2019wsc,Moutafis:2019wbp,Aoude:2019tzn,Brivio:2019myy}.

No obstante, puede construirse un escenario m\'as general. Tomando como referencia s\'olo el contenido de part\'iculas y las simetr\'ias del SM se llega a una nueva clase de teor\'ias efectivas en las que el bos\'on de Higgs es considerado un singlete del SM y puede ser por tanto tratado de forma independiente a los bosones de Goldstone. Estos \'ultimos s\'olo poseen acoplamientos derivativos asociados a nuevas interacciones fuertes. Por lo tanto, en este supuesto, el Higgs y los bosones de Goldstone no forman un doblete de $SU(2)_L$, y la ruptura de la simetr\'ia electrod\'ebil es implementada de manera no lineal~\cite{Appelquist:1980vg,Longhitano:1980iz,Longhitano:1980tm,Chanowitz:1985hj,Cheyette:1987jf,Dobado:1989ax,Dobado:1989ue,Dobado:1990zh,Espriu:1991vm,Dobado:1990jy,Dobado:1990am,Dobado:1995qy,Dobado:1999xb}. Los operadores contenidos en el Lagrangiano est\'an organizados, en este caso, en una expansion en potencias del momento externo debido al car\'acter derviativo de las interacciones entre bosones de Goldstone, siento el Lagrangiano de orden m\'as bajo aquel que contiene t\'erminos con dos derivadas, el de siguiente orden aquel que contiene t\'erminos con cuatro derivadas, y as\'i sucesivamente. El campo de Higgs se introduce normalmente a trav\'es de una funci\'on polin\'omica gen\'erica, ya que sus interacciones no est\'an limitadas a acoplamientos derivativos. Este Lagrangiano no lineal o Lagrangiano quiral electrod\'ebil~\cite{Alonso:2012px,Alonso:2012pz,Buchalla:2013rka,Espriu:2012ih, Delgado:2013loa, Delgado:2013hxa, Brivio:2013pma, Espriu:2013fia, Espriu:2014jya,  Delgado:2014jda, Buchalla:2015qju, Arnan:2015csa,Brivio:2016fzo} (EChL de sus siglas en ingl\'es), es llamado as\'i en analog\'ia al Lagrangiano quiral en cromodin\'amica cu\'antica utilizado para explicar la din\'amica de piones a bajas energ\'ias~\cite{Weinberg:1978kz,Gasser:1984gg, Gasser:1983yg}. Adem\'as, el marco te\'orico lineal es realmente un l\'imite espec\'ifico del caso no lineal, aunque la relaci\'on entre ambos es ciertamente no trivial~\cite{Alonso:2012px,Eboli:2016kko,Brivio:2016fzo}.

En esta Tesis nos centraremos en la descripci\'on de la din\'amica de ruptura de la simetr\'ia electrod\'ebil dada por el EChL. En particular, estudiaremos las predicciones de esta teor\'ia efectiva y la posibilidad de testarlas en experimentos actuales y futuros. Para ello, lo primero que debemos hacer es determinar los observables concretos en los que las desviaciones con respecto al SM deber\'ian observarse con mayor claridad. Idealmente, acceder directamente a las interacciones entre bosones de Goldstone nos dar\'ia la informaci\'on que necesitamos, ya que estos representan el coraz\'on del sector de ruptura de la simetr\'ia electrod\'ebil. Sin embargo estas part\'iculas no son f\'isicas y no pueden ser observadas experimentalmente. No obstante, debido a que conforman el grado de polarizaci\'on longitudinal de los bosones gauge electrod\'ebiles, el {\it scattering} de estos \'ultimos, com\'unmente conocido como scattering de bosones vectoriales~\cite{Szleper:2014xxa,Rauch:2016pai,Anders:2018gfr,Bellan:2019xpr} (VBS de {\it vector boson scattering}), deber\'ia ser el mejor lugar para buscar las se\~nales predichas por el EChL. Adem\'as, ya que los acoplamientos de los bosones de Goldstone son proporcionales al momento externo de las part\'iculas, la regi\'on de altas energ\'ias de los procesos de VBS resultar\'ia ser la m\'as sensible a la nueva f\'isica. Debido a que basamos nuestra hip\'otesis en la idea de que las interacciones puramente bos\'onicas deber\'ian ser aquellas que sufrieran mayores desviaciones provenientes de un sector de ruptura de la simetr\'ia electrod\'ebil m\'as all\'a del SM, nos centraremos s\'olo en ellas. Asumiremos, por lo tanto, que las interacciones que involucren fermiones permanecer\'an id\'enticas a como est\'an descritas en el SM.

Ahora que hemos determinado el observable con mayor potencial para descubrir se\~nales del EChL, la pregunta reside en qu\'e experimento deber\'ia buscar dichas se\~nales. Llegados a esta pregunta la escala de energ\'ia t\'ipica del EChL entra en juego. Esta escala, que controla las contribuciones cu\'anticas de la teor\'ia, resulta ser $4\pi v\sim3$ TeV (como sucede de forma similar en el Lagrangiano quiral de las interacciones fuertes, donde esta escala es $4\pi f_\pi\sim$1 GeV) lo cual motiva claramente la escala del TeV apuntando, por tanto al Gran Colisionador de Hadrones (LHC de {\it Large Hadron Collider}). Adem\'as, la producci\'on de bosones vectoriales desde quarks (protones) iniciales da lugar a part\'iculas finales con unas caracter\'isticas cinem\'aticas muy bien definidas que las hacen f\'aciles de detectar. Esto supone una incre\'ible ventaja a la hora de reconocer de manera eficiente las topolog\'ias del VBS, facilitando su selecci\'on de entre otros procesos generados en el LHC.

Sin embargo, cuando se analizan las predicciones de observables del VBS en el EChL aparece una inconsistencia relacionada con la violaci\'on de unitariedad perturbativa. A pesar de que la teor\'ia ultravioleta completa debe ser unitaria para asegurar la conservaci\'on de la probabilidad, la expansi\'on perturbativa a bajas energ\'ias que da lugar a la descripci\'on de la teor\'ia efectiva sufre, normalmente, problemas de violaci\'on de unitariedad. En particular, esto sucede para los elementos de la matriz S que involucran bosones gauge polarizados longitudinalmente, cuyas amplitudes de probabilidad crecen an\'omalamente con la energ\'ia. Es por esto que la validez de la teor\'ia efectiva se pierde cerca del valor de la energ\'ia al cual se da la violaci\'on de unitariedad. En el escenario no lineal es t\'ipica la aparici\'on de resonancias pesadas generadas din\'amicamente por las interacciones fuertes que curan el problema de la violaci\'on de unitariedad. Las propiedades de estas resonancias, en algunos casos, pueden obtenerse directamente desde las predicciones de la teor\'ia efectiva~\cite{Dobado:1999xb,Alboteanu:2008my,Eboli:2011ye,Ballestrero:2011pe,Espriu:2012ih, Espriu:2013fia,Delgado:2013loa, Delgado:2013hxa, Delgado:2014dxa, Espriu:2014jya, Arnan:2015csa, Dobado:2015hha,Corbett:2015lfa,Reuter:2016kqo,DelgadoLopez:2017ugq,BuarqueFranzosi:2017prc,Delgado:2017cls,BuarqueFranzosi:2017jrj,Delgado:2017cat,Delgado:2019ucx} empleando lo que se conoce como m\'etodos de unitarizaci\'on: prescripciones que permiten tomar la predicci\'on cruda de la teor\'ia efectiva y convertirla en unitaria. Se sabe que estos m\'etodos funcionan tambi\'en en el caso no resonante y que solo algunos de ellos pueden acomodar la aparici\'on de estados pesados en el espectro.

Llegados a este punto, podr\'iamos decir que hemos extendido nuestro mapa del tesoro y que hemos marcado con cruces los lugares donde hay m\'as probabilidad de encontrar el cofre enterrado, por lo que ahora nos toca cavar y agarrar las gemas.

Esta Tesis desarrolla un estudio fenomenol\'ogico de las implicaciones del EChL apoy\'andose en VBS en el LHC como observable principal para ello. Con este prop\'osito, se caracterizan exhaustivamente las generalidades de los procesos de VBS tanto a nivel de subproceso como en el contexto del LHC. La importancia que tuvieron este tipo de configuraciones en el descubrimiento del bos\'on de Higgs motiva la primera parte del an\'alisis. En ella, el potencial de los procesos de VBS es utilizado para estudiar la producci\'on de dos bosones de Higgs con el objetivo de obtener una medida competitiva del autoacoplamiento del Higgs en estos canales, complementar\'ia a la obtenida a trav\'es la fusi\'on de gluones explorada com\'unmente en la literatura~\cite{Glover:1987nx, Dicus:1987ic, Plehn:1996wb, Dawson:1998py, Djouadi:1999rca, Baur:2003gp, Grober:2010yv, Dolan:2012rv, Papaefstathiou:2012qe, Baglio:2012np, deFlorian:2013jea, Dolan:2013rja,  Frederix:2014hta, Liu-Sheng:2014gxa, Goertz:2014qta, Azatov:2015oxa, Dicus:2015yva, Dawson:2015oha, He:2015spf, Dolan:2015zja, Cao:2015oaa, Cao:2015oxx, Behr:2015oqq,  Bishara:2016kjn, Cao:2016zob, deFlorian:2016spz, Adhikary:2017jtu, Banerjee:2018yxy, Goncalves:2018qas}. Para ello se consideran tanto el valor del SM como valores m\'as all\'a del SM, siendo estos \'ultimos explicados por las nuevas interacciones introducidas por el EChL.

En la segunda parte se presta especial atenci\'on al problema de la violaci\'on de unitariedad. Primero, debido a que existen varios m\'etodos de unitarizaci\'on que arreglan el mencionado problema, llevamos a cabo por primera vez un an\'alisis comparativo de sus predicciones para el {\it scattering} el\'astico de WZ en WZ en el LHC. Estudiamos el impacto de estos m\'etodos en las cotas experimentales que pueden imponerse a los par\'ametros del EChL y motivamos su uso combinado en el caso no resonante. Por otro lado, analizamos el escenario resonante, en la que estados pesados se generan din\'amicamente desde las interacciones fuertes, que resulta ser diferente. 

En la parte final de la Tesis empleamos el M\'etodo de la Amplitud Inversa~\cite{Espriu:2012ih, Espriu:2013fia,Delgado:2013loa, Delgado:2013hxa, Delgado:2014dxa, Espriu:2014jya, Arnan:2015csa, Dobado:2015hha, Corbett:2015lfa,BuarqueFranzosi:2017prc}, ampliamente conocido en el contexto de QCD a bajas energ\'ias~\cite{Truong:1988zp, Dobado:1989qm, Dobado:1992ha, Hannah:1995si}, para caracterizar se\~nales de resonancias vectoriales en el WZ {\it scattering} en el LHC. Cuantificamos la sensibilidad presente y futura de este experimento a dichas resonancias en el canal de desintegraci\'on puramente lept\'onico de los bosones gauge finales. Adem\'as, exploramos tambi\'en el  WW {\it scattering} y, dado que en este caso la desintegraci\'on lept\'onica involucra dos neutrinos en el estado final, lo cual no permite reconstruir de forma precisa las propiedades de las resonancias, estudiamos el caso puramente hadr\'onico. Examinamos la regi\'on cinem\'atica en la que los productos de desintegraci\'on hadr\'onicos de los bosones vectoriales son detectados como un \'unico jet de radio considerable empleando t\'ecnicas modernas de reconstrucci\'on de fat jets, como son llamados. De este modo, proporcionamos la sensibilidad correspondiente a las resonancias vectoriales en el WW {\it scattering}.

Esta Tesis esta organizada como sigue: En el Cap\'itulo \ref{EChL} se introducen los conceptos b\'asicos relacionados con la ruptura espont\'anea de la simetr\'ia electrod\'ebil, incluyendo la realizaci\'on lineal y no lineal, as\'i como las diferentes teor\'ias efectivas que describen la din\'amica de dicha ruptura. Se hace especial \'enfasis en el EChL ya que corresponde al marco te\'orico m\'as general que puede construirse en este contexto. Se lleva tambi\'en a cabo una comparaci\'on ilustrativa entre esta teor\'ia y su versi\'on lineal, el SMEFT, y , por \'ultimo, se discute el problema de la violaci\'on de unitariedad presente en este tipo de teor\'ias, as\'i como las posibles maneras de solucionarlo.

El Cap\'itulo \ref{VBS} est\'a dedicado a la caracterizaci\'on del VBS. En \'el se motivan este tipo de observables como uno de los canales m\'as prometedores para descubrir nuevas interacciones entre escalares para despu\'es presentar las predicciones tanto del SM como del EChL de diferentes canales de VBS. La violaci\'on de unitariedad es caracterizada en este contexto en t\'erminos del criterio de unitariedad sobre las ondas parciales. Tras esto, se discute el VBS en el LHC, y en particular su cinem\'atica asociada que ser\'a la clave para seleccionar estos procesos de entre los muchos que tienen lugar en las colisiones del LHC. Finalmente, se resumir\'an brevemente las b\'usquedas actuales y futuras de configuraciones VBS en el LHC.

En el Cap\'itulo \ref{HH}, se estudia la posibilidad de obtener una medida del autoacoplamiento del  Higgs a trav\'es de la producci\'on de dos bosones de  Higgs via VBS en el LHC. Este estudio est\'a motivado por el hecho de que los procesos VBS sufren muchos menos incertidumbres te\'oricas que otros que tambi\'en pueden llevar a medidas de este par\'ametro, tambi\'en llamado acoplamiento trilineal. En este Cap\'itulo estudiamos escenarios en los que el acoplamiento del Higgs toma valores tanto del SM como m\'as all\'a, estando el \'ultimo caso contemplado en la prescripci\'on del EChL. En este contexto, realizamos un an\'alisis exhaustivo de los eventos de la se\~nal y de los posibles fondos en el LHC prestando especial atenci\'on a la identificaci\'on de las configuraciones del VBS y a la reconstrucci\'on de los pares de bosones de Higgs para obtener sensibilidades competitivas a valores del acoplamiento trilineal m\'as all\'a del SM.

El Cap\'itulo \ref{Methods} contiene el estudio del impacto que diferentes m\'etodos de unitarizaci\'on  pueden tener en la interpretaci\'on de los datos experimentales en el caso no resonante. Utilizamos el {\it scattering} el\'astico de WZ en WZ para ilustrar c\'omo las cotas impuestas a los par\'ametros del EChL dependen fuertemente del m\'etodo de unitarizaci\'on empleado. Con este objetivo, realizamos un an\'alisis de canales de helicidad acoplados y seleccionamos los cinco m\'etodos de unitarizaci\'on m\'as com\'unmente empleados en la literatura. Obtenemos as\'i las regiones de exclusi\'on en el espacio de par\'ametros del EChL que corresponden a interpretar los datos experimentales con cada uno de los m\'etodos y con todos ellos combinados. Por lo tanto, proporcionamos la incertidumbre te\'orica asociada a la determinaci\'on experimental de los par\'ametros del EChL debida a la elecci\'on del m\'etodo de unitarizaci\'on.

En cuanto al Cap\'itulo \ref{Resonances}, en \'el se incluyen los resultados concernientes a la sensibilidad del LHC en observables de VBS a resonancias vectoriales cargadas generadas din\'amicamente en el contexto del EChL. Las propiedades de dichas resonancias (masa, anchura y acoplamientos a los bosones W y Z) se derivan del M\'etodo de la Amplitud Inversa. Debido a la dificultad de introducir este procedimiento en un generador de eventos de Monte Carlo, desarrollamos un modelo para MadGraph que reproduce el comportamiento de las resonancias generadas din\'amicamente. Con esta herramienta cuantificamos la sensibilidad del LHC a estas resonancias cargadas para futuras luminosidades propuestas. Para ello, analizamos el {\it scattering} the WZ en WZ y, en particular, su canal de desintegraci\'on puramente lept\'onico. Tambi\'en se presentan algunas estimaciones muy preliminares del caso hadr\'onico.

En el Cap\'itulo \ref{ResonancesWW} extendemos los resultados del Cap\'itulo anterior al caso del {\it scattering} de WW en WW, que da acceso a las resonancias vectoriales neutras. Dado que el canal puramente lept\'onico en este caso involucra dos neutrinos, la eficiencia de reconstrucci\'on de las propiedades de la resonancia se reduce considerablemente. Por esta raz\'on estudiamos el caso puramente hadr\'onico, y, m\'as concretamente, la regi\'on cinem\'atica del mismo en las que cada par de jets provenientes de la desintegraci\'on de un W son detectados como un \'unico fat jet. Con la ayuda de t\'ecnicas de reconstrucci\'on de fat jets obtenemos la sensibilidad del LHC a estas resonancias vectoriales neutras generadas din\'amicamente.

Para finalizar, presentamos las principales Conclusiones de la Tesis al final de este documento.

Los contenidos expuestos en esta Tesis, las Conclusiones y los Ap\'endices corresponden a  trabajos originales que ha sido publicados en los art\'iculos~\cite{Delgado:2017cls,Arganda:2018ftn,Garcia-Garcia:2019oig,Delgado:2019ucx} y en los proceedings de conferencias~\cite{Delgado:2018lvo,Delgado:2018nnh}.

Vamos a buscar el tesoro!


\chapter*{Introduction}\label{Introduction}
\addcontentsline{toc}{chapter}{\scshape\bfseries Introduction}
\chaptermark{Introduction}

There exists quite a resemblance between trying to understand how nature really works and being in the search for a treasure. First things first, one needs a map. A reliable guideline pointing you towards the buried gems, a solid base from which to start. 

In the case of fundamental physics, our map is the Standard Model~\cite{Glashow:1961tr,GellMann:1964nj,Weinberg:1967tq,Salam:1968rm} (SM) of fundamental interactions: perhaps one of the most successful theories in the history of science. This framework, albeit a priori simple, classifies all the known elementary particles and describes three of the four fundamental interactions among them: electromagnetic, weak and strong.

The SM is a quantum field theory invariant under local or gauge transformations of the group $SU(3)_C\times SU(2)_L\times U(1)_Y$. It is also invariant under transformations of the Poincar\'e group (Lorentz invariant), and it is, by construction, renormalizable. These symmetry principles lead to the description of the SM interactions, mediated by their corresponding gauge, spin 1 bosons. The strong and electromagnetic interactions have massless and electrically neutral carriers: the gluon and the photon, respectively. Weak interactions, on the contrary, are mediated by massive vector bosons, the W${}^+$, the W${}^-$, and the the Z, whose charges are, in units of the electron charge, one, minus one and zero. 

The matter content of the SM is organized in three families, each of them containing four fermions. The only existing difference between families concerns the masses of the particles therein. The rest of their properties and quantum numbers remain the same, embedding the SM fermions in a structure of three identical copies of increasing mass. Every family accommodates two quarks, one with positive, and one with negative charge, and two leptons, one with negative charge, and one neutral. This four-fermion pattern in each family is in fact fixed by the gauge principle in order to ensure anomaly cancellation.

The last piece of the theory, also the last one to be experimentally observed, consists of a scalar boson: the Higgs boson, remnant of the SM mass generation process achieved through the Brout-Englert-Higgs (BEH, or just Higgs, as it is often found in the literature) mechanism~\cite{Higgs:1964ia,Higgs:1964pj,Englert:1964et,Guralnik:1964eu, Higgs:1966ev}. This mechanism exemplifies how the spontaneous breaking of the electroweak (EW) symmetry can give rise to the observed masses of the W, the Z, and the fermions, with the exception of the neutrinos, that remain massless in the SM. 

With this particle content and these symmetry principles the Lagrangian of the SM can be constructed. It depends on several parameters that need to be fixed on an empirical basis. These are the three gauge couplings, controlling the strengths of the gauge interactions, the Yukawa coupling matrices, encoding fermion masses and mixings, and the two parameters of the Higgs potential, the Higgs mass and its self-coupling. With these parameters fixed, the SM is able to predict, impressively, most of the known phenomena in particle physics with extreme accuracy. 

Nevertheless, our endeavour to find the precious treasure gets interrupted at this point: the map is incomplete.

Despite the remarkable accuracy of the predictions of the theory, and of the agreement between those and the corresponding experimental measurements, we know that the SM cannot be the ultimate description of nature. It lacks satisfactory explanations for diverse observed phenomena and for some theoretical prejudices. 

Neutrino oscillations and the existence of dark matter are the most important examples of the former, among others. 
Although neutrinos are massless by construction in the SM, it has been experimentally stated that they must have non-zero masses due to the fact that they oscillate in flavour~\cite{Fukuda:1998mi,Fukuda:1998ah,Ahmad:2002jz,Eguchi:2002dm}. Thus, the SM cannot account for neutrino masses, and physics beyond the SM (BSM)  has to be addressed. A common way to give masses to these particles is to extend the SM particle content by adding right handed neutrinos, absent in the SM. With these new fields one can construct Dirac mass terms for the neutrinos through their interaction with the Higgs boson in the same way as for the rest of the SM fermions. Nevertheless, due to the quantum numbers of these new states, neutrinos could be Majorana particles and have another source of mass, independent from the Higgs field. 

On the other hand, astrophysical and cosmological evidences point towards the fact that some {\it invisible} matter, amounting to more than 25\% of the total energy content of our universe, needs to exist~\cite{Bessel:1844ck,Zwicky:1933gu,Zwicky:1937zza,Rubin:1970zza,Begeman:1991iy,Taylor:1998uk,Clowe:2003tk,Ade:2015xua}. This dark matter interacts gravitationally, but the whereabouts of its interactions via the other three fundamental forces are yet unknown. For this reason, there is no suitable candidate for dark matter in the SM, and, so, extensions of the latter are constructed.

Regarding the unexplained theoretical prejudices that the SM suffers, it might be important to mention that they are only that: prejudices. They are often addressed as fine tuning or naturalness~\cite{tHooft:1979rat} problems: measurements of extreme values of SM parameters without any symmetry criterion or explanation behind them. They could be, in any case, just mere coincidences; values that nature has chosen among many possible others. Nevertheless, they might serve us a guidance towards understanding what is beyond our current interpretation of the physical world. 

One of these naturalness problems is the so-called strong charge conjugation-parity (CP) problem. Although the strong interactions allow for a CP violating term in the Lagrangian, the electric dipole moment of the neutron induced by it has not been observed yet. This sets bounds on the parameter that controls CP violation in the strong sector, the $\theta$ parameter, implying that it has to be smaller than $10^{-10}$~\cite{Baluni:1978rf,Crewther:1979pi,Kanaya:1981se,Schnitzer:1983pb,Cea:1984qv,Musakhanov:1984qy,Morgan:1986yy}. Since there is no explanation for this extremely small value, one can interpret it either as a fine tuning or as a hint towards BSM physics. In the latter case the well known Peccei-Quinn symmetry~\cite{Peccei:1977hh}, whose spontaneous breaking gives rise to the yet undiscovered axion particle is commonly employed as a possible solution to this problem.

Another of these problems is the one concerning the peculiar pattern of fermion masses and flavour mixings in the SM, i.e., the structure of the Yukawa matrices. Fermion masses span from the MeV scale to hundreds of GeV and, although the flavour mixings in the quarks sector tend to be very hierarchical, this is not the case in the lepton sector. This mystery, together with the existence of three and (so far) just three families is known as the flavor puzzle.

The last theoretically undesired property of the SM model listed here is the hierarchy problem. It relates to the fact that, in the SM, the Higgs boson is a fundamental scalar, whose mass, unprotected by any symmetry, is introduced {\it ad hoc} in the Lagrangian. 
Should some new physics appear above the EW scale, the Higgs boson would in principle couple to it, and the radiative corrections to the Higgs mass would result to be quadratically dependent on the new physics scale. Under the assumption that there is no new physics until the Planck scale, of the order of $10^{19}$ GeV, where gravity is supposed to participate relevantly in elementary particle interactions, the Higgs mass value would be pulled towards the Planck mass. However, the observed value of the Higgs mass differs from this scale by many orders of magnitude: $m_H = 125.09 \pm 0.21{\rm (stat.)} \pm 0.11{\rm (syst.)}$ GeV~\cite{Aad:2015zhl}. It is, in fact, very close to the EW scale, $v=246$ GeV itself, i.e., very close to the EW gauge boson masses. Thus, an extremely precise cancellation must take place between the bare mass and the quantum corrections in order to obtain a prediction that is compatible with the experimental data.

As a consequence of the existence of all these problems, it seems clear that our map is lacking the part where the treasure is buried, and that we need to explore {\it terra incognita} the best we can to be able to find it.  

When analyzing the SM problems presented above, it is interesting to notice that many of them are or could be related to the electroweak symmetry breaking (EWSB) sector of the theory. In the SM, the Higgs field is interpreted as a complex doublet that triggers the spontaneous EWSB when acquiring a vacuum expectation value (vev), $v=246$ GeV, being this vacuum invariant under the electromagnetic symmetry group. In this process, described by the BEH mechanism, three of the degrees of freedom contained in the scalar doublet, the EW Goldstone bosons (GBs), manifest themselves as the longitudinal polarization of the EW gauge bosons, that acquire a mass. The other degree of freedom appears in the particle spectrum as the Higgs boson. 

This mechanism solved various shortcomings of the primitive SM: it provided an explanation for the EW gauge boson masses, known to be finite due to the short range character of the electroweak interactions, as well as an explanation for the fermion mass generation. Besides, it cured the unitarity violation present in the probability amplitudes of the scattering of massive EW gauge bosons. Nevertheless, as it has been commented in the previous paragraphs, there appears to be something beyond the SM EWSB mechanism that needs to be further explored. Besides, after the discovery of the Higgs boson~\cite{Aad:2012tfa,Chatrchyan:2012xdj}, experimental searches are focused on mustering data that could shed some light into the true nature of the scalar sector of the SM~\cite{Chatrchyan:2012jja,Aad:2013xqa,Aad:2015zhl,Khachatryan:2014jba,Khachatryan:2014iha,Aad:2015xua,Aad:2015pla,Khachatryan:2014qaa,Aad:2015gra,ATLAS:2013rma,Khachatryan:2015lga,Aad:2014xva,Khachatryan:2014aep,Heinemeyer:2013tqa,LHCHiggsCrossSectionWorkingGroup:2012nn}, and, thus, it is very timely to investigate in this direction. 

But we do not know towards which cardinal point we should extend our map, so our aim will be to do it in the most general way as possible. With this criterion as our compass, the most clear option is to use effective field theories (EFT). These theories describe the low energy dynamics of a complete, ultraviolet (UV) theory, being this latter, in the present case, the one that might explain the unknown features of EWSB. Their model independence resides in the fact that the high scale physics remains encoded in low energy constants, that are, a priori, free parameters. Should these parameters be determined experimentally, a hint towards the complete UV theory will be accesible.

There are, however, several ways to use EFTs to describe the EWSB dynamics. The first thing that might come to mind is to use the SM as a starting point. Without extending its particle content but giving up the requirement of renormalizability, new operators can be included in the Lagrangian respecting the symmetries of the theory. These operators, that have an energy dimension larger than four, and that are therefore suppressed by corresponding powers of the new physics scale, are organized as an expansion in canonical dimension. In this sense, new structures motivated by the presence of some new physics at a higher scale, and that can modify the EWSB interactions, come into play. This framework corresponds to a linear realization of the EWSB, in which the Higgs field and the Goldstone bosons form a weakly interacting $SU(2)$ doublet. The leading order (LO) Lagrangian in this scenario is the SM Lagrangian itself, and, so, this EFT is often dubbed SMEFT. Its interesting associated phenomenology has been (and is being) in fact extensively studied. Some examples of recent analyses can be found in~\cite{Eboli:2006wa,Corbett:2013pja,Ellis:2014dva,Ellis:2014jta,Corbett:2014ora,Degrande:2016dqg,Ellis:2018gqa,Barzinji:2018xvu,Grazzini:2018eyk,Perez:2018kav,Dawson:2018liq,Dawson:2018jlg,Dawson:2019xfp,Baglio:2018bkm,Almeida:2018cld,Hartland:2019bjb,Buchalla:2019wsc,Moutafis:2019wbp,Aoude:2019tzn,Brivio:2019myy}. 

Nevertheless, a more general setup can be constructed. Taken as reference only the SM particle content and symmetry principles leads to a new class of EFTs, in which the Higgs boson is considered as a SM singlet so it can be treated independently of the EW Goldstone bosons, the latter having only derivative couplings, and being associated, in principle, to new strong interactions. Thus, in this scenario, the Higgs and the EW Goldstone bosons do not need to form an $SU(2)$ doublet and the EWSB is implemented non-linearly~\cite{Appelquist:1980vg,Longhitano:1980iz,Longhitano:1980tm,Chanowitz:1985hj,Cheyette:1987jf,Dobado:1989ax,Dobado:1989ue,Dobado:1990zh,Espriu:1991vm,Dobado:1990jy,Dobado:1990am,Dobado:1995qy,Dobado:1999xb}. The operators contained in the Lagrangian are organized as an expansion in powers of the external momentum, due to the derivative character of the EW Goldstone boson interactions, being the LO Lagrangian the one containing terms with two derivatives, the NLO Lagrangian the one containing terms with four derivatives, and so on. The Higgs field is often introduced via a generic polynomial function, since their interactions are not limited to derivative couplings. This non-linear Lagrangian is called electroweak chiral Lagrangian~\cite{Alonso:2012px,Alonso:2012pz,Buchalla:2013rka,Espriu:2012ih, Delgado:2013loa, Delgado:2013hxa, Brivio:2013pma, Espriu:2013fia, Espriu:2014jya,  Delgado:2014jda, Buchalla:2015qju, Arnan:2015csa,Brivio:2016fzo} (EChL) is named after the chiral Lagrangian developed in the context of quantum chromodynamics (QCD) describing the low energy dynamics of pions~\cite{Weinberg:1978kz,Gasser:1984gg, Gasser:1983yg}. Moreover, the linear case is, in fact, a specific limit of the EChL, albeit the relation between the two is not trivial~\cite{Alonso:2012px,Eboli:2016kko,Brivio:2016fzo}.

In this Thesis we will focus on the EChL description of the dynamics of the EWSB. In particular, we will study the predictions of such theory and their testability in current and future experiments. To this aim, the first thing to be done is to determine the particular observables in which the deviations from the SM predictions should be best observed. Ideally, we would like to test directly the interactions among EW Goldstone bosons, since they are the heart of the EWSB sector. Nevertheless, these are unphysical particles that are not present in the spectrum, so we need to probe them indirectly. Since they manifest through the longitudinal components of the EW gauge bosons, the scattering of the latter, most commonly known as vector boson scattering~\cite{Szleper:2014xxa,Rauch:2016pai,Anders:2018gfr,Bellan:2019xpr} (VBS), should be the finest place to look for signals of the EChL. Besides, as the EW Goldstone boson couplings are proportional to the external momentum, the large energy region of the VBS processes ought to be the most sensitive one to new physics. Since we base our hypothesis in the idea that the purely bosonic interactions should be the ones suffering from strongest deviations coming from a BSM EWSB sector, we will focus only on them. Thus, by assumption, we will take the fermionic interactions as the ones in the SM, for simplicity.

The question now is at what experiment should we look for these signals. Here, the typical energy scale of the EChL comes into play. This scale is the one controlling the contributions of the quantum corrections of the theory, and results to be $4\pi v\sim 3$~TeV, equivalently than in the low energy QCD case where this scale, arising from chiral perturbation theory, is expected to be $4\pi f_\pi\sim1$ GeV. Thus, the TeV scale is motivated, pointing us towards the Large Hadron Collider (LHC). Furthermore, the production of vector bosons from quark (proton) initial states leads to final state particles with very distinctive kinematics which are also easy to detect. This signature supposes an incredibly powerful tool to recognize the VBS topologies very efficiently from the collision products, facilitating our task of studying these processes. 

But an inconsistency arises when analyzing the predictions obtained from an EFT: that of violation of perturbative unitarity. Although the UV complete theory must be unitary to ensure probability conservation, the perturbative expansion at low energies, giving rise to the EFT description, usually suffers from unitarity violation problems. This happens, in particular, for the S-matrices that contain longitudinally polarized gauge bosons, whose probability amplitudes grow anomalously with energy. Thus, the validity of the EFT framework breaks down near the energy value at which perturbative unitarity is lost. In the non-linear scenario, the typical appearance of heavy resonances in the spectrum, arising from the strongly interacting dynamics, cures the unitarity violation problem. The properties of these resonances can be, in some cases, predicted from the EFT itself by using unitarization methods~\cite{Dobado:1999xb,Alboteanu:2008my,Eboli:2011ye,Ballestrero:2011pe,Espriu:2012ih, Espriu:2013fia,Delgado:2013loa, Delgado:2013hxa, Delgado:2014dxa, Espriu:2014jya, Arnan:2015csa, Dobado:2015hha,Corbett:2015lfa,Reuter:2016kqo,DelgadoLopez:2017ugq,BuarqueFranzosi:2017prc,Delgado:2017cls,BuarqueFranzosi:2017jrj,Delgado:2017cat,Delgado:2019ucx}. These are prescriptions that drive unitary the raw, non-unitary EFT prediction, allowing us to have consistent and testable results. These unitarization schemes are known to work in the non-resonant scenarios as well, and only some of them can accommodate the presence of a new heavy state.

At this point, we have extended the map in the hunt for our treasure. We have marked with crosses the points in which the cache could be most likely buried. Now it is time to dig in the dirt and grasp the gems.

In this Thesis, a phenomenological study of the EChL implications is carried out, relying upon VBS at the LHC as the main observable to do it. To this purpose, the generalities of VBS processes are exhaustively characterized both at the subprocess level, with initial gauge bosons, and at the LHC, with protons as initial state particles. The importance that such configurations had in the discovery of the Higgs boson motivates the first part of the present work. The discovery potential of VBS is used to study double Higgs production signals in order to test the feasibility of obtaining a competitive measurement of the Higgs self-coupling in this channel, as a complementary approach to the usual gluon gluon fusion scenario commonly explored in the literature~\cite{Glover:1987nx, Dicus:1987ic, Plehn:1996wb, Dawson:1998py, Djouadi:1999rca, Baur:2003gp, Grober:2010yv, Dolan:2012rv, Papaefstathiou:2012qe, Baglio:2012np, deFlorian:2013jea, Dolan:2013rja,  Frederix:2014hta, Liu-Sheng:2014gxa, Goertz:2014qta, Azatov:2015oxa, Dicus:2015yva, Dawson:2015oha, He:2015spf, Dolan:2015zja, Cao:2015oaa, Cao:2015oxx, Behr:2015oqq,  Bishara:2016kjn, Cao:2016zob, deFlorian:2016spz, Adhikary:2017jtu, Banerjee:2018yxy, Goncalves:2018qas}. The SM value and BSM choices for such coupling are considered, being the latter potentially explained by new EChL interactions.

In the second part, special attention is payed to the violation of perturbative unitarity. First, since there are many different unitarization methods that can be used to cure the mentioned problem, we perform, for the first time, a comparative analysis of their predictions for the elastic WZ scattering at the LHC. We study the impact these methods have on the experimental constraints that can be imposed on the EChL parameters, and motivate their combined use in the non-resonant case. On the other hand, the resonant scenario, in which a resonance is generated dynamically as it is characteristic of strongly interacting theories, is different. 

In the final part of the Thesis, we use the Inverse Amplitude Method (IAM)~\cite{Truong:1991gv,Oller:1997ng,Dobado:1999xb,GomezNicola:2001as,Espriu:2012ih, Espriu:2013fia,Delgado:2013loa, Delgado:2013hxa, Delgado:2014dxa, Espriu:2014jya, Arnan:2015csa, Dobado:2015hha, Corbett:2015lfa,BuarqueFranzosi:2017prc}, well known in the context of low-energy QCD~\cite{Truong:1988zp, Dobado:1989qm, Dobado:1992ha, Hannah:1995si,Dobado:1996ps}, to characterize vector resonance signatures in elastic WZ scattering at the LHC. We quantify the current and future LHC sensitivity to these resonances in the purely leptonic decay channel of the final gauge bosons. Besides, we explore as well the WW scattering. Since, in this case, the purely leptonic final state involves two neutrinos, disallowing to reconstruct accurately the resonance properties, we study the purely hadronic scenario. We examine the kinematical region in which a gauge boson hadronic decay products are detected as a single, large-radius jet, through the use of modern fat jet reconstruction techniques. In this way we provide the corresponding sensitivity to EChL vector resonances in the WW channel. 

The present Thesis is organized in the following way: In Chapter \ref{EChL} we briefly review the main features of EWSB, including its possible linear and non-linear realizations, and the different EFTs describing new EWSB dynamics. We make special emphasis in the EChL, since it corresponds to the more general framework that can be constructed in this context and since it is the one this Thesis is devoted to. We make an illustrative comparison between this theory and the linear version, the SMEFT, and, finally, address theoretically the unitarity violation problem present in these setups as well as some of the different possible ways to repair it.

Chapter \ref{VBS} is devoted to VBS characterization. We motivate these observables as one of the most likely discovery channels of new scalar interactions. Then, the SM and the EChL predictions for the various VBS channels, VV$\to$VV, are presented and studied. Unitarity violation in this context is quantified as well, in terms of the partial wave unitarity criterion.  The LHC case is also reviewed in this Chapter. The specific kinematics of VBS topologies at the LHC are illustrated, as they should be the key to disentangle these processes from the other ones produced in the LHC collisions. Finally, the current and future experimental searches looking for this kind of configurations are discussed.

In Chapter \ref{HH} the possibility of achieving a measurement of the Higgs self-coupling through double Higgs production in VBS at the LHC is addressed. We motivate this particular channel since it owns moderately large event rates and suffers from small theoretical uncertainties. We study scenarios in which this coupling can take SM or BSM values, being the latter case contemplated in the EChL as a possible modification to the EWSB mechanism. Within this setup, we perform a dedicated analysis of signal and background LHC events, paying special attention to VBS identification and HH reconstruction techniques, in order to obtain competitive sensitivities to BSM values of the Higgs self-coupling. 

Chapter \ref{Methods} contains the study of the impact that different unitarization methods can have on the interpretation of experimental data in the non-resonant scenario. We use the elastic WZ scattering to illustrate how the constraints imposed on EChL parameters strongly depend on the unitarization method that is used. To this aim, we rely on a coupled analysis of the helicity channels involved and select five of the most commonly used of these unitarization schemes. We obtain the exclusion regions in the EChL parameter space that correspond to interpreting the experimental results with each of these methods at a time and with all of them combined. Thus, we provide a theoretical uncertainty in the determination of the EChL parameters due to the choice of the unitarization method.

Regarding Chapter \ref{Resonances}, it includes the results concerning the sensitivity to charged vector resonances from the EChL in VBS at the LHC. These resonances are assumed, as previously mentioned, to be generated dynamically from the self interactions of the strongly interacting longitudinal gauge bosons. The properties of the vector resonances, mass, width and couplings to the W and Z gauge bosons are derived from the IAM. Due to the difficulty that introducing this procedure in a Monte Carlo generator represents, we develop a MadGraph model that mimics the behaviour of the IAM resonances. With this tool, we quantify the sensitivity of the LHC to these charged resonances for expected future luminosities. To that purpose, we analyze the WZ scattering, and, in particular, its purely leptonic decay channel. Some very naive estimates of the hadronic channel are also provided.

In Chapter \ref{ResonancesWW}, we extend the results of the previous Chapter to the WW case, which gives access to the neutral vector resonances. Since the purely leptonic decays of these gauge bosons involve two neutrinos, the efficiency of reconstructing the resonance properties diminishes significantly. This is the reason why we study the purely hadronic case. Specifically, we are interested in the kinematical region in which each of the jet pairs coming from the decay of the gauge bosons are identified as a single, large-radius jet. With the help of up to date reconstruction techniques of these so-called fat jets, we obtain the sensitivity to the dynamically generated neutral vector resonances in this channel at the LHC.

To finalise, we present the main Conclusions of this work at the end of the present document.

The contents presented in this Thesis, the Conclusions and the Appendices, are original works that have been published in the scientific articles~\cite{Delgado:2017cls,Arganda:2018ftn,Garcia-Garcia:2019oig,Delgado:2019ucx} and in the conference proceedings~\cite{Delgado:2018lvo,Delgado:2018nnh}.

Let us dig for the treasure!

%
%
%

    \titleformat
{\chapter} 
[display] 
{\fontsize{21pt}{21pt}\scshape\bfseries} 
{} 
{-6.0ex} 
{
    {\color{col1}\rule{\textwidth}{0.8mm}}
    \raggedleft
    {{\fontsize{30pt}{30pt}\selectfont\textcolor{col1}{\thechapter}}}
} 
[
\vspace{-0.5ex}%
{\color{col1}\rule{\textwidth}{0.8mm}}
] 



\chapter[\bfseries Effective Theories for electroweak symmetry breaking]{Effective Theories for electroweak symmetry breaking}\label{EChL}
\chaptermark{Effective Theories for electroweak symmetry breaking}

Since the discovery of the electroweak force the fact that its carriers should be massive was clear, as it was known to be a short-ranged interaction. However, the EW theory explaining the weakly interacting phenomena, could not account for W and Z masses without spoiling gauge invariance.  The same happened with fermion masses. Their presence in the Lagrangian was just not allowed by the gauge symmetries. Therefore, a breakthrough was needed in order to allow the SM to include the observed particle masses.    

This breakthrough would arise from an idea that only a genius or a fool could have: that of breaking a symmetry without breaking the symmetry. This geniality is known as spontaneous symmetry breaking (SSB), and it takes places when the vacuum state of a theory does not exhibit the same invariance as the complete Lagrangian. Although it may sound strange at first, SSB is a rather common phenomenon in nature. The typical example that is used to illustrate it is that of a ball sitting at the top of a hill. In this case, the system is rotationally invariant, since there is not a privileged direction for the ball to roll down along. Nevertheless, once the ball has fallen down, a particular path has been chosen, and the symmetry is spontaneously broken. In fact, when the ball is placed at the bottom of the hill, it sits in just one of the infinite possible vacuum states, all degenerate in energy and connected through transformations of the broken symmetry (rotations). 

A very similar mechanism to the one described by the ball-in-a-hill example can be implemented in the EW theory, such that masses for the fermions and gauge bosons are generated without giving up gauge invariance. This is the Brout-Englert-Higgs mechanism~\cite{Higgs:1964ia,Higgs:1964pj,Englert:1964et,Guralnik:1964eu, Higgs:1966ev}, in which a scalar $SU(2)_L$ doublet is introduced so that its vacuum state breaks spontaneously the EW symmetry.  

Nevertheless, very specific conditions have to be fulfilled by the Higgs potential in order for it to acquire a vacuum expectation value that can break the  $SU(2)_L\times U(1)_Y$ symmetry in an adequate way. In other words: the Higgs potential has to be similar to a hill-shaped potential. Thus, the BEH mechanism, despite its brilliance, only {\it describes} the spontaneous breaking of the EW symmetry, but does not account for an explanation of the underlying dynamics. It supposes an elegant and clever way to describe the process through which fermions and gauge bosons acquire their masses but lacks an explanation of the true origin of the EWSB mechanism. 
For this reason, there is a necessity of BSM theories that can elucidate the dynamics of the EWSB sector. 

Nowadays, there exist a large number of SM extensions that try to give an explanation for the dynamical origin of EWSB. With the aim of finding the one that really incorporates the correct interpretation of nature, we could just take them all one by one, compute their predictions, and test them at particle experiments. However, this is not likely to be very efficient. It will be equivalent to take a large number of different extensions of your initial map, each of them indicating different treasure locations, and start digging hoping to get a little lucky.

Another possibility is, nonetheless, to extend the map in all directions taking, somehow, all the different maps into account at once. In this way, as you advance in your endeavour you can get hints of where the gems could be buried and redirect your efforts. This would be, more or less, the equivalent to use effective theories: low-energy descriptions of UV complete theories encoding the physical properties of many of them at the same time through a finite set of low-energy parameters. 

In this Chapter a review and description of the basic ingredients commented in the paragraphs above is presented. Specifically, we will revise the generalities of spontaneous symmetry breaking, both in the case of global and local symmetries, to introduce a description of the possible realizations of the EWSB process. The linear and the non-linear scenarios will be described with simple examples in QCD that can be easily translated to the EW case. Once this is done, the effective theory setup is discussed. First we will introduce the effective electroweak chiral Lagrangian with a light Higgs, on which this Thesis is based, establishing its main features as well as the relevant parameters, operators, and counting. Then, we will briefly comment on the SMEFT, and on its relation with the EChL. Finally, the issue of the violation of perturbative unitarity is addressed, being it another key point of this Thesis.

\section{Spontaneous symmetry breaking}

Spontaneous symmetry breaking is a well known phenomenon in many areas of physics. The example of the ball sitting at the top of a hill is just one among many others, but there is a long list of systems showing this behaviour. In high energy physics, its importance is paramount: not is it only key in the mass generation mechanism of fundamental fermions and gauge bosons, but also in the appearance of new particles in the spectrum. This latter case occurs because of the fact that the total number of degrees of freedom before and after the spontaneous breaking is conserved. However, precisely because of this, there could also be no new particles emerging from the breaking. Whether this happens or not depends on the nature of the symmetry and on its global or local character.\\ 

{\noindent\scshape\bfseries Spontaneous breaking of global symmetries: Goldstone's theorem}\\

The spontaneous breaking of a global symmetry is very accurately describe by Goldstone's theorem~\cite{Nambu:1960tm,Goldstone:1961eq,Goldstone:1962es}, first formulated in the context of condensed matter physics. It states that a certain number of massless scalar bosons with the same quantum numbers as the vacuum must appear in the spectrum. Besides, this number corresponds to the number of generators of the broken symmetry. These new scalar states, generated to preserve at all times the total number of degrees of freedom, are the Goldstone bosons.

Perhaps the most important example of a global spontaneous symmetry breaking in the context of particle physics is that of the chiral symmetry in QCD.  The simplified version of the QCD Lagrangian with two flavors, ($u$, $d$),  of massless quarks exhibits an invariance under the global $SU(2)_L\times SU(2)_R$ chiral symmetry. Under this group, the fermions of each chirality transform differently:
\begin{align}
Q_L\equiv \left(\begin{array}{c} u_L\\d_L \end{array}\right) \to g_L\,Q_L~~~~~~~~~ Q_R\equiv \left(\begin{array}{c} u_R\\d_R \end{array}\right) \to g_R\,Q_R\,,
\end{align}
with $g_L\subset SU(2)_L$ and $g_R\subset SU(2)_R$ being the transformations under each corresponding group. Nevertheless, the vacuum of this system is not chiral invariant. The ground state corresponds to quark condensates that form spin-0 states whose vev is non-vanishing,
\begin{align}
\langle \overline{Q}_L Q_R+\overline{Q}_R Q_L \rangle \neq 0\,,
\end{align}
 triggering the spontaneous symmetry breaking. The vacuum is only symmetric under the diagonal or vectorial subgroup contained in $SU(2)_L\times SU(2)_R$, i.e., transformations that require $g_L=g_R$. Thus, the breaking  $SU(2)_L\times SU(2)_R\to SU(2)_{L+R}$ takes place and $\big[ {\rm dim}(SU(2)_L\times SU(2)_R)-{\rm dim}(SU(2)_{L+R})\big]=3 $ Goldstone bosons appear as physical states. These GBs are identified with the QCD pions: composite pseudoscalar states arising from the spontaneous breaking of the chiral symmetry. The fact that they have a mass, unlike what is stated by Goldstone's theorem, can be explained through the non-vanishing masses of the two lightest quarks. Their mass terms break explicitly the chiral symmetry in a small amount and serve as a source for the therefore light mass of  the pion.\\
 
{\noindent\scshape\bfseries Spontaneous breaking of local symmetries: BEH mechanism}\\

 When a local or gauge symmetry is spontaneously broken Goldstone's theorem does not account accurately for the physical properties of the system. This was first shown in 1964, simultaneously, by three independent groups: those of Brought and Englert~\cite{Englert:1964et}, Higgs~\cite{Higgs:1964ia,Higgs:1964pj,Higgs:1966ev}, and Guralnik, Hagen and Kibble~\cite{Guralnik:1964eu}. Interestingly, the correct mechanism describing spontaneously broken local symmetries is most commonly named only after Higgs, and, some times, after Brought, Englert and Higgs.
 
In this case, the Goldstone bosons that emerge from the breaking do not manifest as new physical particles, and are therefore called {\it would be} Goldstone bosons. Instead, they become the longitudinal polarization degree of freedom of the gauge bosons associated to the broken generators. The latter acquire a non-vanishing mass in this process without spoiling the good symmetry properties of the Lagrangian. In this sense, the BEH mechanism was indeed the ideal way to implement the spontaneous breaking of the EW symmetry of the SM.

The minimal setup required to produce the desired breaking is achieved by introducing four real scalar degrees of freedom. These correspond to one complex scalar transforming as an $SU(2)_L$ doublet with hypercharge 1/2:
\begin{align}
\Phi=\left(\begin{array}{c}\Phi^+\\ \Phi^0  \end{array}\right)\,,\label{doublet}
\end{align}
whose dynamics and interactions, compatible with the SM symmetries, are generically described by the Lagrangian:
\begin{align}
\mL_{\Phi}=&(D_\mu \Phi)^\dagger D^\mu \Phi+\mu^2 (\Phi^\dagger \Phi)-\lambda( \Phi^\dagger \Phi)^2\nn\\
&-\Big[Y_u \overline{Q}_L\widetilde\Phi u_R+Y_d \overline{Q}_L\Phi  d_R + Y_e  \overline{L}_L\Phi  e_R + h.c.\Big]\,,\label{lagphi}
\end{align}
where
\begin{align}
 D_\mu\Phi=\partial_\mu\Phi+\frac{ig}{2}(\vec{W}_\mu\cdot\vec{\tau})\Phi+\frac{ig'}{2}B_\mu\Phi\,,
~~~\widetilde\Phi=i\tau_2\Phi^*=\left(\begin{array}{c}\phantom{-}\Phi^{0*}\\ -\Phi^-  \end{array}\right)\,,\label{Dphi}
\end{align}
and with $Q_L$ and $L_L$ being the quark and lepton doublets and $u_R$, $d_R$ and $e_R$  being the right handed type-$u$ quark, type-$d$ quark and charged lepton, respectively. The Pauli matrices are denoted here by $\tau$. $g$ and $g'$ correspond to the $SU(2)_L$ and $U(1)_Y$ gauge couplings, respectively. Their values, together with those of  $\mu$, $\lambda$ and the Yukawa matrices, $Y_{u,d,e}$, have to be determined experimentally, since they are not predicted in the SM.

The Lagrangian of \eqref{lagphi} is manifestly $SU(2)_L\times U(1)_Y$ invariant, but, does the vacuum preserve this symmetry? The answer depends on the values of $\mu$ and $\lambda$. In order to have a potential bounded from below in the SM, $\lambda$ has to take positive values. Now, if $\mu^2$ is negative, the potential has a minimum corresponding to $\langle \Phi \rangle\equiv\langle0|\Phi|0\rangle=0$, and the spontaneous breaking of the EW symmetry simply does not take place. On the other hand, if the opposite condition holds, $\mu^2 >0$, the potential develops an infinite number of degenerate minima sharing the property of the Higgs doublet that ensures the minimization of the potential.

\begin{figure}[t!]
\begin{center}
\includegraphics[width=0.4\textwidth]{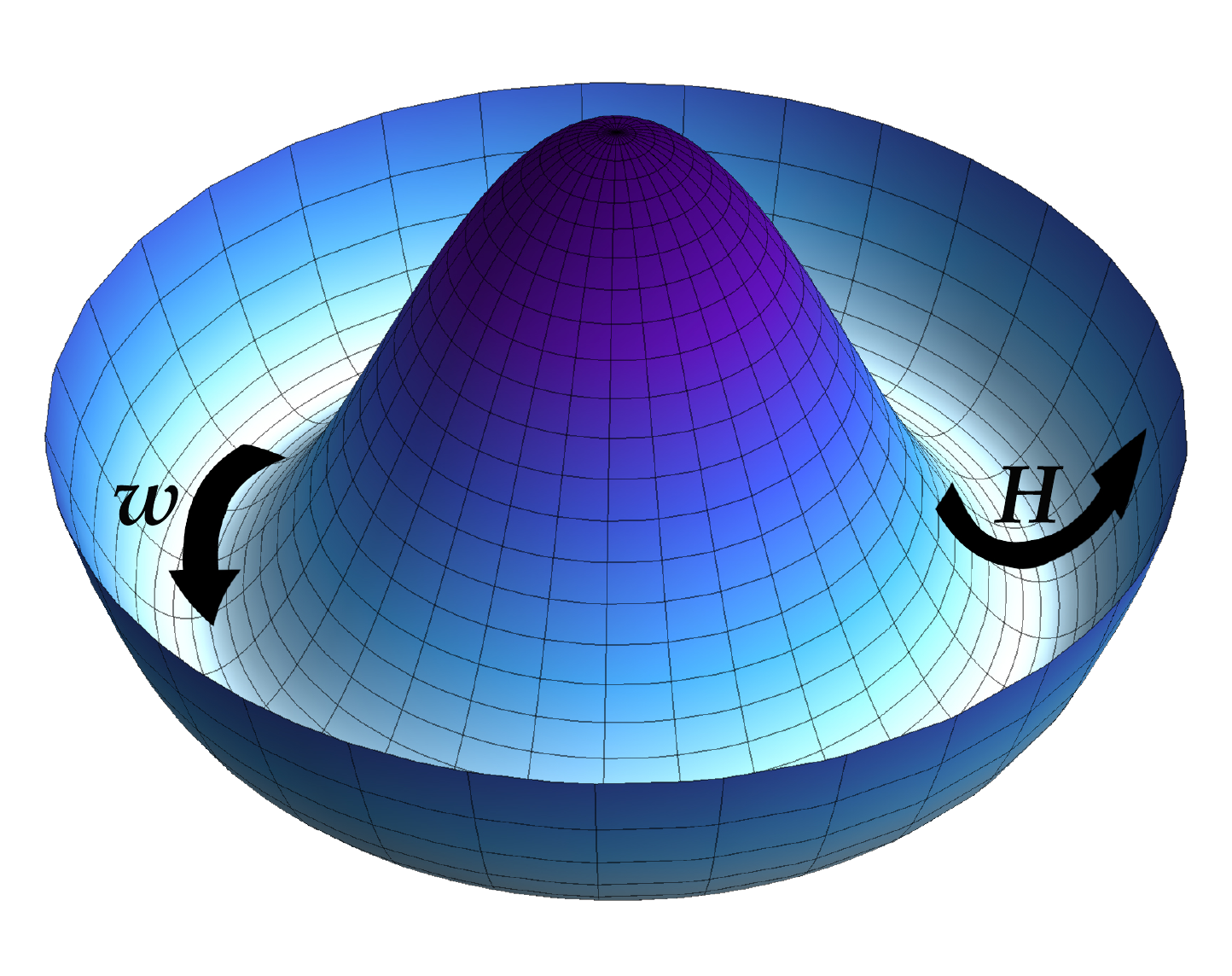}
\caption{Qualitative shape of the SM scalar potential for $\mu^2>0$. Radial excitations around the vacuum correspond to the physical Higgs mode, whereas angular excitations correspond to the EW Goldstone bosons.}
\label{fig:MexicanHat}
\end{center}
\end{figure}

In this scenario, we have a {\it mexican hat}-shaped potential for the scalar fields, as the one shown in \figref{fig:MexicanHat}, whose minima lie on a ring of radius equal to $v$. None of these vacua are invariant under the $SU(2)_L\times U(1)_Y$, but only under a residual $U(1)$. Since the electromagnetic (EM) interactions are known to be a good symmetry of the ground state, it is usual to select the fundamental configuration $\langle \Phi \rangle=\big(0,v/\sqrt{2}\big)^T$, that implies a symmetry breaking pattern of the form $SU(2)_L\times U(1)_Y\to U(1)_{em}$, where
 \begin{align}
v=\sqrt{\frac{\mu^2}{\lambda}}\,, \label{vev}
\end{align}
corresponds to the vacuum expectation value, whose size is not predicted in the SM but has been determined experimentally to be $v=246~{\rm GeV}$.

The implications of \eqref{vev} are of great importance. First, the value of the vev defines the EW scale, in such a way that any dimensionful quantity related to the EW interactions must depend on it. Precisely because of this, the masses of fermions and gauge bosons shall be directly related to this scale, whose precise value is determined from the decay rate of the muon\footnote{In fact, the value of the EW scale is inferred from that of the Fermi constant $G_F$, a low energy parameter of the EFT describing EW interactions at low energies. This is an example of how powerful the use of EFTs can be in determining outstandingly important properties of a physical system.}. To understand how these masses arise from the spontaneous breaking of the EW theory, one can consider small perturbations around the vacuum, parameterized, in polar coordinates, as:
\begin{align}
\Phi=\frac{v+H}{\sqrt{2}}e^{\,(i\vec{w}\cdot\vec{\tau})/v}\left(\begin{array}{c}0\\ 1  \end{array}\right)=\frac{1}{\sqrt{2}}\left(\begin{array}{c} iw_1+i\,w_2    \\ v+H-i\,w_3  \end{array}\right)+\dots
\end{align}
Here, the radial excitations correspond to the physical Higgs boson, $H$. On the other hand, the EW Goldstone bosons that are ``eaten'' (as it is usually referred to in the high energy physics jargon) by the EW gauge bosons, correspond to  the three phases, $\vec{w}=(w_1,w_2,w_3)$. Being these the excitations along the ring of vacua, there is no energy cost for them to move along the circle of minima, and, therefore, the Goldstones posses a shift symmetry and are constrained to have derivative couplings only. Furthermore, since these three scalars are non-physical particles, they can be decoupled via a gauge transformation. In this simple case (unitary gauge) the scalar potential drives a mass and a self-coupling for the Higgs particle:
\begin{align}
\mL_{\Phi}\supset&\,\mu^2(\Phi^\dagger \Phi)-\lambda( \Phi^\dagger \Phi)^2=-\mu^2H^2-v\lambda H^3-\frac{\lambda}{4}\,H^4;~~~~ m_H^2=2\mu^2=2\lambda v^2\,,\label{lagHmass}
\end{align}
while the kinetic term leads to the EW gauge boson mass terms after the symmetry breaking has taken place,
\begin{align}
\mL_{\Phi}\supset&\,(D_\mu \Phi)^\dagger D^\mu \Phi=\frac{1}{2}\partial_\mu H\partial^\mu H+\frac{g^2v^2}{4}W^+_\mu W^{-\mu}+\frac{(g^2+g'^2)v^2}{4}Z_\mu Z^{\mu}+\dots\label{WZmasses}
\end{align}
once they are rotated to the physical basis,
\begin{align}
W^{\pm}_\mu=\frac{1}{\sqrt{2}}\big(W^1_\mu\mp iW^2_\mu\big)\,, ~~~~~~~Z_\mu=\cosw W^3_\mu - \sinw B_\mu\, , ~~~~~~~A_\mu=\sinw W^3_\mu + \cosw B_\mu\,.\label{massrot}
\end{align}
Here, $\cosw$ and $\sinw$ are the cosine and sine of the weak mixing angle $\theta_{\rm w}={\rm arctan}(g'/g)\sim 0.48$, respectively.

The result obtained in \eqref{WZmasses} is remarkable, as it provides masses for the W${}^\pm$ and Z bosons, leaving the photon massless, as observed experimentally. Furthermore, these masses
\begin{align}
m_W=\frac{gv}{2}\sim 80~{\rm GeV},  ~~~~~~~m_Z=\frac{\sqrt{g^2+g'^2}\,v}{2}\sim 90~{\rm GeV}\,,
\end{align}
result to be very close to their experimental value~\cite{Tanabashi:2018oca}. This means that, once the EW scale is determined, the BEH mechanism provides an excellent description for the generation of the EW gauge boson masses.

Besides, it is also able to explain the existence of fermion masses without the explicit breaking of the EW invariance. Reviewing again \eqref{lagphi}, it is plain that when the Higgs doublet acquires a vev, fermionic mass terms are generated involving the two chiralities of the corresponding fermions. These terms are called Dirac mass terms, and predict a mass value related to the EW scale, $m_f= (v Y_f)/\sqrt{2}$. At this point, it is interesting to notice that a neutrino mass term is not present in the Lagrangian even after EWSB, since the right-handed component of these fields is absent in the SM.

A final remark about the BEH mechanism has to be commented at this point. It is that of the equivalence theorem~\cite{Cornwall:1974km,Vayonakis:1976vz, Lee:1977eg,Chanowitz:1985hj,Gounaris:1986cr} (ET), which translates the relation between the longitudinally polarized gauge bosons and the corresponding Goldstone bosons to a relation between scattering amplitudes.  It is somehow intuitive that at sufficiently large energies compared to the EW scale, the longitudinal polarization of the EW gauge bosons should reveal their Goldstone boson condition, since, in the opposite limit, at zero momentum, the polarization state is not well defined. Because of these reason, there is a direct relation between the scattering amplitudes of EW gauge bosons and the scattering amplitudes computed with Goldstone bosons at high energies. They are indeed the same quantity up to a phase and corrections of the order $\mO(m_W/\sqrt{s})$, with $\sqrt{s}$ being the center of mass energy of the scattering. Therefore, the ET enunciates the following connection:
\begin{align}
&A\big(V_L(p_1),...,V_L(p_N)\to V_L(k_1),...,V_L(k_M)\big)=\nn\\&(-i)^{N+M}A\big(w(p_1),...,w(p_N)\to w(k_1),...,w(k_M)\big)+\mO\big(m_W/\sqrt{s}\big)\,,\label{ET}
\end{align}
with $V=$ W${}^\pm$, Z and where $w$ in this case denotes the corresponding Goldstone bosons, i.e., either $w^\pm$ or $z$. This is an outstanding and useful result of the BEH mechanism, which allows to perform computations of spin 1 particle interactions in the simplified scenario in which only scalar degrees of freedom are involved. Nevertheless, it is important to keep in mind that in order to understand the full behaviour of the EW gauge bosons at all energies, one has to take into account the full gauge boson configurations, and not just their scalar counterparts. 

The spontaneous breaking of a local symmetry served to describe very important physical properties of the EW theory. It allowed to generate masses for the elementary fermions of the SM and for the EW gauge bosons through the incorporation of the emergent {\it would be} Goldstone bosons as their longitudinal polarization degrees of freedom. Even more, considering the proper breaking pattern $SU(2)_L\times U(1)_Y\to U(1)_{em}$, the photon, aligned with the preserved generator, remained massless. Besides, as a bonus, a physical scalar should appear in the spectrum as the remnant of the symmetry breaking process, leaving a trace of the underlying breaking if observed experimentally. In 2012, a new particle with properties compatible with those of the SM Higgs boson~\cite{Chatrchyan:2012jja,Aad:2013xqa,Aad:2015zhl,Khachatryan:2014jba,Khachatryan:2014iha,Aad:2015xua,Aad:2015pla,Khachatryan:2014qaa,Aad:2015gra,ATLAS:2013rma,Khachatryan:2015lga,Aad:2014xva,Khachatryan:2014aep,Heinemeyer:2013tqa,LHCHiggsCrossSectionWorkingGroup:2012nn} was discovered~\cite{Aad:2012tfa,Chatrchyan:2012xdj}, with a mass of approximately 125 GeV~\cite{Aad:2015zhl}. However, there is still some room for BSM physics in the EWSB sector, and the characteristics of the observed scalar state are being measured and analyzed to elucidate if we need to keep exploring in this direction.

\section{Realizations of electroweak symmetry breaking}

In the process of building the EW theory, it is clear that spontaneous symmetry breaking has to be an ingredient in order to ensure the generation of fermion and gauge boson masses. However, the way in which this breaking can be parameterized is not such an obvious statement. The scalar sector of the SM is known to be invariant, as we shall see later on, under global transformations of the EW chiral symmetry $SU(2)_L\times SU(2)_R$. From now on we will use indistinctly the terms ``chiral symmetry'' and ``EW chiral symmetry''. It should be understood that we are always referring to the EW chiral symmetry except if specified otherwise. The different forms of parameterizing the EWSB basically differ in the transformation properties under this $SU(2)_L\times SU(2)_R$ symmetry group of the scalar particles: the three EW Goldstone bosons and the Higgs.  

The two main avenues to describe  EWSB in this sense are the so-called linear and non-linear realizations. Whereas the former embeds the four scalars in a bi-doublet transforming linearly under the chiral symmetry, the latter places the EW Goldstone bosons in a triplet that transforms non-linearly and interprets the Higgs boson as a singlet of the $SU(2)_L\times SU(2)_R$ symmetry.

The choice between these two ways of describing the EWSB lies in the physical motivations behind them: the linear case is often related to weakly-coupled theories and the non-linear case to strongly-coupled ones. This implies that in non-linear realizations of EWSB, arbitrary insertions of Goldstone bosons in the scattering amplitudes are far less suppressed than in the linear one. 

Nevertheless, in a certain limit, both implementations can lead to the same physical properties, but using one or the other to describe the relevant physics may seem more or less natural. For example, if in a very tall building you observe that as you go up there are fewer and fewer people on each floor, you might think that the elevator is out of order or that there is no elevator at all. This would be in analogy with the linear scenario, in which you have to pay a large energy price for generating more and more particles. However, if you find more or less the same amount of people on the thirtieth floor than on the second, you would assume that the elevator works properly. This would correspond to a strongly interacting case, in which the scattering probability of many scalar particles is not so suppressed with respect to the probability of having a few-particle scattering. 

Nonetheless, in this latter case, you could just have very sportive neighbours that take the stairs to every floor without a problem, but it just seems more unnatural to believe. In this sense, using one EWSB realization or another would facilitate the interpretation of the physical properties of the system depending on the setup one initially assumes.  

In this section, we will briefly review both realizations with pedagogical examples that are well known in the context of low-energy QCD, to later translate those examples into concrete scenarios of EWSB parametrizations.

\subsection{The linear realization}

As already commented in the lines above, the linear realization of EWSB assumes that the Goldstone bosons and the Higgs are embedded together in a bi-doublet of the chiral symmetry that transforms linearly. This implementation is the most usual one in weakly coupled theories, like Supersymmetry (SUSY)~\cite{Golfand:1971iw,Volkov:1973ix,Wess:1974tw}, and it corresponds to the SM BEH mechanism as well. Although the role of the chiral symmetry in the EWSB sector has not been clarified yet in this Thesis, its importance will become manifest in the following pages.

As an illustrative example, it is worth revisiting the linear $\sigma$-model describing the spontaneous breaking of the chiral symmetry in the QCD case for a simplified scenario with massless pions. 

The simplified version of the linear $\sigma$-model Lagrangian, developed by Gell-Mann and L\'evy~\cite{GellMann:1960np}, contains only four real scalar fields: the sigma particle $\sigma$ and the three pions $\vec{\pi}=(\pi_1,\pi_2,\pi_3)$. We do not include the fermion fields in this section for simplicity. The most general Lagrangian that can be written with this field content respecting the chiral symmetry reads
\begin{align}
\mL_{\text{{\it lin}-}\sigma}=\frac{1}{2}\partial_\mu \sigma \partial^\mu \sigma+\frac{1}{2}\partial_\mu \vec{\pi} \partial^\mu \vec{\pi}+\mu^2(\sigma^2+\vec{\pi}\cdot \vec{\pi})-\lambda(\sigma^2+\vec{\pi}\cdot \vec{\pi})^2\,.\label{linsigma1}
\end{align}
However,  the $SU(2)_L\times SU(2)_R$ invariance of this Lagrangian seems to be a bit obscured. To simplify the argument, one can construct a bi-doublet, $\Sigma$, containing the four scalar fields:
\begin{align}
\Sigma=\mathbb{1}\sigma+i\vec{\pi}\cdot \vec{\pi}=\left(\begin{array}{cc} \sigma+i\pi_3 & i\pi_1+\pi_2 \\i\pi_1-\pi_2 & \sigma-i\pi_3
 \end{array}\right),~~~~~\Sigma \to g_L \,\Sigma \, g_R^\dagger\,,
\end{align}
with $g_L$ and $g_R$ being $SU(2)_L$ and $SU(2)_R$ transformations respectively. Noticing that Tr$(\Sigma^\dagger\Sigma)=2(\sigma^2+\vec{\pi}\cdot \vec{\pi})$ one obtains:
\begin{align}
\mL_{\text{{\it lin}-}\sigma}=\frac{1}{4}{\rm Tr}\big[(\partial_\mu\Sigma)^\dagger\partial^\mu\Sigma\big]+\frac{\mu^2}{2}{\rm Tr}\big[\Sigma^\dagger\Sigma\big]-\frac{\lambda}{4}\big({\rm Tr}\big[\Sigma^\dagger\Sigma\big]\big)^2\,,\label{linsigma2}
\end{align}
which explicitly shows the $SU(2)_L\times SU(2)_R$ chiral invariance. The spontaneous breaking of the chiral symmetry takes place when the scalar fields acquire a vev in the case in which $\mu^2>0$:
\begin{align}
\frac{1}{2}\langle {\rm Tr}\big[\Sigma^\dagger\Sigma\big] \rangle=\langle \sigma^2+\vec{\pi}\cdot \vec{\pi} \rangle = \frac{\mu^2}{2\lambda}\equiv f^2\,.
\end{align}
Here, $f$ denotes the pion decay constant, the scale controlling the destruction of a pion field through its coupling with the broken currents.  Its value is measured in pion decay experiments with good accuracy, leading to a result of $f\equiv f_\pi\sim94$ MeV.

From the infinite vacua, it is customary to choose the one in which the pion fields have a vanishing vev, while the sigma particle's vacuum expectation value is set to $f$: $\langle \sigma \rangle=f;~ \langle \vec{\pi} \rangle=0 $, so that $\langle \Sigma \rangle=\mathbb{1}f$. This vacuum breaks the $SU(2)_L\times SU(2)_R$ symmetry, but is invariant under the diagonal subgroup $SU(2)_{L+R}$, often called strong isospin symmetry. Expanding now in small perturbations around the vacuum, the following Lagrangian is obtained:
\begin{align}
\mL_{\text{{\it lin}-}\sigma}=\frac{1}{2}\partial_\mu \vec{\pi} \partial^\mu \vec{\pi}+\frac{1}{2}\partial_\mu \sigma \partial^\mu \sigma-\frac{m_{\sigma}^2}{2}\sigma^2-4\lambda f\sigma(\sigma^2+\vec{\pi}\cdot \vec{\pi})-\lambda(\sigma^2+\vec{\pi}\cdot \vec{\pi})^2\,,
\end{align}
with $m_{\sigma}^2=4\mu^2$. 

Therefore, in the lineal $\sigma$-model implementation of the spontaneous breaking of the chiral symmetry, a massive particle together with three massless\footnote{The fact that the pions are not exactly massless can be explained by the approximate character of the chiral symmetry in QCD, explicitly broken by the small quark mass terms appearing in the Lagrangian, as it was mentioned in the previous section.} states that manifest also in the spectrum. Moreover, the couplings between the four scalars are deeply connected, since they are all embedded in the same representation of the symmetry group. Furthermore, the model is renormalizable, so quantum corrections can be calculated consistently with the usual prescriptions in perturbation theory.

Regardless of normalization factors, the linear $\sigma$-model results to be very similar to the BEH mechanism presented in the previous section. In fact, the latter can be parameterized via a  linear $\sigma$-model, in which, therefore, EWSB is linearly realized. The sole difference between both scenarios lies in the fact that the broken symmetry in the EW case is local, and not global, and, as a consequence, the Goldstone bosons do not show up in the physical spectrum.

Implementing the BEH mechanism via a  linear $\sigma$-model can be very useful, specially for studying explicitly the chiral invariance of the EWSB sector of the SM. The scalar contribution of the SM Lagrangian shows a global invariance under the chiral symmetry, as it will be shown shortly. After the $SU(2)_L\times U(1)_Y$ group of the chiral symmetry is gauged, the EW gauge invariance becomes clear. Thus, there is a close relation between the chiral and EW gauge invariances of the theory that should become manifest in the following examples.

Starting from the initial doublet of the BEH mechanism, we can place our four scalar degrees of freedom in a bi-doublet of the global chiral symmetry for the EW sector:
\begin{align}
\Sigma=\frac{1}{\sqrt{2}}\Big(\widetilde\Phi\,\Phi\Big)=\frac{1}{\sqrt{2}}\left(\begin{array}{cc} \phantom{-}\Phi^{0*} & \Phi^+ \\-\Phi^- & \Phi^{0}
 \end{array}\right),~~~~~\Sigma \to g_L \,\Sigma \, g_R^\dagger\,,
\end{align}
in such a way that the EWSB SM Lagrangian given in \eqref{lagphi} (excluding fermion operators) reads
\begin{align}
\mL_{\Phi}\to\mL_{\Sigma}=&{\rm Tr}\Big[(D_\mu \Sigma)^\dagger D^\mu \Sigma\Big]+\mu^2  {\rm Tr}\big[\Sigma^\dagger\Sigma\big]-\lambda \big({\rm Tr}\big[\Sigma^\dagger\Sigma\big]\big)^2\,,\label{lagSigma}
\end{align}
where in this case, the covariant derivative is slightly modified with respect to the one presented in the previous section:
\begin{align}
 D_\mu\Sigma=\partial_\mu\Sigma+\frac{ig}{2}(\vec{W}_\mu\cdot\vec{\tau})\,\Sigma-\frac{ig'}{2}B_\mu\,\Sigma\,\tau_3\,.\label{covdev}
\end{align}
The modification in the last term arises from the fact that, while the whole $SU(2)_L$ group is gauged in the EW theory, only a subgroup of $SU(2)_R$ is gauged: the one corresponding to $U(1)_Y$, which is generated by $\tau_3$. In any case, it is fairly easy to check that this expression matches the one given in \eqref{Dphi} in terms of the doublet $\Phi$.

The Lagrangian shown in \eqref{lagSigma} is manifestly EW chiral invariant, but the ground state of the system, if the condition $\mu^2>0$ is fulfilled, is not. In fact, the same breaking pattern of the linear $\sigma$ model is obtained, once the $\Sigma$ field acquires a vev:
\begin{align}
\langle {\rm Tr}\big[\Sigma^\dagger\Sigma\big] \rangle=\langle \Phi^\dagger\,\Phi \rangle = \frac{\mu^2}{2\lambda}\equiv \frac{v^2}{2}\,,
\end{align}
that in this case is identified with the EW scale, $v=246$ GeV. The appropriate vacuum is such that the breaking $SU(2)_L\times SU(2)_R\to SU(2)_{L+R}\equiv SU(2)_c$, to the so-called custodial symmetry, is achieved. Small perturbations around this vacuum lead to the exact same Lagrangians as the ones given in Eqs.~(\ref{lagHmass}) and (\ref{WZmasses}), thus stating that the BEH mechanism is intimately related to the linear $\sigma$-model parameterization.

An important comment has to be made at this point about this custodial symmetry~\cite{Sikivie:1980hm}, which corresponds to the diagonal subgroup of $SU(2)_L\times SU(2)_R$, i.e., $SU(2)_{L+R}$. It is the remnant symmetry under which the EW vacuum is invariant if one neglects the fact that only the $U(1)_Y$ subgroup of $SU(2)_R$ is gauged. This scenario would correspond to the case in which $g'=0$, and, therefore to $\theta_{\rm w}=0$ and to $m_W=m_Z$.  It is usually called the isospin limit, after the other common name of the $SU(2)_{L+R}$ symmetry.  Although the isospin limit can be extremely useful and computationally convenient, this is known not to be the exact case in the SM, specially at low energies compared to the EW gauge boson mass scale, well the approximation $m_W=m_Z$ is poorly justified. In this case, instead, the relation between the W and the Z masses is given by the $\rho$ parameter:
\begin{align}
\rho=\frac{m_W^2}{m_Z^2 \cosw^2}\,,
\end{align}  
that is predicted to be exactly one in the SM at the tree level, and which assumes that the only source of custodial symmetry breaking comes from the gauging of the hypercharge group and from the difference in the fermion masses within the same doublet. This result has very interesting phenomenological implications, since many BSM setups induce sizable corrections to de $\rho$ parameter. In particular, the implications of custodial invariance affect significantly the predictions of the $T$ oblique parameter~\cite{Peskin:1991sw}, and of the Higgs couplings to the EW gauge bosons. From the measurement of these two quantities, specially from the former~\cite{ALEPH:2010aa,Peskin:1990zt,Pich:2012dv,Pich:2013fea,Ciuchini:2014dea}, whose value is known experimentally with good precision, it appears that other sources of custodial breaking rather than those coming from $g'\neq0$ are strongly disfavoured. It seems, so far, that nature has chosen a custodial  preserving EW vacuum. Because of this reason, we will always assume from now on in this Thesis that custodial symmetry is preserved in the bosonic sector\footnote{The difference between the various fermion masses represents another source of custodial breaking that is taken into account in the SM and in the present Thesis, since the fermionic sector remains unchanged with respect to that of the SM.} of the EW interactions, up to corrections of the order $g'$.

\subsection{The non-linear realization}

In the previous subsection the linear realization of spontaneous symmetry breaking through weakly-coupled, renormalizable models was exemplified. In particular, the correspondence between the BEH mechanism described in the pages above and these linear implementations, in which the nature of the four scalars involved in the EWSB is tightly related, was discussed.

In the present subsection the non-linear realization of EWSB is revisited. In this scenario the Higgs particle is decoupled from the EW Goldstone bosons in a consistent way with the SM symmetries. This is achieved by  assigning the Higgs to the singlet representation of the chiral symmetry while placing the EW Goldstone bosons in a triplet, making the transformation of these scalar fields non-linear under the $SU(2)_L\times SU(2)_R$ group. Due to this fact, the interactions among the Goldstone bosons become non-perturbative at high energies, i.e., they become strongly interacting particles. Besides, in this setup, the Higgs is not needed to preserve any global or gauge invariances, since it is a singlet of the EW chiral and gauge symmetries, and can be integrated out from the theory. This feature was specially appealing before the Higgs discovery, where the non-observation of such scalar particle at light masses seemed to point towards a Higgsless model or to a scenario in which the Higgs could arise as a heavy resonance from a strongly interacting system.

To illustrate this non-linear realization we rely once again in its simple QCD counterpart: the non-linear $\sigma$ model~\cite{GellMann:1960np}. Contrary to the linear case, in which the four scalars were necessary to preserve the chiral invariance due to their quantum numbers, in the non-linear setup one can restrain oneself to a more minimal scenario. Since the only truly necessary components for the spontaneous $SU(2)_L\times SU(2)_R\to SU(2)_{L+R}$ breaking are the three Goldstone bosons, the other scalar degree of freedom can be, in principle, removed.

Nevertheless, we need to include the Higgs particle in the EW EFT, so we part from the same particle content and the symmetry principles than in the linear scenario, starting from the same Lagrangian for the scalar fields given in \eqref{linsigma1}, which is the most general one in these conditions. The way in which we organize the fields is, however, different. This time, we define four new fields, $\varphi$ and $\vec{\chi}=(\chi_1,\chi_2,\chi_3)$, related to the $\sigma$ and $\vec{\pi}$ through non-linear transformations, such that
\begin{align}
\Sigma=\varphi\, \Ugold\,, \label{Sigmanl}
\end{align}
with U containing the $\vec{\chi}$ particles. The $\varphi$ in this case represents the singlet of the chiral symmetry and, therefore, the U field must transform in the same way as the $\Sigma$: U$\to g_L\, \Ugold \, g_R$. U can be paremeterized in many different ways, since the physics encoded in it must be independent of the parameterization. Nevertheless, the most common way of describing it is through its exponential form:
\begin{align}
\Ugold=e^{\,(i \vec{\tau}\cdot\vec{\chi})/f}\,,
\end{align}
although others, like, for instance, the spherical form~\cite{Delgado:2014jda}, can be equivalently used~\cite{Haag:1958vt}. Here $f$ is again identified with the pion decay constant and with the vacuum expectation value of the $\varphi$ field, $\langle \varphi \rangle =f$. It is clear from this construction that the roles of the $\varphi$ and the $\vec{\chi}$ are not directly related as it was the case in the linear $\sigma$-model. In fact, the original Lagrangian becomes, employing the definition in \eqref{Sigmanl},
\begin{align}
\mL_{\text{{\it nonlin}-}\sigma}=\frac{1}{2}\partial_\mu\varphi\partial^\mu\varphi+\frac{\varphi}{4}{\rm Tr}\big[(\partial_\mu \Ugold)^\dagger\partial^\mu \Ugold\big]+\mu^2\varphi^2-\lambda\varphi^4\,,
\end{align}
making manifest the separation between the $\varphi$ and the $\vec{\chi}$ fields, which do not even appear in the scalar potential. In this sense, the minimization to obtain the asymmetric vacuum is performed only taking into account the $\varphi$ state:
\begin{align}
\langle \varphi \rangle = \sqrt{\frac{\mu^2}{2\lambda}}=f\,.
\end{align}
Following the same recipe as before and expanding around the $SU(2)_{L+R}$ invariant vacuum one arrives at
\begin{align}
\mL_{\text{{\it nonlin}-}\sigma}=\frac{(f+\varphi)^2}{4}{\rm Tr}\big[(\partial_\mu \Ugold)^\dagger\partial^\mu \Ugold\big]+\frac{1}{2}\partial_\mu\varphi\partial^\mu\varphi-\frac{m_\varphi^2}{2}\varphi^2-4\lambda f \varphi^3-\lambda\varphi^4\,,\label{nllag}
\end{align}
with $m_\varphi^2=4\mu^2$ as in the $\sigma$ case.

Regarding this expression, it is clear that the decoupling limit for the $\varphi$ field is easily achieved by taking $m_\varphi\to\infty$ while keeping $f$ constant. By doing this, a theory describing the pion dynamics for energies below the $\varphi$ mass is obtained, and it is, actually, the most general one that can be constructed:
\begin{align}
\mL_{2}=\frac{f^2}{4}{\rm Tr}\big[(\partial_\mu \Ugold)^\dagger\partial^\mu \Ugold\big]\,.\label{originalchiral}
\end{align}

 This effective theory of low energy QCD is called the QCD chiral Lagrangian\footnote{More specifically, it corresponds to the lowest order Lagrangian in the chiral expansion.} and it can be UV completed with any $SU(2)_L\times SU(2)_R$ symmetric theory. In fact, the linear case is just one of these possible completions. 
 
In the most general case, the terms included in the Lagrangian of this EFT are not organized through their canonical dimension but through their chiral dimension. Due to the unitary character of the matrix U, the canonical dimension of the operator shown in \eqref{originalchiral} is not well defined, and, therefore, another prescription is needed in order to classify the different contributions of the theory. The expansion is performed in powers of the pion external momentum $p$ (or, equivalently, in powers of derivatives) over a scale $\Lambda$ that accounts for the energy at which the EFT  expansion breaks down. The particular value of this scale is not a priory fixed. It is known to be related to the quantity $4\pi f$, being this the scale controlling the quantum corrections at which the Goldstone dynamics become strongly-interacting. In this sense, it can also be related to the mass of the $\varphi$ particle in the case in which the linear model corresponds to the UV completion of the non-linear EFT. This scale $\Lambda$ represents the new physics (NP) scale or the entrance in the non-perturbative regime of the scalar dynamics, and is usually chosen as the minimum among them $\Lambda=\min(\Lambda_{\rm NP},\, 4\pi f)$. 

Having this in mind, the chiral dimension $d$ of an operator is defined as the number of momentum insertions it possesses, independently of the value of $\Lambda$. For example, the operator given in \eqref{originalchiral} has chiral dimension $d=2$, and, so, it represents the leading order term in the chiral expansion. For this reason, the corresponding Lagrangian is often denoted by $\mL_2$. The chiral  , i.e., the classification of the operators by means of their chiral dimension, is indeed crucial for the computation of the quantum corrections in the EFT.  The chiral Lagrangian was built so that higher order terms in the loop expansion could be safely computed providing finite results. However, the Lagrangian displayed in \eqref{originalchiral} is far from being renormalizable. The one-loop computations obtained from $\mL_2$ produce divergences that cannot be absorbed by terms already present in this Lagrangian. Interestingly, these divergences correspond to the next order in the chiral expansion, $\mO(p^4)$, so they can be absorbed by redefining the parameters of the next to leading order (NLO)  Lagrangian:
\begin{align}
\mL_{4}=a_4\, \big({\rm Tr}\big[(\partial_\mu\Ugold)\Ugold^\dagger (\partial^\nu\Ugold)\Ugold^\dagger\big]\big)^2+a_5\, \big({\rm Tr}\big[(\partial_\mu\Ugold)\Ugold^\dagger (\partial^\mu\Ugold)\Ugold^\dagger\big]\big)^2\,.
\end{align}
This means that these new terms of $\mO(p^4)$ can act as counterterms, so that the divergences generated by the $\mL_2$ operators can be absorbed by redefining the $a_4$ and $a_5$ parameters.

In chiral perturbation theory (ChPT) each operator in the Lagrangian contributes to higher order terms when considering loop diagrams due to Weinberg's power counting theorem~\cite{Weinberg:1978kz}, which, in a simplified version, states that the matrix element of each diagram will scale generically with energy as:
\begin{align}
\mathcal{M}\sim E^{~2+\sum_n N_n(n-2)+2N_L}\,,
\end{align}
with $N_L$ being the number of loops and $N_n$ the number of vertices coming from a term with $n$ derivatives. This way, every one-loop diagram coming from the  $\mL_n$ Lagrangian term contributes to a higher order in the expansion than the corresponding tree level one.

The fact that lower chiral dimension operators contribute via loop diagrams to the same order in ChPT than higher dimension ones at tree level is basic to renormalize the theory. For the case that has been discussed, the loop contributions from $\mL_2$ will be of order $\mO(p^4)$ and so will be the possibly generated divergences. Thus, it should be possible to reabsorb them by redefining the parameters of $\mL_{4}$. As these parameters have to be determined phenomenologically the only difference is that after renormalizing the theory our predictions will have to be made in terms of the renormalized parameters, whose running is given by~\cite{Weinberg:1978kz,Gasser:1984gg, Gasser:1983yg}:
\begin{align}
a_4^r(\mu)&=a_4(\mu_0)-\frac{1}{16\pi^2}\frac{1}{12}\log{\frac{\mu^2}{\mu_0^2}}\,,\nn\\
a_5^r(\mu)&=a_5(\mu_0)-\frac{1}{16\pi^2}\frac{1}{24}\log{\frac{\mu^2}{\mu_0^2}}\,,\label{renorma4a5}
\end{align}
as a consequence of the chiral symmetry, and where $\mu_0$ being a reference scale and $\mu$ being the renormalization scale.

With these renormalized parameters, one can compute finite amplitudes of pion-pion scattering processes, up to $\mO(p^4)$ in a consistent way with chiral perturbation theory, so the full EFT prescription in the QCD context is determined.

Once again, as in the linear case, there is an easy translation between the low-energy QCD scenario and the EW theory picture. 
This time, the correspondence between the QCD and the EW case is achieved by defining the $\Sigma$ matrix as a function of the Higgs field and the EW Goldstone bosons in the following way:
\begin{align}
\Sigma=\frac{v+H}{\sqrt{2}}\,\Ugold=\Big(\widetilde\Phi\,\Phi\Big)\,,~~~~~\Ugold=e^{\,(i\vec{\tau}\cdot\vec{w})/v}\,,\label{nlEWsigma}
\end{align}
noticing that now, instead of the pion decay constant, we have $v=246$ GeV, i.e., the EW scale. 
With this definition, there is a complete analogy with the QCD scenario that has just been  described, with the single difference that the spontaneously broken symmetry in this context is local and not global. Therefore, by substituting the partial derivatives by covariant ones of the form
\begin{align}
D_\mu \Ugold=\partial_\mu\Ugold+\frac{ig}{2}(\vec{W}_\mu\cdot\vec{\tau})\,\Ugold-\frac{ig'}{2}B_\mu\,\Ugold\,\tau_3\,,
\end{align}
and employing the definition given in \eqref{nlEWsigma}, the equivalent to the chiral Lagrangian given in \eqref{nllag} for the EW case reads
\begin{align}
\mL_{\text{{\it nonlin}-}H}=\frac{(v+H)^2}{4}{\rm Tr}\big[(D_\mu \Ugold)^\dagger D^\mu \Ugold\big]+\frac{1}{2}\partial_\mu H\partial^\mu H-\frac{m_H^2}{2}H^2-v\lambda H^3-\frac{\lambda}{4}\,H^4\,.
\end{align}

At this point, the Higgs can be integrated out, in the same way as the $\varphi$ particle before, since it is not needed anymore to preserve the EW chiral or gauge invariance. This limit is achieved by taking $m_H\to\infty$, so that the corresponding EFT for the EW Goldstone boson dynamics at low energies, the so-called Higgsless model, is obtained. The corresponding Lagrangian is given by
\begin{align}
\mL_2^{\slashed{H}}=\frac{v^2}{4}{\rm Tr}\big[(D_\mu \Ugold)^\dagger D^\mu \Ugold\big]\,.\label{higgslesslag}
\end{align}

However, this time, since the $SU(2)\times U(1)_Y$ symmetry is local, the Goldstone bosons do not manifest as physical particles but they appear instead as the longitudinal components of the EW gauge bosons. Therefore, recalling the equivalence theorem (\eqref{ET}), the Lagrangian presented in \eqref{higgslesslag} corresponds to an EFT that describes the longitudinally polarized Ws and Zs between the EW scale and the EFT break down scale. Furthermore, when expanding the matrix U in terms of the fields it contains, it is easy to check that the first term in the expansion, $\Ugold=\mathbb{1}$, leads to the correct mass terms for the EW gauge bosons given in \eqref{WZmasses}. Thus, the EWSB is non-linearly realized and the Higgs has been integrated out from the low energy spectrum.

There exists a difference, however, between the QCD and the EW scenarios, due to the character of the spontaneously broken symmetry in each case. Being the EW a local symmetry, in order to ensure gauge invariance, as it has been  previously mentioned, the Lagrangian operators have to be constructed with the covariant derivative. Furthermore, the kinetic terms of the EW gauge bosons and the fermionic operators have to be included as well in the theory\footnote{Notice that this happened also in the linear case, although it was not explicitly mentioned.}. This fact allows to include a larger set of invariants in the Lagrangian apart from those involving only the scalar fields, although the  former are expected to be less relevant than the latter due to the strongly interacting scalar dynamics. In any case, the Goldstone interactions are exactly the same ones as those described by the non-linear $\sigma$-model, and, therefore, all the properties of the Higgsless scenario can be directly extracted from the paragraphs above. Consequently, the next to leading order terms remain unmodified
\begin{align}
\mL_{4}&=\,a_4\, \big({\rm Tr}\big[(D_\mu\Ugold)\Ugold^\dagger (D_\nu\Ugold)\Ugold^\dagger\big]\big)^2+a_5\, \big({\rm Tr}\big[(D_\mu\Ugold)\Ugold^\dagger (D^\mu\Ugold)\Ugold^\dagger\big]\big)^2\nn\\[4pt]
&=\,a_4\,\big({\rm Tr}\big[\Vchi_\mu\Vchi_\nu]\big)^2+a_5\,\big({\rm Tr}\big[\Vchi_\mu\Vchi^\mu]\big)^2\,,
\end{align}
upon defining the chiral vector
\begin{align}
\Vchi_\mu=(D_\mu\Ugold)\Ugold^\dagger,~~~~\Vchi_\mu\to g_L\, \Vchi_\mu \, g_L^\dagger\,.
\end{align}
The same happens with the renormalization of the $\mL_4$ parameters, whose particular expression matches that in \eqref{renorma4a5}. 

It is worth commenting at this point that another chiral structure with scalar properties can be constructed respecting the SM symmetries. It is the so-called T operator defined as
\begin{align}
\Tchi=\Ugold\tau_3\Ugold^\dagger\,,
\end{align}
which clearly supposes an additional source of custodial symmetry breaking. For this reason, since there is good experimental evidence to believe that custodial breaking is driven only by the gauging of the hypercharge symmetry group and by the differences in the masses of the SM fermions, we will not consider this kind of invariants in the present Thesis.

At this point, we have all the ingredients that characterize the so-called Higgsless model: an example of the non-linear realization of EWSB that allows to remove the Higgs particle from the theory in a EW gauge and chiral invariant way. This procedure leads to an EFT for the low energy Goldstone boson dynamics which is the more general one  consistent with the corresponding symmetries, and that can be UV completed by any theory respecting those same symmetry principles. Even the linear case can be retrieved by this construction in a certain limit.

The Higgsless model was particularly motivated several years ago, before ATLAS and CMS discovered a Higgs-like state with a mass of around 125 GeV. The observation of this particle compatible with a light Higgs boson turned the hypothesis of the Higgs being a  potentially arising heavy resonance, appearing as a consequence of the strongly interacting underlying dynamics, to be a very fragile assumption. A consistent way of introducing this new light degree of freedom in the EFT was, therefore, needed. This is how the Electroweak Chiral Lagrangian with a light Higgs boson was formulated, becoming a very complete and model independent way of studying the issue that concerns us: the true nature of the EWSB.

\section{The Effective Electroweak Chiral Lagrangian with a light Higgs}
\label{theEChL}

In this section, the effective electroweak chiral Lagrangian with a light Higgs boson (EChL), also known as Higgs effective field theory (HEFT) in the recent literature, is introduced. This framework, as illustrated in the previous section, is the most general effective description of EW Goldstone bosons dynamics that can be constructed respecting the SM symmetries, and, therefore, it is the appropriate tool to study EWSB in full depth.

The EChL is a gauged non-linear effective field theory that contains as dynamical fields the EW gauge bosons, W${}^{\pm}$, Z and $\gamma$, the corresponding would-be Goldstone-bosons, $w^{\pm}$, $z$, and the Higgs scalar
boson, $H$. For the present Thesis, the fermion sector remains unmodified with respect to its SM description.

In this setup, the Higgs field is therefore a light degree of freedom, in agreement with the recent observation of a light Higgs-like particle at the LHC, and it is introduced as a singlet of the EW chiral and gauge symmetries. For this reason there are no restrictions regarding its implementation in the Lagrangian, and it is consequently introduced in the most generic possible way. This is done through polynomial functions ${\rm F}(H)$, that depend on new parameters that take particular values in specific UV complete models. 

The EW gauge bosons are described in the usual  manner through the covariant derivative of the U field and through the $SU(2)_L$ and $U(1)_Y$ field strength tensors. The EW Goldstone bosons are placed in a matrix field U, in the same way presented in the Higgsless model, that takes values in the $SU(2)_L \times SU(2)_R/SU(2)_{L+R}$ coset, and transforms as $\Ugold \to g_L \,\Ugold\, g_R^\dagger$ under the action of the global EW chiral group  $SU(2)_L \times SU(2)_R$. Given the arguments discussed in the previous sections, we will assume here that the scalar sector of the EChL preserves the custodial symmetry, except for the explicit breaking due to the gauging of the $U(1)_Y$ symmetry. 

The basic building blocks of the $SU(2)_L\times U(1)_Y$ gauge invariant EChL are the following:
\begin{align}
\Ugold(w^\pm,z) &= 1 + i\, (\vec{\tau}\cdot\vec{w})/v + \mO(w^2)\;,\label{Umatrix}\\[4pt]
{\rm F}(H)&= 1+2\,a(H/v)+b(H/v)^2+\dots\,,\label{polynomial}\\[4pt]
D_\mu \Ugold &= \partial_\mu \Ugold + i\hat{W}_\mu \Ugold - i \Ugold\hat{B}_\mu\,, \label{eq.cov-deriv} \\[4pt]
\hat{W}_{\mu\nu} &= \partial_\mu \hat{W}_\nu - \partial_\nu \hat{W}_\mu
 + i  [\hat{W}_\mu,\hat{W}_\nu ],\;\hat{B}_{\mu\nu} = \partial_\mu \hat{B}_\nu -\partial_\nu \hat{B}_\mu\,,
\label{fieldstrength}\\[4pt]
{\Vchi}_\mu &= (D_\mu \Ugold) \Ugold^\dagger\,,
\label{EWfields}
\end{align}
with $\hat{W}_\mu = g (\vec{W}_\mu \cdot\vec{\tau})/2$ , and $\hat{B}_\mu = g'\, B_\mu \tau^3/2$.

According to the usual counting rules, already introduced  in the previous section, the ${SU(2)_L \times U(1)_Y}$ invariant terms in the EChL are organized by means of their chiral dimension, meaning that a term ${\cal L}_d$ with chiral dimension $d$ will contribute to $\mO(p^d)$ in the corresponding power momentum expansion.
The chiral dimension of each term in the EChL can be found out by following the scaling with $p$ of the various contributing basic functions. Derivatives and  masses are considered as soft scales of the EFT and of the same order in the chiral counting, i.e., of ${\cal O}(p)$. The gauge boson masses, $m_W$ and $m_Z$ are examples of these soft masses in the case of the EChL, similarly as the pion masses in the QCD context, which are also considered to be of ${\cal O}(p)$ in the chiral counting. These gauge boson masses are generated from the covariant derivative in
\eqref{eq.cov-deriv} once the U field is expanded in terms of the $\vec{w}$ fields: \begin{align}
D_\mu \Ugold &= i\Frac{( \vec{\tau}\cdot \partial_\mu\vec{w})  }{v} + i\,  \Frac{g v}{2}  \, \Frac{ (\vec{\tau}\cdot\vec{W_\mu})}{v} - i\,  \Frac{g' v}{2}  \, \Frac{B_\mu\,  \tau^3}{v} + \dots
\end{align}
as it happened already in the Higgsless case. The dots represent terms with higher powers of $(w/v)$, and their precise form will depend on the particular parameterization of U.  In this Thesis we will always employ the usual exponential parametrization, although others are equally valid, albeit motivated for different studies. Once the gauge fields are rotated to the physical basis, defined in \eqref{massrot}, they get the usual mass values at lowest order: $m_W=gv/2$ and $m_Z =\sqrt{g^2+g'^{2}}\,v /2$.

Furthermore, in order to have a power counting consistent with the loop expansion, one needs all the terms in the covariant derivative above to be of the same order. Thus, the proper assignment is  $\partial_\mu$, $(g v)$ and  $(g'v) \sim \mO(p)$ or, equivalently, $\partial_\mu$, $m_W$,  $m_Z \sim \mO(p) $. In addition, the Higgs boson mass $m_H$ will be also considered as another soft mass in the EChL with a similar chiral counting as $m_W$ and $m_Z$.  That implies, $m_H \sim \mO(p)$, or equivalently $(\lambda v^2 ) \sim \mO(p^2)$, with $\lambda$ being the SM Higgs self-coupling. 

The typical energy scale that controls the size of the various contributing terms in this chiral expansion is again provided by $4 \pi v\sim 3$ TeV. In the scenarios where there are emerging resonances, a common case in strongly interacting underlying UV theories, then there are additional mass scales given by the masses of the resonances to account for in the EChL.

With these building blocks one can construct the EChL up to a given order in the chiral expansion. This Lagrangian has to fulfil the requirements of being Lorentz invariant and $SU(2)_L \times U(1)_Y$ gauge invariant and custodial preserving. We also assume CP invariance for the present Thesis and include the terms with chiral dimension up to  $\mO(p^4)$. Therefore, the EChL can be generically written as:
\be
\mL_{\rm EChL} = \mL_2 + \mL_4 +\mL_{\rm GF} +\mL_{\rm FP}\, ,
\ee
where  $\mL_2$ refers to the terms with chiral dimension 2, i.e., $\mO(p^2)$,  $\mL_4$ refers
to the terms with chiral dimension 4, i.e., $\mO(p^4)$, and
$\mL_{\rm GF}$ and $\mL_{\rm FP}$ are the gauge-fixing (GF) and the corresponding
non-abelian Fadeev-Popov (FP) terms. As commented in the Introduction, we will be interested in the study of EW gauge boson scattering amplitudes, for whose description the relevant terms are, at leading order:
\begin{align}
\mL_2 =&    -\Frac{1}{2 g^2} {\rm Tr}\big[\hat{W}_{\mu\nu}\hat{W}^{\mu\nu}\big] -\Frac{1}{2 g'^{2}}
{\rm Tr} \big[\hat{B}_{\mu\nu} \hat{B}^{\mu\nu}\big]+ \Frac{1}{2} \partial^\mu H \, \partial_\mu H - V(H)\nn\\
& +\Frac{v^2}{4}\left[%
  1 + 2\,a (H/v) + b (H/v)^2\right] {\rm Tr} \big[D^\mu U^\dagger D_\mu U \big]+\dots
\label{eq.L2}
\end{align}
with $V(H)$ being the usual Higgs potential given in \eqref{lagphi}, whereas the next to leading order terms correspond to\footnote{The notation used is taken from \cite{Herrero:1993nc,Herrero:1994iu} and compares: 1) with~\cite{Longhitano:1980iz} as, $a_1=(g/g') \alpha_1$, $a_2=(g/g') \alpha_2$,
$a_3= -\alpha_3$, $a_4= \alpha_4$, $a_5= \alpha_5$; 2) with~\cite{Gasser:1983yg} as, $\ell_1=4 a_5$, $\ell_2= 4 a_4$, $\ell_5= a_1$, $\ell_6= 2(a_2 -a_3)$; and with~\cite{Gasser:1984gg} as, $L_1=a_5$, $L_2=a_4$, $L_9=a_3-a_2$, $L_{10}=a_1$.}:
\begin{align}
{\mL}_{4} =& %
 ~ a_1\, {\rm Tr}\big[\Ugold \hat{B}_{\mu\nu} \Ugold^\dagger \hat{W}^{\mu\nu}\big]
  + i a_2\,  {\rm Tr}\big [ \Ugold \hat{B}_{\mu\nu} \Ugold^\dagger [\Vchi^\mu, \Vchi^\nu ]\big]
  - i a_3 \,{\rm Tr}\big [\hat{W}_{\mu\nu}[\Vchi^\mu, \Vchi^\nu]\big] \nn \\
&+
  ~ a_4\, \big({\rm Tr}\big[\Vchi_\mu \Vchi_\nu \big]\big) ^2 
  + a_5\,  \big({\rm Tr}\big[\Vchi_\mu \Vchi^\mu\big]\big) ^2
\nn \\[5pt]
  &-
  ~ c_{W} (H/v) {\rm Tr}\big[\hat{W}_{\mu\nu} \hat{W}^{\mu\nu}\big]
  - c_B(H/v)\, {\rm Tr} \big[\hat{B}_{\mu\nu} \hat{B}^{\mu\nu} \big]\, +\dots\label{eq.L4}
\end{align}

The effects of the new physics introduced by these operators will be controlled by the values of the couplings associated to each of them.  These couplings that appear in the Lagrangian are the so called chiral parameters or chiral coefficients of the EChL, and encode the information of the microscopic theory. At this point, it is important to recall that we are mostly interested in vector boson scattering processes, since we believe they are the most sensitive observables to new EWSB physics. Each of the previous Lagrangian terms will contribute differently to these observables and their relative importance will depend on the values of the chiral coefficients. 

To have an insight of how these couplings intervene in the predictions of the EChL, we can inspect the Feynman rules derived from this Lagrangian, collected in Appendix \ref{FR-EChL}. For example, if we take a look at the self interaction vertices of four weak gauge bosons it is clear that for $a_{3,4,5}\neq 0$ the predictions in the ECLh will be different from those in the Standard Model, the latter case corresponding to $a_i=0$. Furthermore, if one focuses on the longitudinally polarized EW gauge boson scattering, the most relevant $\mL_4$ coefficients are $a_4$ and $a_5$, as we shall see quantitatively later on. This can be understood easily by means of the equivalence theorem, since the mentioned process, at energies well above the EW scale, is described by $ww$ scattering and is thus dominated by the derivative Goldstone interactions given by $a_4$ and $a_5$. The other three parameters, $a_1$, $a_2$ and $a_3$, that appear in the Lagrangian, will modify mainly the interaction of the transversely polarized gauge bosons, so their contribution at high energies is expected to be suppressed with respect to that of $a_4$ and $a_5$. However, they are of much importance in, for instance, processes involving photons, which are purely transverse as well as in EW precision observables at LEP energies. The $c_W$ and $c_B$ parameters, also included here, control BSM interactions such as the one relating locally a Higgs with two photon fields that will not be studied in this Thesis. 

With all these considerations in mind, and with the EChL properly introduced, we should be able to compute the relevant observables up to a certain order in chiral perturbation theory. Nevertheless, as discussed in the previous section, a renormalization prescription is needed such that we obtain finite contributions order by order. In the same way as in the preceding examples, the divergences of the one loop contributions from $\mL_2$ will be absorbed in the $\mL_4$ parameters, and, as it is customary, through the renormalization group equations they will acquire a dependence on the renormalization scale, contained in the following equation:
\begin{equation}
\dfrac{da_i^r}{d\log\mu}=-\dfrac{\Gamma_{a_i}}{16\pi^2}\,,\label{eqrunninga4a5}
\end{equation}
where $\Gamma_{a_i}$ gives the running of the particular coupling $a_i$. For completeness we show in Table \ref{tabla}, the running of the most relevant chiral parameters, extracted from the summary in \cite{Delgado:2014jda} and references therein. The Higgsless scenario is also considered, from which the running of $a_4$ and $a_5$ presented in \eqref{renorma4a5} can be easily recovered by setting $a=b=0$. The combinations appearing in Table \ref{tabla} are in general relevant for different processes; the ones containing $a_1$, $a_2$ and $a_3$ will be involved in transverse gauge boson interactions and the ones containing $a_4$ and $a_5$, will be the ones of greater relevance for the present Thesis, since they will contribute to the interactions of longitudinal gauge bosons.
\begin{table}[!t]
\begin{center}
\begin{tabular}{ lcc }
\toprule
\toprule
 & {\bf ECLh  } &  {\bf Higgsless}\\                  
 \midrule
$ \Gamma_{a_1-a_2+a_3}$ & $ 0  $ &  0                
\\[4pt]
$ \Gamma_{a_1}$ & $ -\frac{1}{6}(1-a^2)  $&$ -\frac{1}{6}  $                
\\[4pt]
$ \Gamma_{a_2-a_3} $ & $ -\frac{1}{6}(1-a^2)  $& $ -\, \frac{1}{6}  $                  
\\[4pt]
$ \Gamma_{a_4}$ & $ \frac{1}{6}(1-a^2)^2   $ & $ \frac{1}{6} $                 
\\[4pt]
$ \Gamma_{a_5}$ & $ \frac{1}{8}(b-a^2)^2
+\frac{1}{12}(1-a^2)^2 $ &
$ \frac{1}{12} $                 
\\[4pt]
\bottomrule
\bottomrule
\end{tabular}
\vspace{0.4cm}\caption{
Running of the relevant ECLh parameters and some of their combinations taken from references \cite{Delgado:2014jda,Espriu:2012ih, Espriu:2013fia, Espriu:2014jya, Delgado:2013loa, Delgado:2013hxa}. The third column provides the corresponding running  for the Higgsless case~\cite{Herrero:1993nc}.\label{tabla}}
\end{center}
\end{table}

The relations displayed in \tabref{tabla} are totally model independent, as they are valid for any set of values of the chiral parameters. Therefore, if we estimate measurable observables involving the scattering of longitudinally polarized gauge bosons with these formulas, and compare the results with the experiment, we would be able to discern the properties and structure of the model that is preferred by data, as we could deduce the values of the chiral parameters, which determine the complete ultraviolet theory.

We present here, for completeness, some illustrative examples of values of the EChL coefficients corresponding to specific interesting models. The SM, and therefore a way of recovering the linear EWSB realization, is chosen as one of these representative benchmarks. We also show the Higgsless case, as a connection to the previous section. The well motivated minimal composite Higgs model~\cite{Agashe:2004rs,Contino:2006qr,Contino:2011np,Barducci:2014kxa,Sanz:2017tco} (MCHM), constructed from the original composite Higgs models~\cite{Kaplan:1983fs,Kaplan:1983sm,Banks:1984gj,Georgi:1984ef,Georgi:1984af,Dugan:1984hq} and based on a $SO(5)/SO(4)$ symmetry breaking pattern is presented as well, together with the dilatonic models~\cite{Halyo:1991pc,Goldberger:2008zz} corresponding values:
\begin{center}
\begin{tabular}{ ll }
$a^2=b=1$ &  SM,\\[4pt]                  
$a^2=b=0$ & Higgsless,\\[4pt] 
$a^2=1-\frac{v^2}{f^2},~~b=1-\frac{2v^2}{f^2}$~~~~ &$SO(5)/SO(4)$ MCHM,\\[4pt] 
$a^2=b=\frac{v^2}{f^2}$ & Dilaton.\\
\end{tabular}
\end{center}
The NLO Lagrangian parameters have as well specific values in each of the presented theories. The SM would imply to set all the $a_i$, and the $c_W$ and $c_B$ coefficients to 0, while, for instance, recovering the Higgsless model would require to set all the parameters involving a Higgs particle to 0 in the NLO Lagrangian as well.

The presence of the scale $f$ in the last two examples is explained by the fact that, in these models, the spontaneous breaking of the EW symmetry takes place in two steps instead of one. For instance, in the MCHM, some ultraviolet strong dynamics triggers the spontaneous breaking of the $SO(5)$ group to a $SO(4)$ remnant, such that four Goldstone bosons emerge in the spectrum. The characteristic scale of these scalar modes, that at this point remain massless, corresponds to $f$. Afterwards, the EW symmetry is gauged, and the $SO(5)$-breaking radiative corrections induce a potential for the Higgs particle, interpreted as another Goldstone boson in this scenario. The minima of this potential break the EW symmetry spontaneously and the other three initial Goldstones are {\it eaten} by the EW gauge bosons that acquire a mass. In these scenarios, due to the Goldstone character of the Higgs, the hierarchy problem is automatically solved since the Higgs mass is protected by a larger symmetry. However, although well motivated, the MCHM is just a possible UV complete theory whose low-energy dynamics can be parameterized through the EChL.

From the previous paragraphs it is plain that there is a plethora of ultraviolet completions for the EChL leading to different explanations of the EWSB dynamical generation. But our final aim is to discern which is the BSM setup that describes best the electroweak symmetry breaking, and thus, the experimental determination of the chiral parameters is crucial. In the years to come it is expected that the physics of the vector boson scattering will be tested in such a way that will allow to shed some light on this issue. 

Regarding the present experimental constraints on the previous EW chiral coefficients, it is important to have in mind that the particular value of the bounds imposed on EFT parameters depends enormously on the phenomenological interpretation of the experimental data. As we will discuss throughout this Thesis, especially in Chapter \ref{Methods}, different EFT treatments can lead to different EChL parameter constraints. 

However, it is pertinent to include at this point a general comment on some of the most established current bounds to have a first guiding value with which obtain our predictions later on. We will focus on the EChL parameters that will be the most relevant ones for this Thesis: $a$, $a_4$ and $a_5$. 

Different strategies were used in the literature to obtain an appropriate fit for the Higgs coupling to vector bosons using experimental data~\cite{Khachatryan:2014jba,ATLAS:2014yka,Buchalla:2015qju}, which lead, approximately, to a value of $a$ centered around the SM prediction ($a$=1) with a 10\% deviation allowed, i.e., $0.9<a<1.1$. These bounds, however, have been recently updated  by the ATLAS collaboration, that has provided a constraint for the $HVV$ coupling at 95\% C.L. corresponding to~\cite{ATLAS:2019slw}
\begin{align}
0.97<a<1.13\,.
\end{align}
This constraint seems to be, in principle, consistent with the full EFT picture.

In what concerns the current allowed values of $a_4$ and $a_5$, the most recent constraints, provided by CMS at 95\% confidence level, read~\cite{Sirunyan:2019der}:
\begin{align}
 |a_4| < 6\cdot 10^{-4}\,,~~~|a_5| < 8\cdot 10^{-4}\,.\label{cotasa4a5}
\end{align}
These results correspond to the translation\footnote{This translation will be exemplified in the next section. \eqref{fs0fs1} is the appropriate one to use in this particular comparison.} of the bounds imposed on linear EFT parameters and are obtained analyzing one parameter at a time and without employing any unitarization procedure upon the EFT predictions.

There are, however, other studies regarding the same kind of measurements that might allow to constrain the $a_4$ and $a_5$ parameter space and that are performed with a different theoretical interpretation. For instance, in~\cite{Aad:2019xxo} a maximum total cross section of various VBS processes, and, therefore, a model independent experimental study, is reported, whereas in \cite{Aaboud:2016uuk} an interesting bound on $a_4$ and $a_5$ is provided using a K-matrix unitarization analysis, following the procedure proposed in \cite{Alboteanu:2008my}. The constraints obtained for all these studies can vary significantly, as we will see in forthcoming Chapters of this Thesis, so it is important to have in mind that the values given in \eqref{cotasa4a5} carry an intrinsic theoretical error that will be quantified for the first time in Chapter \ref{Methods}.

We would also like to make a comment about the other parameters presented in Eqs.~(\ref{eq.L2}) and (\ref{eq.L4}). Regarding the $\mL_2$ parameter $b$, it is very interesting to point out that it has been finally constrained to be $-1.02<b<2.71$ at 95\% C.L. via search strategies of VBS double Higgs production in the $b\bar{b}b\bar{b}jj$ final state~\cite{ATLAS:2019dgh}. This represents the first experimental bound imposed on this parameter nowadays.

The coefficients controlling $\mO(p^4)$ operators, $a_1$, $a_2$ and $a_3$ have been  constrained in different setups, and, particularly, $a_1$ results to be quite heavily constrained by LEP data due to its relation to the oblique $S$ parameter~\cite{Herrero:1993nc,Falkowski:2013dza,Brivio:2013pma,Fabbrichesi:2015hsa}. With the combined LEP data, provided in~\cite{Tanabashi:2018oca}, a constraint on this oblique correction of $-0.12<S<0.16$ can be imposed at the 95\% C.L., and with the relations presented in~\cite{Herrero:1993nc,Delgado:2014jda}, which imply $S=-4\pi a_1$, the EChL parameter $a_1$ is restrained to be $-0.013<a_1<0.009$.

 \begin{figure}[t!]
\begin{center}
\includegraphics[width=0.8\textwidth]{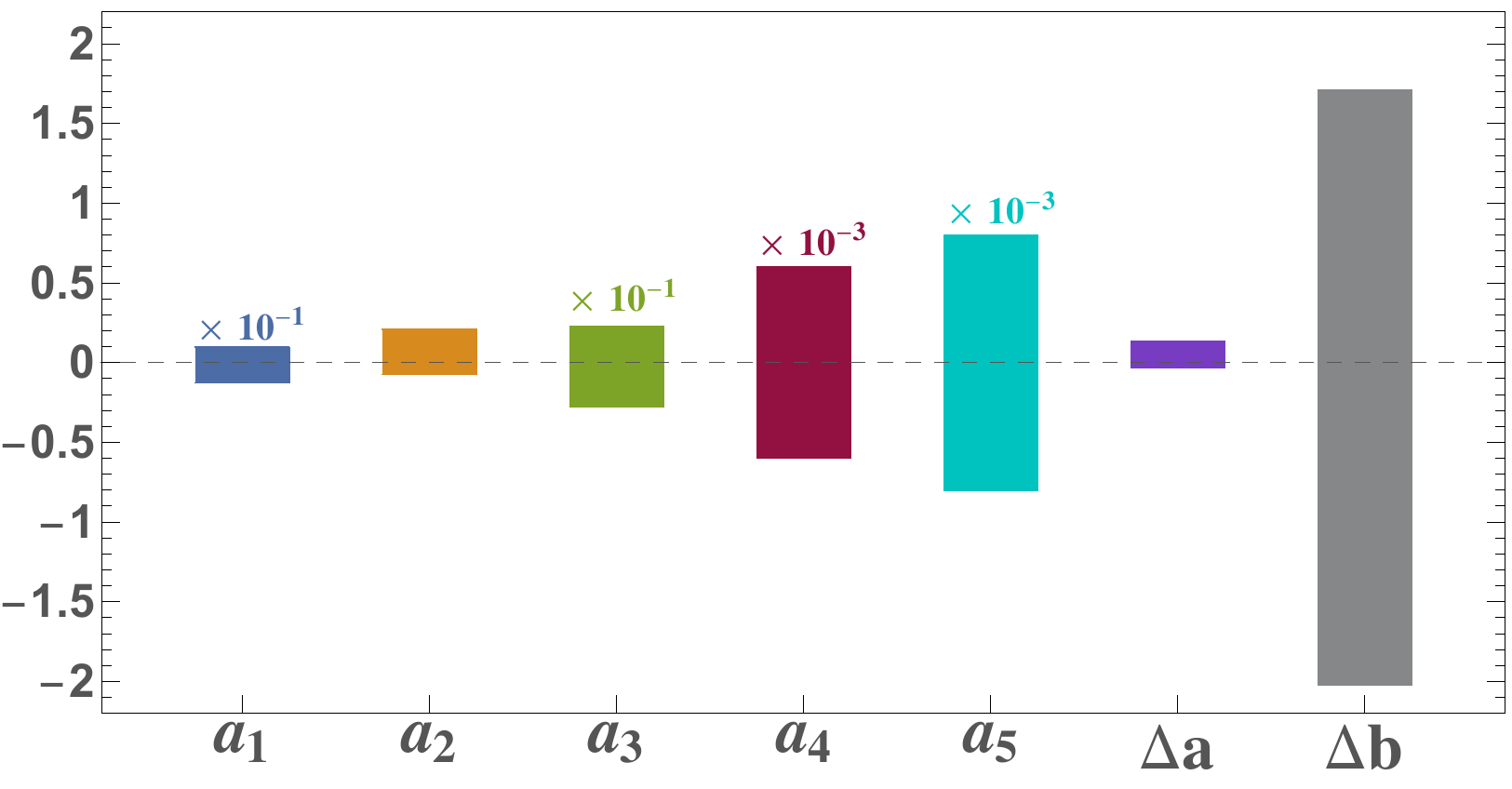}
\caption{Summary of the present experimental constraints imposed on the EChL coefficients.
Bounds for $\Delta a=a-1$~\cite{ATLAS:2019slw} and $\Delta b=b-1$~\cite{ATLAS:2019dgh} are extracted directly from the cited references. Limiting values of $a_4$ and $a_5$~\cite{Sirunyan:2019der} can be derived using \eqref{fs0fs1}. $a_1$ is constrained by LEP data through its relation to the S parameter~\cite{Herrero:1993nc,Falkowski:2013dza,Brivio:2013pma,Fabbrichesi:2015hsa}. Constraints on $a_2$ and $a_3$ are naively obtained using the global fit data from~\cite{Almeida:2018cld} together with Eq.~(\ref{hastaelpapo}) explained in the text.}
\label{fig:constraints}
\end{center}
\end{figure}

The other two, $a_2$ and $a_3$ have been bounded thanks to their relation to triple gauge couplings~\cite{Falkowski:2013dza, Brivio:2013pma, Khachatryan:2014jba, Aad:2014zda, ATLAS:2014yka, Fabbrichesi:2015hsa, Buchalla:2015qju,deFlorian:2016spz,Aaboud:2016uuk}. The most recent constraints on $a_2$ and $a_3$ can be derived from those obtained for their related linear coefficients. As we will see in the next Section, there is not a unique way to translate linear into non-linear parameters: both setups can be compared in a Lagrangian language (i.e., comparing operators), at the level of their repercussion in a particular observable (and, therefore, observable dependent), etc. Thus, the obtention of an accurate bound on $a_2$ and $a_3$ would imply to analyze the data directly in the EChL framework. 

Nevertheless, to give a first insight of the expected order of magnitude of the constraints imposed on these two parameters, one can take the most recent values for the bounds on the SMEFT parameters\footnote{These appear in \cite{Almeida:2018cld} as $C_W$ and $C_B$, not to be confused with the EChL parameters $c_W$ and $c_B$ denoted here in lowercase.} $f_W$ and $f_B$, controlling the operators $\mO_W$ and $\mO_B$ as described in~\cite{Almeida:2018cld}, and naively translate them into bounds for $a_2$ and $a_3$ using the relations given in~\cite{Rauch:2016pai} and~\cite{Herrero:1993nc}. In such a way, with the global fit data provided at the 95\% C.L. in~~\cite{Almeida:2018cld} for the linear coefficients
\begin{align}
-3.0<\frac{f_W}{\Lambda^2}<3.7\,,~~-8.3<\frac{f_B}{\Lambda^2}<26.0\,,
\end{align}
and by applying 
\begin{align}
a_2=\frac{m_W^2}{2g^2}\left(\frac{f_B}{\Lambda^2}\right)+a_1\,,~~a_3=-\frac{m_W^2}{2g^2}\left(\frac{f_W}{\Lambda^2}\right)\,,\label{hastaelpapo}
\end{align}
taking into account the limiting values of $a_1$ derived from LEP data, 
one arrives at
\begin{align}
 -0.07<a_2<0.20\,,~~ -0.03<a_3<0.02\,.
 \end{align}
 It is important to keep in mind that, because of the ambiguous matching between both realizations, these values represent a naive estimate and just serve as an orientation to the order of magnitude we might be dealing with.

Regarding $c_W$ and $c_B$, also coming from $\mL_4$, the most stringent bounds come from the related coefficient $c_{\gamma\gamma}=c_W+c_B$ appearing in the photonic $\frac{e^2}{16 \pi^2}c_{\gamma\gamma}\frac{H}{v}F_{\mu\nu}F^{\mu\nu}$ Lagrangian term, that has been constrained to be $ c_{\gamma\gamma}=-0.24\pm 0.37$~\cite{Buchalla:2015qju}.

A summary of the most relevant constraints explained in the lines above is displayed in \figref{fig:constraints}. In this Figure the parameters $a$ and $b$ are presented as $\Delta a=a-1$ and $\Delta b=b-1$, respectively, in order to show their departure from the SM value $a=b=1$.

In the preceding pages we have thoroughly presented and discussed the effective electroweak chiral Lagrangian with a light Higgs framework that will serve us to explore BSM deviations in the relevant VBS observables throughout the present Thesis. This EFT is the most general one that can be constructed regarding low energy descriptions of the EWSB dynamical origin, and it can be UV completed in an arbitrary number of ways if the SM symmetries are preserved.

Nevertheless, there is another EFT, the SMEFT, in which the EWSB is linearly realized which is also aimed to describe possible BSM physics. This theory, albeit less general, is equally valid as the EChL, and it can be indeed recovered from the latter in a particular limit.  Since the SMEFT is commonly employed nowadays in the search for deviations with respect to the SM predictions for different observables we will briefly discuss its properties in the next section as well as its specific relation to the EChL.  

\section{The Standard Model effective field theory and its relation with the EChL}

The Standard Model effective field theory (SMEFT) is based upon a linear realization of the EWSB as the one presented in the discussion of the BEH mechanism. Therefore, the nature and interactions of the Higgs boson are intimately related to those of the EW Goldstone bosons. In particular, the four scalars are embedded in a bi-doublet that transforms linearly under the EW chiral group $SU(2)_L\times SU(2)_R$, which means that, contrary to the EChL case, the Higgs boson interaction terms have always a polynomial structure in powers of $(v+H)$. 

The Lagrangian of this theory is built such as all the SM symmetries are preserved. Actually, the leading order Lagrangian is identified with the SM one (thus the name SMEFT), and the tower of BSM operators is organized in terms of their canonical dimension:
\begin{align}
\mL_{\rm SMEFT}=\mL_{\rm SM}+\sum_i \frac{f_i}{\Lambda^2}\mO_i^{d=6}+\sum_i \frac{f_i}{\Lambda^4}\mO_i^{d=8}+\dots\,
\end{align}
Here, the scale $\Lambda$, at which the new physics is expected to emerge, is completely unknown, and, in principle, there is no guidance towards a particular value. 

The specific form of the dimension 6, $\mO_i^{d=6}$, and dimension 8, $\mO_i^{d=8}$, operators depends on the choice of basis~\cite{Buchmuller:1985jz,Hagiwara:1993ck,Hagiwara:1996kf,Grzadkowski:2010es,Corbett:2012ja}. There are, nowadays, many different bases in which to write the relevant operators motivated by different types of physics studies~\cite{Eboli:2006wa,Corbett:2013pja,Ellis:2014dva,Ellis:2014jta,Corbett:2014ora,Degrande:2016dqg,Ellis:2018gqa,Barzinji:2018xvu,Grazzini:2018eyk,Perez:2018kav,Dawson:2018liq,Dawson:2018jlg,Dawson:2019xfp,Baglio:2018bkm,Almeida:2018cld,Hartland:2019bjb,Buchalla:2019wsc,Moutafis:2019wbp,Aoude:2019tzn,Brivio:2019myy}. In this section we will consider just a very reduced sample of operators given in a concrete basis to illustrate the relation between this linear EFT and its non-linear version, the EChL.

First of all, since the EChL is a more general theory, the number of invariants contained in the Lagrangian is, in principle, larger. This means that independent operators in the EChL might appear as correlated structures in the SMEFT case. However, this statement strongly depends on the different orders one is considering in each scenario, and it has to be treated carefully.

Secondly, the counting in both EFTs is very different. Whereas in the linear scenario the canonical dimension of each operator is used to organize the expansion, in the non-linear case the chiral counting governs the classification of invariants. In this sense, operators of different canonical dimension can contribute to the same order in the chiral expansion, causing, therefore, a reshuffling of the terms appearing in the Lagrangian between both theories. Thus, an EChL operator of fixed order in the chiral expansion might be comparable with operators of different canonical dimension, and therefore different order, in the linear theory. Since the counting in the non-linear scenario is more involved, there is some controversy on the weight that is assigned to each operator~\cite{Cohen:1997rt,Jenkins:2013sda,Buchalla:2013eza,Gavela:2014vra,Buchalla:2014eca,Gavela:2016bzc}. In this Thesis we will always use the chiral counting described in the previous section, but other options can be implemented. 

Having these considerations in mind, it is possible to match both EFTs at the Lagrangian level by defining the following relation:
\begin{align}
\Phi\leftrightarrow\frac{(v+H)}{\sqrt{2}}\,\Ugold \left(\begin{array}{c}0\\1\end{array}\right)\,,\label{translationlinnonlin}
\end{align}
which translates the doublet structure of the linear theory, that equals the one given in \eqref{doublet}, to the framework of the non-linear theory in terms of the singlet $H$ and the matrix U.

As a concrete example of this matching between both scenarios, we can analyze the following dimension-8 operators of the SMEFT~\cite{Kilian:2014zja}:
\begin{align}
\mL_{\rm SMEFT}=\frac{f_{S,0}}{\Lambda^4}\,\big({\rm Tr}\big[(D_\mu {\rm \bf H})^\dagger D_\nu {\rm \bf H}\big]\big)^2 + \frac{f_{S,1}}{\Lambda^4}\,\big({\rm Tr}\big[(D_\mu {\rm\bf H})^\dagger D^\mu {\rm\bf H}\big]\big)^2\,,\label{Lfs0fs1}
\end{align}
with 
\begin{align}
{\rm\bf H}=\frac{1}{2}\left(\begin{array}{cc} v+H-iw_3 & -i\sqrt{2}w^+ \\-i\sqrt{2}w^- & v+H+iw_3
 \end{array}\right)\,,
\end{align}
and the covariant derivative defined as in \eqref{eq.cov-deriv} in terms of \textbf{H} instead of U. 

If, now, the replacement given in \eqref{translationlinnonlin} is implemented, restraining oneself to the lowest order in the U expansion in the Goldstone fields, it is easy to arrive at the following result:
\begin{align}
D_\mu{\rm\bf H}=\frac{v}{2}\Vchi_\mu\,,
\end{align}
which allows to directly relate the two SMEFT operators presented in \eqref{Lfs0fs1} with the EChL structures controlled by the chiral couplings $a_4$ and $a_5$:
\begin{align}
a_4=v^4\,\frac{f_{S,0}}{16\Lambda^4}\equiv v^4\,\frac{F_{S,0}}{16}\,,~~~~~~~
a_5=v^4\,\frac{f_{S,1}}{16\Lambda^4}\equiv v^4\,\frac{F_{S,1}}{16}\,.\label{fs0fs1}
\end{align}
We have denoted $F_{S,i}=f_{S,i}/\Lambda^4$ for shortness and since it is the true combination that can be probed experimentally, due to the fact that the coefficient $f_i$ cannot be separated from the scale.

With this simple example, the relation between both EFT setups at the Lagrangian level is shown so experimental and theoretical results derived from each theory can be compared. Nevertheless, another, more physically motivated, prescription to match both EFTs is that of comparing on-shell observables, like the VBS cross sections of our interest. Computing a specific observable in each EFT and identifying structures governed by different parameters supposes a more intricate way of relating both scenarios, since the translation of the linear parameters to the EChL ones and viceversa becomes process dependent. However, since theses observables can be tested directly, unlike Lagrangian operators or even Feynman rules, a more physical interpretation between the matching of both EFTs can be performed. This is indeed of paramount importance  as the intercourse of these theories is crucial in order to be able to distinguish between both effective descriptions. Should this be achieved experimentally, the true nature of EWSB would be revealed to be linearly or non-linearly realized.

\section{Violation and restoration of perturbative unitarity}

We have seen, so far, that effective theories can be an incredibly useful tool to study the low-energy dynamics of a system in a very model independent way. Specifically the effective EChL serves as the most general framework to help us understand the true EWSB dynamical origin. Nevertheless, these constructions, and in particular the EChL, can suffer from inconsistency problems such as the one of unitarity violation. 

Although the UV complete theory has to be unitary in order to ensure probability conservation, the EFT that describes it at low energies might violate the unitarity condition at some point. This happens indeed in the EChL due to the structure of the operators it contains. The fact that in the context of this strongly interacting dynamics, operators, and thus, interactions among gauge bosons, scale directly with the external momentum, leads to a scenario in which predictions of observables can behave pathologically with energy from a certain energy scale upwards. This pathology translates into a violation of unitarity of the S matrix, which basically implies an unphysical leak in the interaction probability among EW gauge bosons at energies that can be probed now at the LHC.

However, predictions that are to be tested at colliders must be fully unitary to be consistent with the underlying quantum field theory. Therefore, a prescription is needed to translate these non-unitary predictions into reliable, unitary ones with which interpret the experimental data. These prescriptions are called unitarization methods or procedures, that drive unitary the non-unitary EFT predictions.

In this section the violation of unitarity will be revisited generically and in the specific context of the EChL. Furthermore, several ways of restoring unitarity, i.e., several unitarization methods, will be introduced and described, since they will be a fundamental ingredient for this Thesis in the forthcoming Chapters.

\subsection{Unitarity of the S matrix}

In a physical process, for the probability current to be conserved, the operator that controls the evolution of the system has to be unitary, otherwise unphysical quantities appear in the computation of observables, such as probabilities that exceed unity. 

In quantum field theory, the operator that relates the initial and the final states in a scattering process, and that, consequently, controls the evolution of the system, is the S matrix, which is usually defined as:
\begin{align}
\label{smatrixdef} {\rm S} = \mathbb{1} + i\,{\rm T}\,,
\end{align}
with T being the transition amplitude matrix containing the actual information about the scattering and whose entries are the scattering amplitudes. The analytical properties of the S matrix can be found in~\cite{Eden:1966dnq}.

The unitarity condition for the S matrix reads
 \begin{align}
\label{unitdef} {\rm S} {\rm S}^\dagger = {\rm S}^\dagger {\rm S} = \mathbb{1}\,,
\end{align}
which obviously implies some constraints on the transition matrix so that unitarity is preserved:
\begin{align}
& {\rm S} {\rm S}^\dagger = (\mathbb{1}+i\,{\rm T})(\mathbb{1}-i\,{\rm T}^\dagger) = \mathbb{1}^2-i\mathbb{1}{\rm T}^\dagger+i\mathbb{1}{\rm T}+{\rm T} {\rm T}^\dagger = \mathbb{1}+i({\rm T}-{\rm T}^\dagger) + {\rm T} {\rm T}^\dagger = \mathbb{1}\,,\nn\\
&i({\rm T}-{\rm T}^\dagger) + {\rm T} {\rm T}^\dagger = 0~,~~({\rm T}-{\rm T}^\dagger) = i\,{\rm T} {\rm T}^\dagger\,,
\end{align}
leading to the final condition:
\begin{align}
&  2\,{\rm Im}[{\rm T}]=|{\rm T}|^2 \label{T}\,.
\end{align}

This relation translates into conditions for the scattering amplitudes of the processes parameterized by the S matrix, which, in terms of its matrix elements will be:
\begin{align}
\langle f \,| \,{\rm S} \, |\, i \rangle \equiv {\rm S}_{fi}=\delta_{fi}+i(2\pi)^4\delta^4(p_i-p_f){\rm T}_{fi}  \,,
\end{align}
leading to an equivalent relation to Eq.~(\ref{T}) for the elements of the T matrix:
\begin{align}
{\rm T}^{}_{fi}-{\rm T}^*_{if}=i\sum_n {\rm T}^{}_{fn}{\rm T}_{ni}^*(2\pi)^4\delta^4(p_i-p_n)\delta^4(p_n-p_f)\label{T2},
\end{align}
where $p_i$, $p_f$ and $p_n$  are the momenta of the initial, final and all possible intermediate particles respectively.

A more friendly and convenient expression regarding the unitarity condition can be obtained by performing a partial wave analysis~\cite{Jacob:1959at,Chaichian:2003yu}. As most scattering processes are rotationally invariant, we can choose a basis to express our scattering amplitudes in a way in which this symmetry is manifest. This simplifies greatly the computations when trying to implement the unitarity conditions of the S matrix, and can be made by just projecting the amplitudes into the basis of the eigenstates of the angular momentum operator. For the case we will be interested in, i.e., the $2\to 2$ scattering of EW gauge bosons $V=W,Z$: $V_{\lambda_1}V_{\lambda_2}\to V_{\lambda_3}V_{\lambda_4}$, we have
 \begin{align}
\label{pwexp}&A_{\lambda_1\lambda_2\lambda_3\lambda_4}(s,\cos\theta)= 16 \pi K \sum_J\, (2J+1)\, d_{\lambda,\lambda'}^{J}(\cos\theta) \,a^J_{\lambda_1\lambda_2\lambda_3\lambda_4}(s)\,.
\end{align}
Here, $A(s,\cos\theta)$ is the scattering amplitude for a fixed initial and final polarization state, denoted by $\lambda_1\lambda_2\lambda_3\lambda_4$, being $s$ the squared center of mass energy and $\theta$ the scattering angle. $J$ is the total angular momentum and $d^{J}_{\lambda\lambda'}(\cos\theta)$ are the Wigner functions with $\lambda=\lambda_1-\lambda_2$ and $\lambda'=\lambda_3-\lambda_4$. $K$ is a factor accounting for identical particles in the final state that takes the value of 1 or 2 if the final particles are distinguishable or indistinguishable, respectively. Finally, $a^{J}_{\lambda_1\lambda_2\lambda_3\lambda_4}(s)$ is the corresponding amplitude of the $J$-th partial wave, which by virtue of the orthogonality relations of the Wigner functions can be obtained as:
\begin{align}
\label{pwamp} a^{J}_{\lambda_1\lambda_2\lambda_3\lambda_4}(s)=\dfrac{1}{32 \pi K}\int_{-1}^{1} d(\cos\theta)\,d_{\lambda,\lambda'}^{J}(\cos\theta)\, A_{\lambda_1\lambda_2\lambda_3\lambda_4}(s,\cos\theta)
\,.
\end{align}
In all these considerations, we have factored out the dependence on the azimuthal scattering angle $\phi$, since it enters always as a phase that will be trivial for the cross section computations.

Now, through the relation between the scattering amplitudes and the S matrix it is easy to obtain the following identity
\begin{align}
{\rm Im} \Big[a^J_{\lambda_1\lambda_2\lambda_3\lambda_4}(s)\Big]=&\,\Gamma(s,m_i)\,|a^J_{\lambda_1\lambda_2\lambda_3\lambda_4}(s)|^2\nn\\
=&\,\Gamma(s,m_i)\sum_{\lambda_a,\lambda_b,\lambda_c,\lambda_d} a^J_{\lambda_1\lambda_2\lambda_a\lambda_b}(s) a^{J*}_{\lambda_c\lambda_d\lambda_3\lambda_4}(s) \,, \label{unitarity}
\end{align}
where the $\Gamma (s,m_i)$ factor accounts for the phase space integration. 

The relation presented above will be of great importance for this Thesis. As it can be seen in the second line of \eqref{unitarity}, the unitarization condition of the partial wave amplitude for a given helicity state involves not just the amplitudes of that specific combination of helicities, but many others. Thus, the implementation of the unitarity criterium has to be performed taking into account the whole coupled system of helicity amplitudes. This statement will be very relevant in the discussion developed in Chapter~\ref{Methods}, since, in general, this coupled analysis is neglected in the literature.

Coming back to the unitarization condition, by parameterizing the partial wave amplitude through its modulus, and its complex phase, $\varphi$,  one can extract the maximum value of the partial wave modulus that allows to have an elastic S matrix which is unitary:
\begin{align}
\Big| a^J_{\lambda_1\lambda_2\lambda_3\lambda_4}(s)\Big| \sin(\varphi)=\Gamma(s,m_i) \Big| a^J_{\lambda_1\lambda_2\lambda_3\lambda_4}(s) \Big| ^2\,.\label{cotaiso}\end{align}
Besides, at high energies, where $s \gg m_i^2$, so that $\Gamma(s,m_i)\sim 1$, the preceding expression turns into the very simple result:
\begin{align}
\label{cota2}\Big| a^J_{\lambda_1\lambda_2\lambda_3\lambda_4}(s)\Big|  \leq 1\,.
\end{align}
This unitarity condition implemented at the partial wave level defines a circle in the complex plane, the Argand circle, in which all partial waves have to lie in order to ensure the unitarity of the S matrix. It defines as well the unitarity violation scale, $\Lambda_{\rm U}$, that corresponds to the value of the center of mass energy at which the modulus of a given partial wave becomes one:
\begin{align}
\Big| a^J_{\lambda_1\lambda_2\lambda_3\lambda_4}(\Lambda_{\rm U}^2)\Big|  = 1\,.\label{unitbound}
\end{align}
Furthermore, it can be directly translated into a physical observable, such as the cross section of a determined process, which has to fulfil some requirements in order to ensure the conservation of probability as well.  Using the optical theorem, that relates the total cross section with the scattering amplitude in the forward direction, $\theta =0$, and using the unitarity relations given above, one arrives at the Froissart condition \cite{PhysRev.123.1053}:
\begin{eqnarray}
& A(s,\cos\theta=1) < \mathcal{O}\big(s ~(\log~s)^2\big), \nn\\
& \sigma \sim \dfrac{1}{s} \, {\rm Im}\big[A(s,\cos\theta=1)\big]~ \Rightarrow~ \sigma < {\rm const.}~(\log~s)^2\,,\label{froi}
\end{eqnarray}
which establishes an upper bound for the growth of the cross section with energy. It is clear that, as many interactions coming from the EChL involve derivatives and, therefore, momenta, when taking matrix elements, the observables related to these interactions will grow with energy. Therefore, the predictions from this EFT, mainly for longitudinally polarized vector boson scattering, will violate unitarity at some energy scale, regarded that the corresponding operator coefficients are not 0. This is a typical feature of chiral Lagrangians, and will be vastly exemplified and treated in posterior parts of this Thesis.

\subsection{Restoring Unitarity: Unitarization Methods}

Unitarity violation can be a serious problem when trying to obtain reliable predictions for certain observables with the EChL. As we have seen, the operators contained in the Lagrangian scale with the external momentum of the GBs, so the corresponding cross sections grow anomalously with energy violating the Froissart condition at a given point. This is, however, incompatible with the basic principles of quantum field theory, and non-unitarity predictions cannot be trusted to test them against experimental data.

A possible solution to this unitarity violation problem, that will most importantly arise in the scattering probabilities of longitudinal EW gauge bosons, is to choose a particular UV complete theory that has to be unitary by construction. For instance, in the pure SM limit of the EChL, the unitarity violation of the VBS amplitudes is cured by the light Higgs interactions~\cite{Lee:1977yc,Lee:1977eg,Romao:2016ien}. Taking as an illustrative example the elastic scattering of a W and a Z, WZ$\to$WZ, and studying each diagram contribution to the total cross section in the longitudinal case, it is easy to see how the energy cancellations take place among different diagrams leading to a total unitary result.

There exist four diagrams contributing to this scattering at the tree level in the SM: a contact term, an $s$ and $u$ channels with a W exchange and a $t$-channel with a Higgs exchange. Here $s$, $t$ and $u$ are the usual Mandelstam variables, defined as
\begin{align}
s=(p_{{\rm W}_i}+p_{{\rm Z}_i})^2\,,~~~t=(p_{{\rm W}_i}-p_{{\rm W}_f})^2\,,~~~u=(p_{{\rm W}_i}-p_{{\rm Z}_f})^2\,,
\end{align}
with $p$ being the external particle momenta and where $i$ and $f$ denote initial and final, respectively.

 \begin{figure}[t!]
\begin{center}
\includegraphics[width=.49\textwidth]{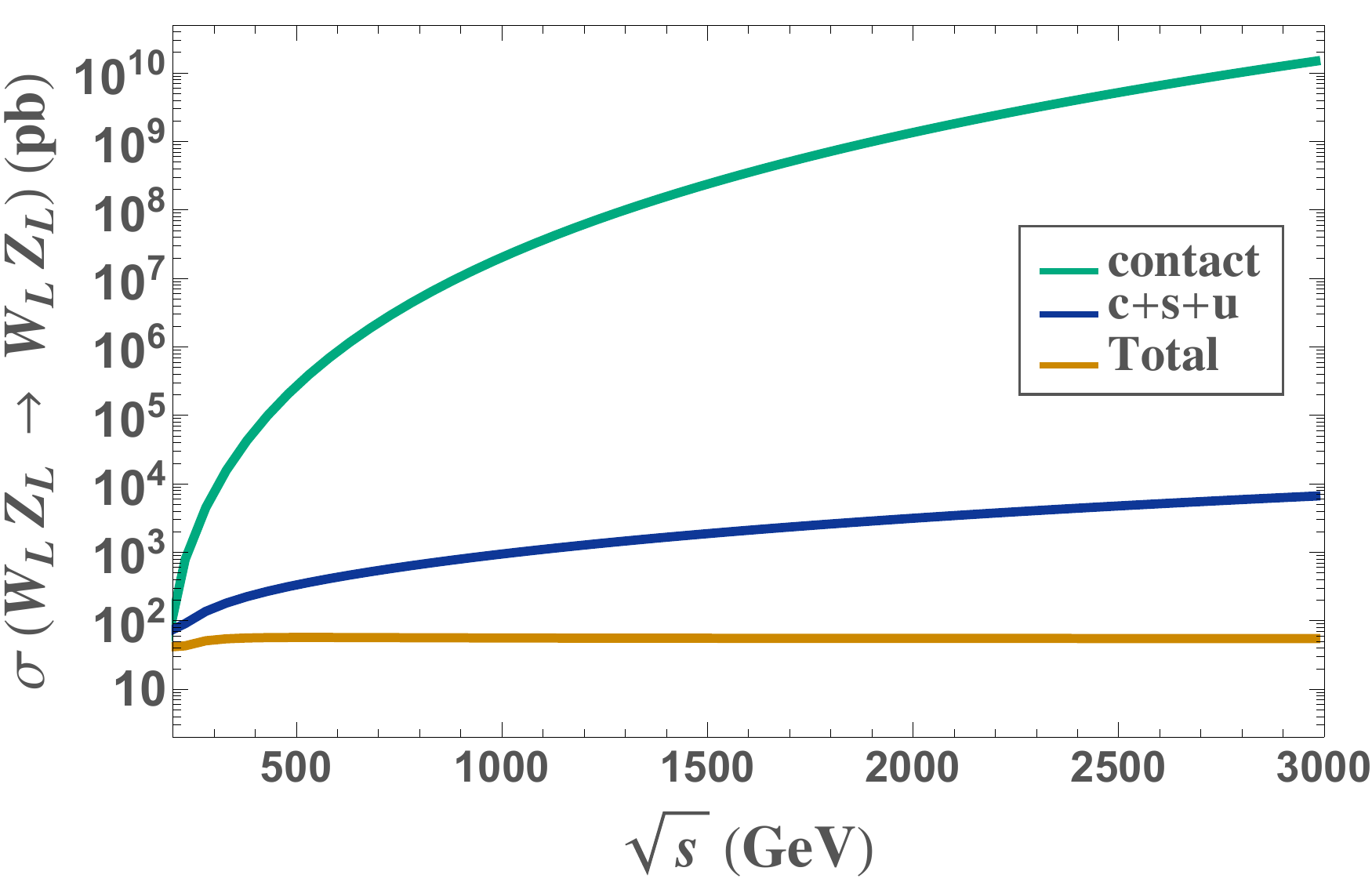}
\includegraphics[width=.49\textwidth]{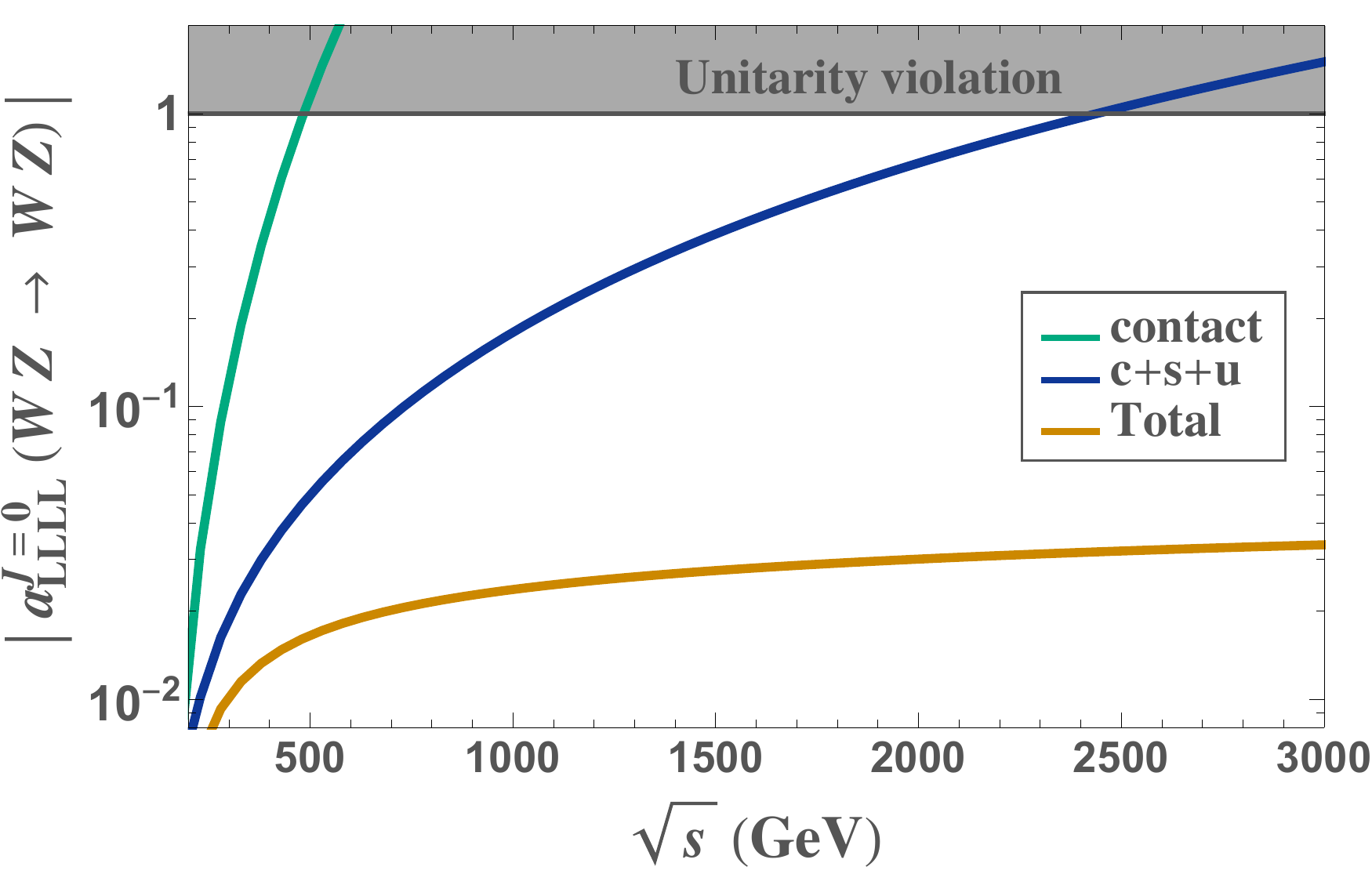}
\caption{SM cross section of the longitudinally polarized WZ$\to$WZ scattering (left) and modulus of the analogous $J=0$ partial wave (right) as a function of the center of mass energy $\sqrt{s}$. Contact, contact (called $c$ for shortness) plus $s$ and $u$ ($c+s+u$), and total ($c+s+t+u$) contributions are displayed in each case, as well as the unitarity violation limit (only right panel). This constitutes an example of the restoration of unitarity in the SM.}
\label{fig:Higgsneed}
\end{center}
\end{figure}

Each of the amplitudes of these diagrams behaves differently with energy. The contact term grows as $s^2$. The $s$ and $u$ channels, due to the gauge invariance of the theory, are such that their contribution exactly cancels the $s^2$ factors from the contact term. This is already key for the S matrix to be unitary, albeit not enough, since the $s$ and $u$ channels leave a remnant linear dependence on $s$ after the cancellation of the $s^2$ terms has taken place. Nevertheless, when the Higgs boson diagram, corresponding to the $t$ channel configuration, is introduced, also this linear term in the energy squared is cancelled. This {\it magic} restores completely the unitarity\footnote{The addition of the Higgs diagrams leads to a constant term with $s$ proportional to $m_H$, and, thus, it lead to unitary observables provided that H is light enough, particularly, below the TeV.} of the scattering amplitudes at all energies. 

The translation from unitarity at the amplitude level to the unitarity of the cross sections is a bit less intuitive to see, since the angular integration of the absolute value squared of the sum of all the diagrams has to be performed. However, one can still track the $s$ dependence of each of the contributions to the cross section to convince oneself that VBS cross sections saturate, in general, to a constant value at high energies and that they are, therefore, unitary. 

Once the 1/$s$ phase space factor has been taken into account, the $s^3$ and $s^2$ terms of the sum of the contact (called $c$ for shortness), $s$ and $u$ diagramas at the cross section level cancel out. This leads to a $c+s+u$ cross section contribution that scales as $s$ plus a constant. Nevertheless, when inspecting the analytical expression of the total cross section, i.e., including all four diagrams, the final result at large energies does not shown any dependence on  $s$. Thus, the unitarity of the cross section is manifest at all energies.

 In this \figref{fig:Higgsneed} the cross section of the longitudinally polarized WZ$\to$WZ scattering and the modulus of the analogous $J=0$ partial wave are displayed as a function of the center of mass energy $\sqrt{s}$. Three curves are shown on each panel, corresponding to the $c$, $c+s+u$, and total ($c+s+t+u$) contributions. In the left panel the cancellations amongst each of the contributing diagrams are manifest, leading to a total cross section that behaves flatly with energy as we have already commented. In the right panel, the consequences of these cancellations in terms of the partial wave modulus can be analyzed. It is clear that, without the consistent inclusion of the Higgs diagram in the SM scenario, unitarity is violated in VBS processes at the few TeV scale.

The previous result supposes a very simple example of how choosing a concrete UV completion for the EChL solves the unitarity violation problem. There are many other examples in this regard, such as strongly interacting scenarios in which heavy resonances dynamically generated by the strong dynamics emerge curing the violation of unitarity at high energies. The MCHM or the now disfavoured Technicolor models~\cite{Weinberg:1975gm,Susskind:1978ms,Dimopoulos:1981xc} correspond to this kind of strongly interacting UV theories.

However, the appeal of using the EChL as our main tool to study possible BSM physics in the EWSB sector lies in its generality, in its model independence. Therefore, we need a prescription to solve the issue of unitarity violation without relying upon choosing a particular UV completion. To this aim, unitarization methods are often addressed to construct unitary scattering amplitudes from the raw, non-unitary, EFT predictions. In the following pages we will introduce and discuss some of these methods that will be relevant for the results presented in Chapters \ref{Methods}, \ref{Resonances} and \ref{ResonancesWW}. Nonetheless, before entering in the specific details of these unitarization methods, some general considerations have to be commented.

First of all, it is important to have in mind that relying on a particular unitarization method implies to make some assumptions about  the behaviour of the scattering amplitudes at higher energies than the typical scale energies controlling the EFT expansion ($4 \pi v$ in the EChL case). In consequence,  choosing a concrete method would somehow be equivalent to make some assumptions about the UV complete theory. There is therefore a trade between obtaining unitary predictions for observables and losing some of the model independence inherent to EFTs.

Nevertheless, there is a caveat in this statement. When the EFT includes by construction the presence of resonant heavy states in the spectrum the various unitarization methods for VBS usually provide comparable results, since the main features of the resonances (mass and width) are present in all cases. However, when the resonances are instead generated dynamically by the unitarization method itself this is not the case anymore and the results may vary substantially from one method to another. Nevertheless, it is important to notice that if the unitarization methods provide amplitudes with the proper analytical structure, they can all accommodate dynamically generated resonances as poles in the unitarized amplitudes whose mass and width are predicted to be more or less the same independently of the employed method, when applicable~\cite{DelgadoLopez:2017ugq}. This means that in the resonant scenario, the model independence of the EFT prediction, which is in principle distorted by the unitarization method, is, in practice, achieved. The resonant scenario will be deeply studied in Chapters \ref{Resonances} and \ref{ResonancesWW}, where estimates of the LHC sensitivity to vector resonances generated dynamically by a strongly interacting EWSB system are provided with the help of different techniques.

Furthermore, in the case of non-resonant scenarios, i.e., when looking for smooth deviations from the SM continuum,  different unitarization methods can lead to outstandingly different predictions for diverse observables. This suggests that, in order not to lose the appealing model independence of EFTs in the non-resonant case, the predictions given from the different unitarization methods available have to be contrasted. This inevitably introduces a theoretical uncertainty in the unitarized EFT predictions, whose quantitative estimation is precisely what will be pursued in Chapter \ref{Methods}.

Second of all, all unitarization schemes have to provide similar predictions in the low energy region. This is a well known feature in the context of ChPT, where the scattering amplitudes from the chiral Lagrangian, unitarized with the various methods, do recover the ChPT prediction at low energies as a consequence of the well known low-energy theorems. This is of outstanding importance for the consistency of the resulting predictions.

Having stated all these considerations, we proceed to briefly explain the unitarization methods that we are going to consider in the present Thesis. We have selected them based on the fact that they are the most commonly employed ones in the literature. They can be classified in two categories: 1) the ones that directly suppress {\it ad-hoc} the pathological energy behaviour of the amplitudes with energy (that we call here, as it is usual in the literature, Cut off, Form Factor~\cite{Baur:1988qt,Figy:2004pt,Hankele:2006ma} and Kink~\cite{Rauch:2016pai}), and  2) the ones that unitarize the first three partial waves ($J=0,1,2$) from which then the unitary total amplitude is reconstructed (K-matrix~\cite{Schwinger:1948yk,Berger:1991uj,Gupta:1993tq,Chanowitz:1999se,Alboteanu:2008my,Kilian:2014zja} and Inverse Amplitude Method~\cite{Truong:1988zp,Truong:1991gv,Dobado:1989qm, Dobado:1992ha,Hannah:1995si,Dobado:1996ps,Oller:1997ng,Dobado:1999xb,GomezNicola:2001as,Espriu:2012ih, Espriu:2013fia,Delgado:2013loa, Delgado:2013hxa, Delgado:2014dxa, Espriu:2014jya, Arnan:2015csa, Dobado:2015hha, Corbett:2015lfa,BuarqueFranzosi:2017prc} (IAM)). 

The reason why only the three lowest order partial waves are unitarized can be understood by means of the ET. In the scattering involving just scalars in the external legs, since the amplitude corresponds to a polynomial expansion in energy up to order $s^2$, once one computes with the EChL at order $\mO(p^4)$, all partial waves with $J>2$ project to 0. Thus, the unitarity violation arising from the strongly interacting character of the interactions among scalars must be encoded in just the three mentioned partial waves even if we consider full gauge bosons in the external legs of our computations.

The various unitarization methods considered in the present Thesis differ in their physical implications and motivation, and in their analytical properties, that we will discuss in the next paragraphs. Despite these differences, and the fact that some of them could be more physically justified than others, there is not a clear prior to choose a particular method with respect to the others. 

We now list the five unitarization prescriptions considered in  this Thesis with a brief explanation of each of them.
Nevertheless, other procedures, such as the improved K-matrix~\cite{Delgado:2015kxa}, the T-matrix~\cite{Kilian:2014zja,Kilian:2015opv} or the N/D method~\cite{Truong:1991gv,Truong:1991ab}, are often used in the literature and are well motivated from the EFT point of view as well. For brevity, since they are not relevant for the work presented in this Thesis, we will omit them in this section. A complete description and comparison among unitarization methods can be found in~\cite{DelgadoLopez:2017ugq}.

\begin{itemize}
\item{\textbf{Cut off:}
The Cut off corresponds to a prescription that allows to obtain unitary amplitudes but it is not a unitarization method per se. It just consists in discarding those predictions given for energy values above the unitarity violation scale $\Lambda_{\rm U}$, defined as the lowest value of $\sqrt{s}$ at which any partial wave crosses the unitarity bound stated in \eqref{unitbound}.
}
\item{\textbf{Form Factor (FF):}
In this case what is done is to suppress the pathological behavior of the amplitudes with energy above the scale at which each of them violate unitarity. To that purpose, a function of the form~\cite{Baur:1988qt,Figy:2004pt,Hankele:2006ma}:
\begin{align}
f_i^{\rm FF}=(1+s/\Lambda_i^2)^{-\xi_i}\,,\label{eqFF}
\end{align}
is employed. In this expression $s$ corresponds to the center of mass energy squared, $\Lambda_i$ to the specific value of $\sqrt{s}$ at which the helicity channel $i$ violates unitarity according to \eqref{unitbound} and $\xi_i$ to the minimum exponent that is sufficient  to fix the energy behavior of the corresponding $i^{th}$ helicity amplitude. Thus, every non-unitary helicity amplitude will be unitarized in the following manner:
\begin{align}
\hat{A}_{\lambda_1,\lambda_2,\lambda_3,\lambda_4}=A_{\lambda_1,\lambda_2,\lambda_3,\lambda_4}\cdot(1+s/\Lambda_{\lambda_1,\lambda_2,\lambda_3,\lambda_4}^2)^{-\xi_{\lambda_1,\lambda_2,\lambda_3,\lambda_4}}\,,\label{FF}
\end{align}
with $\hat{A}$ being the unitary amplitude and $A$ the non-unitary EFT prediction.
}
\item{\textbf{Kink:}
The Kink method is conceptually the same as the Form Factor one. The only difference existing between both prescriptions is that the suppression in the Kink method is  performed with a step function~\cite{Rauch:2016pai}:
\begin{align}
f_i^{\rm Kink}=\left\{\begin{array}{l}1~~~~~~~~~~~~~~~~~~~s\leq\Lambda_i^2 \\(s/\Lambda_i^2)^{-\xi_i} ~~~~~ s>\Lambda_i^2\end{array}\right.\,.\label{eqKink}
\end{align}
However, the rest of the discussion regarding the Form Factor is equally valid for the Kink case.
}

\item{\textbf{K-matrix:}
This method is a prescription applied to the partial wave amplitudes that consists in projecting the non-unitary partial waves into the Argand circle through a stereographic projection. This means that a imaginary part is added {\it ad hoc} such that the unitarity limit is saturated. For each helicity partial wave, this is achieved by using the following expresion:
\begin{align}
\hat{a}^{J;{\rm K-matrix}}_{\lambda_1\lambda_2\lambda_3\lambda_4}=\dfrac{a^J_{\lambda_1\lambda_2\lambda_3\lambda_4}}{1-i\,a^J_{\lambda_1\lambda_2\lambda_3\lambda_4}}\,, \label{eqKmatrix}
\end{align}
that holds for each helicity amplitude at a time.
}
\item{\textbf{Inverse Amplitude Method (IAM):}
The Inverse Amplitude Method is very well known in the context of ChPT in QCD for effective descriptions of pion-pion scattering~\cite{Truong:1988zp, Dobado:1989qm, Dobado:1992ha, Hannah:1995si,Dobado:1996ps}, and its accuracy has been proved in various scenarios, like, for instance, in the prediction of the $\rho$ meson as an emergent resonance in these scattering processes. It is based on the application of dispersion relations (bidirectional mathematical prescriptions allowing to relate the real and imaginary parts of complex functions) to the inverse of the partial wave amplitudes computed in the EFT framework. This unitarization procedure can actually be understood as the result of the first Pad\'e approximant~\cite{Basdevant:1969rw,Basdevant:1969sz,Dicus:1990ew} derived from the chiral expansion series provided by ChPT. In practice, this method implements the re-summation of loops with bubble configurations in the s-channel of the given scattering process. Therefore, in the context of the EChL, it accounts for re-scattering effects that are not taken into account with the other unitarization methods. As we have extensively commented, this makes sense in the context of a strongly interacting theory since these re-scattering contributions are not suppressed as in weakly interacting systems. 

Within ChPT the GB scattering amplitudes are computed as a series expansion at different orders in momenta of the  external scalars, also at the partial wave level, 
\begin{equation}
a^{{J}}_{\lambda_1\lambda_2\lambda_3\lambda_4}(s)=a^{{(2)\,J}}_{\lambda_1\lambda_2\lambda_3\lambda_4}(s)+a^{{(4)\,J}}_{\lambda_1\lambda_2\lambda_3\lambda_4}(s)+\dots\label{expansion}
\end{equation}
where $a^{{(2)\,J}}_{\lambda_1\lambda_2\lambda_3\lambda_4}(s)$ accounts for the contributions that come from the terms in $\mL_2$ at tree level and that are of order $\mO(p^2)$, and $a^{{(4)\,J}}_{\lambda_1\lambda_2\lambda_3\lambda_4}(s)$ for the contributions of order $\mO(p^4)$, coming from $\mL_4$ at tree level and from $\mL_2$ at one loop level. Because of this, $a_{J}(s)$ will behave as a polynomial in $s$ that, when truncated at a given order, will lead to the violation of unitarity. Still, the unitarity condition holds perturbatively,
\begin{align}
{\rm Im}\Big[a_{J}^{(4)}(s)\Big]=\Gamma(s,m_i) \Big|a_{J}^{(2)}(s)\Big|^2\,.\label{unitpert}
\end{align}
Thus, within ChPT, one recovers unitarity perturbatively, meaning that the imaginary part of the NLO (i.e., the contribution of the chiral loops) in the chiral expansion unitarizes the LO part.

We wish, however, to go beyond perturbative unitarity by applying the IAM, which leads to the following fully unitarized helicity partial wave amplitudes:
\begin{align}
\hat{a}^{J;{\rm IAM}}_{\lambda_1\lambda_2\lambda_3\lambda_4}=\dfrac{\left(a^{{(2)\,J}}_{\lambda_1\lambda_2\lambda_3\lambda_4}\right)^2}{a^{(2)\,J}_{\lambda_1\lambda_2\lambda_3\lambda_4}-a^{(4)\,J}_{\lambda_1\lambda_2\lambda_3\lambda_4}}\,.\label{IAM}
\end{align}
This rather simple expression encodes an important property of the strongly interacting EW Goldstone bosons dynamics. It does not provide just unitary predictions, but also the appropriate analytical structure of the VBS amplitudes. This implies that it can accommodate dynamically generated resonances appearing as poles in the second Riemann sheet of the partial wave with the corresponding $J$ quantum number. Furthermore, it is worth commenting that, according to \cite{Delgado:2015kxa}, similar results as those obtained with the IAM regarding the appearance of dynamical resonances  are also provided by other alternative unitarization methods that lead to the proper analytical structure, such as the N/D or the improved K-matrix.    

This is in contrast to the unitarized partial waves with the (original) K-matrix method that do not have poles. These resonances, that appear naturally at high energies in strongly interacting theories, such as in the case of low-energy QCD, can be the key to disentangle the true nature of the EWSB sector. Therefore, the capability of the IAM to accommodate them consistently within the EFT predictions is of paramount importance. 

We have just outlined the main properties of this particular unitarization method in this section, since it will be specifically used in the following chapters in more depth, especially in Chapters \ref{Resonances} and \ref{ResonancesWW}. The pertinent further details regarding the IAM will be commented, therefore, in the forthcoming Chapters of this Thesis.
}
\end{itemize}

In this Chapter we have reviewed the generalities of spontaneous EW symmetry breaking and its possible linear and non-linear realizations. We have also introduced the effective electroweak chiral Lagrangian with a light Higgs boson in full detail, since it corresponds to the theoretical pilar of this Thesis. The Standard Model effective field theory has been briefly discussed as well, and so has its matching to the EChL. Finally, unitarity violation in the context of the effective description of EWSB has been revisited, together with some of the most usual procedures that allow to cure this pathology. 

Being the theoretical bases of this Thesis presented, we now move on to a more phenomenological Chapter, in which we will exhaustively study the interactions between EW gauge bosons, i.e., the vector boson scattering processes.


\chapter[\bfseries Vector boson scattering]{Vector boson scattering}\label{VBS}
\chaptermark{Vector boson scattering}

Ideally, the study of the EWSB sector would imply to scrutinise the EW Goldstone boson dynamics with exquisite detail. Nevertheless, as it has become clear in the previous Chapter, when a local symmetry is spontaneously broken these particles are unphysical, and, therefore, they cannot be produced directly at experiments. Being this so, how can we study EWSB efficiently? Fortunately, despite the absence of the Goldstone bosons in the spectrum, we can profit from the close relation existing between them and the EW gauge bosons. 

Thanks to the equivalence theorem, \eqref{ET}, it is known that, at energies well above the EW gauge boson masses, the Goldstone boson dynamics can be comprehended through the longitudinally polarized EW gauge boson interactions. Interestingly, this high energy region is the one in which the GBs would manifest their strongly interacting character, so it is, indeed, the one we should explore in our quest to understand the EWSB.

All this points towards the fact that, in order to study the EWSB sector, one must look at vector boson scattering processes, since they hide the heart of the strong interactions among scalars we are aiming to probe. In particular, the scattering of longitudinally polarized vector bosons should help us shed some light into this interesting issue. Furthermore, VBS processes are well characterized experimentally, and have even served to achieve very important measurements. They have also helped determining some of its fundamental properties at the LHC. 

For all these reasons, in this Thesis we will base our study on vector boson scattering observables to try to disentangle some relevant aspects of a possibly strongly interacting EWSB sector. To this purpose, in this Chapter we will revisit some of the most important features of these VBS configurations. 

Since we expect the relevant physical properties of these systems to appear already at the subprocess level, we will start by characterizing various interesting VBS processes with EW gauge bosons as external particles. First, we will study the SM predictions for these observables, as the SM itself will be unavoidably one of the irreducible backgrounds we will have to deal with once we move on to the LHC case.  Then we will analyze the properties of VBS in the EChL and we will compare them with the SM values to get an idea of the deviations we would expect to observe if the EChL were the correct description of the true EWSB nature. Afterwards we will explore the violation of unitarity in VBS processes and how it could affect our predictions and their interpretation when compared against experimental data. All this will be done paying special attention to the polarization of the initial and final gauge bosons, since our real interest will be to access the longitudinal components of these particles.

Nevertheless, since, nowadays, there is no W or Z collider, we will ultimately need to study the production of EW gauge bosons that then re-scatter to obtain the full VBS picture. Because of this, we will also present in this Chapter a review of the main properties of VBS processes at the LHC, since through these observables we will be able to test the EChL experimentally. Besides, the LHC characterization of VBS topologies will lead to a well known and very interesting result, which is that the VBS kinematical configurations are really special. Their genuine properties will allow us to disentangle them very efficiently from undesired backgrounds in the forthcoming Chapters of this Thesis, making these observables not only theoretically interesting due to their relation to the EWSB dynamics, but also experimentally due to their remarkable cleanliness. 

Finally, we will conclude this Chapter with a review of the most interesting and up to date VBS experimental searches that the LHC has performed, as well as the future prospects on the study of these observables at colliders.

\section{Scattering of Electroweak gauge bosons}

As it has been already stated, new physics appearing in the EWSB sector should be observable already at the subprocess level. Therefore, it is important to study the scattering of EW gauge bosons directly both in the SM as our starting point, and in the EChL to gain intuition on the deviations we should expect from the new interactions it describes. 

This section is devoted to this task: a profound characterization of VBS at the subprocess level in the SM and in the EChL.

\subsection{Vector boson scattering in the Standard Model}

The SM prediction of VBS observables represents our starting point as it will always correspond to our null hypothesis, i.e., the case in which there were neither new interactions nor new particles in the EWSB sector. Thus, it will always be, somehow, our irreducible background: the value one should compare against in order to determine if there is BSM physics around the corner.

\begin{figure}[t!]
\begin{center}
\includegraphics[width=\textwidth]{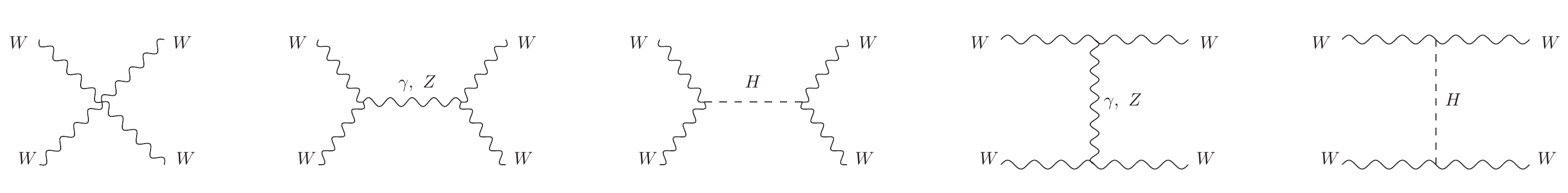}
\caption{SM diagrams contributing to the W${}^+$W${}^-$$\to$ W${}^+$W${}^-$ process in the Unitary gauge.}
\label{fig:diagramsWWWW}
\end{center}
\end{figure}

\begin{figure}[t!]
\begin{center}
\includegraphics[width=\textwidth]{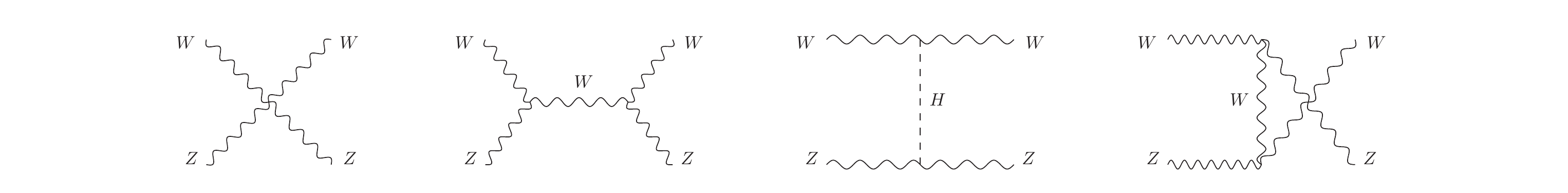}
\caption{SM diagrams contributing to the W${}^+$Z $\to$ W${}^+$Z process in the Unitary gauge.}
\label{fig:diagramsWZWZ}
\end{center}
\end{figure}

\begin{figure}[t!]
\begin{center}
\includegraphics[width=\textwidth]{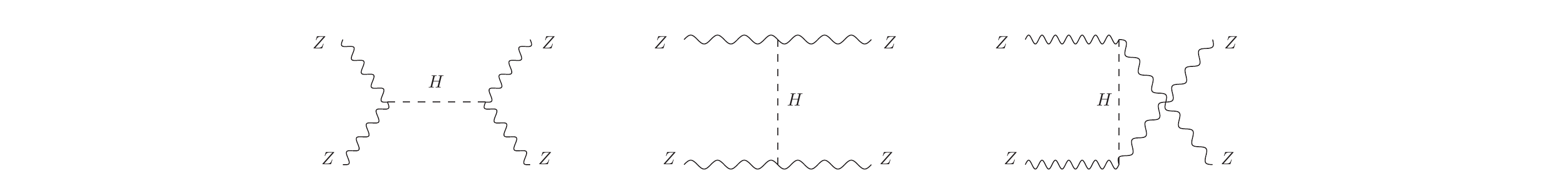}
\caption{SM diagrams contributing to the ZZ $\to$ ZZ process in the Unitary gauge.}
\label{fig:diagramsZZZZ}
\end{center}
\end{figure}

In this subsection we present the main SM features of VBS. We will start by comparing the behaviour of five different VBS channels W${}^+$W${}^+$$\to$ W${}^+$W${}^+$, W${}^+$W${}^-$$\to$ W${}^+$W${}^-$, W${}^+$W${}^-$$\to$ ZZ, W${}^+$Z $\to$ W${}^+$Z and  ZZ $\to$ ZZ as a function of different kinematical variables. We have not included the W${}^-$W${}^-$$\to$ W${}^-$W${}^-$ nor the W${}^-$Z$\to$ W${}^-$Z, since, at the subprocess level, they lead to the same results as their positively charged counterparts due to CP invariance of EW gauge boson self-interactions. The Feynman diagrams that contribute to the W${}^+$W${}^-$$\to$ W${}^+$W${}^-$, W${}^+$Z $\to$ W${}^+$Z and ZZ $\to$ ZZ are displayed in Figs.~\ref{fig:diagramsWWWW}-\ref{fig:diagramsZZZZ}. The rest can be obtained by crossing symmetry.

\begin{figure}[t!]
\begin{center}
\includegraphics[width=.49\textwidth]{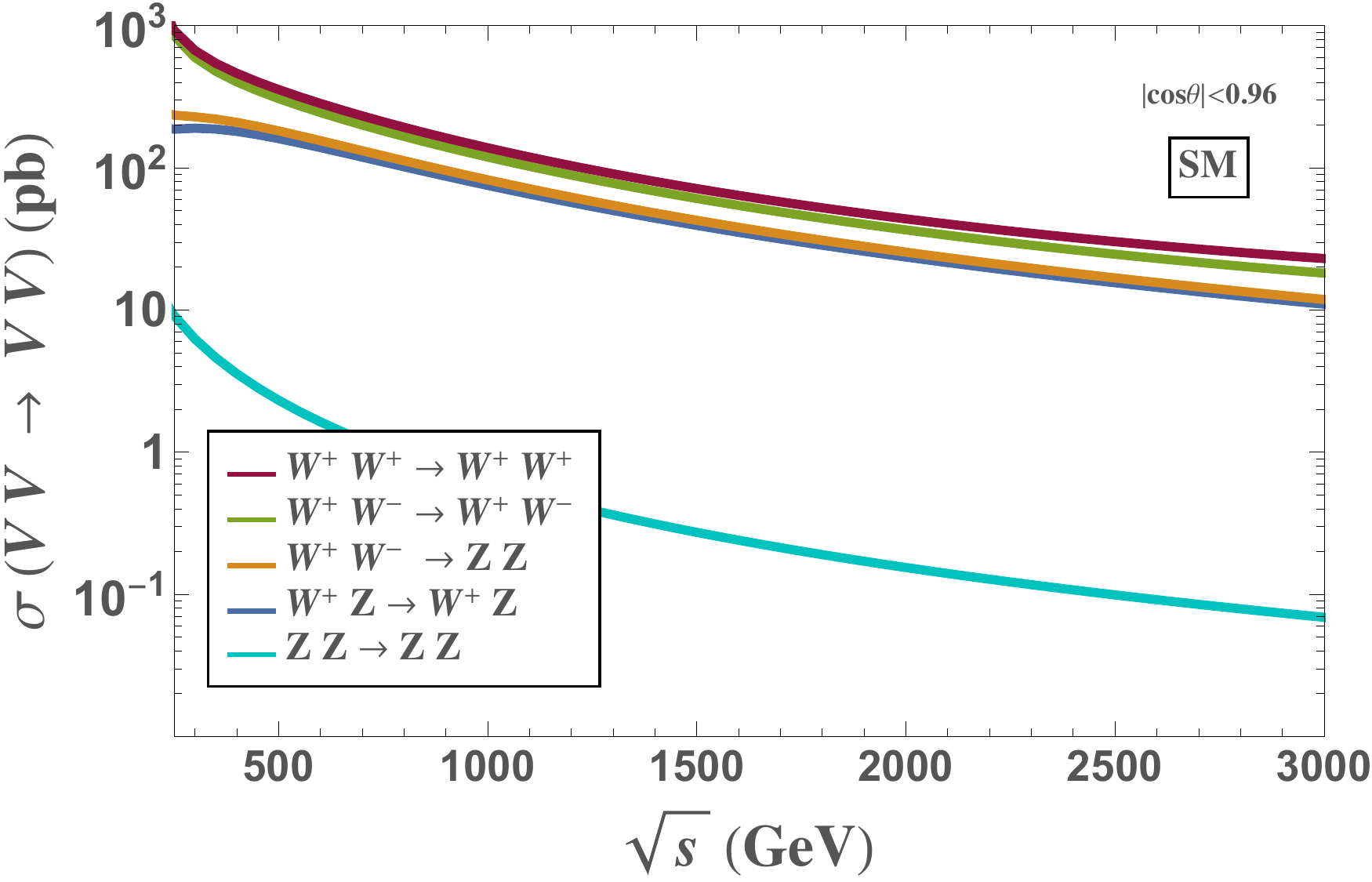}
\includegraphics[width=.49\textwidth]{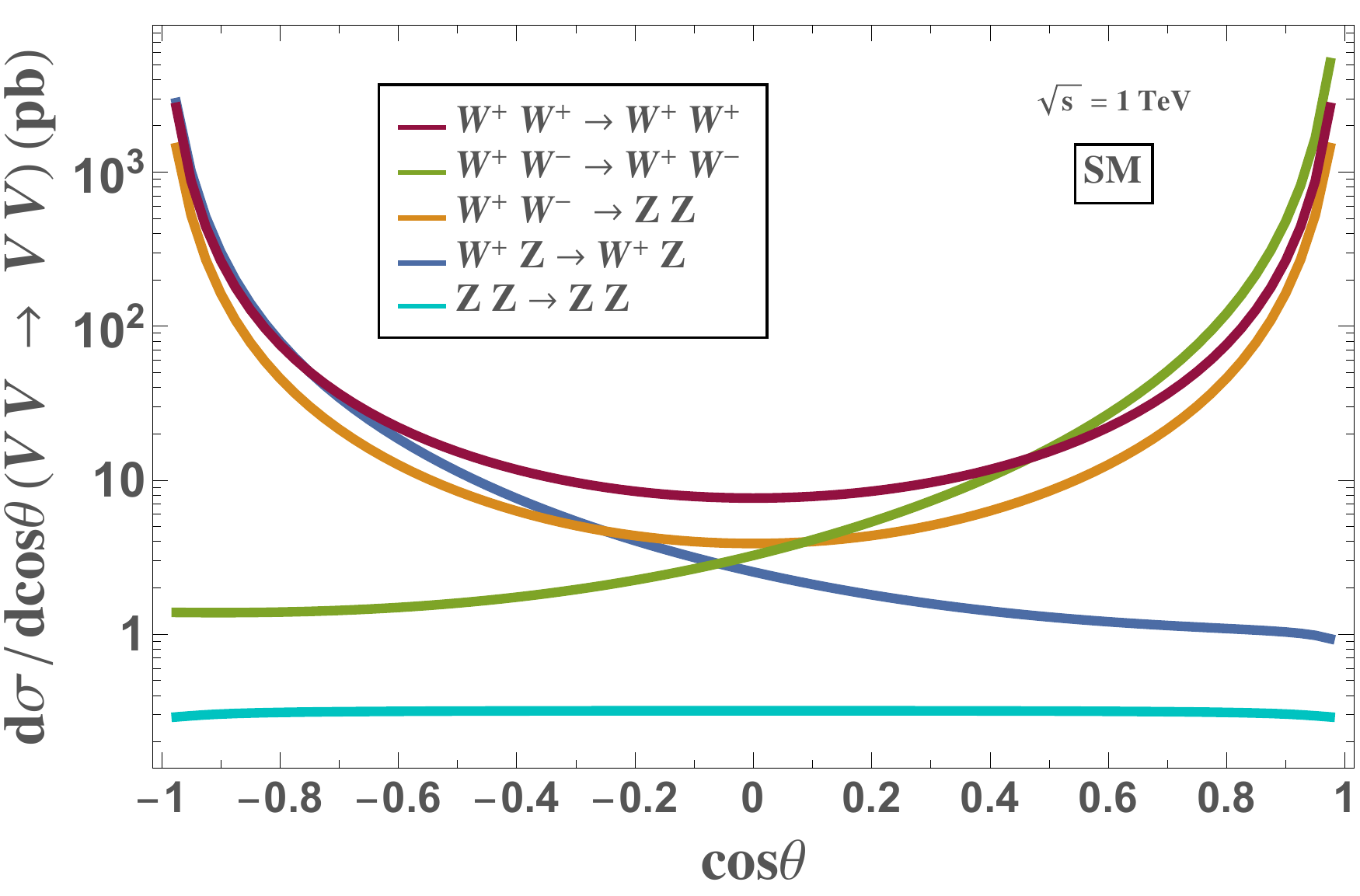}
\caption{SM predictions for the cross sections of five representative VBS processes: W${}^+$W${}^+$$\to$ W${}^+$W${}^+$, W${}^+$W${}^-$$\to$ W${}^+$W${}^-$, W${}^+$W${}^-$$\to$ ZZ, W${}^+$Z $\to$ W${}^+$Z and  ZZ $\to$ ZZ. Left panel shows the total cross section as a function of the center of mass energy integrated in $|\cos\theta|\leq 0.96$, with $\theta$ being the scattering angle as defined in the text. Right panel shows the differential cross section as a function of the cosine of the scattering angle for a fixed energy of  $\sqrt{s}=$1 TeV. }
\label{fig:VBSSM1}
\end{center}
\end{figure}

In \figref{fig:VBSSM1} we present the unpolarized (i.e., adding consistently all the possible polarization states for initial and final gauge bosons and averaging over the initial states) SM cross sections of the five mentioned processes with respect to the two main kinematical variables: center of mass energy $\sqrt{s}$ (left) and scattering angle $\theta$ (right). The latter is defined as the angle formed by the incoming W${}^+$ and the outgoing W${}^+$ or Z, except in the W${}^+$W${}^+$$\to$ W${}^+$W${}^+$ and ZZ $\to$ ZZ cases, where the assignment is made between one of the incoming particles and one of the outgoing particles, since they are all identical.

Very important features of the VBS processes can be extracted from this Figure. First of all, regarding the left panel, and thus the behaviour of VBS observables with the center of mass energy, it is clear that, at high energies, all five channels depend similarly on the energy. It is important to mention that a kinematical cut has been set in the phase space integration, requiring $|\cos\theta|\leq 0.96$, to cure the Coulomb singularity present in some of these processes. This cut is responsible of the fact that cross sections fall slightly as the energy increases. In the cases in which the cross section can be computed safely integrating in all possible angular values, the predictions show a flat behaviour with energy, which is very characteristic of the VBS configurations, as it was already the case in \figref{fig:Higgsneed}. Second of all, although the energy behavior is shared between all the studied processes, the cross section value at a given energy is very different among them. The ZZ$\to$ ZZ case results to be, in general, two orders of magnitude smaller than the others, which differ in less than a factor 2 at high energies.

In the right panel of \figref{fig:VBSSM1} we can see the shape of the differential cross sections as a function of the cosine of the scattering angle for a fixed energy value of  $\sqrt{s}=$1 TeV as a reference. Here, the differences between the various channels are clear.  The ones having identical particles in the final state, W${}^+$W${}^+$$\to$ W${}^+$W${}^+$, W${}^+$W${}^-$$\to$ ZZ and ZZ $\to$ ZZ, are, as expected, symmetrical. The former two peak at $\cos\theta\sim\pm1$ due to the exchange of a vector boson in the $t$ and $u$ channels, related to the scattering angle:
\begin{align}
t=\frac{(\sum m_i^2-s)}{2}(1-\cos\theta)\,,~~~~~u=\frac{(\sum m_i^2-s)}{2}(1+\cos\theta)\,,
\end{align}
where $m_i$ stands for any initial or final particle mass.  The latter ZZ $\to$ ZZ process, however, seems to be nearly independent of the scattering angle as only scalars are exchanged in internal legs.

The remaining two, W${}^+$W${}^-$$\to$ W${}^+$W${}^-$ and W${}^+$Z $\to$ W${}^+$Z are highly asymmetrical: the one involving no Z bosons becomes larger in the forward direction ($\cos\theta\sim1$),  whereas the other does it in the backward direction ($\cos\theta\sim-1$). This can be easily understood by checking the corresponding diagrams contributing to each of them, since in the W${}^+$W${}^-$$\to$ W${}^+$W${}^-$ case a vector boson is exchanged in the $t$ channel and a scalar boson in the $u$ channel. The opposite happens in the W${}^+$Z $\to$ W${}^+$Z process.

Now that the energy and angular behaviour of the SM predictions are settled, we can move on to study the characteristics of the cross sections involving different polarizations of the final and initial gauge bosons. Since each massive vector boson has three possible polarization states, the total number of helicity amplitudes corresponds to 81 (3$\times$3 initial states times 3$\times$3 final states). However, if we group all the transverse modes into a single category, and consider the LT state together with the TL one, defining AB$={\rm V}_{\rm A}{\rm V}_{\rm B}$ and LT$=$(LT$+$TL), this quantity reduces to 9 possibilities: LL $\to$ LL, LL $\to$ LT, LL $\to$ TT, LT $\to$ LL, LT $\to$ LT, LT $\to$ TT, TT $\to$ LL, TT $\to$ LT and TT $\to$ TT.

\begin{figure}[t!]
\begin{center}
\includegraphics[width=.49\textwidth]{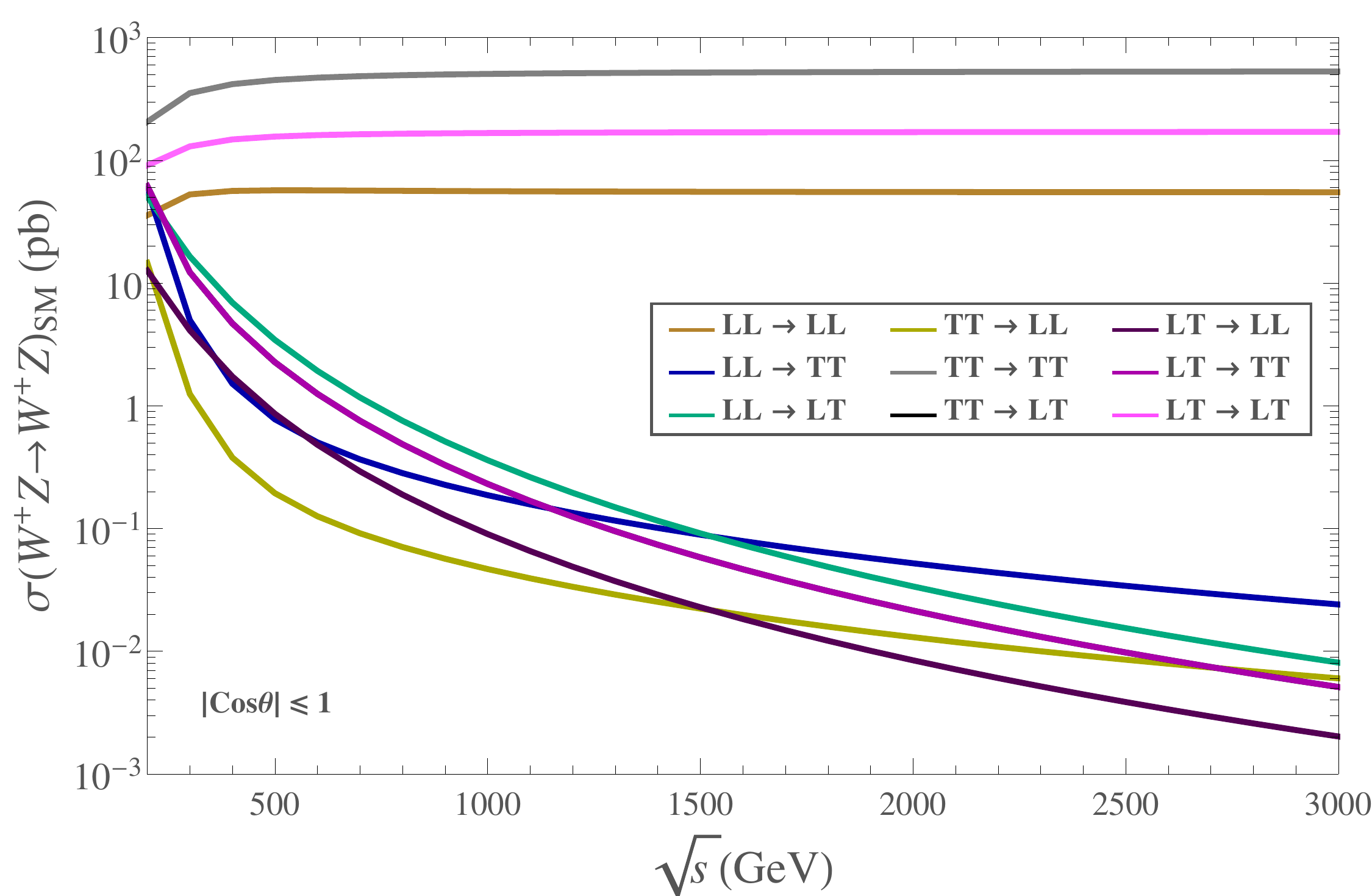}
\includegraphics[width=.49\textwidth]{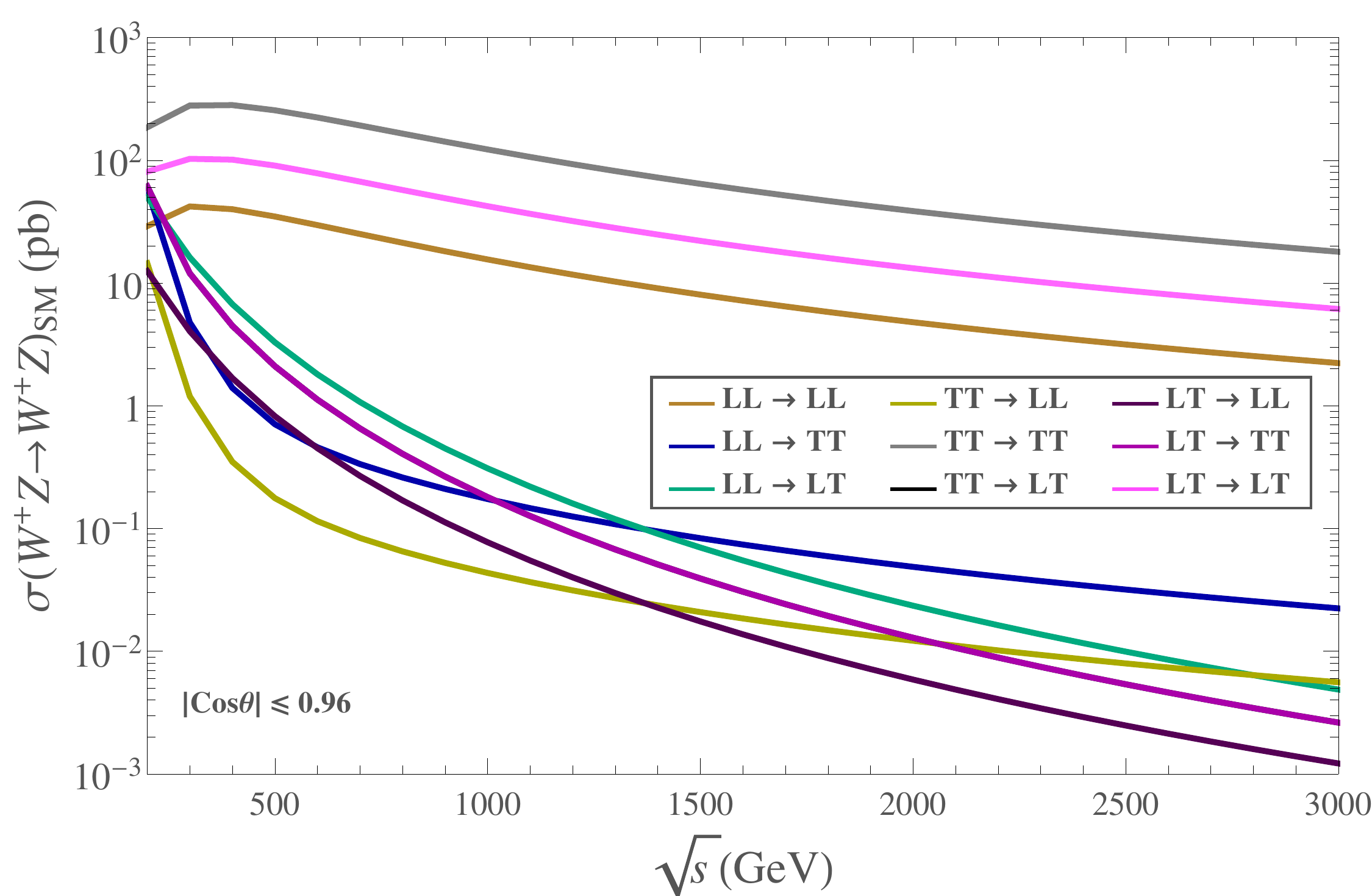}
\caption{Predictions of the SM cross section as a function of the center of mass energy of the process W${}^+$Z $\to$W${}^+$Z for different polarizations of the initial and final  bosons. 
We display the different polarization cross sections integrated in two choices of the center of mass scattering angle, $|\cos\theta|\leq 1$ (left panel) and $|\cos\theta|\leq 0.96$ (right panel), corresponding the latter to $|\eta_{W,Z}|\leq2$.}
\label{fig:VBSSMpols}
\end{center}
\end{figure}

The W${}^+$Z $\to$ W${}^+$Z scattering cross sections corresponding to the different helicity states as a function of the center of mass energy are displayed in \figref{fig:VBSSMpols}, integrating in the whole available phase space (left) and with the same kinematical cut imposed in  \figref{fig:VBSSM1} (right), for comparison. We have selected a representative example among the ones shown in the previous results to illustrate the behaviour of the different polarizations involved in the scattering. The  W${}^+$Z $\to$ W${}^+$Z process serves very well to this purpose, and will be of great relevance in subsequent parts of this Thesis, so we will choose it as our main example in several occasions from now on.

An outstanding conclusion follows from  \figref{fig:VBSSMpols}: the polarization modes of the gauge bosons (meaning longitudinal or transverse) are, to a good approximation, conserved in VBS processes. It is clear that the three amplitudes that preserve these polarization modes, LL $\to$ LL, LT $\to$ LT and TT $\to$ TT, are, especially at high energies, more than three orders of magnitude larger than the others, so they will clearly dominate in the total, unpolarized cross section. Thus, it is fair to say, that if the polarization of the final gauge bosons was measured, it will most likely come from the scattering of two gauge bosons with the same polarization mode. This is a remarkable result, since obtaining a measurement of a longitudinally polarized diboson system would imply to probe directly the heart of the interactions among scalars.

Beyond the conservation of the polarization state, from these images it can also be inferred that the purely transverse scattering dominates, followed by the mixed transverse-longitudinal and by the purely longitudinal one, respectively. In fact, there is roughly an order of magnitude difference between the purely transverse and the purely longitudinal contributions, being the mixed one between both of them. This indicates that the main Goldstone boson interactions will be in general slightly suppressed in VBS since the transverse modes tend to dominate the total cross sections, so tiny deviations in the longitudinally polarized gauge bosons will might not be visible as deviations in the total cross sections. However, sizable deviations, such as the ones currently allowed experimentally in the EChL case, might be observed, as we will see in the forthcoming pages.

The last important comment regarding this figure concerns the typical behaviour of the VBS cross sections with energy commented before. As we have said, when the phase space integration is performed in the whole range of the scattering angle a flat energy dependence is obtained, at least for the polarization-conserving cross sections that then dominate the unpolarized, total one. This feature can be indeed seen in the left panel of  \figref{fig:VBSSMpols}. Nevertheless, when a cut is imposed on the scattering angle, they tend to fall mildly with energy (right panel of \figref{fig:VBSSMpols}). This feature will be relevant in order to identify VBS topologies against possible backgrounds. 

We conclude this subsection with the partial wave analysis of the unpolarized SM  W${}^+$Z $\to$ W${}^+$Z cross section as a reference. Recalling \eqref{pwexp}, every helicity amplitude can be expanded in the base of angular momentum. The coefficients of this expansion are the partial waves of each helicity configuration, and they correspond to different values of the angular momentum $J$. Thus, according to the given expression, the total cross section can be reconstructed from the values of the partial waves.

\begin{figure}[t!]
\begin{center}
\includegraphics[width=0.49\textwidth]{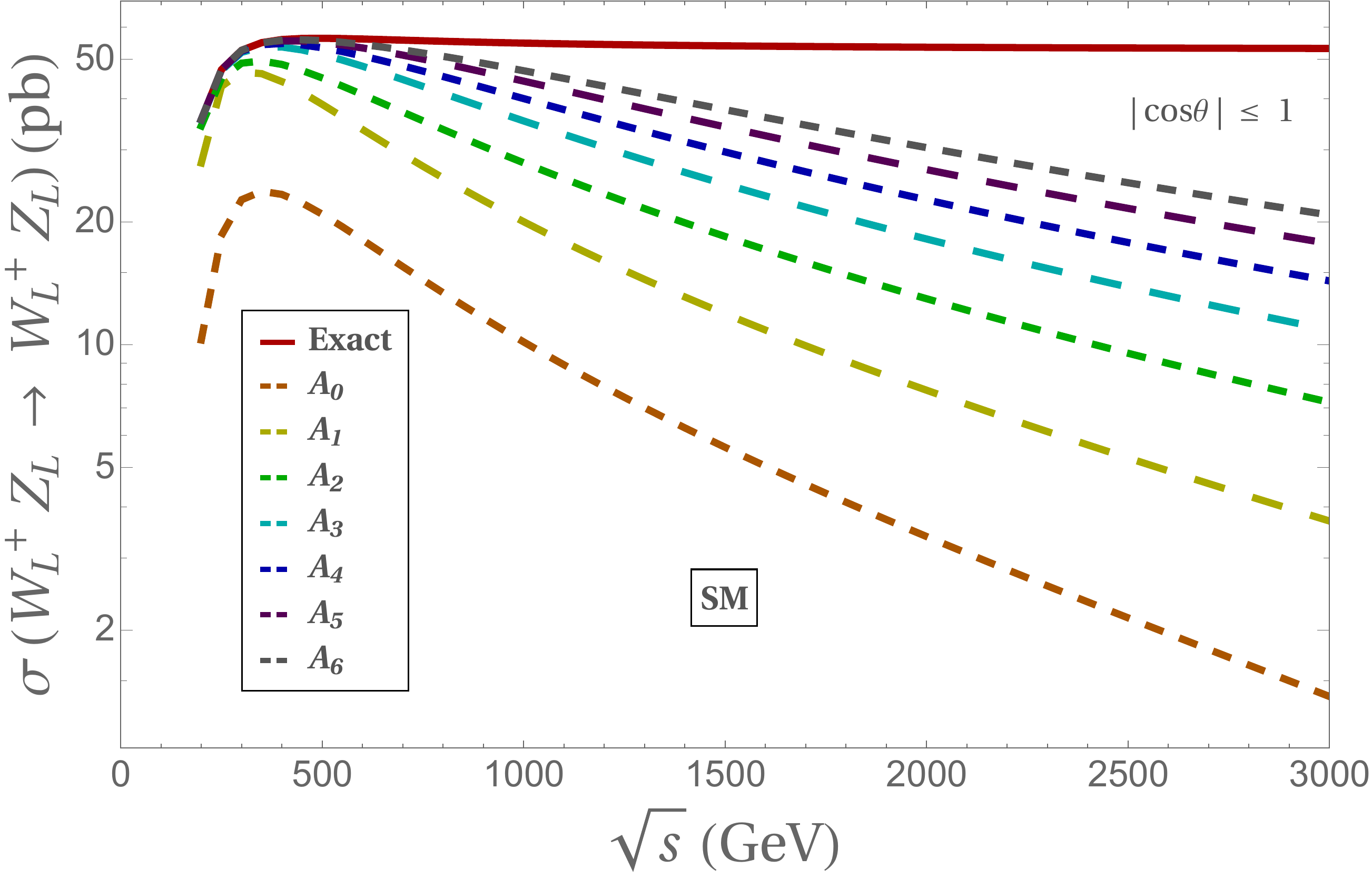}
\includegraphics[width=0.49\textwidth]{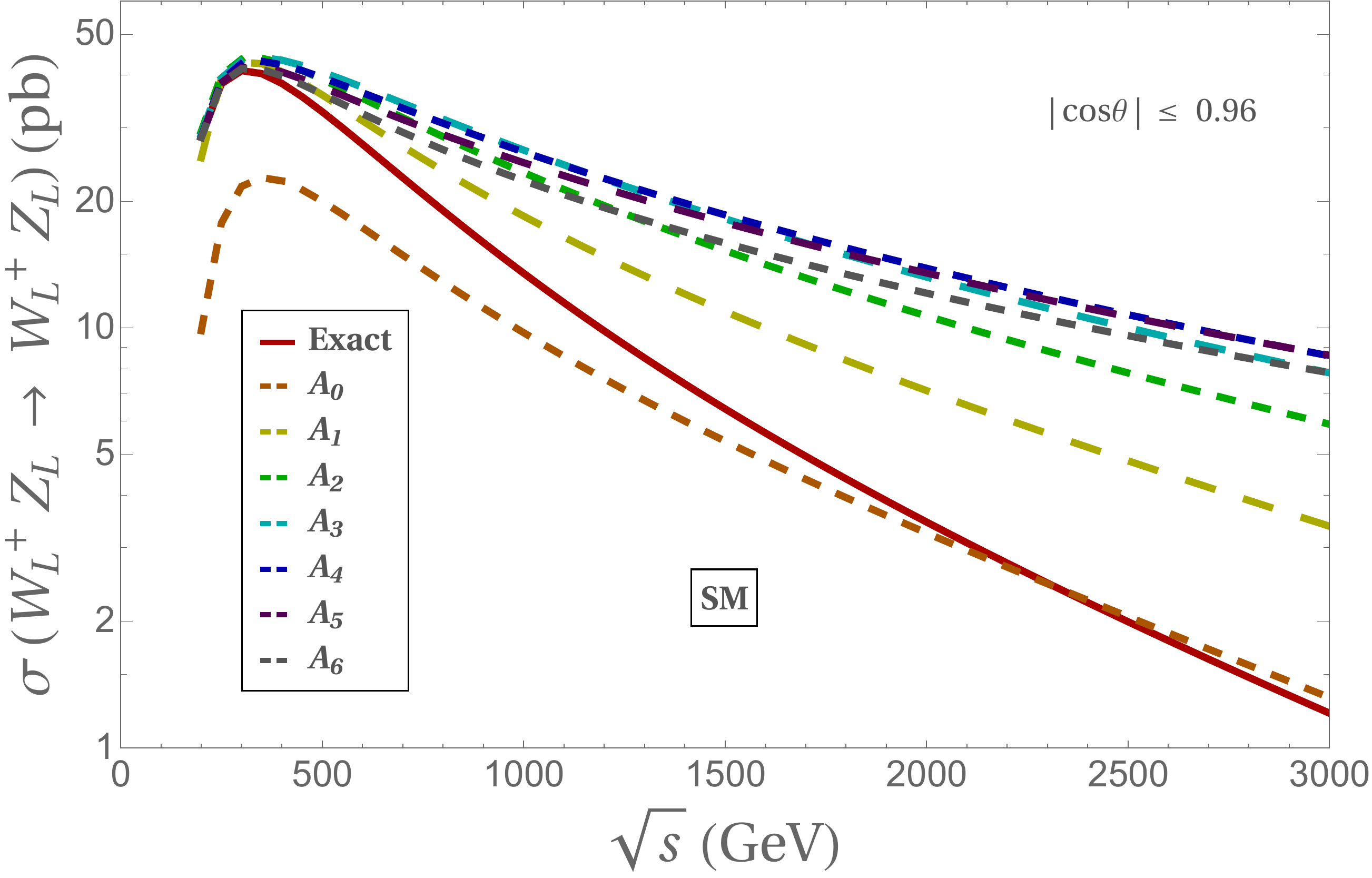}\\
\caption{ Cross section as a function of the center of mass energy of the process W${}^+_L$ Z${}_L\to$ W${}^+_L$ Z${}_L$ computed in the SM integrating in the whole scattering angle (left panel) and with a kinematical cut: $|\cos\theta|\leq0.96$ (right panel). Solid red line corresponds to the cross section computed directly from the original scattering amplitude. Dashed lines show the cross section reconstructed using Eq.~(\ref{pwexp}) truncating the series at different values of the angular momentum $J$. }
\label{fig:pwExcompSM}
\end{center}
\end{figure}

It is customary to find in the literature that it is possible to obtain the total cross section with moderate accuracy by cutting the series at $J=2$. As it has already been commented in this Thesis, this is the exact case in the equivalence theorem, when only scalars are considered in the cross section computations, since all the higher order terms, i.e., $J>2$, project to 0.  However, this assumption fails when considering full gauge bosons in the external legs, as it is shown in  \figref{fig:pwExcompSM}. There, the purely longitudinal SM W${}^+$Z $\to$ W${}^+$Z cross section is displayed as a solid, red line, again integrating in two choices of the scattering angle, $|\cos\theta|\leq 1$ and $|\cos\theta|\leq 0.96$. The dashed lines represent the cross section reconstructed using Eq.(\ref{pwexp}) truncating the partial wave series at different values of the angular momentum $J$. It is plain that, especially at high energies, the partial wave expansion converges very slowly, and many orders in the angular momentum expansion need to be considered to retrieve the correct prediction of the initial cross section. This conclusion will be of paramount importance in posterior Chapters, especially in the ones regarding unitarity violating issues.
 
At this point we have extensively characterized the VBS processes in the SM at the subprocess level. It is time to study the possible deviations from these predictions that will arise from the new EWSB interactions introduced by the EChL.

\subsection{Vector boson scattering in the electroweak chiral Lagrangian}

The electroweak chiral Lagrangian, introduced in the previous Chapter, parameterizes new EWSB physics through operators that are controlled by a finite set of low-energy parameters. In this sense, for determined values of these parameters, deviations with respect to the SM predictions should be observed in a variety of observables. Concretely, VBS should be very sensitive to these new interactions. In this subsection we will explore how the EChL operators affect VBS observables. 

Not all the terms appearing in the EChL are expected to affect the VBS processes in the same way. For instance, taking a look at \eqref{eq.L4}, it can be seen that the operators weighted by $a_4$ and $a_5$ are the most relevant ones in terms of VBS processes. They only involve Goldstone boson interactions, which in the end are the heart of the strongly interacting system, and they are the only ones that remain present in the Lagrangian if the gauge interactions are {\it switched off}. Since their contribution is expected to be the dominant one in terms of deviations with respect to the SM in VBS observables we will use them as reference in the forthcoming pages in order to illustrate the different characteristics of VBS in the EChL. Other parameters will become relevant in posterior parts of this Thesis, as we shall see.

\begin{figure}[t!]
\begin{center}
\includegraphics[width=.49\textwidth]{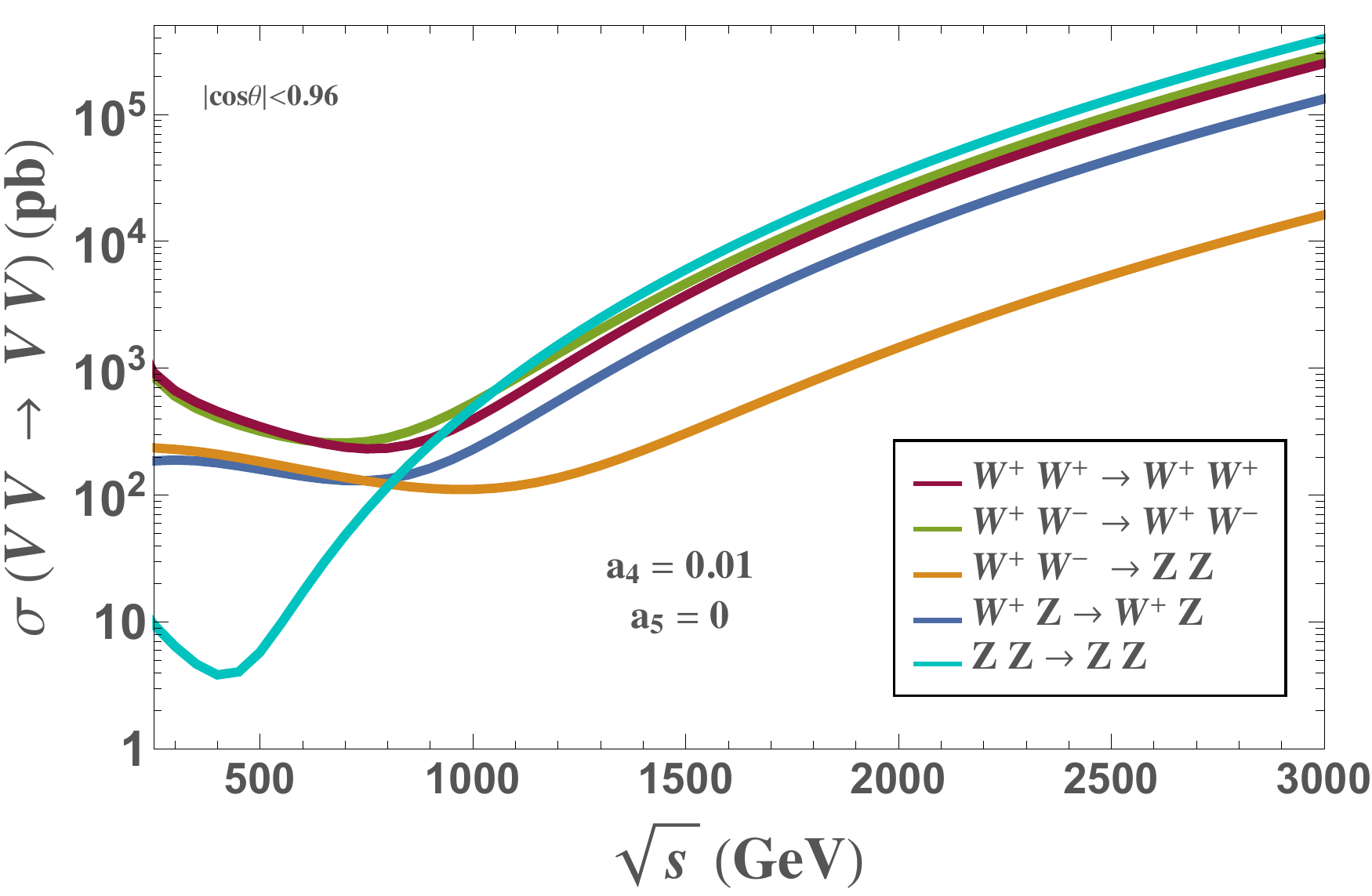}
\includegraphics[width=.49\textwidth]{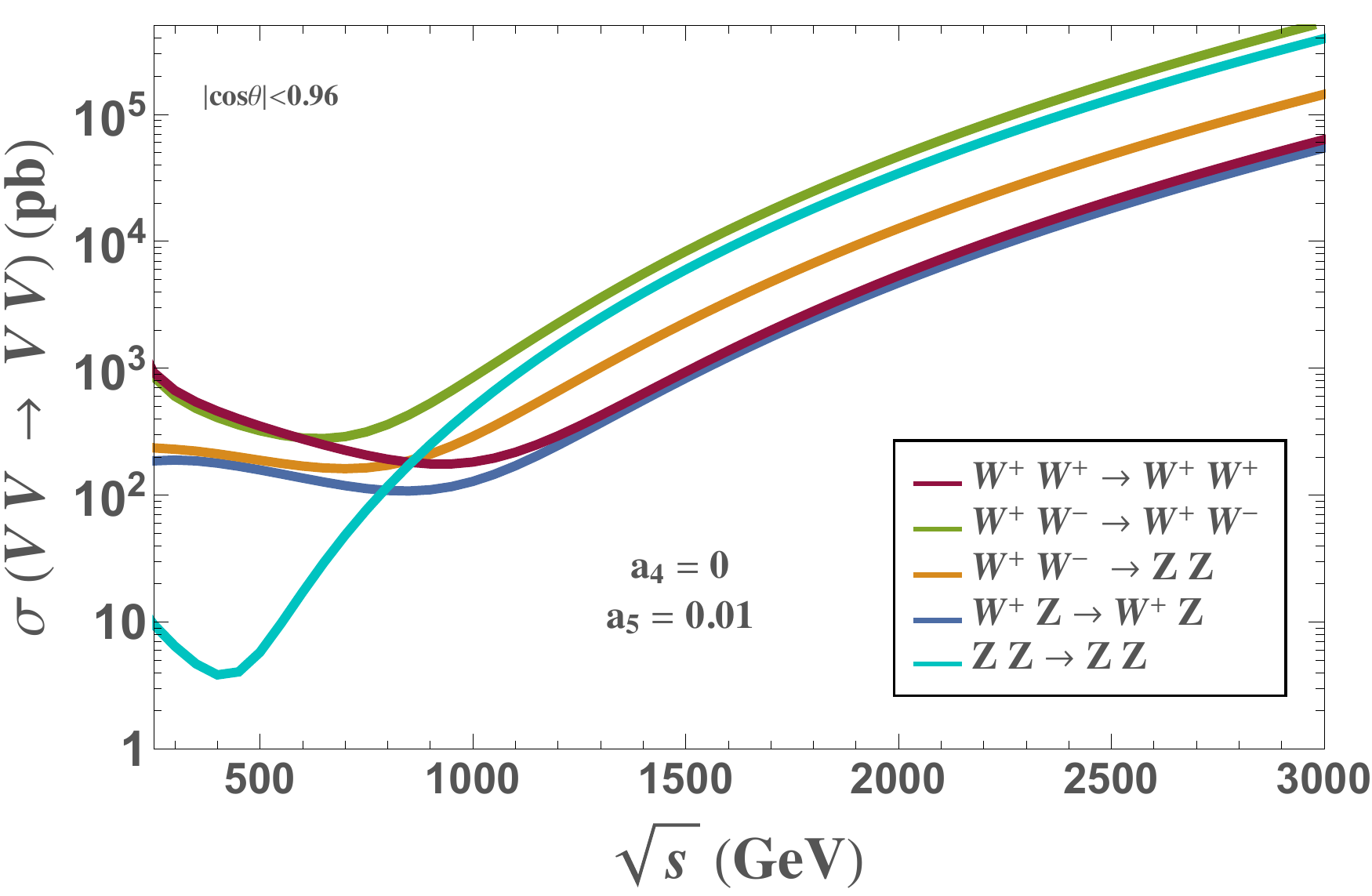}\\
\includegraphics[width=.49\textwidth]{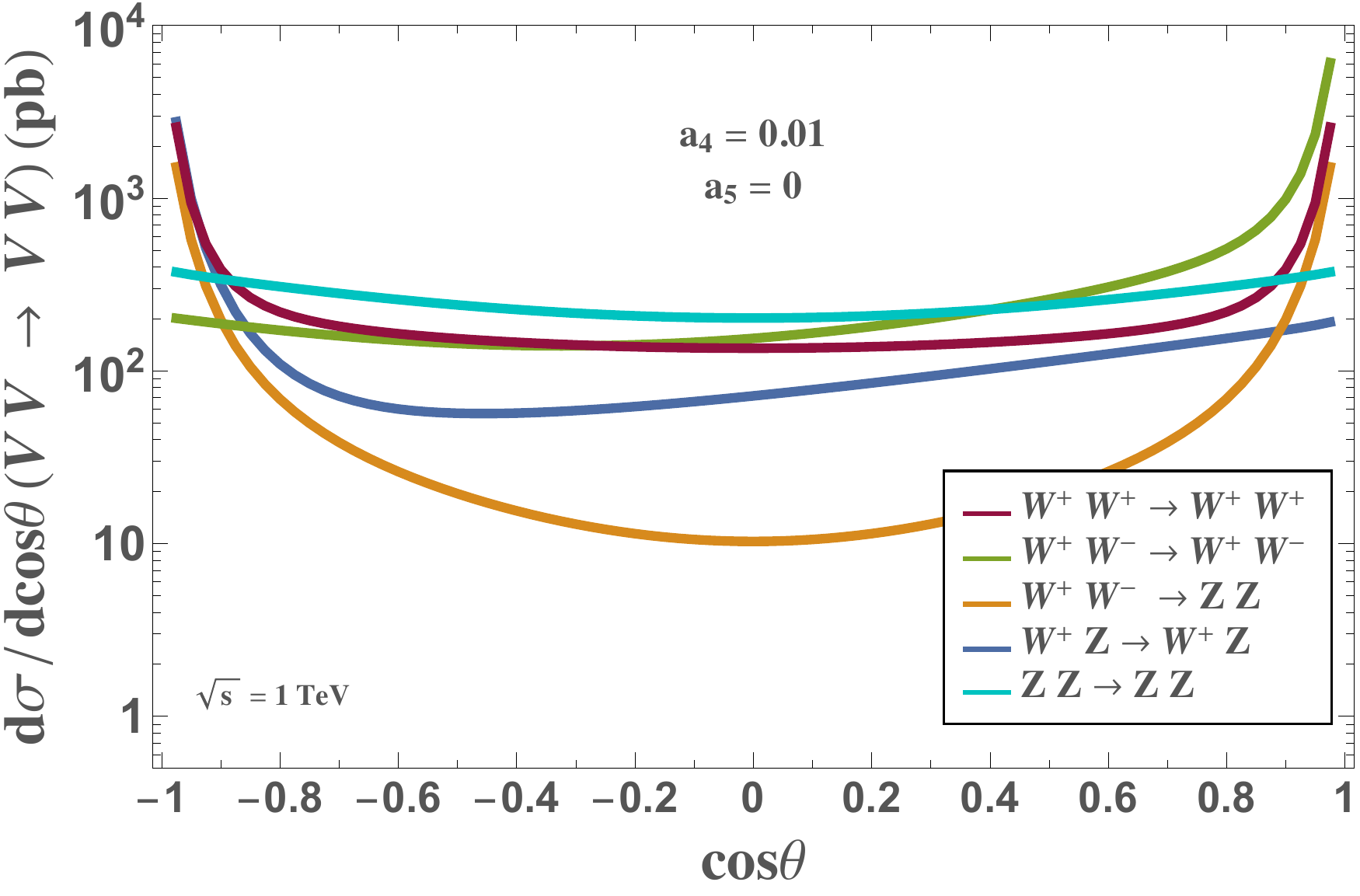}
\includegraphics[width=.49\textwidth]{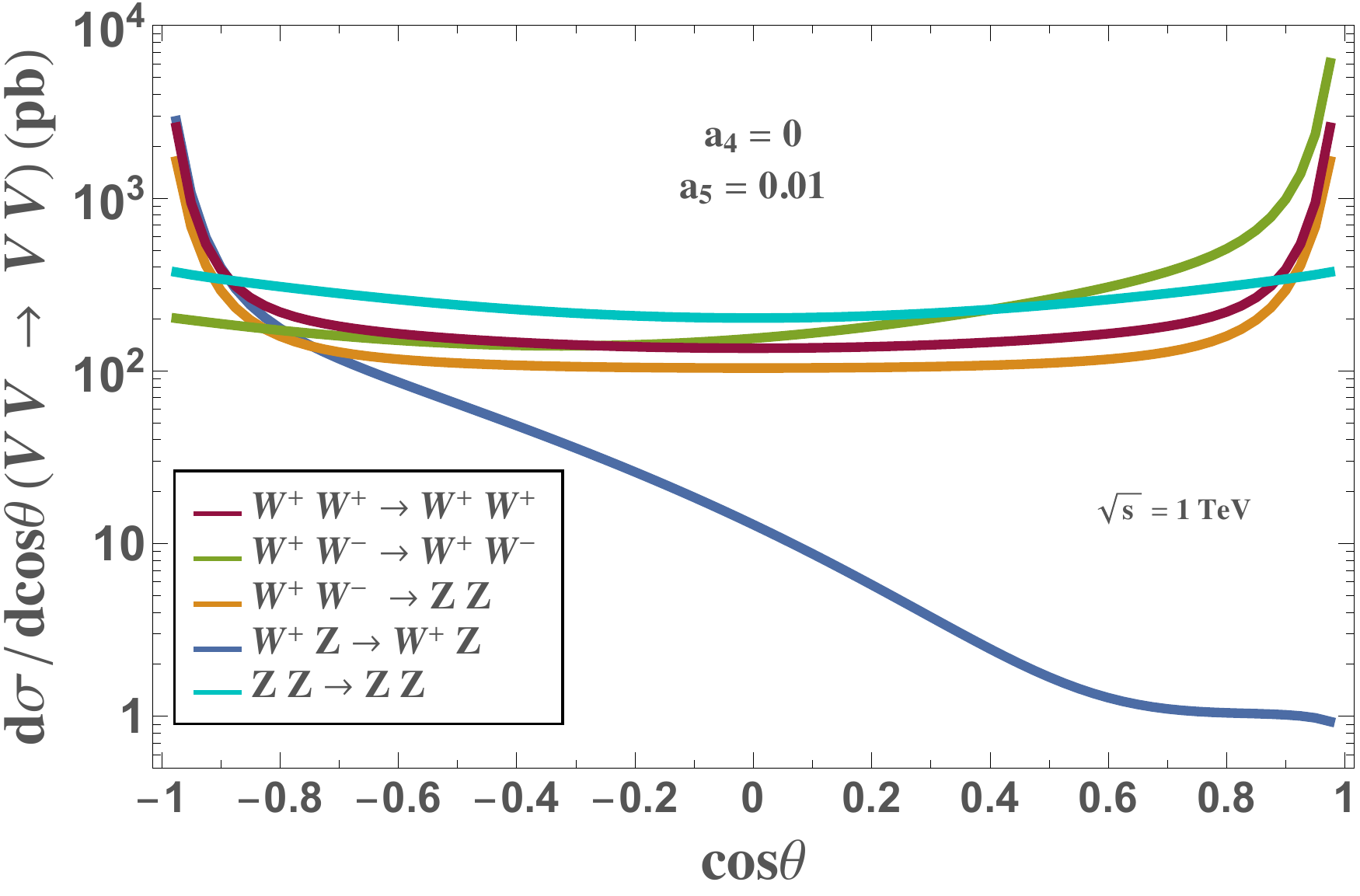}
\caption{EChL predictions for the cross sections of the five chosen VBS processes: W${}^+$W${}^+$$\to$ W${}^+$W${}^+$, W${}^+$W${}^-$$\to$ W${}^+$W${}^-$, W${}^+$W${}^-$$\to$ ZZ, W${}^+$Z $\to$ W${}^+$Z and  ZZ $\to$ ZZ. Upper panels show the total cross section as a function of the center of mass energy integrated in $|\cos\theta|\leq 0.96$, with $\theta$ being the scattering angle as defined in the text. Lower panels show the differential cross section as a function of the cosine of the scattering angle for a fixed energy of $\sqrt{s}=$1 TeV. Left panels correspond to $a_4=0.01$ and $a_5=0$, whereas right panels display the opposite scenario: $a_4=0$ and $a_5=0.01$.}
\label{fig:VBSa4a5scos}
\end{center}
\end{figure}

In order to characterize the EChL predictions for VBS observables, since our final aim is to compare them against the SM values, we may start by scrutinizing their energy and angular behaviour, as we did for the SM case. In  \figref{fig:VBSa4a5scos} we present the results of the previously commented processes (W${}^+$W${}^+$$\to$ W${}^+$W${}^+$, W${}^+$W${}^-$$\to$ W${}^+$W${}^-$, W${}^+$W${}^-$$\to$ ZZ, W${}^+$Z $\to$ W${}^+$Z and  ZZ $\to$ ZZ) computed within the EChL with non-vanishing values of $a_4$, $a_5$ or both parameters simultaneously. Upper panels show the total cross section as a function of the center of mass energy integrated in $|\cos\theta|\leq 0.96$, with $\theta$ being the scattering angle as defined before. Lower panels show the differential cross section as a function of the cosine of the scattering angle for a fixed energy of $\sqrt{s}=$1 TeV. Left panels correspond to $a_4=0.01$ and $a_5=0$, whereas right panels display the opposite scenario $a_4=0$ and $a_5=0.01$.

It is clear that these results differ very much from those obtained within the SM (\figref{fig:VBSSM1}), except at low energies where both predictions match as it was expected from the low-energy theorems. However, regarding the behaviour at high energies, the cross sections do not fall down with energy anymore. Instead, they grow, as it was expected due to the derivative character of the Goldstone boson interactions. This happens in all channels and for both $a_4$ and $a_5$ different from 0. Furthermore, the net size of each of the studied VBS channels changes with respect to the SM case. Whereas before, in the SM computation, the ZZ $\to$ ZZ cross section was the smallest by several orders of magnitude, in the EChL it grows to be, at high energies, the largest or next-to largest one. This might be a consequence of the fact that in the SM there is no quartic term between four Zs, i.e., no contact diagram, whereas in the EChL this four vector interaction exists. The present results indicate that the deviations from the EChL with respect to the SM will be larger in the ZZ $\to$ ZZ case than in the rest, that are of comparable size up to a factor 10.

From the upper panels it is also plain that $a_4$ and $a_5$ have different effects in each of the studied VBS channels. For instance, the cross section obtained for the W${}^+$W${}^+$$\to$ W${}^+$W${}^+$ scenario is significantly larger, at high energies, for $a_4=0.01$ and $a_5=0$ than for the opposite case: $a_4=0$ and $a_5=0.01$. In the ZZ $\to$ ZZ channel, on the contrary, both choices of the EChL parameters lead to the same results. This facts can be understood from the Feynman rules collected in Appendices \ref{FR-SM} and \ref{FR-EChL} in terms of the $a_4$ and $a_5$ dependence of the different contributions.

In what concerns the angular behaviour of these cross sections, i.e., the results shown in the lower panels of \figref{fig:VBSa4a5scos}, many differences with the SM prediction can be seen as well, and the difference between the  $a_4$ and $a_5$ contribution to each of the VBS processes is even clearer than in the upper panels. First of all, although their precise shape changes, the symmetry or asymmetry of each differential cross section is trivially maintained with respect to the SM case: W${}^+$W${}^+$$\to$ W${}^+$W${}^+$, W${}^+$W${}^-$$\to$ ZZ and  ZZ $\to$ ZZ are symmetric in the cosine of the scattering angle whereas W${}^+$W${}^-$$\to$ W${}^+$W${}^-$ and W${}^+$Z $\to$ W${}^+$Z are asymmetric. Second of all, the relative size of each cross section changes, as it was expected from the commented results presented in the upper panels.

However, the most surprising results is that of the W${}^+$Z $\to$ W${}^+$Z channel, that shows an outstanding different behaviour for $a_4\neq0$ than for $a_5\neq0$. In the former case, the diminishment of the cross section in the forward direction that took place in the SM computation is highly alleviated, whereas in the latter it is maintained even to its SM value. Taking a look at the interaction vertex of two Ws and two Zs in the EChL, (see Appendix \ref{FR-EChL}), for the kinematical configuration of the W${}^+$Z $\to$ W${}^+$Z process, $a_4$ will control the $(s^2+u^2)$ behaviour of the amplitude and $a_5$ the $t^2$ one. Since $t$ becomes minimal in the forward direction, when only $a_5$ is switched on, the $t^2$ factor diminishes the cross section for $\cos\theta\sim1$, pulling it towards the SM one. On the other hand, $u$ diminishes in the opposite direction and becomes maximal at $\cos\theta\sim1$. Therefore, if $a_4$ is different from 0 the cross section rises for values close to $\cos\theta\sim1$.  This feature is indeed very interesting since it could be used to disentangle the values of $a_4$ and $a_5$ independently by looking at different kinematical regions of the same VBS process. The same arguments can be used in the analysis of the other VBS channels, that suffer, however, from less severe modifications due to their concrete kinematical configurations.

\begin{figure}[t!]
\begin{center}
\includegraphics[width=.49\textwidth]{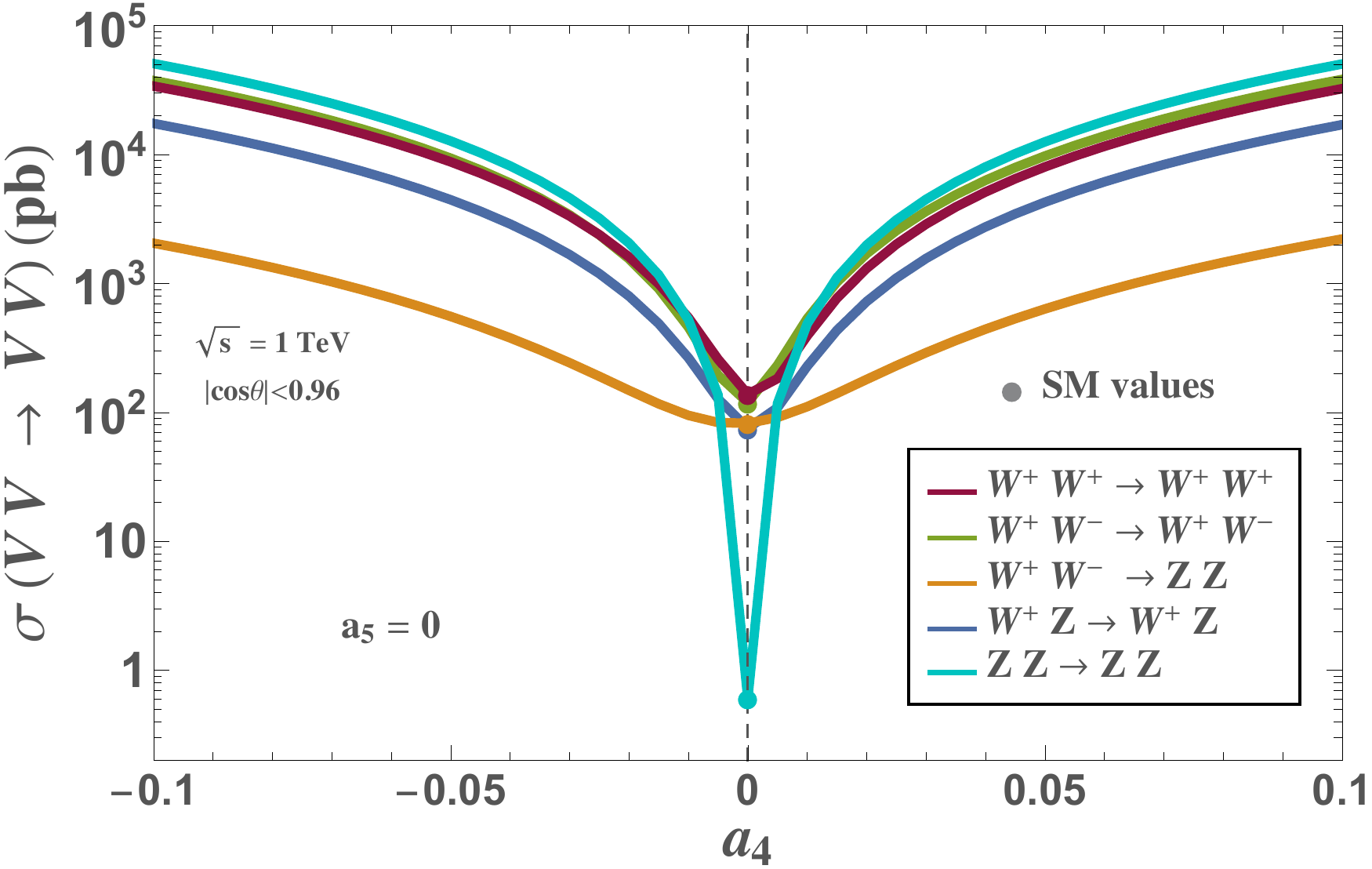}
\includegraphics[width=.49\textwidth]{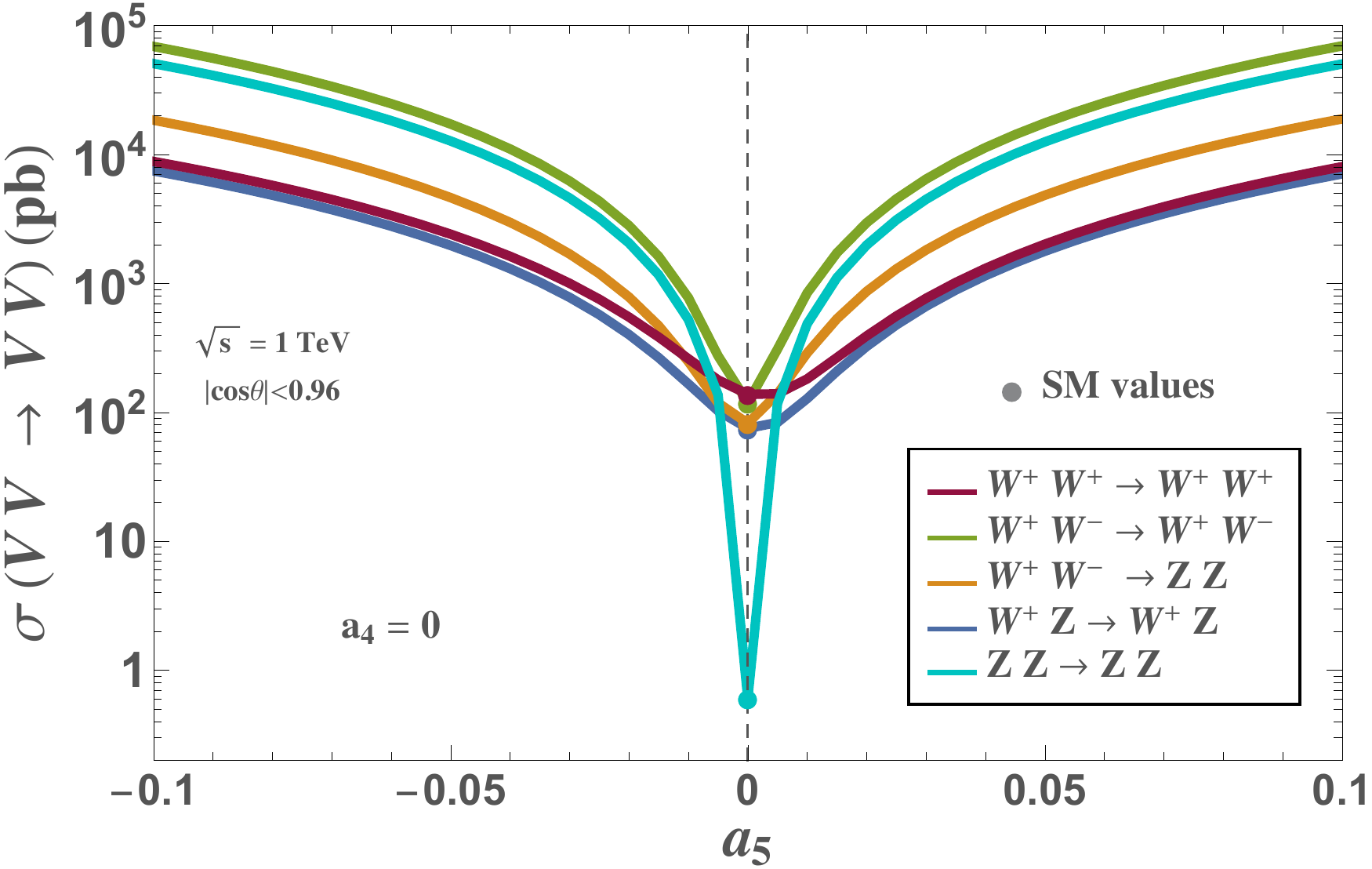}
\caption{Predictions for the total cross section of W${}^+$W${}^+$$\to$ W${}^+$W${}^+$, W${}^+$W${}^-$$\to$ W${}^+$W${}^-$, W${}^+$W${}^-$$\to$ ZZ, W${}^+$Z $\to$ W${}^+$Z and  ZZ $\to$ ZZ computed within the EChL as a function of $a_4$ with $a_5$ set to 0 (left) and as a function of $a_5$ with $a_4$ set to 0 (right). The center of mass energy has been fixed to $\sqrt{s}=$1 TeV and $|\cos\theta|\leq 0.96$ has been required. SM cross section values are marked with solid circles.}
\label{fig:VBSa4a50}
\end{center}
\end{figure}

It is manifest that our two main chiral parameters, $a_4$ and $a_5$, have different impacts in the various VBS channels. Therefore, it is interesting to study the EChL total cross section predictions as a function of these two coefficients. This is precisely what is displayed in \figref{fig:VBSa4a50}, for $\sqrt{s}=$1 TeV, $|\cos\theta|\leq 0.96$ and $a_5=0$ (left) and for $\sqrt{s}=$1 TeV, $|\cos\theta|\leq 0.96$ and $a_4=0$ (right). The SM reference values are marked with a solid circle for comparison. The most important feature that becomes clear in these Figures is that the cross section values do not practically depend on the sign of each of the chiral parameters, since the results are almost symmetric with respect to $a_i=0$.

\begin{figure}[t!]
\begin{center}
\includegraphics[width=.49\textwidth]{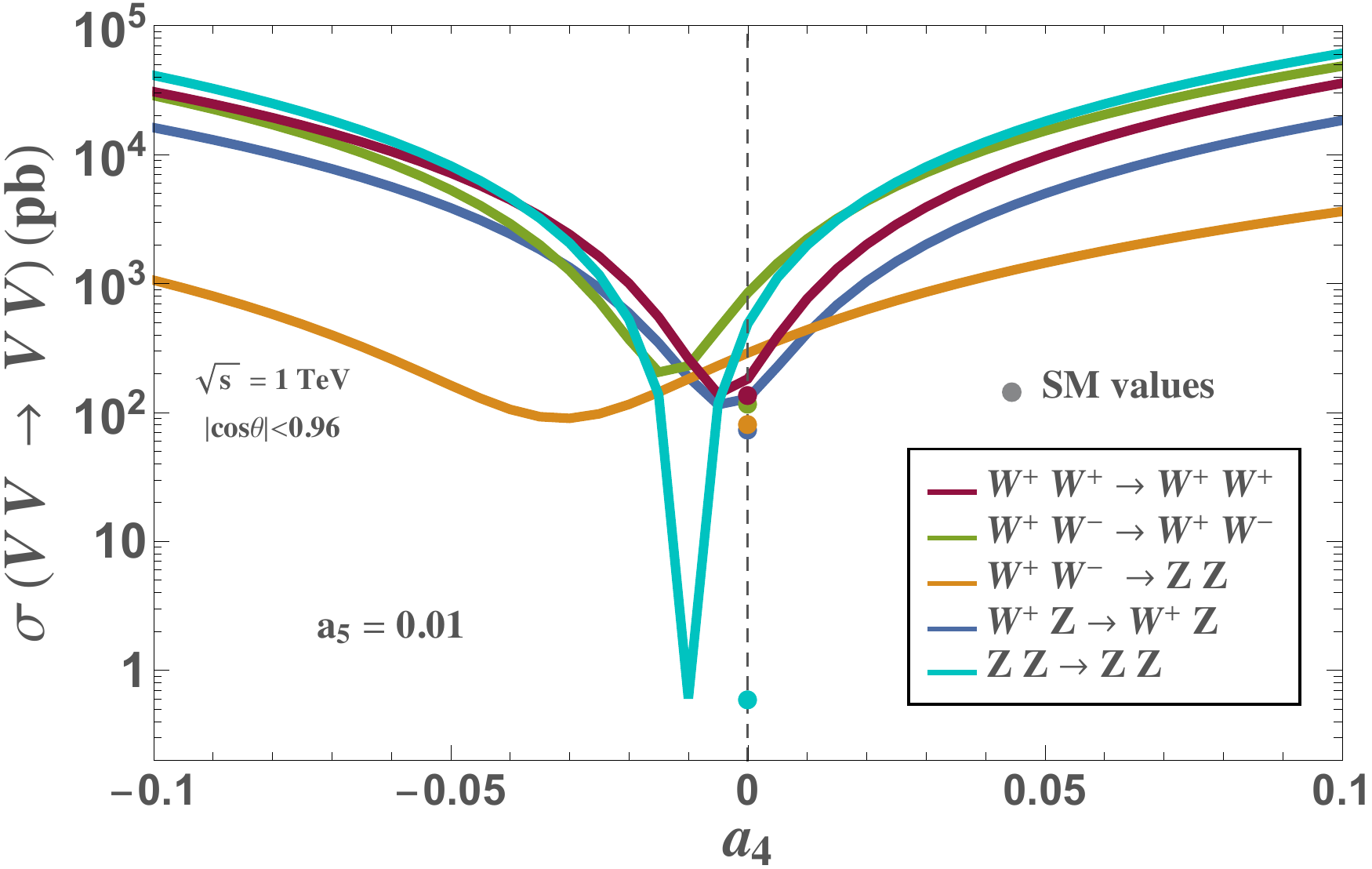}
\includegraphics[width=.49\textwidth]{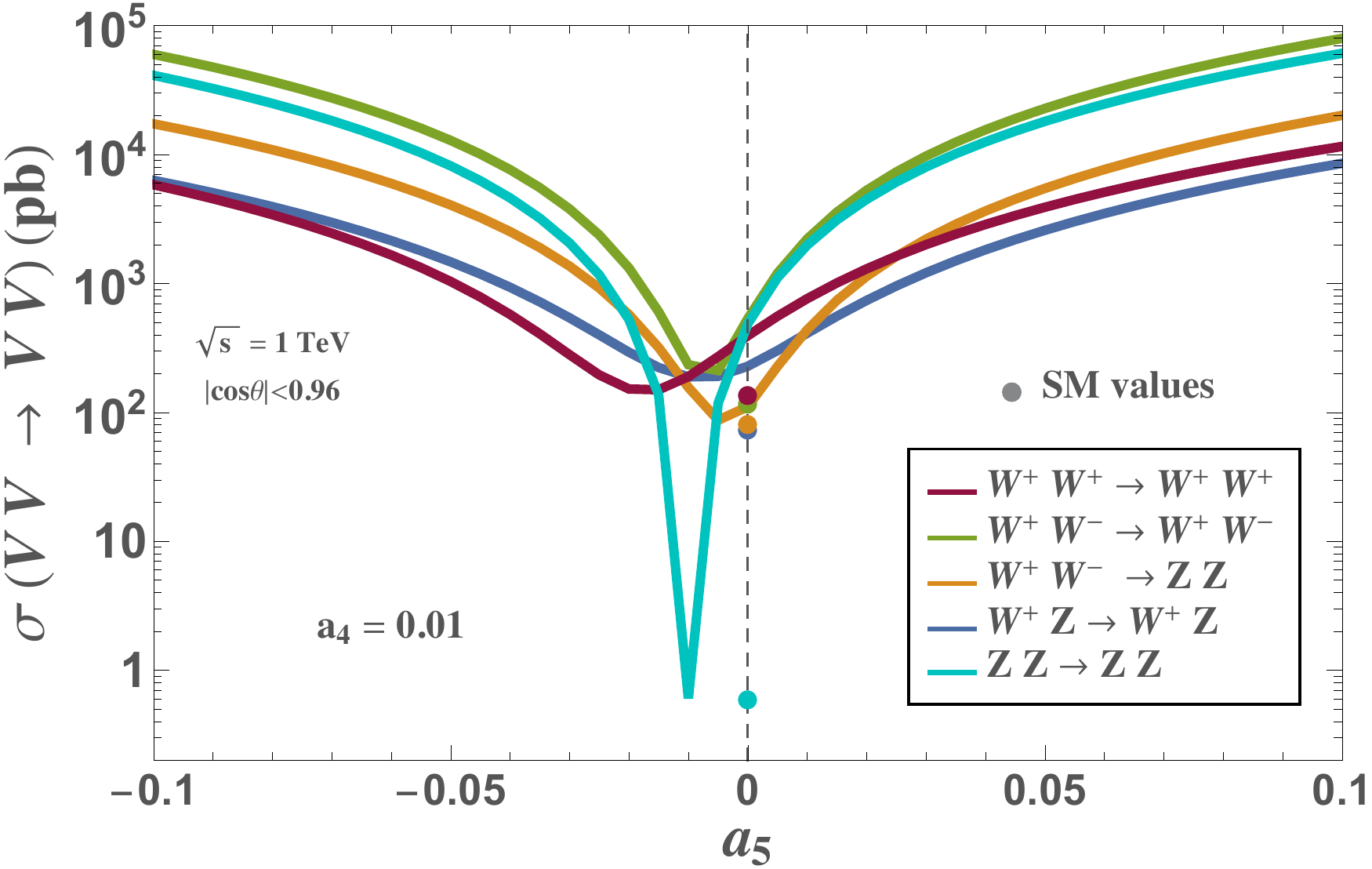}\\
\includegraphics[width=.49\textwidth]{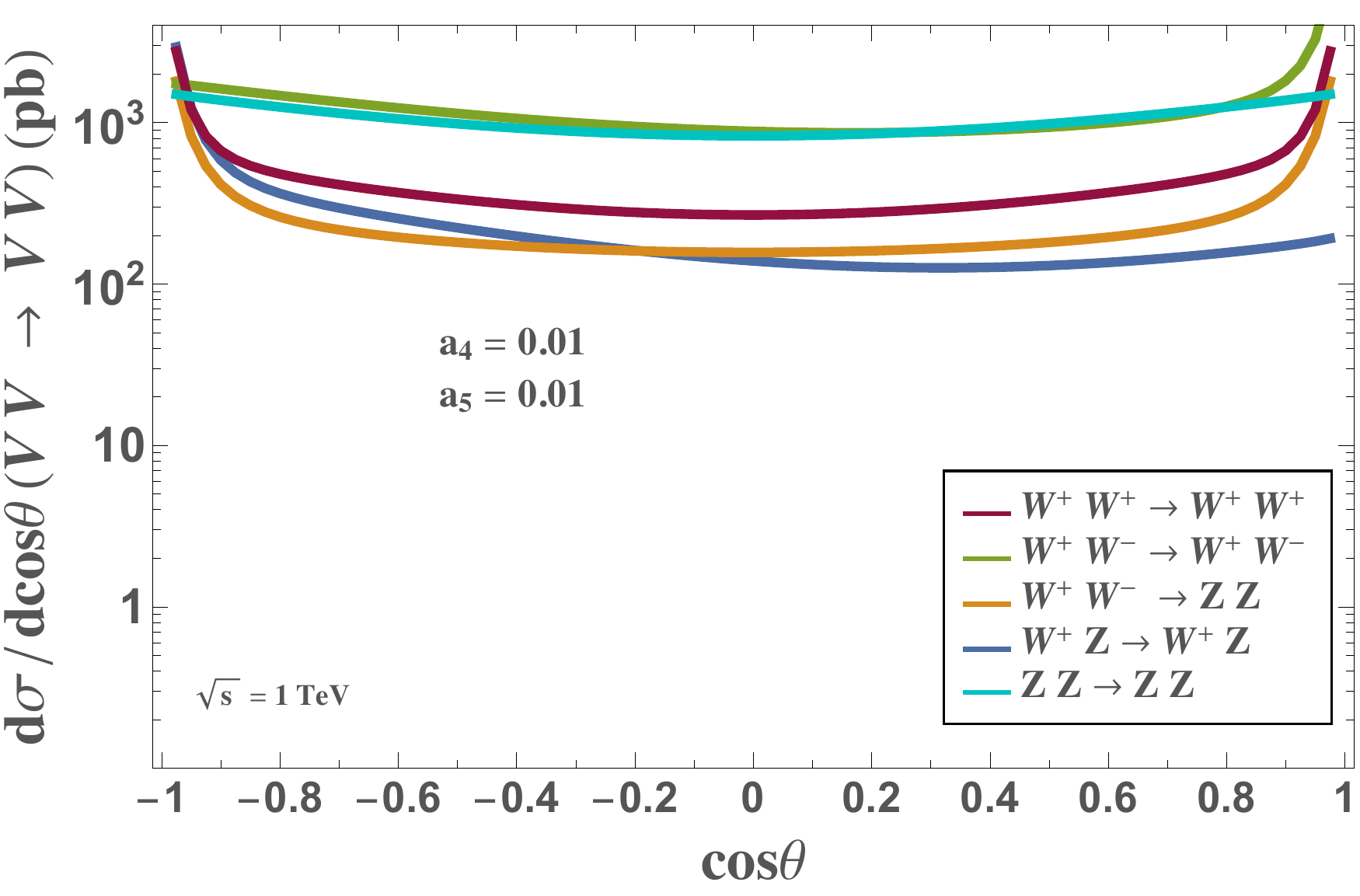}
\includegraphics[width=.49\textwidth]{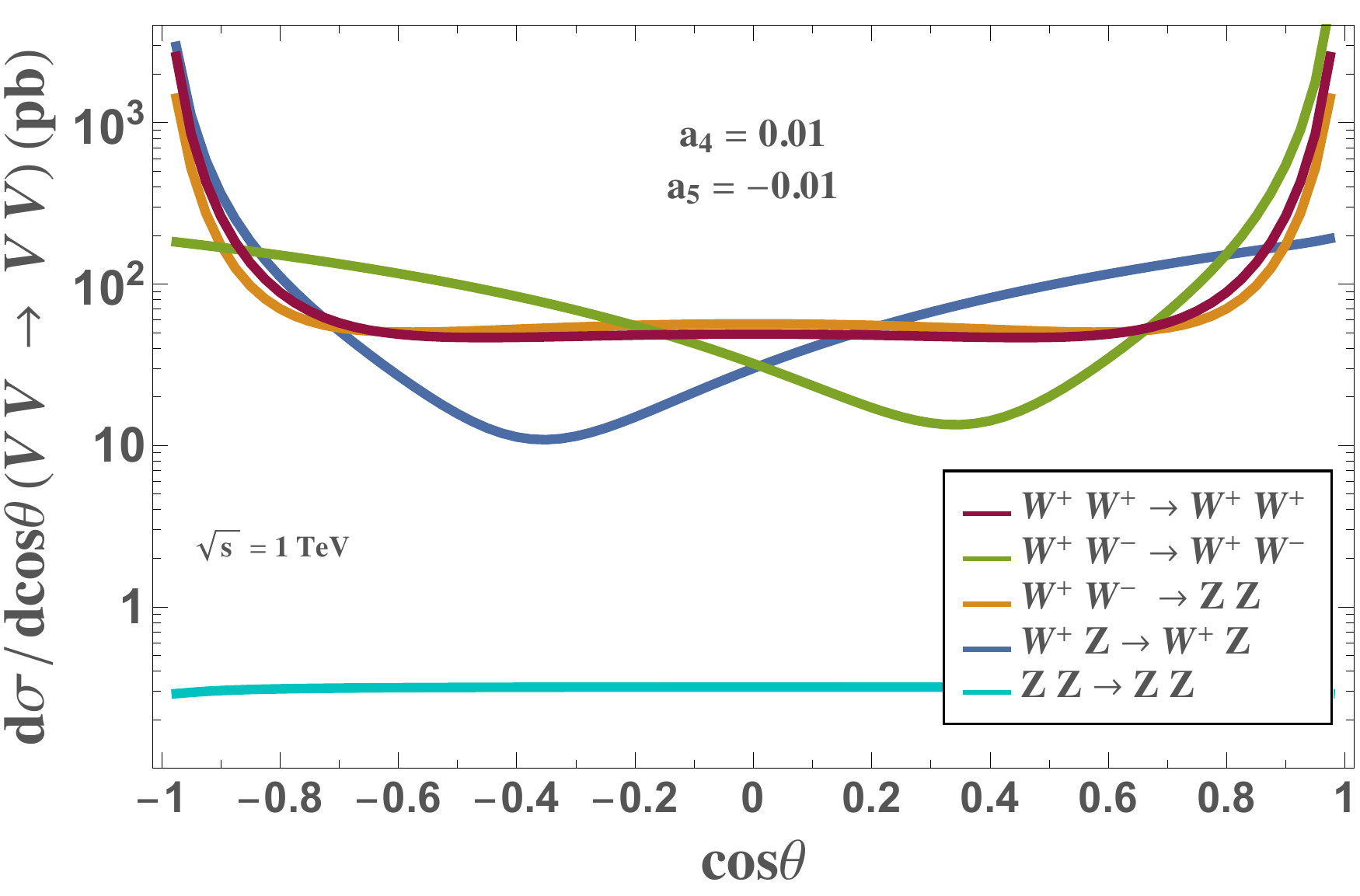}
\caption{Total cross section of the five VBS processes introduced in the text computed within the EChL  as a function of $a_4$ with $a_5=0.01$ (upper left), and as a function of $a_5$ with $a_4=0.01$ (upper right). Lower panels show the differential cross section as a function of the cosine of the scattering angle for $a_4=a_5=0.01$ (left) and for $a_4=-a_5=0.01$ (right). In all cases the center of mass energy has been fixed to $\sqrt{s}=$1 TeV. In the upper panels $|\cos\theta|\leq 0.96$ has additionally been required and the SM prediction has been marked with a solid circle. }
\label{fig:VBSa4a5}
\end{center}
\end{figure}

But, what happens if both parameters are taken into account at the same time? To answer this question we present in \figref{fig:VBSa4a5} the total cross section of the five VBS processes computed within the EChL  as a function of $a_4$ with $a_5=0.01$ (upper left), and as a function of $a_5$ with $a_4=0.01$ (upper right). The lower panels show the differential cross section as a function of the cosine of the scattering angle for $a_4=a_5=0.01$ (left) and for $a_4=-a_5=0.01$ (right). In all cases the center of mass energy has been fixed to $\sqrt{s}=$1 TeV. In the upper panels the phase space integration has been performed in $|\cos\theta|\leq 0.96$. Again, the SM cross section is signalled with a solid circle.

The first clear conclusion that one can extract from these Figures is that, paying attention to the upper panels first, the different VBS channels are governed by different combination of $a_4$ and $a_5$ when they are considered simultaneously. For instance, the process ZZ $\to$ ZZ shows a clear overall dependence on the combination $(a_4+a_5)$, since for $a_4=-a_5$ the SM value is recovered. In the other cases the concrete dependence is not as obvious, since in most of them the $a_4$ and $a_5$ effect cannot be factored out directly, like in the W${}^+$Z $\to$ W${}^+$Z, where the chiral parameter dependence of the amplitude is, as we already saw, of the generic form $a_4(s^2+u^2)+a_5\,t^2$. All these dependences can be, nevertheless, inferred from the inspection of the EChL Feynman rules. It is worth commenting at this point that we have checked that fixing the opposite sign values for $a_5$ and $a_4$ in the upper left and right panels, respectively, one obtains almost the mirror image with respect to $a_i=0$ of these plots. 

Regarding the angular distributions obtained when both parameters are considered together, shown in \figref{fig:VBSa4a5}, the general behaviours of the achieved results for the cases $a_4=a_5=0.01$ (lower left) and for $a_4=-a_5=0.01$ (lower right) are very similar to those presented in   \figref{fig:VBSa4a5scos}. The main difference lies in the ZZ $\to$ ZZ process once again, since, in the case in which both parameters are equal but of opposite sign, the BSM contributions cancel out and the SM prediction is retrieved. In the other cases, the kinematical configurations of the specific contact diagram lead to the explanation of the curves appearing in the Figure, following the same kind of arguments presented above in these pages. 

 \begin{figure}[t!]
\begin{center}
\includegraphics[width=.49\textwidth]{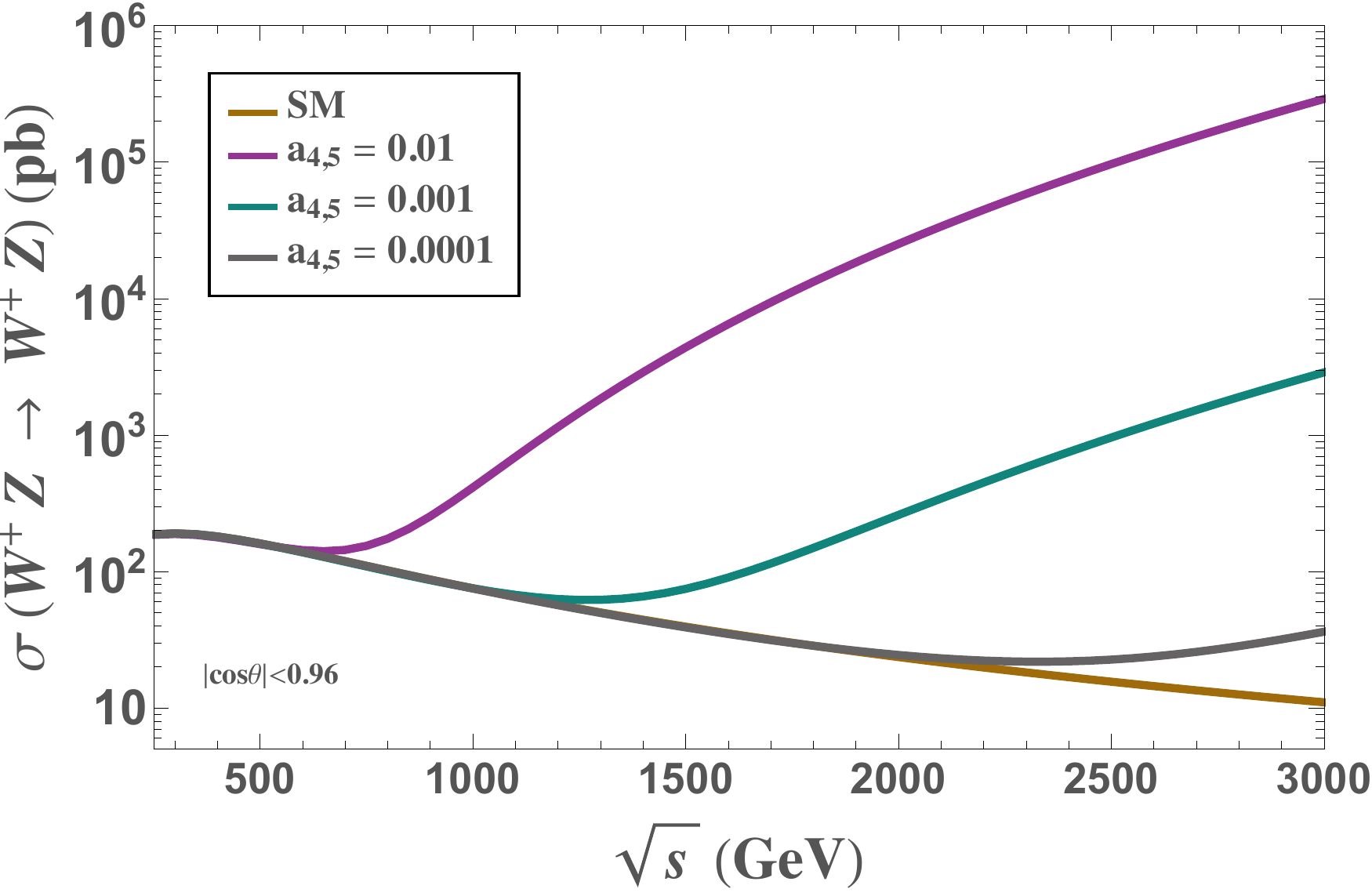}
\includegraphics[width=.49\textwidth]{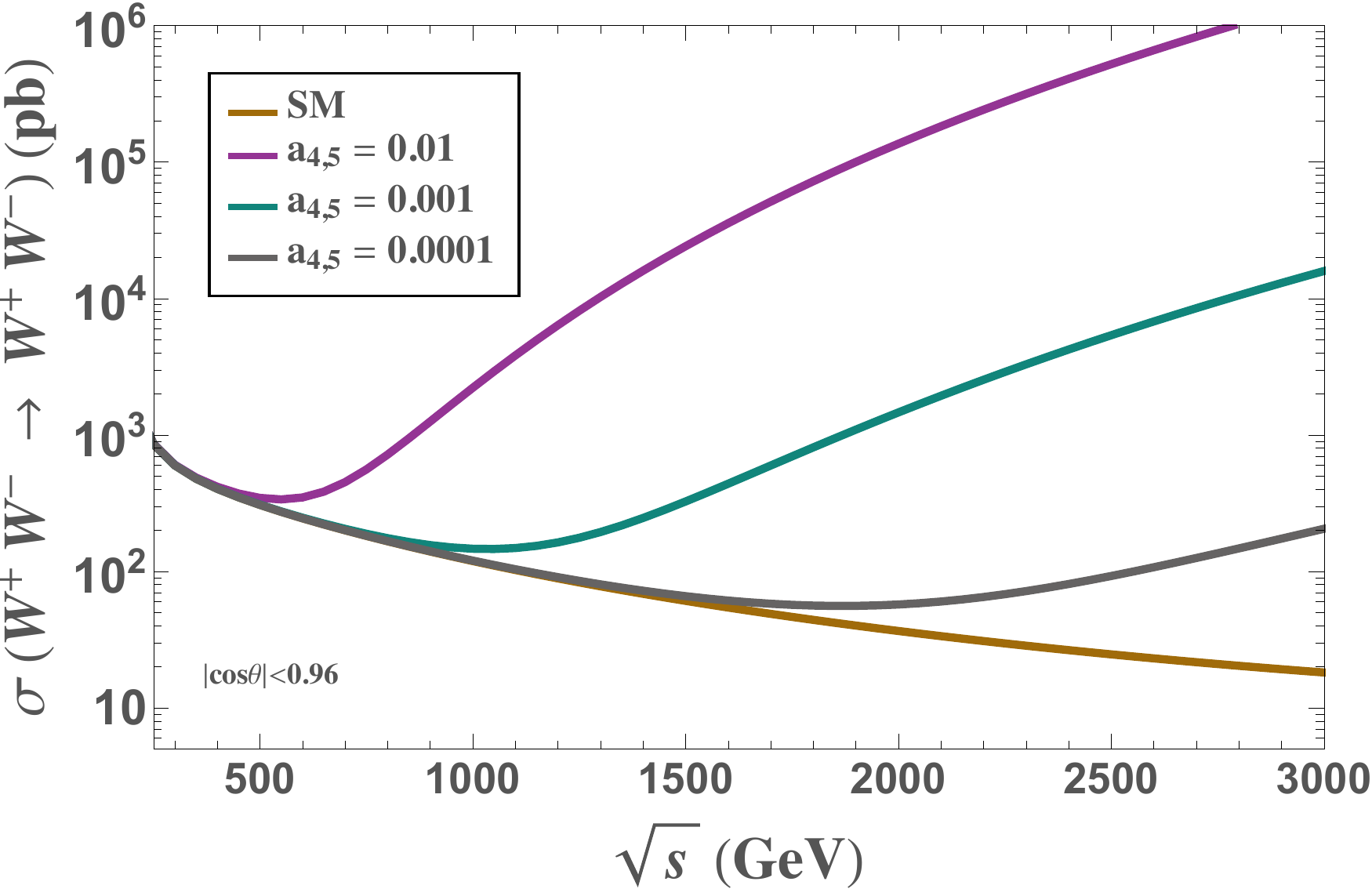}
\caption{EChL total cross sections of W${}^+$Z $\to$ W${}^+$Z (left panel) and W${}^+$W${}^-$$\to$ W${}^+$W${}^-$ (right panel) as a function of the center os mass energy $\sqrt{s}=$ for different values of the chiral parameters: $a_4=a_5=0.01,~0.001,~0.0001$. The SM prediction is shown as well for comparison.  Phase space integration has been performed in $|\cos\theta|\leq 0.96$.}
\label{fig:VBSEChLcomp}
\end{center}
\end{figure}

To conclude the characterization of the VBS processes in the EChL framework, we present \figref{fig:VBSEChLcomp}, where a clearer comparison between the EChL and the SM predictions can be made. To this purpose we have chosen two examples, the W${}^+$Z $\to$ W${}^+$Z and the W${}^+$W${}^-$$\to$ W${}^+$W${}^-$ to illustrate the different energy behaviour of the EChL total cross sections comparatively to the SM one for  $|\cos\theta|\leq 0.96$. In this Figure, three different values of the chiral parameters are considered: $a_4=a_5=0.01,~0.001,~0.0001$. At low energies, near the EW scale, these three cross sections match exactly the SM one, as they should as a consequence of the well known features of chiral perturbation theory. At high energies, on the contrary, they depart from the SM value significantly. The larger the values of the parameters, the larger the enhancement with respect to the SM, that, for $\sqrt{s}=$3 TeV, goes from a factor 4 in the W${}^+$Z $\to$ W${}^+$Z for $a_4=a_5=0.0001$ to five orders of magnitude in the W${}^+$W${}^-$$\to$ W${}^+$W${}^-$ for $a_4=a_5=0.01$. This indicates that if the true dynamics of the EWSB sector can be described by the EChL with parameter values of the order of $0.01$, the deviation with respect to the SM in this VBS observable could be clearly visible. 

At this point, when we have stablished the characteristics of the various VBS  channels in the context of the EChL, two features have to be pointed out. The first one concerns the behaviour of the different polarizations in the EChL case. We have not shown it here for brevity, but we have explicitly checked that the polarization modes are approximately conserved even beyond the SM predictions. The EChL cross sections manifest a similar behaviour  than those presented in \figref{fig:VBSSMpols} for the SM case when considering different polarizations. Obviously, the channels involving longitudinally polarized gauge bosons change with respect to the SM prediction due to the new interactions driven by the EChL operators while the transverse modes are much less affected. In this sense, the predictions of both setups are not comparable, but we wanted to state that even in BSM physics scenarios, such as in the EChL, the polarization modes are conserved especially at high energies in VBS processes.

The other important comment to be made at this stage concerns the violation of unitarity discussed in the previous Chapter. Since the EChL cross sections grow with energy, as we have just concluded, for each fixed value of $a_4$ and $a_5$ (and the rest of the chiral parameters) there exists an energy value at which unitarity is violated. Intuitively, the larger the values of the chiral coefficients, the smaller the energy at which the unitarity violation takes place. If this occurs for very high energies, larger than those accesible now at experiments, the EFT framework can be safely used. If not, the unitarity violation problem has to be cured in some way to be consistent with the underlying quantum field theory. Being in one scenario or the other would depend on the values of the chiral parameters, so a study of the violation of unitarity in the EChL as a function of these coefficients is now required. The next subsection is devoted precisely to such an analysis.

\subsection{Violation of unitarity in vector boson scattering observables}

The violation of unitarity arising from the EChL Goldstone interactions is one of the central issues of this Thesis. It will be extensively studied in posterior Chapters for different setups and scenarios so this subsection is aimed to be just a brief introduction to this phenomenon in VBS observables.

As we have seen, the EChL predictions of VBS processes grow with energy as a consequence of the momentum dependence of the interactions among the longitudinally polarized EW gauge bosons. At a given energy, for a determined value of the chiral parameters, the partial waves corresponding to those predictions surpass the unitarity limit given in \eqref{unitbound}. If the scale of unitarity violation lies within the energy interval that can be probed experimentally, the interpretation of the data through the non-unitary predictions falls in an inconsistency. If this were the case, a prescription would be needed to solve the unitarity violation problem. Chapters \ref{Methods}, \ref{Resonances} and \ref{ResonancesWW} will be devoted to this issue.

But, first, we need to get an idea of how the unitarity violation scale changes as a function of the chiral parameters. In other words, we need to characterize the violation of unitarity in our particular framework: VBS within the EChL. This first estimate can be easily done by studying the partial wave amplitudes of the different VBS channels in the isospin limit, i.e., when $g'=0$, $\sinw=0$ and $m_W=m_Z$. Due to the smallness of the value of $g'$ and $\sinw$, and to the similarity of the EW gauge boson masses, this assumption, in which custodial symmetry is exactly conserved, is well justified, and, therefore, it is  a good first approximation to the full result (as we will discuss later on) especially at high energies. 

In the isospin limit the Ws and the Z form an exact triplet of the $SU(2)_L$ symmetry, so the total isospin value of a gauge boson is 1. In a system of two vector bosons, like the one we are dealing with in VBS, the isospin values of each of them add up following the usual angular momentum sum rules. This means that the possible isospin quantum numbers of the diboson system can only be 0, 1 or 2.

For this reason we can construct three amplitudes of fixed isospin from the combinations of the different VBS processes we have been studying in the previous paragraphs. In this sense, defining
\begin{equation}
A^{abcd}=A({\rm V}^a{\rm V}^b \rightarrow {\rm V}^c {\rm V}^d)\,,
\end{equation}
with $a,~b,~c$ and $d$ denoting the concrete vector boson we are referring to, and by virtue of the crossing symmetry, we can construct a generic VV$\to$ VV amplitude of the total isospin-coupled process as
\begin{align}
A^{abcd}(p^a,p^b,p^c,p^d)= \delta^{ab} \delta^{cd} A(s,t,u)+\delta^{ac} \delta^{bd} A(t,s,u)+\delta^{ad} \delta^{bc} A(u,t,s)\,, \label{isospingold}
\end{align}
where $p^i$, $i=a,b,c,d$ are the momenta of the corresponding EW gauge bosons. 

We now introduce a more intuitive notation, which allows to relate the amplitudes labeled by the $s$, $t$ and $u$ variables with the amplitudes of each of the processes that contribute to the scattering we are looking at. These relations are the following, in which, for instance, $A^{+-00}$ corresponds to $A(W^+W^-\to ZZ)$:
\begin{align}
A^{+-00} &= A(s,t,u)\,,\label{ampwwzz}  \\
A^{+-+-} &= A(s,t,u)+A(t,s,u)\,,  \\
A^{++++} &= A(t,s,u)+A(u,t,s)\,,
\end{align}
and with them it is easy to build the amplitudes with a fixed value of the isospin quantum number, given by:
\begin{align}
A_0&=3A^{+-00}+A^{++++} = 3A(s,t,u)+A(t,s,u)+A(u,t,s)\,,\\
A_1&=2A^{+-+-}-2A^{+-00}-A^{++++} = A(t,s,u)-A(u,t,s)\,,  \\
A_2&=A^{++++} = A(t,s,u)+A(u,t,s)\,.\label{isodef}
\end{align}
Here, the subindices 0, 1 and 2 denote isospin quantum number associated to each amplitude.

The partial waves corresponding to these isospin amplitudes can be computed using \eqref{pwamp}, substituting the scattering amplitude $A$ by its corresponding $A^I$:
\begin{align}
\label{pwampiso} a^{IJ}_{\lambda_1\lambda_2\lambda_3\lambda_4}(s)=\dfrac{1}{32 \pi K}\int_{-1}^{1} d(\cos\theta)\,d_{\lambda,\lambda'}^{J}(\theta)\, A^I_{\lambda_1\lambda_2\lambda_3\lambda_4}(s,\theta)
\,.
\end{align}
They represent the coefficients of the expansion of the scattering amplitude of fixed isospin $I$ in the base of the angular momentum $J$. The condition of the violation of unitarity remains the same as before, i.e., unitarity is no longer preserved when the modulus of the corresponding $IJ^{th}$ partial wave becomes one.

With these considerations in mind we can study the partial wave behaviour of the fixed isospin channels that encode the characteristics of the varios VBS processes studied in the previous subsections. We have chosen to analyze in this first simple exercise the two lowest partial waves in terms of angular momentum $J$, since they are expected to be the most sensitive ones to this issue. Furthermore, we will simplify our study by considering only the purely longitudinally polarized scattering, from which the violation of unitarity arises, mainly. 

\begin{figure}[t!]
\begin{center}
\includegraphics[width=.49\textwidth]{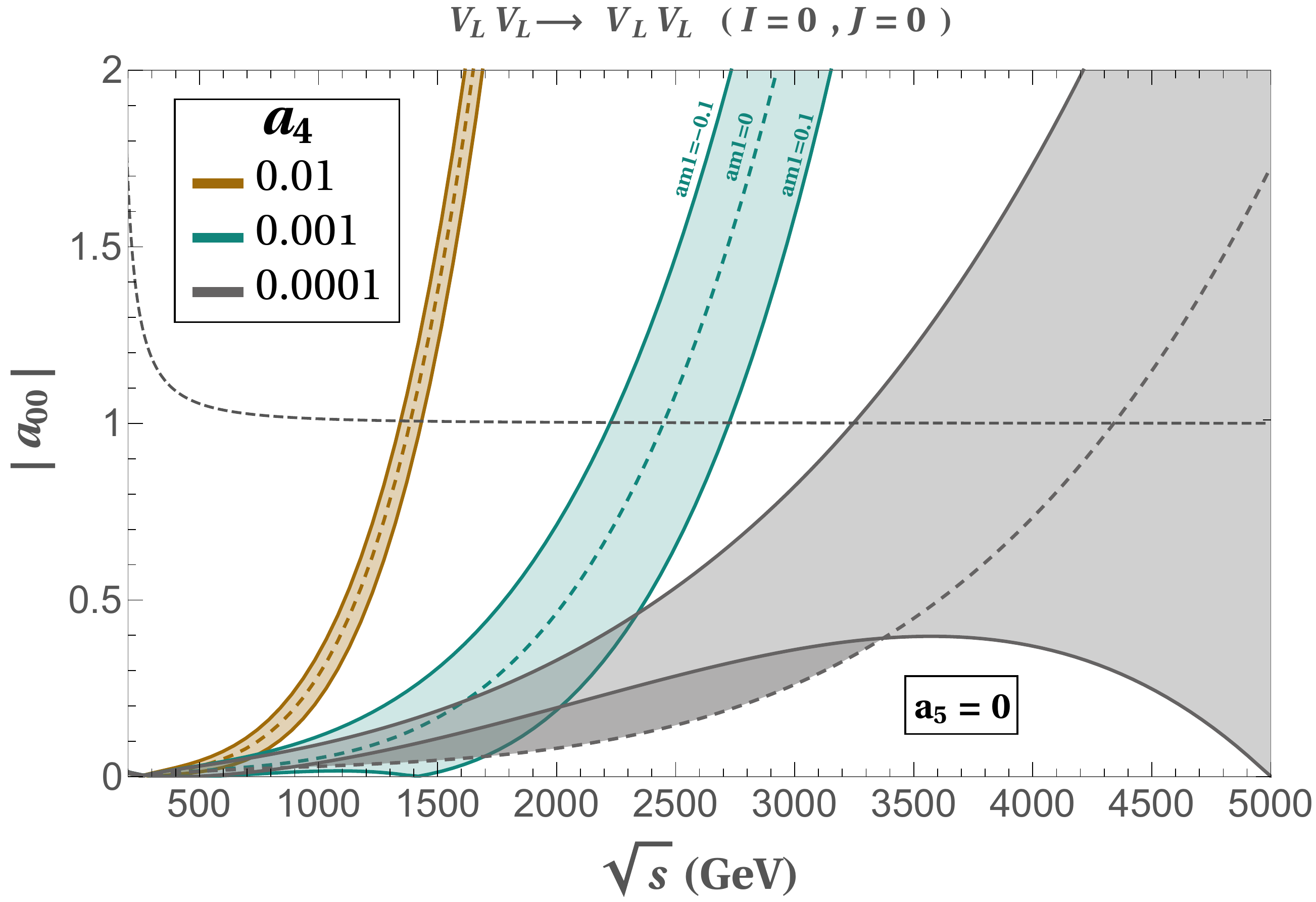}
\includegraphics[width=.49\textwidth]{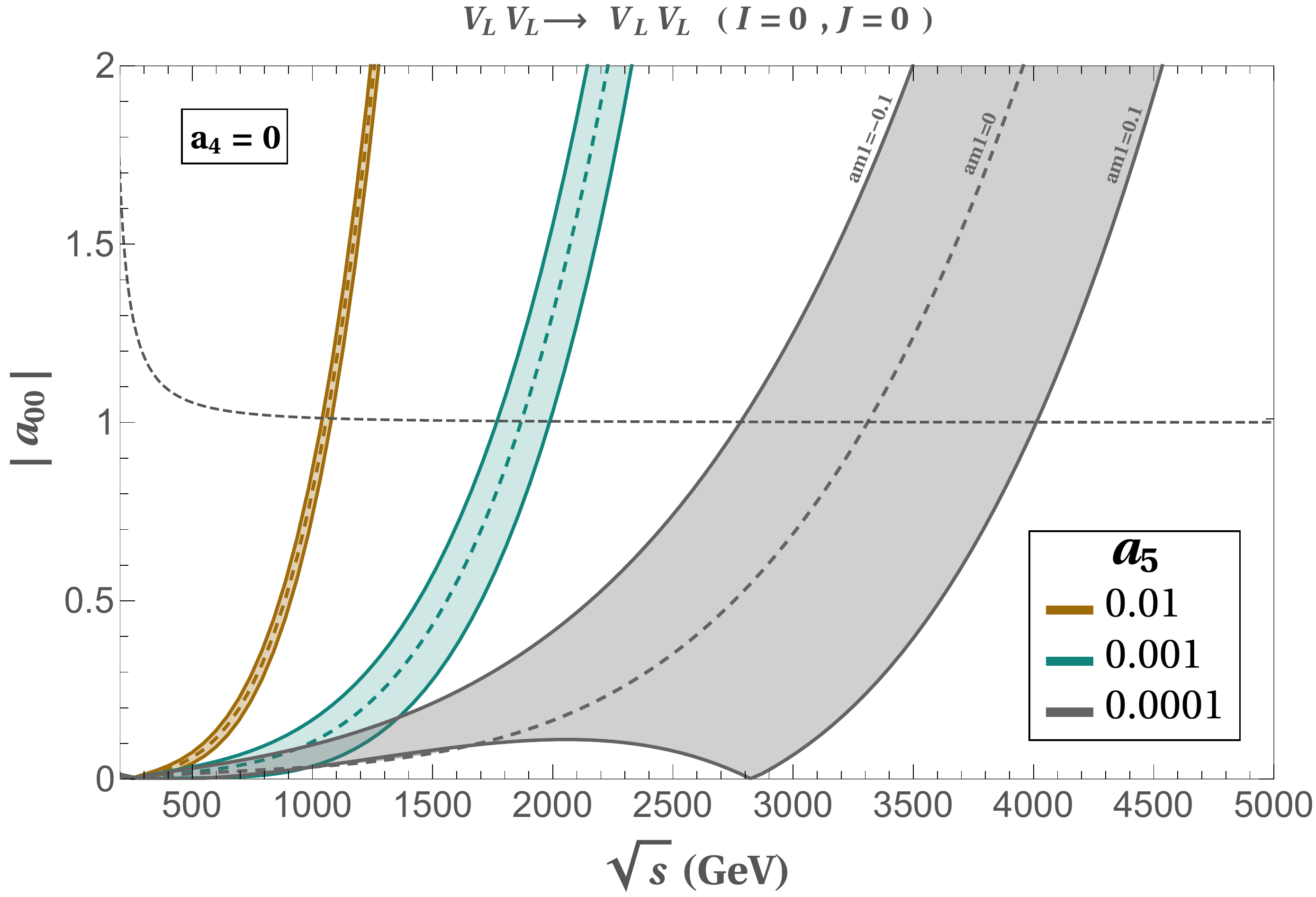}\\
\includegraphics[width=.49\textwidth]{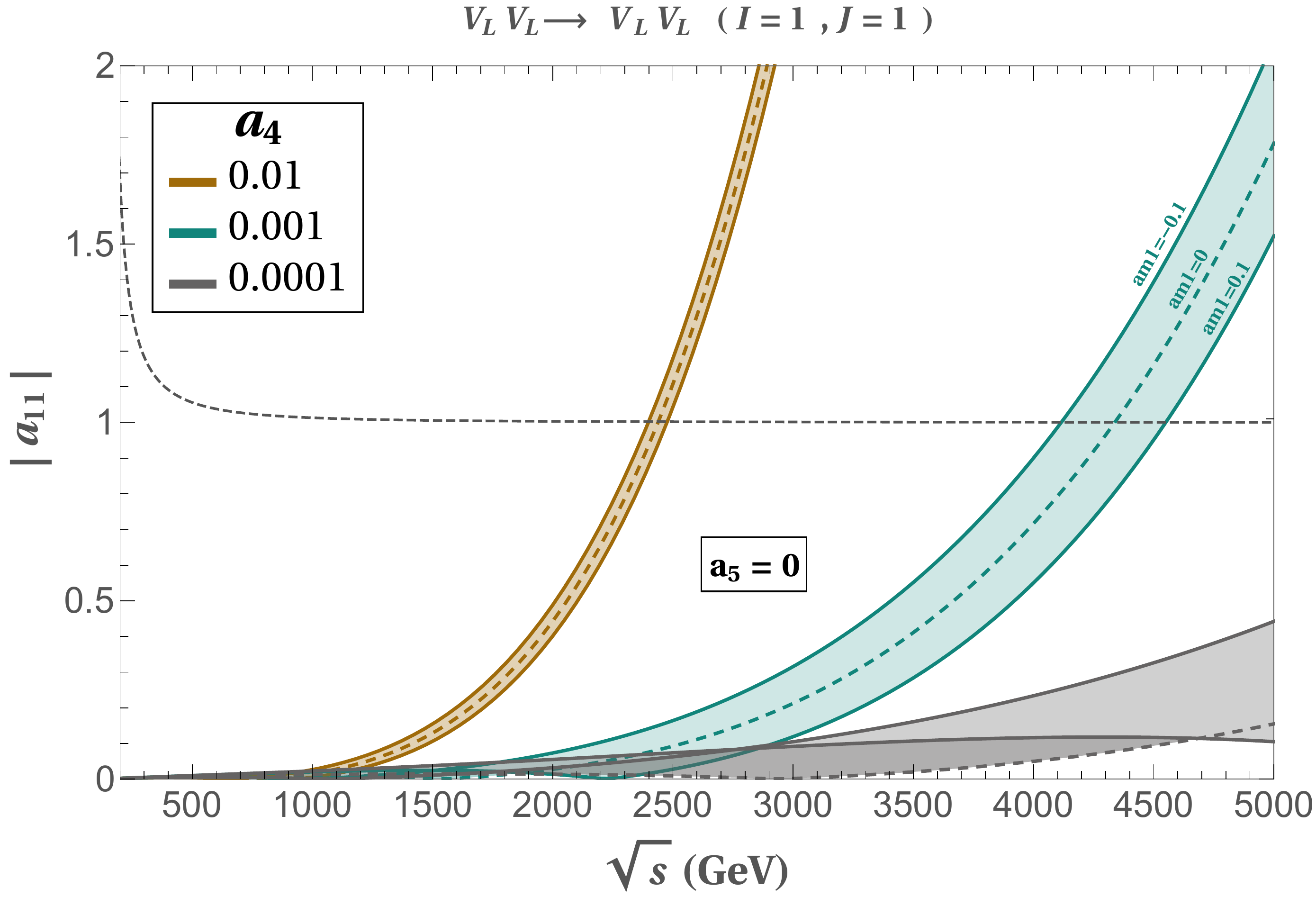}
\includegraphics[width=.49\textwidth]{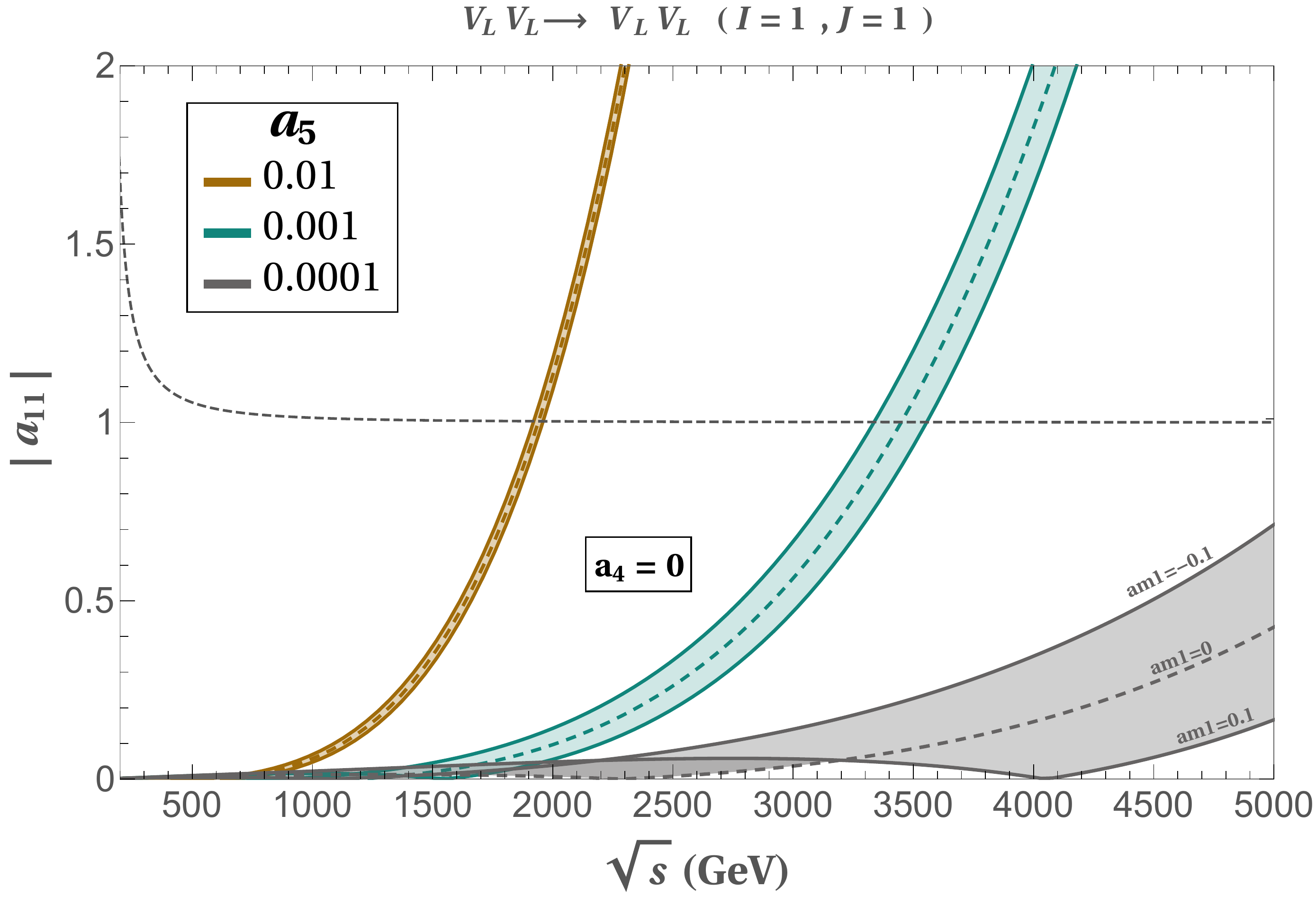}
\caption{Predictions of the modulus of the $a_{00}$ (upper panels) and $a_{11}$ (lower panels) partial waves as a function of the center of mass energy for the isospin-coupled process $V_L V_L \to V_L V_L$.  Computations have been performed in the EChL for different values of the chiral coefficients $a_4$ (left) and $a_5$ (right), shown in different colours, and for different values of the $am1 \equiv a-1$ parameter, depicted in the shadowed areas. The unitarity bound is shown as a dashed line.}
\label{fig:violationunitarity}
\end{center}
\end{figure}
 
We have selected three values of $a_4$ and $a_5$ as before: 0.01, 0.001 and 0.0001, {\it switching on} each of them at a time, and we have also varied the parameter $am1=a-1$\footnote{This definition simplifies the comparison with the SM predictions since by setting $a=1$ and therefore $am1=0$ the SM is recovered. Along this Thesis we will use the different notations: $am1=\Delta a=a-1.$} that controls the BSM coupling between the Higgs boson and two EW gauge bosons. The values for this parameter have been set to -0.1, 0 and 0.1. The results are displayed in \figref{fig:violationunitarity}, where the modulus of the partial wave amplitudes corresponding to $IJ=00$ (upper panels) and $IJ=11$ (lower panels) are shown as a function of the center of mass energy. The violation of unitarity limit is represented with a dashed line so it is easier to check the point at which the unitarity bound is crossed.
 
Unitarity violation is manifest in the ECLh for the given values of the parameters, all of them different from those in the SM, whose predictions just saturate to a constant value below the unitarity violation bound. It is clear as well that the bigger the values of the chiral coefficients $a_4$ and $a_5$, the lower the energies at which these violations occur. However, the behaviour with $a$ is different, as the higher and positive $(a-1)$ becomes, the later is unitarity violated.  Another interesting conclusion extracted from these plots is that in the scalar channel, $IJ=00$, the unitarity bound is trespassed at lower energies than in the vector channel, $IJ=11$, for the same values of the chiral parameters.

The values of the energies for which the unitarity condition does not longer hold for each set of parameters correspond to a limit of validity of the effective theory. As we have seen, they strongly depend on the choice of the coefficients and on the particular $IJ$ channel that we are considering. For instance, if we choose to look at the $IJ=11$ channel, for values of $a_4$ (with $a_5=0$) of 0.01, the violation of unitarity happens at energies around 2.5 TeV, whereas if we take $a_4=0.001$ it happens much later, between 4.1 and 4.5 TeV depending on the values of $a$. In the latter case, the EFT breakdown would take place when the energy reaches $4\pi v\sim 3$ TeV, i.e., prior to the violation of unitarity.

\begin{figure}[t!]
\begin{center}
\includegraphics[width=0.49\textwidth]{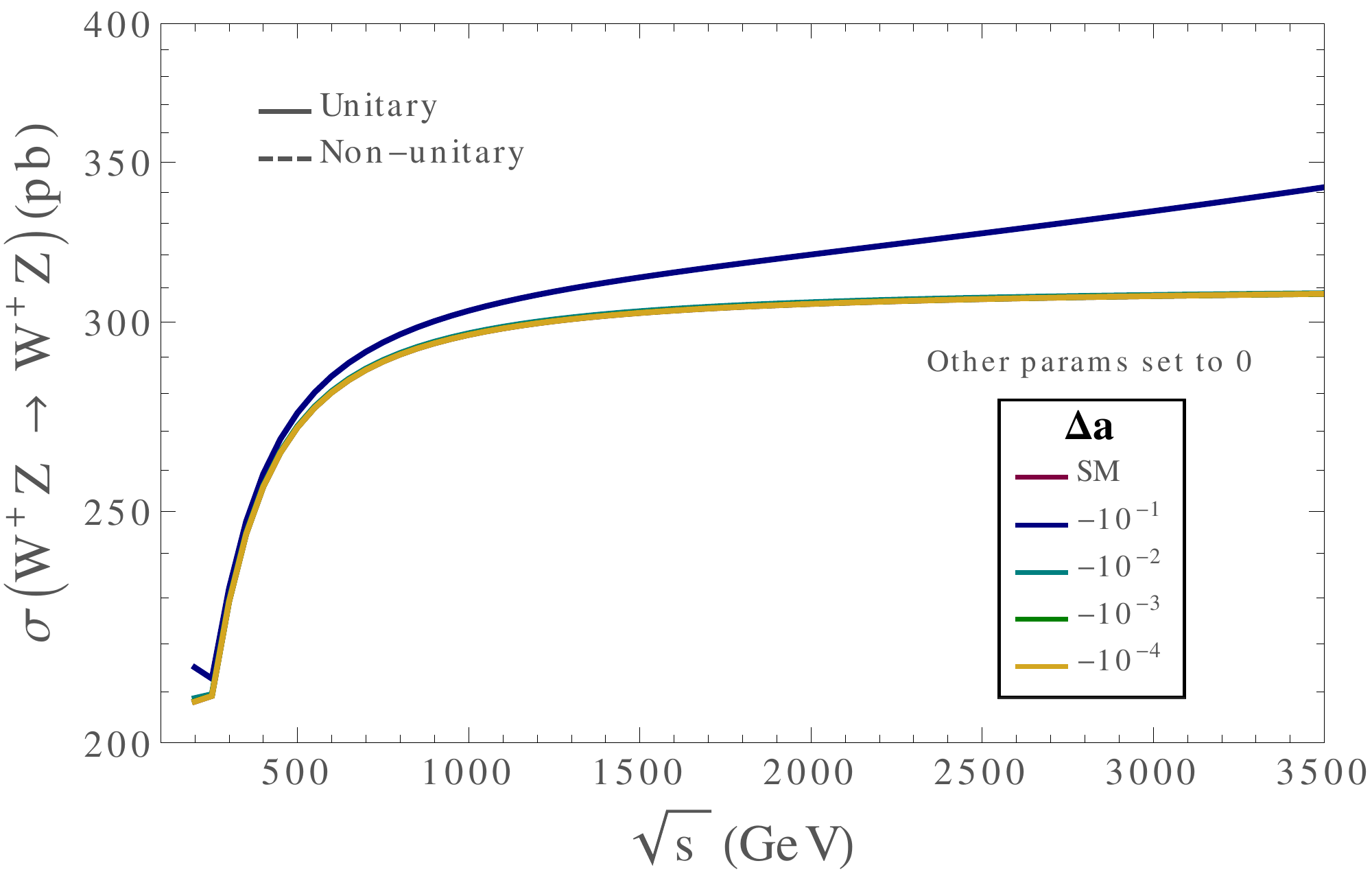}
\includegraphics[width=0.49\textwidth]{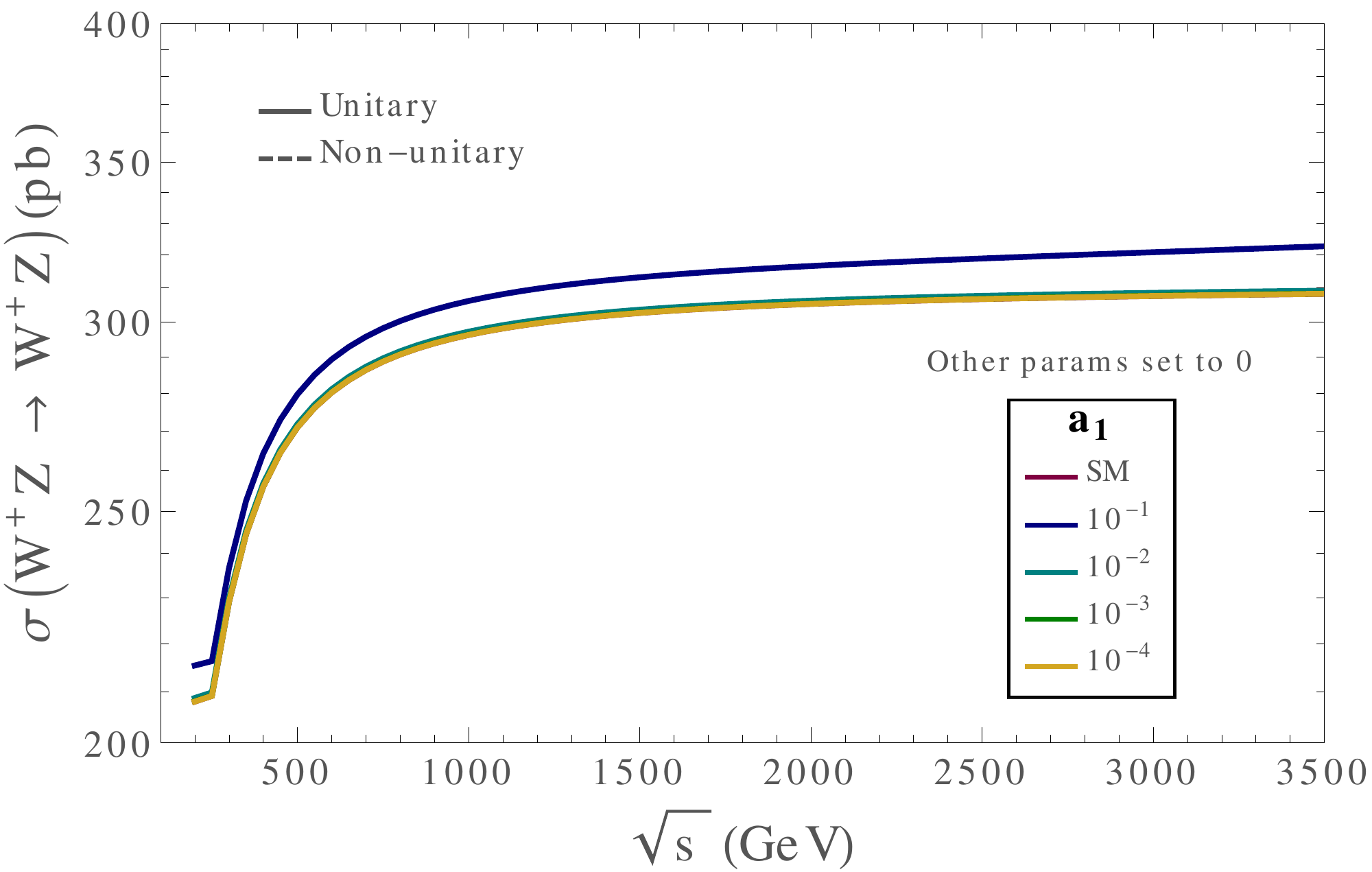}\\
\includegraphics[width=0.49\textwidth]{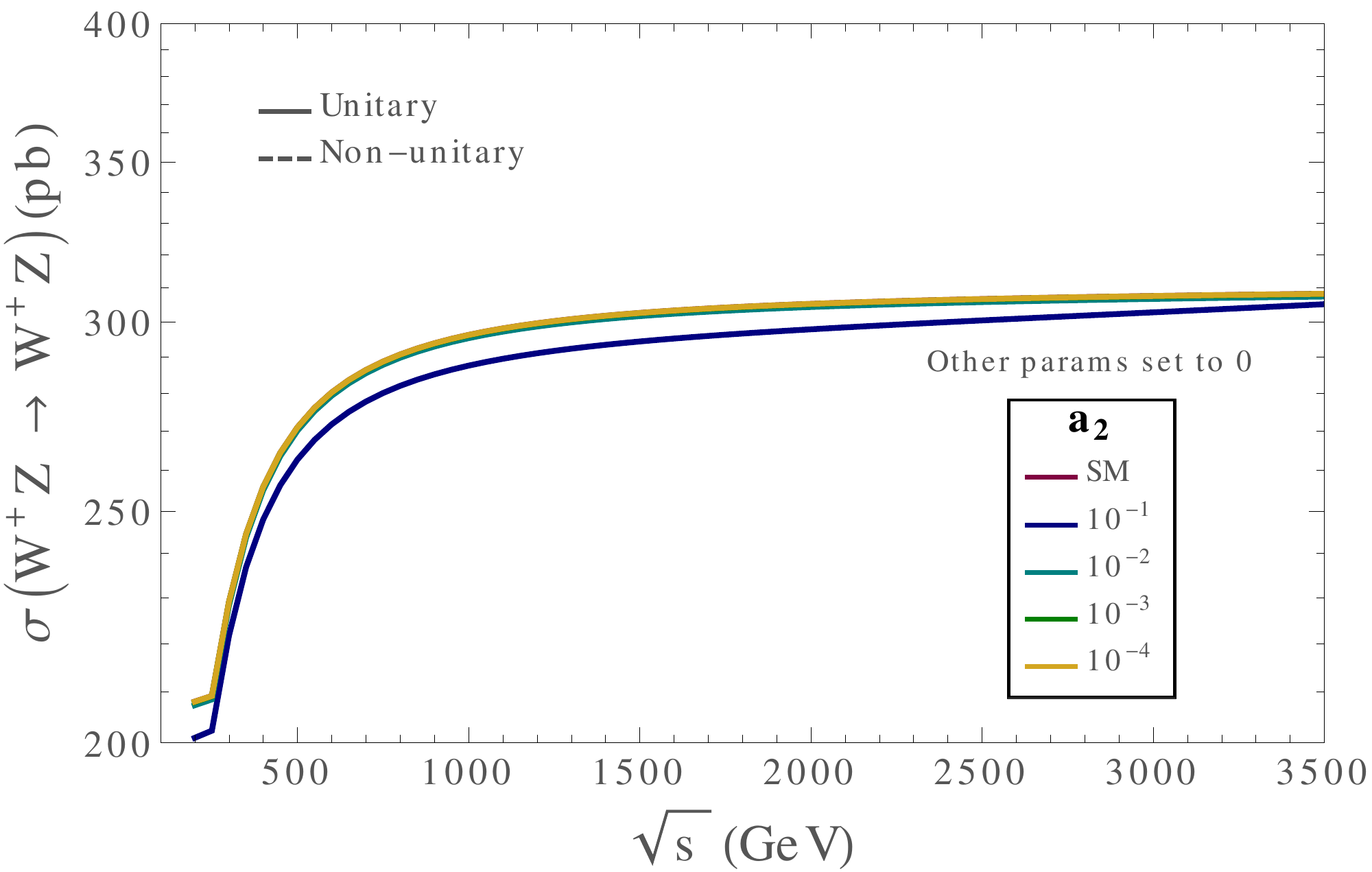}
\includegraphics[width=0.49\textwidth]{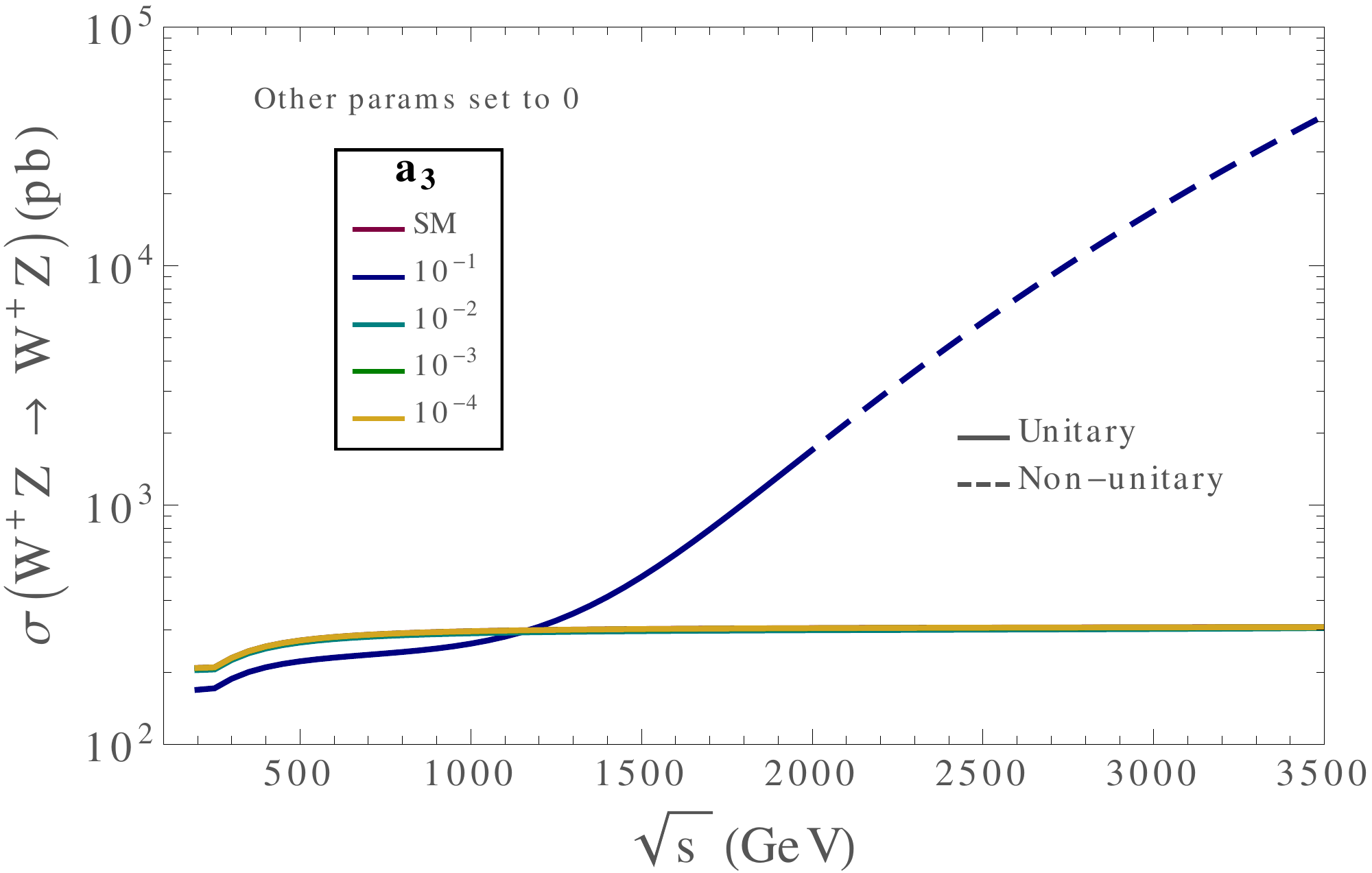}\\
\includegraphics[width=0.49\textwidth]{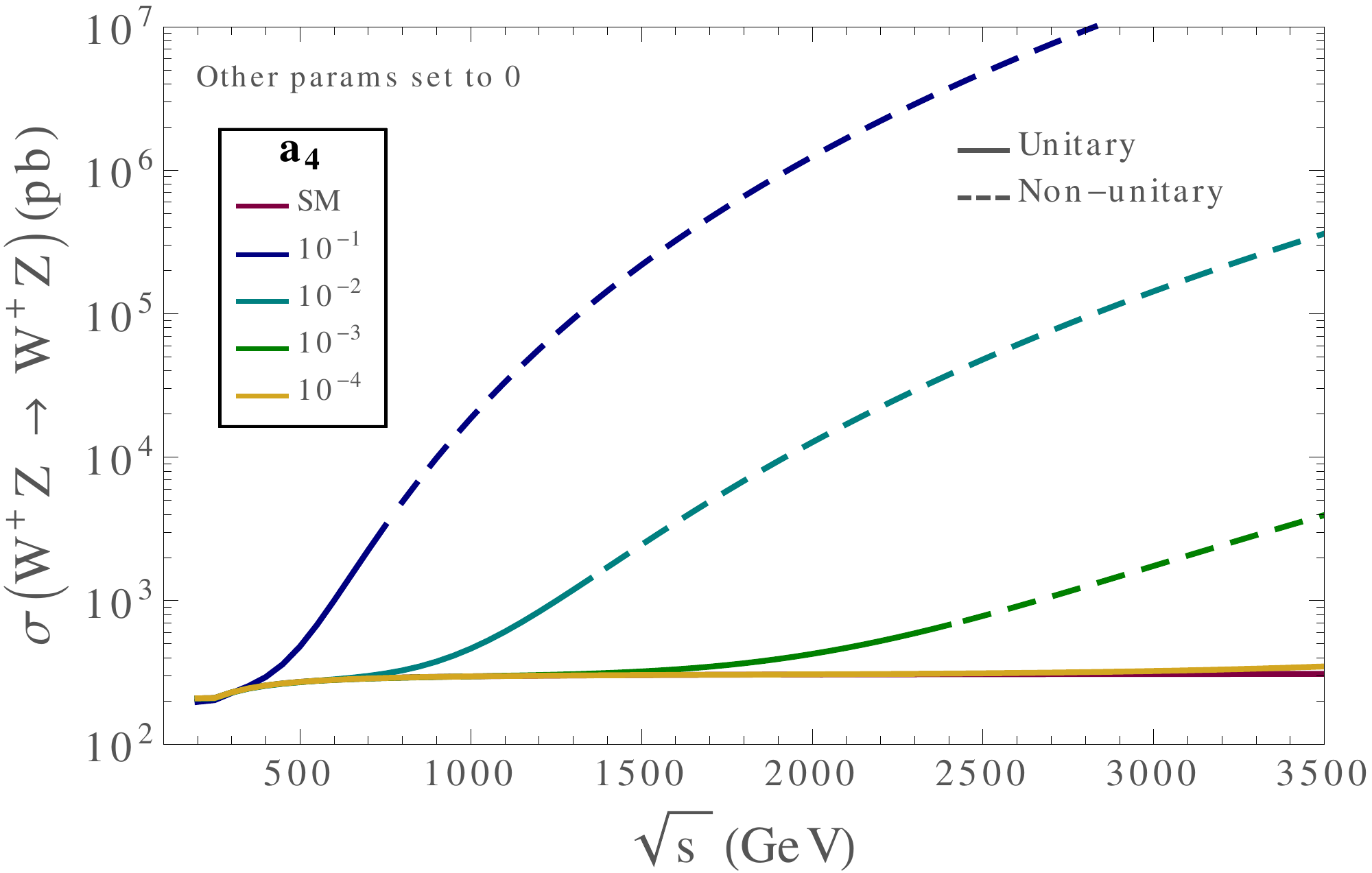}
\includegraphics[width=0.49\textwidth]{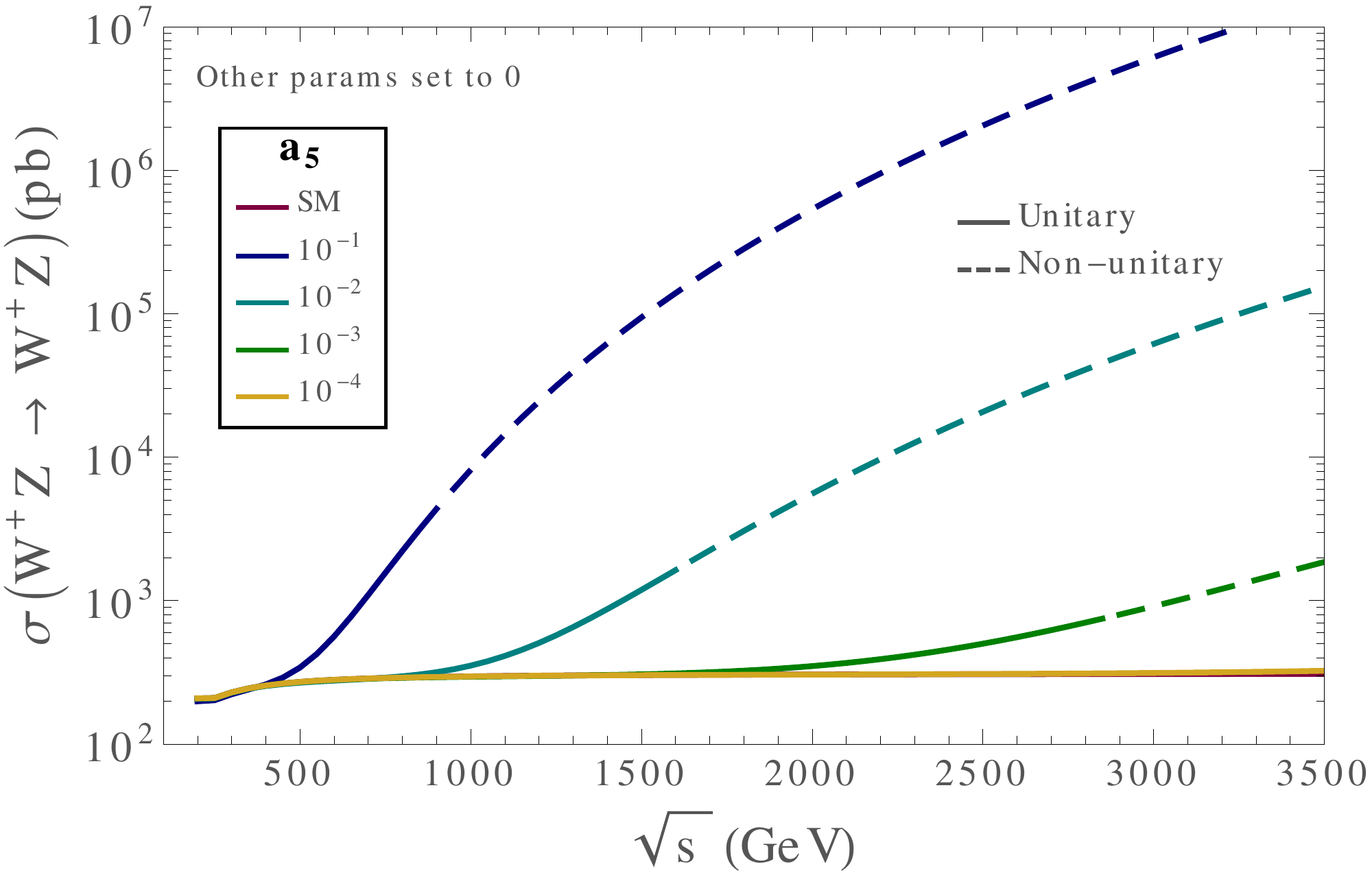}\\
\caption{Predictions of the total cross section of the process W${}^+$Z $\to$ W${}^+$Z as a function of the center of mass energy computed in the EChL framework for different values of one of the chiral parameters at a time. The rest are set to their SM value for a simpler comparison. From top to bottom and left to right $\Delta a=am1=a-1,~ a_1,~a_2,~a_3,~a_4$ and $a_5$ are varied, respectively. Solid lines represent a unitary prediction whereas dashed lines denote unitarity violating values.}
\label{fig:unitviolai}
\end{center}
\end{figure}

However, when any $IJ$ partial wave crosses the unitarity limit for a given set of values of the chiral coefficients, the EFT description cannot be trusted anymore. In this sense, from \figref{fig:violationunitarity}, it is easy to gain some intuition regarding the relation between the unitarity violation scale and the overall values of $a_4$ and $a_5$ (and, more mildly, of $am1$). For values of the chiral parameters restricted to $|a_{4,5}|<10^{-4}$ and $|a-1|<10^{-1}$ unitarity is respected for energies below 3 TeV, so we will reach the intrinsic energy scale of the EFT before the violation of unitarity takes place. Nevertheless, for larger values of $a_4$ and $a_5$, the picture changes. If one chooses $a_4,~a_5\sim 10^{-3}$, the partial wave moduli trespass the unitarity limit between 1.8 and 2.7 TeV. For values of $a_4,~a_5\sim 10^{-2}$, it happens even at lower energies; between 1 and 1.3 TeV. 

This can be seen even more clearly in terms of total cross sections instead of analyzing the partial wave behaviour. Selecting a particular VBS process, like, for instance, the W${}^+$Z $\to$ W${}^+$Z one, we can see the impact of each of the chiral parameters in the violation of unitarity at the cross section level.

To this aim we present in \figref{fig:unitviolai} the total cross section of W${}^+$Z $\to$W${}^+$Z scattering in the EChL at the tree level for different representative values of one parameter at a time, setting the rest of them to their SM values. In this case we have not considered the isospin limit but the full result. We show these cross sections as a function of the center of mass energy of the process and we mark the unitarity-violating predictions with dashed lines. The value of the energy at which each cross section overcomes the unitarity limit is chosen as the lowest one at which any of the corresponding $J$ and/or helicity  partial wave crosses the unitarity bound.

 In this figure it can be clearly seen that in this scattering process the parameters $\Delta a=am1=a-1,~ a_1$ and $a_2$ (upper left, upper right and middle left panels respectively) do not play a relevant role in the violation of unitarity, since there is no unitarity violation driven from these coefficients in the energy range that has been explored. Notice that the $b$ parameter, which controls the interaction between two EW gauge bosons and two Higgs bosons, does not appear in this scattering at tree level.

When the parameter $a_3$ is considered (middle right panel), however, cross sections show a unitarity violating behaviour in this same energy range. This happens only for large values of $a_3$, of the order of $10^{-1}$, for which unitarity is violated at around 2 TeV. Nevertheless, this size of $10^{-1}$ is already at the border of being in conflict with the EW precision data, so no unitarity violation driven from $a_3$ is expected below 3 TeV in this channel.

Moreover, it is clear that $a_4$ and $a_5$ are the most relevant parameters regarding the issue of the violation of unitarity, as we expected. If one takes a look at the two lower panels of \figref{fig:unitviolai} it is manifest that for values of these two parameters between $10^{-1}$ and $10^{-3}$, the violation of unitarity occurs well bellow 3 TeV. Actually, the crossing of the unitarity limit occurs, approximately, at the same corresponding energy values shown in \figref{fig:violationunitarity}, where the isospin limit was considered. 

Unitarity violation in VBS observables supposes a serious problem in the framework of the EChL, as we have seen. Non-unitary predictions cannot be contrasted consistently against experimental data since they do not respect the premise of probability conservation. For this reason, a prescription to cure this problem is needed. We leave the exhaustive analysis of various ways of solving the unitarity problem and their consequences to Chapters \ref{Methods}, \ref{Resonances} and \ref{ResonancesWW}, characterizing here only the main features of the violation of unitarity in VBS processes.
 
To summarize, in this section, it has been stated how the EChL predictions affect the various VBS observables at the subprocesses level. Depending on the values of the chiral parameters, especially of $a_4$ and $a_5$, significant deviations appear in the VBS cross sections within the EChL with respect to those obtained in the SM. Furthermore, the violation of unitarity can take place at energies below 3 TeV for experimentally allowed values of the chiral couplings. 

Nevertheless, in order to really understand the implication of the EChL new interactions, a subprocess level analysis is not enough. Since our main tool to study the scattering of vector bosons is the LHC, where they are produced and then re-scattered, a comprehensive exploration of VBS configurations at the LHC is on demand. We devote the next section of this Thesis to this interesting study.

\section{Vector Boson Scattering at the LHC}

In the previous Chapters we have stated our hypothesis that vector boson scattering observables should be the most sensitive ones to new physics in the EWSB sector. Besides, the TeV scale is motivated in the EChL, the effective theory we use to parameterize these new physics contributions. For these reasons VBS at the LHC is without a doubt one of the most fundamental parts of this Thesis, so its correct characterization is of great importance.

This section is aimed to introduce the basic notions of the VBS configurations at the LHC, paying special attention to their kinematical features. After all, these will be the ones that will allow us to disentangle the VBS topologies from the undesired backgrounds that also populate the LHC events. To this aim, we will employ the SM predictions since they should be enough to illustrate the main VBS features. Besides, they will be our main irreducible background, so it is important to have good control of them. Further on, in the forthcoming Chapters, the EChL predictions for VBS observables at the LHC will be studied in depth for each particular case, so we will show here just a preliminar example of how BSM physics can be distinguished from the SM in the context of VBS observables. For a more extensive review on VBS physics at the LHC see~\cite{Szleper:2014xxa,Rauch:2016pai,Anders:2018gfr,Bellan:2019xpr}.

\begin{figure}[t!]
\begin{center}
\includegraphics[width=.3\textwidth]{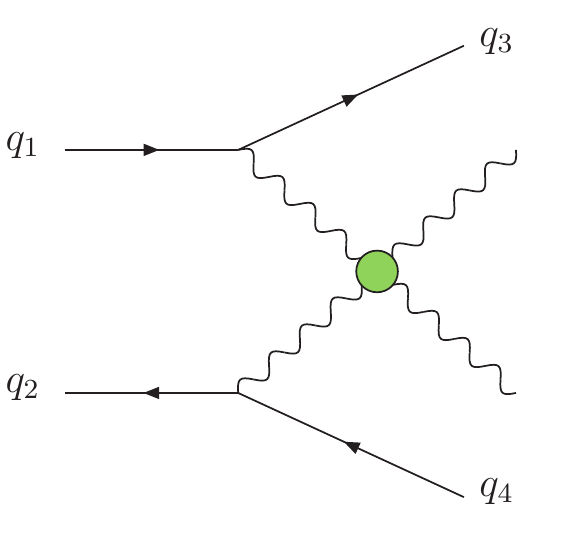}
\caption{Generic diagram of a vector boson scattering process at the LHC: $pp\to$ VV$jj$, more concretely  $q_1q_2\to$ VV$q_3q_4$. The green circle represents all the possible interactions among the four vector bosons. For instance, it shall contain the contributions depicted in Figs.~\ref{fig:diagramsWWWW}-\ref{fig:diagramsZZZZ}.}
\label{fig:DiagramVBS}
\end{center}
\end{figure}

VBS processes take place at the LHC as depicted, generically, in \figref{fig:DiagramVBS}. From the constituent quarks of the initial protons two EW gauge bosons are radiated. These vector bosons re-scatter leading to a final state containing two EW gauge bosons and two jets. The characteristics of the two final-state jets will be the key to select the VBS topologies from the multiple types of events that occur at the LHC. 

Although the various channels of VBS processes are interesting, we will mainly focus our attention on the W${}^+$Z channel, i.e.,  W${}^+$Z $\to$W${}^+$Z at the subprocess level. We have chosen this specific channel because of three main reasons: the first one is that when performing the LHC study, this channel, $pp \to$ W${}^+$Z$jj$, will suffer from less severe backgrounds than, for instance, the  $pp \to$ W${}^+$W${}^-jj$ one, as we will discuss in subsequent parts of this Thesis. In particular, the background process coming from gluon gluon one-loop contributions is absent in the WZ case.  The second one concerns the results provided in Chapters \ref{Resonances} and \ref{ResonancesWW}, where an analysis of the production at the LHC of vector resonances dynamically generated from the EChL strong interactions is performed. In order to do so, the WZ channel is a very promising window, as, due to its quantum numbers, it will have a $s$-channel contribution from the charged vector resonances, such as the ones we aim to study. Furthermore, it will not have any other contribution from other resonances, like for instance a scalar one. The third reason is that positively charged channels benefit from larger rates than negatively charged ones, in general. The probability of radiating a W${^+}$ from a proton is higher than that of radiating a W${}^-$ and the latter is also slightly higher than that of radiating a Z. Thus, taking into account the subprocess cross sections, the $pp \to$ W${}^+$W${}^+jj$ channel will be the one with the largest LHC cross section, whereas the $pp \to$ ZZ$jj$ will be the one with the smallest LHC cross section. 

In this section we will, therefore, revisit the main properties of the $pp \to$ W${}^+$Z$jj$ process at the LHC as an example to illustrate the generic features of the VBS topologies. We will introduce its specific kinematical properties with the aim of recognizing this type of events from the ones that will conform the backgrounds, commenting on the different polarization channels of the EW gauge bosons involved. Finally, we will devote the last part of this section to briefly review the current experimental status of VBS searches at the Large Hadron Collider. 

\subsection{Kinematics of vector boson scattering processes}

Apart from the fact that, theoretically speaking, VBS observables should be the most relevant ones to look for BSM physics in the EWSB sector, these configurations have another important advantage. Their kinematical properties result to be very characteristic, especially those of the two final jets.

\begin{figure}[t!]
\begin{center}
\includegraphics[width=0.5\textwidth]{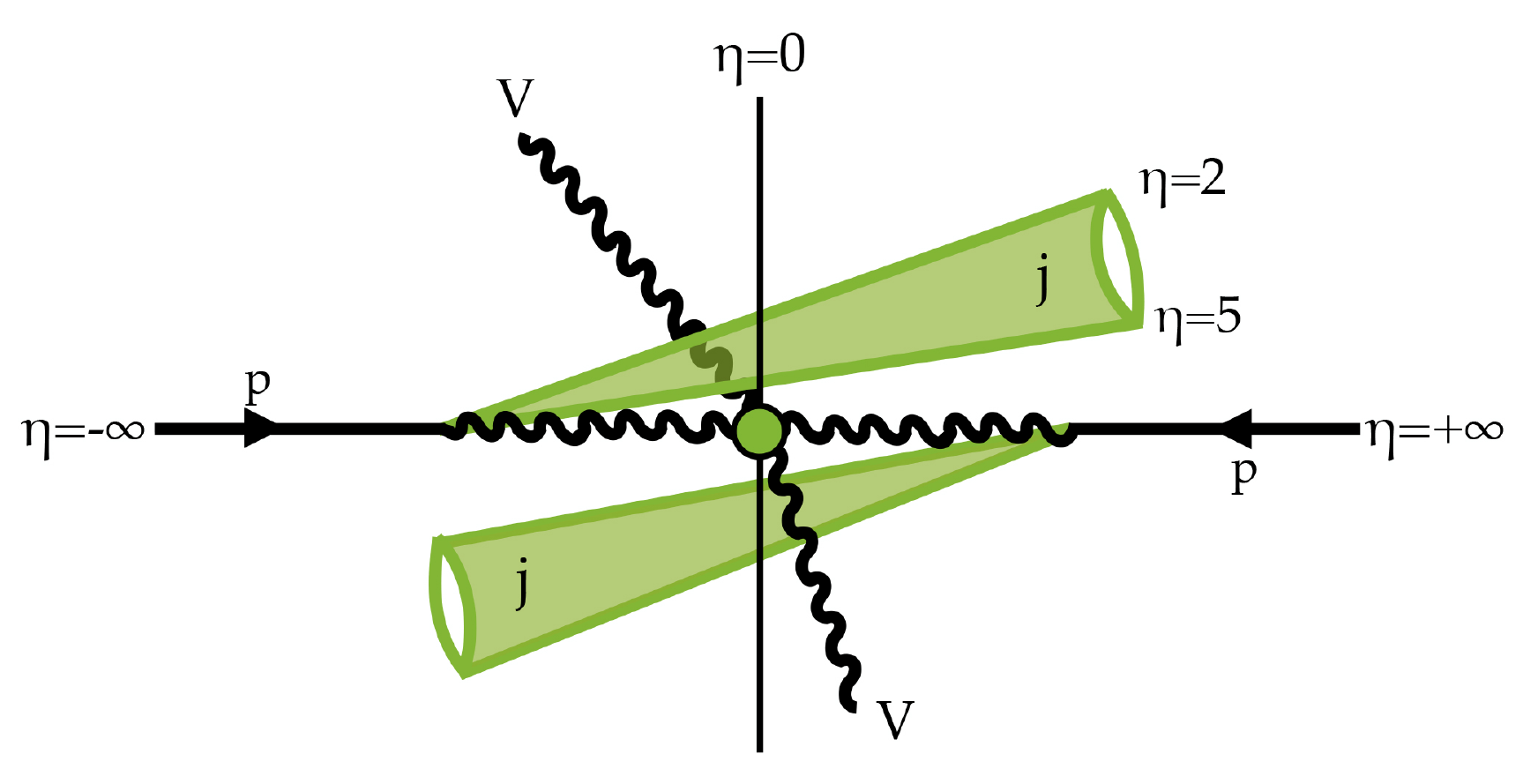}
\caption{Charateristic kinematical configuration of the final state particles of a VBS process $pp\to$ VV$jj$ at the LHC. The Figure represents the longitudinal section of the detector.}
\label{fig:VBSconfig}
\end{center}
\end{figure}

Due to the radiation of the EW gauge bosons, the final-state jets result to be very forward/backward. They lie typically in two opposite-sign pseudorapidity cones of $\eta$ values between 2 and 5, as shown in \figref{fig:VBSconfig}. In contrast, the produced vector bosons tend to populate the central region of the detector, i.e., small values of the pseudorapidity. Furthermore, the invariant mass of the two extra jets is usually larger than in other processes. This means that the behaviour of the differential cross section as a function of the invariant mass of the dijet system does not show a steep behaviour, but a rather flat one. 
 
These two features, the presence of the two final jets in determined cones of opposite-sign pseudorapidities and the fact that their invariant mass tends to be large, define the most characteristic kinematics of the VBS topologies, since no other subprocess shares these same exact properties. 

\begin{figure}[t!]
\begin{center}
\includegraphics[width=.49\textwidth]{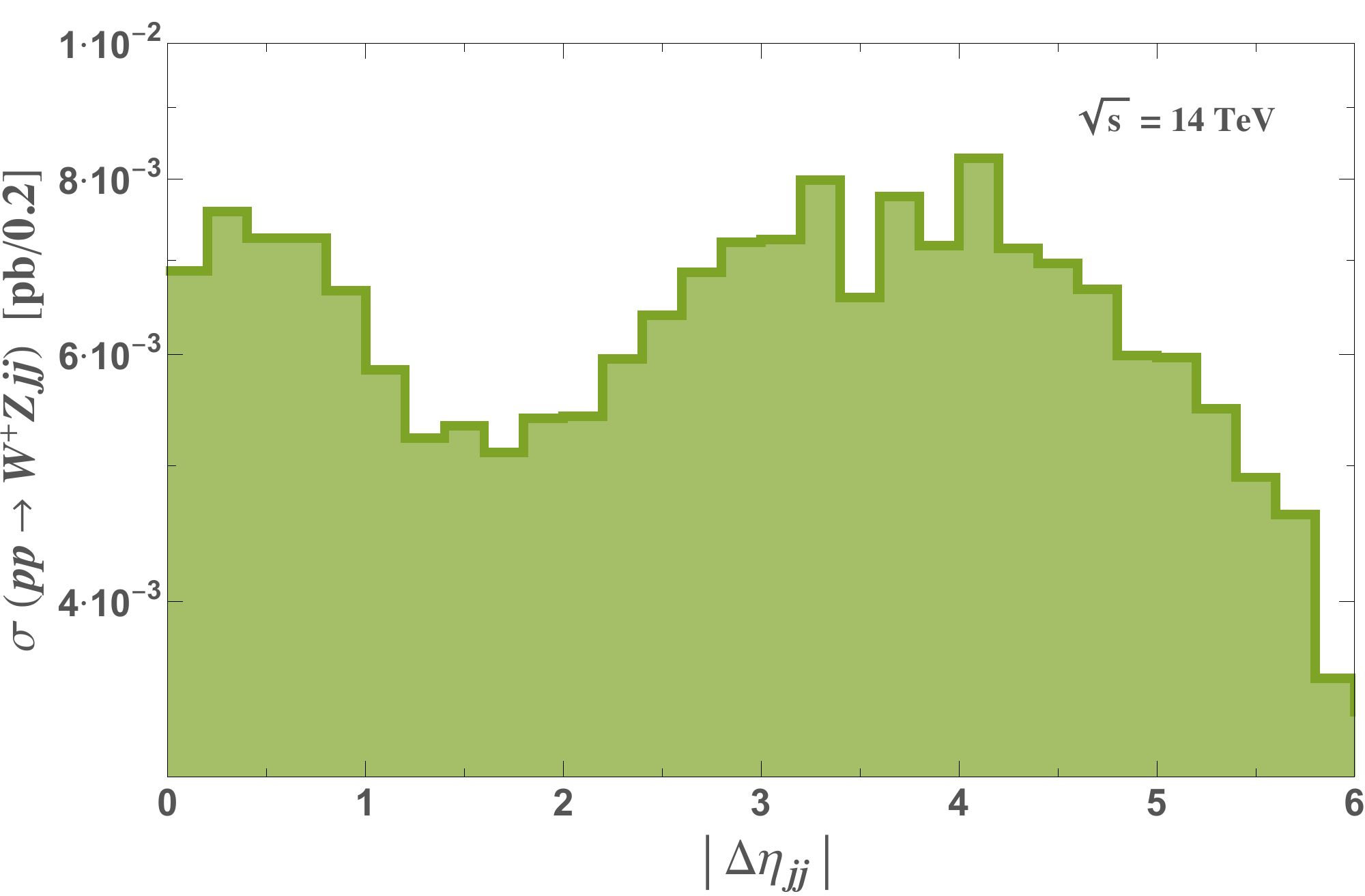}
\includegraphics[width=.49\textwidth]{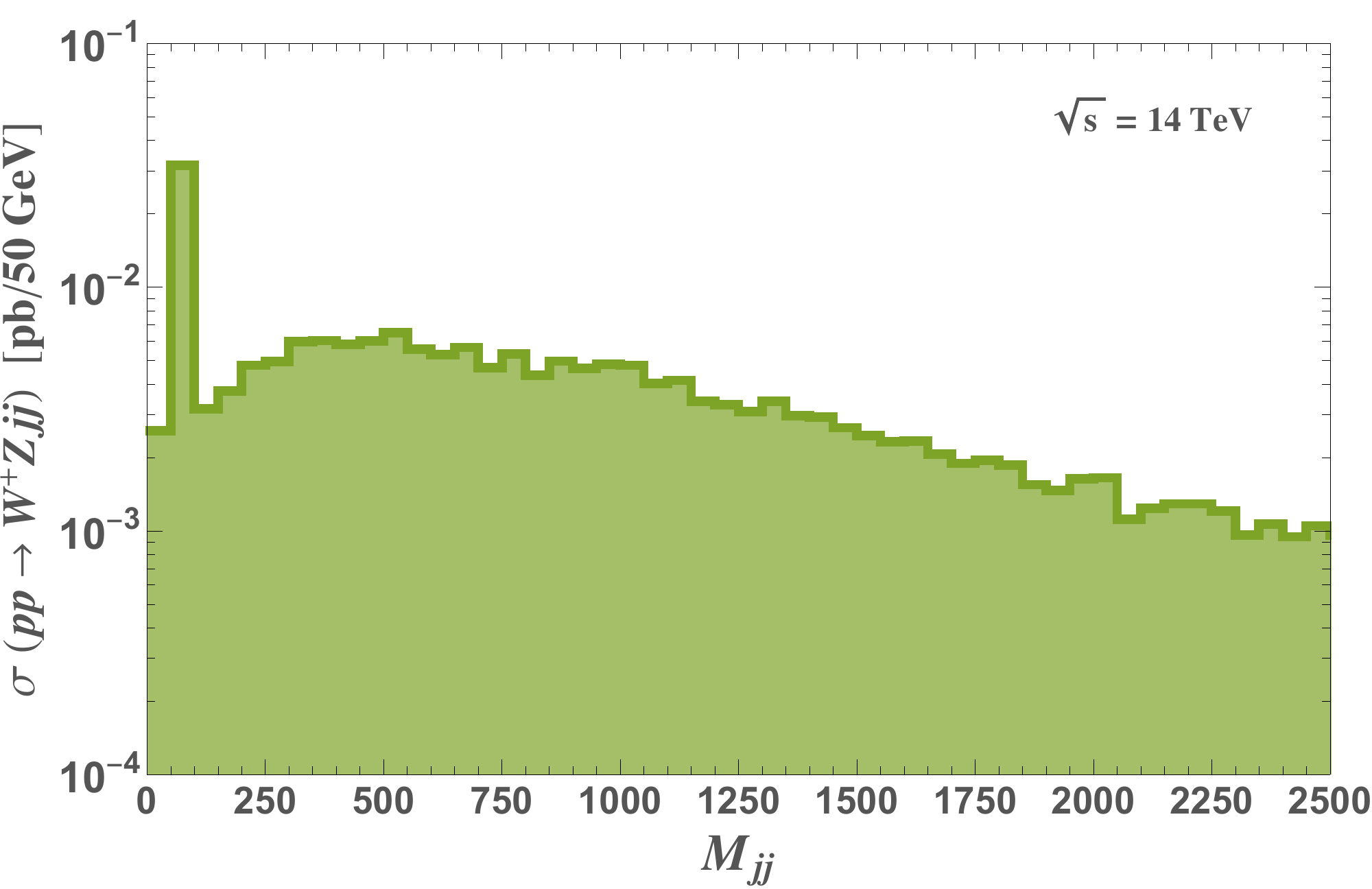}
\caption{SM distributions of $pp\to$ W${}^+$Z$jj$ events with the pseudorapidity separation of the two jets, $\Delta\eta_{jj}$ (left) and with the invariant mass of the $jj$ pair, $M_{\rm jj}$ (right). Basic kinematical cuts, $|\eta_{j_1,j_2}|<5\,,~|\eta_{W,Z}| < 2\,,~p_{T_{j,W,Z}}>20$ GeV, have been imposed. }
\label{fig:VBSkin}
\end{center}
\end{figure}

In \figref{fig:VBSkin} these peculiarities can be seen in terms of the cross section distributions of the EW production, i.e., $\mO(\alpha^2)$ at the amplitude level, of the $pp\to$ W${}^+$Z$jj$ process. Other production modes, such as the ones involving QCD interactions, of the order of $\mO(\alpha_s\alpha)$ will be treated as backgrounds, so we will not discuss them in this subsection. We will comment on them in posterior parts of this Thesis. In the left panel of this Figure, the distribution of 10000 Monte Carlo events generated with MadGraph5 (MG5)~\cite{Alwall:2014hca,Frederix:2018nkq} is shown as a function of the difference in pseudorapidity of the two final jets, defined as 
\begin{align}
\Delta\eta_{jj}\equiv|\eta_{j_1}-\eta_{j_2}|\,.
\end{align}
Since these particles are expected to be in opposite sides of the detector, very localized in the two mentioned pseudorapidity cones, the distribution should peak at large differences of pseudorapidity, which it does indeed (see left panel in \figref{fig:VBSkin}). There is a peak at low $\Delta\eta_{jj}$ values, corresponding to the central part of the detector, that comes mostly from the non-VBS diagrams that contribute to this process. These have to be taken into account in the computation since the VBS contributions alone are not gauge invariant by themselves. However, as we say, a second, larger peak appears at $\Delta\eta_{jj}\sim[2.5,5]$, where the bulk of the events lie, meaning that a large amount of the total cross section come from the VBS topologies. 

In the right panel of \figref{fig:VBSkin} the distribution of the same MG5 events is presented, this time as a function of the invariant mass of the final two jets, $M_{jj}$. The enhancement of the cross section in the second bin corresponds to the W and Z resonances, i.e., to the contributing diagrams in which the two extra jets come from the decay of one of these gauge bosons. Apart from that, it is clear that the cross section falls very mildly with this variable, making it a useful discriminant to select this kind of topologies. 

In all the plots presented in this Section, the center of mass energy of the proton-proton system has been set to $\sqrt{s}=14$ TeV and a set of minimal cuts
\begin{align}
|\eta_{j_1,j_2}|<5\,,~~~|\eta_{W,Z}| < 2\,,~~~p_{T_{j,W,Z}}>20 ~{\rm GeV}\,,
\end{align}
 has been imposed to the final state particles to ensure their detection.
 
The particular selection cuts that will allow us to disentangle the VBS processes from the undesired backgrounds would inevitably depend on the kinematics of the latter. For this reason, when studying the concrete scenarios of the following Chapters the VBS selection cuts will be discussed again for each concrete case, since different searches will be performed in different decay channels of the final EW gauge bosons. 
 
 The characteristics of the final jets are very important from the point of view of discriminating between the VBS contributions to an observable and the rest. However, the BSM signals will be related to the EW gauge boson subsystem. In the previous section the most clear deviations with respect to the SM were seen, mainly, in the  behaviour of the cross sections with the center os mass energy. At the LHC, this points out towards the invariant mass of the final diboson system as the proper variable to look for these deviations.

Furthermore, the study of different polarizations of the final gauge bosons (and, thus, of the initial ones due to the approximate polarization conservation shown in the previous section) would suppose an amazing strategy to get an insight of the EW Goldstone boson dynamics at the LHC. Unfortunately, it is still very challenging to measure experimentally the polarization of the vector bosons, but theoretical developments have been carried out in this regard, see, for instance~\cite{Fabbrichesi:2015hsa,Ballestrero:2019qoy}. Besides, measurements of EW gauge boson polarization fractions have been also achieved at the LHC~\cite{Aaboud:2019gxl}, albeit not in VBS configurations. This latter are, however, expected  in the near future of the LHC.

\begin{figure}[t!]
\begin{center}
\includegraphics[width=.49\textwidth]{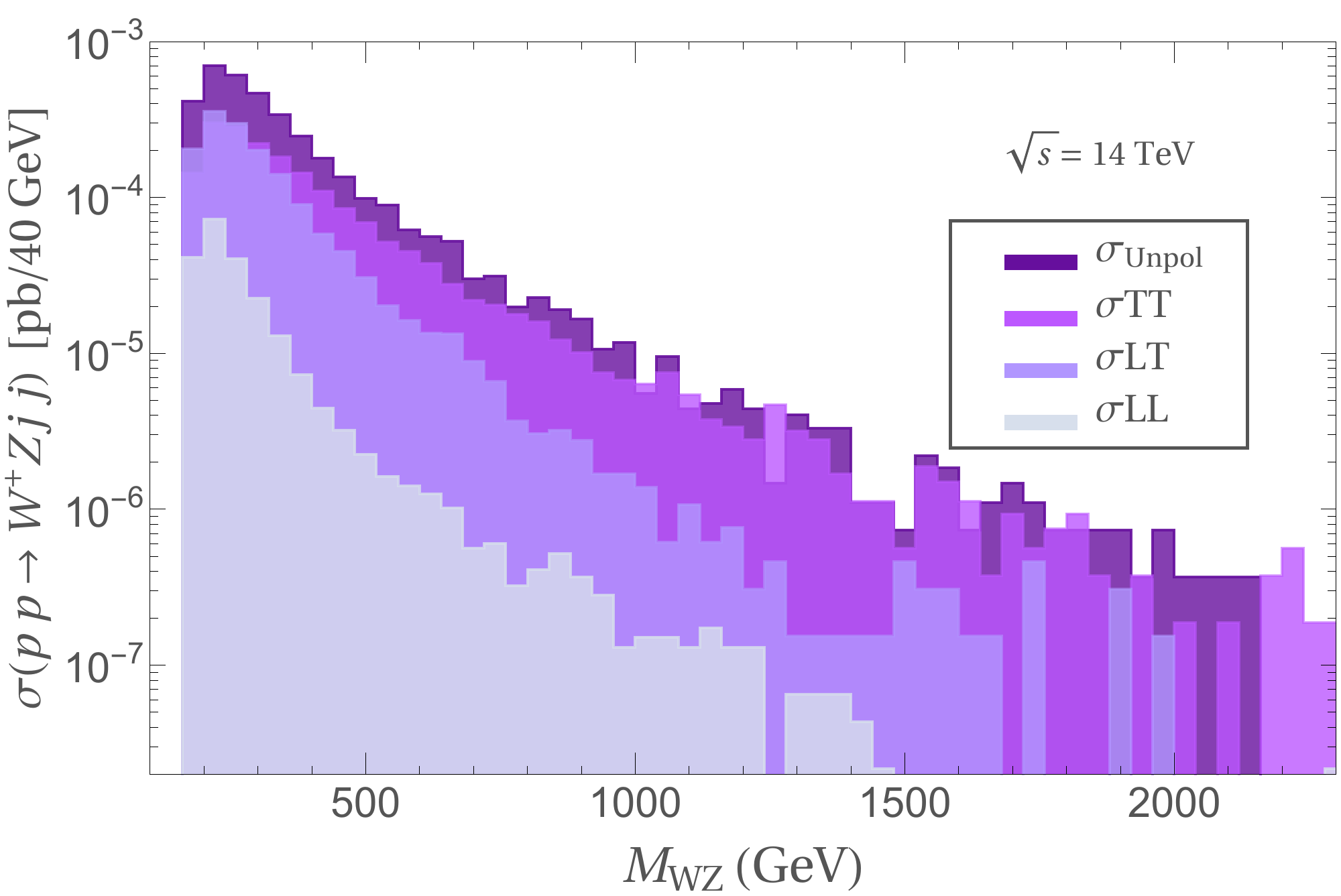}
\caption{SM distributions of $pp\to$ W${}^+$Z$jj$ events with the invariant mass of the WZ pair, $M_{\rm WZ}$. The predictions for the various polarizations $\sigma_{AB}$ of the final $\text{W}^{}_A\text{Z}^{}_B$ pair as well as the total unpolarized result, $\sigma_{\rm Unpol}$, are displayed separately, for comparison. The following kinematical cuts have been imposed: $|\eta_{j_1,j_2}|<5\,,~\eta_{j_1} \cdot \eta_{j_2} < 0\, {\rm and}~|\eta_{W,Z}| < 2$.}
\label{fig:MWZgen}
\end{center}
\end{figure}

Having this in mind, we present in \figref{fig:MWZgen} the cross section distributions of 10000 $pp\to$ W${}^+$Z$jj$  events simulated with MG5 as a function of the invariant mass of the WZ system, $M_{WZ}$ for different polarizations of the final\footnote{The polarizations of the intermediate weak bosons are summed over in MadGraph since the intermediate states are generally considered off shell in this event generator. } EW gauge bosons. A minimal set of cuts that select mainly events which have two opposite-sided large pseudorapidity jets together with two gauge bosons, $W$ and $Z$ within the acceptance of the detector, corresponding to 
\begin{align}
|\eta_{j_1,j_2}|<5\,,~~~\eta_{j_1} \cdot \eta_{j_2} < 0\,,~~~~|\eta_{W,Z}| < 2\,,\label{VBScutsgen}
\end{align}
 has been imposed in this Figure.

One can observe that the fact that the transverse modes largely dominate in the total cross section is still true in the case in which we have protons in the initial state. Also, although it is a very mild effect, the slope of the tails in the distributions with $M_{WZ}$ is different for each polarization, being slightly steeper when having two transversely polarized final gauge bosons. Regarding the falling, we can understand its behaviour due to the fact that the total cross section for the VBS contribution to $pp\to W^+Zjj$, can be approximately estimated as the one of the subprocess convoluted with the probability functions of the W${}^+$ and the Z, that, in a similar way as the Parton Distribution Functions (PDFs), give the probability of radiating a weak boson from a proton at a certain momentum fraction. Thus, since even the unpolarized, total cross section does not show a very steep behaviour with energy, the distribution with $M_{WZ}$ in VBS configurations is expected to be flatter than in other processes, due to its inheritance from the VBS subprocesses properties. 

\begin{figure}[t!]
\begin{center}
\includegraphics[width=.49\textwidth]{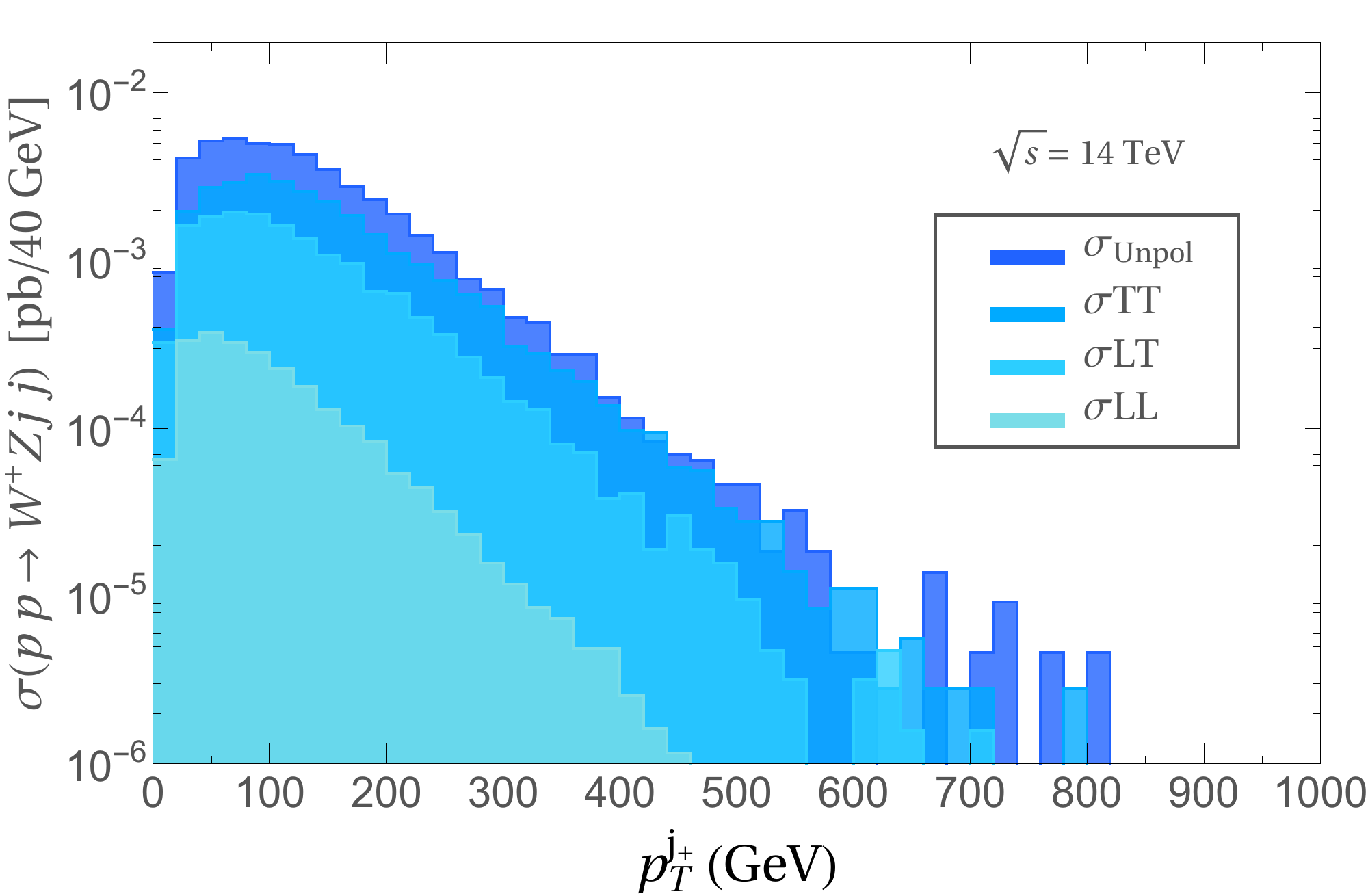}
\includegraphics[width=.49\textwidth]{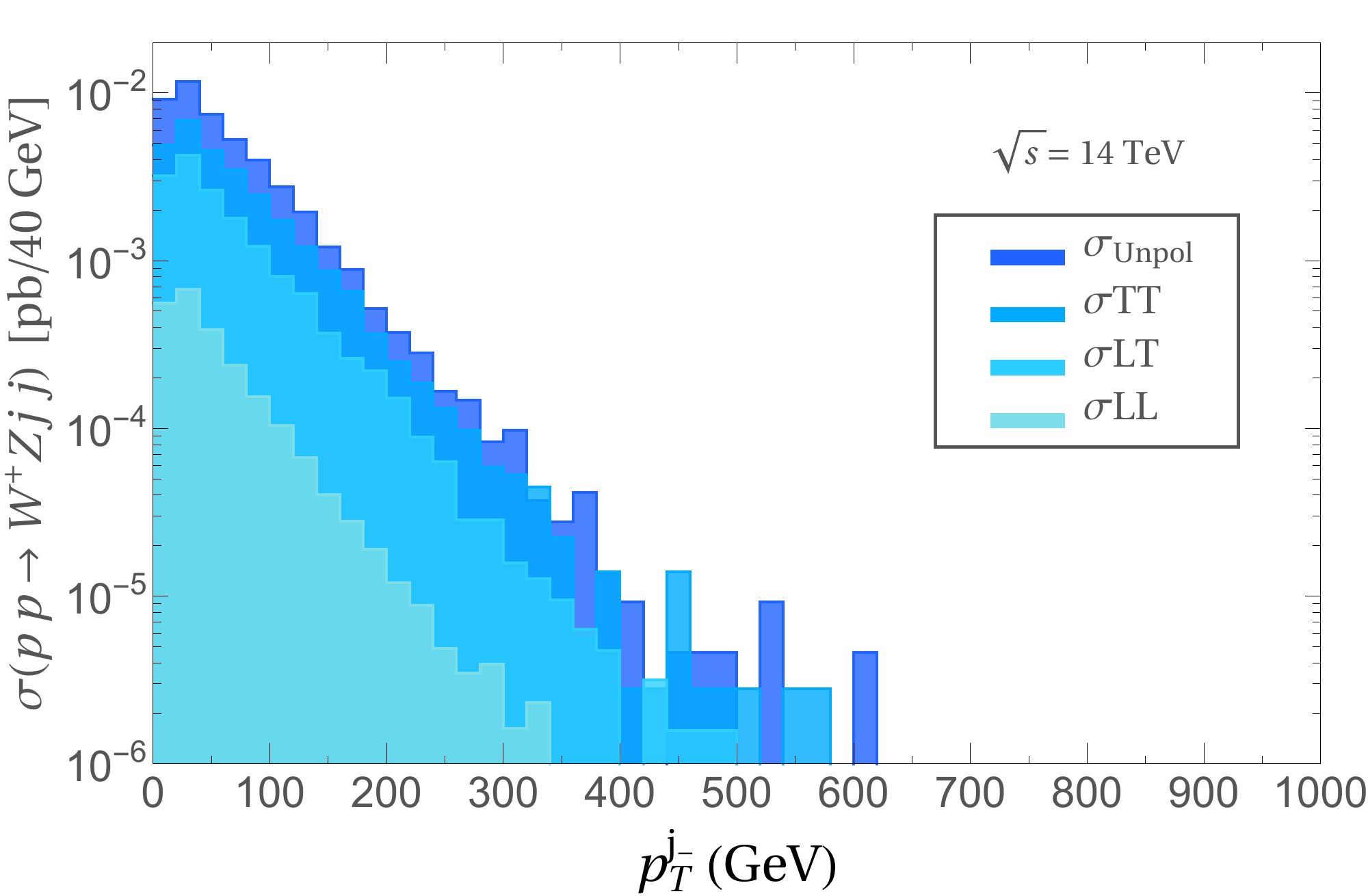}
\caption{SM distributions of $pp\to$ W${}^+$Z$jj$ events with the transverse momentum of the most energetic jet, $p_T^{j_+}$ (left panel) and with the transverse momentum of the least energetic jet $p_T^{j_-}$ (right panel). The predictions for the various polarizations $\sigma_{AB}$ of the final $\text{W}^{}_A\text{Z}^{}_B$ pair as well as the total unpolarized result, $\sigma_{\rm Unpol}$, are displayed separately, for comparison. The following kinematical cuts have been imposed: $|\eta_{j_1,j_2}|<5\,,~\eta_{j_1} \cdot \eta_{j_2} < 0\, {\rm and}~|\eta_{W,Z}| < 2$.}
\label{fig:ptjgen}
\end{center}
\end{figure}

 It is interesting to comment as well on the two distributions in the transverse momentum of the jets, $p_T^j$ (also denoted by $p_{T_j}$ in this Thesis), shown in \figref{fig:ptjgen} for the different polarization states of the W and the Z. The left panel shows the Monte Carlo event distribution of the W${}^+$Z process as a function of the leading jet, $p_T^{j_+}$, defined as the one with larger modulus of the transverse momentum, whereas the right panel corresponds to the distribution in the subleading jet, $p_T^{j_-}$, the one with less $p_T$.  It is then understandable that the distributions with respect to $p_T^{j_+}$ peak at a higher value than those with respect to $p_T^{j_-}$. The same cuts given in \eqref{VBScutsgen} have been applied and the center of mass energy has been again fixed to $\sqrt{s}=14$ TeV.
 
From \figref{fig:ptjgen} it is plain that the various polarization channels behave differently. The longitudinal distributions tend to peak at lower $p_T$ values than the transverse ones, whose dominance remains clear. This can be understood by the fact that longitudinally polarized vector bosons tend to be emitted at a smaller angle with respect to the quark that has radiated them, and hence at smaller transverse momentum with respect to the proton beam, than the transversely polarized ones. As a consequence, the final quark (and thus the final jet) accompanying a longitudinal gauge boson is more forward than the one accompanying a transverse W or Z. This translates into different $p_T^j$ distributions. Whereas the ones coming from events with transverse gauge bosons tend to peak closer to the EW boson mass, the ones with longitudinally polarized W or Z  peak normally around half of the EW boson mass.

These features are very interesting regarding future prospects of polarization studies. As we have argued, being able to disentangle the polarization of the gauge bosons in the final state will be enormously helpful to discriminate signal versus background in these scenarios and to access more directly the interactions among Goldstone bosons. Indeed, a more detailed study of the relevant kinematical variables to perform this kind of discrimination deserves some future development, although there are already some analysis in this direction, as it has already been pointed out. However, as sophisticated techniques to distinguish  the polarizations of the final W and Z are not yet well established, we are not going to use a polarization analysis as a discriminant in this Thesis. 

With this comment about the issue of measuring the different polarizations of the EW we finalise the charaterization of the VBS processes at the LHC. Nevertheless we have explored only the SM predictions to get some intuition about the properties of the VBS topologies. To conclude this section, we find pertinent to include an example of a BSM scenario in this context, although specific EChL signals will be presented through more devoted studies in the forthcoming Chapters.

We have chosen the previously introduced Higgsless model to be this example. Although this model is currently ruled out by experiments it is illustrative to control its prediction for the VV scattering, as it will be somehow an extreme reference model among the possible strongly interacting scenarios for EWSB.

\begin{figure}[t!]
\begin{center}
\includegraphics[width=0.49\textwidth]{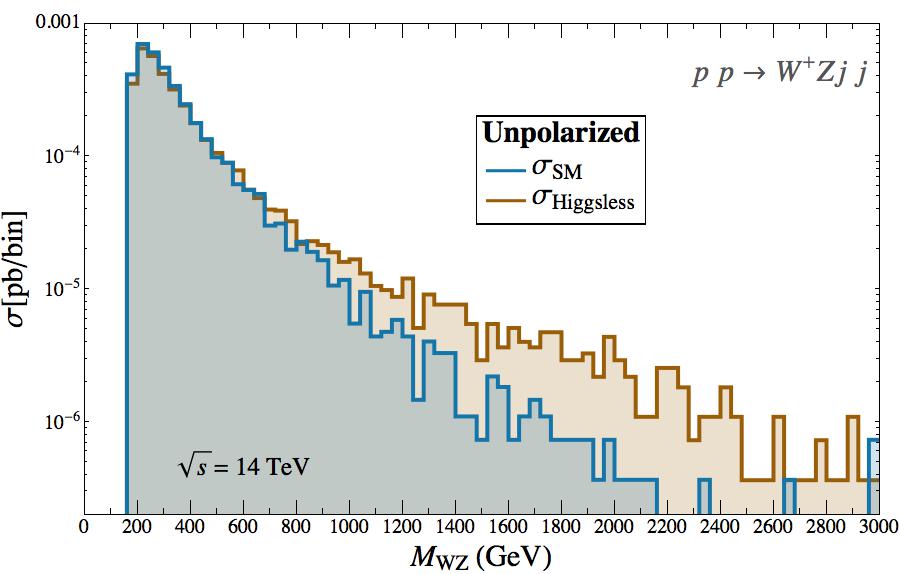}
\includegraphics[width=0.49\textwidth]{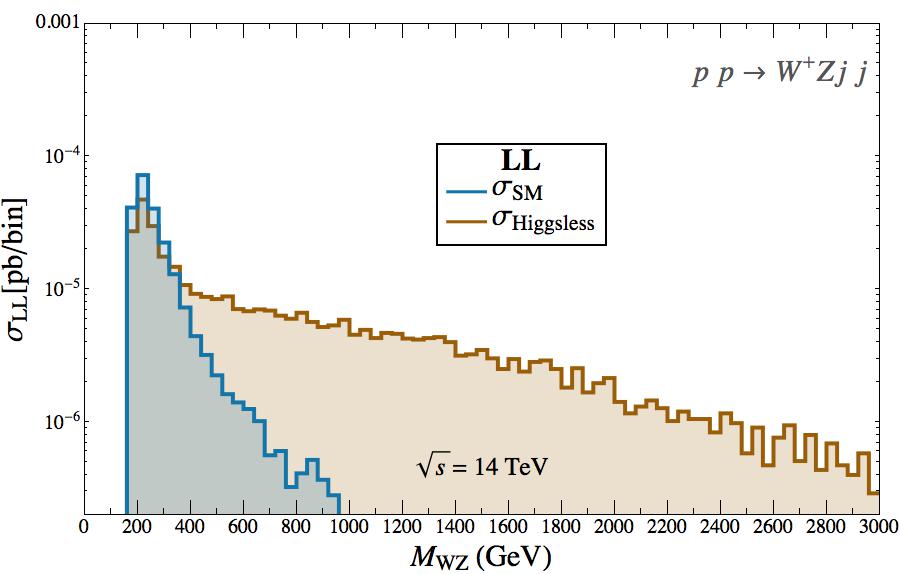}
\includegraphics[width=0.49\textwidth]{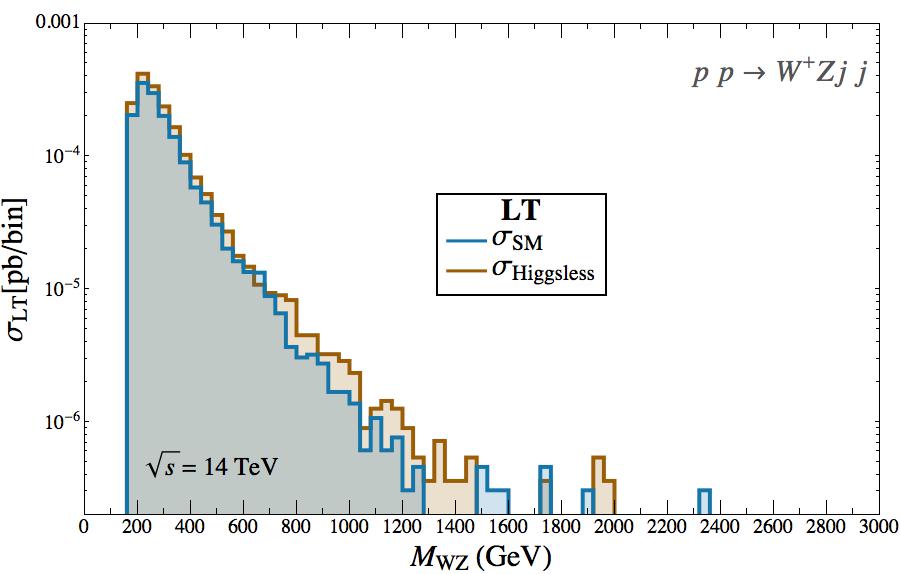}
\includegraphics[width=0.49\textwidth]{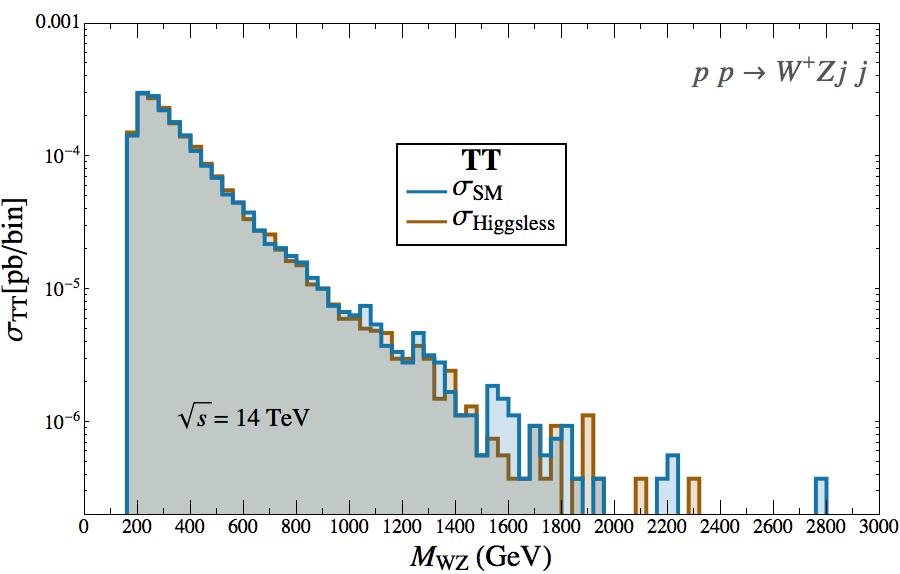}
\caption{ Predictions of the cross section per bin of the process $pp\to$ W${}^+$Z$jj$ from the Higgsless model, i.e., the EChL with $a=b=0$, $a_i=0$, and for the SM. They are computed for the unpolarized (upper left), longitudinally polarized (upper right), longitudinally and transversely polarized (lower left) and transversely polarized (lower right) final gauge bosons as a function of the invariant mass of the WZ pair, $M_{WZ}$. Kinematical cuts have been implemented: $|\eta_{j_1,j_2}|<5\,,~\eta_{j_1} \cdot \eta_{j_2} < 0\, {\rm and}~|\eta_{W,Z}| < 2$ . The center of mass energy is set to 14 TeV. Binsize of 40 GeV.}\label{fig:higgsless}
\end{center}
\end{figure}

Using MG5, we have computed the cross section per bin of the process $pp\to$ W${}^+$Z$jj$ as a function of the invariant mass of the WZ pair for the different polarization states of the weak gauge bosons within the Higgsless model and within the SM, for comparison. The results are displayed in \figref{fig:higgsless}. It is very interesting to notice that, indeed, the greater modifications with respect to the SM appear in the longitudinal modes, as expected. Moreover, they emerge in the total cross section as well, despite the dominance of the transverse modes that do not suffer any significant modification. 

This means that, at high energies, deviations with respect to the SM predictions introduced by the EChL operators might be seen at the LHC in VBS observables, even if the polarizations of the final gauge bosons cannot be precisely determined. This is indeed a very motivating result that will be exploited throughout posterior parts of this Thesis. 

Some last comments have to be made at this point, in order to conclude a complete review of VBS at the LHC. The first one concerns the size of the VBS cross sections within the SM. They are in general smaller than those of other (even EW mediated) processes taking place as well at the LHC like Drell-Yan configurations or single top production. This is due to the 
large number of powers of the weak coupling $g$ involved in the VBS diagrams. The total production cross section of two vector bosons and two jets amounts to, roughly, 0.1-1 pb. 

Nevertheless, since the W and Z are unstable particles they are detected through their decay products. Their decays suppose another source of suppression related, again, to insertions of the weak coupling $g$. For instance, in the purely leptonic case, which is the cleanest final state in a hadronic collider, an extra factor of ${\rm BR}({\rm W}\to \ell\nu)\times {\rm BR}({\rm Z}\to \ell\ell)\sim 0.03$ has to be added to the rates shown in the present section. In the purely hadronic case, on the contrary, the rates are larger, since the cross sections diminish only a factor ${\rm BR}({\rm W}\to {\rm hadrons})\times {\rm BR}({\rm Z}\to {\rm hadrons})\sim 0.5$, but the final jets coming from the EW gauge boson decays are more difficult to select from the unwanted backgrounds. This trade between both search strategies depending on the decays of the Ws and Zs, as well as the semileptonic channel, will be discussed in the next subsection.

The second comment is that of next to leading order corrections to VBS processes, both from QCD and from EW interactions. These higher order contributions are of great importance in precision physics, since small theoretical uncertainties are needed to extract accurate information when comparing the data with the corresponding predictions. Recent works have obtained the NLO corrections~\cite{Biedermann:2017bss}, both QCD and EW, of some VBS processes, relying, particularly, in the purely leptonic scenario. They have found that in the most extreme case, and for some kinematical regions, these contributions can amount to as much as 30\% of the total LO cross section in the SM. The NNLO cross section of diboson production processes (not particularizing, however, in VBS topologies) has been also calculated. In any case, these computations result to be a difficult task, due to the six fermion final state of these processes. Because of this, a factorization process is often employed.

It is well known that VBS can be simulated with good accuracy by assuming that the initial gauge bosons are radiated collinearly from the proton-proton system to posteriorly re-scatter on-shell. This is known as the effective W approximation (EWA), developed for the first time in \cite{Dawson:1984gx,Johnson:1987tj}, that translates the Weisz\"{a}cker-Williams~\cite{vonWeizsacker:1934nji,Brodsky:1971ud} approximation for photons to the case os massive EW gauge bosons. This framework has two important advantages: the first one is that it has the intuitive physical interpretation of the distribution functions of the W and the Z as the PDFs in the parton model, and the second one is that it is  computationally simple and, as we will see in forthcoming parts of this Thesis, leads to very good results. 

The EWA provides  probability  functions, $f_{W,Z}(x)$, for the W and the Z that describe the probability of the EW gauge boson to be radiated collinearly from a fermion carrying a  fraction $x$ of its total momentum. In order to get the total cross section at the LHC for the full process that starts with protons, these functions, taking quarks as the mentioned fermions, are then convoluted with the PDFs of the quarks and with the corresponding subprocess cross section for the scattering of on-shell EW gauge bosons. Thus, the total process $pp\to$ VV$jj\to f_1f_2f_3f_4jj$ can be computed in three steps: first, the initial protons radiate the two initial gauge bosons on-shell, then, these re-scatter, and, finally, using the narrow width approximation, they decay. Factorizing the process in such a way allows for easier computations that contain the main features of the VBS observables. This procedure will be studied in Chapter \ref{Methods}, as well as the accuracy of its predictions when compared with the full result from MG5.  

With the revision of the main relevant features of VBS at the LHC we finalise this subsection. The next one is aimed to complement the information presented here by reviewing the experimental status of VBS searches at the LHC. 

\subsection{Experimental searches on vector boson scattering observables}

Measuring accurately the properties of VBS observables is one of the main tasks of the ATLAS and CMS experiments at the LHC. Since the cross sections of these processes are somewhat small compared to others, of the order of a few femtobarns once the gauge bosons have decayed, high energies are required to access this kind of physics. Thus, the LHC is the ideal place to look for deviations in VBS observables coming from BSM physics in the EWSB.

Other kinds of colliders, such as leptonic $e^+e^-$ ones, can also be useful to test the EW gauge boson self-interactions. This was, for instance, the case of LEP, that could access this kind of physics through precision observables~\cite{Falkowski:2013dza,Fabbrichesi:2015hsa}. Future linear colliders such as CLIC~\cite{Abramowicz:2016zbo} or the ILC~\cite{Baer:2013cma} might as well contribute to the determination of the precise VBS characteristics. They have the advantage of being very clean environments, but the center of mass energies they can reach are much smaller than those at which the LHC is currently running. Because of this, and since we will aim to probe directly the VBS observables in the high diboson invariant mass region, where the new physics is expected to emerge clearer, we will focus this subsection on the LHC searches for VBS.

As we have said, measuring VBS event rates is quite challenging due to their small associated cross sections. Currently, only three of the VBS channels have been observed with a statistical significance of more than  5$\sigma$. These measurements were performed in  $pp\to{\rm W}^+{\rm W}^+jj\to\ell^+\ell^+E_T^{\rm miss}jj$, $pp\to{\rm W}^+{\rm Z}jj\to\ell^+\ell^+\ell^-E_T^{\rm miss}jj$ and $pp\to{\rm Z}{\rm Z}jj\to(\ell^+\ell^-\ell^+\ell^-)jj+(\ell^+\ell^-E_T^{\rm miss})jj$ events, for a center of mass energy of $\sqrt{s}=13$ TeV and with 36.1 fb${}^{-1}$, 36.1 fb${}^{-1}$ and 139 fb${}^{-1}$ of accumulated data by the ATLAS experiment, respectively. Other VBS channels have also been observed in different decay modes of the EW gauge bosons, although with smaller statistical significances. A summary of these measurements for both CMS and ATLAS and for their corresponding center of mass energies and luminosities is presented in \tabref{Tablaexp}.

 \begin{table}[t!]
\begin{center}
\begin{tabular}{c|cc|cc|cc}
\toprule
\toprule
\multicolumn{1}{c|}{Channel}  & \multicolumn{2}{|c|}{$\sqrt{s}$ [TeV]} & \multicolumn{2}{|c|}{Lum. [fb${}^{-1}$] }& \multicolumn{2}{|c}{Obs. (exp.) significance [$\sigma$]} \\
\midrule
  & ATLAS & CMS & ATLAS & CMS & ATLAS & CMS \\
\midrule
W${}^\pm$W${}^\pm$  & 8  & 8  & 20.3 & 19.4 & 3.6 (2.3)\cite{Aad:2014zda,Aaboud:2016ffv} & 2.0 (3.1)\cite{Khachatryan:2014sta} \\
W${}^\pm$W${}^\pm$  & 13  & 13  & 36.1& 35.9 & 6.5 (4.6)\cite{Aaboud:2019nmv} & 5.5 (5.7)\cite{Sirunyan:2017ret} \\
ZZ  & 13  & 13  & 139 & 35.9 & 5.5 (4.3)\cite{ATLAS:2019vrv} & 2.7 (1.6)\cite{Sirunyan:2017fvv} \\
WZ  & 8  & 8  & 20.2 & 19.4 & aQGC lim.\cite{Aad:2016ett} & Unreported\cite{Khachatryan:2014sta} \\
WZ  & 13  & - & 36.1 & - & 5.3 (3.2)\cite{Aaboud:2018ddq} &- \\
WV  & 8  & - & 20.2 & - &aQGC lim.\cite{Aaboud:2016uuk} & - \\
VV  & 13  & - & 35.5& - & 2.7 (2.5)\cite{Aad:2019xxo} & - \\
\bottomrule
\bottomrule
\end{tabular}
\vspace{0.4cm}\caption{
Summary of the experimental measurements (observed and expected statistical significances in terms of standard deviations, $\sigma$) of various VBS processes involving Ws and Zs decaying to different final state particles. VV represent the combined study of WW, WZ and ZZ. In most cases, the leptonic decays for both gauge bosons are considered. Specification of each search can be found in the corresponding references. The center of mass energy and luminosities employed in each ATLAS and CMS  search are shown as well. In some studies the significances are provided directly. Others present their results in term of anomalous quartic gauge couplings (aQGC) only. }
 \label{Tablaexp}
\end{center}
\end{table}


The CMS and ATLAS searches on VBS in the various available channels allow to constraint the parameter space of EFTs  (being these parameters often dubbed anomalous quartic gauge coupling (aQGC)), since, so, far, no significant deviation from the SM prediction has been observed. The concrete values of the different limits have already been presented in the previous section and will be discussed in posterior parts of this Thesis, so we will not include them here for brevity. The two experiments, ATLAS and CMS, perform different studies that can lead to different interpretations of the experimental data from the EFT point of view. Apart from the fact that some analyses rely on the linear EFT and others in the non-linear one, the main difference concerns the treatment of the unitarity violation issue. 

Usually, when the non-linear formalism is adopted, the K-matrix method explained in the previous Chapter implemented following \cite{Alboteanu:2008my}, is used to compare the data against a unitarized EChL prediction. Nonetheless, when the linear EFT is used, different approaches are used by CMS and ATLAS. In general, ATLAS employs a Form Factor technique that suppressed by hand the pathological high energy behaviour of the EFT cross sections. CMS, on the contrary, often provides a {\it validity bound}: the energy (or invariant mass of the diboson system) at which the observed limits would violate unitarity. In some cases no unitarization method is considered.

These differences introduce an unavoidable degeneracy in the information about the EFT parameters extracted from the experimental LHC results. A first approach to the quantification of this uncertainty arising in the determination of the chiral coefficients due to the different unitarity restoration treatments is presented in Chapter \ref{Methods}, so we will exhaustively review this issue in forthcoming sections.

\begin{figure}[t!]
\begin{center}
\includegraphics[width=0.49\textwidth]{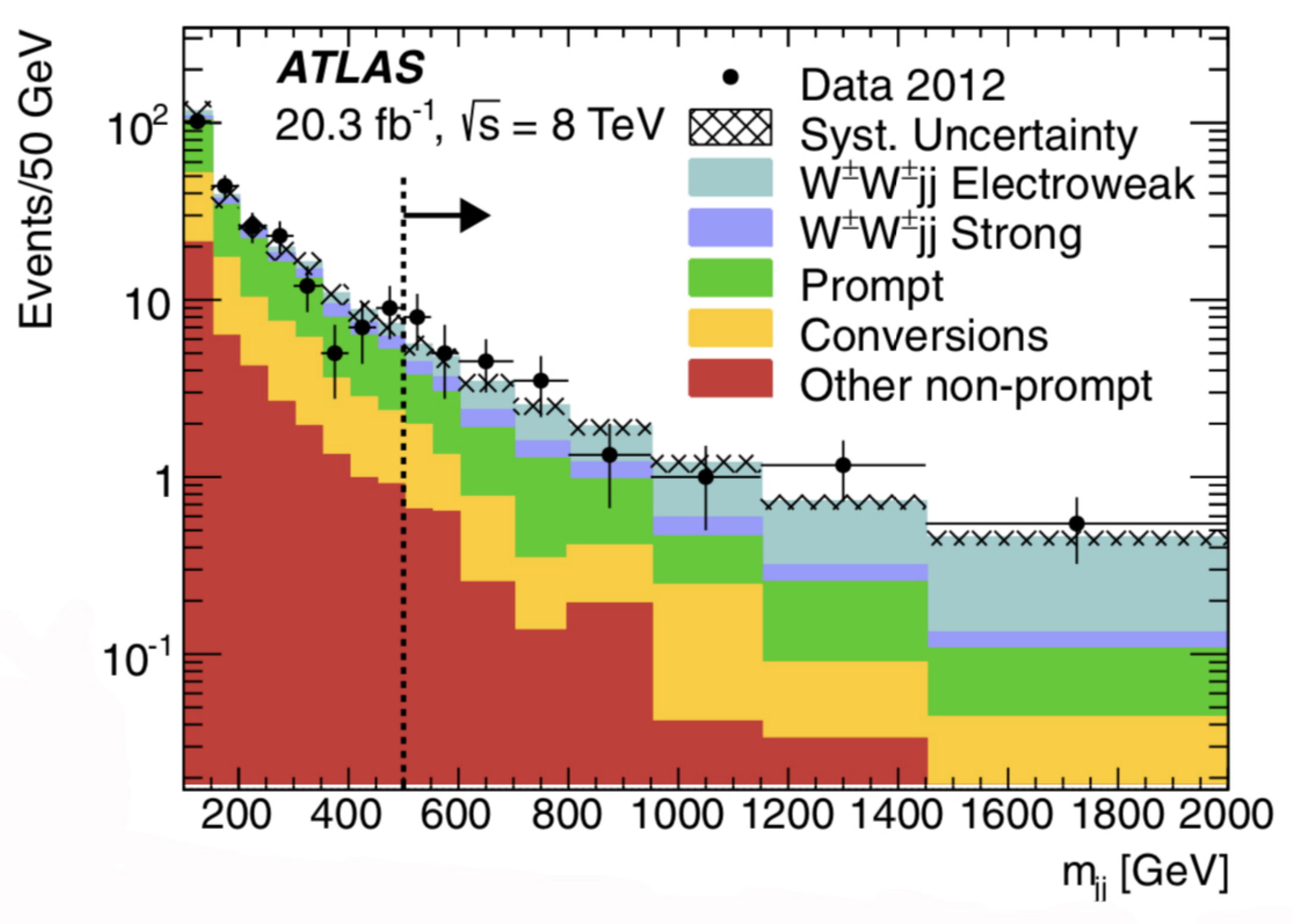}
\includegraphics[width=0.49\textwidth]{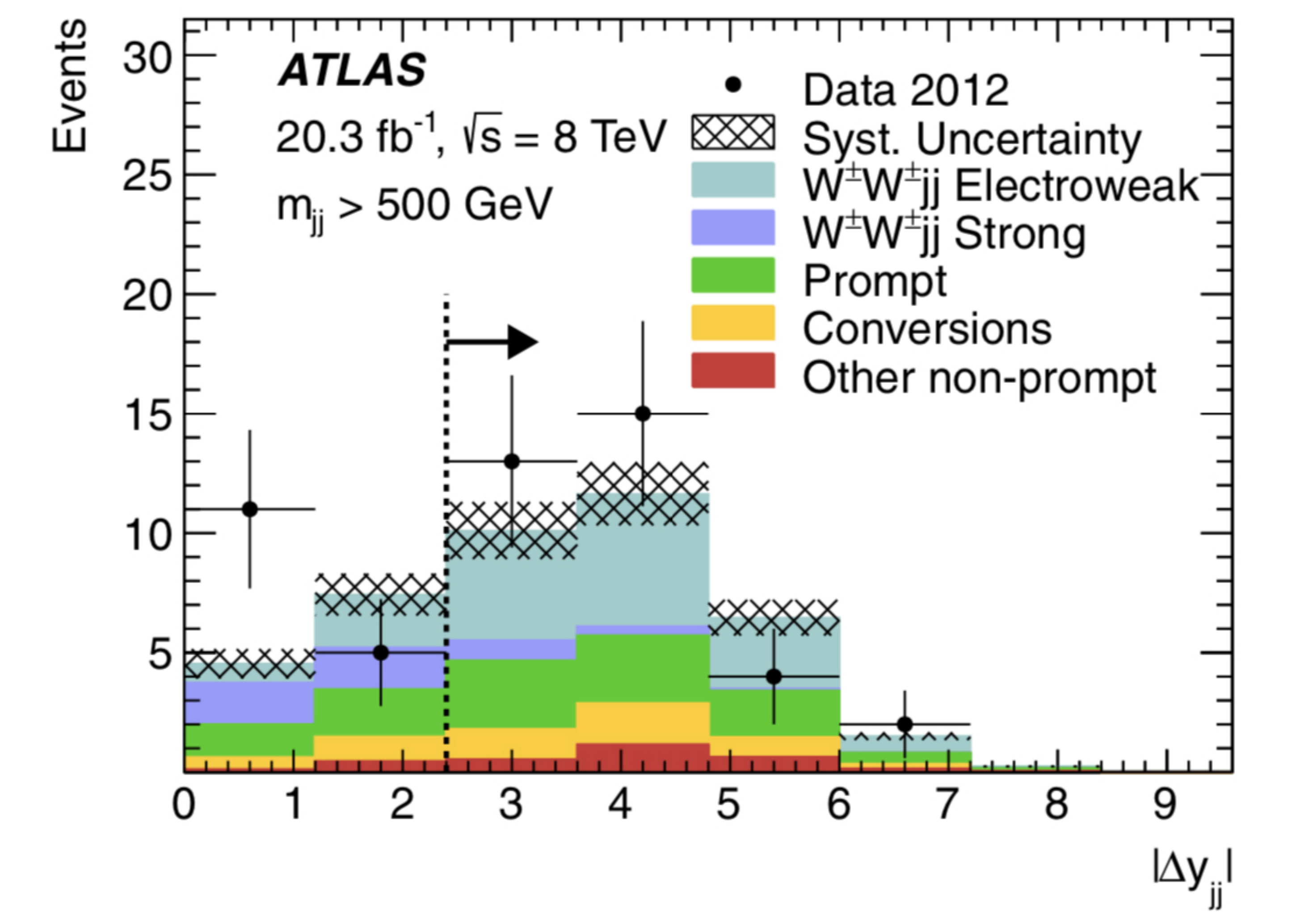}\\
\includegraphics[width=\textwidth]{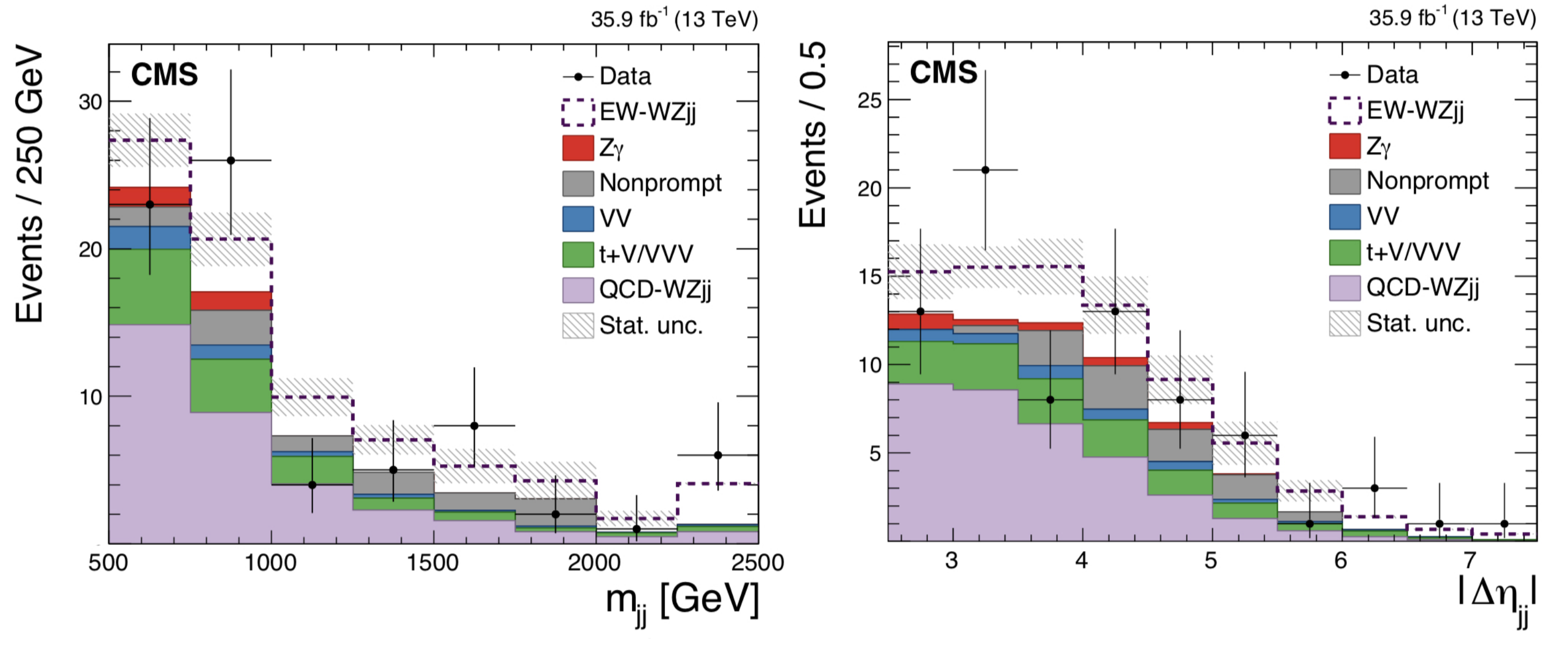}
\caption{Distributions of $pp\to$ W${}^\pm$W${}^\pm jj$ and $pp\to$ WZ$ jj$ observed and expected events in the two principal VBS variables: $m_{jj}$ (left panels)  and $|\Delta\eta_{jj}|$ (equivalent to $|\Delta y_{jj}|$ in the ATLAS case, right panels). Upper panels correspond to ATLAS studies whereas lower panels correspond to CMS studies. 
The main backgrounds to these observables are included as well in different colours. Luminosity and center of mass energies are indicated in the figures. In both cases, the purely leptonic decay channel is considered. Plots borrowed from \cite{Aaboud:2016ffv} and \cite{Sirunyan:2019ksz}.}
\label{fig:VBSATLASMCS}
\end{center}
\end{figure}

\begin{figure}[t!]
\begin{center}
\includegraphics[width=0.49\textwidth]{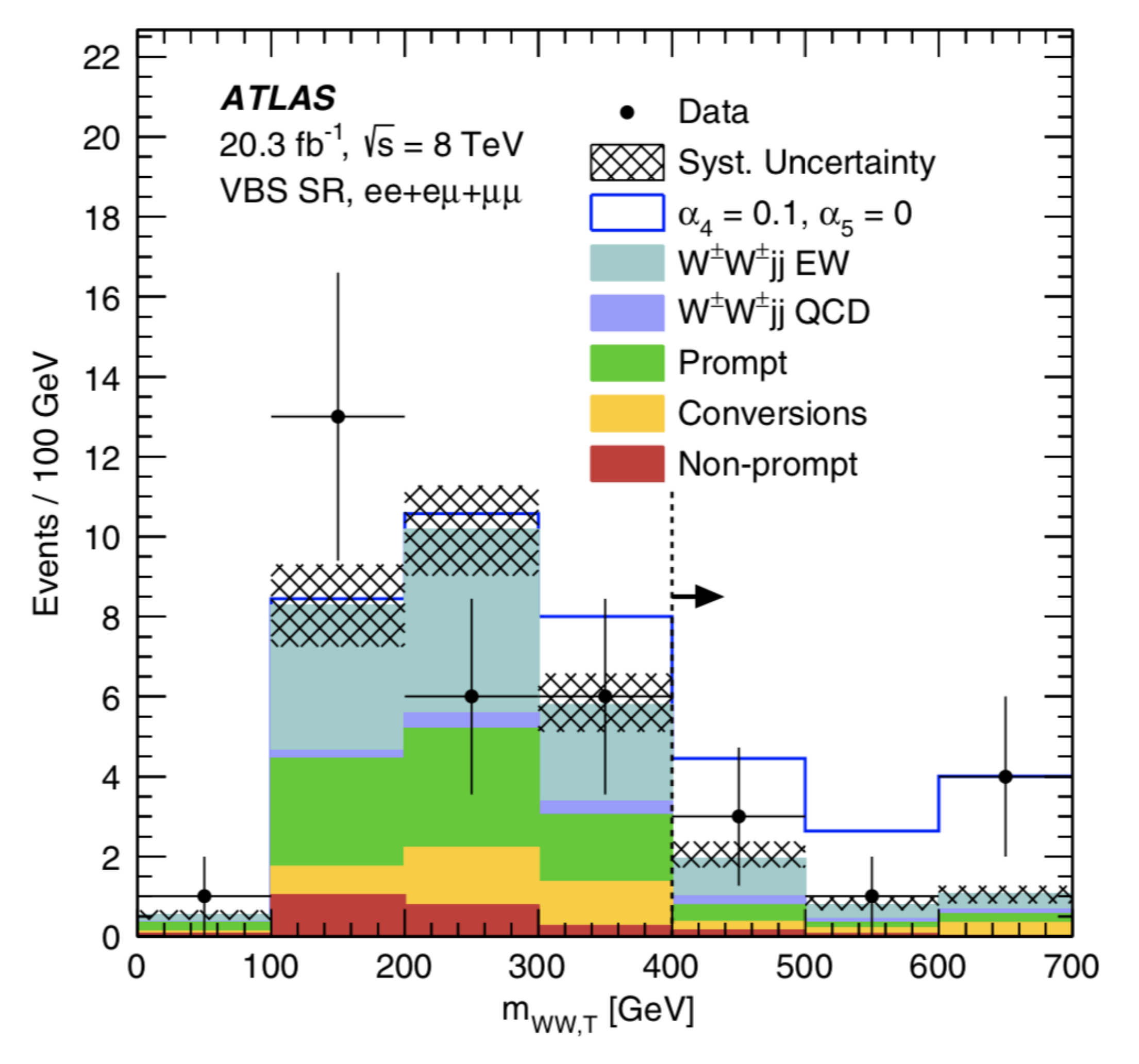}
\caption{Distribution of $pp\to$ W${}^\pm$W${}^\pm jj$ observed and expected events in the transverse invariant mass of the two Ws (for a precise definition of this variable see~\cite{Haywood:1999qg} and~\cite{Aaboud:2016ffv}) provided by ATLAS. The prediction of the (non-unitarized) EChL for $a_4\equiv\alpha_4=0.1$ and $a_5\equiv\alpha_5=0$ is shown for comparison. The main backgrounds to these observable are included as well in different colours. Luminosity and center of mass energies are indicated in the figure. The purely leptonic decay channel is considered. Plot borrowed from \cite{Aaboud:2016ffv}.}
\label{fig:VBSMWWT}
\end{center}
\end{figure}

Even if there exist some disparities between both experiments, the ATLAS and CMS searches share a quite similar set of the VBS selection criteria. In general the VBS topology is well established and, despite the differences between channels and decay products which lead obviously to different selection and identification requirements, and the fact that the different experiments have different geometrical acceptances and efficiencies, there is a common set of 
cuts that allow to tag the extra jets and to select the VBS topologies quite efficiently. These are usually defined by 
\begin{align}
m_{jj} > 500~{\rm GeV}\,,~~~~~~~|\Delta\eta_{jj}|>2.5\,,\label{VBScutsexp}
\end{align}
 as it can be seen in \figref{fig:VBSATLASMCS}. There, the $m_{jj}$  and $|\Delta\eta_{jj}|$ (equivalent to $|\Delta y_{jj}|$ in the ATLAS case) distributions of the VBS processes $pp\to$ W${}^\pm$W${}^\pm jj$ and $pp\to$ WZ$ jj$ obtained by ATLAS (upper panels) and CMS (lower panels), respectively, are shown. The main backgrounds to these observables are included as well in different colours. From these Figures it is clear that the bulk of the VBS events lie at large invariant masses of the final dijet system and at large angular separations of the two jets. In fact, especially in the ATLAS plots, it is plain that the VBS cuts given in \eqref{VBScutsexp} select the phase space region in which the VBS topologies are maximal with respect to the other processes.  With these cuts imposed invariant mass distributions\footnote{In this case, due to the presence of neutrinos in the final state, the so-called transverse invariant mass is used as discriminant. For more information on this variable see~\cite{Haywood:1999qg} and~\cite{Aaboud:2016ffv}.} can be presented and compared to different EFT scenarios, such as the one given in \figref{fig:VBSMWWT} as an example.

Remarkable progress is being made in the VBS characterization at the LHC but there are still some challenges to overcome in order to fully explore this kind of observables. The most important of these challenges  is that of  the correct evaluation of the backgrounds, especially the ones arising from QCD events. Depending on the VBS channel one considers different backgrounds will pollute the signal. For instance, the charged scenarios, and, especially, the doubly charged channel  $pp\to$ W${}^\pm$W${}^\pm jj$ suffer from less severe backgrounds than the neutral ones. The proper characterization of the backgrounds is, therefore, a key ingredient to be able to access all the diverse VBS channels at the LHC experiments.

The reduction of the statistical  and theoretical uncertainties is another challenge of the VBS searches. Due to the small VBS cross sections, high luminosity is needed to perform accurate observations. Besides, precise NLO computations, PDF interpretations and scale determinations are needed from the theoretical point of view to interpret the data correctly. Furthermore, the exploration of different decay modes of the final gauge bosons is also very important to extract as much information as possible from the experimental data. 

The various final states related to VBS processes present different advantages and disadvantages. The purely leptonic case is rather clean, since electron and muons can be identified and reconstructed with good precision both in ATLAS and CMS. Besides, the backgrounds associated to these final states are, in general, small. However, due to the modest branching fractions of the gauge bosons to leptons, these channels have small event rates and high luminosities are required to observed them. Moreover, the usual presence of neutrinos, (not) detected as missing energy, complicates the reconstruction of the intermediate EW gauge bosons properties.

On the other hand, the purely hadronic final state benefits from larger cross sections compared to the leptonic one. Furthermore, the scenarios in which the decay products of the gauge bosons are identified as  single, large radius jets, usually called fat jets, are very appealing since higher energy regions of the VBS phase space can be probed. However the backgrounds associated to this state are quite difficult to overcome. Especially those coming from QCD multijet events, from $t\bar{t}$ decays and for W+jets processes really complicate the task of selecting VBS configurations from the rest. In this sense, the semileptonic decay modes of the EW gauge bosons are often used since they represent the intermediate point between the leptonic and the hadronic cases.  

Concerning the most recent results on VBS measurements, a good summary can be found in \tabref{Tablaexp}. For the complete results of other channels involving photons see \cite{Bellan:2019xpr} and references therein. These studies provide the experimental data in different formats: some of them report the observed cross section measurements of determined processes directly whereas others implement in their analyses the determination of EFT constraints already. 

\begin{figure}[t!]
\begin{center}
\includegraphics[width=0.6\textwidth]{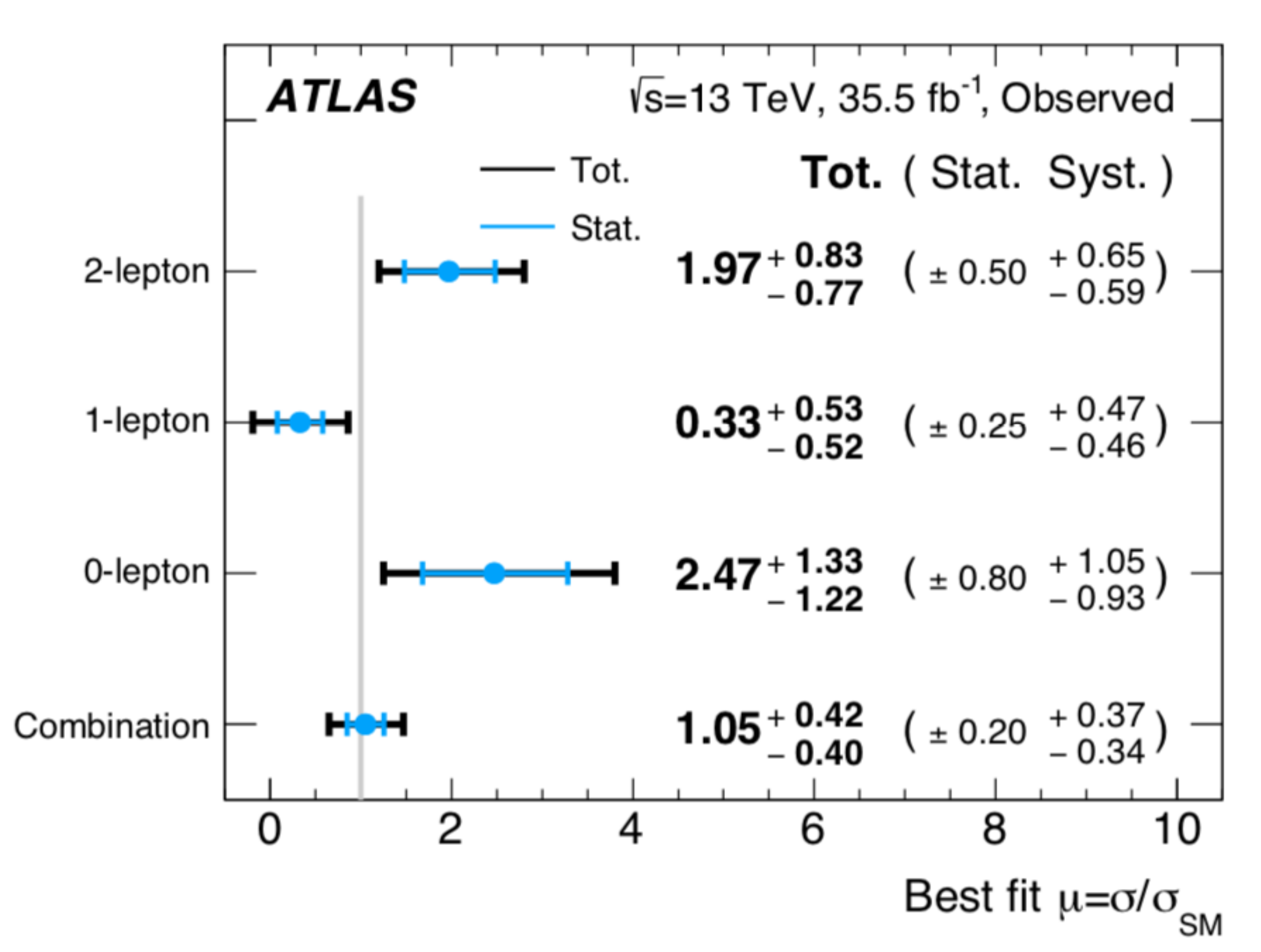}
\caption{Measurements of the signal-strength $\mu=\sigma/\sigma_{\rm SM}$ corresponding to $pp\to$(WW/WZ/ZZ)$jj$ events for the 0, 1 and 2 lepton channels and their combination. The center of mass energy, luminosity and interpretation of the error bars are explicitly given in the plot. Figure borrowed from \cite{Aad:2019xxo}.}
\label{fig:exampleXSVBS}
\end{center}
\end{figure}

An example of the former studies can be seen in \figref{fig:exampleXSVBS}, where the cross section measurements of the joint WW/WZ/ZZ  processes are given por different final state configurations (with 2, 1 or 0 leptons). The combined prediction is presented as well. There, one can see that the various measurements are not definitely compatible with the SM prediction when considering the error bars, and that their central values differ from the expected SM cross section. However, the combined result matches quite well the SM prediction. This points towards the fact that a deeper analysis and more data taking of this kind of processes are needed.

The latter type of analyses, i.e., those that provide their results directly as bounds on EFT parameters, will be of great relevance for this Thesis. The most recent works in this direction correspond to references \cite{Sirunyan:2019der,Aad:2019xxo,Sirunyan:2019ksz}. Precisely because of the mentioned differences between the experimental treatment of EFT predictions  (and particularly of the issue of unitarity violation) those constraints have to be interpreted carefully. In Chapter \ref{Methods} we will discuss this interpretation exhaustively, so we leave the presentation of these results for these forthcoming sections.

It is important to make a comment as well about another type of search strategies that will also be a fundamental part of this Thesis: the resonant VBS searches. We will not get into much detail in this section since the last two Chapters will be completely devoted to the characterization of these resonances and more information will be given, therefore, later on.

 Both ATLAS and CMS perform analyses dedicated to observe the appearance of resonances motivated by different theoretical frameworks. Since no positive observation of such states has been done so far they impose constraints on the masses and couplings to the SM particles of these new states. However, most of these searches are carried out assuming that the new heavy degrees of freedom are produced via Drell-Yan, although they might decay posteriorly into a pair of vector bosons. Thus, if the resonance is very weakly coupled to fermions and manifests only in VBS interactions these bounds are overestimated. There are some works that consider the VBS production channel in the resonant scenario as well \cite{Aaboud:2017fgj}, but the invariant masses they can reach are, in general, smaller.

Now that we have briefly reviewed the experimental status of VBS searches, it is pertinent to comment on some of the future prospects of such measurements. The first thing that comes to mind in this sense is that of the separation of the different gauge boson polarizations. A lot of efforts and improvements are being made both theoretically and experimentally in this direction, and it is likely that in the near future the accurate distinction between polarization modes will be achieved.

 Proposals such as looking for specific angular distributions of the decay products of W and Z bosons or new kinematical variables sensitive to the initial vector boson polarization~\cite{Fabbrichesi:2015hsa} are the most promising avenues to this purpose. In the experimental side, the use of these distributions has led to the determination of polarization fractions of W and Z bosons \cite{Aaboud:2019gxl}, albeit not in VBS configurations. Being able to distinguish among polarizations, especially in VBS observables, will be the key to really access possible new physics in the EWSB sector. 
 
Another important improvement being currently developed concerns the reconstruction techniques of final state jets. 
As we commented before, the case in which the decay products of a gauge boson are identified as a single, large radius jet allows to probe the very boosted regime of VBS processes. Nonetheless, the correct reconstruction and characterization of these fat jets supposes a difficult task and its current efficiency is not very high. In the near future this efficiency is expected to rise significantly so that boosted topologies of the vector boson scattering could be accessed through fat jet measurements. Machine learning techniques such as boost decision trees will be very useful in this sense and also to disentangle signal and background events.

Nevertheless, the most direct future prospect one expects is the increase of the experimental luminosity that will allow to obtain statistically accurate measurements of all the VBS channels in their different decay modes. The High Luminosity LHC (HL-LHC) will thus be the stage at which VBS is expected to be deeply understood, and, luckily, so will be the  true nature of EWSB.

In summary, in the two previous Chapters we have revisited the theoretical bases of EWSB as well as the effective theory description of new physics explaining its yet unknown dynamical generation. We have discussed the properties of the electroweak chiral Lagrangian, which will be our main tool in the forthcoming Chapters, and of vector boson scattering observables in this framework, especially in the context of the LHC experiment. This means that our map is ready and that we shall go get the treasure!



\chapter[\bfseries Probing the Higgs self-coupling in double Higgs  production \newline through  vector  boson scattering  at the LHC]{Probing the Higgs self-coupling in double Higgs production through  vector  boson scattering  at the LHC}\label{HH}

\chaptermark{Probing the Higgs self-coupling in HH production via VBS at the LHC}

As a first example of the power of VBS observables to access new physics in the EWSB sector at the LHC, we will study in this Chapter the feasibility of probing the Higgs self coupling in VBS double Higgs production.  As we have said in previous parts of this Thesis, the observation of the Higgs boson by the ATLAS and CMS experiments \cite{Aad:2012tfa, Chatrchyan:2012xdj} in 2012 confirmed the prediction of the last particle of the Standard Model (SM) of fundamental interactions. But, although this discovery allowed us to answer many important and well established questions about elementary particle physics, it also posed a lot of new mysteries concerning the scalar sector of the SM.

One of these mysteries is that of the true value of the Higgs self-coupling $\lambda$, involved in trilinear and quartic Higgs self-interactions appearing in the Higgs potential (\eqref{lagphi}), as well as its relation to other parameters of the SM.  Particularly, understanding and testing experimentally the relation between $\lambda$ and the Higgs boson mass, $m_H$, will provide an excellent insight into the real nature of the Higgs particle. This relation, given in the SM at the tree level by $m_H^2=2 v^2 \lambda$, with $v=246$ GeV, arises from the BEH mechanism, as we have seen, so to really test this theoretical framework one needs to measure $\lambda$ independently of the Higgs mass. Unfortunately, the  value of the Higgs self-coupling has not been established yet with precision at colliders, but there is (and will be in the future) a very intense experimental program focused on the realization of this measurement (for a review, see for instance \cite{Simon:2012ik, Dawson:2013bba, Baer:2013cma, Abramowicz:2016zbo, deFlorian:2016spz}).

The Higgs trilinear coupling can be probed in double Higgs production processes at the LHC, process that have been extensively studied both theoretically 
in \cite{Glover:1987nx, Dicus:1987ic, Plehn:1996wb, Dawson:1998py, Djouadi:1999rca, Baur:2003gp, Grober:2010yv, Dolan:2012rv, Papaefstathiou:2012qe, Baglio:2012np, Yao:2013ika, deFlorian:2013jea, Dolan:2013rja, Barger:2013jfa,  Frederix:2014hta, Liu-Sheng:2014gxa, Goertz:2014qta, Azatov:2015oxa, Dicus:2015yva, Dawson:2015oha, He:2015spf, Dolan:2015zja, Cao:2015oaa, Cao:2015oxx, Huang:2015tdv, Behr:2015oqq, Kling:2016lay, Borowka:2016ypz, Bishara:2016kjn, Cao:2016zob, deFlorian:2016spz, Adhikary:2017jtu, Kim:2018uty, Banerjee:2018yxy, Goncalves:2018qas, Bizon:2018syu, Borowka:2018pxx, Gorbahn:2019lwq}, and experimentally in \cite{Aaboud:2016xco, CMS:2016foy,  CMS:2017ihs, Aaboud:2018knk,  Sirunyan:2018iwt, Aaboud:2018ftw, Aad:2019uzh}. At hadron colliders, these processes can take place through a variety of production channels, being gluon-gluon fusion (GGF) and vector boson scattering, also called vector boson fusion (VBF) in the literature, the main ones regarding the sensitivity to the Higgs self-coupling. Focusing on the LHC case, on which we will base our study in this Chapter, the dominant contribution to double Higgs production comes from GGF. It associated production rate for $\sqrt{s}=14$ TeV is about a factor 17 larger than that of VBS \cite{Frederix:2014hta}. Because of this, most of the works present nowadays in the literature focus on this particular HH  production channel, GGF, to study the sensitivity to $\lambda$. In fact, all these works and the best present combined measurement at the LHC have made possible to constraint this parameter in the range $\lambda\in[-5.0,12.0]\cdot \lambda_{\rm SM}$ at the 95\% CL \cite{Aad:2019uzh}.

Although GGF benefits from the highest statistics and rates, it suffers the inconveniences of having large uncertainties, being a one loop process initiated by gluons, and being dependent of the top Yukawa coupling. Double Higgs production via VBS~\cite{Grober:2010yv, Dawson:2013bba, Baglio:2012np, Frederix:2014hta, Liu-Sheng:2014gxa, Dicus:2015yva, He:2015spf, Bishara:2016kjn, deFlorian:2016spz} is, in contrast, a tree level process not initiated by gluons and it is independent of top physics features, leading therefore to smaller uncertainties. Also, at a fundamental level, since VBS processes involving longitudinally polarized gauge bosons, like the process V${}_L$V${}_L\to$ HH that we are interested in, probe genuinely the self interactions of the scalar sector of the SM, it results interesting to study this kind of scenarios. This would happen specially at high energies, such as those available at the LHC, since, in this regime, each V${}_L$ behaves as its corresponding would-be-Goldstone boson. Therefore, testing V${}_L$V${}_L\to$ HH is closely related to testing $w w$HH interactions. In this way, a new window, qualitatively different than GGF, would be open with VBS to test $\lambda$, meaning that being able to measure these processes for the first time will be a formidable test of the SM itself, and it could even lead to the discovery of physics beyond the Standard Model (BSM). Moreover, the VBS production channel is the second largest contribution to Higgs pair production, and the VBS topologies have very characteristic kinematics, which allow us to select these processes very efficiently as well as to reject undesired backgrounds, as we just introduced in the previous Chapter.  Thus, in summary, VBS double Higgs production might be very relevant to study the sensitivity to the Higgs self-coupling, despite the fact that it is considerably smaller in size than GGF, since it could lead to a cleaner experimental signal. Besides, it will be a complementary measurement to that of GGF and will, in any case, help to improve the determination of the $\lambda$ coupling with better precision.  

In this Chapter, motivated by the previously commented advantages, we will analyze in full detail Higgs pair production via VBS at the LHC to probe the Higgs self-coupling. To this end, we will first explore and characterize the subprocesses of our interest, VV $\to$ HH with V=W,Z, both for the SM with $\lambda= \lambda_{SM}$ and for BSM scenarios with $\lambda=\kappa\,\lambda_{SM}$. We will consider values of $\kappa$ between 10 and -10. For this study, we fix $m_H$ to its experimental value, $m_H=125.18\pm 0.16$ GeV  \cite{Tanabashi:2018oca}, and set the Higgs vacuum expectation value (vev) to $v=246$ GeV. In this way, studying the sensitivity to $\lambda$ in VBS will provide the desired independent test of this coupling. 

Once we have deeply studied double Higgs production at the subprocess level, we will then explore the LHC scenario. First we will analyze the process $pp\to$  HH$jj$, to fully understand the properties of this scattering, and then we will study and give quantitative results for the sensitivity to the Higgs self-coupling after the Higgs decays. The production of HH$jj$ events at the LHC, including VBS and GGF, has been studied previously in \cite{Dolan:2013rja, Dolan:2015zja}, where they focus on $b\bar{b}\tau\bar{\tau} jj$ final states. Our main study is performed, in contrast, in the four bottoms and two jets final state, $pp\to b\bar{b}b\bar{b}jj$, since it benefits from the highest rates. We also make predictions for the interesting $pp\to b\bar{b}\gamma\gamma j j$ process which, although with lower rates, leads to cleaner signatures. We would like to point out that all computations and simulations are performed at the parton level with no hadronization or detector response simulation taken into account, since the work is aimed to be a first and simple approximation to the sensitivity to $\lambda$ in VBS processes at the LHC. A discussion on the implication of these additional considerations will be performed at the end of this Chapter.


\section{Double Higgs production in vector boson scattering}
\label{Subprocess}
As already stated in the paragraphs above, we are interested in exploring the sensitivity to the Higgs self-coupling, $\lambda$, through VBS processes, in particular at the LHC. For that purpose, we have to study and characterize first the subprocess that leads to the specific signal we will be dealing with once we perform the full collider analysis. This subprocess will be, in our case, the production of two Higgs bosons in the final state from the scattering of two EW gauge bosons, VV $\to$ HH, with V=W,Z\footnote{Notice that this signal was not introduced in Chapter \ref{VBS} due to the presence of Higgs bosons in the external legs that makes it a non-standard VBS channel. It will be, however, exhaustively review in this section.}. Within this context, in this section we aim to understand the role of the Higgs trilinear coupling in the SM and beyond, as well as the generic characteristics of the scattering processes W${}^+$W${}^-\to$ HH and ZZ $\to$ HH.

\begin{figure}[t!]
\begin{center}
\includegraphics[width=\textwidth]{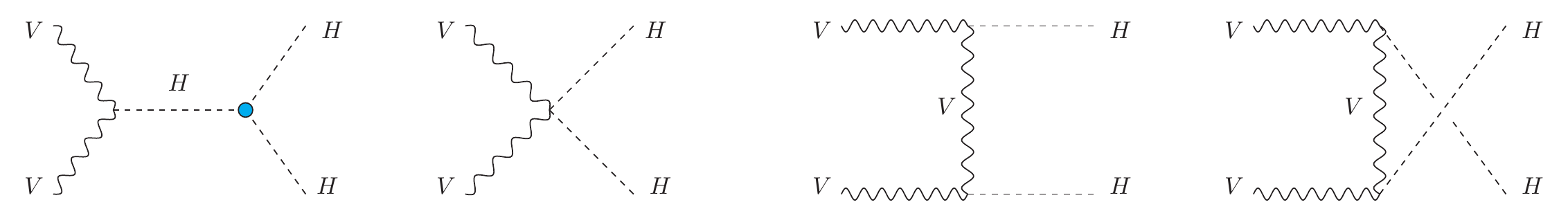}
\caption{Tree level diagrams that contribute to double Higgs production in vector boson scattering in the Unitary gauge. The cyan circle represents the presence of the Higgs self-coupling in the interaction vertex.}
\label{fig:subprocess_diagrams}
\end{center}
\end{figure}

The Higgs self-coupling is only present, at the tree level and in the Unitary gauge, in the $s-$channel diagram of  the studied processes, so the sensitivity to $\lambda$ will only depend on this particular configuration. However, a contact diagram, a $t-$channel diagram and a $u-$channel diagram have to be taken into account too as shown in \figref{fig:subprocess_diagrams}, in which we display all the possible tree level contributions to the mentioned scattering processes in the Unitary gauge. Each of these diagrams has its own energy dependence and its own relative size, so they participate differently in the total amplitude $A\big(V_1(p_1,\varepsilon_1) V_2(p_2,\varepsilon_2)  \to H_1(k_1) H_2(k_2)\big)$. This can be seen in \eqrefs{amplitudeHHs}{amplitudeHHu}, where we show the amplitude of each diagram of the process $W^+W^-\to HH$, $A_d$, with $d=s,c,t,u$ from $s$, contact, $t$ and $u$ channels respectively, computed consistently in the Unitary gauge:
\begin{align}
A_s(W^+W^-\to HH)=&~3 g^2v^2\dfrac{ \lambda}{s-m_H^2}(\varepsilon_1 \cdot \varepsilon_2)\, ,\label{amplitudeHHs} \\
A_c(W^+W^-\to HH)=& ~\dfrac{g^2}{2}(\varepsilon_1 \cdot \varepsilon_2)\, , \label{amplitudeHHc}\\
A_t(W^+W^-\to HH)=&~\dfrac{g^2}{t-m_W^2}(m_W^2(\varepsilon_1 \cdot \varepsilon_2) +(\varepsilon_1 \cdot k_1) (\varepsilon_2 \cdot k_2) )\, , \label{amplitudeHHt}\\
 A_u(W^+W^-\to HH)=&~\dfrac{g^2}{u-m_W^2}(m_W^2(\varepsilon_1 \cdot \varepsilon_2) +(\varepsilon_1 \cdot k_2) (\varepsilon_2 \cdot k_1) )\, .\label{amplitudeHHu}
\end{align}
Here, $g$ is the EW coupling constant, $m_W$ is the mass of the W boson, and $s,t$ and $u$ are the usual Mandelstam variables. The amplitudes for the $ZZ\to HH$ case are identical except for a global factor $1/c_{\rm w}^2$ (with $c_{\rm w}=\cos\theta_{\rm w}$ and with $\theta_{\rm w}$ being the weak angle) and the substitution of $m_W^2$ by $m_Z^2$ in the $t$ and $u$ channel expressions.

\begin{figure}[t!]
\begin{center}
\includegraphics[width=0.49\textwidth]{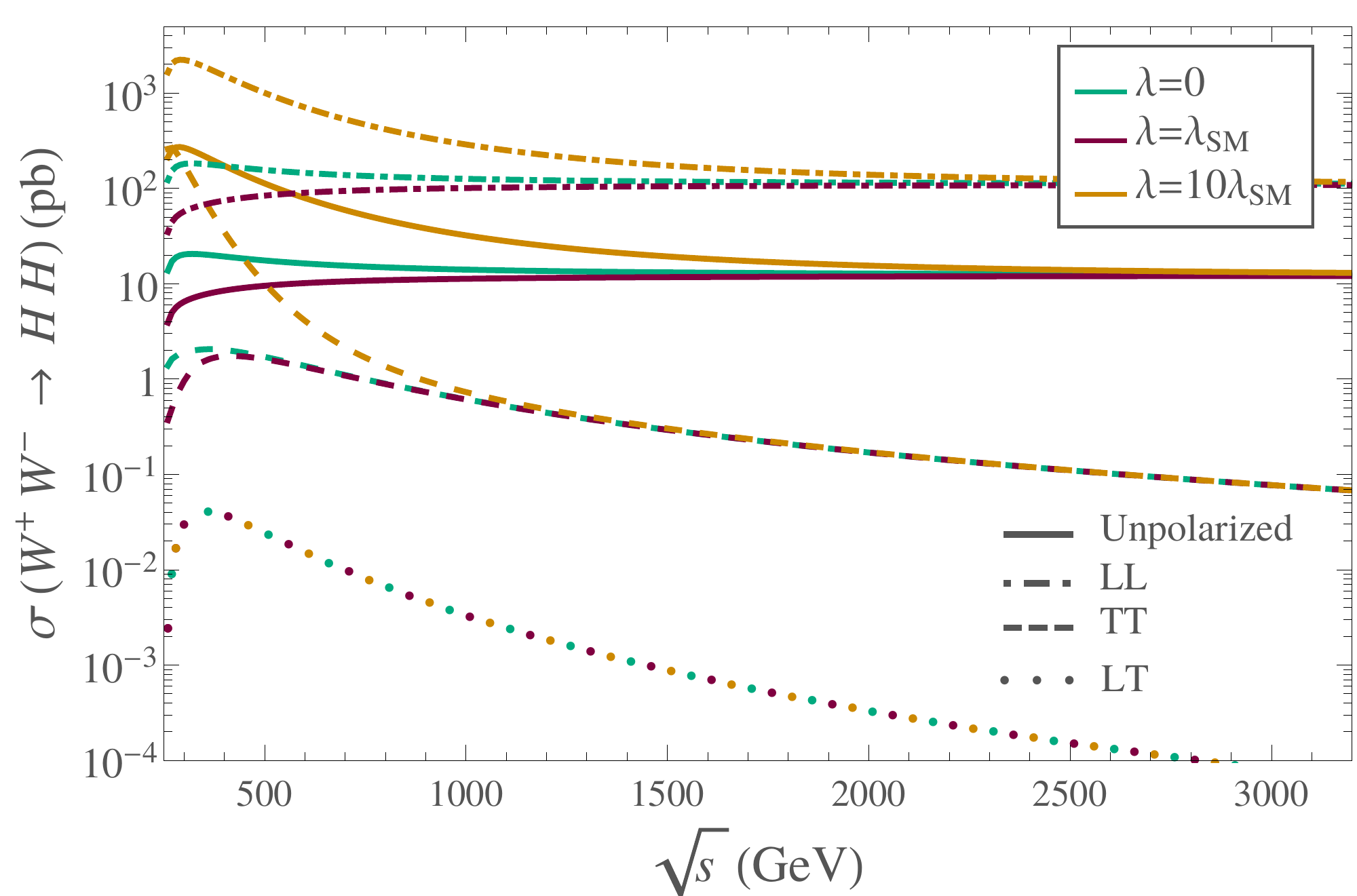}
\includegraphics[width=0.49\textwidth]{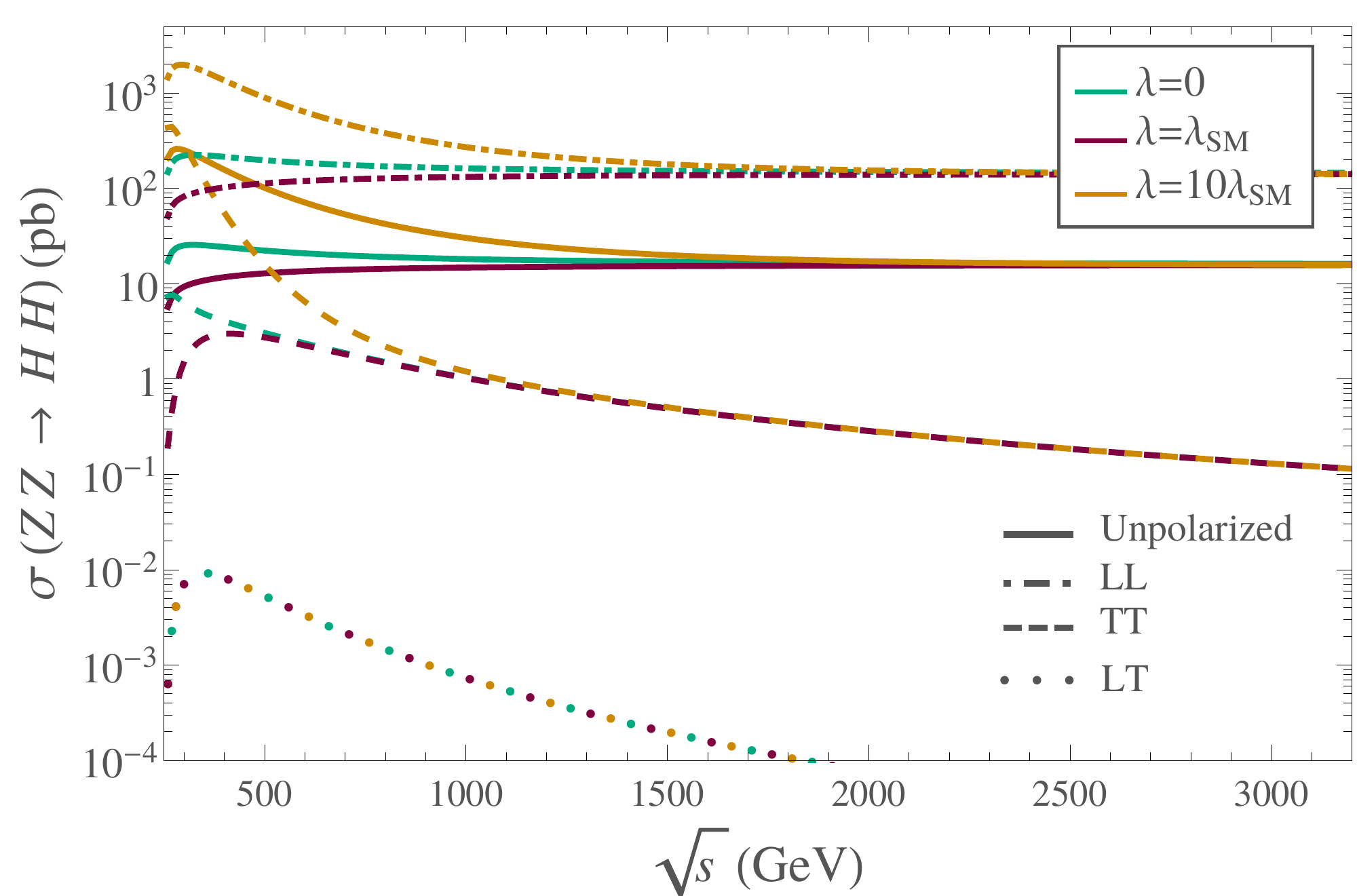}
\caption{Predictions of the cross sections of $W^+W^-\to HH$ (left panel) and $ZZ\to HH$ (right panel) as a function of the center of mass energy $\sqrt{s}$ for three different values of $\lambda$ and for different polarizations of the initial gauge bosons: $V_LV_L$ (upper dot-dashed lines), $V_TV_T$ (middle dashed lines) and $V_LV_T+V_TV_L$ (lower dotted lines). The unpolarized cross section is also included (solid lines). Each polarized cross section contributes with a factor 1/9 to the unpolarized (averaged) cross section.}
\label{fig:subprocess_polarizations}
\end{center}
\end{figure}

On the other hand, the contribution of each polarization state of the initial EW gauge bosons behaves differently, not only energetically, but also in what concerns to the sensitivity to $\lambda$. There are only two polarization channels that do depend on $\lambda$: the purely longitudinal, $V_LV_L$, and the purely transverse in which both vector bosons have the same polarization, $V_{T^+}V_{T^+}$ and $V_{T^-}V_{T^-}$. All the other channels have vanishing $s$-channel contributions and will not actively participate, therefore, in the study of the Higgs trilinear coupling, although all polarization states contribute to the total cross section. Moreover, this total cross section is dominated, especially at high energies, by the purely longitudinal $V_LV_L$ configuration, and so is each diagram contribution. All these features can be seen in \figref{fig:subprocess_polarizations}, where we display the predictions for the cross sections of $W^+W^-\to HH$ and $ZZ\to HH$ as a function of the center of mass energy for three different values of $\lambda$ separated by polarizations of the gauge bosons, including, also, the unpolarized cross section. In this figure two things are manifest: the first one is that the $V_LV_T$ configuration is indeed independent of $\lambda$. The second one is that the total cross section is clearly strongly dominated by the purely longitudinal contribution at all energies. This is a very interesting result, since it means that, if this process was measured, we would be being sensitive to the purely longitudinal configurations of the gauge bosons, and therefore to the heart of the self-interactions of the SM scalar sector.

\begin{figure}[t!]
\begin{center}
\includegraphics[width=0.49\textwidth]{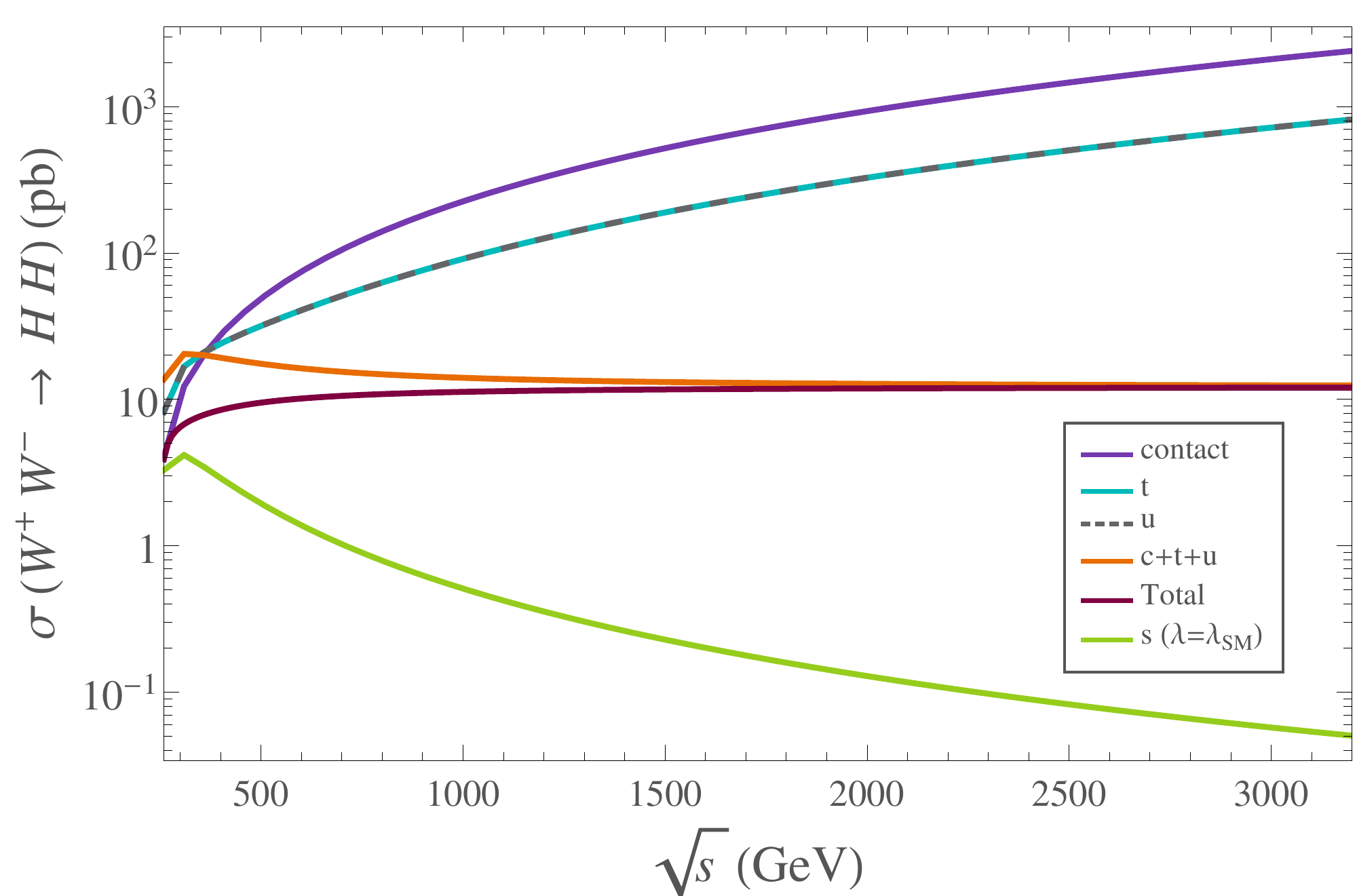}
\includegraphics[width=0.49\textwidth]{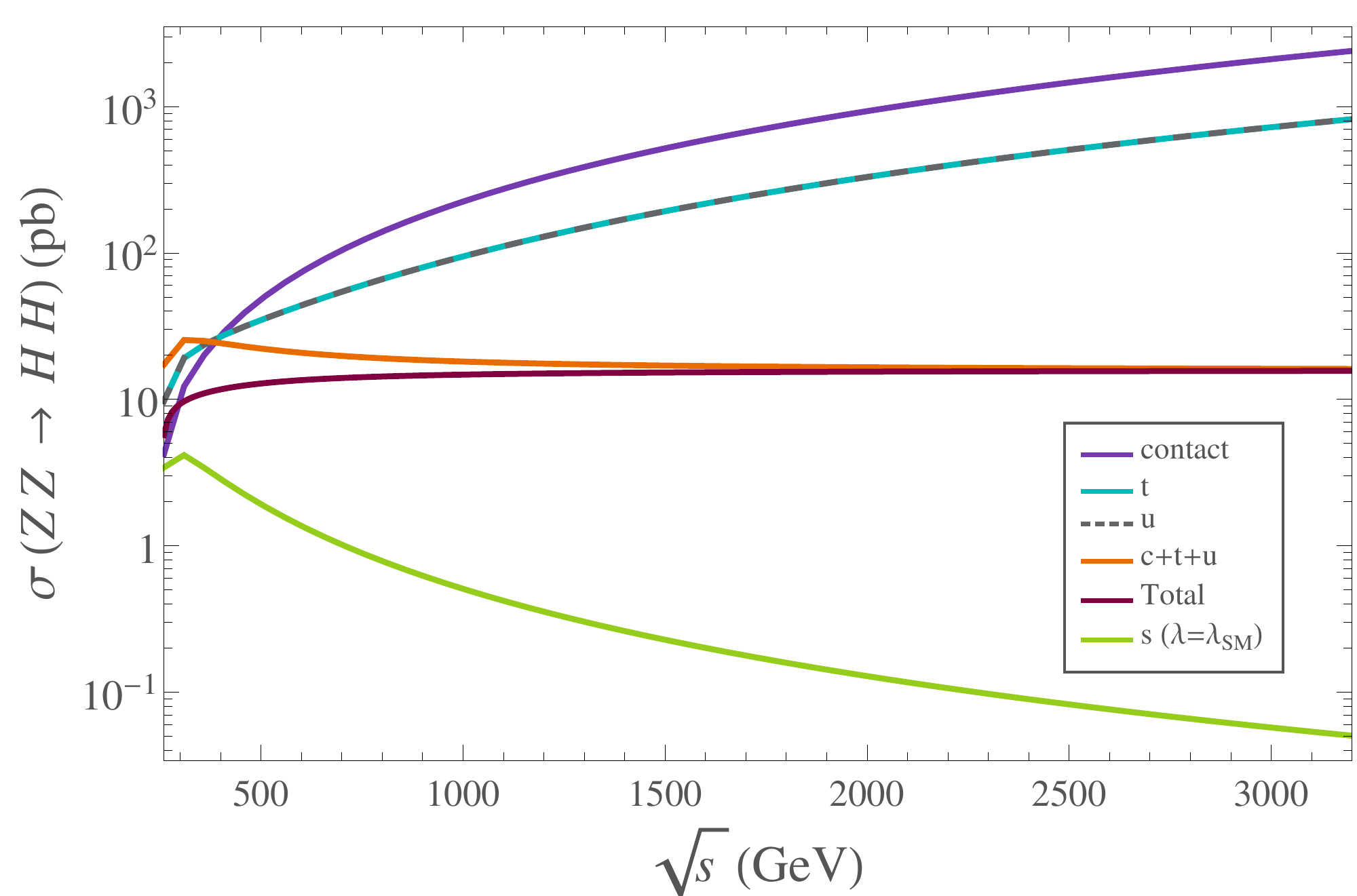}
\caption{Contribution to the total cross section of $W^+W^-\to HH$ (left panel), and of $ZZ\to HH$ (right panel) in the SM, i.e., $\lambda=\lambda_{SM}$, of each diagram displayed in \figref{fig:subprocess_diagrams} as a function of the center of mass energy $\sqrt{s}$. The sum of the contributions of the contact, $t$-channel and $u$-channel diagrams as well as the sum of all diagrams that contribute are also presented.}
\label{fig:subprocess_diagrams_effect}
\end{center}
\end{figure}

The $V_LV_L$ dominance can be understood through the inspection of the energy dependence of the longitudinal polarization vectors, $\varepsilon_V$, at high energies. They are all proportional, for $\sqrt{s}\gg m_V$, to a power of the energy over the mass, $E_V/m_V$. This leads to a behavior of the amplitudes, presented in Eqs.(\ref{amplitudeHHc})-(\ref{amplitudeHHu}), for the contact, $t$ and $u$ channels respectively, proportional to $s$,  and to a constant behavior with energy of the $s$-channel amplitude given in Eq.(\ref{amplitudeHHs}). Including the extra $1/s$ suppression factor to compute the cross section from the squared amplitude, one obtains the energy dependence seen in \figref{fig:subprocess_diagrams_effect}, where we present the contribution of each diagram to the total cross sections of $W^+W^-\to HH$ and $ZZ\to HH$ in the SM, as well as the sum of the contact, $t$-channel and $u$-channel diagrams, $(c+t+u)$, and the total cross section taking all diagrams into account. In this figure, we see clearly that the sum of the contact, $t$ and $u$ channels tends at high energies to a constant value. This happens because in the SM there is a cancellation among the linear terms in $s$ corresponding to these three channels, similarly as in the case presented in \figref{fig:Higgsneed}. In contrast, the $s$-channel contribution decreases as $1/s$ and is subleading numerically in the SM with respect to the other $(c+t+u)$ contributions. It is only at low energies, near the production threshold of two Higgs bosons, where the s-channel contribution is numerically comparable to the other channels. In fact, a mild cancellation occurs between this $s$-channel and the rest $(c+u+t)$. Therefore, the s-channel and in consequence $\lambda$, do not effectively participate in  the constant behavior at high energies of the total cross section in the SM. At this point, it is worth recalling once again that these constant behaviors of the cross sections with energy are characteristic of VBS processes at high energies.

When going beyond the SM by taking $\lambda \neq \lambda_{SM}$, the previously described dependence with energy and the delicate cancellations commented above among the various contributing diagrams may change drastically. In fact, varying the size of the Higgs trilinear coupling could modify the relative importance of the contributing diagrams and, in particular, it could allow for the s-channel contribution to be very relevant or even dominate the scattering. This could happen not only at low energies close to the threshold of $HH$ production, but also at larger energies, where the pattern of cancellations among diagrams could be strongly modified. This may lead to a different high energy behavior, and, hence, to a different experimental signature. The crucial point is that such a large  deviation in $\lambda$ with respect to the SM value is still experimentally possible, as the present bounds on the trilinear coupling are not yet very tight. The best bounds at present set $\kappa = \lambda/\lambda_{SM}\in[-5.0,12.0]$ \cite{Aad:2019uzh}, so values of order 10 times the SM coupling are still allowed by LHC data. Then, if in the future the LHC could improve this sensitivity to lower values of $\lambda$ it would be a formidable test of the presence of new physics beyond the SM. We will show next that this sensitivity can be indeed reached in the future by means of VBS. 

\begin{figure}[t!]
\begin{center}
\includegraphics[width=0.49\textwidth]{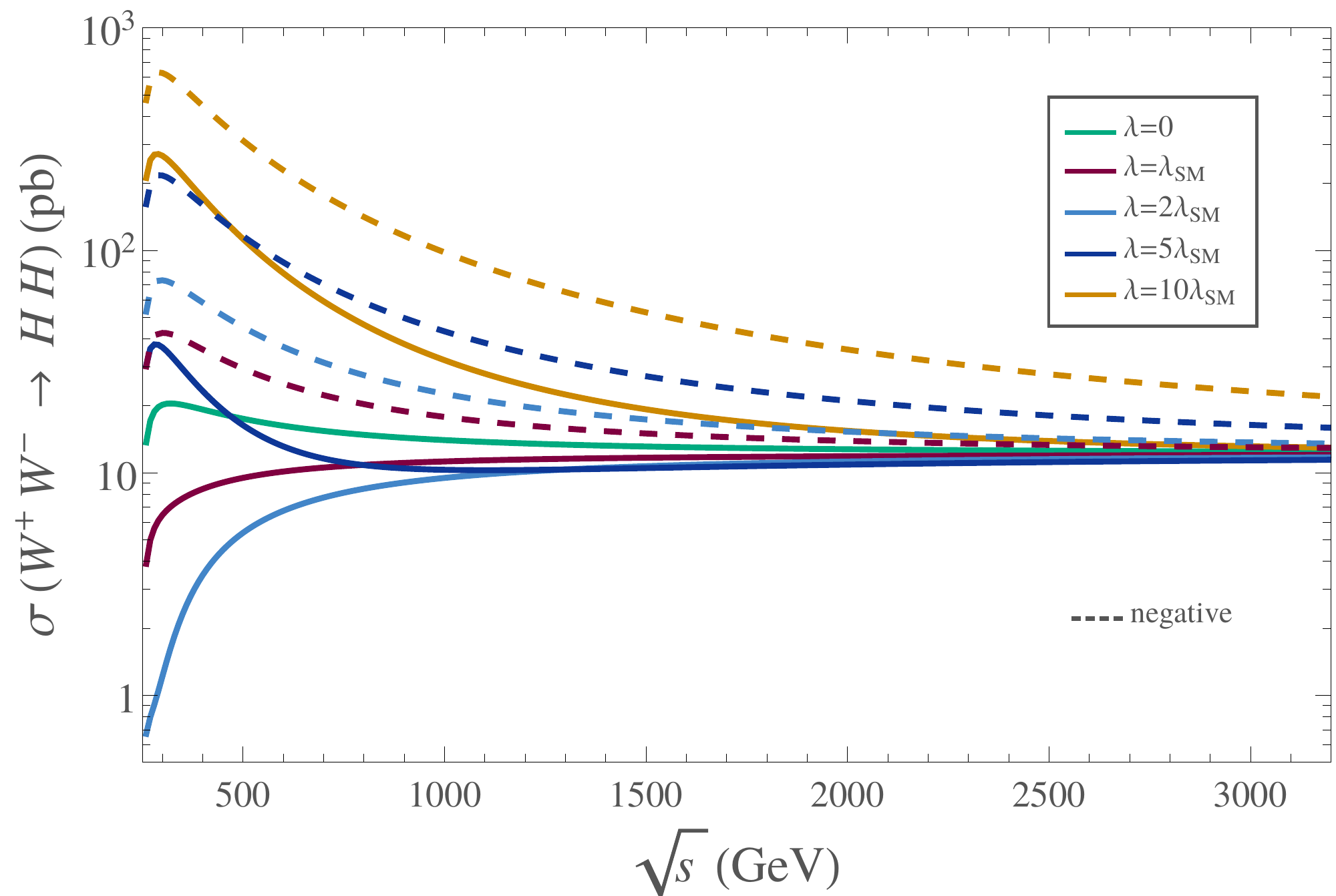}
\includegraphics[width=0.49\textwidth]{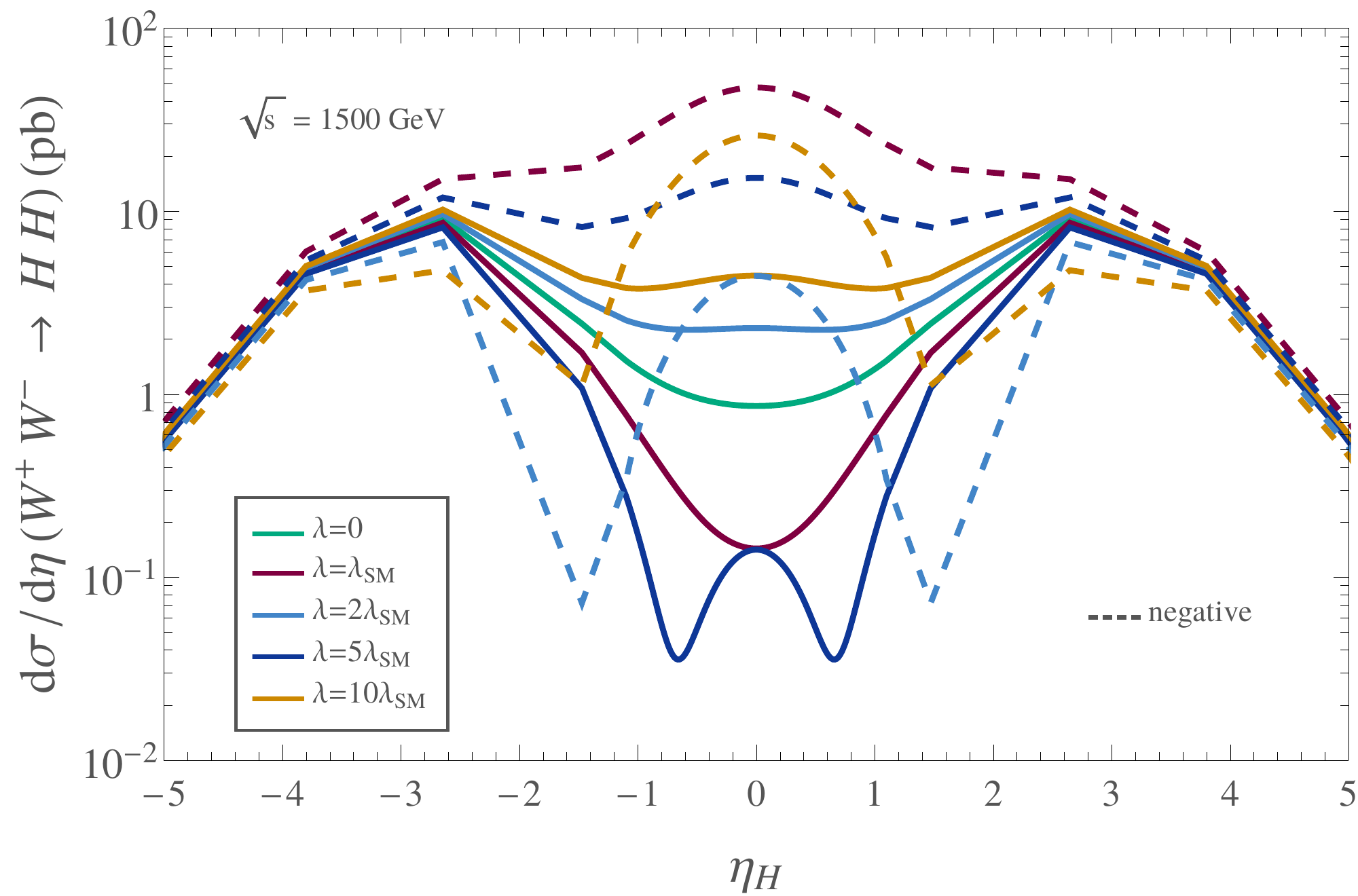}
\caption{Predictions for the total cross section of the process $W^+W^-\to HH$,   as a function of the center of mass energy $\sqrt{s}$ (left panel) and as a function of the pesudorapidity of one of the final H at a fixed center of mass energy of $\sqrt{s}=1500$ GeV (right panel) for different values of the Higgs self-coupling $\lambda$. Solid (dashed) lines correspond to positive (negative) values of $\lambda$.}
\label{fig:subprocess_varylambda}
\end{center}
\end{figure}

It is important to understand in more detail at this point the implications of setting $\lambda$ to a different value than $\lambda_{SM}$ in the kinematical properties of the VBS processes we are studying here. For this purpose, we present in \figref{fig:subprocess_varylambda} the total cross section of the process $W^+W^-\to HH$ as a function of the center of mass energy $\sqrt{s}$  and the differential cross section with respect to the pseudorapidity $\eta_H$ of one of the final Higgs bosons (notice that the distribution with respect to the pseudorapidity of the other Higgs particle is the same) for different values of positive, vanishing and negative $\lambda$. We assume here a phenomenological approach when setting $\lambda\neq\lambda_{SM}$ , meaning that it is not our aim to understand the theoretical implications of such a result like potential instabilities for negative values of $\lambda$, etc. We understand that the deviations in this coupling would come together with other BSM Lagrangian terms that would make the whole framework consistent. For instance, such BSM values for the Higgs self-coupling can be accommodated in the EChL description. Since in this context the Higgs boson is a singlet of the chiral symmetry, as we have seen, a term of the form $(H/v)^3 \,{\rm Tr} \big[D^\mu U^\dagger D_\mu U \big]$ (among others) can be included in the Lagrangian presented in \eqref{eq.L2}, accounting for BSM deviations in the Higgs trilinear coupling. 

Back to the results in \figref{fig:subprocess_varylambda} it can be seen that, first and most evidently, the total cross section changes in magnitude and in energy dependence with respect to the SM one, as already announced. This happens especially near the $HH$ production threshold, confirming that the sensitivity to deviations in $\lambda$ with respect to the SM value is larger in this region. For the case of positive $\lambda$ the total BSM cross section can be larger or lower than that in the SM, depending on the size of the deviations in $\lambda$ with respect to $\lambda_{SM}$, since in this case there is a destructive interference between the $s$ channel contribution and the rest $(c+t+u)$. In contrast, for the case of negative $\lambda$ values, the sum of diagrams is always constructive and one obtains bigger cross sections than the SM one independently of the absolute value of the coupling. The details of these features will be extended when commenting the next figure. As a final comment, it is important to mention that he results for $ZZ\to HH$ (not shown) are very similar to those of $W^+W^-\to HH$.

Regarding the angular dependence of the differential cross section, or, correspondingly, the distribution respect to $\eta _H$ also shown in \figref{fig:subprocess_varylambda}, we see clearly that it also changes in the BSM scenarios respect to the SM one. We particularly learn from this figure that for central values of the Higgs pseudorapidity, concretely for $|\eta_H|<2.5$, it is much easier to distinguish between different values of $\lambda$. Therefore, this suggests the kind of optimal cuts in this variable $\eta _H$ (or the equivalent one in terms of the final particles from the Higgs decays) we should be imposing to enhance the sensitivity to the signal when moving to the realistic case of the $pp$ collisions at the LHC.

\begin{figure}[t!]
\begin{center}
\includegraphics[width=0.49\textwidth]{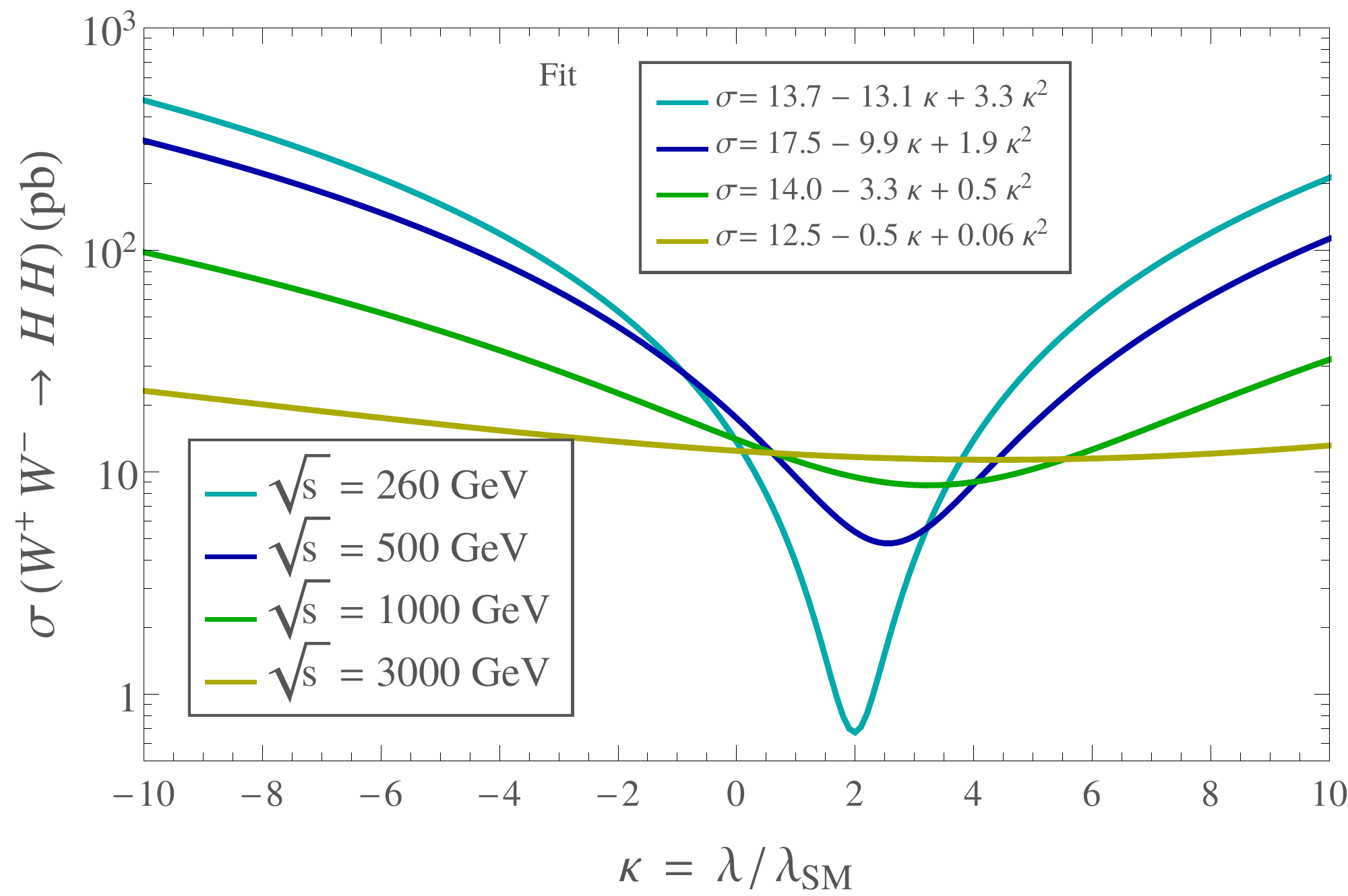}
\includegraphics[width=0.49\textwidth]{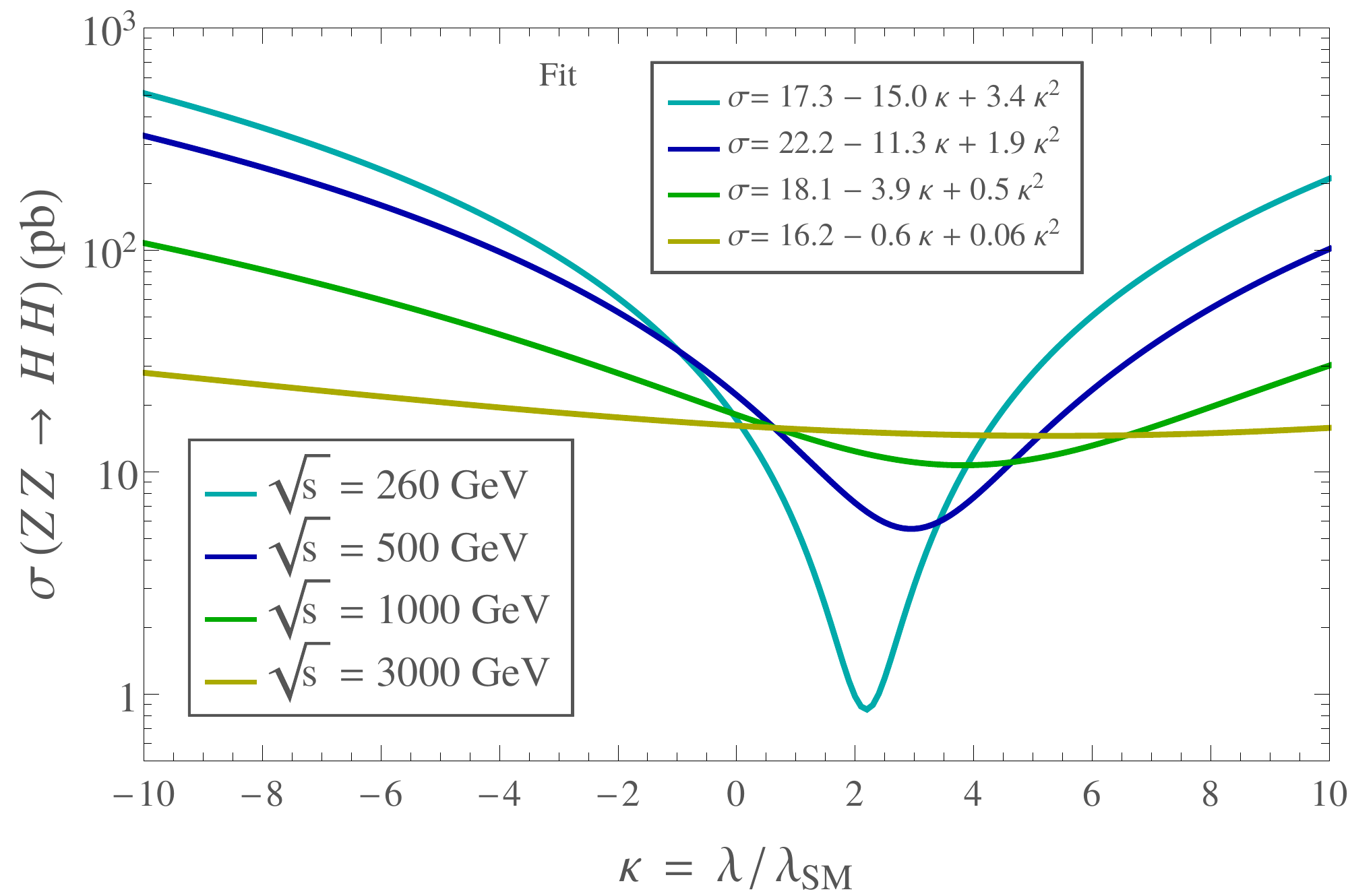}
\caption{Prediction for the total cross section of the VBS process $W^+W^- \to HH$ (left panel) and of $ZZ\to HH$ (right panel) as a function of the ratio of a generic $\lambda$ value over the SM value for four different center of mass energies: $\sqrt{s}=$ 260, 500, 1000 and 3000 GeV.}
\label{fig:subprocess_cancellation_energy}
\end{center}
\end{figure}

In \figref{fig:subprocess_cancellation_energy} we display our predictions for the total cross section of the two relevant VBS processes as a function of $\kappa$ for four different values of fixed center of mass energy $\sqrt{s}=260,\,500,\,1000,\,3000$ GeV. We also display the parabolic fits that allow us to describe each of the curves to have a more analytical insight into the details of how the above commented cancellations among diagrams do actually occur. The formulas of the fits in this figure manifest that, in general, the cross section has a quadratic, a constant and a linear term in $\kappa$, coming, respectively, from the $s$-channel contribution, from the $(c+t+u)$ contribution and from the interference between them. The sign of the interference is negative for positive values of $\kappa$ and positive for  negative values of $\kappa$. This destructive interference for $\lambda>0$ produces that the minima of these lines are placed at $\lambda > \lambda_{SM}$. Besides, depending on the energy and on the size of $\kappa$, the behavior of the cross section will be dominantly constant, linear or quadratic in $\lambda$, and therefore the sensitivity to $\lambda$ will vary accordingly.

 Near the production threshold, i.e., at energies around 250 GeV, two issues can be seen. The first one is that, as we already saw in \figref{fig:subprocess_varylambda}, the differences in the cross section when we vary $\lambda$ are maximal, and so will be the sensitivity to differences in this coupling. The second one is that, at these low energies, the SM, corresponding to $\kappa=1$, suffers, as already said, a mild cancellation between the linear and the constant terms, and therefore the sensitivity to $\lambda$ will be mainly quadratic. We can also see that the minima of the parabolas soften (in the sense that the variations in the cross section when we vary $\lambda$ become smaller) and that their position moves from $\lambda/\lambda_{SM}$ close to 2 to larger values as the energy is increased. Because of this, the bigger the energy, the bigger the value of $\lambda$ that maximizes the cancellations. Thus, as a first conclusion at this point, we will have to keep in mind, once we perform the full collider analysis, that the sensitivity to different values of the trilinear coupling and the issue of delicate cancellations among diagrams in VBS are clearly correlated and this will affect  the final results at the LHC.

A final comment has to be made in this section, and it is that of a potential unitarity violation problem for large $|\lambda|$ values in the processes of our interest here, $VV \to HH$. To check this unitarity issue, we have evaluated the partial waves $a_J$ of the dominant polarization channels for this scattering, which, as we have said, are the longitudinal ones, i.e., $V_LV_L \to HH$. These $a_J$ of fixed angular momentum $J$ are evaluated following \eqref{pwamp}\footnote{Notice that in this case only two polarizations intervene, i.e., those of the initial gauge bosons, since the final Higgs particles are scalars.}.
By doing this exercise, we find that all the partial waves $|a_J|$ that we have computed are below 0.1 for values of $\lambda$ between -10 and 10 times the SM value at all energies. So, for the present study, we are safe from unitarity violation problems.

 For completeness, we have also made a fast estimate of the value of $\lambda$ that would be required to violate unitarity in this process. For large values of $|\lambda|$, the dominant contribution to the total amplitude comes from the $s$-channel. This contribution, as we mentioned before, behaves, at high energies and for the purely longitudinal case, as a constant. In particular, one obtains that $A_s(V_LV_L\to HH)\sim 6\,\lambda$ for $\sqrt{s}\gg m_H$. With this amplitude, one can compute the value of $\lambda$ for which the biggest partial wave (in this case we have checked that it is the one corresponding to $J=0$) becomes one. We obtain $\lambda_{\rm unit}\sim 17$. Notice that this upper limit of $\lambda$ is above the perturbativity limit given naively by $\lambda_{\rm pert}\sim\sqrt{16\pi}\sim 7$. 
  
With all these features in mind, we can move on from the subprocess level to the full process at the LHC to study the sensitivity of this collider to the Higgs self coupling in VBS processes.

 
 \section{Sensitivity to the Higgs self-coupling at the LHC}
 \label{LHC}
 
Once we have characterized completely the scattering $VV \to HH$, it is time to explore the full process at the LHC to quantify how sensitive this machine could be to the Higgs trilinear coupling in VBS processes. This is precisely the aim of this section, in which we first promote the previous analysis at the subprocess level to that of its LHC signal, $pp\to HHjj$, so that we can fully understand its behaviour and properties, and then we give more quantitative and realistic results for the sensitivity to $\lambda$ once the Higgs bosons have decayed. Specifically, we will focus first on the dominant Higgs decays to bottoms, leading to the process \bbbbjj. This process benefits from having more statistics due to the large branching ratios involved, and, because of this, it is presumably the one that will lead to the best sensitivities. We will also present results on other channels, concretely for $p p \to b \bar{b} \gamma\gamma jj$, where one of the two Higgs bosons has decayed into two photons, that, despite its smaller number of events, might also provide interesting results since it suffers from less severe backgrounds.

For all computations and results of the signal events we use MadGraph5 \cite{Alwall:2014hca}, setting the factorization scale to $Q^2=m_Z^2$ and using the set of PDF's NNPDF2.3~\cite{Ball:2013hta}. We have found that changing the chosen value of $Q^2$ does not lead to relevant changes in the signal rates. Concerning the backgrounds, all of them are simulated with the same settings and PDF's as the signal, using  MadGraph5 as well. 
For the case of the multijet QCD background in the $pp \to b \bar{b} b \bar{b} jj$ channel, due to its complexity, we have simulated events using both MadGraph5 with the previous mentioned settings and PDF's, and AlpGen \cite{Mangano:2002ea}, this time choosing $Q^2=(p_{T_b}^2+p_{T_{\bar{b}}}^2+\sum p_{T_j}^2)/6$ and selecting the set of PDF's CTEQ5L \cite{Lai:1999wy}. We have found agreement between the results of these two Monte Carlos in the total normalization of the cross section with the basic cuts and in the shape of the relevant distributions within the provided errors. All results are presented for a center of mass energy of $\sqrt{s}= 14$ TeV.

\begin{figure}[t!]
\begin{center}
\includegraphics[width=0.3\textwidth]{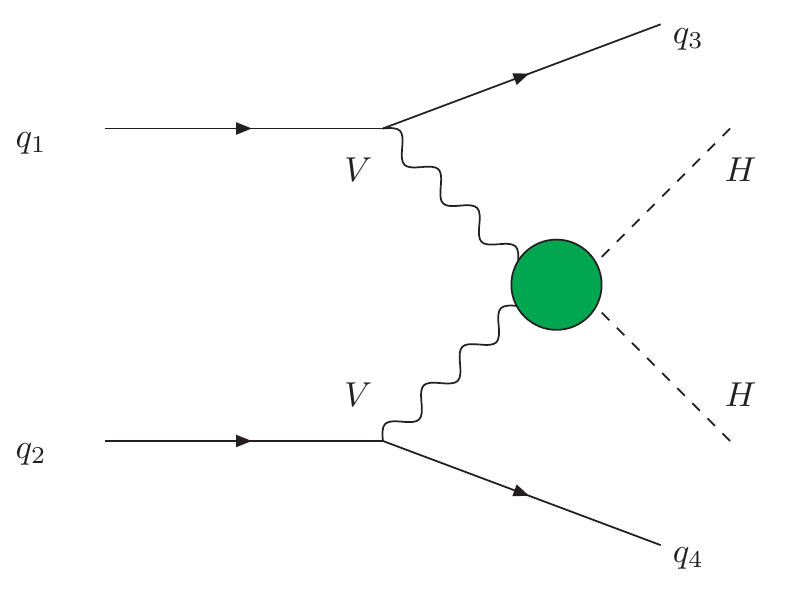}
\caption{Schematic representation of  partonic double Higgs production though VBS at the LHC. The green blob represents the presence of the Higgs self-coupling $\lambda$ in the process, although all diagrams in Fig.(\ref{fig:subprocess_diagrams}) are considered.}
\label{fig:LHC_diagrams}
\end{center}
\end{figure}

 
 \subsection[Study and characterization of  $pp\to HHjj$ signal events]{Study and characterization of  $\boldsymbol{pp\to HHjj}$ signal events}
 \label{HHjj}
 
In order to be able to estimate the sensitivity to the Higgs self-coupling in VBS at the LHC, we need to understand how the results of the previous section translate into the full process when we start with protons as initial particles. This full process, $pp\to HHjj$,  can be produced via many different channels, and not only in VBS configurations. In fact, it is well known that this VBS subset of diagrams contributing to $q_1 q_2 \to q_3 q_4 H H$ is not gauge invariant by itself and all kinds of contributing diagrams have to be included to get  a gauge invariant result. This is indeed what we are doing here, since when we use MadGraph to compute the signal all kinds of diagrams are included. 

The crucial point regarding the phenomenological interest of VBS, that indeed motivates this work, is that the specific VBS configuration can be very efficiently selected by choosing the appropriate kinematic regions of the two extra jets variables, as it has already been discussed in the previous Chapter, as well as in the recent literature \cite{Goncalves:2018qas, Doroba:2012pd, Szleper:2014xxa, Delgado:2017cls}. We recall here that  the VBS topologies are characterized by large separations in pseudorapidity of the jets, $|\Delta\eta_{jj}|=|\eta_{j_1}-\eta_{j_2}|$, and by large invariant masses of the dijet system, $M_{jj}$. Imposing proper cuts over these two variables makes possible to obtain events that come dominantly from VBS processes and, as we will see later on, also to reject many background events.

\begin{figure}[t!]
\begin{center}
\includegraphics[width=0.49\textwidth]{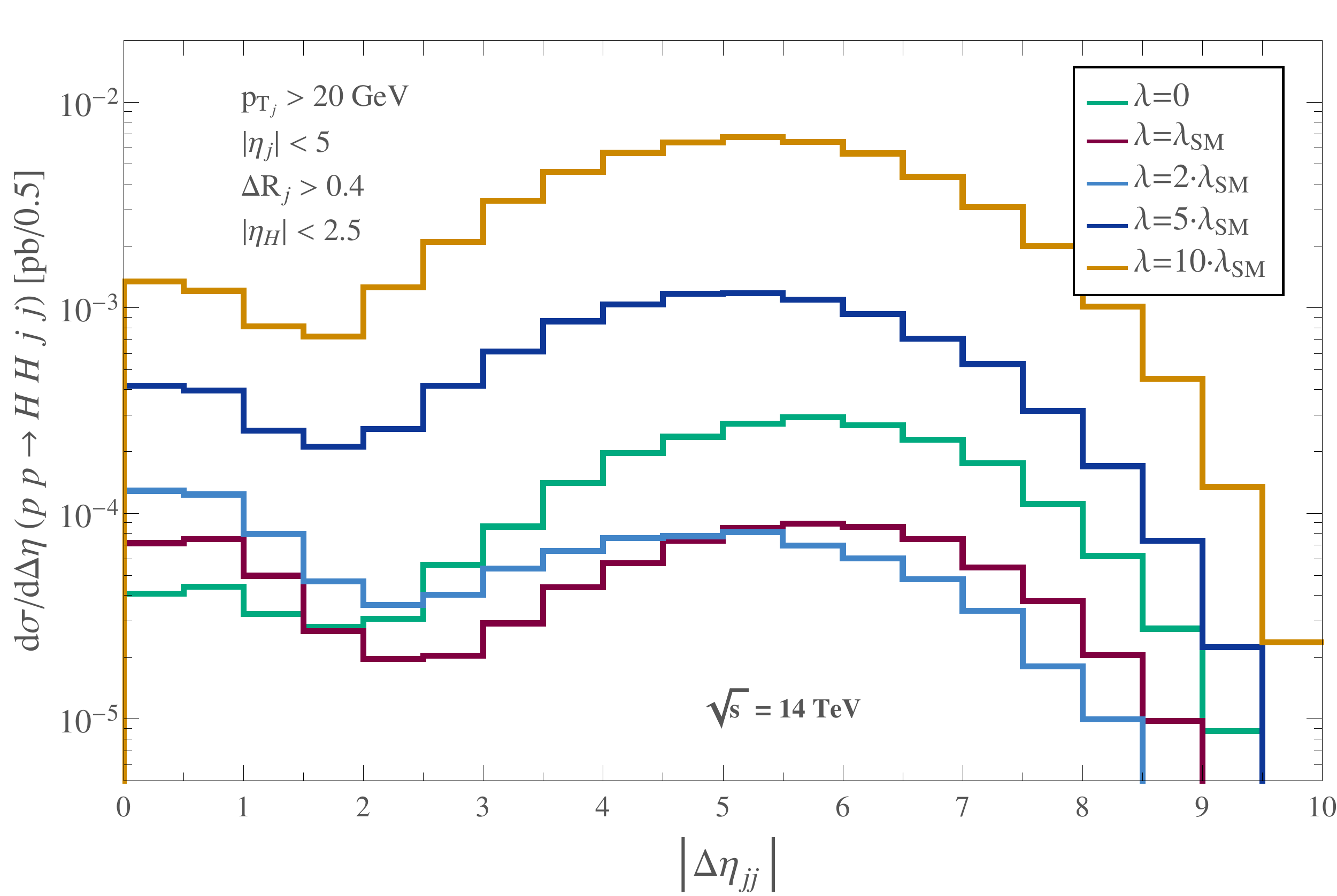}
\includegraphics[width=0.49\textwidth]{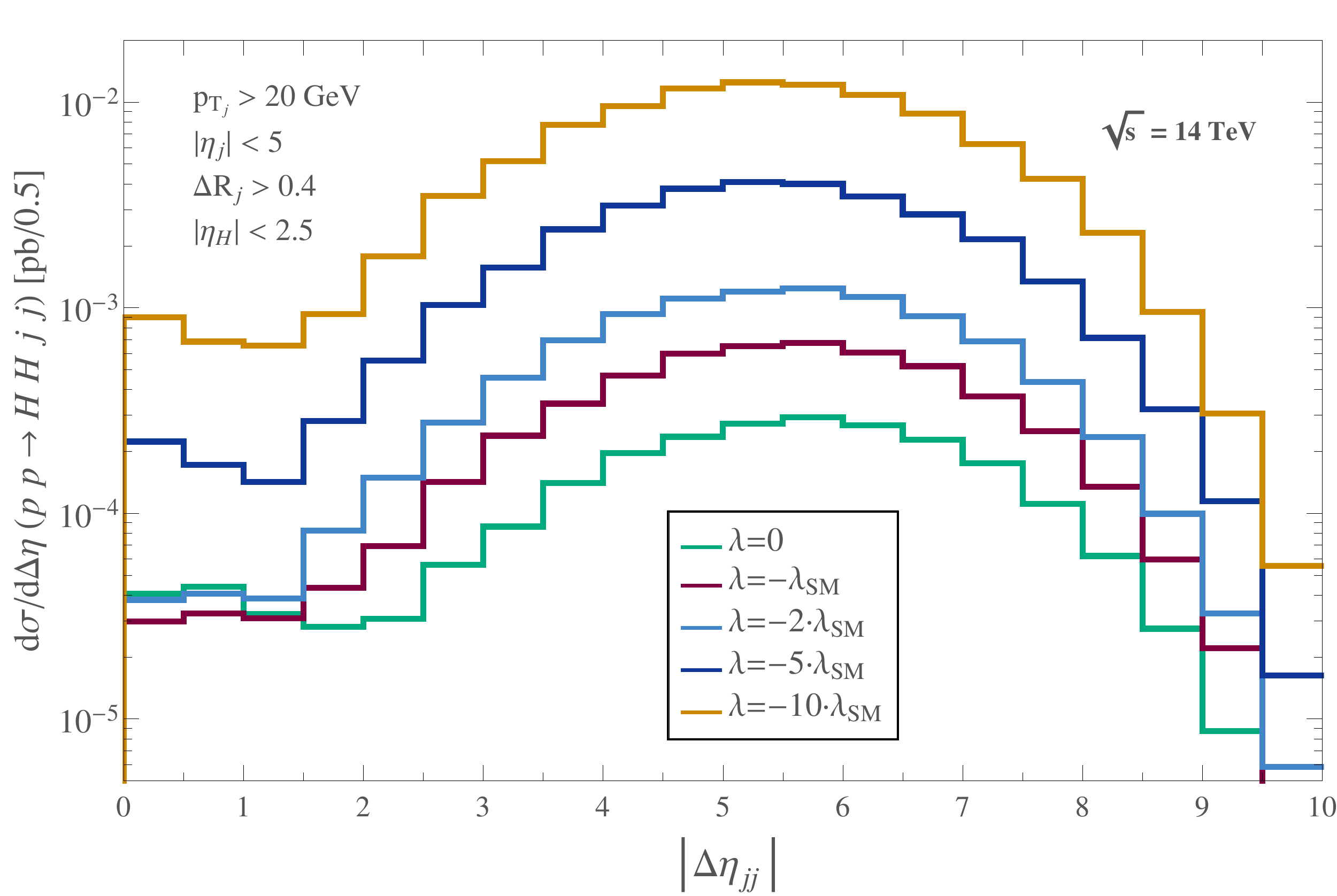}
\includegraphics[width=0.49\textwidth]{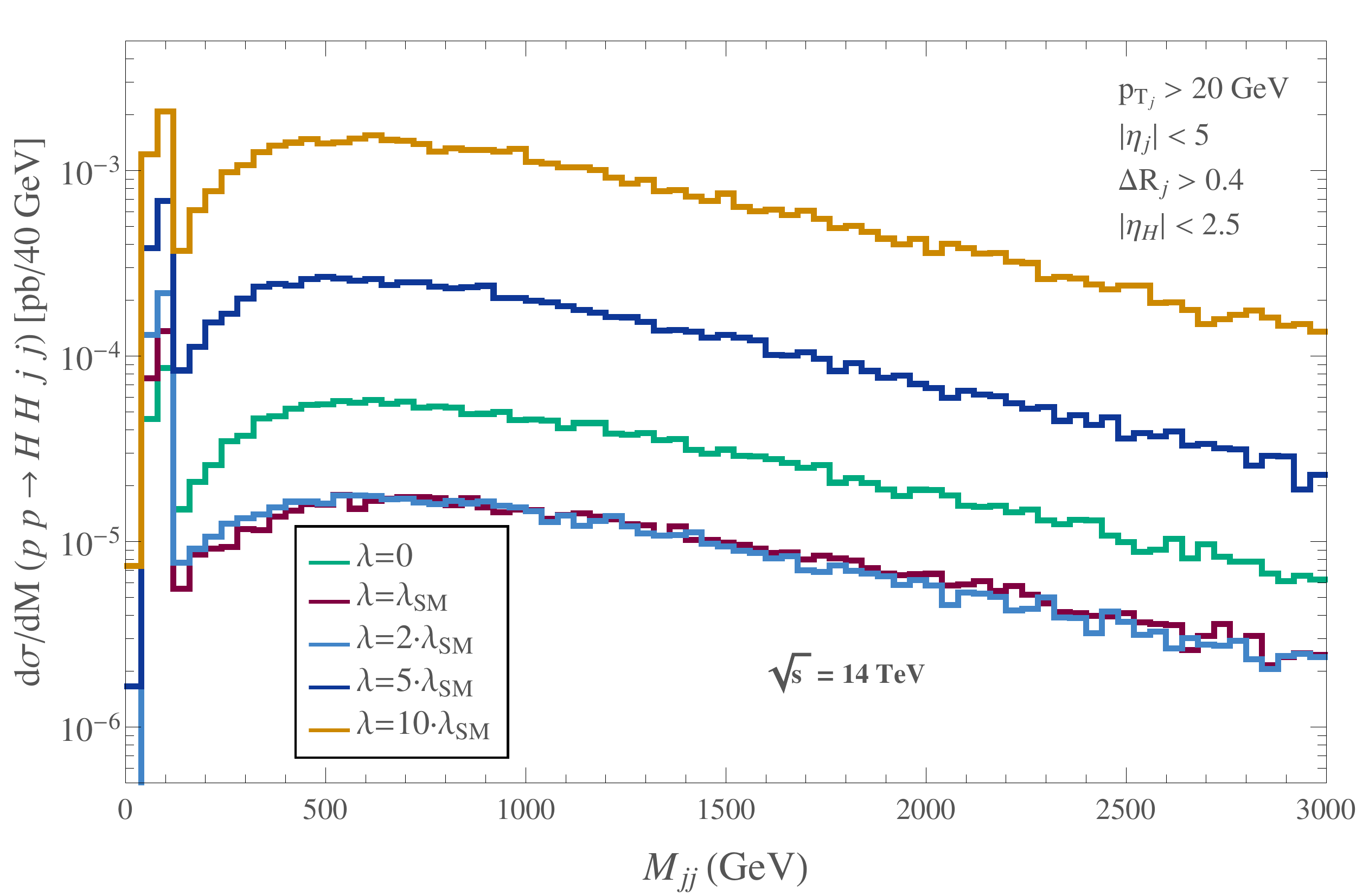}
\includegraphics[width=0.49\textwidth]{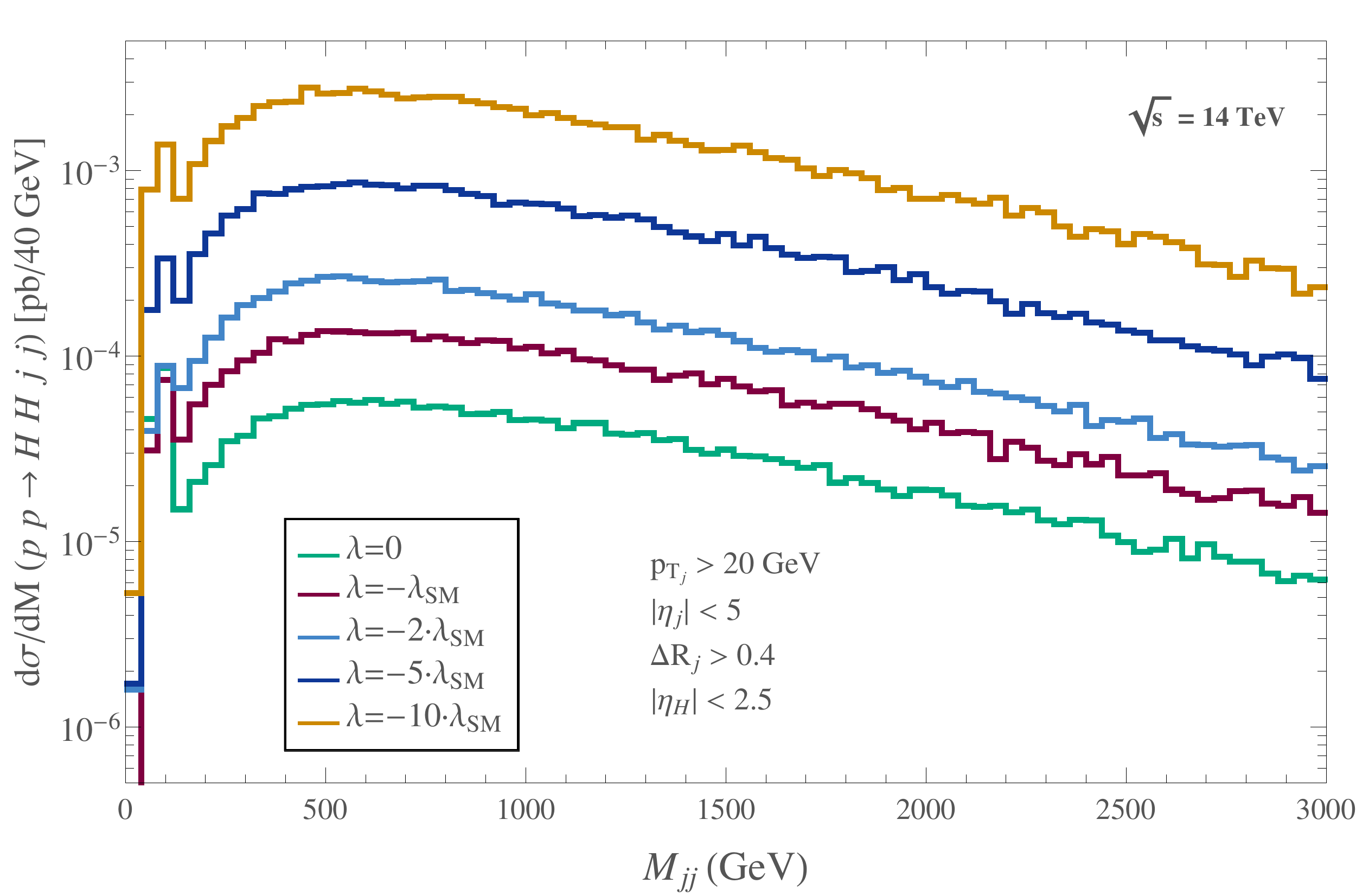}
\caption{Predictions for the total cross section of the process $pp\to HHjj$ as a function of the absolute value of the difference between pseudorapidities of the two jets $|\Delta\eta_{jj}|$ (upper panels) and as a function of the invariant mass of the two jets $M_{jj}$ (lower panels) for different values of the Higgs self-coupling $\lambda$. We display positive (left panels) and negative (right panels) values of $\lambda$ for comparison. We also include the case $\lambda=0$. Cuts in Eq.(\ref{basiccutsHHjj}) have been applied and the center of mass energy has been set to $\sqrt{s}=14$ TeV.}
\label{fig:HHjj_VBSdistributions}
\end{center}
\end{figure}

 The VBS processes involved in $pp\to HHjj$ can be seen schematicaly in \figref{fig:LHC_diagrams}, where the green blob represents all diagrams in \figref{fig:subprocess_diagrams}, including the presence of the $s$-channel with the generic Higgs trilinear coupling $\lambda$. This kind of processes will inherit the properties of the sub-scatterings we have studied, but will also have differences with respect to them due to the fact that we now have protons in the initial state. Then, it is important to know at this stage how close to the ``pure'' VBS configuration our $pp\to HHjj$ signal is. To this end, we have generated with MadGraph5 $pp\to HHjj$ signal events for this process for different values of $\lambda$ with a set of basic cuts that allow for the detection of the final particles, given by: 
\begin{align}
p_{T_j} & > 20~ {\rm GeV}\,,~~~ |\eta_j|<5\,,~~~ \Delta R_{jj} > 0.4\,,~~~ |\eta_H|<2.5\,,\label{basiccutsHHjj} 
\end{align}
where $p_{T_j}$ is the transverse momentum of the jets, $\eta_{j,H}$ is the pseudorapidity of the jets or of the Higgs bosons, and $\Delta R_{jj}$ is the angular separation between two jets defined as $\Delta R_{jj}=\sqrt{\Delta\eta_{jj}^2+\Delta\phi_{jj}^2}$, with $\Delta\eta_{jj}$ and $\Delta\phi_{jj}$ being the angular separation in the longitudinal and transverse planes, respectively.

 In \figref{fig:HHjj_VBSdistributions} we present the predictions for the cross section of the process $pp\to HHjj$ for different values of $\lambda$ as a function of the separation in pseudorapidity of the final jets $|\Delta\eta_{jj}|$ and as a function of the invariant mass of these two jets $M_{jj}$. In these plots we can see that our signal is indeed dominated by the VBS configuration, since a very large fraction of the events populates the kinematic regions that correspond to VBS topologies. To have a quantitative estimation, we can take, for instance, the VBS selection cuts proposed in \cite{Delgado:2017cls} and impose them to the events shown in \figref{fig:HHjj_VBSdistributions}. Thus, by imposing these cuts:
\begin{align}
{\rm VBS\,\, CUTS :\,}\,\,\,\,\,|\Delta\eta_{jj}|&>4\,,~~~ M_{jj}>500~ {\rm GeV}\,,\label{VBSselectioncuts}
\end{align}
 we obtain that between 50\% and 75\% (depending on the value of $\lambda$, since the larger the value of the Higgs self-coupling the larger this percentage) of the events are accepted within them, which means that the VBS topologies amount\footnote{In the sense of the fraction of events that pass the VBS cuts with respect to the total number of events.}, at least, to half of the total cross section of $pp\to HHjj$. This is indeed a very interesting result, since, as we will see in the forthcoming section, the VBS cuts allow us to reduce some backgrounds even in two orders of magnitude. The fact that the signal is practically left unaffected by these cuts is an excellent outcome as the signal to background ratio will favor a better sensitivity to $\lambda$.  
 
 \begin{figure}[t!]
\begin{center}
\includegraphics[width=0.49\textwidth]{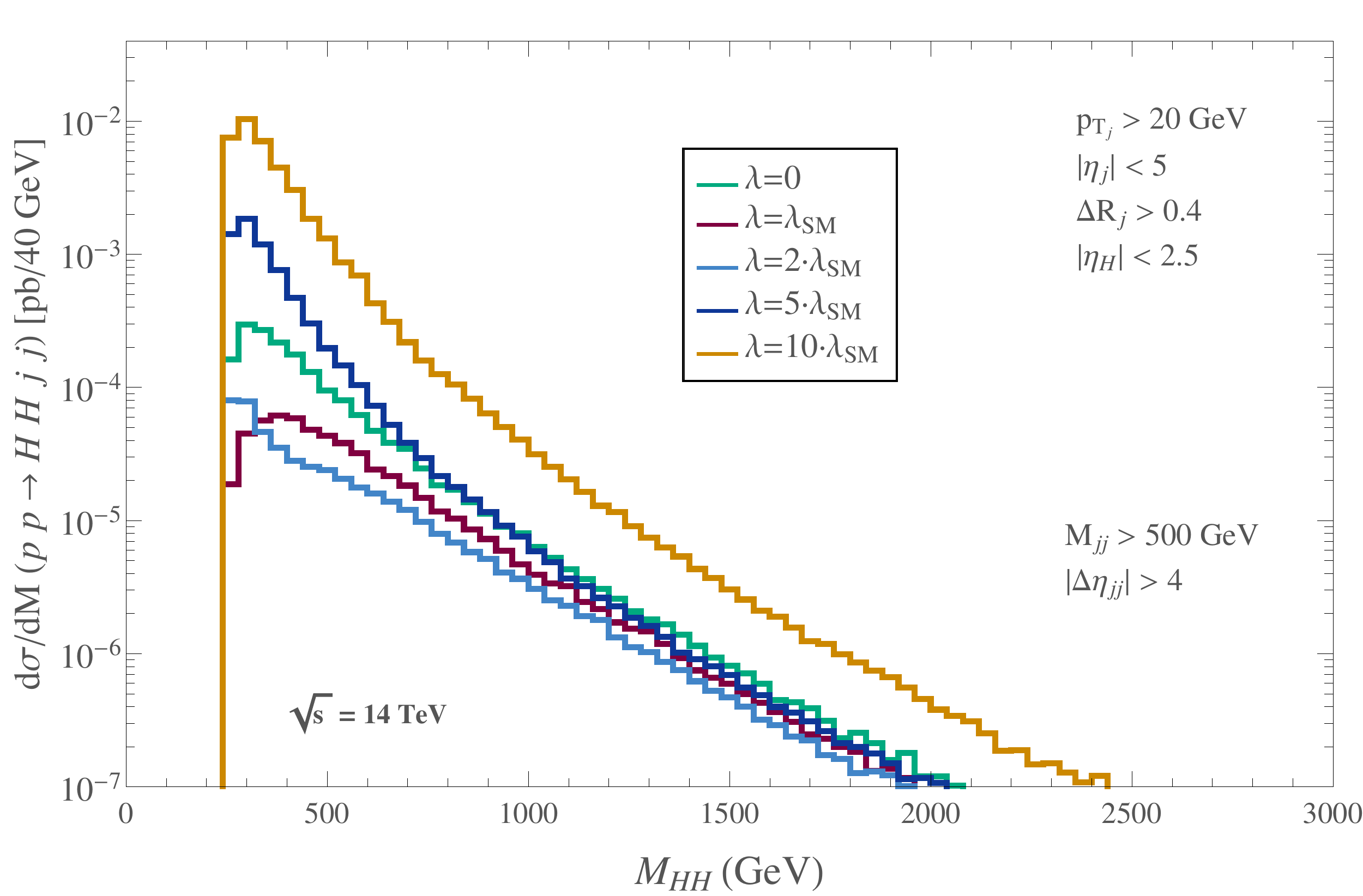}
\includegraphics[width=0.49\textwidth]{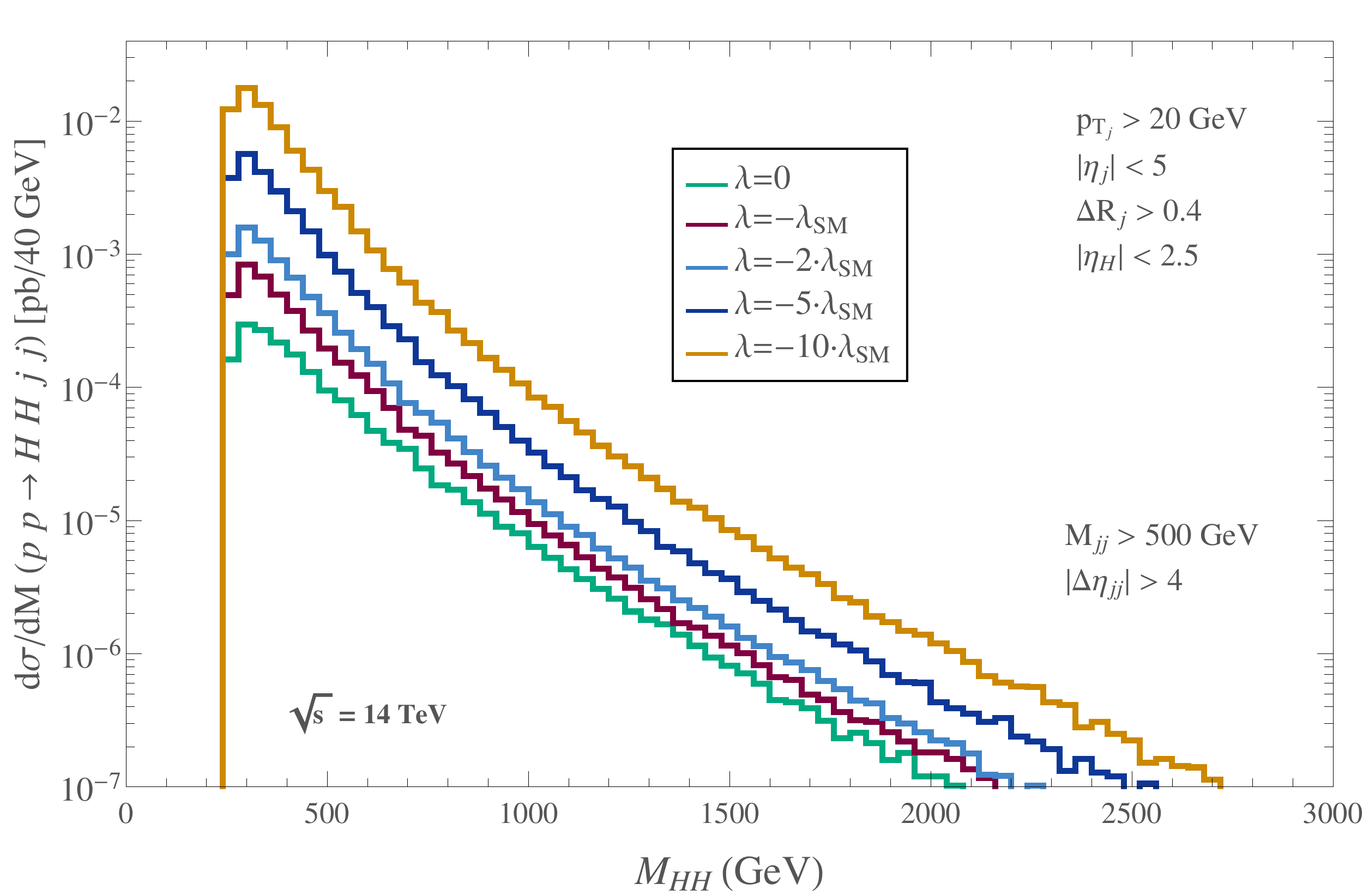}
\caption{Predictions for the total cross section of the process $pp\to HHjj$ as a function of the invariant mass of the di-Higgs system $M_{HH}$ for different values of the Higgs self-coupling $\lambda$. We display positive (left panel) and negative (right panel) values of $\lambda$ for comparison. We also include the case $\lambda=0$. Cuts in Eq.(\ref{basiccutsHHjj}) and VBS selection cuts presented in Eq.(\ref{VBSselectioncuts}) have been applied. The center of mass energy has been set to $\sqrt{s}=14$ TeV.}
\label{fig:MHH_distributions}
\end{center}
\end{figure}

Furthermore, knowing that the process of our interest at the LHC has a dominant VBS configuration, we would expect  the translation from the subprocess results to the complete ones at this level to be straightforward. This appears to be the case, as shown in \figref{fig:MHH_distributions}, where we display the predictions for the total cross section of the process $pp\to HHjj$ as a function of the invariant mass of the diHiggs system, $M_{HH}$, for different values of the Higgs self-coupling after imposing the cuts given in Eqs. (\ref{basiccutsHHjj}) and (\ref{VBSselectioncuts}). In these plots, it is manifest that the curves follow the same tendency as the subprocess when we vary $\lambda$. Near the $HH$ production threshold the difference in the cross sections for different values of the coupling is more pronounced, and one can see again that the cancellations play a role in the same way we learnt at the subprocess level. The SM cross section ($\kappa=1$, in red) lies between the $\kappa=0$ (in green) one, which is bigger, and the $\kappa=2$ (in light blue) one, which is smaller. Again, for negative values of $\kappa$ the cross section is always larger than the SM one, so we will have, for the same absolute value of the coupling, better sensitivities for negative $\lambda$ values.

\begin{figure}[t!]
\begin{center}
\includegraphics[width=0.49\textwidth]{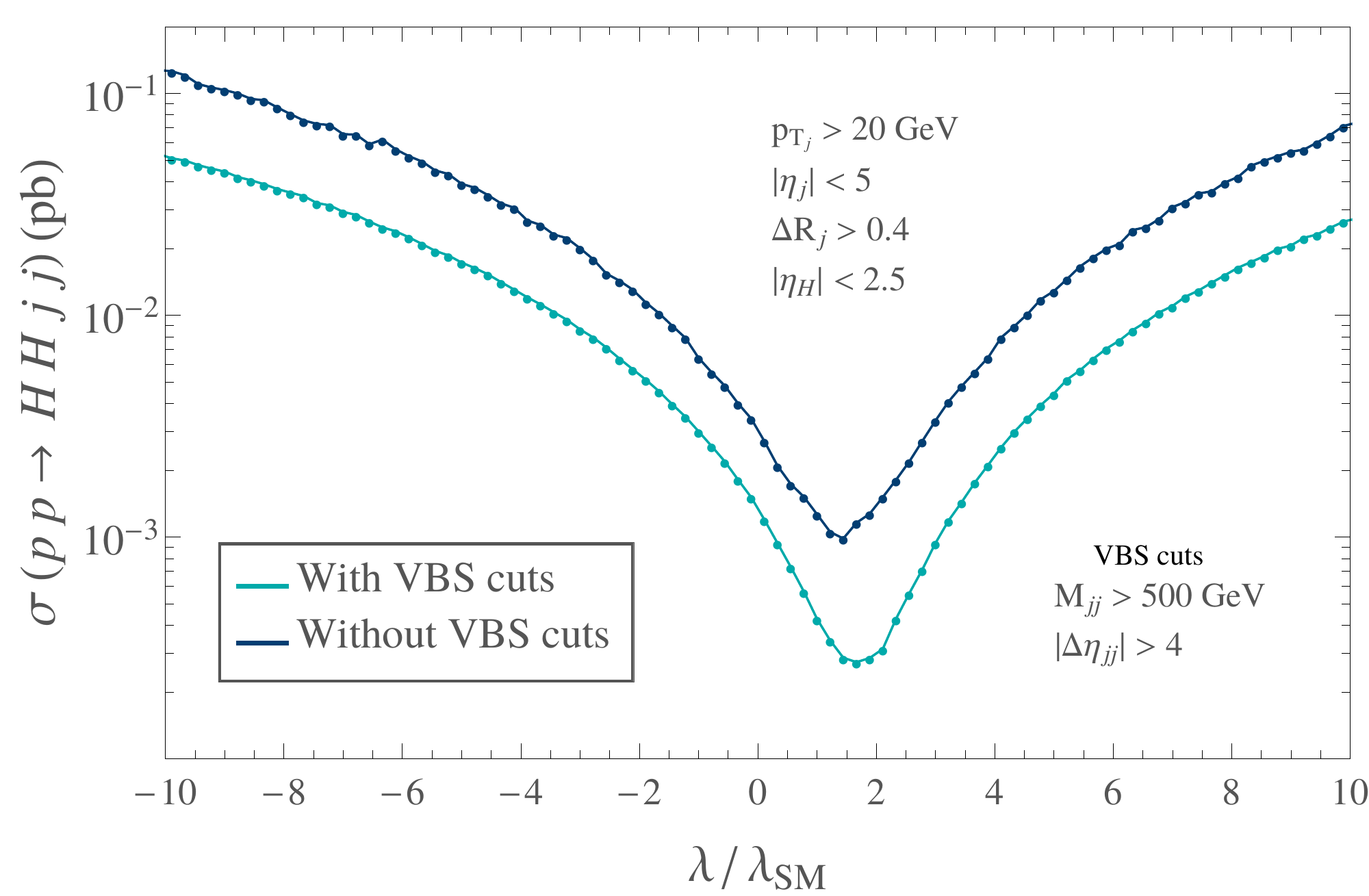}
\includegraphics[width=0.49\textwidth]{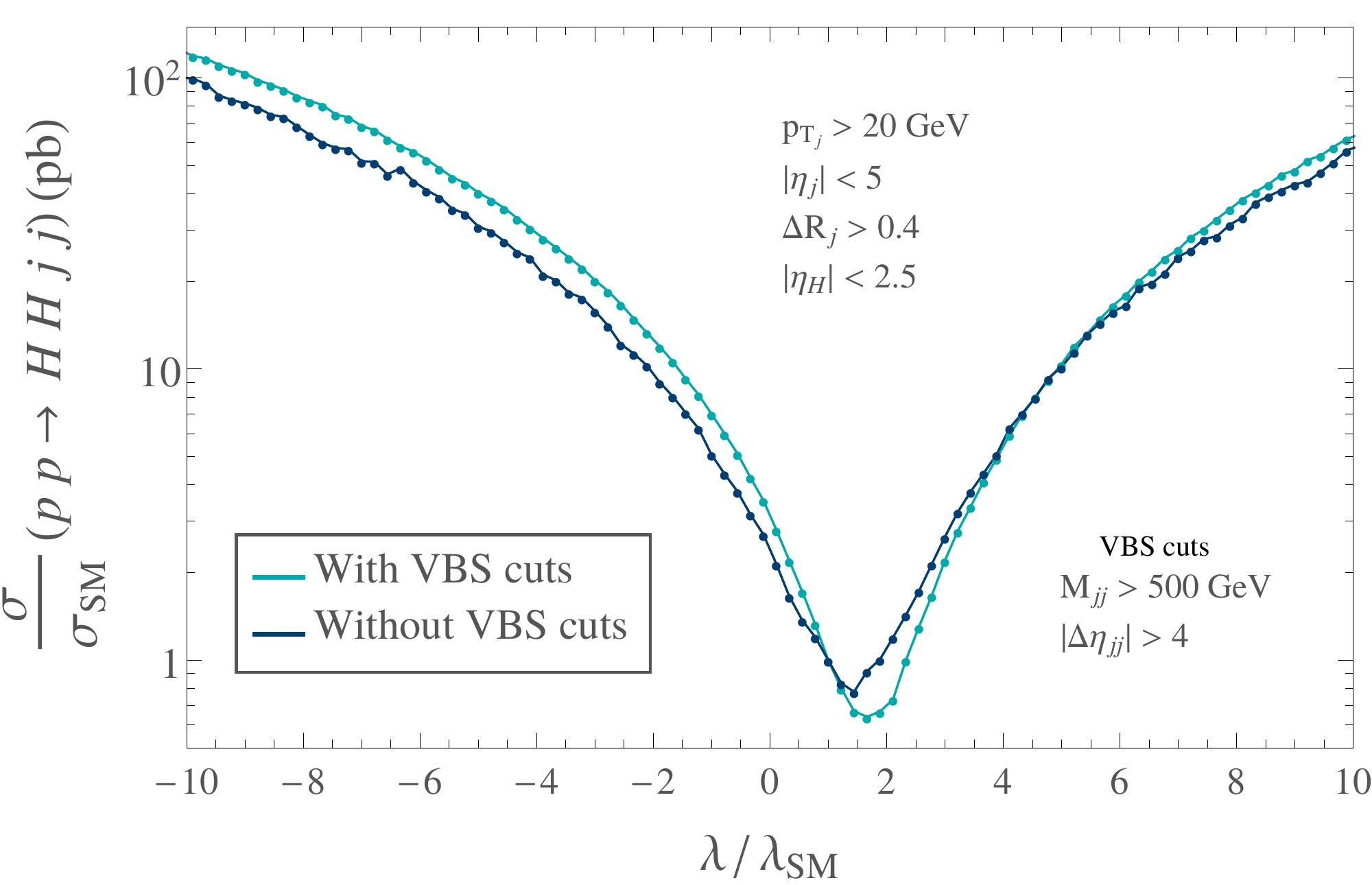}
\caption{Predictions for the total cross section (left panel) and for the ratio of the total cross section over its SM value (right panel) as a function of the Higgs self-coupling $\lambda$ with and without imposing the VBS selection cuts given in Eq.(\ref{VBSselectioncuts}). Cuts in Eq.(\ref{basiccutsHHjj}) have been applied and the center of mass energy has been set to $\sqrt{s}=14$ TeV. }
\label{fig:cancelations}
\end{center}
\end{figure}

The issue of the cancellations that take place between the $\lambda$-dependent diagram and the rest is shown in more detail in \figref{fig:cancelations}. In this figure, we present the predictions for the total cross section for $pp\to HHjj$, and for the ratio of the total cross section over its SM value as a function of the Higgs self-coupling. We also compare the results with and without imposing the VBS cuts given in Eq.(\ref{VBSselectioncuts}) to explore how the cancellation happens at the LHC, and how it depends on the selection of the VBS topologies. We learn again, that, for the same absolute value of $\lambda$, negative values give rise to larger cross sections, and therefore to better sensitivities. The smallest cross section corresponds roughly to $\kappa\sim1.6$, which is the value that will be harder to reach at the LHC.
One may notice that this value does not coincide exactly with that in \figref{fig:subprocess_cancellation_energy}, even for the dominant contribution close to the threshold. This slight displacement of the minimum is due to the fact that many different topologies in addition to those of VBS contribute to this final state, in contrast with the results in  \figref{fig:subprocess_cancellation_energy} that took into account only VBS configurations. In fact, once we apply the VBS cuts the minimum gets closer to that of \figref{fig:subprocess_cancellation_energy}.
 Besides, and interestingly, the effect of imposing the VBS selection cuts can ameliorate the sensitivity to $\lambda$. Although the cross sections reduce in value after applying the cuts, the ratio of the total cross section for a given trilinear coupling over the SM cross section increases when we are away from the region in which the cancellations are relevant, i.e., for $\kappa>3$ and $\kappa<1$.
 
The last issue we would like to point out in this section refers to the kinematical behavior of the VBS subsystem, that is then translated to the kinematics of the final Higgs bosons. Usually, in vector boson scattering processes at the LHC, most of the energy of the initial $pp$ state is transmitted to the radiated EW gauge bosons. This leads, as a consequence, to a very boosted system of final $HH$ pairs, which can be profitable to select these kind of events against backgrounds. If the final Higgs particles are very boosted, their decay products, will have, in general, small angular separations. This, together with the fact that the invariant mass of the two particles that come from the Higgs decay has to lie near the Higgs mass, will allow us to characterize very efficiently the Higgs boson candidates as we will see in the next subsection. With this and the VBS topologies under control, we can study the full processes in which the Higgs bosons have decayed, and compute the sensitivities to $\lambda$ in these realistic BSM scenarios.


 \subsection[Analysis after Higgs boson decays: sensitivity to $\lambda$ in $pp\to b\bar{b}b\bar{b}jj$]{Analysis after Higgs boson decays: sensitivity to $\boldsymbol{\lambda}$ in $\boldsymbol{pp\to b\bar{b}b\bar{b}jj}$}
 \label{4b2j}

 As previously mentioned, once we have fully characterized our most basic process, $pp\to HHjj$, we need to take into account the Higgs decays to perform a realistic analysis at the LHC. The channel we are going to focus on is $p p\to b\bar{b}b\bar{b}jj$, since the decay of the Higgs boson to a bottom-antibottom pair benefits from the biggest branching ratio, BR($H\to b\bar{b}$) $\sim$ 60 \%. Because of this, we will obtain the largest possible rates for our signal, which will allow us to probe the broadest interval of deviations in the Higgs self-coupling. 
 
 Although this process is really interesting because of its large statistics, it is important to mention that it also suffers from having a severe background: the one coming from pure multijet QCD events. This QCD background, of  $\mathcal{O}(\alpha_S^6)$ at the cross section level, leads to the same final state as our signal, $p p\to b\bar{b}b\bar{b}jj$, and, although in general they have very different kinematics, its rates are so high that some of the events can mimic the signal coming from the decay of two Higgs particles. For this reason, we need to be very efficient when applying selection cuts and criteria to be able to reject this particular background.
 
 We learnt in the previous sections that our signal is very dominated by the VBS configuration. Oppositely, the multijet QCD background is composed primarily by topologies that do not share kinematical properties with VBS processes. This is the reason why we will first select those QCD events that can be misidentified as  signal events coming from VBS, and take them as a starting point to perform our more refined study of the signal and background.

To have a first insight on how efficient the VBS selection criteria are, we have generated with MadGraph5 ten thousand events for our signal, $p p\to HHjj\to b\bar{b}b\bar{b}jj$ in the SM, i.e., $\kappa=1$, and for the multijet QCD background with a set of basic cuts that ensure the detection of the final state particles:
 \begin{align}
p_{T_{j,b}}>20~ {\rm GeV}\,;~|\eta_j|<5\,;~|\eta_b|<2.5\,;~\Delta R_{jj,jb}>0.4\,;~\Delta R_{bb}>0.2\,.\label{basiccuts4b2j}
 \end{align}
where $p_{T_j,b}$ is the transverse momentum of the jets and bottoms, $\eta_{j,b}$ are the pseudorapidities of the jets or of the bottom particles, and $\Delta R_{ij}$ is the angular separation between the $i$ and $j$ particles.

\begin{figure}[t!]
\begin{center}
\includegraphics[width=0.49\textwidth]{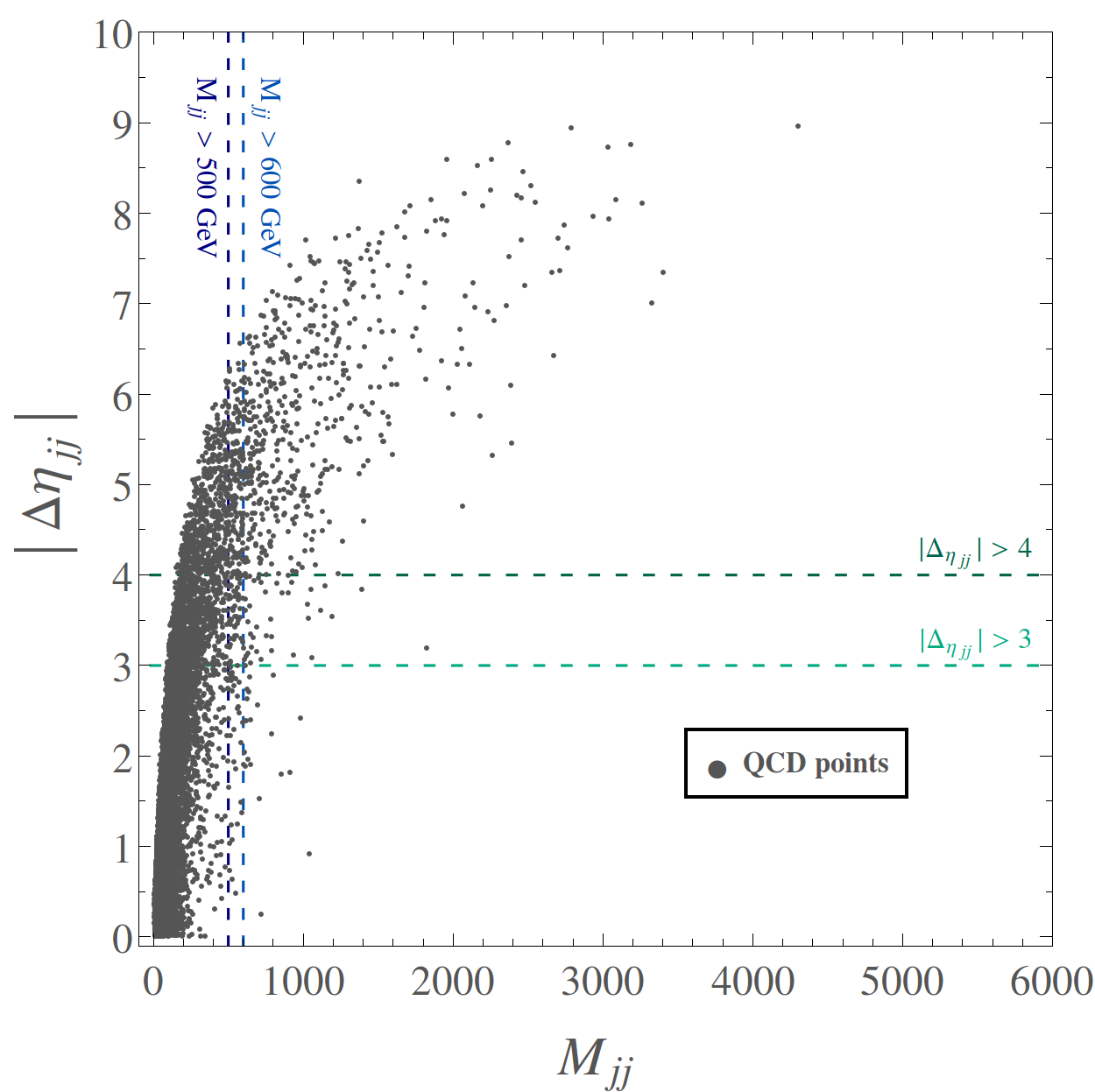}
\includegraphics[width=0.49\textwidth]{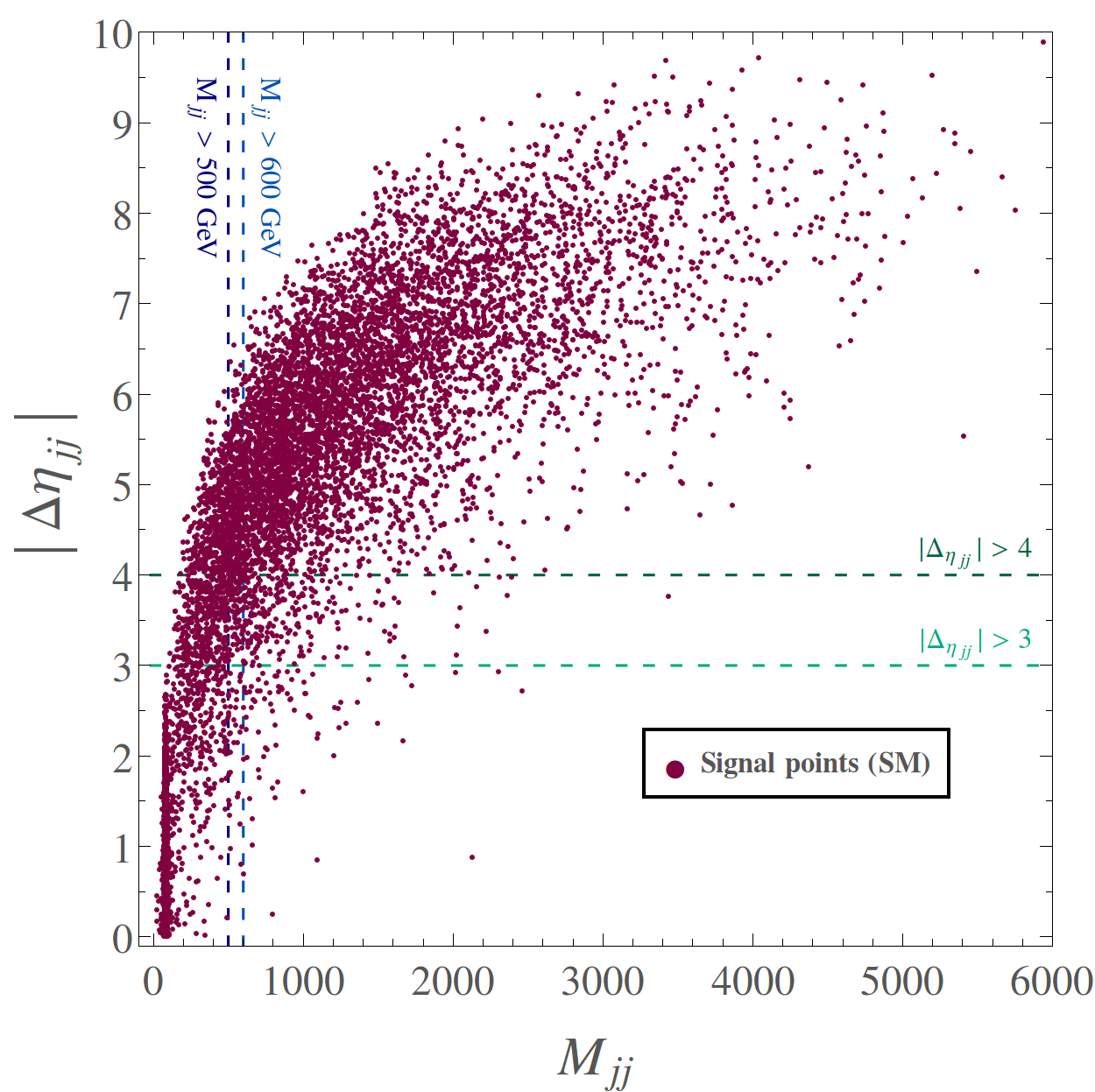}
\caption{Distribution of 10000 Monte Carlo events of multijet QCD background $pp\to b\bar{b}b\bar{b}jj$ (left panel) and signal $pp\to HHjj\to b\bar{b}b\bar{b}jj$ (right panel) in the plane of the absolute value of the difference between pseudorapidities of the two jets $|\Delta\eta_{jj}|$ versus the invariant mass of the two jets $M_{jj}$. Cuts in \eqref{basiccuts4b2j} have been implemented. The center of mass energy has been set to $\sqrt{s}=$14 TeV.}
\label{fig:4b2j_VBSvariables2D}
\end{center}
\end{figure}

In \figref{fig:4b2j_VBSvariables2D} we display the localization of these events in the $|\Delta\eta_{jj}|-M_{jj}$ plane, the two variables that better characterize the VBS processes. One can see, indeed, that the QCD events populate mostly the region of small invariant masses of the dijet system and of small differences in pseudorapidity of the jets, as opposed, precisely, to the signal events. Thus, imposing the proper VBS cuts, like those in \eqref{VBSselectioncuts}, should relevantly reduce the QCD background leaving the signal nearly unaffected.

\begin{figure}[t!]
\begin{center}
\includegraphics[width=0.49\textwidth]{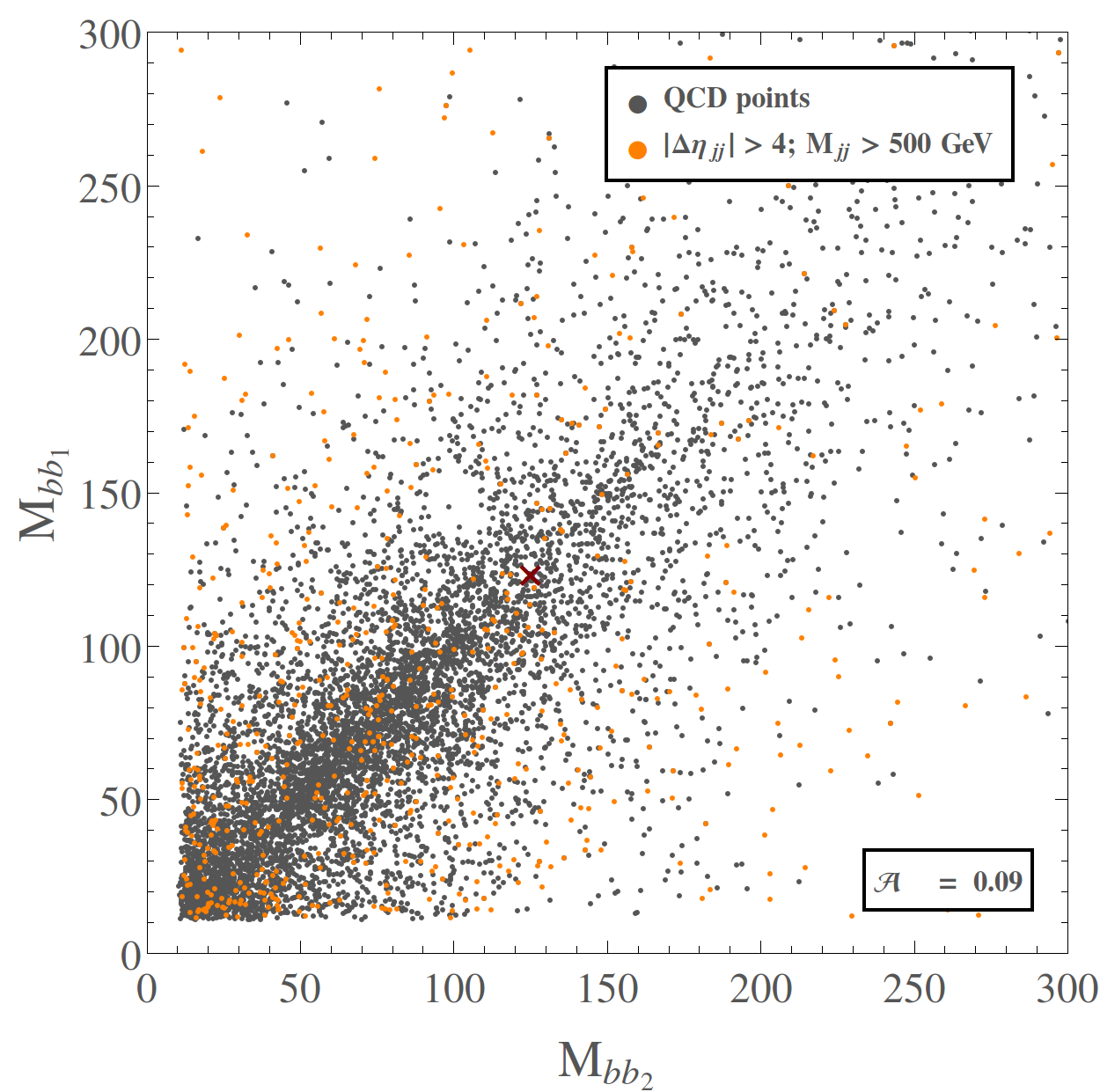}
\includegraphics[width=0.49\textwidth]{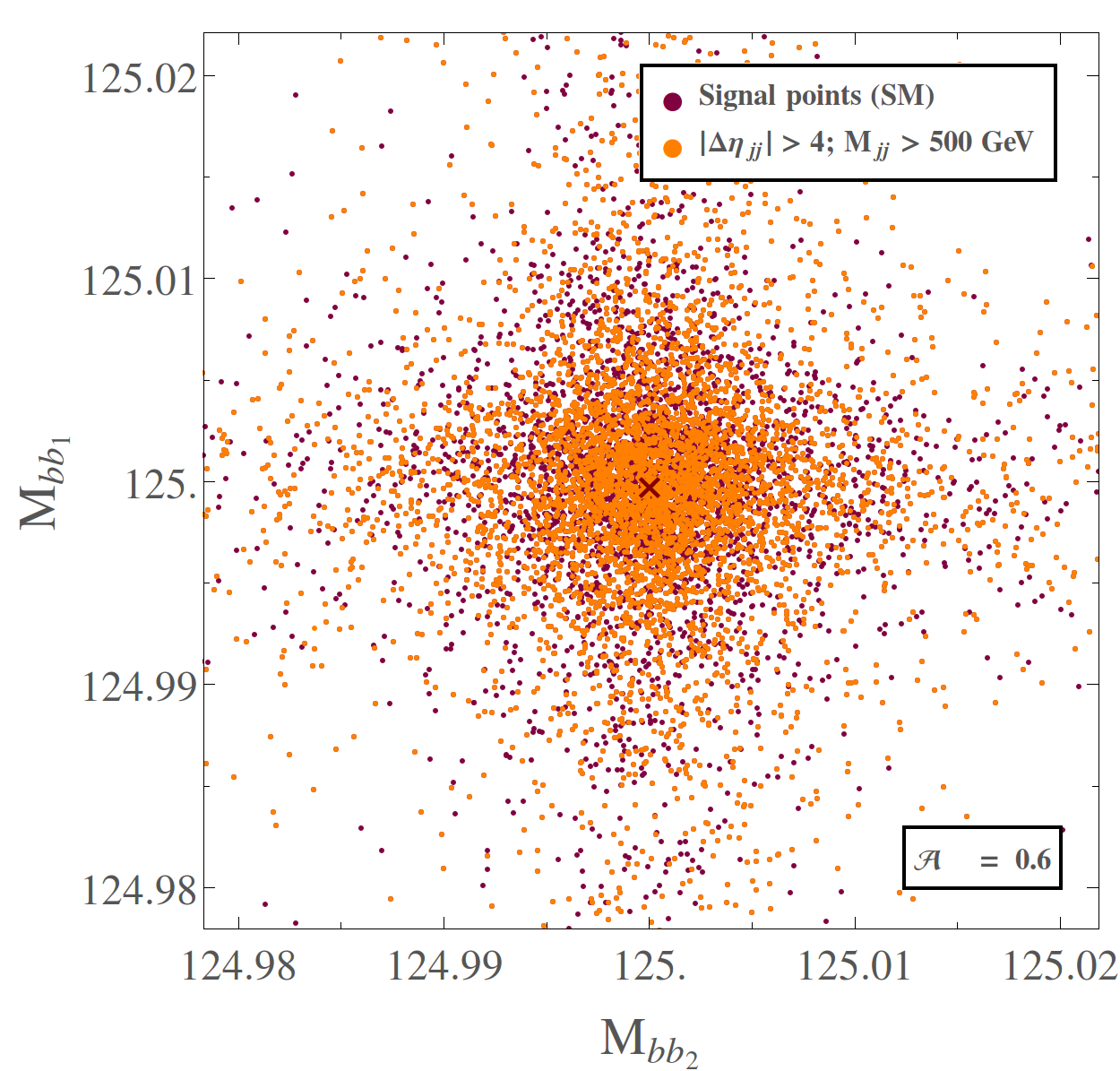}
\caption{Distribution of 10000 Monte Carlo events of multijet QCD background $pp\to b\bar{b}b\bar{b}jj$ (left panel) and signal $pp\to HHjj\to b\bar{b}b\bar{b}jj$ (right panel) in the plane of the invariant mass of one bottom pair identified as a Higgs candidate following the criteria presented in the text $M_{bb_1}$ versus the invariant mass of the other bottom pair identified as the other Higgs candidate $M_{bb_2}$. Orange dots correspond to those events that pass the implemented VBS selection cuts given in Eq.(\ref{VBSselectioncuts}). Cuts in \eqref{basiccuts4b2j} have been implemented. The value of the acceptance $\mathcal{A}$ is also included. The red cross represents the value of the Higgs mass The center of mass energy has been set to $\sqrt{s}=$14 TeV.}
\label{fig:4b2j_2D_Mbb}
\end{center}
\end{figure}

In \figref{fig:4b2j_2D_Mbb} we aim precisely to see this effect, since we present the same set of events as in \figref{fig:4b2j_VBSvariables2D} for the QCD background and for the signal highlighting in orange those events that fulfill the VBS selection criteria given in \eqref{VBSselectioncuts} as an example. This time we show the results in the $M_{bb_1}-M_{bb_2}$ plane, where $M_{bb_{1,2}}$ are the corresponding invariant masses of the two bottom pairs that are the best candidates to come from the decay of a Higgs boson, as we will see later.

The first thing one can observe in both plots of \figref{fig:4b2j_2D_Mbb} is that very few QCD events survive the imposition of the VBS cuts, whereas practically all events of the signal do. The concrete fraction of the events ($\mathcal{A}$) that survive in both cases is also presented in the plots. We call $\mathcal{A_{\rm VBS}}$ the acceptance of the VBS cuts, defined as
\begin{equation}
\mathcal{A}_{\rm VBS}\equiv\dfrac{\sigma(pp\to b\bar{b}b\bar{b}jj)|_{\rm VBS}}{\sigma(pp\to b\bar{b}b\bar{b}jj)}\,,
\end{equation}
i.e., the ratio between the cross section of the process after applying the VBS cuts like those in \eqref{VBSselectioncuts} over the cross section of the process without having applied them. The basic cuts are imposed in all cases. Taking a look at these numbers, we see that 60\% of the signal events pass these cuts while only 9\% of the QCD events do. At this point, one might wonder wether these results are very dependent on the specific VBS cuts we impose or not. In \tabref{table:VBS} we show the predictions for the acceptances, $\mathcal{A}_{\rm VBS}$, of different sets of VBS selection cuts, i.e., different cuts in $|\Delta\eta_{jj}|$ and in $M_{jj}$, for both the multijet QCD background and the signal with $\kappa=1$. From those predictions we can see that all the sets of cuts considered lead to very similar results: around 60\% of the signal fulfills the VBS selection criteria whereas a 5-10\% of the multijet QCD background does. We have checked that for other values of $\kappa$ the acceptance for the signal varies between a 55\% and a 75\%. From now on we will apply the VBS selection cuts given in \eqref{VBSselectioncuts}, since this set is well explored in the literature and qualitatively provides the same results as the other sets of cuts that we have analyzed.

The second issue that we can notice about \figref{fig:4b2j_2D_Mbb} is that, again, the QCD events populate a very different region of this plane than those of the signal. QCD events tend to lie at low values of ${M_{bb_i}}$, somehow away from the  region close to the $[M_{bb_1}=m_H,M_{bb_2}=m_H]$ point in the $M_{bb_1}-M_{bb_2}$ plane, in which most of our signal settles. Notice, however, that the illustration in \figref{fig:4b2j_2D_Mbb} is ideal since the signal events populate a region which is very close to the $[M_{bb_1}=m_H,M_{bb_2}=m_H]$ point. We will comment on the effects of having a larger, more realistic dispersion of the signal points later on. In any case, it is evident that two particles coming from the decay of a Higgs boson should have a total invariant mass value near the Higgs boson mass, $m_H$, as our signal does. This motivates the next selection criteria we are going to apply, following the search strategies of ATLAS \cite{Aaboud:2018knk} and CMS \cite{CMS:2016foy} for double Higgs production, that are aimed to efficiently identify the $HH$ candidates.

The $HH$ candidate identification criteria are also based on what we have learned in the previous sections. Logically, each H candidate corresponds to a $b$-quark pair, and therefore we first need to define how we are going to pair the final $b$-quarks. From now on, it is worth mentioning that we will not distinguish between bottom and anti-bottom, similarly to what is done in experimental analyses. Therefore, with four bottom-like particles in the final state we have three possible double pairings. Among these three possibilities, we select the one in which the values of the invariant masses of the pairs are closer, i.e., the one that minimizes $|M_{bb_1}-M_{bb_2}|$, where $M_{bb_1}$ is the invariant mass of one of the $bb$ pairs and $M_{bb_2}$ is the invariant mass of the other pair. Once we have defined the $b$-quark pairing, we can profit from the fact that, as mentioned before, if two $b$-quarks come from the decay of a boosted Higgs boson, as it happens in VBS processes, the angular separation between them should be small. Thus, we should look for pairs of $b$-quarks with small (and yet  enough to resolve the particles) $\Delta R_{bb}$. Furthermore, we have already discussed that our signal is characterized by the fact that the invariant mass of each $b$-quark pair should be around the Higgs mass, $m_H$. Therefore, imposing this criterion will ensure that we are maximizing the selection of events that come from the decays of two Higgs bosons.

\begin{table}[t!]
\begin{center}
\vspace{.2cm}
\begin{tabular}{ c @{\extracolsep{1cm}} c  @{\extracolsep{1cm}} c }
\toprule
\toprule
Set of VBS cuts & $\mathcal{A}_{\rm VBS}^{\rm QCD}$  & $\mathcal{A}_{\rm VBS}^{{\rm Signal};\kappa=1}$\\
\midrule
$|\Delta\eta_{jj}|>4,~M_{jj} > 500$ GeV &  0.086 & 0.631\\
\rule{0pt}{1ex}
$|\Delta\eta_{jj}|>4,~M_{jj} > 600$ GeV   &  0.066  &   0.597\\
\rule{0pt}{1ex}
$|\Delta\eta_{jj}|>4,~M_{jj} > 700$ GeV & 0.054   &     0.558\\
\rule{0pt}{1ex}
 $|\Delta\eta_{jj}|>3,~M_{jj} > 500$ GeV&  0.098  &     0.669\\
\rule{0pt}{1ex}
$|\Delta\eta_{jj}|>3,~M_{jj} > 600$ GeV &  0.071  & 0.626\\
\rule{0pt}{1ex}
$|\Delta\eta_{jj}|>3,~M_{jj} > 700$ GeV&   0.057 &     0.580\\
\bottomrule
\bottomrule
\end{tabular}
\vspace{0.4cm}
\caption{Predictions for the acceptance of different sets of VBS cuts, including those in \eqref{VBSselectioncuts}, for the multijet QCD background and for the signal with $\kappa=1$. Signal acceptances for the other values of $\kappa$ considered in the present work, $\kappa\in[-10,10]$, vary between 0.5 and 0.75.}
\label{table:VBS}
\end{center}
\end{table}


With all these features in mind, and guided by the ATLAS search strategies \cite{Aaboud:2018knk}, we define the following set of cuts as the requirements to efficiently select the  Higgs boson pair candidates: 
\begin{align}
&HH \,\,\, {\rm CANDIDATE\,\,\, CUTS \,\,\,:} \nonumber \\[0.3em]
&p_{T_b}> 35~ {\rm GeV}\,,\label{cutsHH1}\\[4pt]
&\hat{\Delta}R_{bb}\equiv\left\{ \begin{array}{l}
M_{b\bar{b}b\bar{b}}<1250~{\rm GeV} \left\{ \begin{array}{l}
 0.2 < \Delta R_{bb^l} < (653/M_{b\bar{b}b\bar{b}})+0.475\,,\\[4pt]0.2 < \Delta R_{bb^s} < (875/M_{b\bar{b}b\bar{b}})+0.35\,,  \end{array} \right.\\[20pt]
M_{b\bar{b}b\bar{b}}>1250~{\rm GeV}\, \left\{ \begin{array}{l} 0.2 < \Delta R_{bb^l} < 1\,,\\[4pt]0.2 < \Delta R_{bb^s} < 1\,,\,  \end{array} \right.
 \end{array} \right.\label{cutsHH2}\\[4pt]
&\hat{p}_{T_{bb}}\equiv p_{T_{bb^l}}>M_{4b}/2-103 {\rm GeV}\,;~ p_{T_{bb^s}}>M_{4b}/3-73{\rm GeV}\,,\label{cutsHH3}\\[4pt]
&\chi_{HH}\equiv\sqrt{\left(\dfrac{M_{bb^l}-m_H}{0.05\,m_H}\right)^2+\left(\dfrac{M_{bb^s}-m_H}{0.05\,m_H}\right)^2}<1\,,\label{cutsHH}
\end{align}
where the super-indices $l$ and $s$ denote, respectively, leading and subleading, defining the leading $b$-quark pair as the one with largest scalar sum of $p_T$. $M_{b\bar{b}b\bar{b}}$ designates the invariant mass of the four final $b$-quarks. One might notice that the requirement of small angular separation between the two $b$-quarks of a pair, and the fact that the invariant mass of each $b$-quark pair has to lie near the mass of the Higgs, are encoded in the $\hat{\Delta}R_{bb}$ and in the $\chi_{HH}$ cuts, respectively. The latter is equivalent to impose that the events in the $M_{bb_1}-M_{bb_2}$ plane have to lie inside a circle of radius $0.05\,m_H=6.25$ GeV centered in the point $[M_{bb_1}=m_H,M_{bb_2}=m_H]$.

Nevertheless, although multijet QCD events represent the most severe background, there are other processes that can fake our signal. One of them is the $t\bar{t}$ background, with the subsequent decays of the top quarks and W bosons, $t\bar{t}\to b W^+ \bar{b} W^-\to   b\bar{b}b\bar{b}jj$. This is, however, a very controlled background, since it is well suppressed by non-diagonal CKM matrix elements and its kinematics are radically different than those of VBS. Starting from a cross section of $5.4\cdot 10^{-5}$ pb with all the basic cuts in \eqref{basiccuts4b2j} applied, one ends up in $1.7\cdot 10^{-7}$ pb after applying the $HH$ candidate cuts, and in $2.0 \cdot 10^{-10}$ pb after applying the VBS cuts afterwards. Therefore, since this background is five orders of magnitude smaller than the smallest of our signals, we will neglect it from now onwards.
Finally, we still have to deal with other potentially important backgrounds corresponding to $pp\to HZjj \to b\bar{b}b\bar{b}jj$ and $pp\to ZZjj \to b\bar{b}b\bar{b}jj$. These two $HZ$ and $ZZ$ production processes, receiving contributions of order ($\alpha^2\cdot\alpha_S^2$) and  ($\alpha^4$) at the cross section level, also drive to the same final state as our signal and may give rise to similar kinematics, since they can also take place through VBS configurations. In fact, their rates are very close to those of our signal after applying the VBS selection cuts, that reduce these backgrounds less efficiently than the multijet QCD one. However, we can again take advantage of the fact that the $b$-quark pairs have to come from a Higgs boson with a well defined mass. Therefore the $HH$ candidate cuts should allow us to reject these backgrounds.

\begin{table}[t!]
\begin{center}
\vspace{.2cm}
\begin{tabular}{ c @{\extracolsep{0.8cm}} c @{\extracolsep{0.8cm}} c @{\extracolsep{0.8cm}} c }
\toprule
\toprule
Cut & $\sigma_{\rm QCD}$  [pb]& $\sigma_{ ZHjj,ZZjj}$ [pb] & $\sigma_{{\rm Signal};\kappa=1}$ [pb]\\
\midrule
Basic detection cuts in \eqref{basiccuts4b2j} & 602.72  & 0.028 &5.1$\cdot 10^{-4}$\\
$p_{T_b}>$ 35 GeV, \eqref{cutsHH1}  &  98.31  &  0.01 &3.0$\cdot 10^{-4}$\\
$\hat{\Delta}R_{bb}$, \eqref{cutsHH2}  & 33.80   &  $6.3\cdot 10^{-3}$   &1.1$\cdot 10^{-4}$\\
$\hat{p}_{T_{bb}}$, \eqref{cutsHH3}&  29.77  &  5.8$\cdot 10^{-3}$   &9.0$\cdot 10^{-5}$\\
$\chi_{HH}<1$, \eqref{cutsHH} &  $7.9\cdot10^{-2}$  &  8.6$\cdot 10^{-6}$   &9.0$\cdot 10^{-5}$\\
VBS cuts in \eqref{VBSselectioncuts} &  $6.8\cdot10^{-3}$  &   5.5$\cdot 10^{-6}$  &4.1$\cdot 10^{-5}$\\
\bottomrule
\bottomrule
\end{tabular}
\vspace{0.4cm}
\caption{Predictions for the total cross section of the multijet QCD background, of the combined $pp\to HZjj \to b\bar{b}b\bar{b}jj$ and $pp\to ZZjj \to b\bar{b}b\bar{b}jj$ background and of the signal with $\kappa=1$ after imposing each of the cuts given in \eqref{basiccuts4b2j} and in Eqs. (\ref{cutsHH1})-(\ref{cutsHH}) subsequently. We show as well the total cross section after applying, afterwards, the VBS selection cuts in \eqref{VBSselectioncuts}. }
\label{table:cutflow}
\end{center}
\end{table}


In \tabref{table:cutflow} we present the cross sections of the multijet QCD background, of the combined $pp\to HZjj \to b\bar{b}b\bar{b}jj$ and $pp\to ZZjj \to b\bar{b}b\bar{b}jj$ background and of the signal with $\kappa=1$ after applying each of the cuts in Eqs.(\ref{cutsHH1})-(\ref{cutsHH}) subsequently.  The basic cuts had already been applied in all cases. This way, we see the reduction factor after each cut, and the total cross section of both signal and background once we have performed our complete $HH$ candidate selection. We show as well the effect of applying the VBS cuts given in \eqref{VBSselectioncuts} afterwards, since we have checked that both sets of cuts ($HH$ candidate cuts and VBS cuts) are practically independent. Thus, we have the total cross sections of the two main backgrounds and of our SM signal after applying all the selection criteria. In \tabref{Tab:allSig_4b2j} we provide the total cross sections of the signal for all the values of $\lambda$ considered in this work,  again after applying all the selection criteria, for comparison.

From the results in \tabref{table:cutflow} we can learn that the sum of the two backgrounds, $ZHjj$+$ZZjj$,  is under control after applying the $HH$ candidate cuts, since its cross section lies an order of magnitud below the SM signal. On the other hand, the multijet QCD background remains being very relevant even after imposing all the selection criteria. However, as we will see later, the total reduction that it suffers still allows to be sensitive to interesting values of $\kappa$ even for low luminosities. This reduction, along with that suffered by the $ZHjj$+$ZZjj$ backgrounds and with that suffered by the SM signal, is presented in \tabref{table:acceptances}. There we show the acceptances of the VBS cuts and the $HH$ candidate cuts separately and together for the multijet QCD background, for the combined $pp\to HZjj \to b\bar{b}b\bar{b}jj$ and $pp\to ZZjj \to b\bar{b}b\bar{b}jj$ background and for the SM signal, for comparison.

\begin{table}[b!]
\begin{center}
\begin{tabular}{c @{\extracolsep{0.6cm}} c @{\extracolsep{0.6cm}} c @{\extracolsep{0.6cm}} c @{\extracolsep{0.6cm}} c @{\extracolsep{0.6cm}} c @{\extracolsep{0.6cm}} c @{\extracolsep{0.6cm}} c @{\extracolsep{0.6cm}} c @{\extracolsep{0.6cm}} c}
\toprule
\toprule
$\kappa$&  $0$& $1$ & $-1$ & $2$& $-2$& $5$& $-5$& $10$& $-10$\\
\midrule
 $\sigma_{\rm Signal}\cdot 10^4$ [pb] &$1.9$ & $0.4$ & $5.0$ & $0.4$ & $9.7$&$10.1$ & $33.2$ & $56.4$ & $102.6$ \\
\bottomrule
\bottomrule
\end{tabular}
\vspace{0.4cm}
\caption{Predictions for the total cross section of the signal $pp\to b\bar{b}b\bar{b} j j$ after imposing all the selection criteria, VBS cuts given in in \eqref{VBSselectioncuts} and $HH$ candidate cuts given in Eqs. (\ref{cutsHH1})-(\ref{cutsHH}) for all the values of $\kappa$ considered in this work: $\kappa=0,\pm 1, \pm 2, \pm 5, \pm 10$. Basic cuts in \eqref{basiccuts4b2j} are also applied.}
\label{Tab:allSig_4b2j}
\end{center}
\end{table}

\begin{table}[t!]
\begin{center}
\vspace{.2cm}
\begin{tabular}{ c @{\extracolsep{0.8cm}} c @{\extracolsep{0.8cm}} c @{\extracolsep{0.8cm}} c }
\toprule
\toprule
Cut & $\mathcal{A}^{\rm QCD}$& $\mathcal{A}^{ ZHjj,ZZjj}$  & $\mathcal{A}^{{\rm Signal};\kappa=1}$
\\
\midrule
VBS cuts in \eqref{VBSselectioncuts} & 0.086  & 0.630 & 0.631
\\
$HH$ candidate cuts in Eqs. (\ref{cutsHH1})-(\ref{cutsHH})   &  1.3$\cdot 10^{-4}$  &  3.1$\cdot 10^{-4}$ & 0.17
\\
VBS cuts $+$ HH candidate cuts   &  1.1$\cdot 10^{-5}$  &  2.0$\cdot 10^{-4}$ & 0.081
\\
\bottomrule
\bottomrule
\end{tabular}
\vspace{0.4cm}
\caption{Predictions for acceptances of the VBS cuts given in \eqref{VBSselectioncuts}, of the $HH$ candidate cuts given in Eqs. (\ref{cutsHH1})-(\ref{cutsHH}), and of both sets of cuts combined  for the multijet QCD background, for the combined $pp\to HZjj \to b\bar{b}b\bar{b}jj$ and $pp\to ZZjj \to b\bar{b}b\bar{b}jj$ background and for the signal with $\kappa=1$. All the results are computed with the basic cuts in \eqref{basiccuts4b2j} already applied.} 
\label{table:acceptances}
\end{center}
\end{table}


 It must be noticed that other backgrounds apart from those having the same final particle content as our signal can contribute relevantly. This would  be the case if some final state particles were misidentified, leading to a ``fake'' $ b\bar{b}b\bar{b}jj$ state.  The most important of these backgrounds is the production of a $t\bar{t}$ pair decaying into two $b$ quarks and four light jets,  $t\bar{t}\to b W^+ \bar{b} W^-\to   b\bar{b}jjjj$ with two of these final light jets being misidentified as two $b$ jets.  In order to estimate the contribution of this background, we have generated with MadGraph5 $t\bar{t}\to b W^+ \bar{b} W^-\to   b\bar{b}jjjj$ events applying first the minimal cuts $|p_{T_{j,b}}|>20$ GeV, $|\eta_{j,b}|<5$ and $\Delta R_{jj,bj,bb}>0.2$, obtaining a total cross section of 246 pb. Applying a mistagging rate of 1\% per each light jet misidentified as a $b$ jet, we obtain $246\cdot(0.01)^2=2.5\cdot 10^{-2}$ pb as starting point to compare to our main multijet QCD $ b\bar{b}b\bar{b}jj$ background. Now we need to apply our selection cuts described in Eqs. (\ref{VBSselectioncuts}), and (\ref{cutsHH1})-(\ref{cutsHH}) to see their impact on this particular background. 
 
 We apply first the VBS selection cuts demanding at least one pair of light jets fulfilling the criteria in \eqref{VBSselectioncuts}. These cuts reduces the cross section to $1.3\cdot 10^{-4}$ pb. Now analyzing the events that pass the VBS cuts, if there is only one pair of ``VBS-like'' light jets, the other two light jets are identified as $b$ quarks. If there is more than one, we select as $b$ quarks those that minimize $|M_{pp_1}-M_{pp_2}|$, with $p=b,j$ among all possibilities. Once we have characterized our two light jet candidates and our four $b$-quark candidates, we proceed with the $HH$ candidate selection cuts. This way, applying subsequently the criteria explained in \eqrefs{cutsHH1}{cutsHH}, we obtain the following cross sections: $1.2\cdot 10^{-5}$ pb $\left(p_{T_b}\right)$, $2.5\cdot 10^{-6}$ pb $\left(\hat{\Delta}R_{bb}\right)$, $4.4\cdot 10^{-7}$ pb $\left(\hat{p}_{T_{bb}}\right)$ and finally $2.1\cdot 10^{-8}$ pb $\left(\chi_{HH}\right)$. Therefore, since this $t\bar{t}$ background is five orders of magnitude below our main considered background, whose final cross section given in \tabref{table:cutflow} is $6.8\cdot10^{-3}$ pb, we conclude that it can be safely neglected.

We have also considered the possible backgrounds coming from multijet QCD processes leading to different final states than that of $b\bar{b}b\bar{b}jj$, such as $6j$ and $b\bar{b}jjjj$, in which some of the final state light jets are again misidentified as $b$ jets. To estimate their contribution to the background we have used the total cross sections of these processes given in \cite{Mangano:2002ea}. These are, for a center of mass energy of 14 TeV, $1.3\cdot10^5$ pb and $7.5\cdot 10^3$ pb, respectively. If we apply now the corresponding misidentification rates we end up with  $1.3\cdot10^5 \cdot (0.01)^4=1.3\cdot10^{-3}$ pb for the case in which we have six light jets, and $7.5\cdot 10^3 \cdot (0.01)^2=7.5\cdot10^{-1}$ pb for the case in which we have two $b$ jets and four light jets. We now assume that the selection cuts we specify in Eqs. (\ref{VBSselectioncuts}) -(\ref{cutsHH}) will have a similar impact on these backgrounds as they do on the multijet QCD production of four $b$ jets and two light jets, since they all take place through similar QCD configurations. Thus, applying the corresponding acceptance factor of these cuts we obtain the following total cross sections: $1.3\cdot10^{-3}\cdot 1.1 \cdot 10^{-5}=1.4 \cdot 10^{-8}$ pb for the six light jets case and  $7.5\cdot10^{-1}\cdot 1.1 \cdot 10^{-5}=8.2 \cdot 10^{-6}$ pb for the two $b$ jets and four light jets case. Both of these cross sections are more than three orders of magnitude below that  of the  $b\bar{b}b\bar{b}jj$ background, so we conclude that they can also be safely neglected without introducing big uncertainties.

\begin{figure}[t!]
\begin{center}
\includegraphics[width=0.49\textwidth]{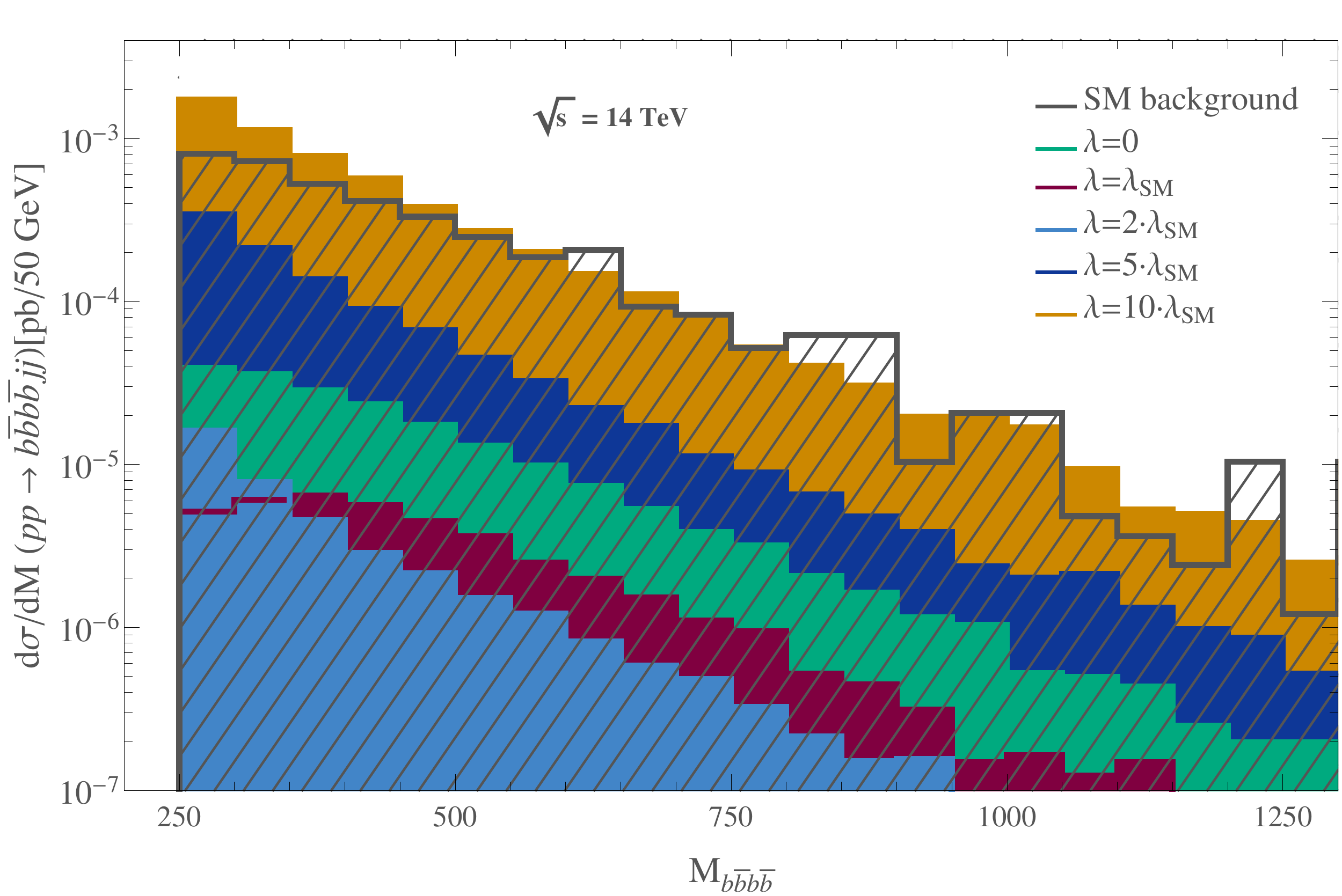}
\includegraphics[width=0.49\textwidth]{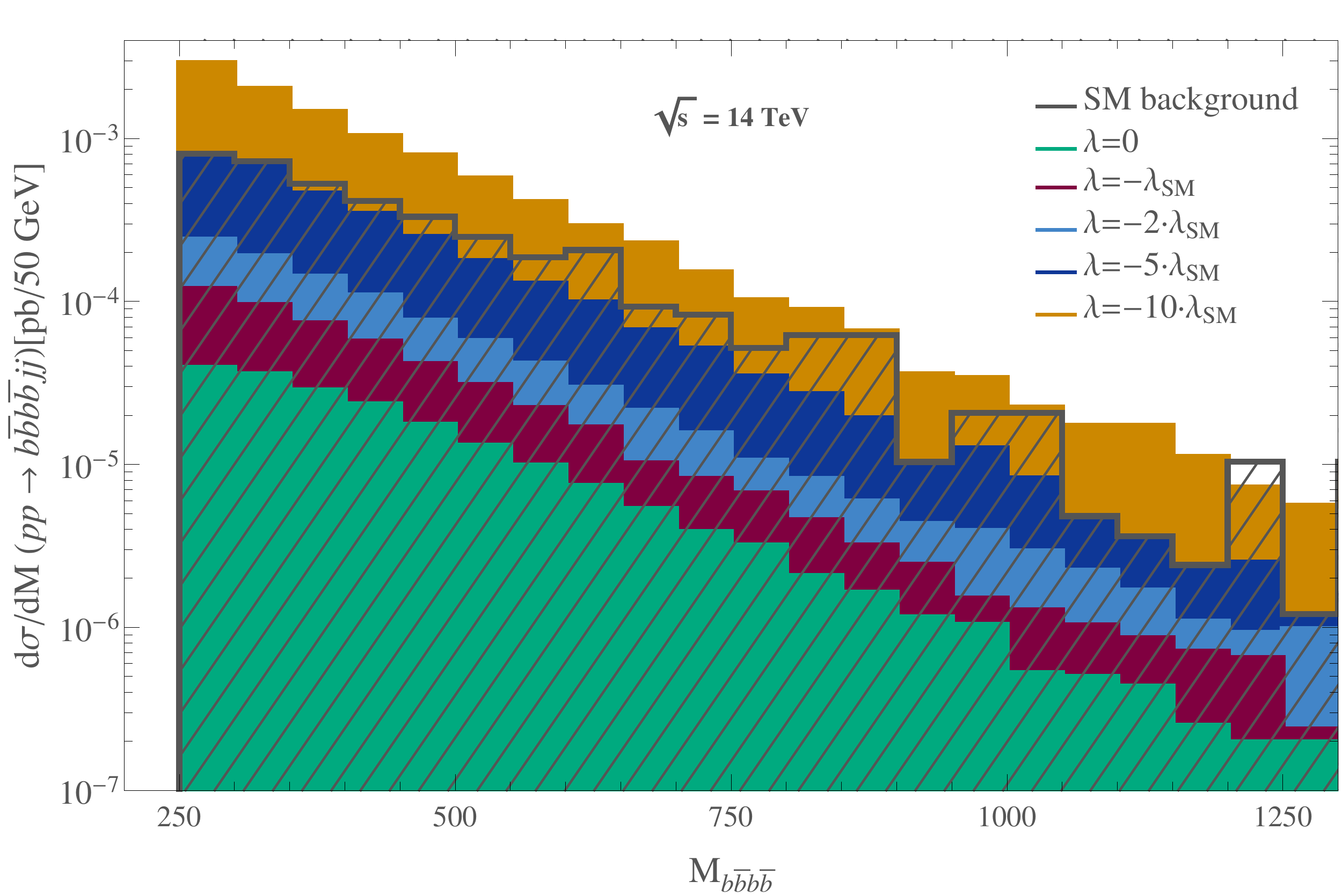}
\caption{Predictions for the total cross section of the process $pp\to b\bar{b}b\bar{b}jj$ as a function of the invariant mass of the four-bottom system $M_{b\bar{b}b\bar{b}}$ for different values of the Higgs self-coupling $\lambda$. We display the predictions for the signal with positive (left panel) and negative (right panel) values of $\lambda$ for comparison, as well as the total SM background given by the sum of $ZHjj$, $ZZjj$ and the multijet QCD background. Cuts in Eq.(\ref{basiccutsHHjj}) and VBS selection cuts presented in Eq.(\ref{VBSselectioncuts}) have been applied. The center of mass energy has been set to $\sqrt{s}=14$ TeV.}
\label{fig:M4b_distributions}
\end{center}
\end{figure}

Once we have the possible backgrounds under control, we can move on to fully explore the sensitivity to the Higgs self-coupling $\lambda$ in $p p\to b\bar{b}b\bar{b}jj$ events. In \figref{fig:M4b_distributions} we display the predictions for the cross section of the total SM background (the sum of multijet QCD background and  the combined $ZHjj+ZZjj$ backgrounds) and of the signal for different values of $\lambda$ as a function of the invariant mass of the four-bottom system $M_{b\bar{b}b\bar{b}}$. These distributions should be the analogous to those in \figref{fig:MHH_distributions} after the Higgs boson decays, as it is manifest since the signal curves follow the same tendency and are very similar except for the global factor of the Higgs-to-bottoms branching ratio. In this figure we can also see that the total SM background is of the same order of magnitude than the $\kappa=10$ and $\kappa=-5$ signals, and it is even below the $\kappa=-10$ signal prediction. This is a very interesting result, since it means that if, for example, the true value of $\lambda$ was minus five times that of the SM, the LHC should be able to measure twice as many events as those expected from the SM background only in this VBS configuration. Similar conclusions can be extracted for other values of $\kappa$.

Given the encouraging previous results, our last step is to give quantitative predictions for the sensitivity to $\lambda$ in $p p\to b\bar{b}b\bar{b}jj$ processes at the LHC. To this end, we compute the statistical significance $\mathcal{S}_{\rm stat}$, as defined in \cite{Cowan:2010js} by: 
\begin{equation}
\mathcal{S}_{\rm stat}=\sqrt{-2\left((N_S+N_B)\log\left(\dfrac{N_B}{N_S+N_B}\right)+N_S\right)}\,,\label{signieq}
\end{equation}
where $N_S$ and $N_B$ are the number of events of signal and background, respectively. Notice that for  $N_S/N_B \ll 1$, this definition of $\mathcal{S}_{\rm stat}$ tends to the usual $N_S/\sqrt{N_B}$ expression. This computation is going to be performed for four different values of the luminosity: $L=50,300,1000,3000$ fb${}^{-1}$, that correspond to a near-future LHC value for the current run (50 fb${}^{-1}$), and to planned luminosities for the third run (300 fb${}^{-1}$) and the High-Luminosity LHC (1000 and 3000 fb${}^{-1}$) \cite{Barachetti:2016chu}.

\begin{figure}[t!]
\begin{center}
\includegraphics[width=0.49\textwidth]{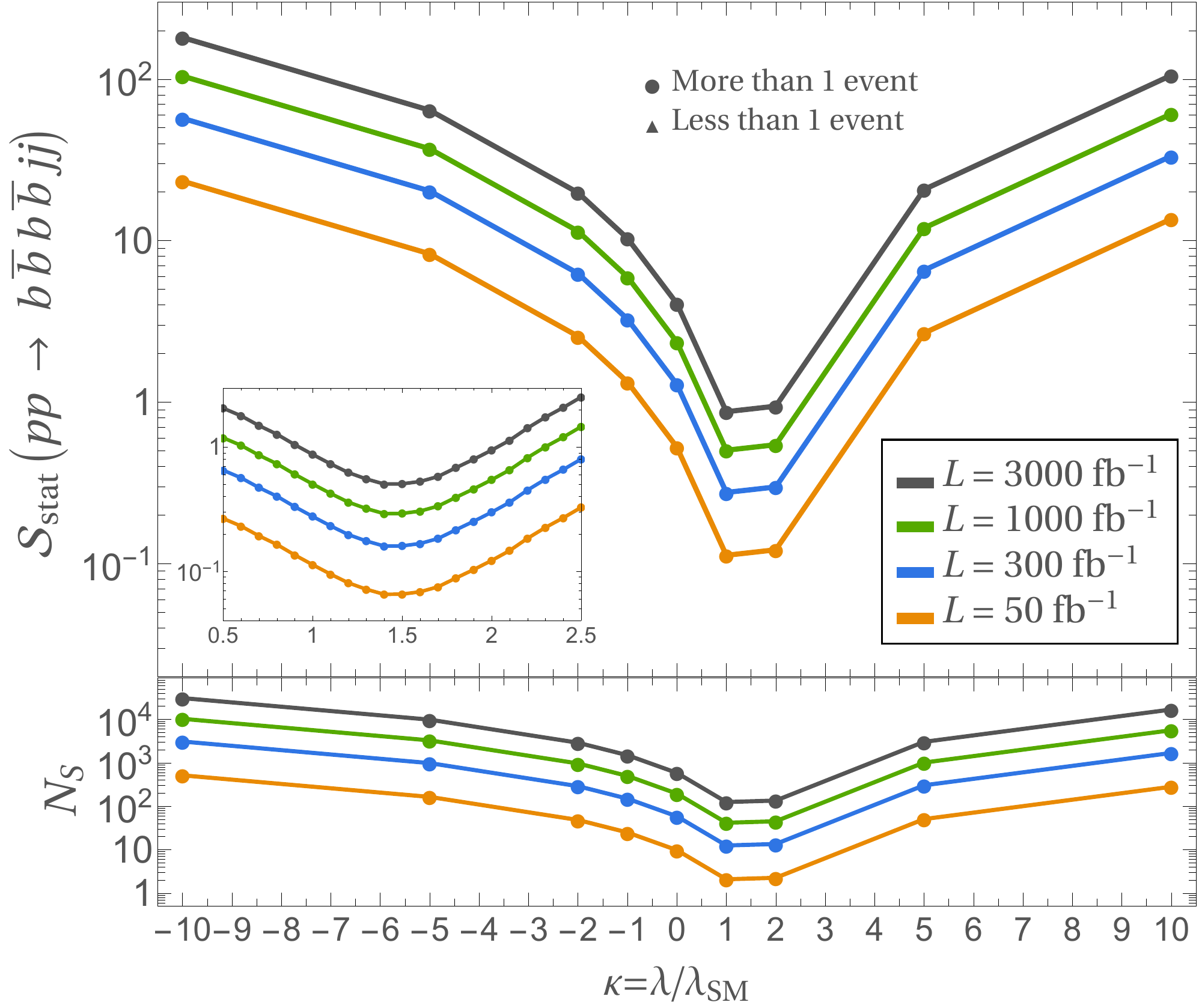}
\includegraphics[width=0.49\textwidth]{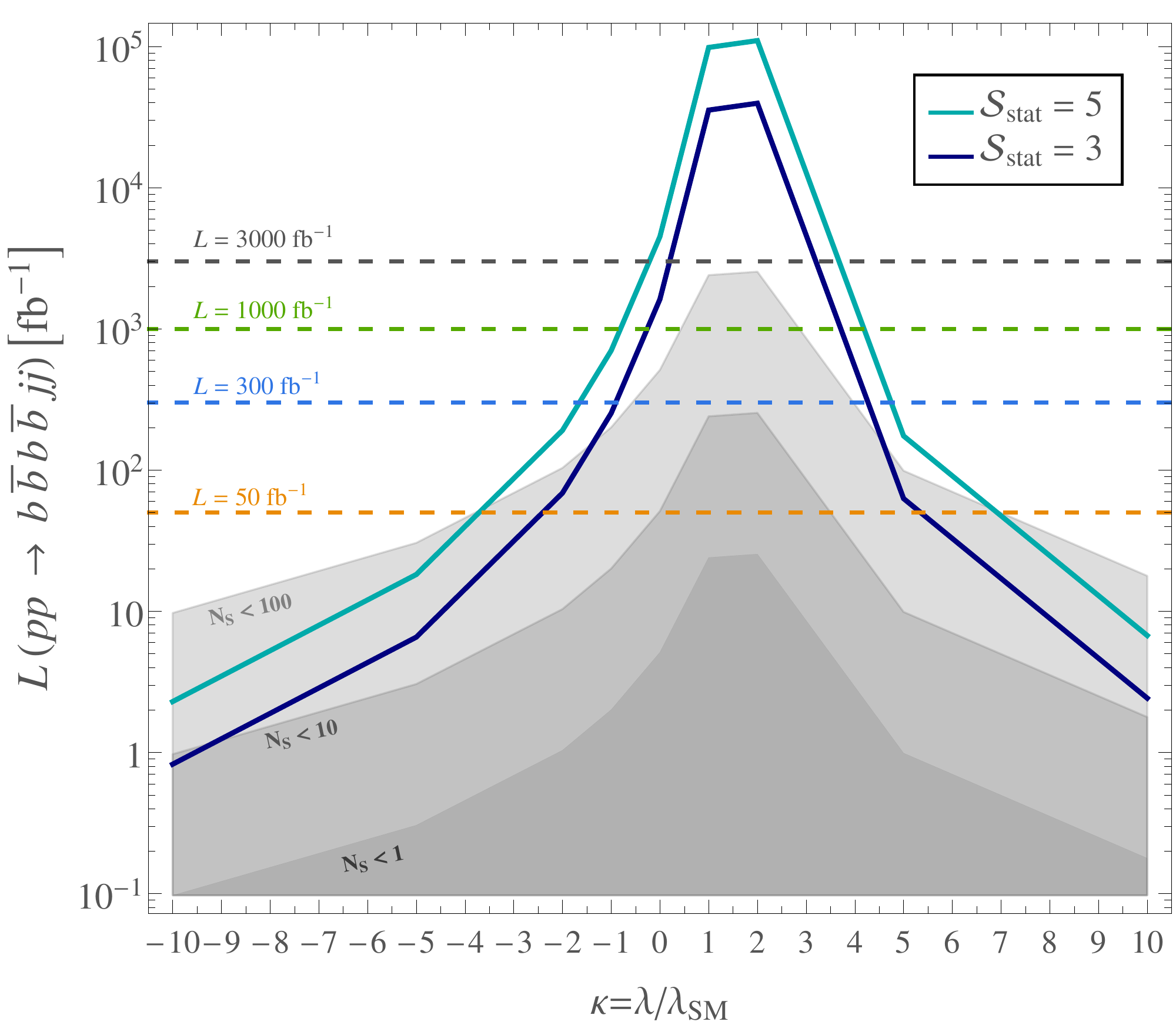}
\caption{Prediction of the statistical significance, $\mathcal{S}_{\rm stat}$, of the process $p p\to b\bar{b}b\bar{b} jj$ for the four luminosities considered $L=50,300,1000,3000$ fb${}^{-1}$ (left panel) and  of the value of the luminosity that will be required to probe a given $\kappa$  at the LHC at  3$\sigma$ and at 5$\sigma$ (right panel),  as a function of the value of $\kappa$. The marked points represent our evaluations.  In the left panel, a zoom is performed on the interesting values of $\kappa$ ranging between 0.5 and 2.5. The shadowed areas in the right panel correspond to the regions where the number of predicted signal events $N_S$ is below 1, 10 and 100. The center of mass energy has been set to $\sqrt{s}=14$ TeV. }
\label{fig:significances_4b2j}
\end{center}
\end{figure}

In \figref{fig:significances_4b2j} we present the results of the statistical significance of our signal, $\mathcal{S}_{\rm stat}$, in $pp\to b\bar{b}b\bar{b}jj$ events as a function of the value of $\kappa$, for the four luminosities considered. We  display as well a closer look for the values of $\kappa$ ranging between 0.5 and 2.5, interesting for an elevated number of well motivated BSM models. In the lower part of the left panel we also present the corresponding predictions for the total number of signal events, $N_S$, as a function of $\kappa$. The marked points correspond to the predictions we have directly evaluated. We show as well, in the right panel of this figure, our predictions for the value of the total integrated luminosity, $L$, as a function of the value of $\kappa$ as well, that will be required to obtain a sensitivity to a 
given $\kappa$ in $pp\to b\bar{b}b\bar{b} jj$ events at the 3$\sigma$ and  5$\sigma$ level. In this plot,  we have also marked the areas in luminosity where the number of predicted signal events $N_S$ is below 1, 10 and 100, respectively, to get a reference of the statistics obtained. 

From these plots, we can extract directly the conclusions on the sensitivity to $\lambda$ in VBS processes at the LHC in $pp\to b\bar{b}b\bar{b}jj$ events. The first thing one might observe is the high statistics and significances of the signal  for most of the studied cases, except for the region close to the SM value, say for $\kappa$ between 1 and 2. Studying carefully this particular region of the parameter space, we conclude that it is the most challenging one to access at the LHC, since all the predicted statistical significances given for $\kappa\in[0.5,2]$ are below 2$\sigma$ even for the highest luminosity considered. The second one is that, for the same absolute value of the coupling, the sensitivities to negative values of $\kappa$ are higher than to positive values of $\kappa$.  The third conclusion is that the LHC should be sensitive to very broad intervals of $\kappa$, even for the lowest luminosity considered, $L=50$ fb${}^{-1}$, with high statistical significance. These means that VBS processes could allow us to probe the value of $\lambda$ with very good accuracy in the near future. More specifically, in \tabref{Tab:range_4b2j} we show the summary of the predictions for the values of $\kappa\equiv\lambda/\lambda_{SM}$ that the LHC would be able to probe in  $pp\to b\bar{b}b\bar{b}jj$ events, with a sensitivity equal or better than  $3\sigma$ ($5\sigma$) for the four luminosities considered: $L=50,300,1000,3000$ fb${}^{-1}$.

\begin{table}[t!]
\begin{center}
\begin{tabular}{c @{\extracolsep{0.4cm}} c @{\extracolsep{0.4cm}} c @{\extracolsep{0.4cm}} c @{\extracolsep{0.4cm}} c}
\toprule
\toprule
$L$ \big[fb${}^{-1}$\big]& 50  & 300  & 1000  & 3000\\
\midrule
$\kappa > 0$ & $\kappa > 5.4 ~(7.0)$ &$\kappa >  4.3  ~(4.8)$ & $\kappa > 3.7 ~ (4.2)$ & $\kappa> 3.2 ~ (3.7)$
\\
$\kappa < 0$ & $\kappa < -2.4 ~ (-3.8)$ &$\kappa <  -1.0  ~(-1.7)$ & $\kappa < -0.3  ~(-0.8)$ & $\kappa < 0 ~ (-0.2)$
\\
\bottomrule
\bottomrule
\end{tabular}
\vspace{0.4cm}
\caption{Predictions for the values of $\kappa\equiv\lambda/\lambda_{SM}$ that the LHC would be able to probe in  $pp\to b\bar{b}b\bar{b}jj$ events, with a sensitivity equal or better than  $3\sigma$ ($5\sigma$) for the four luminosities considered: $L=50,300,1000,3000$ fb${}^{-1}$.}
\label{Tab:range_4b2j}
\end{center}
\end{table}
 
These results are indeed very interesting, since the sensitivities to $\lambda$ that one can obtain from studying VBS double Higgs production are very promising even for the lowest luminosity considered $50$ fb${}^{-1}$. The ranges of $\lambda$ that the LHC could be able to probe in this kind of processes indicate that it is worth to study VBS as a viable and useful production mechanism to measure the Higgs trilinear coupling. On the other hand, it can be seen that the HL-LHC should be able to test
very small deviations in the value of the Higgs self-coupling and that it should be sensitive to all the explored negative values for $\kappa$. Although the present work is a naive study, since it is performed at the parton level and does not take into account hadronization and detector response simulation, the results in  \tabref{Tab:range_4b2j} show that the VBS production channel could be very promising to measure the true value of $\lambda$, and, therefore, to understand the nature of the Higgs mechanism.


 \subsection[Analysis after Higgs boson decays: sensitivity to $\lambda$ in $pp\to b\bar{b}\gamma\gamma jj$]{Analysis after Higgs boson decays: sensitivity to $\boldsymbol{\lambda}$ in $\boldsymbol{pp\to b\bar{b}\gamma\gamma jj}$}
 \label{2b2a2j}
 
The $p p\to b\bar{b}b\bar{b} jj$ process is, as we have seen, a very promising channel to study the Higgs self-coupling at the LHC due to its large event rates. However, it is clear that it suffers from quite severe backgrounds, coming specially from multijet QCD events, so one could think of studying complementary channels with smaller rates but with a cleaner experimental signature. This is the reason why we would like to explore the case in which one of the Higgs bosons decays to a $b$-quark pair, as before, while the other one decays to two photons through gauge bosons and fermion loops. This implies a large reduction in statistics due to the comparative low branching ratio BR$(H \to \gamma \gamma)\sim 0.2\%$, a factor 0.003 smaller than that of $H \to b\bar{b}$.  

The analysis of the process $pp\to b\bar{b}\gamma\gamma jj$ implies to go through its main backgrounds as well. We will consider in this section the same background $ZH$ of the previous case, since the $ZH$ final state can also lead to processes with two photons and two bottoms, $pp\to HZjj \to b\bar{b}\gamma\gamma jj$, coming from the decays of the H and the Z. In addition, we also consider the mixed QCD-EW $pp\to b\bar{b}\gamma\gamma jj$ background, of $\mO(\alpha^2\cdot\alpha_S^4)$ at the cross section level, that should be the most severe one.

As we did before, to study signal and background, we first need to establish a set of cuts that ensure particle detection, so we apply the following basic cuts:
\begin{align}
 &p_{T_{j,b}}>20~ {\rm GeV}\,,~p_{T_{\gamma}}>18 ~{\rm GeV}\,,\nonumber\\&|\eta_j|<5\,,~|\eta_{b,\gamma}|<2.5\,, \nonumber\\&\Delta R_{jj,jb,\gamma\gamma,\gamma b,\gamma j}>0.4\,,~\Delta R_{bb}>0.2\,,\label{basiccuts2b2a2j}
 \end{align}
and afterwards, to reject the QCD-EW and the $ZHjj$ backgrounds we will apply first the VBS cuts given in \eqref{VBSselectioncuts} and subsequently the following kinematical cuts given by CMS in \cite{Sirunyan:2018iwt}:
\begin{equation}
p_{T_{\gamma^l}}/M_{\gamma\gamma}>1/3\,;~~~p_{T_{\gamma^s}}/M_{\gamma\gamma}>1/4\,,\label{bkgcuts2b2a2j}
 \end{equation}
where $l$ and $s$ stand for leading (highest $p_T$ value) and subleading photons, and where $M_{\gamma\gamma}$ is the invariant mass of the photon pair. The final ingredient is to apply the $\chi_{HH}$ cut, taking now into account that we have a $b$-quark pair and a photon pair in the final state:
\begin{equation}
\chi_{HH}=\sqrt{\left(\dfrac{M_{bb}-m_H}{0.05\,m_H}\right)^2+\left(\dfrac{M_{\gamma\gamma}-m_H}{0.05\,m_H}\right)^2}<1\label{chiHHphotons}\,.
\end{equation}
 This ensures that the two $b$-quarks and the two photons come from the decay of a Higgs particle.

  Once again, there might be important background contributions from multijet QCD processes leading to different final states than that of $b\bar{b}\gamma\gamma jj$, such as $6j$ and $b\bar{b}jjjj$, in which some of the final state light jets are again misidentified as $b$ jets and some are misidentified as photons. Taking again as the presumably leading QCD background processes the production of six light jets and of two $b$ jets and four light jets, applying a misidentification rate of 0.1\% per each jet misidentified as a photon, and considering a similar reduction factor after our selection cuts as before, since the selection cuts are very similar, we obtain:   $1.3\cdot10^5 \cdot (0.01)^2\cdot (0.001)^2\cdot 1.1 \cdot 10^{-5}= 1.4 \cdot 10^{-10}$ pb for the six light jets case and $7.5\cdot 10^3 \cdot (0.001)^2\cdot 1.1 \cdot 10^{-5}= 8.2 \cdot 10^{-8}$ pb for the $2b4j$ case. Again in both cases the final cross sections are more than one order of magnitude smaller than the main background we have considered, being of order $10^{-6}$ pb, concluding again that they can be neglected as well.
  
    \begin{figure}[t!]
\begin{center}
\includegraphics[width=0.49\textwidth]{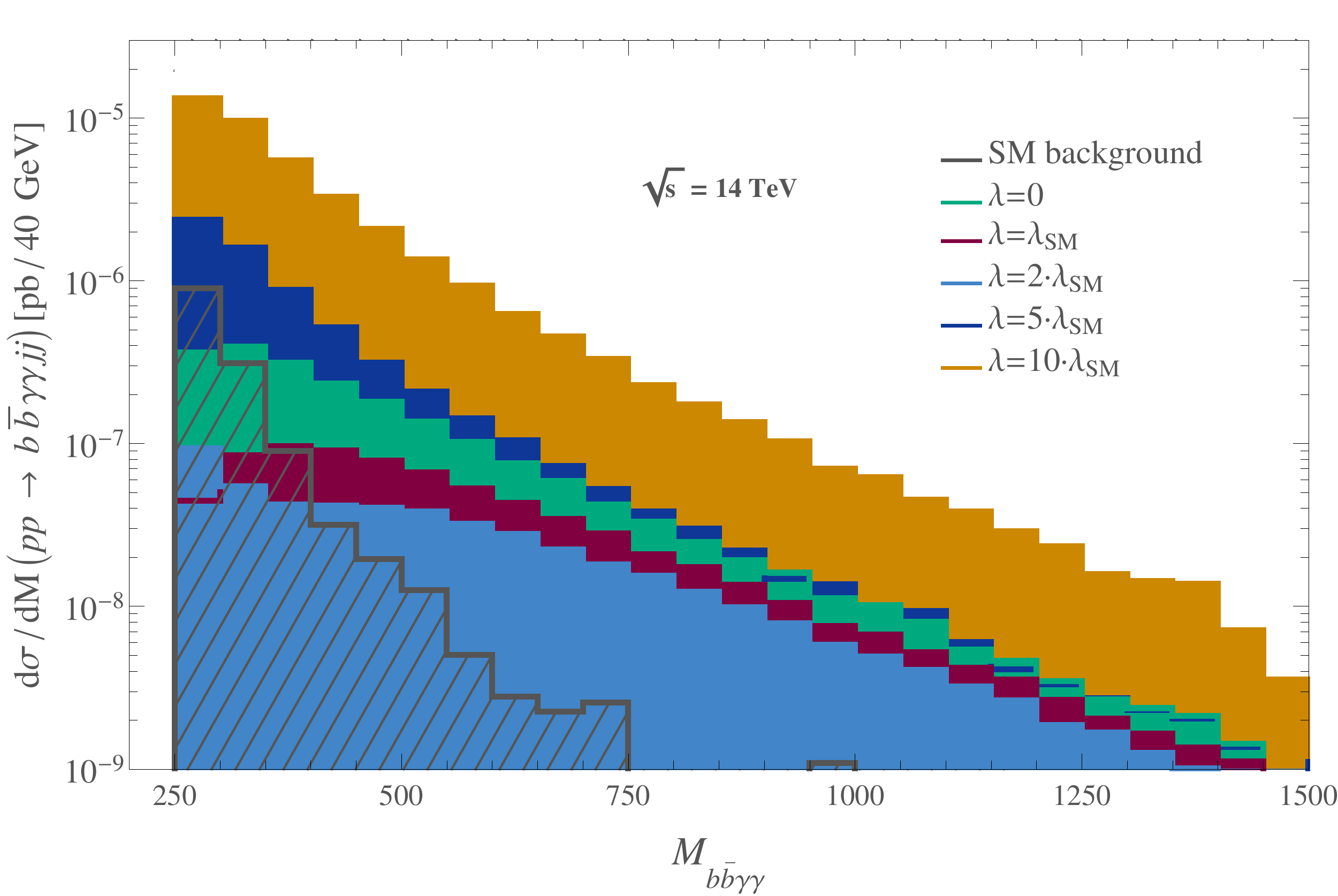}
\includegraphics[width=0.49\textwidth]{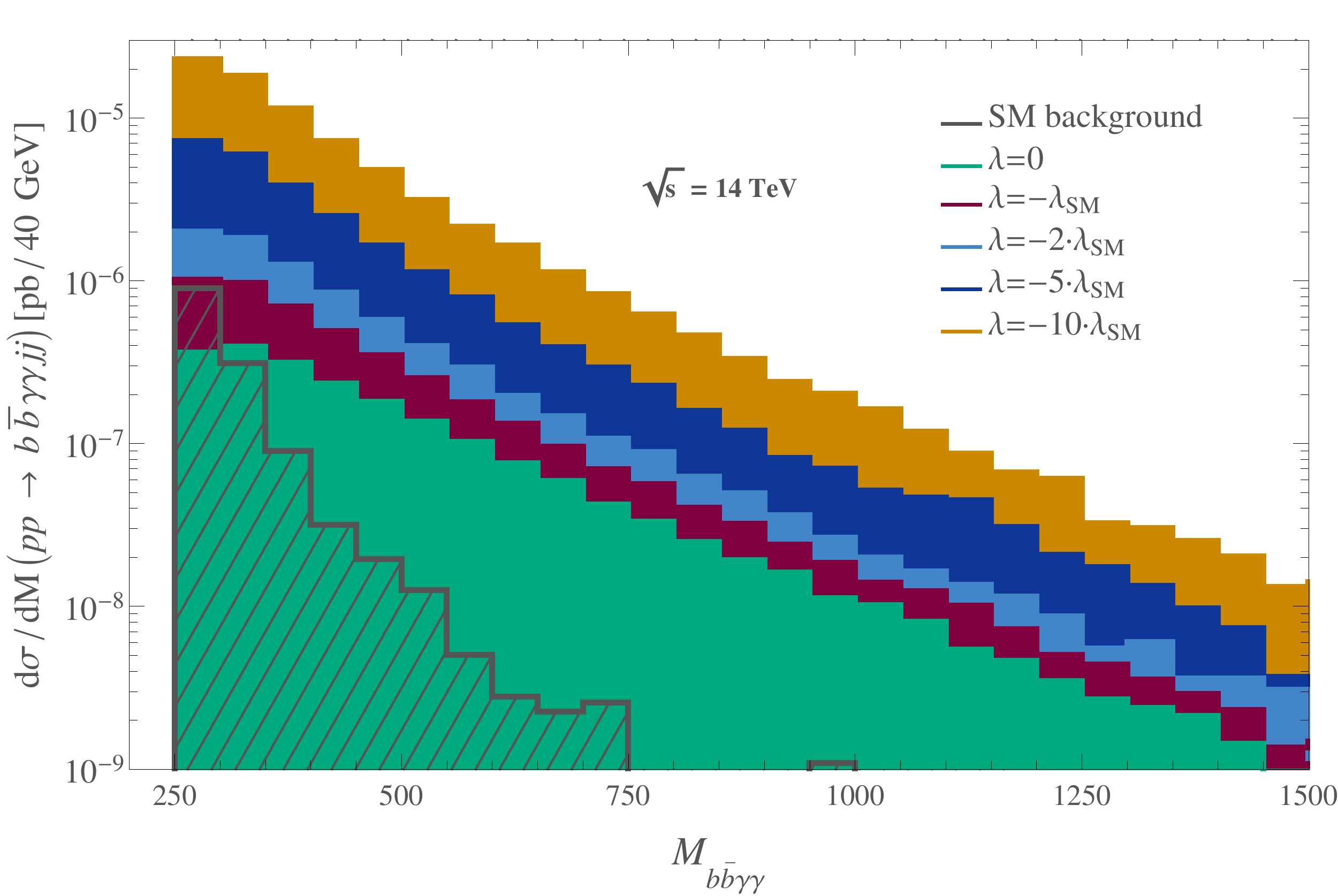}
\caption{Predictions for the total cross section of the process $pp\to b\bar{b}\gamma\gamma jj$ as a function of the invariant mass of the $b\bar{b}\gamma\gamma$ system $M_{b\bar{b}\gamma\gamma}$ for different values of the Higgs self-coupling $\lambda$. We display the predictions for the signal with positive (left panel) and negative (right panel) values of $\lambda$ for comparison, as well as the total SM background. Cuts in Eqs.(\ref{basiccuts2b2a2j})-(\ref{chiHHphotons}) and VBS selection cuts presented in Eq.(\ref{VBSselectioncuts}) have been applied. The center of mass energy is set to $\sqrt{s}=14$ TeV.}
\label{fig:M2b2a_distributions}
\end{center}
\end{figure}

Having all this in mind, we present in \figref{fig:M2b2a_distributions} the predictions for the total cross section of the process $pp\to b\bar{b}\gamma\gamma jj$ as a function of the invariant mass of the $b\bar{b}\gamma\gamma$ system $M_{b\bar{b}\gamma\gamma}$, for different values of the Higgs self-coupling $\lambda$. We also display the prediction for the total SM background (sum of the QCD-EW and the $ZHjj$ background) for comparison. Once again, one can see that the signal distributions for different values of $\kappa$ are very similar to those shown in  \figref{fig:MHH_distributions}, and that the main difference is due to the reduction factor of the branching ratios into photons and into $b$-quarks. They are very similar, too, to the results of the  $b\bar{b}b\bar{b}jj$ final state, in \figref{fig:M4b_distributions}, although two-three orders of magnitude smaller.  The background is, however, very different with respect to the one for  $b\bar{b}b\bar{b}jj$ events. It is smaller in comparison with the signal, specially at high  $M_{b\bar{b}\gamma\gamma}$, since it decreases much more steeply. Therefore, we would expect to have good sensitivities to the Higgs self-coupling despite the lower rates of the process involving photons. For completeness, we display in \tabref{Tab:allSig_2b2a2j} the predictions for the total cross section of the signal, for the set of $\kappa$ values considered, and  after applying all cuts given in Eq. (\ref{VBSselectioncuts}) and in Eqs. (\ref{basiccuts2b2a2j})-(\ref{chiHHphotons}). The prediction for the cross section of the total SM background for this same cuts amounts to $\sigma_{\rm Background}=1.4\cdot 10^{-6}$ pb.

\begin{table}[b!]
\begin{center}
\begin{tabular}{c @{\extracolsep{0.6cm}} c @{\extracolsep{0.6cm}} c @{\extracolsep{0.6cm}} c @{\extracolsep{0.6cm}} c @{\extracolsep{0.6cm}} c @{\extracolsep{0.6cm}} c @{\extracolsep{0.6cm}} c @{\extracolsep{0.6cm}} c @{\extracolsep{0.6cm}} c}
\toprule
\toprule
$\kappa $&  0& 1 & -1 & 2& -2& 5& -5& 10& -10\\
\midrule
 $\sigma_{\rm Signal}\cdot 10^6$ [pb] & 2.0 & 0.7 & 4.5 & 0.5 & 8.0 & 6.4 & 25.2 & 38.4 & 76.0  \\
\bottomrule
\bottomrule
\end{tabular}
\vspace{0.4cm}
\caption{Predictions for the total cross section of the signal $pp\to b\bar{b}\gamma\gamma j j$ after imposing all the selection criteria, VBS cuts given in \eqref{VBSselectioncuts} and cuts given in Eqs. (\ref{bkgcuts2b2a2j}) and (\ref{chiHHphotons}) for all the values of $\kappa$ considered in this work: $\kappa=0,\pm 1, \pm 2, \pm 5, \pm 10$. The cross section of the SM background for this same cuts amounts to $\sigma_{\rm Background}=1.4\cdot 10^{-6}$ pb. Basic cuts in \eqref{basiccuts2b2a2j} are also applied.}
\label{Tab:allSig_2b2a2j}
\end{center}
\end{table}

  \begin{figure}[t!]
\begin{center}
\includegraphics[width=0.49\textwidth]{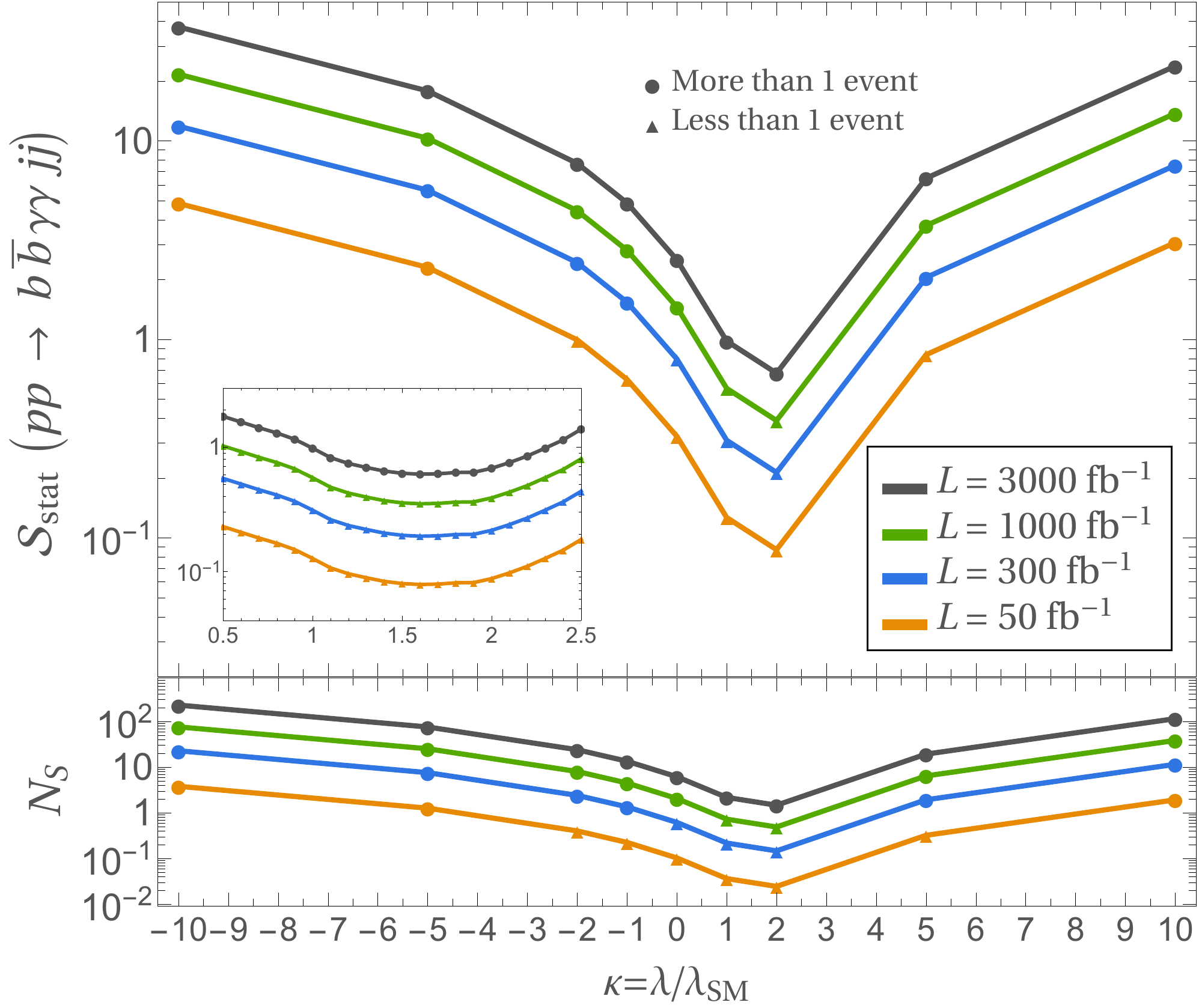}
\includegraphics[width=0.49\textwidth]{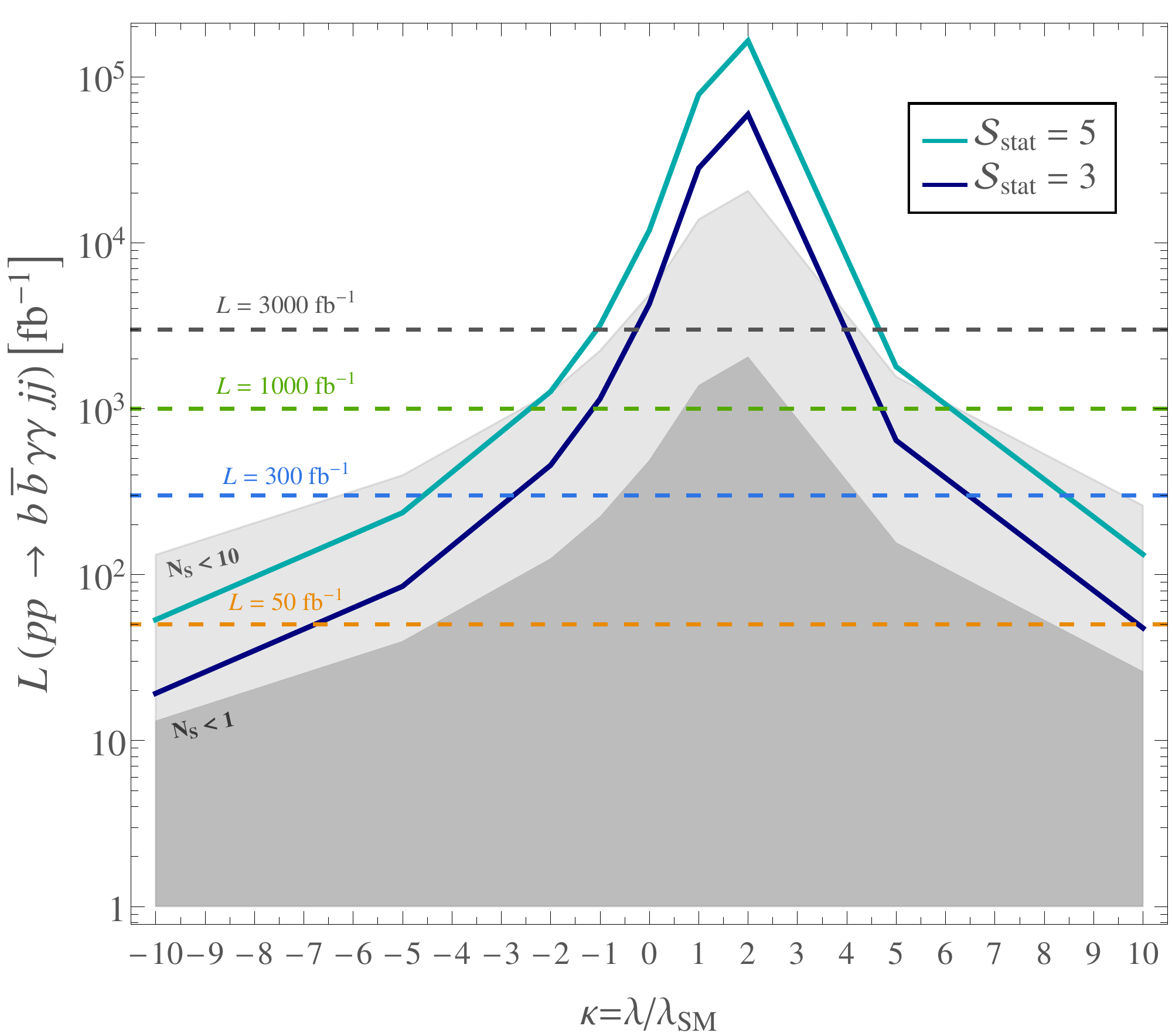}
\caption{Prediction of the statistical significance, $\mathcal{S}_{\rm stat}$, of the process $p p\to b\bar{b}\gamma\gamma jj$ for the four luminosities considered $L=50,300,1000,3000$ fb${}^{-1}$ (left panel) and  of the value of the luminosity that will be required to probe a given $\kappa$  at the LHC at  3$\sigma$ and at 5$\sigma$ (right panel),  as a function of the value of $\kappa$. The marked points represent our evaluations.  In the left panel, a zoom is performed on the interesting values of $\kappa$ ranging between 0.5 and 2.5. The shadowed areas in the right panel correspond to the regions where the number of predicted signal events $N_S$ is below 1, and 10.
The center of mass energy has been set to $\sqrt{s}=14$ TeV.}
\label{fig:significances_2b2a2j}
\end{center}
\end{figure}

In \figref{fig:significances_2b2a2j} we show the predictions for the statistical significance $\mathcal{S}_{\rm stat}$, computed in the same way as in the previous section, making use of \eqref{signieq}, for the four luminosities considered previously, $L=50,300,1000,3000$ fb${}^{-1}$  and taking again a closer look for the values of $\kappa$ ranging between 0.5 and 2.5. We also show the predictions of the final number of signal events, $N_S$ as a function of $\kappa$, for these same luminosities.  On the right panel of this figure we present the prediction for  the value of the luminosity that will be required to probe a given $\kappa$ value with sensitivities at 3$\sigma$ and 5$\sigma$, as a function of the value of $\kappa$. In these plots, due to the lower statistics of this process, some of the computed significances correspond to scenarios in which there is not even one signal event. The concrete predictions for these signal event rates can be read from the lower plot of the left panel.

Taking a look at these figures, we can again extract the conclusions on the sensitivity to the Higgs self-coupling at the LHC in VBS processes, this time in $pp\to b\bar{b}\gamma\gamma j j$ events. One might notice that, although the results are less encouraging than those of  $pp\to b\bar{b}b\bar{b} j j$ events, this channel could also be very useful to measure the value of $\lambda$. Analogously to the previous section, in \tabref{Tab:range_2b2a2j} we present the values of $\kappa\equiv\lambda/\lambda_{SM}$ that would  be accessible at the LHC in these type of events, $pp\to b\bar{b}\gamma\gamma jj$, with a statistical significance equal or better than $3\sigma$($5\sigma$), for the four luminosities considered.

These results show again that the values of $\kappa$ that can be probed in the future at LHC  through the study of VBS processes leading to the final state $ b\bar{b}\gamma\gamma jj$ could be very competitive as well.  Except for the lowest luminosity considered, $L=50$ fb${}^{-1}$, where the signal rates found at the parton level are too low as to survive the extra suppression due to the missing detector efficiencies, hadronization effects etc, the sensitivities found point towards the potential of VBS processes in order to obtain a precise measurement of $\lambda$. The values close to the SM value, are, again, very challenging to reach at the LHC, since the statistical significances of $\kappa\in[0.5,2]$ are always below 2$\sigma$ for this case as well. However, the HL-LHC should be able to probe deviations in $\lambda$ very efficiently in this channel.

\begin{table}[b!]
\begin{center}
\begin{tabular}{c@{\extracolsep{0.3cm}}c@{\extracolsep{0.3cm}}c@{\extracolsep{0.3cm}}c@{\extracolsep{0.3cm}}c}
\toprule
\toprule
$L$ \big[fb${}^{-1}$\big]&  50 & 300 & 1000 & 3000\\
\midrule
$\kappa> 0$ & $\kappa >  9.9~(14.2)$ &$\kappa >  6.4~(8.4)$ & $\kappa > 4.6 ~(6.0)$ & $\kappa >  3.8~(4.7)$
\\
$\kappa <0$ & $\kappa <  -6.7~(-10.0)$ & $\kappa <  -2.7~(-4.6)$ & $\kappa <  -1.1~(-2.3)$ & $\kappa <  -0.2~(-1.0)$ 
\\
\bottomrule
\bottomrule
\end{tabular}
\vspace{0.4cm}
\caption{Predictions for the values of $\kappa\equiv\lambda/\lambda_{SM}$ that the LHC would be able to probe in  $pp\to b\bar{b}\gamma\gamma jj$ events, with a sensitivity equal or better than $3\sigma$ ($5\sigma$) for the four luminosities considered: $L=50,300,1000,3000$ fb${}^{-1}$. }
\label{Tab:range_2b2a2j}
\end{center}
\end{table}

\subsection{Beyond parton level estimates}
\label{discussion}
Finally, to close this section of results, we find pertinent to discuss on how the precision of our predictions could be improved by including additional considerations. We comment here just on those that we consider are the most relevant ones. 
\begin{itemize}
\item[1.-] Our computation of the $HHjj$ signal rates incorporates just those coming from the subprocess $qq \to HH jj$, which includes VBS, but this is not the only contributing channel. It is well known that also the subprocess $gg \to HH jj$, initiated by gluons, does contribute to these signal rates, and it is also sensitive to large BSM $\lambda$ values~\cite{Dolan:2013rja}. Although it is a one-loop process, mediated mainly by top quark loops, it provides  a sizable contribution to the total $HHjj$ signal cross section. For instance, for the case of $\lambda=\lambda_{SM}$, the total cross section at the LHC with $\sqrt{s}$ = 14 TeV is,  according to \cite{Dolan:2015zja},  $5.5$ fb from $gg \to HH jj$ to be compared with $2$ fb from VBS. Therefore, when  considering both contributions to the signal, the sensitivity to $\lambda$ presumably increases. However, we have explicitly checked that once we apply our optimized VBS selection cuts summarized in Eq.~(\ref{VBSselectioncuts}) and in \tabref{table:VBS}, we get a notably reduced cross section for this $gg$ subprocess. In particular, our estimate of the signal rates at the LHC with $\sqrt{s}$ = 14 TeV from $gg \to HH jj$, after applying the stringent $M_{jj}>500$ GeV cut and using the results in \cite{Dolan:2015zja} for the $M_{jj}$ distribution, gives a strong reduction in the corresponding cross section, and leads to smaller rates for $gg$ than those from VBS  by about a factor of 20. Therefore its contribution to the signal rates studied here can be safely neglected, and no much better precision will be obtained by including this new contribution in the signal rates. We have also checked that this finding is true for other BSM values of $\lambda$.   
\item[2.-] When considering next to leading order (NLO) QCD corrections in our estimates of both the signal and background rates, we expect some modifications in our results. These can be very easily estimated, as usual, by using the corresponding $K$-factors. Thus, for instance, for the leading $b \bar b b \bar b j j$ final state,  we can include these NLO corrections by taking into account the $K$-factors for the VBS signal and for the main background from multijet QCD. For the signal we take the $K$-factor  from \cite{Frederix:2014hta}, given  by $K_{\rm VBS}=1.09$. For the QCD-multijet background the corresponding $K$-factor is, to our knowledge, not available in the literature, and different choices are usually assumed. We consider here two choices: $K_{\rm QCD}=1.5$, and another, more conservative one, of $K_{\rm QCD}=3$. This implies that our predictions for the signal rates are practically unchanged, but those for the background rates are increased by a factor of 1.5 and 3 respectively. This modifies our predictions for the statistical significance  of the $b \bar b b \bar b j j$ signal, from the $\mathcal{S}_{\rm stat}$ results given \figref{fig:significances_4b2j} to $\mathcal{S}_{\rm stat}^{\rm NLO}\sim K_{\rm VBS}/\sqrt{K_{\rm QCD}}\,\,\mathcal{S}_{\rm stat} \sim 0.9\,\, \mathcal{S}_{\rm stat}$ ($0.6\,\, \mathcal{S}_{\rm stat}$ ) for $K_{\rm QCD}=1.5$ ($K_{\rm QCD}=3$).  For instance, for the high luminosity considered of 
$1000 \,{\rm fb}^{-1}$ we get sensitivities of $\kappa >3.8 (4.3)$ for $K_{\rm QCD}=1.5$, and of $\kappa >4.5 (4.8)$ for $K_{\rm QCD}=3$, both at the $3\sigma$ ($5\sigma$) level, to be compared with our benchmark result in  \tabref{Tab:range_4b2j} of $\kappa >3.7 (4.2)$. Therefore ignoring these NLO corrections does not provide large uncertainties either.      
\item[3.-] When including $b$-tagging efficiencies in our estimates of the $b \bar b b \bar b j j$ signal and background rates, our predictions of the statistical significance do also change. However, an estimate of this change can be easily done by adding the corresponding modifying factors. Assuming well known $b$-tagging efficiencies of $70\%$, that apply to both the signal and background, the two rates are reduced by a factor of ${0.7}^ 4 \sim 0.24$ . Therefore we get a reduced  statistical significance of $\mathcal{S}_{\rm stat}^{b-{\rm tag}}\sim 0.24/\sqrt{0.24}\,\, \mathcal{S}_{\rm stat} \sim 0.5 \,\,\mathcal{S}_{\rm stat}$ with respect to the ones that we have reported previously. This factor of $0.5$ will change our predicted sensitivities to BSM $\lambda$ values. Again, as an example, for the considered luminosity of
$1000 \,{\rm fb}^{-1}$ we get sensitivities of $\kappa >4.3 (4.9)$ at the $3\sigma$ ($5\sigma$) level to be compared with our benchmark result in \tabref{Tab:range_4b2j} of $\kappa >3.7 (4.2)$. 

Similarly, considering also photon-identification efficiencies (also called in this work $\gamma$- tagging) of $95\%$, as presented in the literature, we get reduced signal and background rates for the $b \bar b \gamma \gamma j j$ final state by a factor of ${0.7}^ 2 \times {0.95}^ 2 \sim 0.44$. Accordingly, we obtain a reduction in the statistical significance of the $b \bar b \gamma \gamma j j$ events, given by  $\mathcal{S}_{\rm stat}^{b, \gamma-{\rm tag}}\sim 0.44/\sqrt{0.44} \,\,\mathcal{S}_{\rm stat} \sim 0.7 \,\,\mathcal{S}_{\rm stat}$ with respect to our results reported in the pages above. The changes in the sensitivities to $\kappa$ can be easily derived. using the same illustrative example, for $1000 \,{\rm fb}^{-1}$ of luminosity, we get sensitivities of $\kappa > 6.0 (8.0)$ at the $3\sigma$($5\sigma$) level to be compared with our benchmark result in \tabref{Tab:range_2b2a2j} of $\kappa >4.6 (6.0)$. 
\item[4.-] One of the largest uncertainties comes from the choice of the energy resolution needed for the reconstruction of the $HHjj$ signal events from the corresponding final state. This basically can be translated into the choice for the particular definition of the $\chi_{HH}$ variable which is very relevant for the selection of the $HH$ candidates. Thus, for the  
$b \bar b b \bar b j j$ final state, in our benchmark scenario we have taken $0.05 \times m_H$ around $m_H$ in the definition of $X_{HH}$ in Eq.~(\ref{cutsHH}), i.e.\footnote{Equivalently in the case of $pp\to b\bar{b}\gamma\gamma jj$ substituting $M_{bb^s}$ by $M_{\gamma\gamma}$.},
\begin{align}
\chi_{HH}\equiv\sqrt{\left(\dfrac{M_{bb^l}-m_H}{ \Delta_{E}\,m_H}\right)^2+\left(\dfrac{M_{bb^s}-m_H}{  \Delta_{E}\,m_H}\right)^2}<1\,,
\label{cutXHH}
\end{align}
with $ \Delta_{E}$ being the energy resolution, which in this case was set to 0.05, leading to a mass resolution of 5\% of the Higgs mass. We chose this value since it optimizes the selection efficiency and could be useful for future experiments with better energy resolution. However a more realistic choice, given the current energy resolution at the LHC experiments, could rather be $\Delta_E\cdot m_H = 0.1 \times m_H$ GeV $\sim 12.5$ GeV. We have redone the analysis with this alternative and more conservative choice and we have obtained, as expected,  a reduced statistical significance. The signal rates do not  change (we still get $4.1 \times 10^{-5}$ pb) , but the main QCD-background does (we get 
$1.8 \times 10^{-2}$ pb  instead of our benchmark value of $6.8 \times 10^{-3}$ pb). This translates in a reduction of the significance given by a factor $\mathcal{S}_{\rm stat}^{\chi_{HH}}\sim 1/\sqrt{18/6.8}\,\, \mathcal{S}_{\rm stat} \sim 0.7 \,\,\mathcal{S}_{\rm stat}$.  The implication of this reduction can directly be seen as a modification of the sensitivity to $\kappa$. Once again, for our benchmark case of $1000 \,{\rm fb}^{-1}$, we obtain sensitivities of $\kappa > 4.0 (4.5)$ at the $3\sigma$ ($5\sigma$) level to be compared with our benchmark result in \tabref{Tab:range_4b2j} of $\kappa >3.7 (4.2)$.

Apart from redoing the analysis for this 10\% resolution\footnote{When mentioning a percentage for the energy resolution we refer to that percentage of the Higgs mass.}, we have also studied other possible and realistic values such as $\Delta_E=20\%$ and $\Delta_E=30\%$, to have a better idea of the implications of the value of the mass determination uncertainty in our predictions. The results for both of our signals are shown in \figref{fig:XHHcut} by the green lines and green shaded areas, where we present the values for the statistical significance at $1000 \,{\rm fb}^{-1}$ as a function of the value of $\kappa$ for different energy resolutions of $\Delta_E=$ 5\% (original scenario throughout the work), 10\%, 20\% and 30\% (the purple lines and purple areas of this figure will be discussed next).  One can see that, as expected, the statistical significance decreases as the energy resolution worsens, but in any case, from the most optimistic case ($\Delta_E=5\%$) to the less optimistic one ($\Delta_E=30\%$), we only obtain a reduction factor of at most  0.4 in the statistical significance.

 \begin{figure}[t!]
\begin{center}
\includegraphics[width=0.49\textwidth]{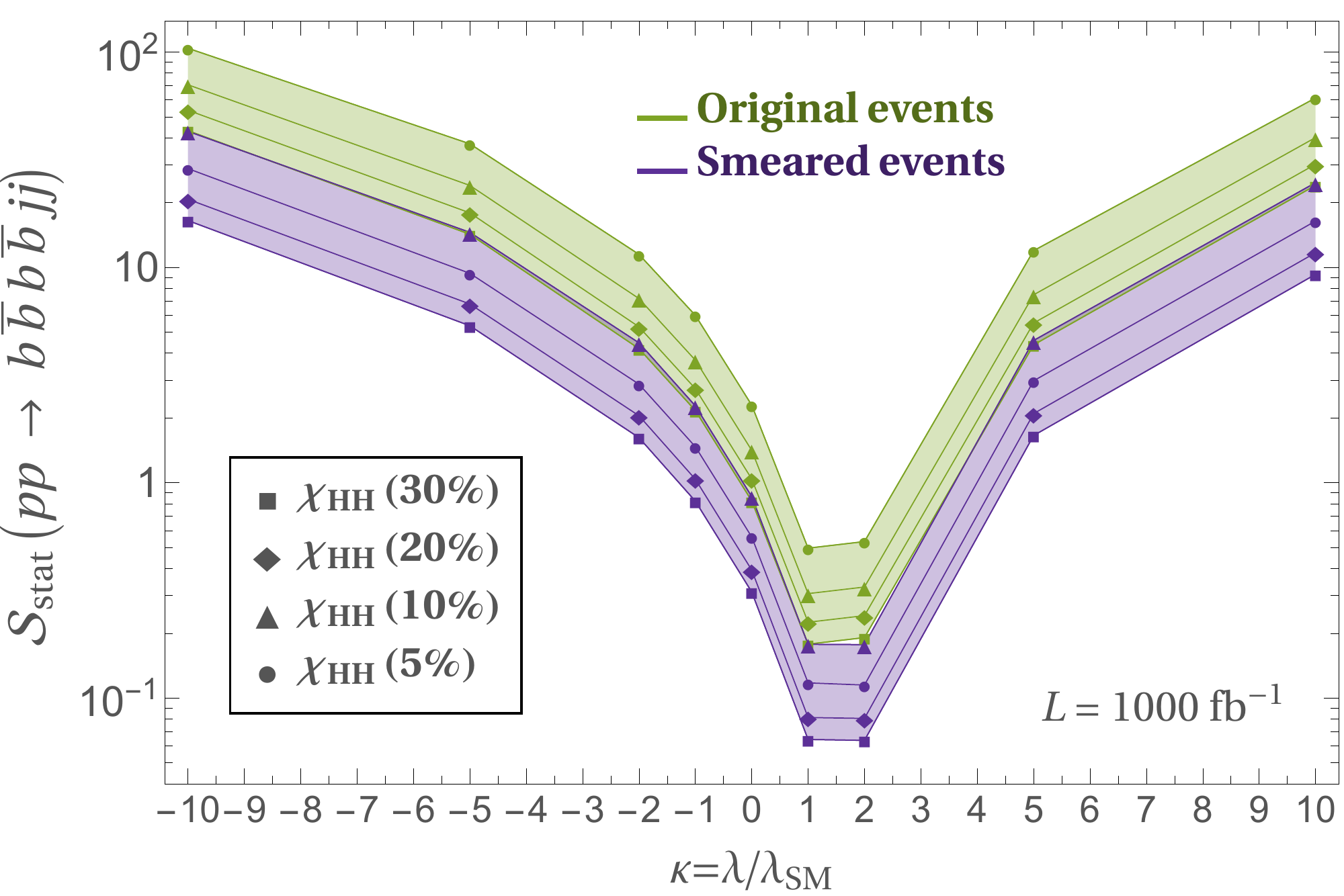}
\includegraphics[width=0.49\textwidth]{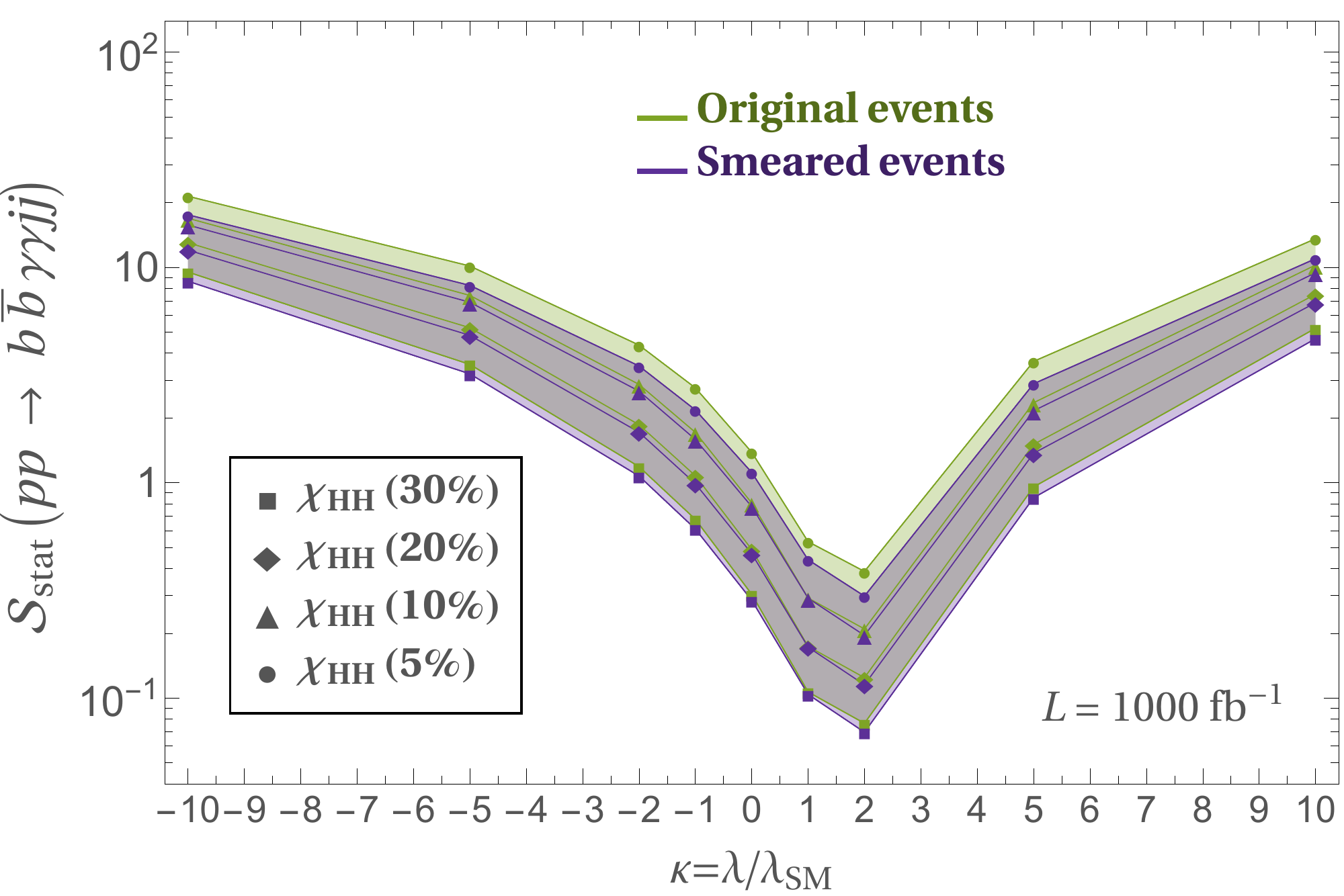}
\caption{Prediction of the statistical significance, $\mathcal{S}_{\rm stat}$, of the process $p p\to b\bar{b} b\bar{b} jj$ (left panel) and of the process $p p\to b\bar{b}\gamma\gamma jj$ (right panel) for $L=1000$ fb${}^{-1}$  as a function of the value of $\kappa$ for different values of the energy resolution, ($\Delta_E$\%), applied through the variable $\chi_{HH}$ defined in \eqref{cutXHH}. These different values are marked with different symbols. We show the predictions for the original events (green lines and green shaded areas; notice that the upper green line corresponds to the green line presented in \figref{fig:significances_4b2j} (left panel) and \figref{fig:significances_2b2a2j} (right panel)), and for the events with a Gaussian smearing applied in order to account for detector effects (purple lines and purple shaded area). The marked points represent our evaluations. The center of mass energy has been set to $\sqrt{s}=14$ TeV.}
\label{fig:XHHcut}
\end{center}
\end{figure}

\begin{figure}[t!]
\begin{center}
\includegraphics[width=0.49\textwidth]{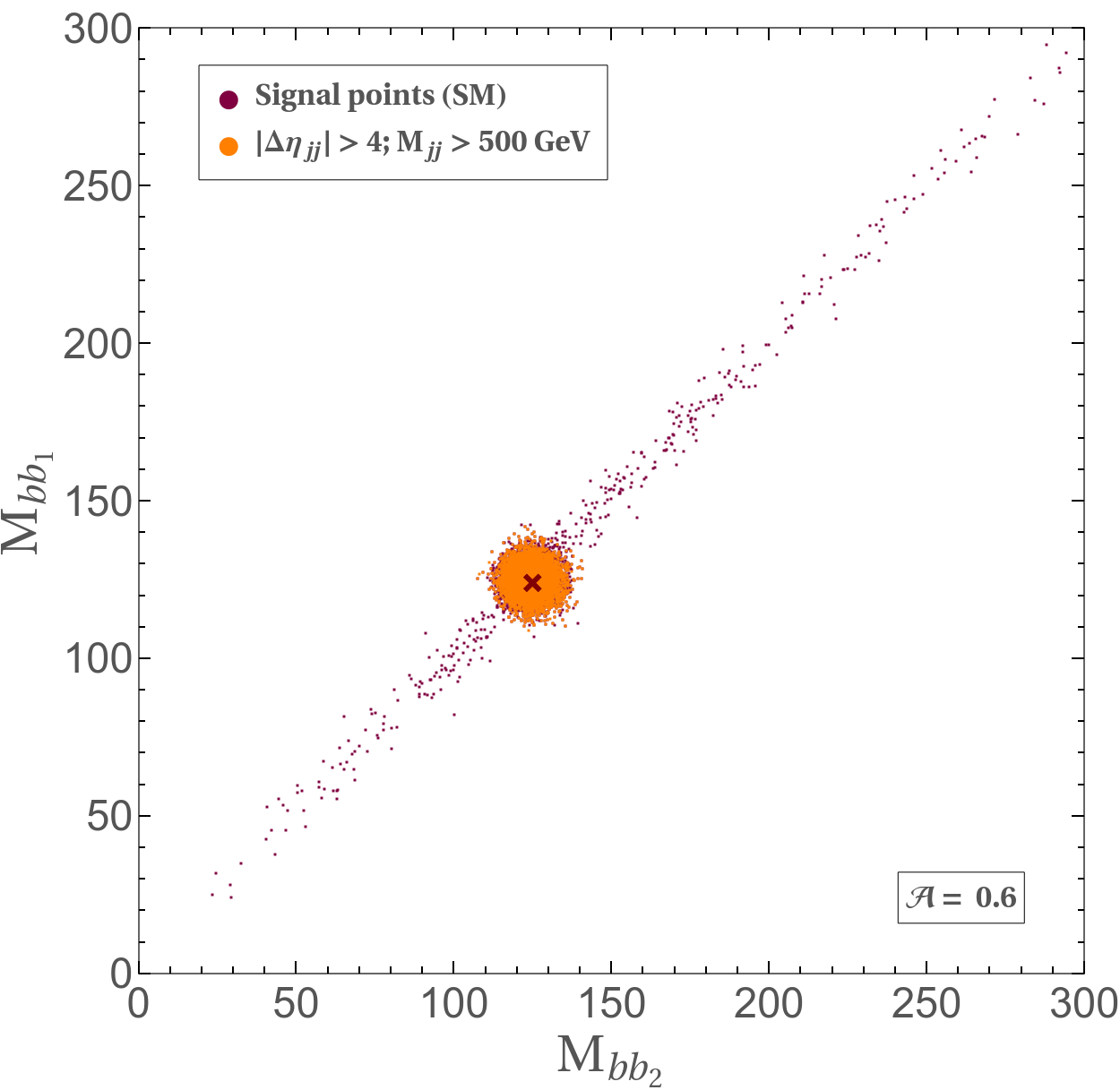}
\includegraphics[width=0.49\textwidth]{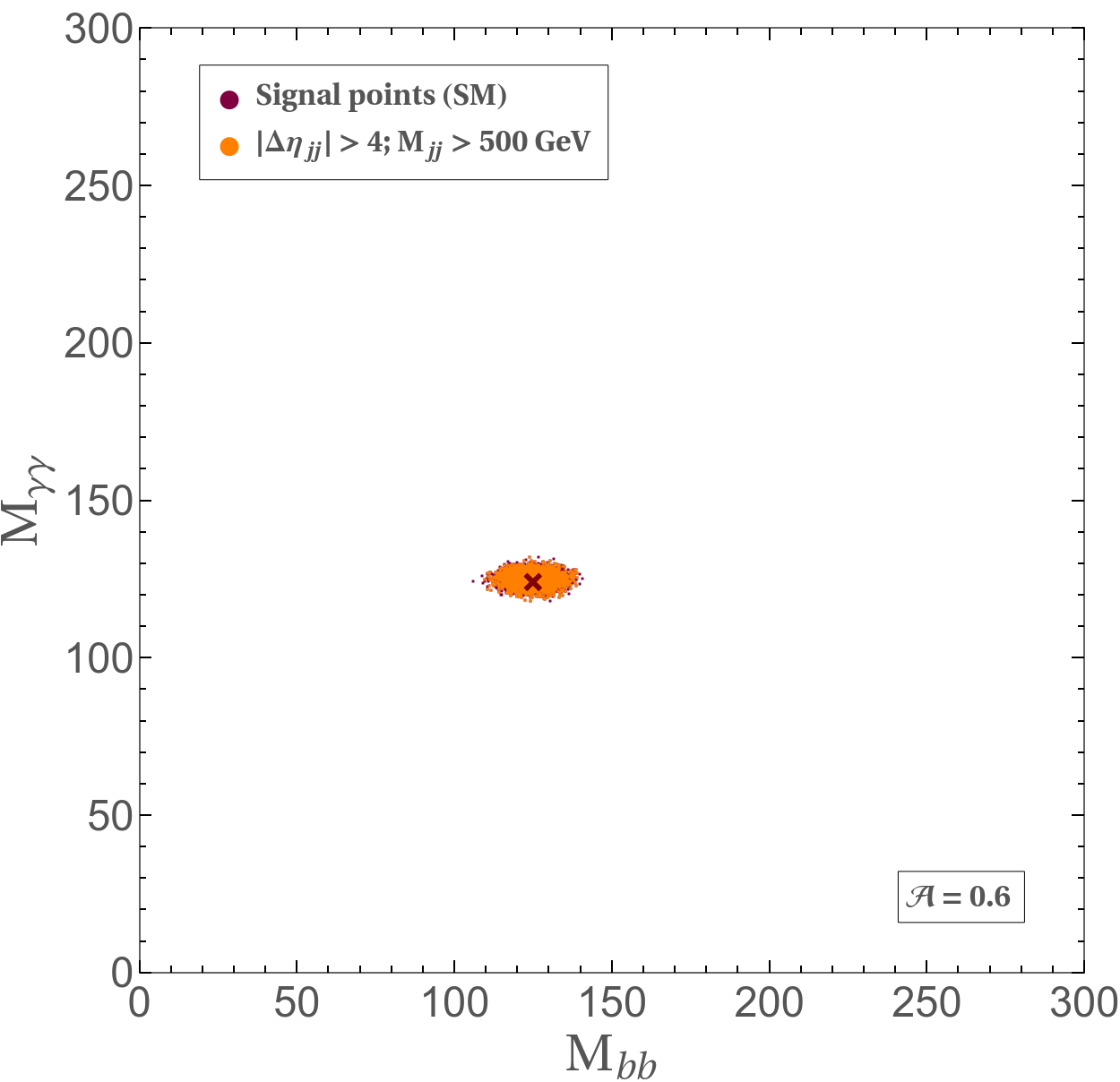}
\caption{Distribution of 10000 Monte Carlo signal events  of $pp\to HHjj\to b\bar{b}b\bar{b}jj$ (left panel) and of $pp\to HHjj\to b\bar{b}\gamma \gamma jj$ (right panel) in the plane of the invariant mass of one the Higgs candidates ($M_{bb_1}$ in the left panel and $M_{bb}$ in the right panel) versus the invariant mass of the other Higgs candidate ($M_{bb_2}$ in the left panel and $M_{\gamma\gamma}$ in the right panel) after applying a Gaussian smearing to the energy of all final state partons as explained in the text. See details of $HH$ candidate selection in the text.  Orange dots correspond to those events that pass the implemented VBS selection cuts given in Eq.(\ref{VBSselectioncuts}). Cuts in \eqref{basiccuts4b2j} (left panel) and in  \eqref{basiccuts2b2a2j} (right panel) have been implemented. The value of the acceptance $\mathcal{A}$ of the VBS cuts is also included. The red cross represents the value of the Higgs mass. The center of mass energy has been set to $\sqrt{s}=$14 TeV.}
\label{fig:MMplanesmearing}
\end{center}
\end{figure}

\item[5.-]  
Another important point that might change significantly our predictions is that introduced by the Higgs mass reconstruction uncertainty coming from detector effects. To estimate this uncertainty, we have applied a Gaussian smearing to the energy of all final state partons. Following~\cite{Pascoli:2018heg}. This gaussian dispersion has been introduced as $1/\sqrt{2\pi\sigma}\cdot e^{-x^2/(2\sigma^2)}$, with $\sigma=0.05\cdot E_{j,b}$ for the energy dispersion of the final light and $b$ jets and with $\sigma=0.02\cdot E_\gamma$ for the energy dispersion of the final photons. We have performed this for each studied signal and for their corresponding backgrounds in order to characterize the impact that these detector effects have regarding the distribution of our events on the relevant kinematical variables. In \figref{fig:MMplanesmearing} we show the distribution of  10000 Monte Carlo signal events  of $pp\to HHjj\to b\bar{b}b\bar{b}jj$ (left panel) and of $pp\to HHjj\to b\bar{b}\gamma \gamma jj$ (right panel) in the plane of the invariant mass of one the Higgs candidates ($M_{bb_2}$ in the left panel and $M_{bb}$ in the right panel) versus the invariant mass of the other Higgs candidate ($M_{bb_1}$ in the left panel and $M_{\gamma\gamma}$ in the right panel). No other cuts than those of the basic selection, given in \eqref{basiccuts4b2j} (left panel) and in  \eqref{basiccuts2b2a2j} (right panel) have been implemented. The orange points correspond to those events that fulfill the VBS selection criteria. The impact of these VBS cuts does not change appreciably after the smearing, not on the signal nor on the background events. The selection of the Higgs candidates in the case of the $b\bar{b}b\bar{b}jj$ signal is performed as explained in the text, following the minimization of $|M_{bb_1}-M_{bb_2}|$. This is the reason why we obtain several points distributed in the diagonal of the left panel. As expected, the detector effects translate into a dispersion of the signal points from the Higgs mass point outwards. In the $b\bar{b}b\bar{b} jj$ case, the dispersion is isotropic, since the smearing affects all four $b$-quarks in the same way, whereas in the $b\bar{b}b\gamma\gamma jj$ case, the dispersion in the $M_{bb}$ direction is bigger with respect to that in the $M_{\gamma\gamma}$ direction, accordingly to the difference in the energy resolution of $b$-quarks and photons in the detector. These results are compatible to those obtained in reference \cite{Kling:2016lay}. In any case, both signals seem to lie inside a circle of radius around 12 GeV, which corresponds to a 10\% of the Higgs mass value. This suggests that the effects of the smearing on our predictions of the statistical significance will severely depend on the $\Delta_E$ we use in the $\chi_{HH}$ selection cut, and, in principle, we expect that for $\Delta_E=10\%$ we will obtain the best sensitivities. This is so because, for this $\Delta_E=10\%$, we select the minimum possible number of background events compatible with selecting all of our signal events simultaneously. 

In order to better understand the impact of the $\Delta_E$ value once the detector effects have been taken into account, we present in the purple lines and purple shaded areas of \figref{fig:XHHcut}  the values for the statistical significance at $1000 \,{\rm fb}^{-1}$ as a function of the value of $\kappa$ for different energy resolutions of 5\% (original scenario throughout the work), 10\%, 20\% and 30\%  after the smearing on the energy of all final state partons has been applied. Is it clear from this figure that, indeed, taking $\Delta_E=10\%$ in the $b\bar{b}b\bar{b}jj$ case maximizes the statistical significance once the detector effects are included. In the $b\bar{b}\gamma\gamma jj $ case (notice that the purple area overlaps with the green one) the $\Delta_E=5\%$ is still the value that gives the best sensitivities, since the signal to background ratio is larger. In any case, from the upper green line to the upper purple line, there is at most a reduction factor of 0.4 in the statistical significance.

\item[6.-]  Considering in addition the effects from showering and clustering of the final jets  will presumably change our naive parton level predictions. However, their estimation will require a more sophisticated and devoted analysis with full computing power and the use of additional techniques like Boost Decision Trees (BDT) and others. This is particularly involved if we wish to control efficiently the background form QCD-multijets and, consequently, we have not performed such an analysis in this Thesis. Nevertheless, to get a first indication of the importance of these effects in the signal rates, we have performed a computation of the $b \bar b b \bar b j j$ signal events after showering with PYTHIA8 \cite{Sjostrand:2014zea} and  clustering with MadAnalysis5 \cite{Conte:2012fm,Conte:2014zja,Dumont:2014tja,Conte:2018vmg} using the anti-$k_t$ algorithm with $R=0.4$, for the BSM example of $\kappa=5$. We have obtained that the cross-section after applying our basic and VBS cuts is $3.0. 10^{-3}$ pb if we include showering+clustering, which should be compared with our parton level estimate of $3.7.10^{-3}$ pb . Therefore, the effect from showering+clustering at this signal level is not very relevant. However, it is expected that it could be relevant in the $HH$ selection candidates and, as we have said, in the reduction efficiency of the QCD-multijet background. 
\end{itemize}

  \begin{figure}[t!]
\begin{center}
\includegraphics[width=0.49\textwidth]{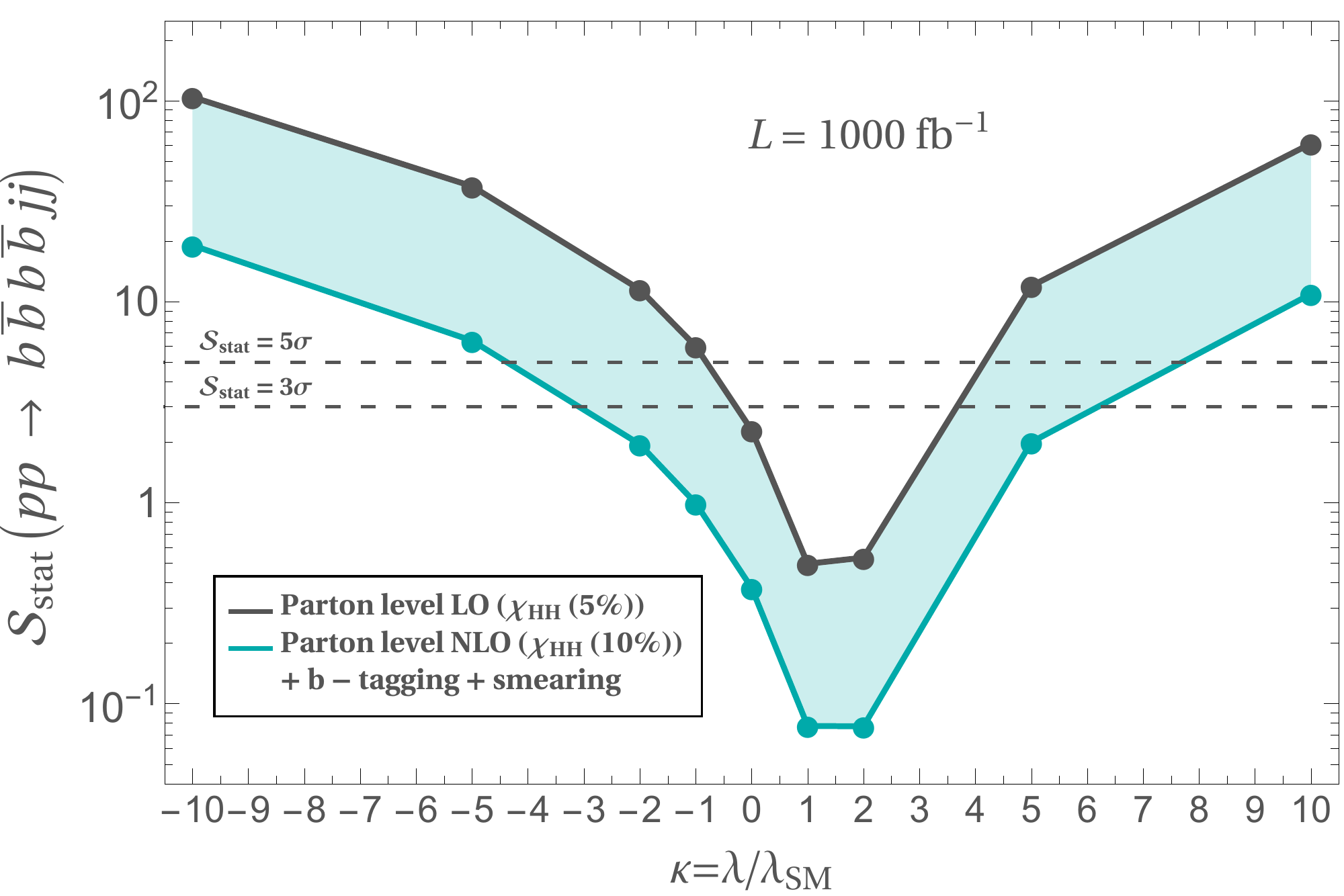}
\includegraphics[width=0.49\textwidth]{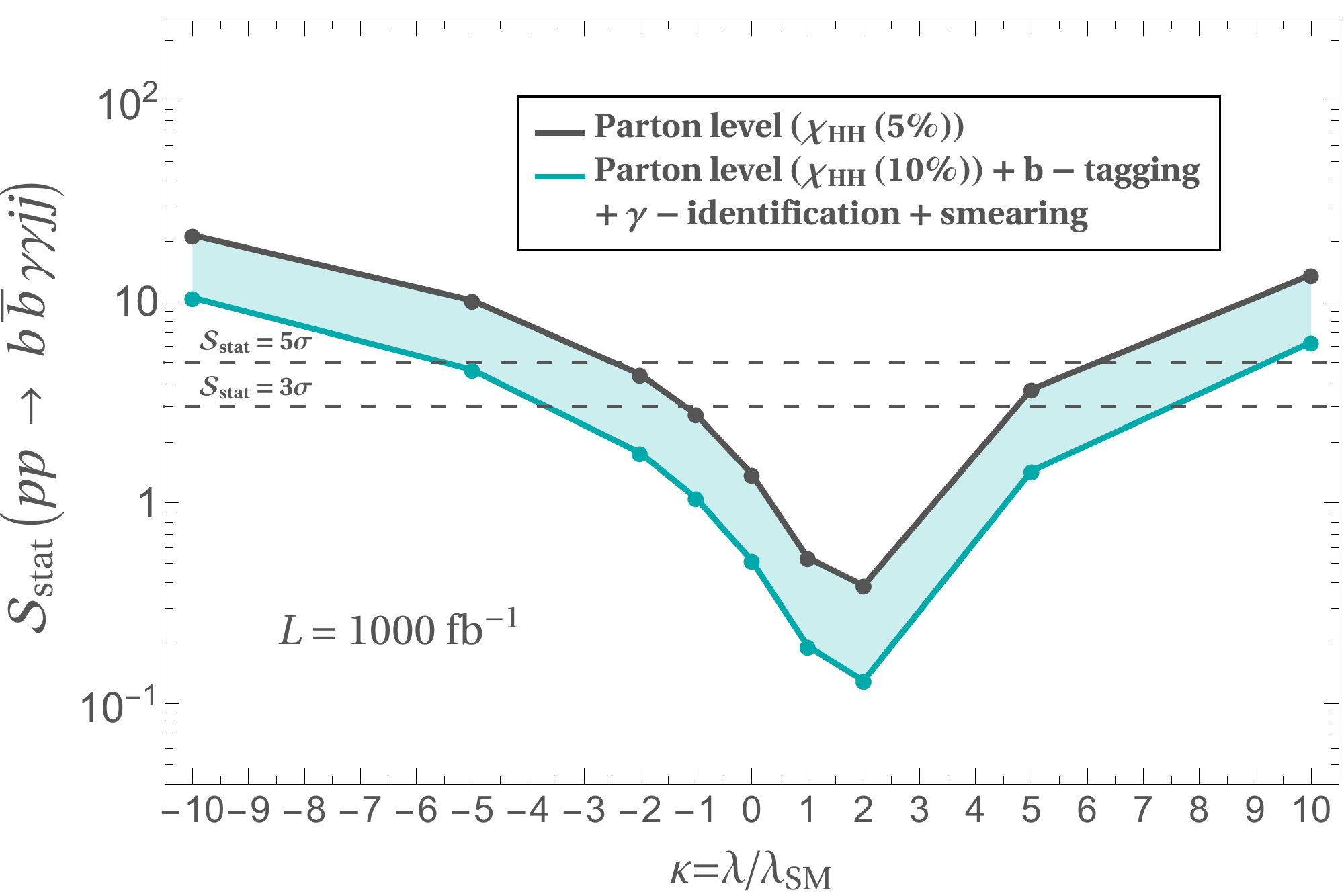}
\caption{Prediction of the statistical significance, $\mathcal{S}_{\rm stat}$, of the process $p p\to b\bar{b} b\bar{b} jj$ (left panel) and of the process $p p\to b\bar{b}\gamma\gamma jj$ (right panel) for $L=1000$ fb${}^{-1}$ as a function of the value of $\kappa$ for two different scenarios: the original parton level analysis (dark grey line, corresponding to the green lines in \figref{fig:significances_4b2j} (left panel) and \figref{fig:significances_2b2a2j} (right panel)) and the analysis performed taking into account the tagging efficiencies of the final state particles, the NLO corrections, the estimation of the detector effects via a Gaussian smearing on the energy of all final state partons and with a 10\% Higgs mass determination uncertainty (blue line, see details of these considerations in the text). The marked points represent our evaluations. The center of mass energy has been set to $\sqrt{s}=14$ TeV.}
\label{fig:errorband}
\end{center}
\end{figure}

Finally, to conclude this subsection and in order to give a more accurate and realistic prediction, all the above mentioned considerations must be taken into account simultaneously. To this aim, we present, in \figref{fig:errorband}, the predictions of the statistical significance as a function of the value of $\kappa$ at $1000~ {\rm fb}^{-1}$ for two comparative scenarios: the original analysis from LO parton level predictions (dark grey line) and the analysis performed after taking into account the main distorting effects which are the tagging efficiencies of the final state particles, described in point 3.- of this discussion, the NLO corrections described in point 2.- and the estimation of the detector effects, introduced in point 5.- with a 10\% Higgs mass determination uncertainty. We give these predictions for both of the studied signals: $p p\to b\bar{b} b\bar{b} jj$ (left panel) and $p p\to b\bar{b}\gamma\gamma jj$ (right panel). The main conclusion is that the biggest uncertainty in our predictions comes from the fact that we are not taking into account, a priori, detector effects.  We have already seen that this can reduce the statistical significance by a factor of 0.4. The second biggest source of uncertainty is the choice of the value of the Higgs mass resolution, $\Delta_E$. Taking a 10\% mass resolution instead of a 5\% can account for a reduction of 0.7 in the statistical significance. Similarly, the $b$-tagging efficiencies in the $b\bar{b}b\bar{b}jj$ case can lead to a similar reduction factor of 0.7. Finally, the NLO corrections play the least relevant role when estimating the uncertainties of the calculation. All the main effects together lead to a reduction factor of at most 0.2 in the statistical significance for $pp\to b\bar{b}b\bar{b}jj$ and of at most 0.5 for $pp\to b\bar{b}\gamma\gamma jj$.
The corresponding changes in the sensitivities to $\kappa$ can be easily derived from \figref{fig:errorband}. Using the same illustrative example, for $1000 \,{\rm fb}^{-1}$ of luminosity, we get sensitivities to $\kappa > 6.2 (7.7)$ at the $3\sigma$ ($5\sigma$) level to be compared with our benchmark result in \tabref{Tab:range_4b2j} of $\kappa >3.7 (4.2)$ for the $b\bar{b}b\bar{b} jj$ case and of $\kappa > 7.7 (9.4)$ at the $3\sigma$ ($5\sigma$) level to be compared with our benchmark result in \tabref{Tab:range_2b2a2j} of $\kappa >4.6 (6.0)$ for the $b\bar{b}\gamma\gamma jj$ one. 

Based on the discussion above we believe that a more dedicated analysis, including more accurately all the considerations above with showering, clustering, and detector effects, and optimizing the selection criteria accordingly, might lead to a sensitivity to the Higgs self-coupling of the same order of magnitude, although a bit smaller, than the one obtained with our naive original analysis. We believe that our findings indicate that double Higgs production via vector boson scattering is a viable and promising observable to measure the Higgs self-coupling in BSM scenarios.

In this Chapter, we have analyzed the sensitivity to $\lambda$ in double Higgs production via vector boson scattering at the LHC, taking advantage of the fact that these processes have very characteristic kinematics that allow us to select them very efficiently against competing SM backgrounds. We have seen how VBS processes can be useful to test BSM scenarios at the LHC, since the two signals we have considered, $pp\to b\bar{b}b\bar{b}jj$ and $pp\to b\bar{b}\gamma\gamma jj$, lead to scenarios in which competitive sensitivities to BSM values of the Higgs self-coupling could be reached at different luminosity stages of the LHC, even beyond our parton level estimates. Because of this, we believe that the vector boson scattering $HH$ production channel will lead to very interesting (and complementary to those of gluon-gluon fusion) findings about the true nature of the Higgs boson.

Having set an example of the VBS potential in the search for new physics, we can move on to the next Chapter, in which we will rely again in this kind of observables to understand the implication of unitarization effects in the information about the EFT characteristics that we can extract from the LHC data.


\chapter[\bfseries Unitarization effects in EFT predictions of WZ scattering at the LHC]{Unitarization effects in EFT predictions of WZ\\ scattering at the LHC}\label{Methods}

\chaptermark{Unitarization effects in EFT predictions of VBS at the LHC}

Throughout the different Chapters of this Thesis it has become manifest that VBS observables at the LHC should be the most promising window to look directly for information about the EWSB dynamics. Concretely, the measurements performed up to this date by the ATLAS and CMS searches, as well as others, have allowed to constraint the parameter space of the EFTs that aim to describe these dynamics at low energies. However, as we have argued already in the previous pages, these measurements can lead to some discrepancies in the interpretation of the experimental data in terms of bounds imposed on the EFT parameters.

The way in which the different experiments treat the issue of the violation of unitarity in VBS measurements is what really complicates the task of achieving a combined constraint on the EFT (in this case EChL) coefficients. As we have introduced, the violation of perturbative unitarity is a common feature of these effective descriptions, and it can take place at values of the center of mass energy that the LHC is currently exploring. Nevertheless, since it is understood as a non-conservation of the probability, unitarity violating predictions shall not be used to compare against experimental data, since they suppose an inconsistency of the underlying EFT.

We wish, however, to rely on the EChL since it is the most general effective theory devoted to explain the EWSB nature, and, in order to do that, the unitarity violation problem must be solved. For this reason, unitarization methods are used to construct unitary scattering amplitudes from the raw, non-unitary EFT predictions, as we already explained in Chapter \ref{EChL}. There, we introduced various of the most commonly employed unitarization methods nowadays, stressing an important fact: that choosing a particular method introduces some model dependence that cannot be avoided, especially in the non-resonant regime. In the resonant case, those methods that are able to reproduce the correct analytical behaviour of the scattering amplitudes (like the IAM or the N/D method) lead to very similar results regarding the resonance properties. On the contrary, in a non-resonant scenario, i.e., when looking for smooth deviations from the continuum, the various manners of unitarizing the computation of an observable lead to very different final results \cite{Alboteanu:2008my,Espriu:2012ih,Delgado:2013loa,Espriu:2014jya,Kilian:2014zja,Delgado:2015kxa,Corbett:2014ora,Corbett:2015lfa,Rauch:2016pai,DelgadoLopez:2017ugq,Perez:2018kav,Kozow:2019txg}. Thus, a theoretical uncertainty arises when computing unitarized EFT  predictions due to the fact that there is a variety of ways of achieving such a unitary outcome.  

Current constraints imposed on some of the mentioned low-energy parameters by LHC experiments do not take this theoretical uncertainty into account, as we have already mentioned throughout several parts of this Thesis (see Chapter \ref{VBS}). They interpret the experimental data using the theoretical EFT predictions in different ways, i.e., using different unitarization methods or no unitarization method at all. 
For instance, the most recent constraints for $a_4$ and $a_5$ (or, equivalently, for their linear counterparts defined in \eqref{fs0fs1}) given in \cite{Aaboud:2016uuk, Sirunyan:2019der,Aad:2019xxo,Sirunyan:2019ksz} provide different analyses. Whereas in \cite{Aad:2019xxo} the measured cross section is directly reported, in \cite{Aaboud:2016uuk} the K-matrix method is used to impose bounds on $a_4$ and $a_5$ and in \cite{Sirunyan:2019der,Sirunyan:2019ksz} the pure EFT predictions are used to obtain such constraints. Therefore, a prescription is needed in order to obtain a unique constraint that can be reliable, and, in some sense, independent of the unitarization method we use.

In this Chapter, we quantify the uncertainty due to the choice of unitarization scheme present in the determination of some of the most relevant low-energy constants for VBS processes, namely, $a_4$ and $a_5$. To this aim, we chose to study the particular VBS process given by the WZ channel in the EChL as a first approach and since it will be a remarkable relevant channel for the next Chapters, as we will see. Within this framework, we characterize the unitarity violation that arises in the predictions of the WZ $\to$ WZ cross sections, and we analyze the impact that a variety of well stablished unitarization methods have on them. We pay special attention to the fact that all helicity states of the incoming and outgoing gauge bosons might play a relevant role in the unitarization process, and consider them all at once as a coupled system. 

Then, we move on to the LHC scenario.  We use the Effective W Approximation \cite{Dawson:1984gx,Johnson:1987tj} to give predictions of $pp\to$WZ$+X$ events at the LHC for different unitarization schemes. In order to check that the EWA works for our purpose here, we compare its predictions for the cases of the SM and the EChL with the corresponding  full results from MG5 \cite{Alwall:2014hca,Frederix:2018nkq}, and we find very good agreement in both cases.  Finally, in order to provide a quantitative analysis of the implications of our study on the LHC searches, we choose to compare our results with those in  \cite{Aaboud:2016uuk}. Concretely, we translate the ATLAS constraints from \cite{Aaboud:2016uuk} to construct the 95\% exclusion regions in some of the EChL parameter space for each of the considered unitarization methods, giving, at the same time, the total theoretical uncertainty driven by the  variety of these methods.

\section{Unitarity Violation and experimental status of WZ scattering}

In Chapter \ref{EChL} we reviewed precisely the example of the WZ scattering in what concerns to the violation of unitarity coming from the EChL operators. Specifically, in \figref{fig:unitviolai}, we studied the relevance of the chiral coefficients in the unitarity violation of this observable at the subprocess level. The main conclusion extracted from this Figure was that only $a_4$ and $a_5$ play a relevant role regarding the violation of unitarity in this context. Therefore, in this Chapter we will consider only these two chiral parameters to present our main conclusions, although some discussion about other coefficients will be provided. 

Since our final aim here is to reinterpret the given experimental constraints on these two parameters, it is important to revisit the current bounds imposed on them. In Chapter \ref{EChL}, we presented already the most stringent bounds available, extracted from the combination of searches in different VBS channels. Furthermore, in Chapter \ref{VBS}, we reviewed the experimental status of VBS observables at the LHC. 
There we pointed out two experimental works, one by ATLAS \cite{Aad:2019xxo} and another one by CMS  \cite{Sirunyan:2019der}, that correspond to the most recent VBS searches available, carried out with $\sqrt{s}=13$ TeV data.  In the former a maximum total cross section of various VBS processes, and, therefore, a model independent experimental study is reported, whereas in the latter direct bounds on the linear counterparts of some EChL parameters are provided. 

These bounds  are obtained without unitarizing the EChL (or, in those references the linear EFT) predictions at all, through a combined study of different VBS channels and analyzing the effect of each parameter at a time. One should keep in mind that these values for the $a_4$ and $a_5$ bounds might be overestimated, since the issue of the violation of unitarity has been neglected in the corresponding study. 

\begin{figure}[t!]
\begin{center}
\includegraphics[width=0.5\textwidth]{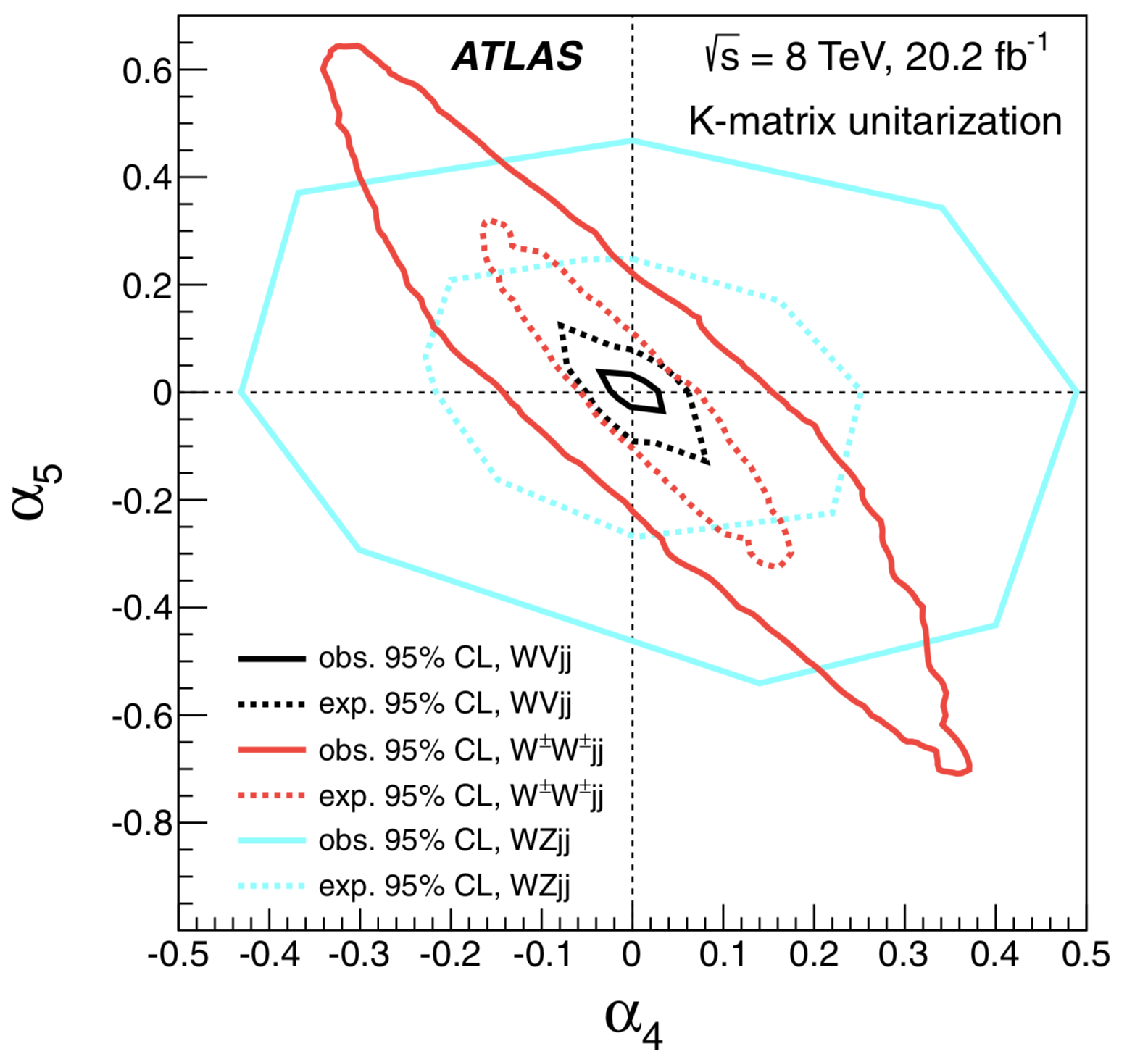}
\caption{The observed $[a_4,a_5]$ 95\% C.L. region for W${}^{\pm}$W${}^{\pm}$ final state (solid red contour), for WZ final state (solid cyan contour) and for the combined analysis (solid black contour) observed by the ATLAS collaboration interpreting the data using the K-matrix unitarization at $\sqrt{s}=8$ TeV and $L=20.2~{\rm fb}^{-1}$. The expected confidence regions are shown as well. Figure borrowed from \cite{Aaboud:2016uuk}, where the notation of  
$[\alpha_4,\alpha_5]$ is used instead of $[a_4,a_5]$.}
\label{fig:elipseATLAS}
\end{center}
\end{figure}

Nevertheless, we have chosen to study the WZ channel only as a first approach to the issue of extracting information from the LHC data about the EFT characteristics. Therefore, we need specific analyses devoted to this channel. In this sense, the result provided in  \cite{Aaboud:2016uuk, Sirunyan:2019ksz} represent the most up to date constraints on $a_4$ and $a_5$ in WZ scattering. In the former, performed with $\sqrt{s}=8$ TeV data, a K-matrix unitarization analysis, following the procedure proposed in \cite{Alboteanu:2008my}, is carried out. With this prescription, the EChL $[a_4,a_5]$ parameter space is constrained, as it is shown in \figref{fig:elipseATLAS}, borrowed from \cite{Aaboud:2016uuk}. We will rely mainly upon this experimental search of \cite{Aaboud:2016uuk} as a first example since it already involves a specific treatment of the violation of unitarity in WZ observables. Besides, as the overall constraints imposed in the EChL parameters in this study are of the order of $a_4\sim a_5\sim0.01$, we will use these values as reference to illustrate different VBS features without loss of generality.

 With all these considerations in mind, we now move on to characterize the restoration of unitarity in WZ scattering. In the next subsection we will comment on the way in which we will implement the various unitarization methods presented in Chapter \ref{EChL} paying special attention at the fact that the unitarization condition couples the different helicity channels. We will present the different predictions these methods provide at the subprocess level in order to have a first insight of their impact in the WZ cross sections.  

\newpage


\section{Restoring Unitarity in WZ scattering}
\label{Unitarity}

In the previous pages we have stated that the EChL, and especially the operators governed by $a_4$ and $a_5$, lead to unitarity violatiing predictions for WZ $\to$ WZ scattering cross sections in the energy range accesible by the LHC. We have also discussed that in order to make the EFT testable at colliders, we need to solve this problem and obtain fully unitary results for the relevant observables. To this aim, unitarization methods are addressed: prescriptions to construct unitary scattering amplitudes from the raw, non-unitary, EFT predictions. A summary of these methods and their generalities was already presented in Chapter \ref{EChL}. In this subsection, therefore, we will just recall the basic aspects of these prescriptions and the concrete way in which we implement them. 

First of all, it is important to have in mind that in the case of non-resonant scenarios different unitarization methods can lead to outstandingly different predictions for diverse observables. This reinforces our hypothesis that, in order not to lose the appealing model independence of EFTs in the non-resonant case, the predictions given from the different unitarization methods available have to be contrasted, and a quantitative estimate of their differences should be provided. This inevitably introduces a theoretical uncertainty in the unitarized EFT predictions, which is precisely the one we want to quantify. Therefore, we will focus in the case in which new resonant states do not manifest in the energies we are going to explore at the LHC via VBS. Besides, if present,  they would also suppose a completely different experimental setup and search strategy, as we will see in the next Chapters.

Second of all, if we recall the unitarity condition given in \eqref{unitarity} that all unitarized amplitudes must fulfil, we see once again that the unitarity of a particular helicity channel does not depend just on itself but in other helicity amplitudes as well. This implies that considering only the most pathological of these amplitudes in terms of the violation of unitarity, could mean that we are neglecting important effects given the fact that the helicity system is coupled. In general, the most worrying helicity channel regarding the violation of unitarity will be the purely longitudinal one ${\rm W}_L{\rm Z}_L\to{\rm W}_L{\rm Z}_L$, due to its close relation with the strongly interacting Goldstone bosons. Thus, the larger the number of longitudinally polarized gauge bosons involved in the scattering, the lower the energy at which unitarity will be violated. In any case, the fact that the purely longitudinal helicity channel dominates at high energies and dominates the violation of unitarity depends on the particular setup that one is considering.

By studying the partial waves with the lowest values of angular momenta, $J=0,1,2$, for the 81 helicity channels independently and for different values of $a_4$ and $a_5$ and at different center of mass energies, one can disentangle the relevance of the purely longitudinal case with respect to the other helicity channels. These three lowest-order partial waves are the ones that should contain all the unitarity violating effects as it has already been explained in previous sections of this Thesis.

\begin{figure}[t!]
\begin{center}
\includegraphics[width=0.324\textwidth]{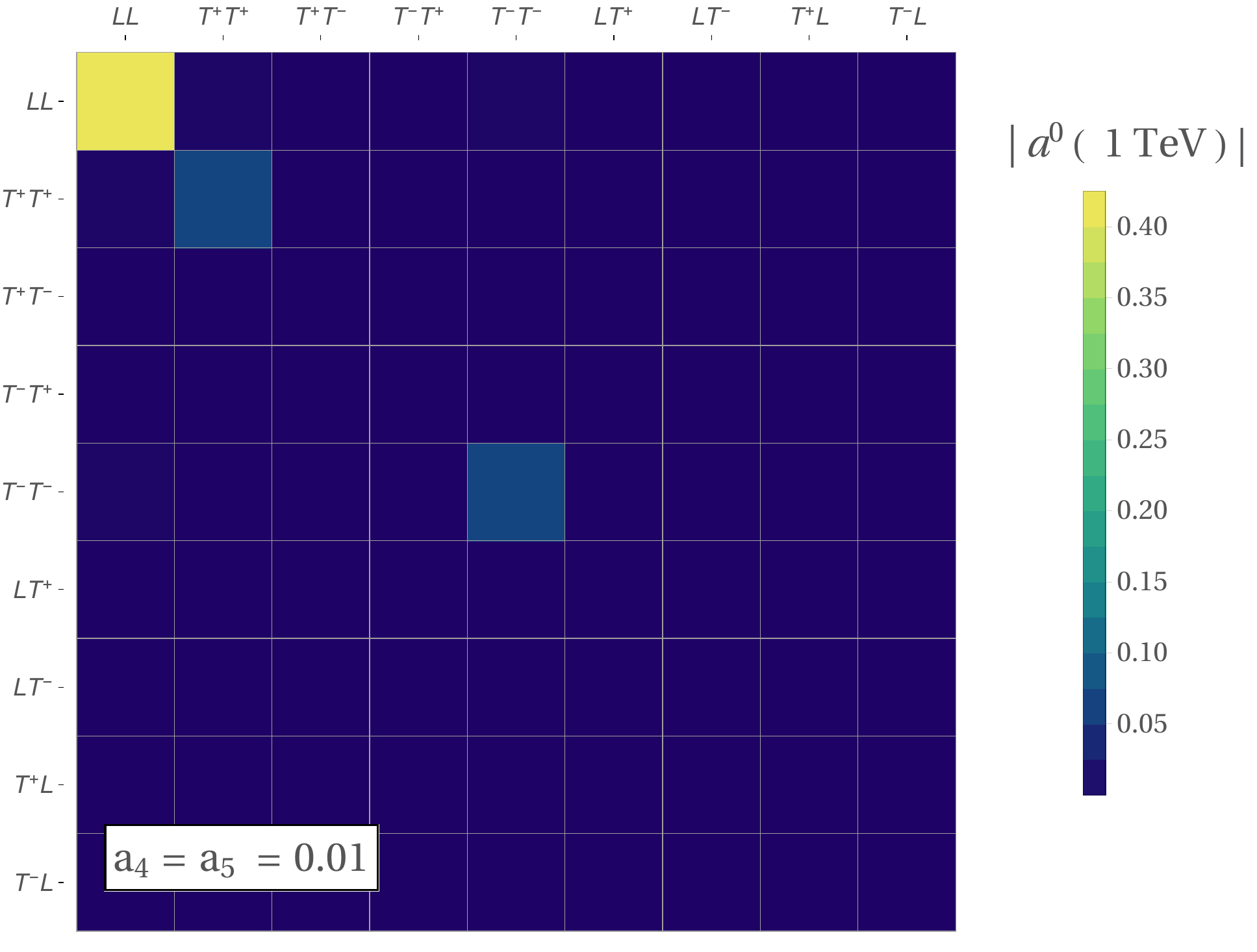}
\includegraphics[width=0.324\textwidth]{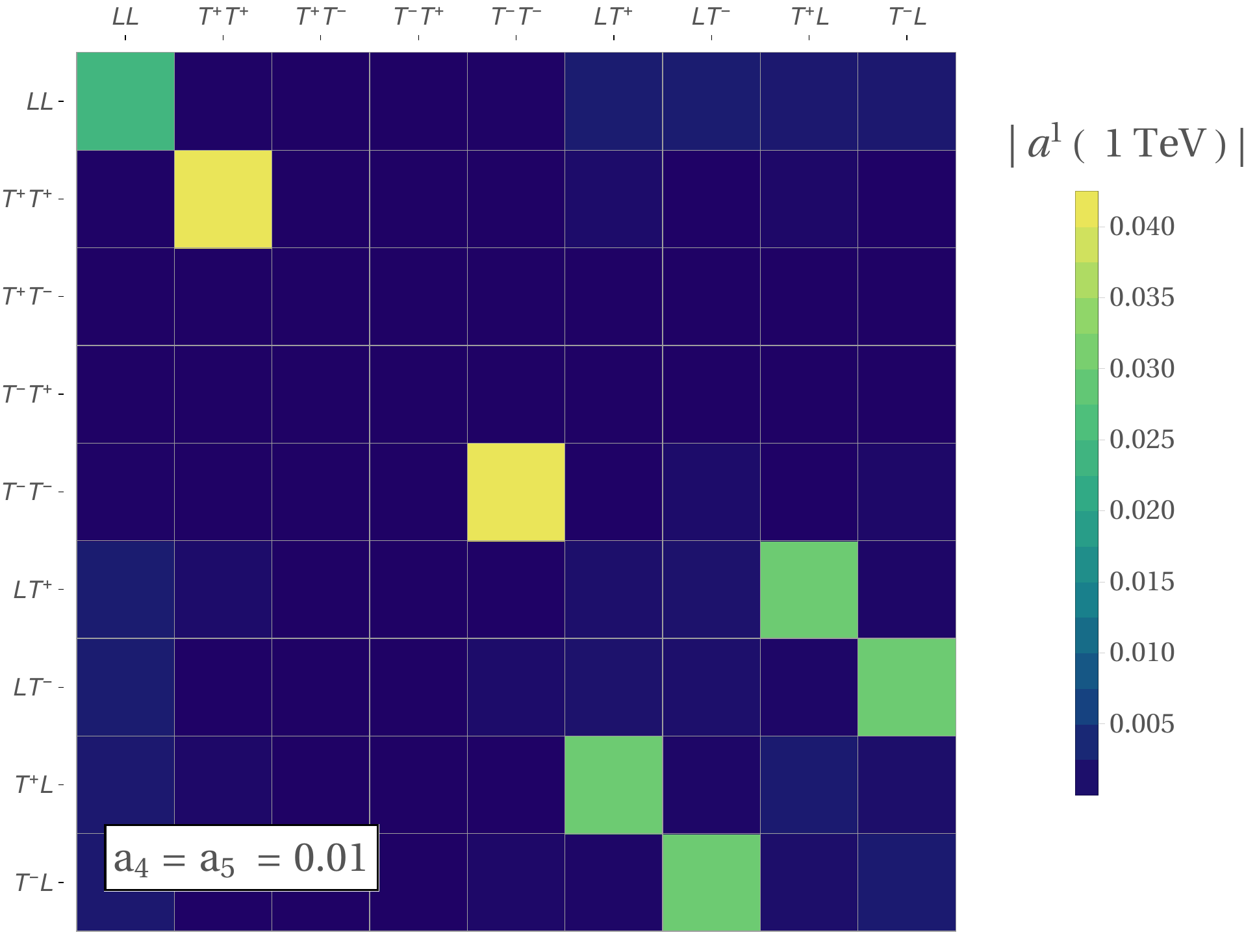}
\includegraphics[width=0.324\textwidth]{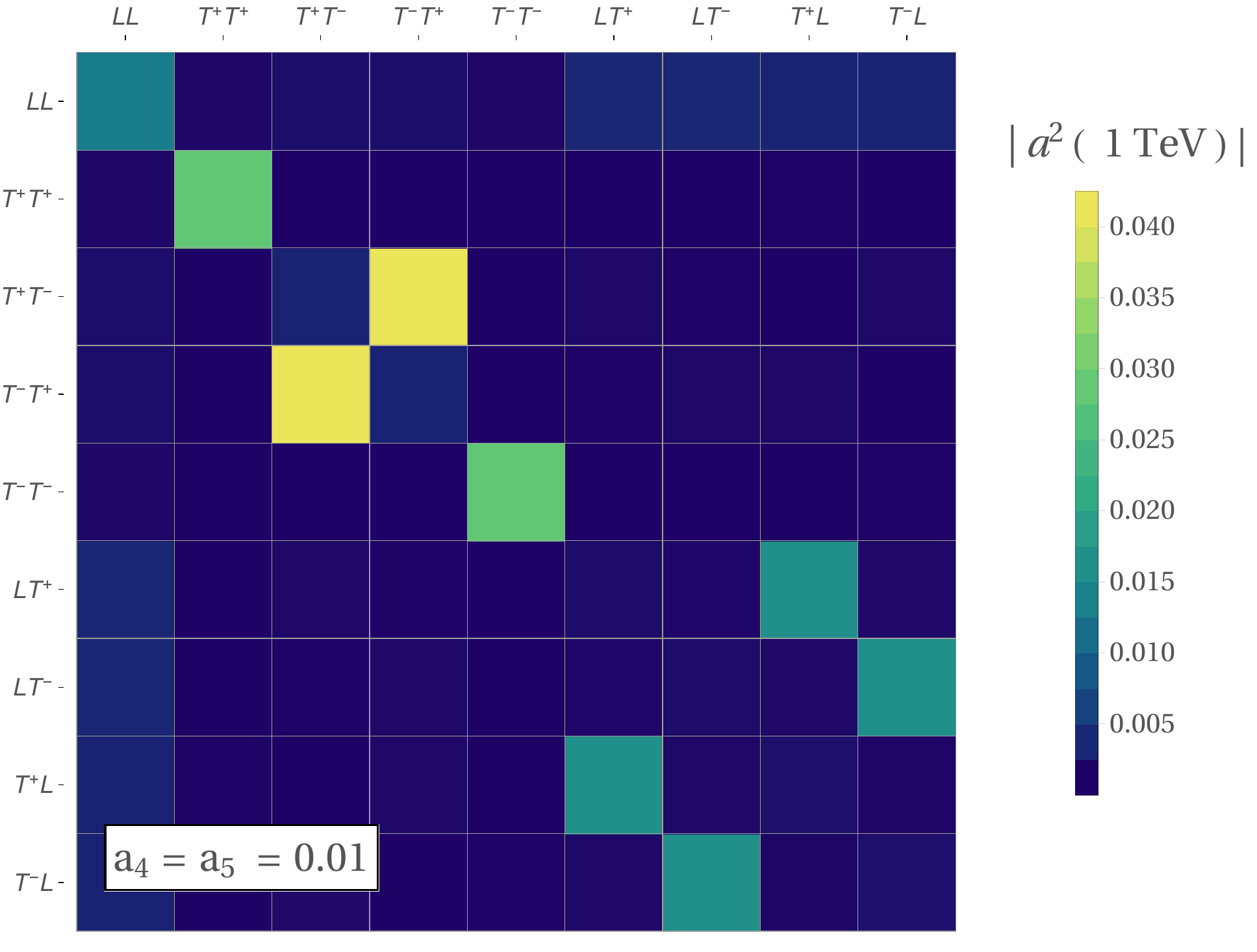}
\caption{Numerical values of the three lowest angular momentum partial waves $a^{J}(\sqrt{s})$ with $J=0$ (left), $J=1$ (middle), and $J=2$ (right), of the 81 helicity combinations of $W^+Z\to W^+Z$ scattering. Predictions are shown for a fixed center of mass energy of $\sqrt{s}=1$ TeV and for $a_4=a_5=0.01$ (with the other parameters set to their SM value) as reference. Incoming and outgoing states can be interpreted indistinctly since the results are presented in a symmetric way due to time-reversal invariance. The included labels of these 9 incoming WZ and 9 outgoing WZ states with two polarized gauge bosons, longitudinal ($L$) and/or transverse ($T^{+,-}$), are ordered and denoted here correspondingly by: $LL$, $T^+T^+$,  $T^+T^-$, $T^-T^+$ , $T^-T^-$, $LT^+$, $LT^-$, $T^+L$ and $T^-L$.}
\label{fig:teselas}
\end{center}
\end{figure}

With this in mind, we have calculated the absolute value of the three lowest-order partial waves for all the helicity channels at a certain center of mass energy and for a particular value of $a_4$ and $a_5$, in order to understand the implication of the different helicity amplitudes in the total cross section and in the unitarization process. In \figref{fig:teselas} we present an example of this for the reference values of $a_4=a_5=0.01$ and for a representative center of mass energy of 1 TeV.

 Looking at this figure, one can observe various interesting features. The first one is that, in general terms, the $J=0$ partial wave (left panel) is around one order of magnitude bigger than the other two, $J=1,2$ (middle and right panels, respectively), as we expected from the results in \figref{fig:violationunitarity}. The second one is that only for that same value of the angular momentum, $J=0$, the purely longitudinal scattering (displayed in the (1,1) entry of these ``matrices'', where incoming and outgoing states can be interpreted indistinctly since the results are presented in a symmetric way due to time-reversal invariance) dominates, being it a factor 5 larger than the next contributing helicity channel and thus becoming practically the only relevant amplitude to take into account.
 
 In the other two cases, $J=1,2$, the $LL\to LL$ amplitude is no longer dominating the picture and other helicity channels become important. In particular we see in this figure that $T^+T^+\to T^+T^+$ and $T^-T^-\to T^-T^-$ play a relevant role in $J=1$ and $T^+T^-\to T^-T^+$ and $T^-T^+\to T^+T^-$ do it in $J=2$. This points towards the fact that, in some setups and for determined values of the relevant chiral parameters, neglecting the unitarity-violating effects of other channels apart from the purely longitudinal one could lead to incomplete predictions. This is the reason why we will consider the whole coupled system of the 81 helicity amplitudes when applying the mentioned unitarization methods.
 
 The different unitarization prescriptions we are going to study in this Chapter are those presented in Chapter \ref{EChL}. There, we introduced five unitarization methods that could be classified in two categories: 1) the ones that directly suppress {\it by hand} the pathological energy behaviour of the amplitudes with energy (Cut off, Form Factor and Kink), and 2) the ones that unitarize the first three partial waves from which then the total unitary amplitude is reconstructed (K-matrix and IAM). It is important to recall that they differ in their physical implications and motivation and in their analytical properties, as we already saw in  Chapter \ref{EChL}, but, since they are the most commonly used ones in the literature nowadays we find pertinent to contrast their predictions. Nevertheless, and despite these differences and the fact that some of them could be more physically justified than others, there is in principle no prior to choose a particular method with respect to the others.  
 
Albeit already introduced in the first Chapter, it is worth reviewing again the five unitarization prescriptions considered in this Chapter with a brief explanation of each of them (for some illustrative reviews on different unitarization methods in the context of VBS see, for instance, \cite{Kilian:2014zja, Rauch:2016pai,DelgadoLopez:2017ugq}). Besides, we also present here the particular way in which we implement them.

\begin{itemize}
\item{\textbf{Cut off:}
The Cut off is not a unitarization method per se but a way to obtain unitary amplitudes by just discarding those predictions given for energy values above the unitarity violation scale $\Lambda$, defined in the previous section as the lowest value of $\sqrt{s}$ at which any partial wave crosses the unitarity bound stated in \eqref{unitlim}. This would mean to reject the predictions  of the cross sections marked with dashed lines in \figref{fig:unitviolai}, sticking only to those that respect the unitarity condition (i.e., solid lines in these figures).
}
\item{\textbf{Form Factor (FF):}
In this case, instead of obviating part of the results computed from the raw EFT, what is done is to suppress the pathological behavior of the amplitudes with energy above the scale at which each of them violate unitarity. To that purpose, a smooth, continuous function of the form:
\begin{align}
f_i^{FF}=(1+s/\Lambda_i^2)^{-\xi_i}\,,
\end{align}
is employed. Here $s$ is the center of mass energy squared, $\Lambda_i$ is the specific value of $\sqrt{s}$ at which the helicity channel $i$ violates unitarity according to \eqref{unitbound} and $\xi_i$ is the minimum exponent that is sufficient  to fix the pathological behavior of the corresponding $i^{th}$ helicity amplitude with energy. Thus, every non-unitary helicity amplitude will be unitarized in the following manner:
\begin{align}
\hat{A}_{\lambda_1,\lambda_2,\lambda_3,\lambda_4}=A_{\lambda_1,\lambda_2,\lambda_3,\lambda_4}\cdot(1+s/\Lambda_{\lambda_1,\lambda_2,\lambda_3,\lambda_4}^2)^{-\xi_{\lambda_1,\lambda_2,\lambda_3,\lambda_4}}\,,\label{FF}
\end{align}
with $\hat{A}$ being the unitary amplitude and $A$ the non-unitary EFT prediction. With all these unitarized amplitudes, then, one would be able to recover a unitary unpolarized, total cross section. In the present case, and for the values of the chiral parameters that are going to be probed in this work, the scales at which unitarity is violated for all helicity channels are above the maximum center of mass energy considered, except in the purely longitudinal case. We have checked that including the Form Factor suppression given in \eqref{FF} for all helicity channels (notice that not only the scale is different in each channel, but also the exponent since they depend differently with energy) is equivalent to do it just in the $LL\to LL$ one for the energies and parameters we are considering, so, for simplicity, from now one we will apply \eqref{FF} to the scattering of longitudinally polarized gauge bosons leaving the rest unchanged. In this way, our prescription to apply the Form Factor unitarization method can be summarized as:
\begin{align}
\hat{A}_{LLLL}=A_{LLLL}\cdot(1+s/\Lambda_{LLLL}^2)^{-2}\,,\label{FFLLLL}
\end{align}
recalling that any other helicity amplitude is left unaffected. The exponent has been set to $\xi_{LLLL}=2$ since it is the minimum value necessary to repair the anomalous growth with energy of the $LL\to LL$ amplitude. The scale $\Lambda_{LLLL}$ has been computed with the VBFNLO utility to calculate Form 
Factors~\cite{Arnold:2008rz,Arnold:2011wj,Baglio:2014uba}.
}
\item{\textbf{Kink:}
The so called Kink unitarization method is very similar to the Form Factor. Conceptually, it is the same, and the only difference present between both prescriptions is that the suppression in the Kink method is not performed smoothly, but with a step function:
\begin{align}
f_i^{Kink}=\left\{\begin{array}{l}1~~~~~~~~~~~~~s\leq\Lambda_i^2 \\(s/\Lambda_i^2)^{-\xi_i} ~~ s>\Lambda_i^2\end{array}\right.\,.
\end{align}
Except for this fact, the rest of the discussion regarding the Form Factor is equally valid for the Kink, so, in this case, we will also apply the method only to the $LL\to LL$ amplitude with an exponent of $\xi_{LLLL}=2$.
}
\item{\textbf{K-matrix:}
The K-matrix unitarization method has been extensively studied and implemented in the context of ChPT  in QCD. This method is a prescription applied to the partial wave amplitudes and basically projects the non-unitary ones into the Argand circle through a stereographic projection. This means that it takes a real, non unitary partial wave amplitude to which an imaginary part is added {\it ad hoc} such that the unitarity limit is saturated. For each helicity partial wave amplitude, this is achieved by using the following simple formula:
\begin{align}
\hat{a}^{J;{\rm K-matrix}}_{\lambda_1\lambda_2\lambda_3\lambda_4}=\dfrac{a^J_{\lambda_1\lambda_2\lambda_3\lambda_4}}{1-i\,a^J_{\lambda_1\lambda_2\lambda_3\lambda_4}}\,.
\end{align}
However, as we have already commented throughout the text, the unitarity condition implies that the whole coupled system of helicities has to be taken into account in our unitarization procedures. Thus, we solve this coupled system in terms of matrices, for which we construct a 9$\times$9 matrix, whose entries correspond to the 81 possible helicity amplitudes of the elastic WZ scattering we are studying, and we unitarize it using the K-matrix method. This way we have:
\begin{align}
\hat{\alpha}^{J;{\rm K-matrix}}=\alpha^J\cdot[1-i\,\alpha^J]^{-1}\,,
\end{align}
being $\alpha$ the 9$\times$9 matrix containing the whole system of helicity partial wave amplitudes. Now, what we need is to reconstruct, from these unitary partial waves, the complete scattering amplitude. To this aim, we substitute from the initial, non-unitary amplitude, the unitarity violating partial waves by their unitarized versions. As we have already explained in the text, these partial waves are those that correspond to $J=0,1,2$, so, what we do is to subtract these three partial waves from the total amplitude to then add the same partial waves after the K-matrix unitarization has been performed:
\begin{align}
\hat{A}_{\lambda_1\lambda_2\lambda_3\lambda_4}(s,\cos\theta)=&A_{\lambda_1\lambda_2\lambda_3\lambda_4}(s,\cos\theta)-16\pi\sum_{J=0}^2 (2J+1)\,d^J_{\lambda,\lambda'}(\cos\theta)\,a^J_{\lambda_1\lambda_2\lambda_3\lambda_4}(s)\,\nn\\
&+16\pi\sum_{J=0}^2 (2J+1)\,d^J_{\lambda,\lambda'}(\cos\theta)\,\hat{\alpha}^{J;{\rm K-matrix}}_{[\lambda_1\lambda_2\lambda_3\lambda_4]}(s)\,.\label{amprec}
\end{align}
Here we denote as $\hat{\alpha}^{J;{\rm K-matrix}}_{[\lambda_1\lambda_2\lambda_3\lambda_4]}(s)$ (in the rest of the formulas it is implicit that all partial waves depend solely on $s$) the element of the 9$\times$9 matrix that corresponds to the $\lambda_1\lambda_2\lambda_3\lambda_4$ polarization state.  In this way, we obtain a unitary amplitude in which we maintain all the fundamental properties introduced by all the  partial wave amplitudes, including those with higher $J>2$ that, since are not involved in the violation of unitarity, remain unaffected. The numerical computations in this K-matrix case and the next one, IAM,  have been performed with a private mathematica code developed  by us.  
}
\item{\textbf{Inverse Amplitude Method (IAM):}
The Inverse Amplitude Method is, probably, the most profoundly studied unitarization prescription considered in this work. It is very well known in the context of ChPT for pion-pion scattering, and its accuracy has been proved in various scenarios, like, for instance, in the prediction of the $\rho$ meson as an emergent resonance in these scattering processes}. It is based on the application of dispersion relations (bidirectional mathematical prescriptions allowing to relate the real and imaginary parts of complex functions) to the inverse of the partial wave amplitudes computed in the EFT framework. This unitarization procedure can be actually understood as the result of the first Pad\'e approximant derived from the chiral expansion series provided by ChPT. In practice, this method implements an approximate re-summation of loops with bubbles in the s-channel of the given scattering process. Therefore in the present context of the EChL it accounts for re-scattering effects in the scattering of the two EW gauge bosons, i.e., WZ in our chosen example,  which are not taken into account with the other unitarization methods. Notice that this makes sense in the context of a strongly interacting theory since these re-scattering contributions are not suppressed as in weakly interacting systems. 

In summary, if one starts with the typical result for a given partial wave amplitude from the chiral Lagrangian, given by the sum of the two contributions in the chiral expansion, one of order ${\cal O}(p^2)$ and the other one of order ${\cal O}(p^4)$ , the corresponding prediction of the IAM  leads to the following unitarized helicity partial wave amplitudes:
\begin{align}
\hat{a}^{J;{\rm IAM}}_{\lambda_1\lambda_2\lambda_3\lambda_4}=\dfrac{(a^{{(2)\,J}}_{\lambda_1\lambda_2\lambda_3\lambda_4})^2}
{a^{(2)\,J}_{\lambda_1\lambda_2\lambda_3\lambda_4}-a^{(4)\,J}_{\lambda_1\lambda_2\lambda_3\lambda_4}}\,,
\end{align}
where $a^{(2)}$ is the contribution to the partial wave amplitude computed with the operators from the $\mL_2$ Lagrangian (\eqref{eq.L2}) at the tree level, which is of order ${\cal O}(p^2)$ and $a^{(4)}$ is the contribution to the partial wave amplitude computed with the operators from the $\mL_4$ (\eqref{eq.L4}) at the tree level, plus the contribution computed with the operators from the  $\mL_2$ Lagrangian at one loop level, which are both of ${\cal O}(p^4)$. In the present work, since the computation of the complete one loop level amplitudes that enter in $a^{(4)}$ has not been performed yet due to the difficulty of the task, we will evaluate this here in an approximate way. Following the usual features in ChPT, we take 
the imaginary part of this contribution to be $\left|a^{(2)}\right|^2$ so that the unitarity condition is fulfilled perturbatively, and neglect the real contribution of the loops which are expected to provide a very small contribution, not being relevant for the present computation. 

Once again we encounter ourselves in the scenario in which we have a prescription to unitarize each helicity amplitude independently. However, we want to take the whole coupled system of helicities in full generality, as explained above. We construct once more the 9$\times$9 matrix $\alpha$, this time splitting it into its $\mO(p^2)$ and $\mO(p^4)$ contributions, that contains the 81 helicity amplitudes, and we unitarize it using the IAM in the following matricial manner:
\begin{align}
\hat{\alpha}^{J;{\rm IAM}}=\alpha^{(2)\,J}\cdot[\alpha^{(2)\,J}-\alpha^{(4)\,J}]^{-1}\cdot\alpha^{(2)\,J}\,.\label{matrixIAM}
\end{align}
At this point, to obtain a fully unitary amplitude, we use the same trick as in the K-matrix case, i.e., we replace the unitarity violating partial waves of the total amplitude by their IAM-unitarized version, following \eqref{amprec} with the only change of K-matrix$\to$IAM. It is pertinent to make now some comments regarding important differences between the IAM unitarization method and the rest we have considered. The IAM does not provide just unitary predictions, but also succeeds to get partial wave amplitudes with the appropriate analytical structure (for more details on this, see, for instance, \cite{DelgadoLopez:2017ugq}). This implies that it is the only method, amongst the ones studied in this work, that can accommodate dynamically generated resonances, since these appear as complex poles in the second Riemann sheet of the partial wave with the corresponding $J$ quantum number. This is in contrast to the unitarized partial waves with the K-matrix method that do not have such poles. These resonances are characteristic of strongly interacting theories, and appear naturally at high energies, such as in the case of low-energy QCD. Furthermore, it is worth commenting that, according to \cite{Delgado:2015kxa}, similar results as those obtained with the IAM regarding the appearance of dynamical resonances  are also provided by other alternative unitarization methods that lead to the proper analytical structure. Example of such methods are the N/D or the improved K-matrix, which for shortness, we have decided not to include here.    
Nevertheless, for the forthcoming study at the LHC, as we have already said, we are interested in studying the non-resonant case of the unitarized theory, so the differences among the various unitarization methods will come in terms of smooth deviations from the SM continuum {\it via} WZ scattering rather than from the appearance of peaks due to the emergence of resonances. It is important, though, to keep in mind that the IAM has some peculiarities regarding its structure and physical motivation, that differentiates it from the others.
\end{itemize}

\begin{figure}[t!]
\begin{center}
\includegraphics[width=0.49\textwidth]{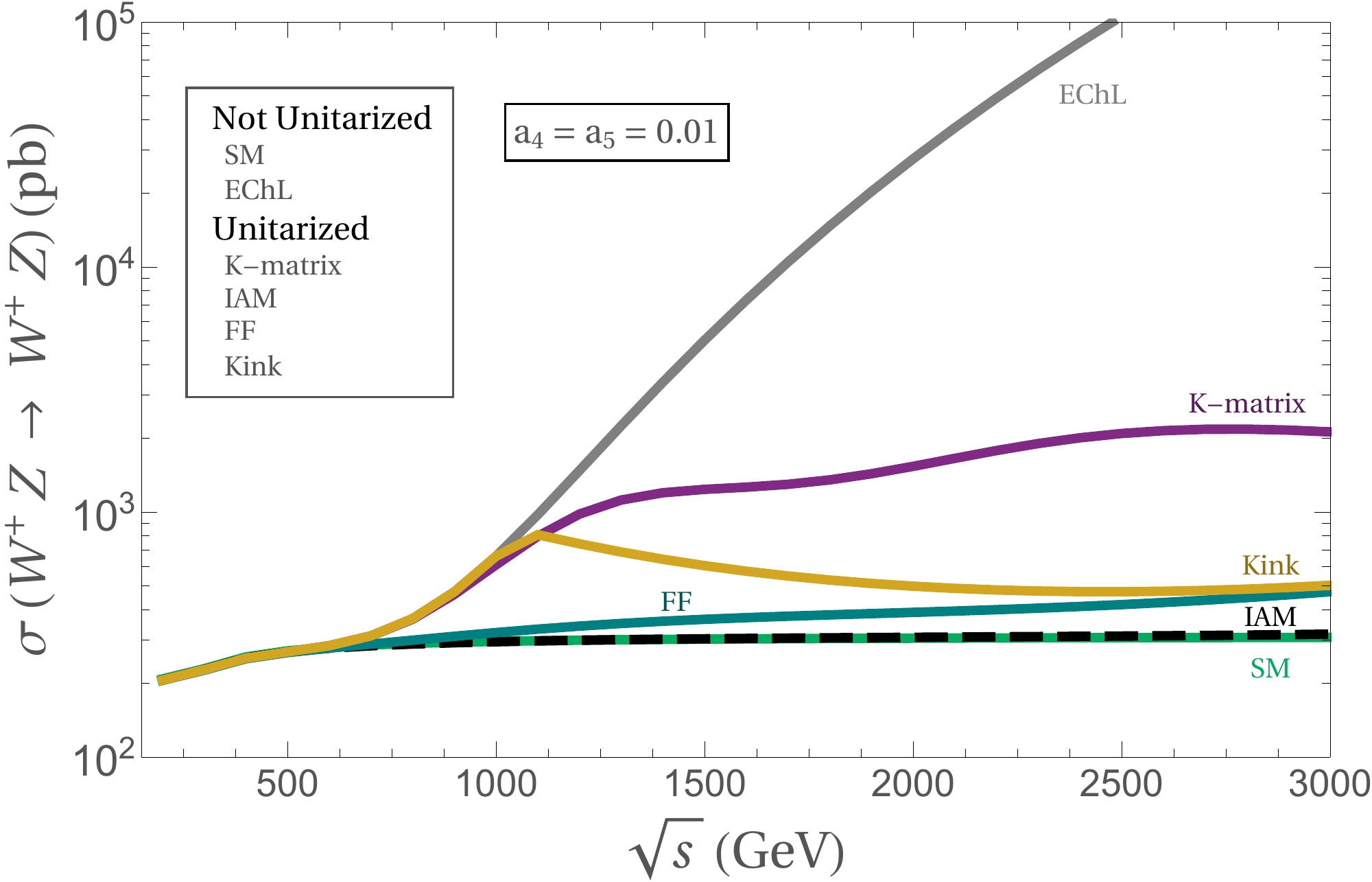}
\includegraphics[width=0.49\textwidth]{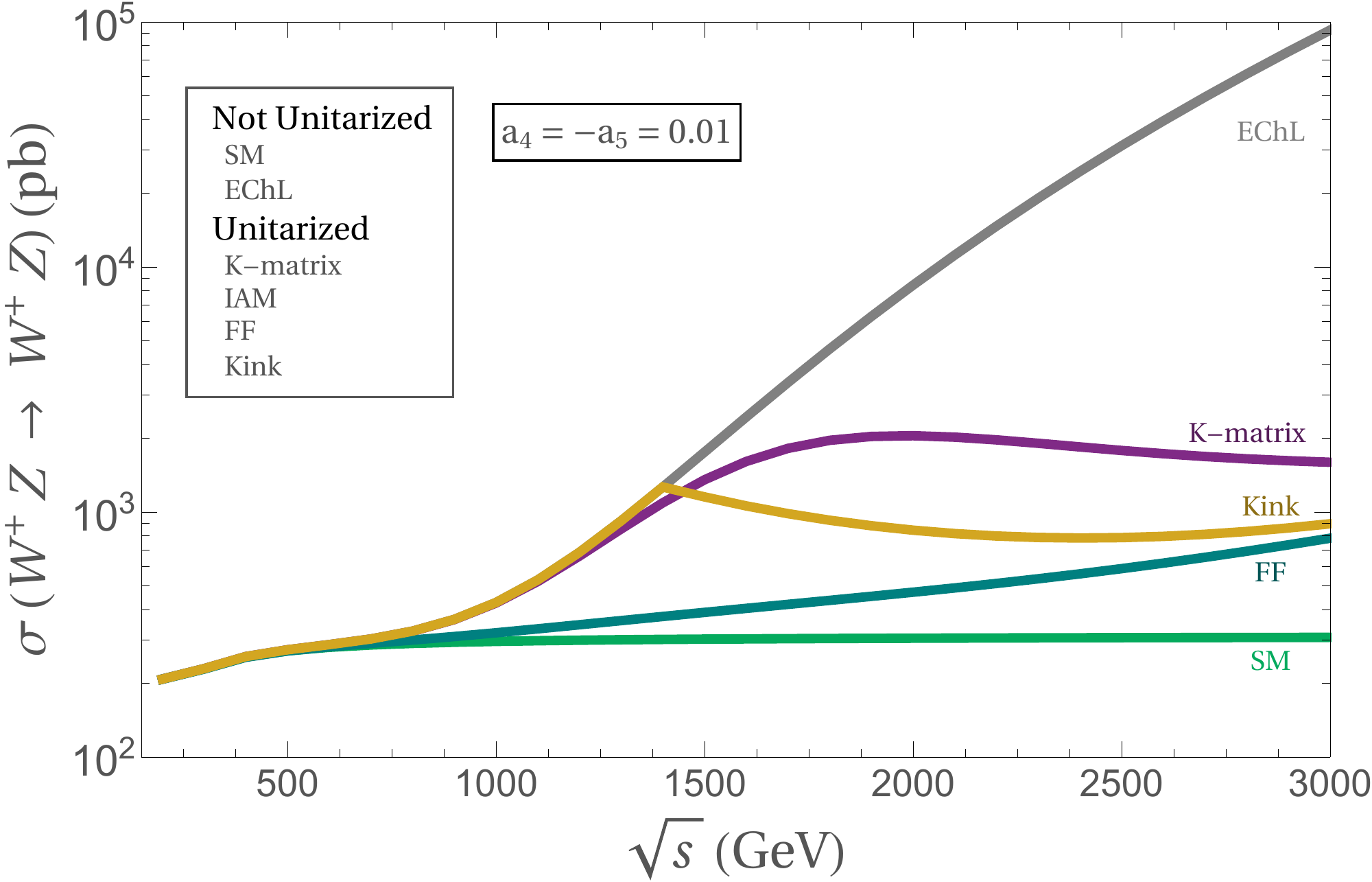}
\caption{Predictions of the total cross section of the process $W^+Z\to W^+Z$ as a function of the center of mass energy for the different unitarization procedures explained in the text: K matrix (purple), Kink (yellow), Form Factor (FF, blue) and IAM(dashed black). Non-unitarized EChL and SM are also displayed. Two benchmark $a_4,a_5$ values are displayed: $a_4=a_5=0.01$ (left) and $a_4=-a_5=0.01$ (right). In all plots $a=1$ (or, equivalently, $\Delta a=0)$. }
\label{fig:methodssubprocess}
\end{center}
\end{figure}

We have already discussed briefly each of the unitarization procedures that we consider in this Chapter and the specific way in which we implement them. Now, what we need is to study the different predictions they provide in regard of VBS observables. To that purpose, we apply each of them, in the way explained before, to the WZ$\to$WZ scattering amplitudes for different values of $a_4$ and $a_5$. The final results of this computation can be seen in \figref{fig:methodssubprocess}, where we present the total, unpolarized cross section of the WZ$\to$WZ scattering as a function of the center of mass energy for the different unitarization methods used. We also display the SM prediction and the non-unitarized EChL prediction for comparison. We consider, moreover, two scenarios for the values of $a_4$ and $a_5$. We set their absolute values to 0.01, as reference, and analyze two cases: the one in which both have the same sign $a_4=a_5=0.01$ (left) and the one in which they have opposite sign $a_4=-a_5=0.01$ (right).

 A great number of interesting facts can be extracted from these plots. Firstly, and most clearly, one can see that different unitarization methods lead, indeed, to very different predictions for this observable. These predictions differ, in general, from those of the raw EChL and the SM, as well. Therefore, one can expect that these differences might be very well seen experimentally. 
 
Secondly, another interesting issue arises from the comparison of both panels in the figure. It appears plainly that the value of the EChL prediction and of the  K-matrix prediction are, in general, smaller in the case in which both parameters,  $a_4$ and $a_5$, have opposite sign. On the contrary, the Kink and the Form Factor provide larger results in this same case. This means that for the non-unitarized prediction and for the K-matrix, the regions of the parameter space in which $a_4$ and $a_5$ have the same sign will be more constrained, whereas for the Kink and the Form Factor the opposite-sign regions will be the most constrained ones. We have checked that the predictions for the scenario in which both parameters are negative give the same results as the ones in which they are both positive, and that in the case in which they have opposite sign the same result is obtained when either of the parameters is positive/negative. 

Thirdly, a comment has to be made regarding the Cut off procedure. The unitarity violation scale is not explicitly shown in these plots, but it can be inferred from the position of the ``knee'' in the Kink prediction. As it is clear, discarding the values of the cross section above this scale will imply to lose a lot of sensitivity, and will of course correspond to a very different prediction with respect to the other studied cases. 
 
Regarding the IAM, we can clearly see that for the particular choice of parameters in the left panel of \figref{fig:methodssubprocess} its prediction lies very close to the SM one. In this case the IAM does not provide an emergent resonance in WZ scattering, since for these particular values of the EChL parameters there are not poles in the reconstructed total amplitude. As a consequence, the outcome provided by the IAM when applied to the LHC context will not show any departure from the SM continuum, whereas for the particular choice of parameters in the right panel there is indeed an emergent resonance below 1 TeV,  which we have decided not to include in this plot since it is already excluded by the present searches at the LHC.

\begin{figure}[t!]
\begin{center}
\includegraphics[width=0.49\textwidth]{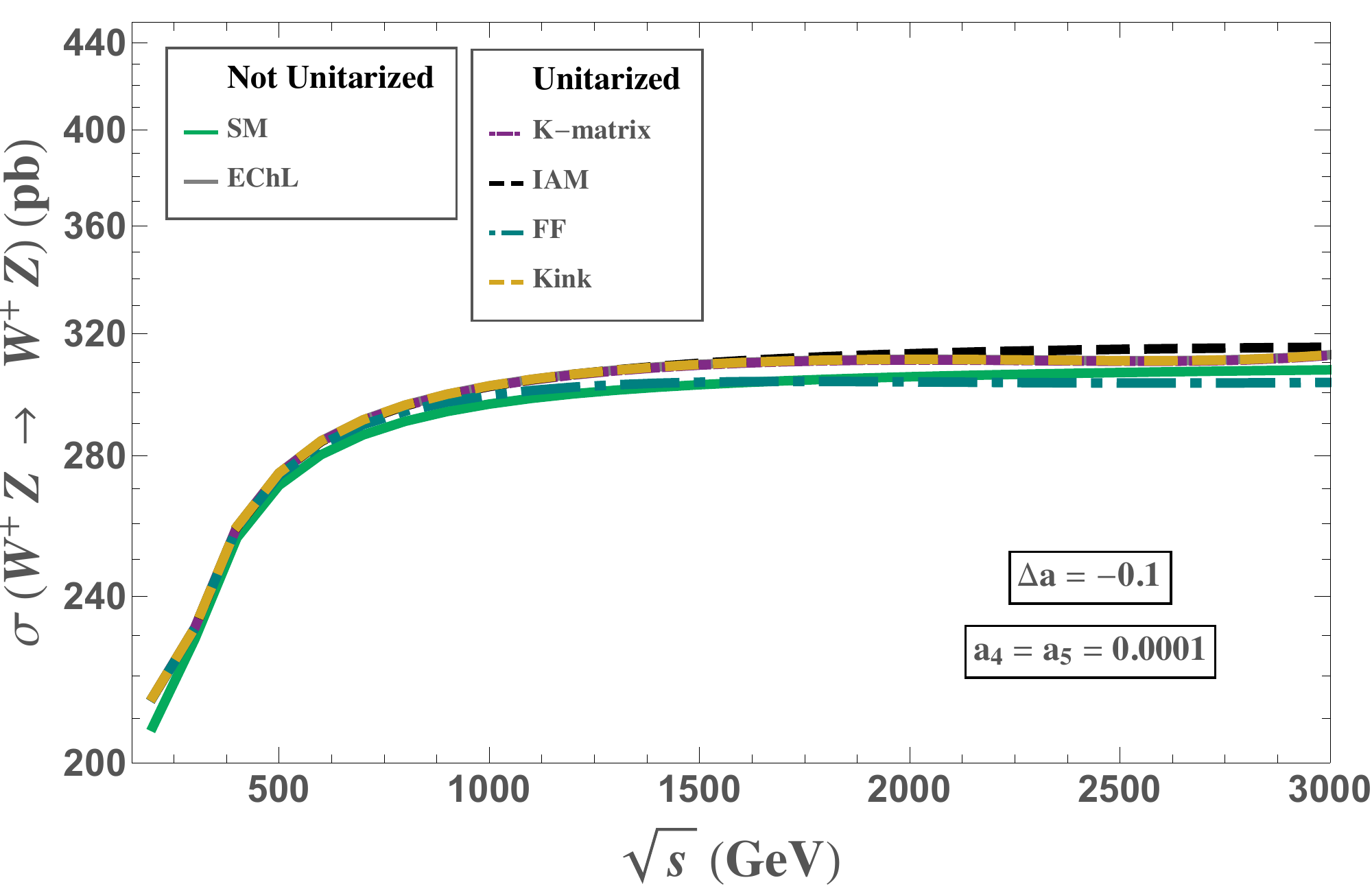}
\includegraphics[width=0.49\textwidth]{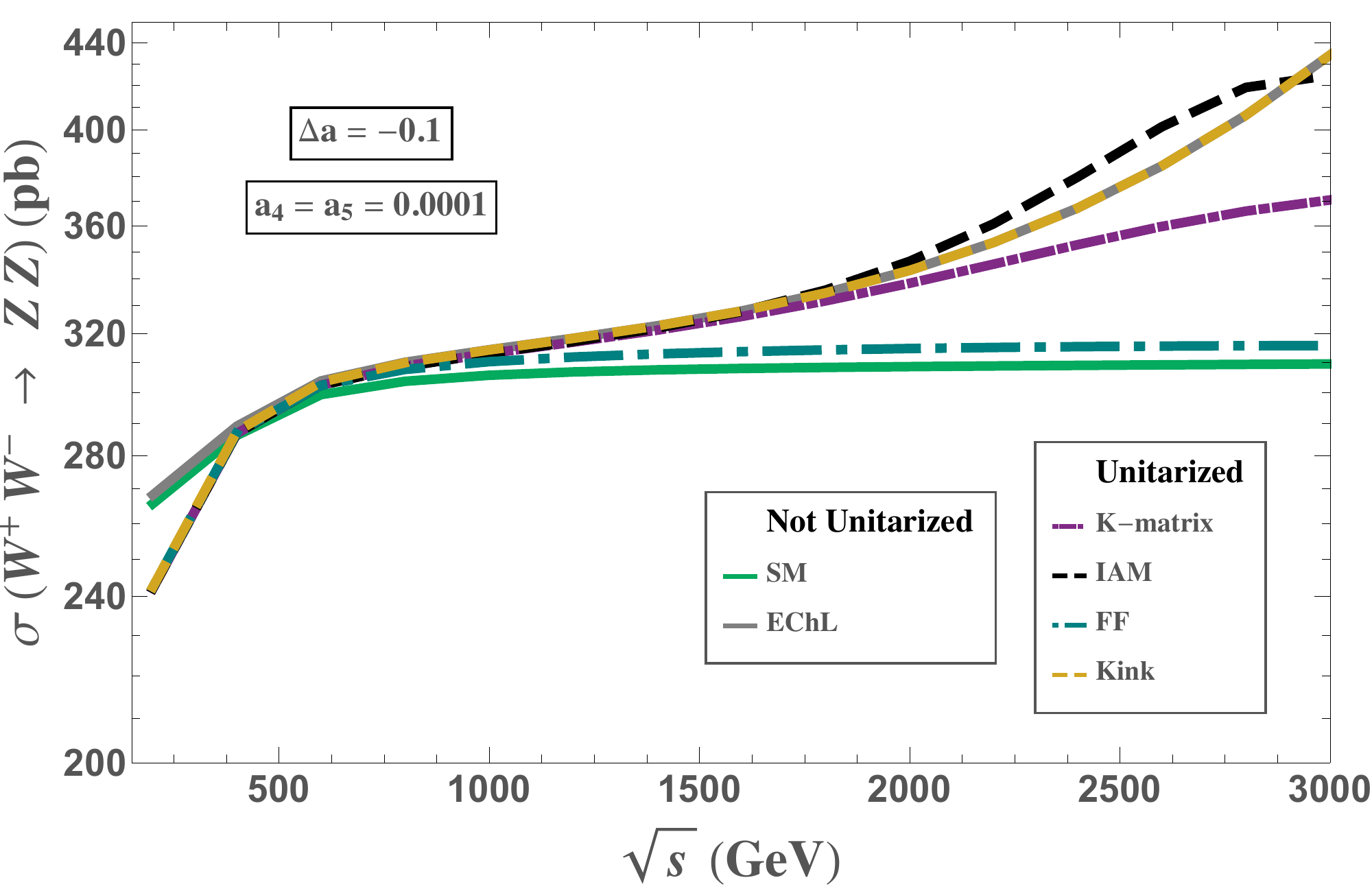}\\
\includegraphics[width=0.49\textwidth]{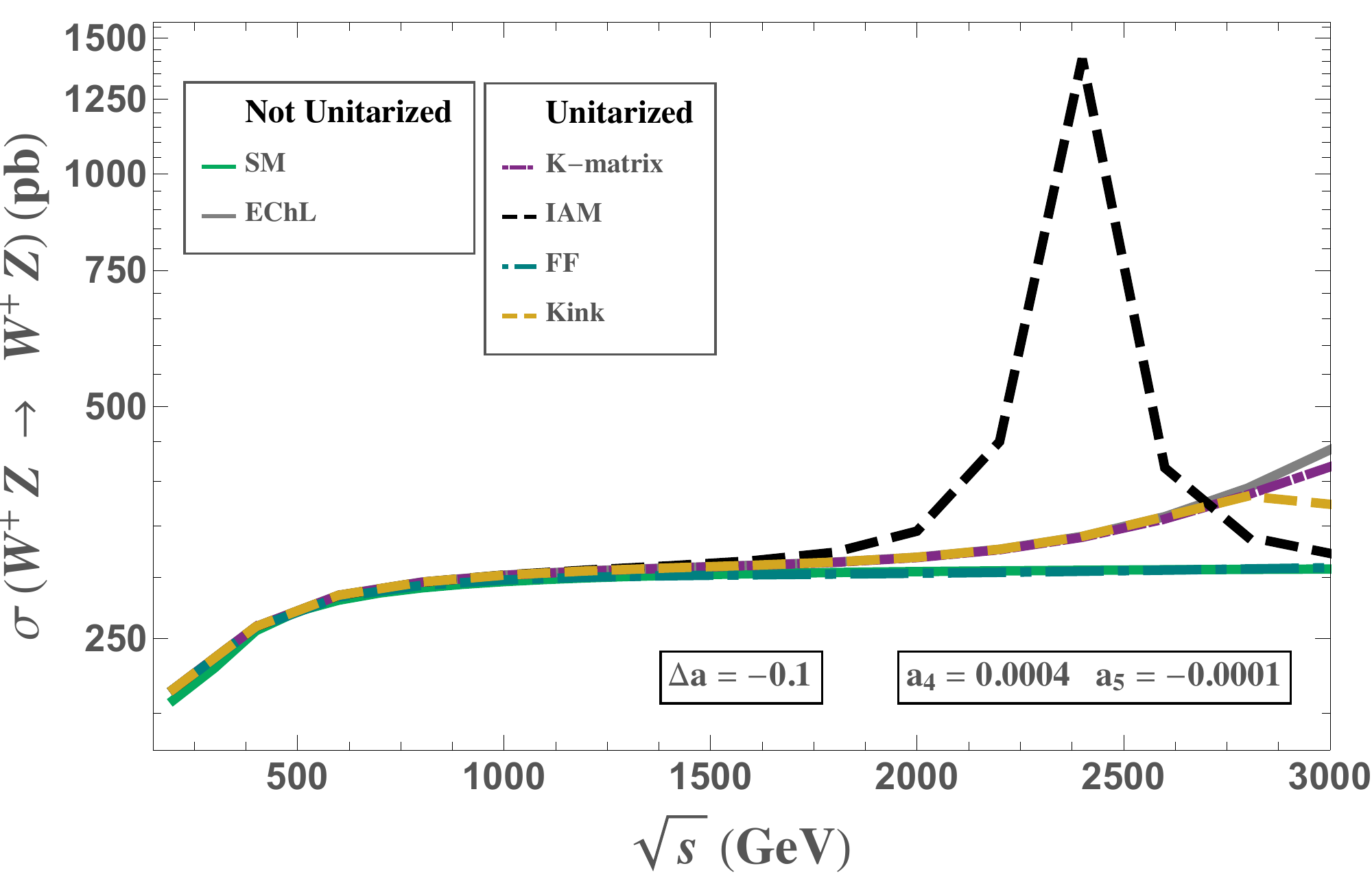}
\includegraphics[width=0.49\textwidth]{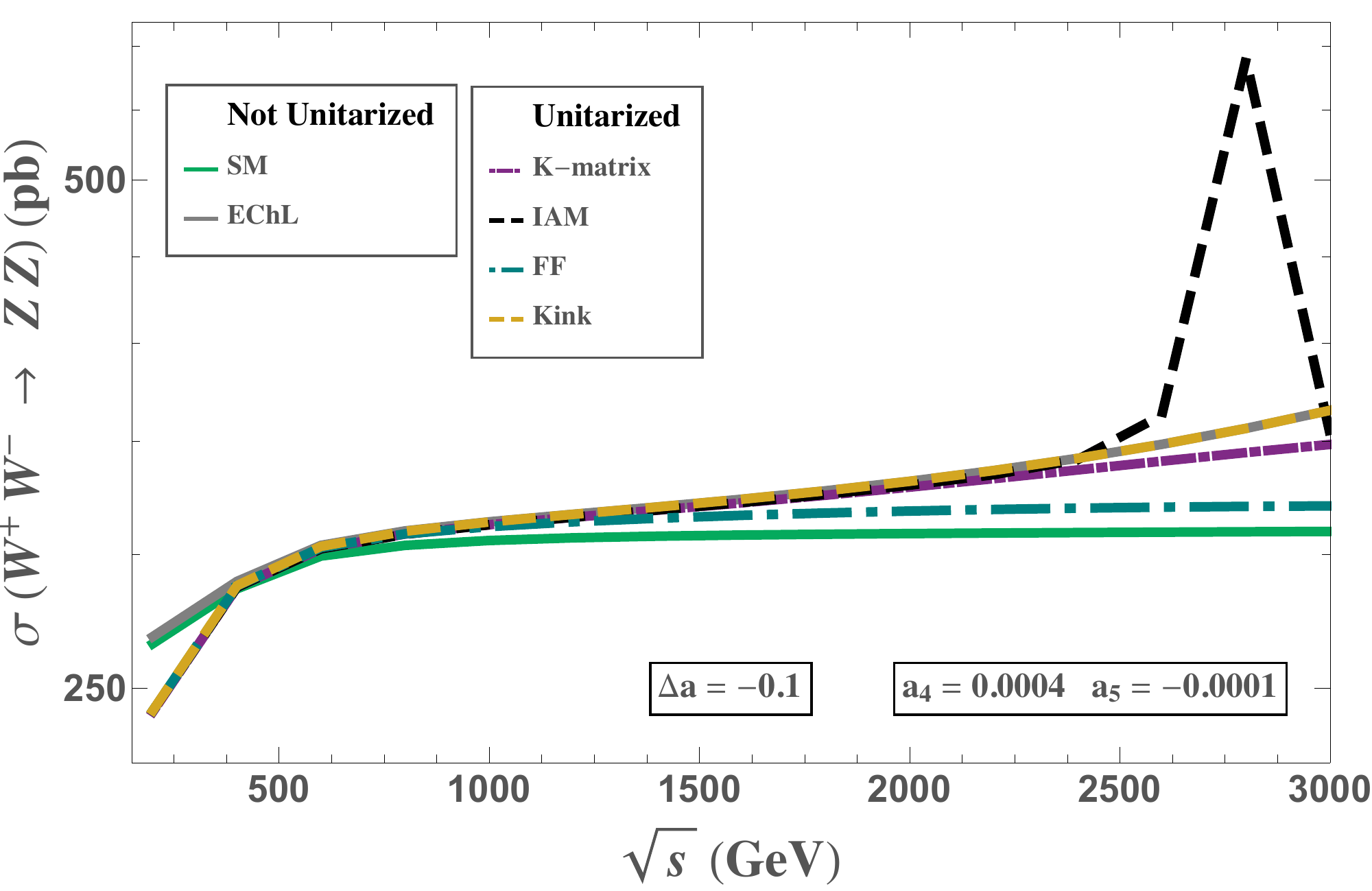}
\caption{Predictions of the total cross section of the process W${}^+$Z$\to$ W${}^+$Z (left panels) as a function of the center of mass energy for the different unitarization procedures explained in the text: K-matrix (purple), Kink (yellow), Form Factor (FF, blue) and IAM(dashed black). Non-unitarized EChL (gray) and SM (green) are also displayed. Two benchmark $a_4,a_5$ values are displayed: $a_4=a_5=0.0001$ (upper) and $a_4=0.0004,~ a_5=-0.0001$ (lower). For comparison, we include the plots corresponding to our predictions for same choice of the parameters but for the channel WW $\to$ ZZ (right panels). In all plots, $a=0.9$ (or, equivalently, $\Delta a=-0.1$.)}
\label{fig:comparisonwithwwzz}
\end{center}
\end{figure}

When other particular values of the EChL parameters are chosen,  different patterns in the predictions of the VBS cross sections from the various unitarization methods are obtained. In general, the choice of smaller values of $|a_4|$ and $|a_5|$ than those in  \figref{fig:methodssubprocess} typically leads, in the non-resonant case, to more similar predictions for the various unitarization methods in the studied energy range which are also closer to the SM prediction. This can be clearly seen in the upper left panel in \figref{fig:comparisonwithwwzz} where the parameters have been set to $a_4=a_5=0.0001$ and $a=0.9$ (or equivalently, $\Delta a= a-1=-0.1$). 

We chose to show these predictions as a complementary result for other EChL parameter values, since it can be interesting to see the roles of different coefficients in the present framework. For the particular choice we have made,  a scalar resonance emerges close to 3 TeV in the IAM unitarized predictions, which does not manifest in the channel of our interest here $WZ \to WZ$ but in the $WW \to ZZ$ channel. This can be seen clearly in the plot of the upper right panel  in   \figref{fig:comparisonwithwwzz}. In this case, studying this VBS channel, $WW \to ZZ$, at the LHC seems more appropriate in order to  analyze the distortions with respect to the SM predictions due to BSM physics represented by this particular choice of parameters. This means that a complete, combined study of the different VBS channels is on demand in order to fully explore the whole parameter space of the EChL properly. In this Thesis, however, to present a first approach to the issue that is aimed to be complemented in the future. 

The other example included in \figref{fig:comparisonwithwwzz}, where the parameters are set to $a_4=0.0004$, $a_5=-0.0001$ and again $a=0.9$, displays the emergence of a vector resonance in the IAM prediction for $WZ \to WZ$ (lower left panel) close to 2500 GeV, and a scalar resonance close to 2800 GeV in the IAM prediction for $WW \to ZZ$ (lower right panel). This resonant behaviour is only found in the predictions with the IAM but not in the predictions with the other unitarization methods, as it was expected due to the special IAM characteristics discussed previously.   

In summary, regarding the IAM, the appearance of dynamically generated resonances in the energy range of a few TeV occurs indeed for a continuum set of $a_4$ and $a_5$ values of the order of ${\cal O}(10^{-3}-10^{-4})$ and its properties, mass and width, also depend on the other relevant parameters, particularly on $a$. These features of the IAM have been studied extensively in the literature and are not the main focus of the present Chapter which, as we have said, is mainly devoted to the non-resonant case. Thus, for the rest of this work we will focus on the other unitarization methods which will produce instead smooth distortions from the SM continuum. The resonant scenario will be studied in depth in the forthcoming Chapters.
 
A significant point has to be made at this stage concerning the K-matrix cross sections, since they are the ones we will use in the next section as a link to the experimental results. We have compared our estimates, obtained with the K-matrix procedure explained in the pages above, with the ones provided by the Wizard group~\cite{Kilian:2014zja,Alboteanu:2008my}. In the given references, the authors construct unitarized four point functions that can be introduced in a Monte Carlo event generator. Their prescription is based in the T-matrix unitarization method, that they implement in a similar way than us: replacing the unitarity violating partial wave amplitudes of the total amplitude by their T-matrix unitarized version\footnote{In the present case, their T-matrix unitarization is equivalent to our K-matrix unitarization, as we have explicitly checked.}. This prescription is used, actually, by the ATLAS collaboration in order to constrain the EChL parameter space~\cite{Aaboud:2016uuk}. Nevertheless, their work is based on the ET, and they unitarize all the helicity amplitudes using the ET calculation, valid only to describe the longitudinally polarized gauge bosons at high energies. Thus, given this difference between their method and ours, we consider pertinent to make some comments about the discrepancies we have found.

Our predictions match those of the Wizard group for all the $LL\to LL$ amplitudes we have considered, i.e., for all the studied energies and values of the chiral parameters. However, there are some regions of the parameter space in which the cross sections of the other helicity channels differ. In the case in which the purely longitudinal scattering dominates at high energies, both procedures give rise to the same values for the cross sections. If other helicity channels have important contributions to the total cross section, we obtain different predictions. This can be the case if the values of $a_4$ and $a_5$ are very small, of the order of, for instance, $10^{-4}$. The authors in~\cite{Kilian:2014zja,Alboteanu:2008my} themselves comment on the limitations of their approach in this regime, so we are proposing here a way to avoid these limitations.

 In this subsection we have seen that different unitarization methods lead to very different predictions for the values of the cross section of the elastic WZ scattering. For this reason one can expect that the translation of these results to the LHC scenario would also show the different behaviours present at the subprocess level. Precisely because of this, the experimental measurements and constraints interpreted using one method or another will be different, and this difference can be understood as a theoretical uncertainty which is precisely the one that we want to quantify in this Chapter. Thus, in the next subsection, we will present our results for the LHC, and we will give an estimate of this uncertainty in the experimental determination of $a_4$ and $a_5$ due to the unitarization scheme choice.

%
 
 \section[Parameter determination uncertainties due to unitarization choice]{Parameter determination uncertainties at the LHC due to unitarization scheme choice}
 \label{MethodsLHC}
 
 In the previous subsection we learnt that the predictions for WZ scattering observables computed in the EChL framework can be very different depending on the unitarization method we apply to them.  This was manifest at the subprocess level, but now we want to study and quantify these deviations as they would be seen at the LHC. 
 
 In order to compute the total cross section at the LHC we have used the simple tool provided by the Effective W Approximation ~\cite{Dawson:1984gx,Johnson:1987tj} and we have compared this approximate result with the full result from MG5~\cite{Alwall:2014hca,Frederix:2018nkq}. The EWA, as we have already said, is the translation to the massive EW gauge bosons case of the Weisz\"{a}cker- Williams  approximation for photons \cite{vonWeizsacker:1934nji,Brodsky:1971ud}, and has the advantages of having the intuitive physical interpretation of the distribution of probability functions of the W and the Z as the PDFs in the parton model, and of being very simple to implement computationally. 
 
There are several studies in the literature that use the EWA to obtain reliable estimates. However, not all of them employ the same  probability functions. For this Thesis, we have considered and compared four of these implementations of the EWA: 1) the original EWA functions given in \cite{Dawson:1984gx}, selecting first the Leading Log Approximation (LLA) formulas (eqs. 2.19 and 2.29 in \cite{Dawson:1984gx}); and second 2) the improved ones which go beyond the LLA by keeping ${\cal O}(M_V^2/E^ 2)$ corrections, with $M_V$ the EW gauge boson mass and $E$ the energy of the initial quark (eqs. 2.18 and 2.28 in  \cite{Dawson:1984gx});  3) the EWA functions derived from \cite{Johnson:1987tj}; and 4) the simplified functions of the beyond LLA given in \cite{Alboteanu:2008my}.

 In principle, all should lead to similar results for the $pp\to$WZ+X process, and they do at high invariant masses of the final diboson system. Nevertheless, they differ quite a lot at lower energies. It is worth mentioning that to compute the  $pp\to$WZ+X rates with the EWA, one has to consider the contributions from two different subprocesses: the intermediate state with a W and a Z radiated from the initial protons that then scatter, and, in addition, the case in which a W and a photon are radiated and then scatter. The latter is of great importance in the low energy region where it dominates indeed over the other one. For the photon case we have used the well established probability function of the Weisz\"{a}cker- Williams approximation~\cite{vonWeizsacker:1934nji,Brodsky:1971ud}.
 
 In order to select the most accurate  probability function for the EW gauge boson case among the ones available in the literature, we have compared the results of the above four mentioned approaches  to the full results for the complete process $pp\to$WZ+X  obtained using MG5. Notice that for this comparison  we have generated MG5 events of the exclusive process $pp\to$WZ$jj$, which automatically contain all the topologies, i.e., the VBS topologies and all the others contributing to the same order in perturbation theory. Besides, in order to compare properly both results, the MG5 one and the EWA one, one has to set particular kinematical cuts on the final state particles.  In particular, as it is well known, in order to regularize  the Coulomb singularity produced by the diagrams with a photon interchanged in a  t-channel, some minimal cuts have to be imposed on the final particles.  Concretely, for this quantitative comparison of the total cross sections we give the following cuts on the transverse momentum and pseudorapidity of the final gauge bosons V and jets $j$, and the angular separation among the jets:
\begin{align}
|p_{T_V}|&> 20 ~{\rm GeV};~~|\eta_V|<2\,,\nn\\
|p_{T_j}|&> 5 ~{\rm GeV};~~|\eta_j|<10\,,~~\Delta R_{jj}>0.1\,,\label{cuts}
\end{align}
both in the EWA and MG5 for the cuts concerning the gauge bosons, and in MG5 events only for the ones concerning the extra jets. 

\begin{figure}[t!]
\begin{center}
\includegraphics[width=0.6\textwidth]{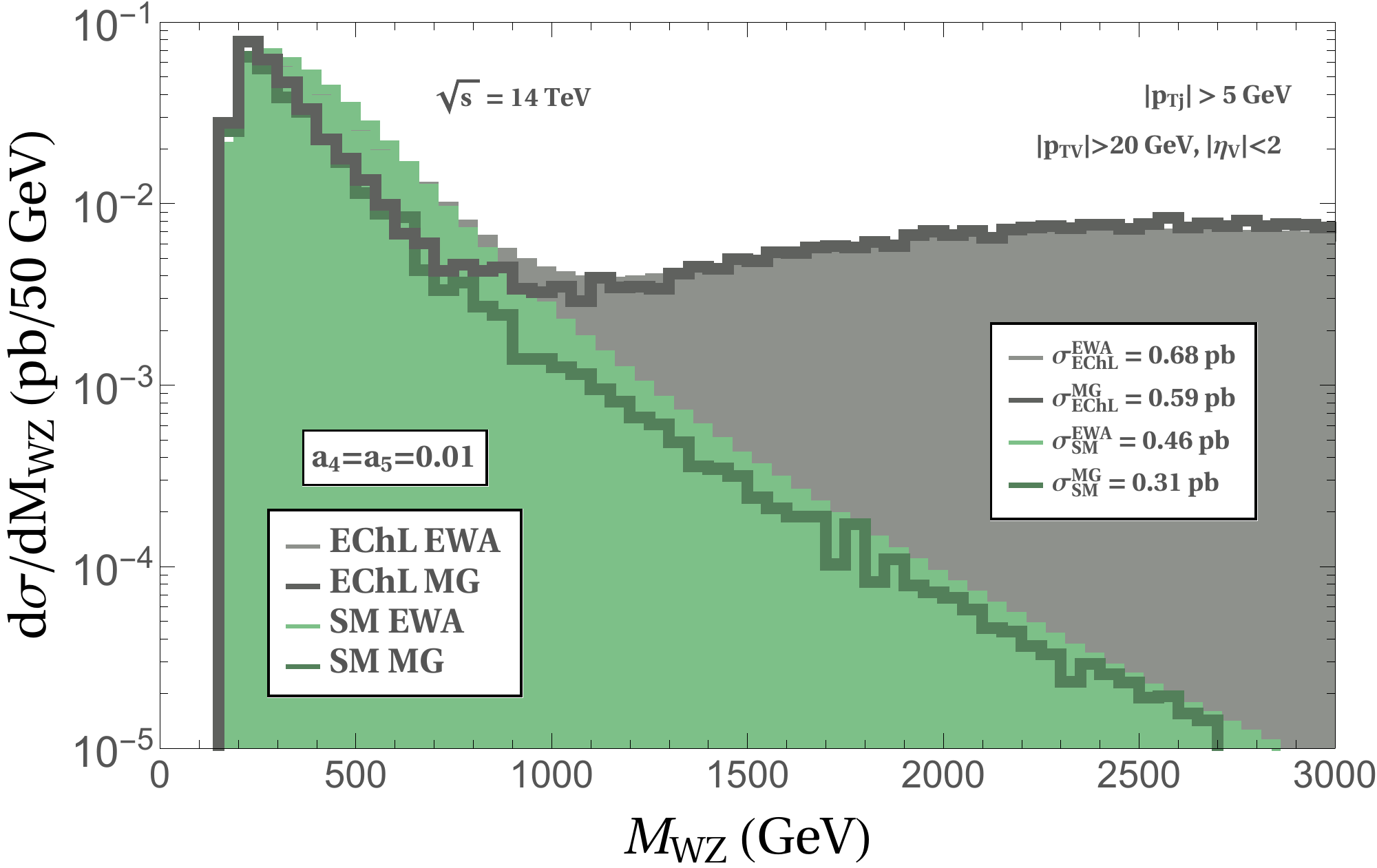}
\caption{Predictions of the differential unpolarized cross section of the process $pp\to$ WZ+X as a function of the invariant mass of the final WZ computed with the EWA (eqs. 2.18 and 2.28 in  \cite{Dawson:1984gx}). SM values (green) and EChL values for $a_4=a_5=0.01$ (gray) are shown. The other chiral parameters are set to their SM value. The MadGraph prediction of $pp\to $WZ$jj$ events with $|p_{T_j}|>5$ GeV is included as a solid line of each corresponding color for reference. All predictions are computed applying the cuts in \eqref{cuts}  and at $\sqrt{s}=14$ TeV.}
\label{fig:EWAcheck}
\end{center}
\end{figure}

With these considerations in mind, we have compared quantitatively the predictions of the $pp\to$WZ+X processes in the SM and in the EChL for $a_4=a_5=0.01$ within the EWA for the four probability functions considered, against the MG5 computation of the $pp\to$WZ$jj$ events. From our numerical comparison we have reached the conclusion that the original,  improved probability functions given in  \cite{Dawson:1984gx} are the ones that better match the MG5 prediction. The others overestimate the probability of radiating a EW gauge boson at low fractions of momentum of the initial quarks, thus missing the correct prediction of the cross section at low energies where most of events lie.

In \figref{fig:EWAcheck}, we display the results of the differential cross section distribution with respect to the invariant mass of the final gauge bosons, computed in the SM (green) and in the EChL (gray) for $a_4=a_5=0.01$ using the EWA and employing these  improved probability functions. We also show the MG5 prediction for these same distributions as solid, darker lines of each corresponding color, as well as the total cross sections obtained with both procedures. Cuts in \eqref{cuts} have been required, if applicable, and center of mass energy has been set to $\sqrt{s}=14$ TeV, as it will be considered for the rest of the Thesis. Regarding the comparison shown in this figure, it is manifest that the EWA works remarkably well, specially at high invariant masses. Not only the total MG5 cross section is recovered within a factor 1.5 at the worst in the SM case and 1.15 in the EChL case, but also the invariant mass distributions match considerably well. 

Now that we have checked that our computations obtained with the EWA employing the  improved probability functions provide reliable predictions of $pp\to$WZ+X observables, we move on to characterize the behaviour of the different unitarization methods at the LHC. To that purpose, we have convoluted the subprocess cross sections of each of the studied unitarization methods, corresponding to the different curves in \figref{fig:methodssubprocess}, with the EW gauge bosons probability functions and with the CT10 set of PDFs  \cite{Lai:2010vv}, evaluated at $Q^2=M_W^2$.

\begin{figure}[t!]
\begin{center}
\includegraphics[width=0.49\textwidth]{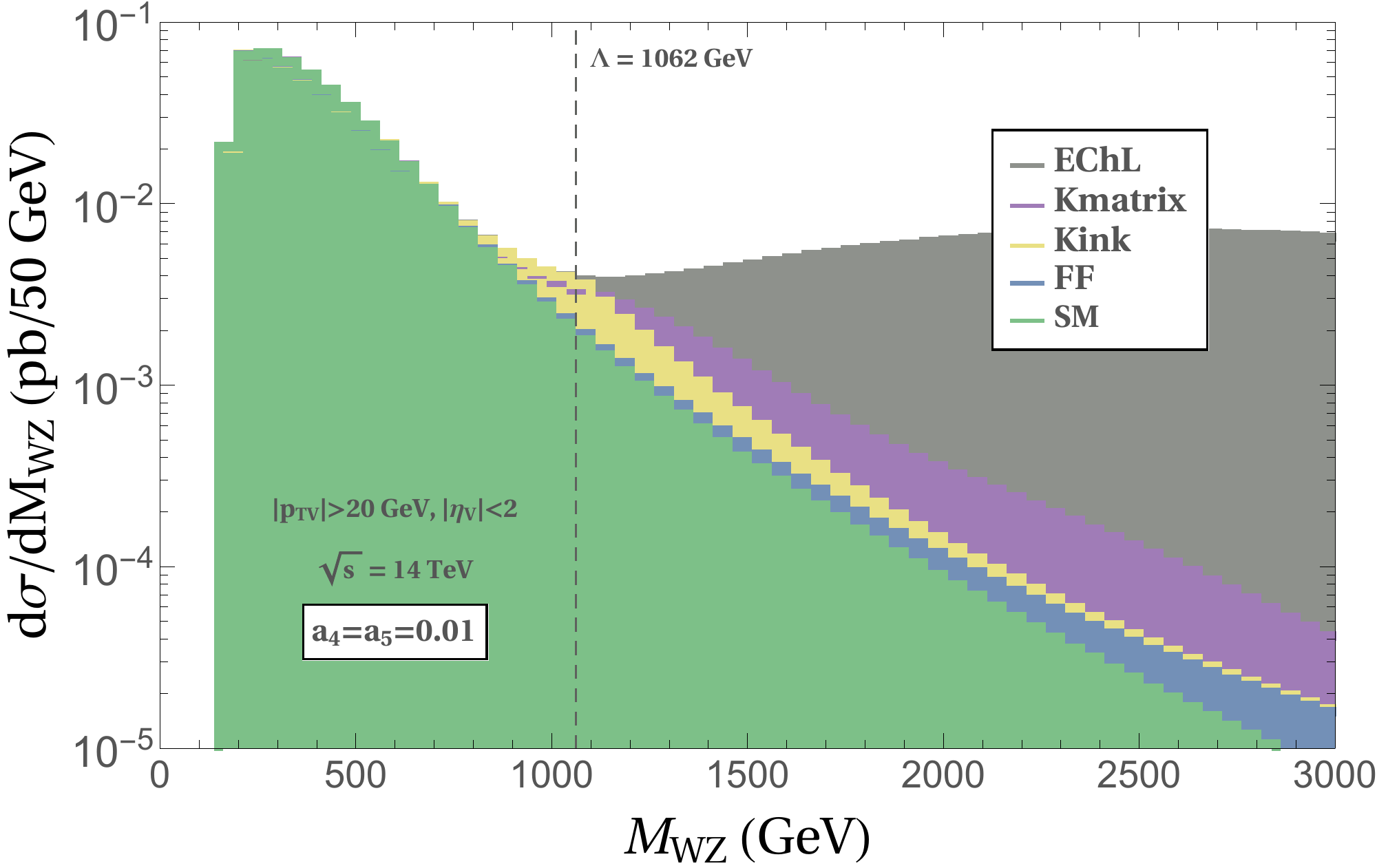}
\includegraphics[width=0.49\textwidth]{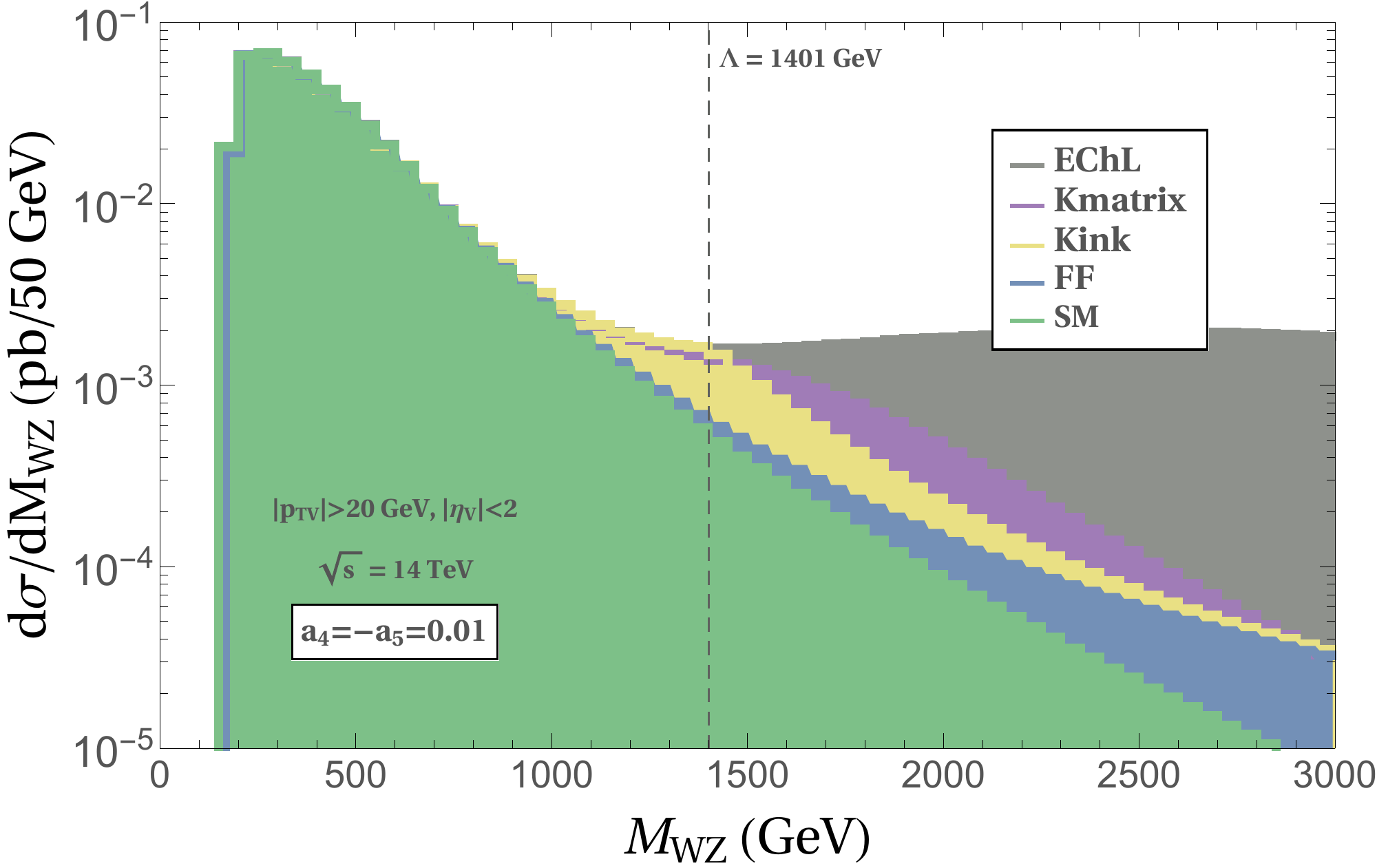}
\caption{Predictions of the differential cross section of the process $pp\to $WZ+X as a function of the invariant mass of the final WZ pair computed with different unitarization methods using the EWA. Predictions of not-unitarized EChL (gray), K matrix unitarization (purple), Kink (yellow), Form Factor (FF, blue) and the SM (green) are shown for two referencial values of the relevant chiral parameters $a_4=a_5=0.01$ (left) and $a_4=-a_5=0.01$ (right). The other chiral parameters are set to their SM value. The unitarity violation scale is also displayed for each case. Predictions are given for $|\eta_V| < 2$ and $|p_{T_V}|>20$ GeV and at $\sqrt{s}=14$ TeV. }
\label{fig:methodsprocess}
\end{center}
\end{figure}

The results are displayed in \figref{fig:methodsprocess}, where we present the invariant mass distributions of the differential cross section of the process $pp\to $WZ+X computed with the EChL for $a_4=a_5=0.01$ (left) and $a_4=-a_5=0.01$ (right) and unitarized with the diverse procedures we have described in the previous section. The non-unitarized EChL and the SM predictions are also shown, for comparison. The unitarity violation scale is marked with a dashed line in each case. The final gauge bosons are required to have $|\eta_V| < 2$ and $|p_{T_V}|>20$ GeV and the evaluation is performed at $\sqrt{s}=14$ TeV.

From these curves we can see that the translation of the subprocess results to the LHC is direct, and the conclusions regarding the diverse curves are very similar. The different predictions among the various unitarization methods are still manifest, which clearly indicates that the experimental constraints imposed on the EChL parameters will strongly depend on the unitarization method used to analyze the data. Besides, the same pattern of the predictions concerning the relative sign of the chiral parameters is encountered: in the EChL and the K-matrix case, same sign $a_4$ and $a_5$ lead to larger predictions than in the opposite sign case. For the Form Factor and the Kink, the reverse setup is recovered. This still points towards the fact that same sign values of $a_4$ and $a_5$ will be more constrained in the EChL and the K-matrix case, opposite to the Form Factor and the Kink case. 

The IAM is not shown in these plots since, as we mentioned, it is more suitable for the resonant case. Besides, as we have seen before, in the present non-resonant case, for the chosen particular channel $WZ \to WZ$, and with the simplified setup of just two non-vanishing chiral coefficients, $a_4$ and $a_5$, the IAM predictions are very close to the SM ones. Notice that it will not be the case if other channels were considered and other chiral coefficients (in particular $a$) were also non-vanishing.

Regarding the Cut off, it is clear that integrating only up to the unitarity violation scale to obtain the total cross section will lead to much smaller predictions than in the rest of the cases. Finally, it is worth commenting that, as it should be, again all predictions match the EChL one at low invariant masses.

We have now characterized the different predictions of the studied unitarization methods at the LHC. The next step should be to translate these predictions into uncertainties in the extracted constraints on the parameter space of the EChL. In order to do that, we will base our approach upon the ATLAS results for $\sqrt{s}=8$ TeV given in ref.~\cite{Aaboud:2016uuk}. In the mentioned reference, a very sophisticated experimental analysis is performed, specially regarding triggers, background estimations, and event selection. Then, using the K-matrix (or T-matrix) unitarization prescription proposed in~\cite{Kilian:2014zja,Alboteanu:2008my} , the 95\% C.L. exclusion regions in the $[a_4,a_5]$ (called $[\alpha_4,\alpha_5]$ in \cite{Aaboud:2016uuk}) parameter space are obtained. It is beyond the scope of this work to reproduce accurately the experimental analysis of the ATLAS searches. However, there is a consistent way in which we can use their results to obtain the experimental constraints corresponding to other unitarization methods apart from the K-matrix one.

Our approach is the following: first, we take the $a_4$ and $a_5$ values lying on the contour of the WZ observed ``elipse'' provided by the ATLAS study, i.e., the solid, cyan line shown in \figref{fig:elipseATLAS}. With those values, we evaluate the total cross section following our K-matrix unitarization procedure for the LHC case, that is, indeed, constant over the mentioned contour. This should be equivalent to what ATLAS has performed, since we have checked that for these values of the parameters our prescription matches the one given by the Wizard group. The cross section that we obtain represents the equivalent cross section in our framework to the one that ATLAS has measured experimentally. It is, so to say, a translation between the experimental results and our naive results. Now, what we do is to find the values of $a_4$ and $a_5$ that lead to the same cross section for the other unitarization methods considered. In this way, we construct the 95\% exclusion regions in the $[a_4,a_5]$ plane for the various unitarization schemes presented in the previous subsection, to see how they differ in magnitude and shape. By applying this procedure, we are assuming that the selection cuts required to be fulfilled by the ATLAS search affect all our predictions equivalently. This could not be the case, but we expect the differences to be small, so our prescription should be a good first approximation to the issue. Furthermore, it is worth commenting that, regarding the backgrounds, since they are the same to all of our signals, it is well justified to proceed in this way.

 \begin{figure}[t!]
\begin{center}
\includegraphics[width=0.8\textwidth]{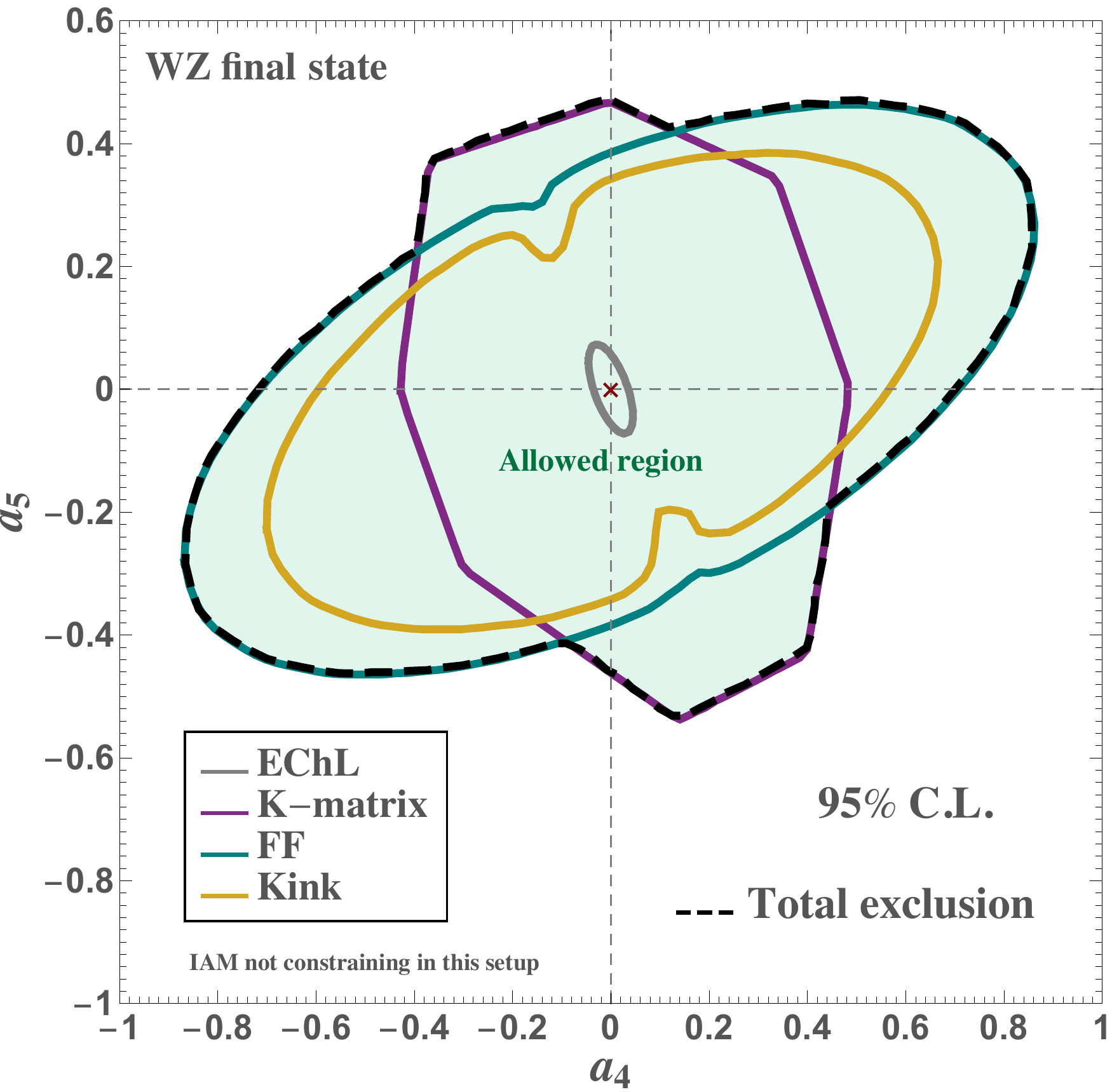}
\caption{Predictions of the 95\% confidence level exclusion regions in the $[a_4,a_5]$ plane for the WZ final state at the LHC with $\sqrt{s}=8$ TeV,  and for the not-unitarized EChL (gray) and the different unitarization methods described in the text: K matrix (purple, corresponding to the solid cyan line in \figref{fig:elipseATLAS}), Kink (yellow) and Form Factor (FF, blue). The total overall exclusion region is the one outside the boundary denoted with a dashed black line. The SM point is marked with a red cross. To obtain this figure we have used the WZ results in \figref{fig:elipseATLAS} as reference.}
\label{fig:elipse}
\end{center}
\end{figure}

The 95\% C.L. exclusion regions in the $[a_4,a_5]$ plane for different unitarization scheme choices, are presented in \figref{fig:elipse}. There we show the corresponding limits for the case in which no unitarization is performed at all (EChL, in gray), for the K-matrix unitarization, matching, of course, the ATLAS results (purple), for the Form Factor prediction (blue) and for the Kink (yellow). We also show the total exclusion region, obtained by the overlap of the former ones.

Many interesting features can be extracted from this figure. First of all, and most importantly, it is indeed very clear that using one unitarization method at a time to interpret experimental data does not consider the full EFT picture. Since there are many unitarization prescriptions that lead to very different constraints, one should take them all into account in order to provide a reliable bound on the EFT parameters. These different constraints can  vary even in an order of magnitude, as it is manifest in \figref{fig:elipse}. For instance, the Form Factor prescription leads to bounds on $[a_4,a_5]$\footnote{This is just a handy estimate, since the extraction of one-dimensional bounds from \figref{fig:elipse} cannot be done directly without losing the statistical interpretation of the 95\% C.L. regions.} of the order of [0.8,0.4], roughly speaking, whereas the case in which there is no unitarization performed leads to constraints of the order of [0.04,0.08]. Notice that these latter bounds, i.e., those obtained from the raw, non-unitarized EChL, are not directly comparable to those given in \cite{Sirunyan:2019ksz}, since our results correspond to $\sqrt{s}=8$ TeV and the ones reported in the mentioned reference to $\sqrt{s}=13$ TeV. We leave the precise computation of the 13 TeV results for a future work.

It is also obvious from this figure that the Kink method leads to more stringent constraints than the FF method (the corresponding pseudo-ellipse is smaller and oriented similarly to the FF one). Also the K-matrix method leads to more stringent constraints than the FF one and, in this case, with a different orientation of the pseudo-ellipse (indeed, similar to the EChL one). Interestingly, for the present studied case of non-resonant  $pp \to$ WZ +X events, there is not just a difference in the magnitudes of the bounds of the EChL parameters, but also in the role of each of them, $a_4$ and $a_5$. This feature was already stated before, since in \figref{fig:methodsprocess} we had already seen that points lying in the region of the plane in which $a_4$ and $a_5$ have the same sign should be more constrained in the EChL and the K-matrix case, just in the opposite direction to the Form Factor and the Kink case.

At this point, two further comments have to be made. The first concerns the IAM, whose prediction is not present in this figure. It is due to the same argument we have been commenting throughout the text, that can be summarized in the fact that, for our particular setup (non-resonant case with deviation with respect to the SM coming only from the two considered $\mO(p^4)$ operators) this method is not suitable to impose reliable constraints on the EChL parameters. Nevertheless, the IAM can be extremely useful when looking for new physics signals at the LHC in the resonant case, as we will see in the next Chapters.

The second concerns the Cut off, also not present in the figure. Since this procedure implies to sum events only up a to a determined invariant mass of the diboson system to obtain the total cross section, a problem arises concerning the backgrounds. In our approach, we are always integrating over the whole studied energy region, for all the unitarization method predictions. This means that the background is considered to be the same for all of our signals and we can use the translation from the ATLAS results safely. However, if we now change the picture and integrate over a smaller invariant mass region, such as in the Cut off procedure, we should take into account this same integration over the background, and the pure translation form the ATLAS results fails, since we don't know the background scaling with energy. For this reason, we have not included the Cut off prediction in our final results, but we really do believe that it should be also considered in proper experimental searches. 

In this Chapter, we have have concluded that the various unitarization methods considered provide very different predictions not only for at the subprocess level but also for the total process at the LHC, that we have estimated making use of the EWA. We have constructed, based on the ATLAS results for $\sqrt{s}=8$ TeV given in~\cite{Aaboud:2016uuk}, the 95\% exclusion regions in the $[a_4,a_5]$ plane for the various unitarization schemes. The main results are contained in \figref{fig:elipse}, from which very interesting features can be extracted. The most important of them is that it is indeed very clear that using one unitarization method at a time to interpret experimental data does not consider the full effective theory picture. Since there are many unitarization prescriptions that lead to very different constraints, one should take them all into account in order to provide a reliable bound on the EFT parameters. There is, therefore, a theoretical uncertainty present in the experimental determination of effective theory parameters due to the unitarization scheme choice. A first approximation to this uncertainty has been quantified in the present work analyzing the predictions of $pp\to$ WZ+X events at the LHC from the EChL in terms of $a_4$ and $a_5$ and with different unitarization methods. We believe that it is important to take these uncertainties into account when relying upon experimental values of the constraints of effective theory parameters, in order to consider the full effective theory properties correctly.

Nevertheless, these conclusions correspond to the non-resonant scenario. We have seen throughout several parts of this Thesis, that, however, the presence of resonant states generated by the strong interaction of the EW Goldstone bosons will be a clear signal of BSM physics in the EWSB sector. For this reason, in the next two Chapters, we will study the production of a type of these resonances in various VBS channels, estimating their LHC sensitivity.


\chapter[\bfseries Vector resonance production in WZ scattering at the  LHC:\newline  a unitarized EChL analysis]{Vector resonance production in WZ scattering at the LHC: a EChL analysis}\label{Resonances}

\chaptermark{Vector resonance production in VBS at the LHC: a EChL analysis}

One of the most likely indications of the existence of physics beyond the standard model could be the appearance of resonances in the scattering of longitudinally polarized $W$ and $Z$ electroweak gauge bosons. This would be a formidable hint of the existence of new interactions involving the EWSB sector of the SM. This possibility is indeed contemplated in all composite Higgs scenarios, characterized by the existence of a scale $f \gg v = 246$ GeV where some new strong interactions trigger the dynamical breaking of a global symmetry group $G$ to a certain subgroup $H$. 
The Goldstone bosons that appear provide the longitudinal degrees of freedom of the weak gauge bosons, while the Higgs boson would be one of the leftover Goldstone bosons, as we have explained in previous Chapters. A non-zero mass for the latter is often provided by electroweak radiative corrections, e.g., via some misalignment mechanism between the gauge group and the global unbroken subgroup~\cite{Agashe:2004rs}.   

The use of the EChL for the study of this strong dynamics suggests that the scale associated to these resonances is related to the parameter with dimension of energy controlling the perturbative expansion within this chiral effective field theory, given typically, in the minimal scenario that we work with, by $4 \pi v$.  Therefore, one expects resonances to appear with masses typically of a few TeV,  clearly in the range covered at the LHC.  The theoretical framework  for the description of such resonances is,  however, not universal and one has to rely on a particular approach. Once one chooses, as we do, the approach provided by the EChL, there are basically two main paths to proceed.  Either the resonances are introduced explicitly at the Lagrangian level or they are instead dynamically generated from the EChL itself. Notice that, in the former case, the new terms added to the EChL are required to share the EChL symmetries. The first approach has been followed in several works~\cite{Pich:2012jv,Pich:2012dv,Pich:2013fea,Pich:2015kwa,Pich:2016lew} essentially along the lines of previous works within the context of low energy QCD \cite{Ecker:1989yg}.
This type of chiral resonances have also been studied at the LHC  \cite{Alboteanu:2008my}.

The second approach has been followed in a number of works that use
the inverse amplitude method  to impose the unitarity of the amplitudes predicted with the EChL~\cite{Espriu:2012ih, Espriu:2013fia,Delgado:2013loa, Delgado:2013hxa, Delgado:2014dxa, Espriu:2014jya, Arnan:2015csa, Dobado:2015hha, Corbett:2015lfa,BuarqueFranzosi:2017prc}.
Within this approach,  the self-interactions of the longitudinal  EW bosons, which are assumed to be strong, are the responsible of the dynamical generation of the resonances,  and these are expected to show up in the scattering of the longitudinal modes, $W_L$ and $Z_L$, as we just saw briefly  in the previous Chapter. The IAM was indeed used long ago in the context of a strongly interacting EWSB sector but without the Higgs particle, and the production of these IAM resonances at the LHC was also addressed \cite{Dobado:1989gr, Dobado:1990jy, Dobado:1990am}. The advantage of this second approach is that it provides unitary amplitudes, which are absolutely needed for a realistic analysis at the LHC, and it predicts the properties of the resonances, masses, widths and couplings, in terms of the chiral parameters of the EChL. The disadvantage of this method is that it does not deal with full amplitudes but with partial waves, which are not very convenient for a Monte Carlo analysis at the LHC.

The present Chapter addresses the question of whether these IAM dynamically generated resonances of the EWSB sector could be visible at the LHC by means of the study of VBS observables. VBS is the most relevant channel to explore at the LHC if the longitudinal gauge modes are really strongly interacting, since they involve the four point self-interactions of the EW gauge bosons, as we have seen. Moreover, the resonances should emerge more clearly in VBS processes as they are generated from this strong dynamics. Our study aims to quantify the visibility of these resonances and also to determine the integrated luminosities that would be required to this end.

 More concretely, our purpose here is to estimate the event rates at the LHC of the production of a $SU(2)_{L+R}$  triplet vector resonance, $V$,  via $WZ\to WZ$ scattering, and the subsequent decays of the final $W$ and $Z$. We have selected this particular subprocess because it has several appealing features in comparison with other VBS channels.  In the presence of such dynamical vector resonances, these  emerge/resonate (in particular, the charged $V^{\pm }$ones) in the s-channel of $W Z  \to W Z$, whereas in other subprocesses like $W^+W^+ \to W^+ W^+$, $W^+W^- \to ZZ$, $ZZ \to W^+W^-$ and $ZZ \to ZZ$ do not. Other interesting cases like $W^+W^- \to W^+ W^-$ where the neutral resonance, $V^0$, could similarly emerge in the s-channel have, however, severe backgrounds.  For this reason it is known to be very difficult to disentangle the signal from the SM irreducible background at the LHC. In particular, the SM one-loop gluon initiated subprocess, $g g \to W^+W^-$, turns out to be a very important background in this case due to the huge gluon density in the proton at the LHC energies. Our selected process $WZ\to WZ$, in contrast,
does not suffer from this background, and therefore it provides one of the cleanest windows to look for these vector resonances at the LHC. This is mainly the reason why we have chosen this particular channel throughout the whole Thesis as our illustrative example of the VBS configurations. 

We work with EW gauge bosons in the external legs of the VBS amplitudes and not with Goldstone bosons. This means that we go beyond the
simpler predictions provided by the equivalence theorem (ET)~\cite{Cornwall:1974km, Vayonakis:1976vz, Lee:1977eg, Gounaris:1986cr}. Furthermore, in order to introduce in a Lagrangian language the resonances that are dynamically generated by the IAM we shall construct an effective Lagrangian including vector resonances, based on the Proca 4-vector formalism~\cite{Pich:2012jv,Pich:2012dv,Pich:2013fea,Pich:2015kwa,Pich:2016lew}. This effective Lagrangian includes the proper resonance couplings to the $W$ and $Z$ and has the symmetries of the EChL, in particular the EW Chiral symmetry.

With this Lagrangian we will mimic the resonant behaviour of the IAM amplitudes, having the resonance masses and widths as predicted by the IAM. Indeed, we will make use of this vector Lagrangian to extract the Lorentz structure of the $WZ$ scattering vertex to be coded in the MonteCarlo. 
The coupling itself will turn out to be a momentum-dependent function that will be derived from the IAM unitarization process
in the $IJ=11$ channel. This IAM-MC model presented here is proper for a MonteCarlo analysis and it is included in MadGraph5 \cite{Alwall:2014hca}
for this work. The corresponding UFO file for the present IAM-MC model can be provided on demand. We would like to emphasize that our IAM-MC model provides full $A(WZ\to WZ)$ amplitudes with massive external EW gauge bosons. The corresponding cross section $\sigma(WZ\to WZ)$ is computed from these full amplitudes and not from the first partial waves that do not provide a sufficiently accurate result, as we have seen in previous Chapters. 

Finally, a careful study of the signal versus backgrounds for the full process $pp \to WZ jj$, leading to events with two jets plus one $W^+$ and one $Z$ will be performed. We will first discuss on the potential of the hadronic and semileptonic channels of the final $WZ$. Then we will explore the cleanest channels leading to events with two jets and the three leptons and missing energy which come from the leptonic decays of the final $W^+$ and $Z$, which will allow us to extract the emergent vector resonances from the SM background in this kind of $\ell {\bar \ell} \ell \nu j j $ events at the LHC.


\section{Selection of scenarios with vector resonances in WZ scattering} 
\label{scenarios}
In this section we present the specific EChL scenarios that will be explored in our forthcoming study at the LHC,  having dynamical vector resonances $V$ emerging in  $WZ$ scattering. First we show the results of the cross-sections for $WZ \to WZ$ from the EChL, which are compared with the SM predictions. Then we unitarize these EChL results, and finally, within these unitarized results, we select the scenarios with emergent vector resonances $V$. This should be complementary to the studies performed on the $WZ \to WZ$ in previous part of this Thesis. 

We know by means of the ET~\cite{Cornwall:1974km, Vayonakis:1976vz, Lee:1977eg, Gounaris:1986cr}, which applies to renormalizable gauges and is valid also for the EChL~\cite{Dobado:1993dg,Dobado:1994vr,Dobado:1997fv,He:1993qa} as we have seen, that the scattering amplitude for this subprocess $W_L Z_L \to W_L Z_L$ can be approximated, at large energies compared to the gauge boson masses, by the scattering amplitude of the corresponding would-be Goldstone bosons,
\begin{equation}
A(W_L Z_L \to W_L Z_L) \simeq A(wz \to wz)\,.
\label{ETRes}
\end{equation}
Since the relevant EW chiral coefficients in the amplitude $A(wz \to wz)$ (i.e., those that remain even switching off the gauge interactions, $g=g'=0$),  are just $a$, $b$, $a_4$ and $a_5$, we conclude again that for our purpose of describing the most relevant departures  from the SM in $A(W_L Z_L \to W_L Z_L)$ it will be sufficient to work with just this subset of EChL parameters. We include now $a$ and $b$ since, although they did not play a relevant role in the context of non-resonant scenarios, in the case in which a resonance emerges in the spectrum some of its most relevant properties will be driven directly by these coefficients. 

Nevertheless, as we have said, in the present work we deal with massive gauge bosons in the external legs of the VBS amplitudes and not with their corresponding Goldstone bosons. The various contributing terms from the EChL to the EW gauge boson scattering amplitude of our interest  are the following:
\begin{align}
A(W_L Z_L \to W_L Z_L)^{{\rm EChL}} = A^{(0)}(W_L Z_L \to W_L Z_L)  + A^{(1)}(W_L Z_L \to W_L Z_L)\,,
\label{WLZL}
\end{align}
where the leading order (LO), ${\cal O}(p^2)$, and next to leading order contributions (NLO), ${\cal O}(p^4)$, are
denoted as  $A^{(2)}$ and $A^{(4)}$ respectively, and are given by:
 \begin{align}
 A^{(2)}(W_L Z_L \to W_L Z_L)&=  A^{{\rm EChL}^{(2)}_{\rm tree}}\,,    \nn\\
 A^{(4)}(W_L Z_L \to W_L Z_L)&=  A^{{\rm EChL}^{(4)}_{\rm tree}}+ A^{{\rm EChL}^{(2)}_{\rm loop}}\,.
\label{LOandNLO}
\end{align}

For completeness, in the Appendices \ref{FR-SM} and \ref{FR-EChL} the necessary Feynman rules for the tree level computation,
\begin{equation}
A(W_L Z_L \to W_L Z_L)^{{\rm EChL}^{(2+4)}_{\rm tree}}=A^{{\rm EChL}^{(2)}_{\rm tree}}+ A^{{\rm EChL}^{(4)}_{\rm tree}}\,.
\label{EChL-tree}
\end{equation}
are presented. 
It should be noticed that, to our knowledge, a full one-loop EChL computation is not available in the literature for this process,  i.e., the full analytical result
of $A^{{\rm EChL}^{(2)}_{\rm loop}}$ is unknown. However, we will use an approximation to estimate the size of this one-loop contribution,
following~\cite{Espriu:2012ih,Espriu:2013fia,Espriu:2014jya}. Concretely, the real part of the loop diagrams is computed 
using the ET (but keeping $m_H\neq 0$) and the imaginary part of the loops is   calculated exactly through the tree-level result by
making use of the optical theorem. In the following, we will refer to this NLO computation, ${\rm EChL}^{(2+4)}_{\rm loop}$,  as quasi exact one-loop EChL result.

\begin{figure}[t!]
\begin{center}
\includegraphics[width=.6\textwidth]{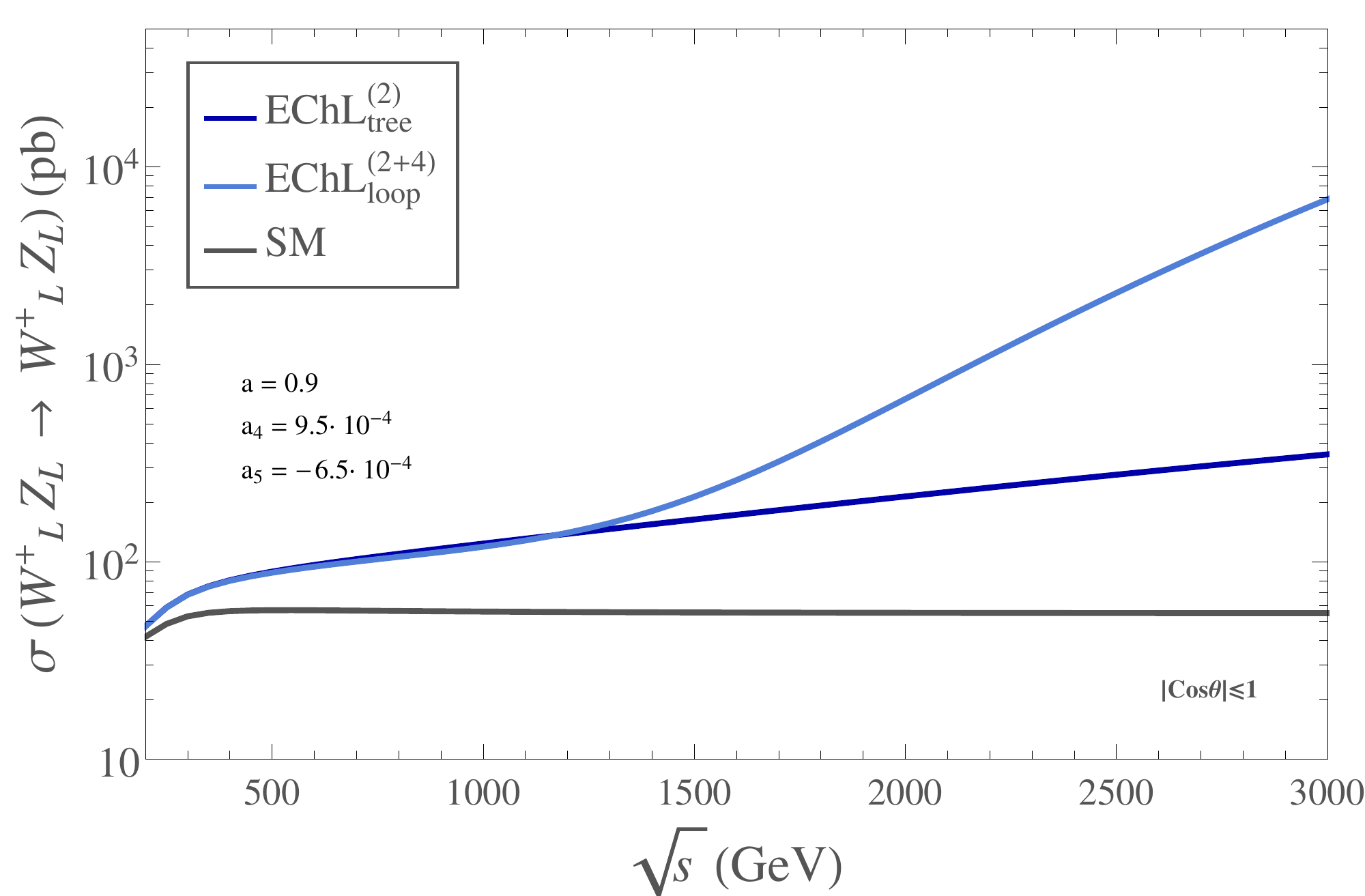}
\caption{Predictions of the cross section $\sigma(W_LZ_L \to W_LZ_L)$ as a function of the center of mass
energy $\sqrt{s}$ from the EChL. The predictions at leading order, ${\rm EChL}^{(2)}_{\rm tree}$,
and next to leading order, ${\rm EChL}^{(2+4)}_{\rm loop}$,
are displayed separately. The EChL coefficients
are set here to $a=0.9$, $b=a^2$, $a_4=9.5 \times 10^{-4}$ and $a_5=-6.5 \times 10^{-4}$. Here the integration is done in the whole $|\cos \theta| \leq 1$ interval of the centre of mass scattering angle $\theta$.
The prediction of the SM cross
section is also included, for comparison. All predictions have been obtained using FormCalc and our private Mathematica code and checked with MadGraph5.} 
\label{fig:xsecEChLLLLL}
\end{center}
\end{figure}

We have chosen one example to illustrate numerically and graphically
the energy behavior of the EChL cross section and the comparison with the SM prediction.
This is displayed in \figref{fig:xsecEChLLLLL}, where the chiral parameters have been
set to $a=0.9$, $b=a^2$, $a_4=9.5 \times 10^{-4}$ and $a_5=-6.5 \times 10^{-4}$. We have taken $a_4$ and $a_5$ values of the order of $10^{-4}$ because this is the overall order of magnitude that will provide resonances with masses of few TeV in the context of the IAM unitarization prescription. 

At this stage, it is also interesting to comment on the accuracy of our assumption of neglecting other loop contributions in our computation of $WZ$ scattering. In particular, as we have said, we are ignoring in this work the contributions from fermions. Since the fermions would only contribute via loops to this $WZ$ scattering process, and since the dominant contributions would come from the third generation-quark loops, we have performed an estimate of the size of these loop contributions to be sure that they are indeed negligible. For this estimate we have assumed that all the fermion interactions are the same as in the SM and we have used the analytical results of  \cite{Dawson:1990cp} which are provided for the SM within the ET. Our numerical estimate of the heavy fermion loops indicates that for the high energies of our interest here, say between 1 and 3 TeV, the contributions from the top loops to $\sigma(w z \to wz)$ decrease with   $\sqrt{s}$, in contrast to the contributions from the EChL loops which increase with energy. Furthermore, they are indeed very small, between $10^{-1}$ pb and $10^{-2}$ pb. These are more than three orders of magnitude below the prediction of $\sigma(W_LZ_L \to W_LZ_L)$ from the EChL (specifically, from our quasi exact prediction   ${\rm EChL}^{(2+4)}_{\rm loop}$ in \figref{fig:xsecEChLLLLL}). Therefore we conclude that our assumption in this work of ignoring the fermion loops is well justified.

The previously studied violation of unitarity of the EChL scattering amplitudes leads to our
major concern in this work: the need of an unitarization method in order to provide realistic predictions at the
LHC. We choose here one of the most used unitarization methods for the partial waves, the IAM,  
since it has the advantage over other methods of being able to
generate dynamically the vector resonances that we are interested in. 

Other unitarization procedures such as N/D and the improved K-matrix (IK) were also studied and compared with the IAM in the present context in detail in Ref.~\cite{Delgado:2015kxa}. In this reference the IAM, N/D and the IK unitarization methods are implemented in a particular way compatible with the electroweak chiral expansion.   All of these three methods turn out to be acceptable, since they produce partial waves which are: IR and UV finite, renormalization scale $\mu$ independent, elastically unitary, have the proper analytical structure (they feature a right and a left cut) and they reproduce the expected low energy results of the EChL up to the one-loop level. Thus the three methods can provide an UV completion of the low-energy chiral amplitudes. Moreover, for some region of the chiral couplings parameter space, they can have a pole in the second Riemann sheet with similar properties. These poles have a natural interpretation as dynamically generated resonances with the quantum numbers of the corresponding channel. 

 By comparing the three methods for different values of the chiral couplings it is possible to realize that all of them normally produce the same qualitative results and, in many cases, the agreement is also quantitative up to high energies. This is particularly  true  for the $I=J=0$ channel. However, as it is explained in detail in Ref.~\cite{Delgado:2015kxa}, the N/D and the IK method cannot be applied to the $I=J=1$ channel considered in this work in the particular case of $b=a^2$, since it leads to contributions from the left and right cuts which cannot be separated in a $\mu$-invariant way, as required by these two methods. Because of this reason, and due to the fact that the other methods considered in this Thesis (Cut off, Form Factor, Kink and K-matrix) do not accommodate the presence of resonances,  in the following we will use only the IAM method.  
 
Contrary to the perturbative expansion of the EChL, the IAM amplitudes fulfil all the analyticity and elastic unitarity requirements. In addition, $a^{\rm IAM}_{IJ}$ may or may not exhibit a pole as discussed before. If present, it can be interpreted as a dynamically generated resonance. In that case we use here the usual convention for the position of the pole in terms of the mass, $M_R$,  and width, $\Gamma_R$, of the corresponding resonance $R$:
$s_{\rm pole}=(M_R-\frac{i}{2}\Gamma_R)^2$.  

\begin{figure}[t!]
\begin{center}
\includegraphics[width=.49\textwidth]{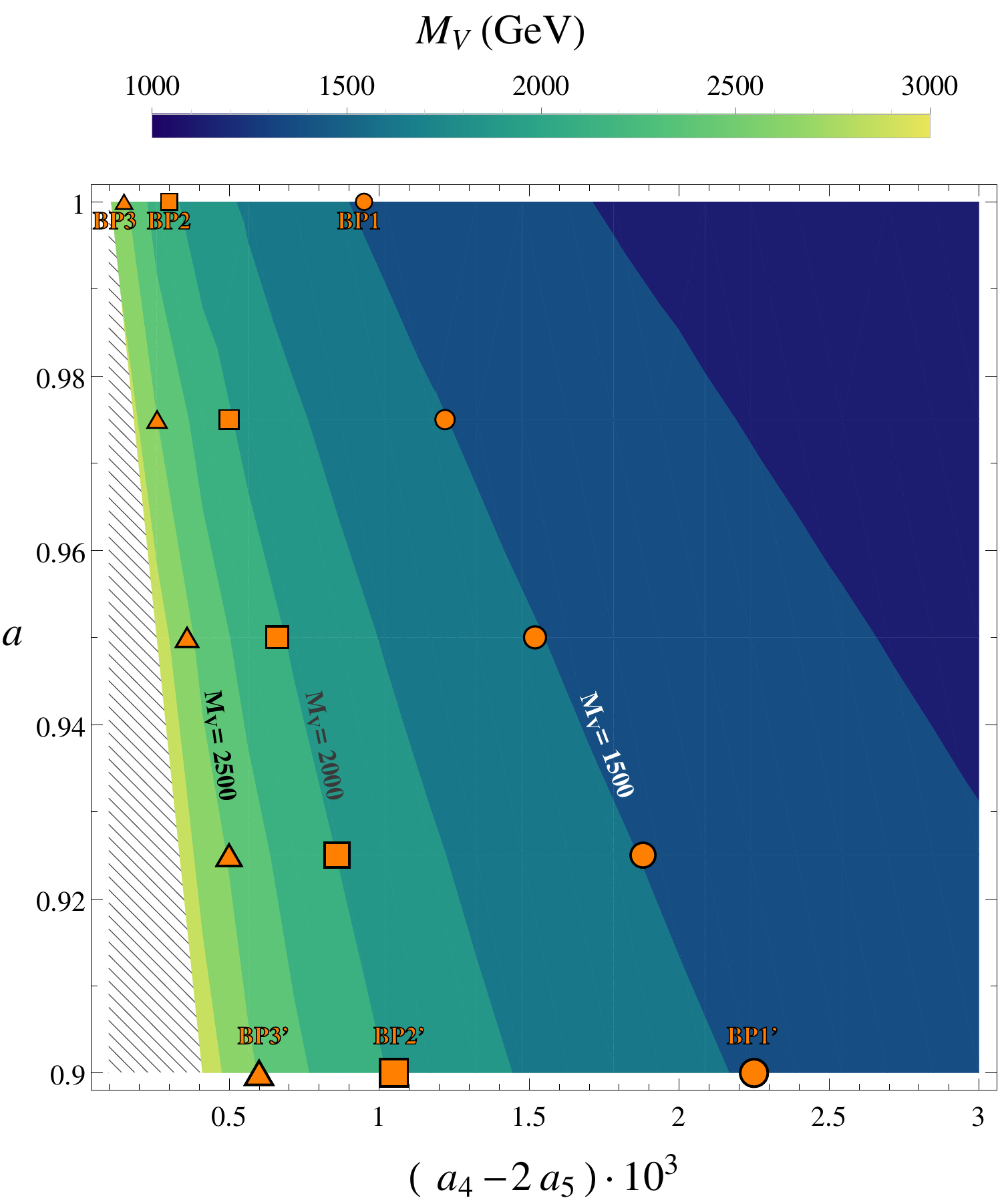}
\includegraphics[width=.49\textwidth]{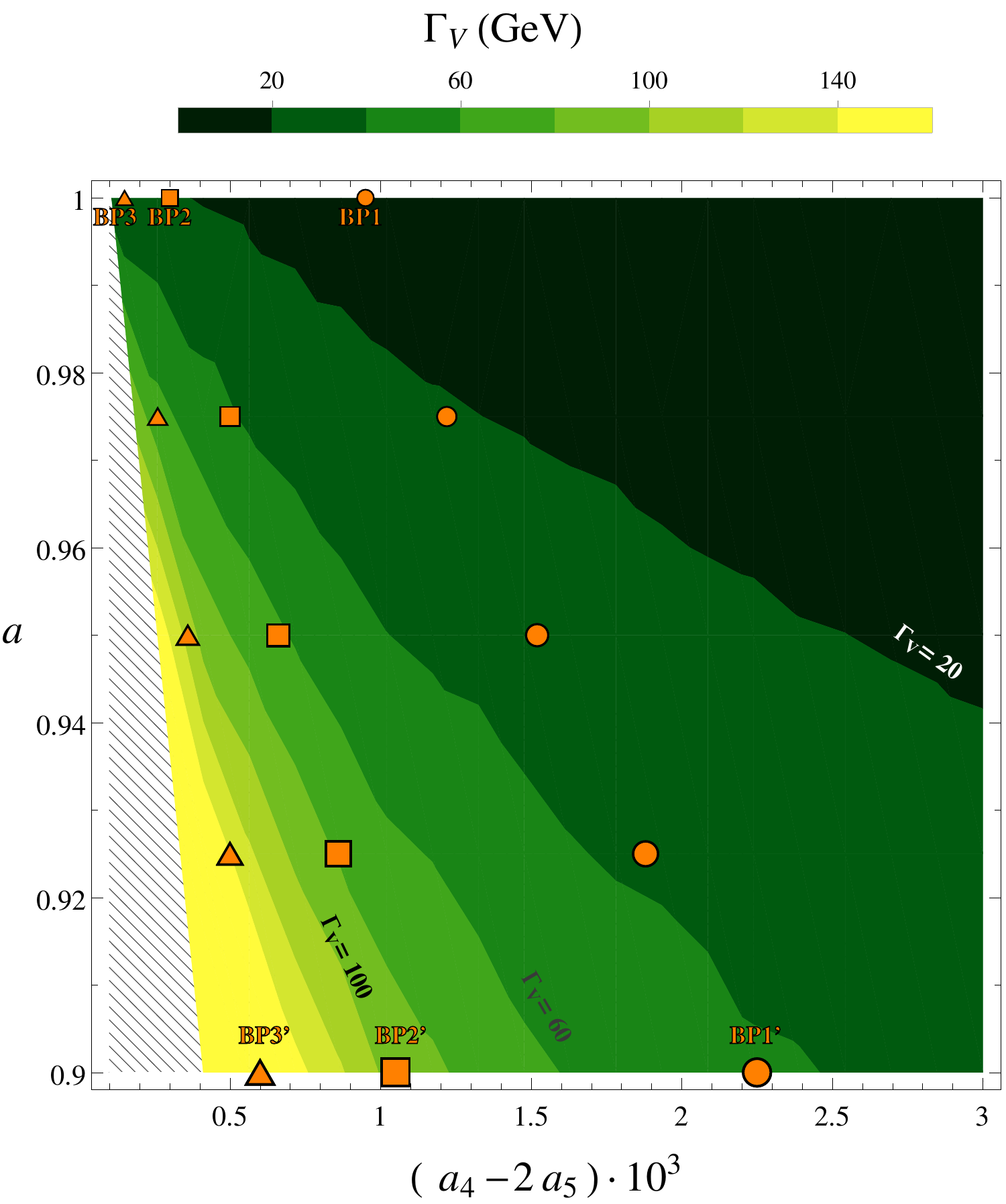}\\
\caption{Predictions for masses (left panel) and widths (right panel) of vector resonances as a function of $a$ and the combination $(a_4-2a_5)$ in the EChL+IAM. Our fifteen selected scenarios lay approximately over the contour lines of fixed $M_V$, 1500 GeV (circles), 2000 GeV (squares), and 2500 GeV (triangles), and have values for $a$ fixed, respectively,  to 0.9 (biggest symbols, corresponding to BP1', BP2' and BP3'), 0.925, 0.95, 0.975 and 1 (smallest symbols, corresponding to BP1, BP2, and BP3). All studied cases with vector resonances are such that no corresponding scalar or tensor resonances appear. The stripped area denotes the region with resonances heavier than 3000 GeV.}
\label{fig:contourMW}
\end{center}
\end{figure}

The solution to the position of the pole in the case of $a^{\rm IAM}_{11}$ is very simple if the ET is used, and gives  simple predictions for the mass and the width of the dinamicaly generated vector resonances in terms of the EChL parameters, $a$, $b$, $a_4$ and $a_5$,  given by~\cite{Delgado:2013loa,Delgado:2013hxa}: 
\begin{align}
(M_V^2)_{\rm ET} &=\dfrac{1152 \pi^2v^2 (1-a^2)}{8(1-a^2)^2-75(a^2-b)^2+4608\pi^2(a_4(\mu)-2a_5(\mu))}\,,\\[4pt]
(\Gamma_V)_{\rm ET}&=\dfrac{(1-a^2)}{96\pi v^2}M_V^3\,\left[1+\dfrac{(a^2-b)^2}{32\pi^2 v^2 (1-a^2)}M_V^2\right]^{-1},
\label{MVGVET}
\end{align}
with $a_4(\mu)$ and $a_5(\mu)$ the scale dependent parameters whose running  equations for arbitrary $a$ and $b$ can be found in \eqref{eqrunninga4a5} complemented with \tabref{tabla} . These solutions apply to narrow resonances, i.e., for $\Gamma_V \ll M_V$, which is indeed our case. It should be noticed that, as it is well known, the case with $a=1$ cannot be treated in the IAM within the ET framework. This will not be the case in our quasi-exact predictions, as we will see in the following.

The solution to the position of the $a^{\rm IAM}_{11}$ pole in the quasi-exact case
with $m_{W,Z} \neq 0$ is more involved~\cite{Espriu:2012ih,Espriu:2013fia,Espriu:2014jya}, but it basically shares the main qualitative features of the previous ET results. First, the main contribution from the parameters $a_4$ and $a_5$ appears also in the particular combination $(a_4 - 2 a_5)$ which is $\mu$-scale independent if $b=a^2$. We have checked explicitly that other contributions from $a_4$ and $a_5$ vanish in the isospin limit where $m_W=m_Z$.

 Second, the main dependence with $a$ also comes in the combination $(1-a^2)$, and the main dependence with $b$ also appears as $(a^2-b)^2$. All these generic features can also be seen in our numerical results, displayed in \figref{fig:contourMW}, which we have generated with the FORTRAN code that implements
the quasi-exact EChL+IAM framework, borrowed from the authors in Refs.~\cite{Espriu:2012ih,Espriu:2013fia,Espriu:2014jya}.

The plots in \figref{fig:contourMW} show the contour lines of fixed $M_V$ and $\Gamma_V$ in the $\left[(a_4-2a_5), a\right]$ EChL parameter space plane. The particular contour lines with $M_V=1500,\, 2000,\, 2500$ GeV are highlighted since they will be chosen as our reference mass values in our next study at the LHC framework. This figure assumes $b=a^2$, but we have checked explicitly that other choices for the $b$ parameter with $b\neq a^2$ do not change appreciably these results. In fact, the contour lines of $M_V$ and $\Gamma_V$ in the  $\left[(a_4-2a_5), b\right]$ plane with $a$ fixed in the interval $a \in (0.9,1)$ (not included here),  do not show any appreciable dependence with $b$ if this parameter is varied in the interval
$b \in (0.8,1)$. The distortions due to $b\neq a^2$ are clearly subleading in comparison to the leading effects from $(1-a^2)$ and $(a_4-2a_5)$, as explicitly shown in the ET formulas of \eqref{MVGVET}, and will be neglected from now on. The main reason of this secondary role of $b$, versus $a$, $a_4$ and $a_5$ is that in the $a^{11}$ amplitude $b$ enters only via loops, whereas $a$, $a_4$ and $a_5$ enter already at the tree level. Therefore our selection of scenarios will be done in terms of $a$, $a_4$ and $a_5$, and $b$ will be fixed to $b=a^2$, for simplicity. This choice of $b=a^2$ is also motivated in several theoretical models~\cite{Coleman:1985,Halyo:1991pc,Goldberger:2008zz}.     
 
\begin{table}[t!]
\begin{center}
\vspace{.2cm}
\begin{tabular}{ ccccccc }
\toprule
\toprule
{\footnotesize {\bf BP}} & {\footnotesize {\bf $M_V ({\rm GeV})$}}  & {\footnotesize  {\bf $\Gamma_V ({\rm GeV)}$}}  & {\footnotesize {\bf $g_V(M_V^2)$}}  & {\footnotesize {\bf $a$}}  & {\footnotesize {\bf $a_4 \cdot 10^{4}$}}  & {\footnotesize {\bf $a_5\cdot 10^{4}$}}
\\
\midrule
\rule{0pt}{1ex}
BP1  & $\quad 1476 \quad $ & $\quad 14 \quad $ & $ \quad 0.033  \quad $ & $ \quad 1 \quad $ & $ \quad 3.5 \quad $ & $ \quad -3 \quad $
\\
\rule{0pt}{1ex}
BP2  & $\quad 2039 \quad $ & $\quad 21 \quad $ & $ \quad 0.018  \quad $ & $ \quad 1 \quad $ & $ \quad 1 \quad $ & $ \quad -1 \quad $
\\
\rule{0pt}{1ex}
BP3  & $\quad 2472 \quad $ & $\quad 27 \quad $ & $ \quad 0.013  \quad $ & $ \quad 1 \quad $ & $ \quad 0.5 \quad $ & $ \quad -0.5 \quad $
\\
\rule{0pt}{1ex}
BP1' & $\quad 1479 \quad $ & $\quad 42 \quad $ & $ \quad 0.058  \quad $ & $ \quad 0.9 \quad $ & $ \quad 9.5 \quad $ & $ \quad -6.5 \quad $
\\
\rule{0pt}{1ex}
BP2'  & $\quad 1980 \quad $ & $\quad 97 \quad $ & $ \quad 0.042  \quad $ & $ \quad 0.9 \quad $ & $ \quad 5.5 \quad $ & $ \quad -2.5\quad $
\\
\rule{0pt}{1ex}
BP3'  & $\quad 2480 \quad $ & $\quad 183 \quad $ & $ \quad 0.033  \quad $ & $ \quad 0.9 \quad $ & $ \quad 4\quad $ & $ \quad -1 \quad $
\\
\bottomrule
\bottomrule
\end{tabular}
\vspace{0.4cm}
\caption{ Selected benchmark points (BP) of dynamically generated vector resonances.
The mass, $M_V$, width, $\Gamma_V$, coupling to gauge bosons, $g_V(M_V)$,
and relevant chiral parameters, $a$, $a_4$ and $a_5$ are given for each of them.
$b$ is fixed to $b=a^2$. This table is generated using the FORTRAN code that implements
the EChL+IAM framework, borrowed from the authors in Refs.~\cite{Espriu:2012ih,Espriu:2013fia,Espriu:2014jya}. 
The effective coupling $g_V(M_V^2)$ is defined in section~\ref{sec-model}.}
\label{tablaBMP}
\end{center}
\end{table}

In  \tabref{tablaBMP} we present a number of selected benchmark points (BP):  specific sets of
values for the relevant parameters
$a, a_4$ and $a_5$ that lead to dynamically generated vector resonances emerging in the $IJ=11$ channel with masses around 1.5, 2 and  2.5 TeV. We also require that resonances in the $IJ=00$ (isoscalar) and $IJ=20$ (isotensor) channels are not present in the spectrum, since we do not consider them in this work. These particular mass values for the vector resonances, belonging to the interval (1000, 3000) GeV have been chosen on purpose as illustrative examples of the a priori expected reachable masses at the LHC.  In the following sections we will use these benchmark points to predict the visibility of vector resonances that may exist in the $IJ=11$ channel, and therefore resonate in the process $WZ \to WZ$ at the LHC. For the $IJ=00$ channel there are recent alternative studies of the IAM scalar resonances and their production at the LHC, see for instance~\cite{BuarqueFranzosi:2017prc}. 

The selected points in \tabref{tablaBMP} are also included in our previous contour plots in Fig.~\ref{fig:contourMW}. They are placed at the upper and lower horizontal axes in these plots, and are chosen on purpose at the two boundary values of the $a$ parameter: 1) $a=1$ for BP1, BP2 and BP3 and 2) $a=0.9$ for  BP1', BP2' and BP3'. These will be our main reference scenarios to which we will devote most of our LHC analysis. However, in order to provide a complementary study of the sensitivity to the $a$ parameter we have also defined a family of additional scenarios belonging to these contour lines of fixed $M_V=1500$, $2000$ and $2500$ GeV, respectively,   but with different values of $a$ in the interval $(0.9,1)$. These BP points are specified by circles, squares and triangles in Fig.~\ref{fig:contourMW} and will also be discussed in the final section.

\section{Dealing with IAM vector resonances in WZ scattering: the IAM-MC}
\label{sec-model}
In order to study how the vector resonances that are predicted in the IAM could be seen at the LHC with a Monte Carlo analysis, we need first to
establish a diagrammatic procedure for $WZ \to WZ$ scattering to implement the basic ingredients of these IAM
resonances in a Lagrangian framework. The use of Monte Carlo event generators like MadGraph requires the
model ingredients to be implemented in a Lagrangian language, which means in our case that we have to specify the
interactions of the emergent vector resonances with the gauge bosons (and Goldstone bosons). Thus, instead
of implementing the $A(W_L Z_L \to W_L Z_L)$ scattering amplitude in terms of the predicted IAM
partial waves, we simulate this scattering amplitude with a simple model that contains the
basic ingredients of the emergent vector resonances. Namely, the mass, the width and the proper couplings to
the gauge bosons $W$ and $Z$.  The simplest Lagrangian to include these vector resonances, $V$, that shares the
chiral and gauge symmetries of the EChL is provided in Refs.~\cite{Ecker:1989yg,DAmbrosio:2006xmn,Pich:2015kwa,Pich:2016lew}.
In the Proca 4-vector formalism,  the corresponding $P$-even Lagrangian is given by:
\begin{align}
\mL_V =& -\Frac{1}{4}{\rm Tr}( {\hat V}_{\mu\nu} {\hat V}^{\mu\nu}) +
 \Frac{1}{2} M_V^2 {\rm Tr}( {\hat V}_\mu {\hat V}^\mu )
 \, + \,\Frac{f_V}{2\sqrt{2}} {\rm Tr}( {\hat V}_{\mu\nu} f_+^{\mu\nu})
 \,\nn\\ &+ \,
\Frac{i g_V}{2\sqrt{2} } {\rm Tr}( {\hat V}_{\mu\nu} \, [u^\mu,u^\nu ] )\, ,
\label{Proca}
\end{align}
which includes the isotriplet vector resonances, $V^{\pm}$ and $V^0$, via the ${\hat V}_\mu$ field as well as their mass $M_V$, and couplings, $f_V$ and $g_V$, and where \cite{Pich:2012dv,Pich:2013fea,Pich:2012jv}:
\begin{align}
&u_\mu  =    \,
i\, u\, \Big(D_\mu \Ugold\Big)^\dagger u, {\rm with}\,\, u^2=\Ugold\,,\\[5pt]
&f_+^{\mu\nu}  =
\, -\, \left(
 u^\dagger \hat{W}^{\mu\nu}  u +   u \hat{B}^{\mu\nu} u^\dagger
\right)\, ,
\\[5pt]
&\nabla_\mu \mX \, =\, \partial_\mu \mX \, +\, [\Gamma_\mu , \mX ] , {\rm with}\,\, \Gamma_\mu =
\Frac{1}{2} \Big(\Gamma_\mu^{L} +\Gamma_\mu^{R}\Big)\, ,\\
&\Gamma_\mu^{L} = u^\dagger \left(\partial_\mu + i\,\frac{g}{2} \vec{\tau}\vec{W}_\mu
\right) u^{\phantom{\dagger}} 
\, , \quad\;
\Gamma_\mu^{R} = u^{\phantom{\dagger}}  \left(\partial_\mu + i\, \Frac{g'}{2} \tau^3 B_\mu\right)u^\dagger 
\,,
\end{align}
with the resonance fields are introduced as follows:
\begin{align}
&{\hat V}_\mu =\Frac{ \tau^a {V}_\mu^a}{\sqrt{2} } \,=\,
\left(\begin{array}{cc}
\Frac{V^0_\mu }{\sqrt{2}}  & V^+_\mu \\ V^-_\mu  &-\Frac{V_\mu^0}{\sqrt{2}}
\end{array}
\right)\,,
\\[5pt]
&{\hat V}_{\mu\nu} = \nabla_\mu {\hat V}_\nu -\nabla_\nu {\hat V}_\mu\, .
\end{align}

In the unitary gauge (convenient for tree-level collider analyses)
we have $u=\Ugold=\mathbb{1}$, and one finds a simpler result. In particular, after rotating to the mass eigenstate basis,
where the mixing terms between the $V$'s and the gauge bosons (introduced by $f_V \neq 0$)
are removed, and after bringing the kinetic and mass terms into
the canonical form, we find:
\begin{align}
\mL_V &=    -\Frac{1}{4} \Big( 2V^+_{\mu\nu}  V^{-\mu\nu}+V^0_{\mu\nu}‚ V^{0\mu\nu}\Big) +
 \Frac{1}{2} M_V^2 \Big( 2V^+_{\mu} V^{-\mu}+V^{0}_{\mu} V^{0\mu} \Big)
\nn\\[5pt]
&-\,\Frac{i f_V }{v^2}
\bigg[m_W^2 V^0_\nu  (W^+_\mu W^{-\,\mu\nu} -W^-_\mu W^{+\,\mu\nu} )
+ m_W m_Z V^+_\nu (W^-_\mu Z^{\mu\nu} -Z_\mu W^{-\, \mu\nu})
\nn\\
&\hspace{1.2cm}+ m_W m_Z V^-_\nu (Z_\mu W^{+\, \mu\nu}- W^+_\mu Z^{\mu\nu} )
\bigg]
\nn\\
&
+ \Frac{i 2g_V }{v^2} \bigg[ m_W^2 V^{0\,\, \mu\nu} W_\mu^+ W_\nu^-
+  m_W \, m_Z\,  V^{+\,\, \mu\nu} W_\mu^- Z_\nu
+  m_W \, m_Z\,  V^{-\,\, \mu\nu} Z_\mu W_\nu^+   \bigg]\, ,
\label{LVugauge}
\end{align}
where we have used the short-hand notation
$V^a_{\mu\nu}= \partial_\mu V^a_\nu - \partial_\nu V^a_\mu$ (for $a=\pm ,0$),
$W^a_{\mu\nu}= \partial_\mu W^a_\nu - \partial_\nu W^a_\mu$  (for $a=\pm$), and
$Z_{\mu\nu}= \partial_\mu Z_\nu - \partial_\nu Z_\mu$.

It should be noticed that in the previous Lagrangian of \eqref{LVugauge} there are not interaction terms between the
vector resonances and two neutral gauge bosons, $VZZ$, (as there are not either  $Vzz$ interactions in
\eqref{Proca} of $V$ with two neutral Goldstones) and this explains why the vector resonances cannot
emerge in the s-channel of $WW \to ZZ$ nor
$ZZ \to ZZ$\footnote{Notice that scalar resonances could resonate in these channels, but we do not considered them here.}. This is a clear consequence of exact custodial invariance and it also confirms that $W^{\pm}Z \to W^{\pm}Z$ are the proper channels to look for emergent
signals from the charged vector resonances $V^{\pm}$. The relevant set of Feynman rules extracted from the
above Lagrangian in \eqref{LVugauge} are collected in Appendix \ref{FR-IAMMC}, for completeness.

Since we are mostly interested here in the deviations with respect to the SM predictions in the case of the
longitudinal modes,  we will mainly focus on their scattering amplitudes.
Therefore, from now on we will simplify our study by setting $f_V=0$. This is well justified since
this $f_V$ predominantly affects the couplings of the resonances to transverse gauge bosons and,
in consequence, $g_V$ is the most relevant coupling to the longitudinal modes.
Some additional comments on the behavior of the scattering amplitudes for the other modes will be made at the end of this section. 

Our aim here is to use the Lagrangian $\mL_V$ in \eqref{LVugauge} as a practical tool to
mimic the main features of
the vector resonances found with the IAM. Specifically, we wish to introduce all these features by means of a tree level computation of 
$A(WZ \to WZ)$ with
$\mL_{\rm model}=\mathcal{L}_2 +\mathcal{L}_V$.
This leads us to the issue of relating $g_V$, $M_V$ and  $\Gamma_V$ to the properties of the
IAM vector resonances found from $a^{11}_{\rm IAM}$.

On one hand, the mass and the width are obviously related to
the position of the pole, $s_{\rm pole}=(M_V-\frac{i}{2}\Gamma_V)^2$,  of $a^{11}_{\rm IAM}(s)$. On the other hand,
the coupling $g_V$ should also be related to the properties of $a^{11}_{\rm IAM}(s)$ in the resonant region.
For instance, one could extract a value of $g_V$ by identifying the residues of $a^{11}_{\rm model}(s)$ and
$a^{11}_{\rm IAM}(s)$ at $s_{\rm pole}$. If, for simplicity, we had used the ET version of the relevant amplitudes,
this would have led to the simple relation $g_V^2=2(a_4-2a_5)$. Alternatively,
one could follow the approach of
 \cite{Pich:2015kwa,Pich:2016lew}
where close to the resonance mass shell, they find $\mL_{\rm model}$
to be equivalent to a more general Lagrangian\footnote{
The Lagrangian in Refs.~\cite{Pich:2015kwa,Pich:2016lew} considers the antisymmetric tensor representation
for the spin--1 resonances, which is fully equivalent to the Proca four-vector representation
provided appropriate non-resonant operators are added to the Lagrangian.}
in which the on-shell vector coupling $g_V$ is related to the $\mathcal O(p^4)$ low-energy chiral parameters
in the form $a_4=-a_5=g_V^2/4$.

\begin{figure}[t!]
\begin{center}
\includegraphics[width=.6\textwidth]{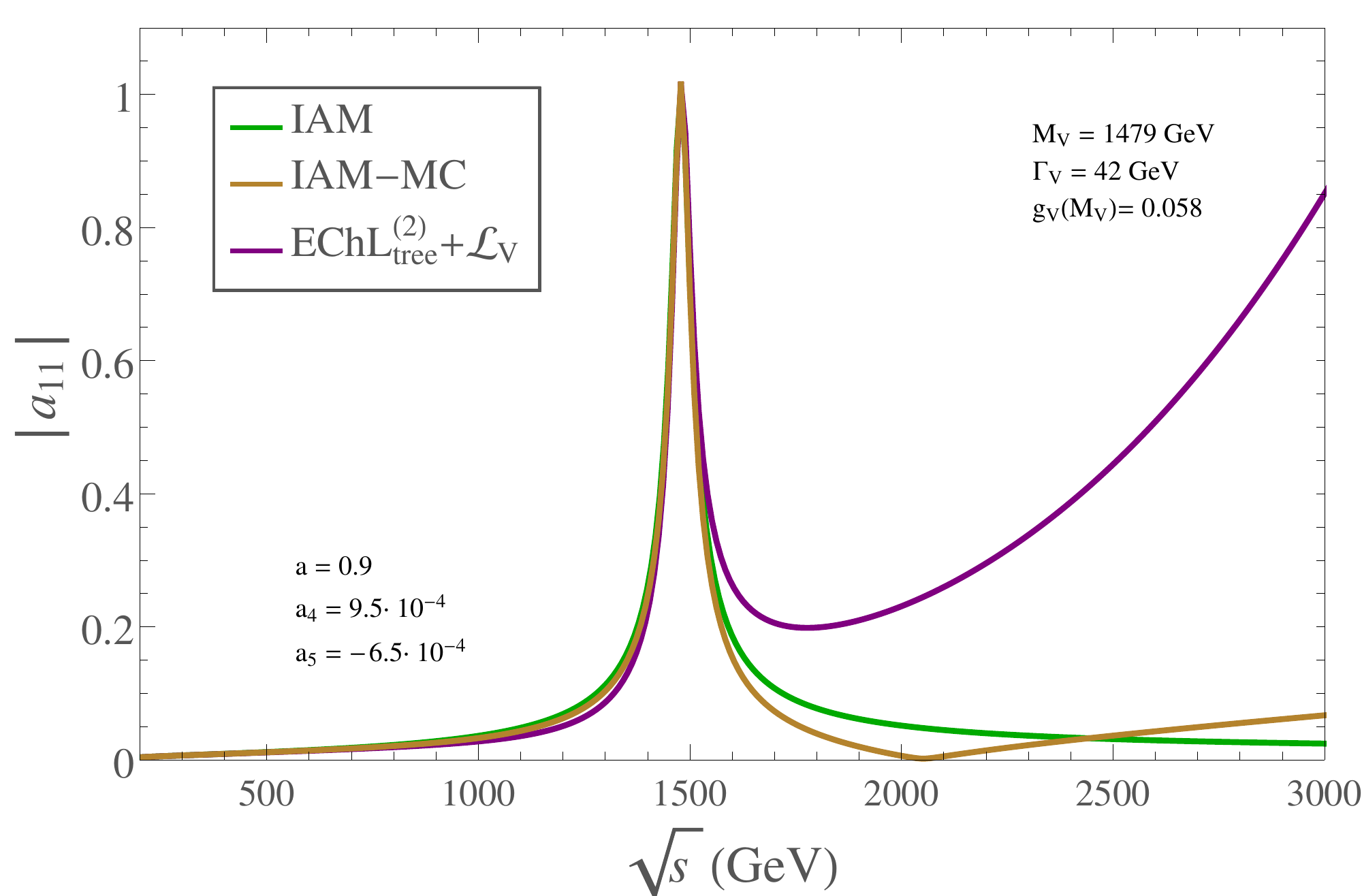}\\
\caption{Prediction of the $|a^{11}|$ partial wave as a function of the center of mass energy $\sqrt{s}$ in the three models explained in the text: IAM (green), IAM-MC (orange) and $\mL_2+\mL_V$ with constant $g_V$ (purple). The values of the parameters are those of BP1' in \tabref{tablaBMP}.}
\label{fig:procagvconst}
\end{center}
\end{figure}

However, this Lagrangian $\mL_2+\mL_V$ leads to problems if a  constant $g_V$ is assumed.
Even though it gives a reasonable estimate of the partial wave at $s\sim M_V^2$,
it does not work satisfactorily away from the resonance region. Indeed, it yields to a bad high energy
behavior for $s>M_V^2$: the subsequent partial wave $a^{11}(s)$ grows too fast with energy and crosses
the unitary bound at energies of a few TeV. This unwanted violation of unitarity happens, indeed, for any choice of the
constant $g_V$ in the Lagrangian $\mL_2+\mL_V$.

We depict this failure in \figref{fig:procagvconst} for one particular example with $a=0.9$, $a_4=9.5\times 10^{-4}$
and $a_5=-6.5\times 10^{-4}$ that produces a IAM vector pole at $M_V=1479$ GeV with
$\Gamma_V=42$ GeV, and where we have assumed a constant value of $g_V=0.058$.  In this case we have found that the crossing over the
unitarity bound occurs at around 3 TeV. From this study, we conclude then that the $a^{11}(s)$ resulting
from $\mL_2+\mL_V$ with constant $g_V$ does not simulate correctly the behaviour of $a^{11}_{\rm IAM}$, which is
by construction unitary and therefore we will not take $g_V$ as a constant coupling.

We will define in the following the specific model that we choose to mimic with a chiral Lagrangian the IAM amplitude, which is referred in \figref{fig:procagvconst} as IAM-MC. 
This will obviously lead us to consider again $\mL_2+\mL_V$ but with a momentum dependent $g_V$.

We work with the Lagrangian  $\mL_2+\mL_V$, first introduced in the EW interaction basis in Eqs~(\ref{eq.L2}) and (\ref{Proca}),
 to mimic the IAM amplitude of $WZ$ scattering but with an energy dependent coupling $g_V(s)$ (remember that we are setting $f_V=0$ in all our numerical estimates), which leads to unitary results in the way that will be described in this subsection. 
 
 Firstly, our $A(W_L Z_L \to W_L Z_L)$ amplitudes have by construction the resonant behavior of the IAM amplitudes at $s_{\rm pole}=(M_{V}-\frac{i}{2}\Gamma_{V })^2$, as commented above. Secondly, it is illustrative to notice that
the effective coupling $g_V(s)$ is in fact related to a form factor, as can be seen for instance using a current algebra language. Concretely, the matrix element of a vector current between two longitudinal $W$ bosons
and the vacuum is described by an energy dependent form factor  $G_V(s)$ given by \cite{Arnan:2015csa}:
\be
\langle W_L^i(k_1)W_L^j(k_2)|J^k_\mu|0\rangle=(k_1-k_2)_\mu G_V(s)\epsilon^{ijk},
\label{genformf}
\ee
\begin{figure}[t!]
\begin{center}
\includegraphics[width=.49\textwidth]{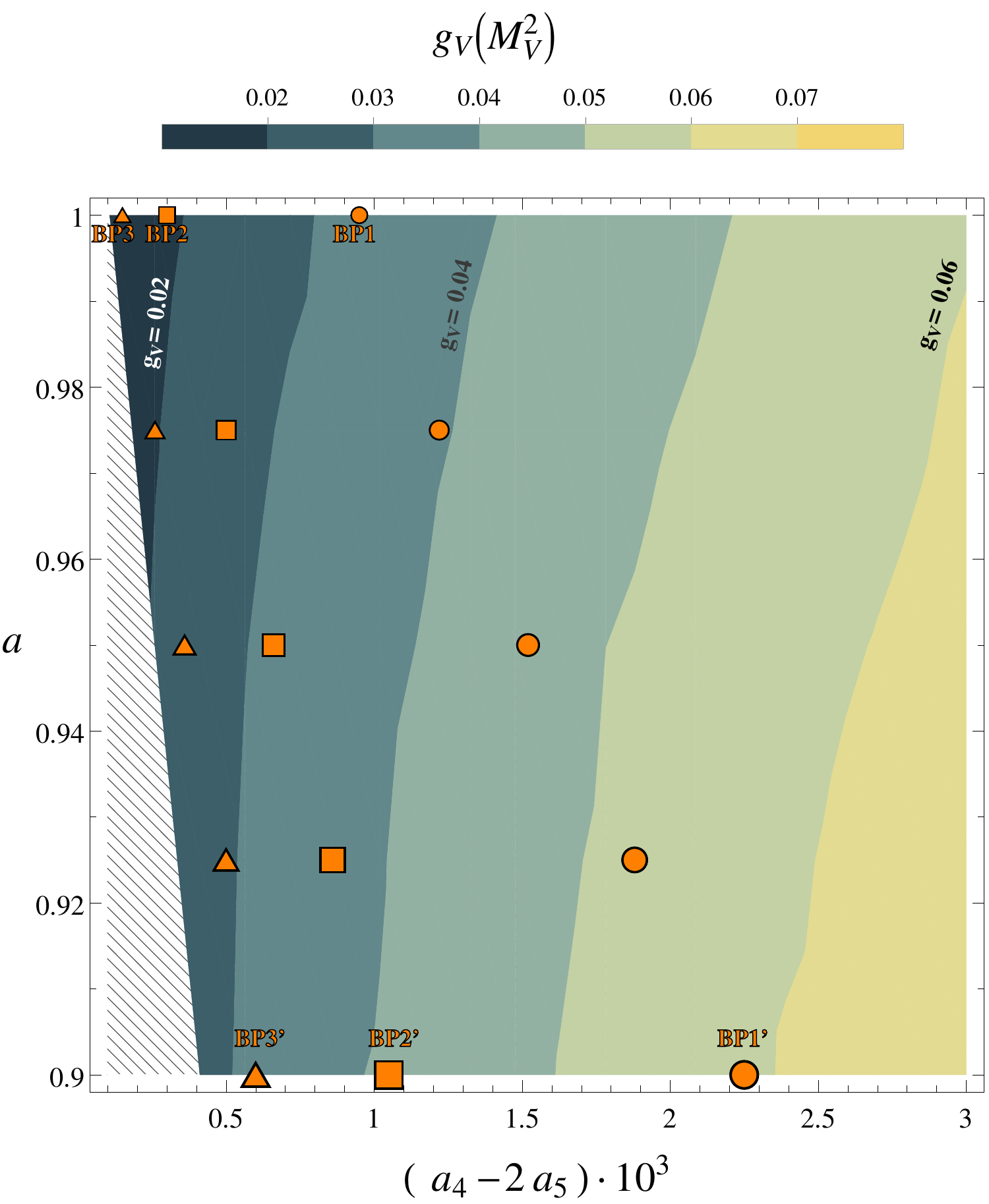}
\caption{Predictions of $g_V(M_V^2)$ as a function of $a$ and $(a_4-2a_5)$   computed from Eq.(\ref{a11MV}), as discussed in the text. The benchmark points specified with geometric symbols correspond respectively to those in \figref{fig:contourMW}.}
\label{fig:contourgv}
\end{center}
\end{figure}
where $J^k_\mu$ is the interpolating vector current with isospin index $k$ that creates a resonance $V$. This form factor $G_V(s)$ can be easily related to $g_V(s)$ at $s=M_V^2$ by $G_V(M_V^2)=\sqrt{2}M_V^2g_V(M_V^2)/v^2$. In practice, $g_V(M_V^2)$ is determined by the matching procedure described next.

In order to build our resonant
$A(W_L Z_L \to W_L Z_L)$ amplitudes we use the following prescription. First, we impose the matching at the partial waves level. Concretely, it is performed by identifying the tree level predictions
from $\mL_2+\mL_V$ with the predictions from the IAM at $M_V$, i.e:
\begin{equation}
\Big|a^{11}_{{\rm EChL}_{\rm tree}^{(2)}+\mL_V}(s=M_V^2)\Big|=\Big|a^{11}_{\rm IAM}(s=M_V^2)\Big|\,,
\label{a11MV}
\end{equation}
where $a^{11}_{{\rm EChL}_{\rm tree}^{(2)}+\mL_V}$ is the partial wave amplitude computed from $\mL_2+\mL_V$.

Solving (numerically) this  \eqref{a11MV} for the given values of $(a,a_4,a_5)$
and the corresponding values of $(M_V,\Gamma_V)$ leads to the wanted solution for $g_V=g_V(M_V^2)$.
For instance, in the previous example of $a=0.9$, $a_4=9.5\times 10^{-4}$ and $a_5=-6.5\times 10^{-4}$ (our benchmark point BP1' in \tabref{tablaBMP}) with
corresponding $M_V=1479$ GeV and $\Gamma_V=42$ GeV, we found $g_V(M_V^2)=0.058$. For the other selected benchmark
points the corresponding values found for $g_V(M_V^2)$ are collected in  \tabref{tablaBMP} and in \figref{fig:contourgv}.  Interestingly, these numerical results in \figref{fig:contourgv} for $g_V(M_V^2)$ show a clear correlation with the previously predicted $M_V$ and $\Gamma_V$ values in \figref{fig:contourMW}, which fulfill approximately:  $\Gamma_V \simeq M_V^5 g_V^2/(48 \pi v^4)$, as naively expected from the Proca Lagrangian for $f_V=0$. 
 
One may notice at this point that the computation of the IAM partial waves has been done with electroweak gauge bosons in the external legs and not with Goldstone bosons. The ET has only been used to compute the real part of the loops involved, as explained before in the previous section.

Away from the resonance we consider an energy dependence in $g_V(s)$ with the following requirements:
\begin{itemize}
\item[i)] Below the resonance, at low energies, one should find compatibility with the result from ${\rm EChL}^{(2+4)}_{\rm loop}$, which implies that the predictions from $\mL_V$ should match those from $\mL_4$ at these energies.
This is what happens indeed to $a^{11}_{\rm IAM}$ below the resonance, by construction.
\item[ii)] Above the resonance, at large energies, we require the cross section not to grow faster than the Froissart bound~\cite{PhysRev.123.1053}, which can be written as:
\begin{equation}
\sigma(s)  \le \sigma_0 \log^2\bigg(\frac{s}{s_0}\bigg)\,,\label{froisbound}
\end{equation}
with $\sigma_0$ and $s_0$ being energy independent quantities. Notice that this definition is equivalent to the one presented in \eqref{froi}.
When using this bound we are implicitly assuming that there are no other resonances (in addition to $V$) emerging in the spectrum, at least until very high energies.
\end{itemize}
We have found that these requirements above are well approximated by setting the following simple function:
\begin{align}
g_V^2(s)&=g_V^2(M_V^2) \frac{M_V^2}{s} \,\,\, {\rm for} \,\, s< M_V^2 \nn\,, \\
g_V^2(s)&=g_V^2(M_V^2) \frac{M_V^4}{s^2} \,\,\, {\rm for} \,\, s> M_V^2\,.
\label{gvenergy}
\end{align}
This $g_V(s)$ coupling should be used when $V$ is propagating in the $s$-channel. In the other channels where the resonance could also propagate, $t$ and/or $u$ channels, the coupling should be the same described in \eqref{gvenergy} in terms of the corresponding $t$ or $u$ variables to be fully crossing symmetric. Nevertheless, we have checked that a completely crossing symmetric energy-dependent coupling, given by $g_V^2(z)=\theta(M_V^2-z)g_V^2(M_V^2)\frac{M_V^2}{z}+\theta(z-M_V^2)g_V^2(M_V^2)\frac{M_V^4}{z^2}$, leads to a moderate violation of the Froissart bound in \eqref{froisbound} at energies in the TeV range. To avoid this violation of unitarity, we propose the following expression for the coupling in terms of the $t$ and $u$ variables:
\begin{align}
g_V^2(z)&=g_V^2(M_V^2) \frac{M_V^2}{z} \,\,\, {\rm for} \,\, s< M_V^2 \nn\,, \\
g_V^2(z)&=g_V^2(M_V^2) \frac{M_V^4}{z^2} \,\,\, {\rm for} \,\, s> M_V^2\,,
\label{gvenergytu}
\end{align}
with $z=t,u$ corresponding to the $t,u$ channels, respectively, in which the resonance is propagating.

The accuracy of the result with this choice of energy dependent coupling in comparison with the previous constant coupling can be seen in \figref{fig:procagvconst}. It is clear from this figure that the result for $a^{11}$ using this energy dependent coupling simulates much better the IAM result than that with a constant $g_V$, and it also provides a good low and high energy behaviors.  It is worth commenting that we have tried other choices for the dependence with energy of this $g_V(s)$ coupling, but none of these alternative tries has passed all the above required conditions.
We have also checked explicitly that our hypothesis in \eqrefs{gvenergy}{gvenergytu} leads to a high-energy behavior of the cross section that is always below and close to the saturation of this Froissart bound.

The above described method, which will be called from now on IAM-MC (named after IAM for MonteCarlo), is the one we choose
to simulate the IAM with a Lagrangian formalism. We find that it is the most appropriate one for the forthcoming MonteCarlo
analysis with MadGraph5 of LHC generated events.

In summary, we follow the subsequent steps to get $A(W_LZ_L \to W_LZ_L)_{\rm IAM-MC}$ for each of the
given $(a,a_4,a_5)$ input values:
\begin{itemize}
\item[1)] Compute the amplitude from the tree level diagrams with the Feynman rules from $\mL_2+\mL_V$.
This gives a result in terms of $a,M_V,g_V$ and $\Gamma_V$.
\item[2)] For the given values of $(a,a_4,a_5)$, then set $M_V$ and $\Gamma_V$ to the corresponding values found from the
poles of $a^{11}_{\rm IAM}$.
\item[3)] Extract the value of $g_V(M_V^2)$ by solving numerically \eqref{a11MV}.
\item[4)] Substitute $g_V$ by $g_V(s)$ in the $s$-channel and by $g_V(u)$ in the $u$-channel (for the process of study, $WZ\to WZ$, the charged vector resonance only propagates in these two channels) and use \eqrefs{gvenergy} {gvenergytu}.
\item[5)] Above the resonance we assume that the deviations with respect to the SM come dominantly from $\mL_V$, which means in practice that the proper Lagrangian for the computation
of the IAM simulated amplitude is
$\mL_{\rm SM}+\mL_V$ rather than $\mL_2+\mL_V$. This is obviously equivalent to use $\mL_2+\mL_V$ with $a=1$ at energies above the resonance.
\end{itemize}

\begin{figure}[t!]
\begin{center}
\includegraphics[width=.49\textwidth]{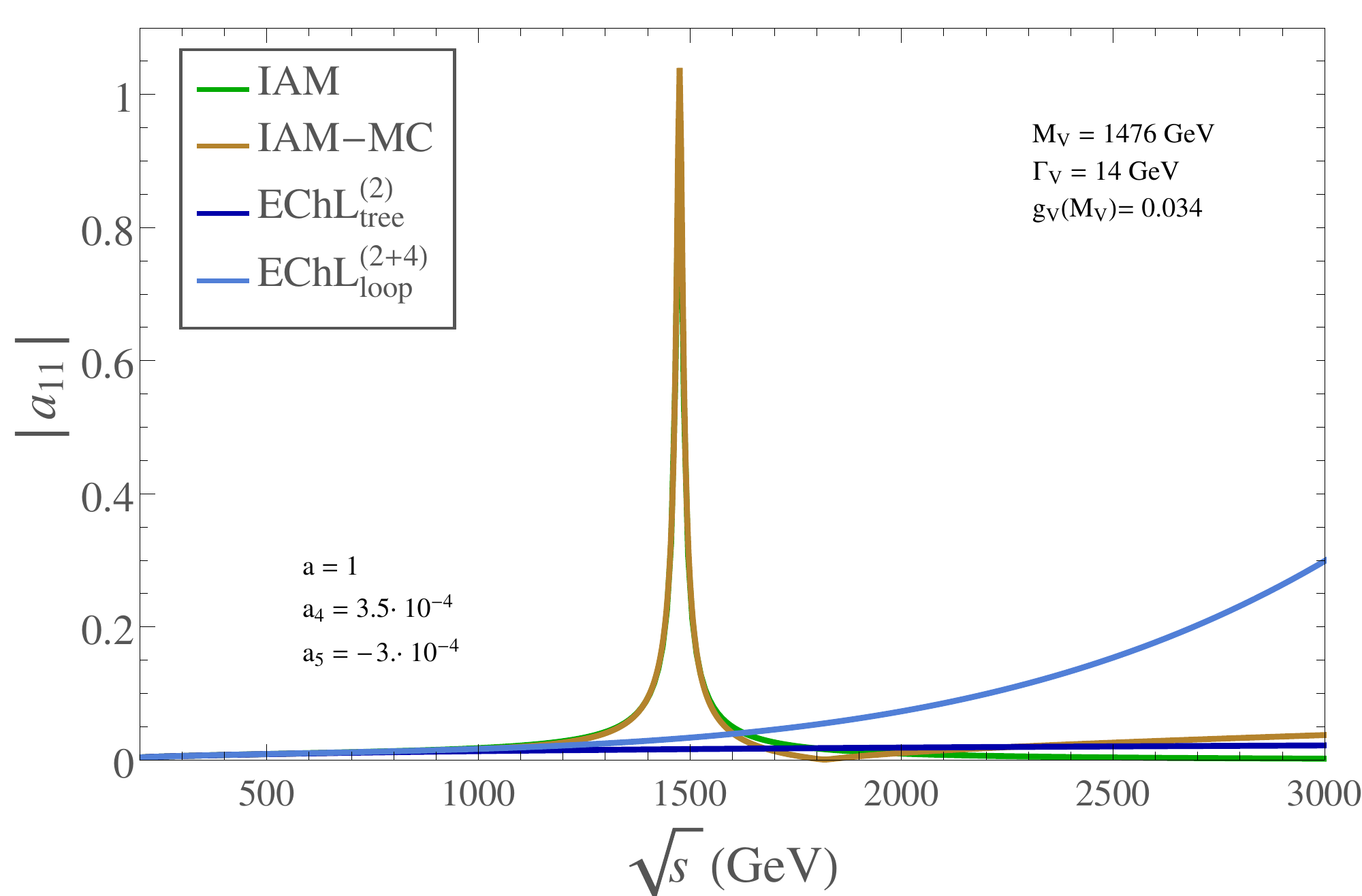}
\includegraphics[width=.49\textwidth]{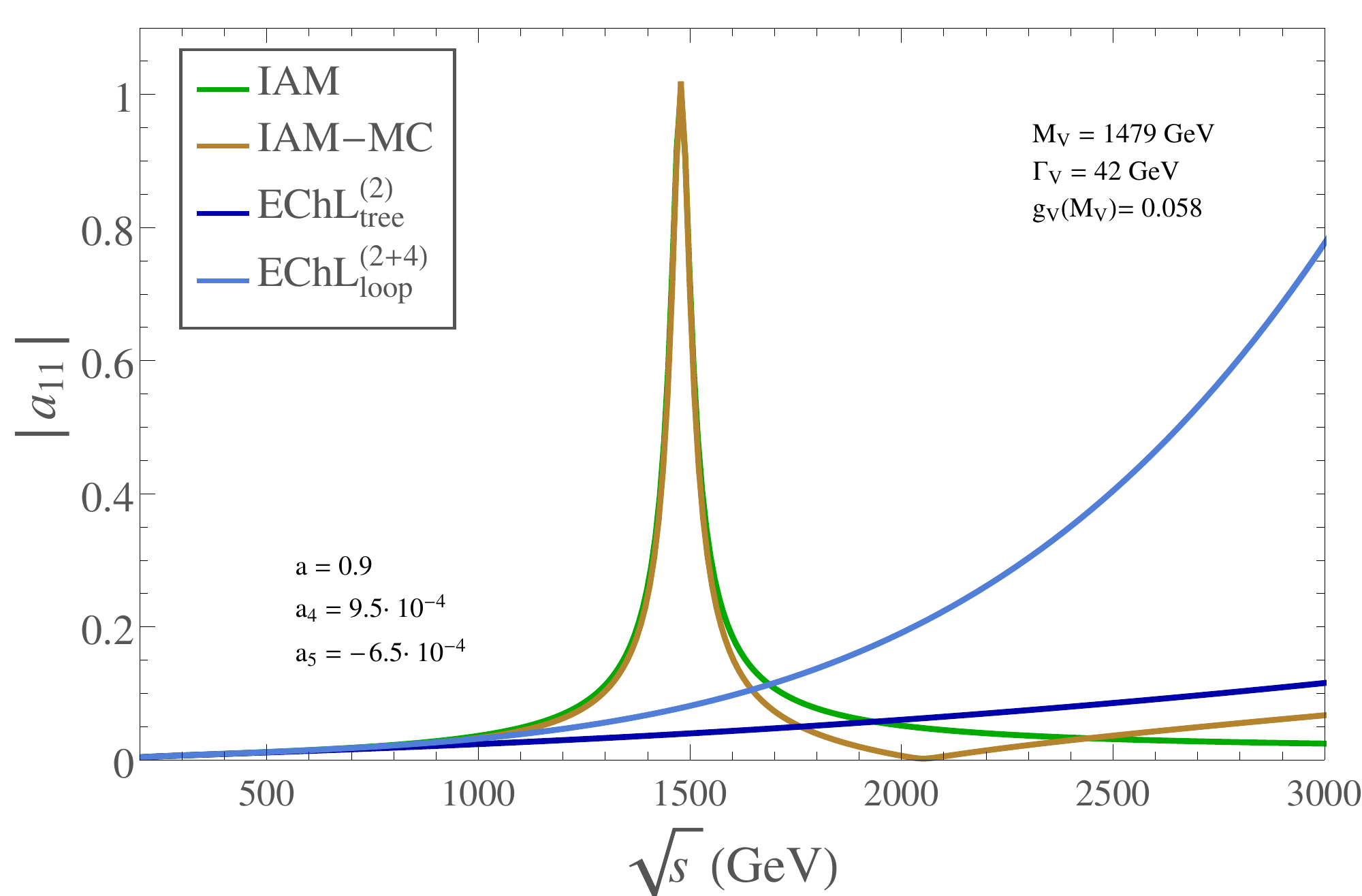}\\
\includegraphics[width=.49\textwidth]{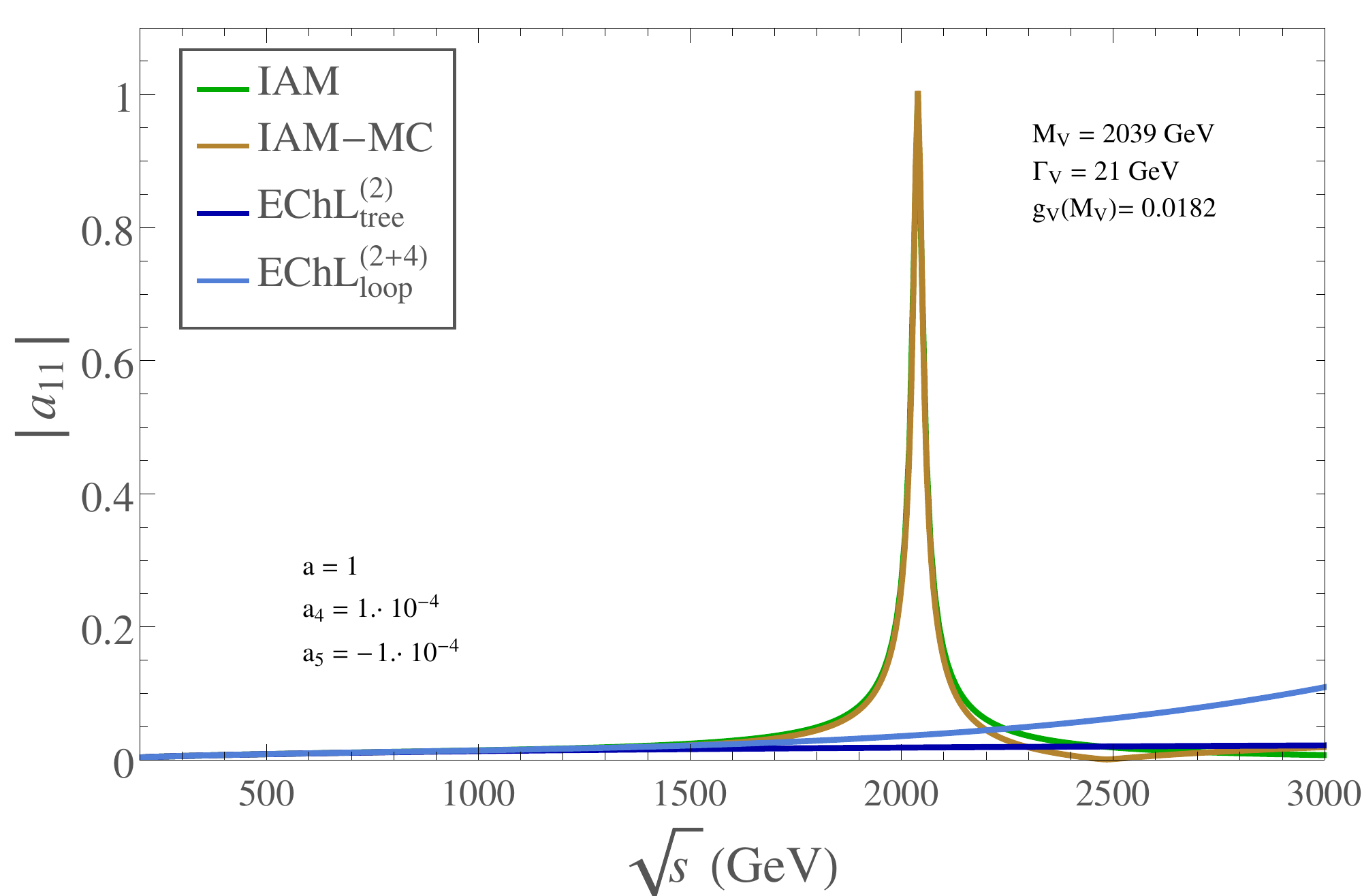}
\includegraphics[width=.49\textwidth]{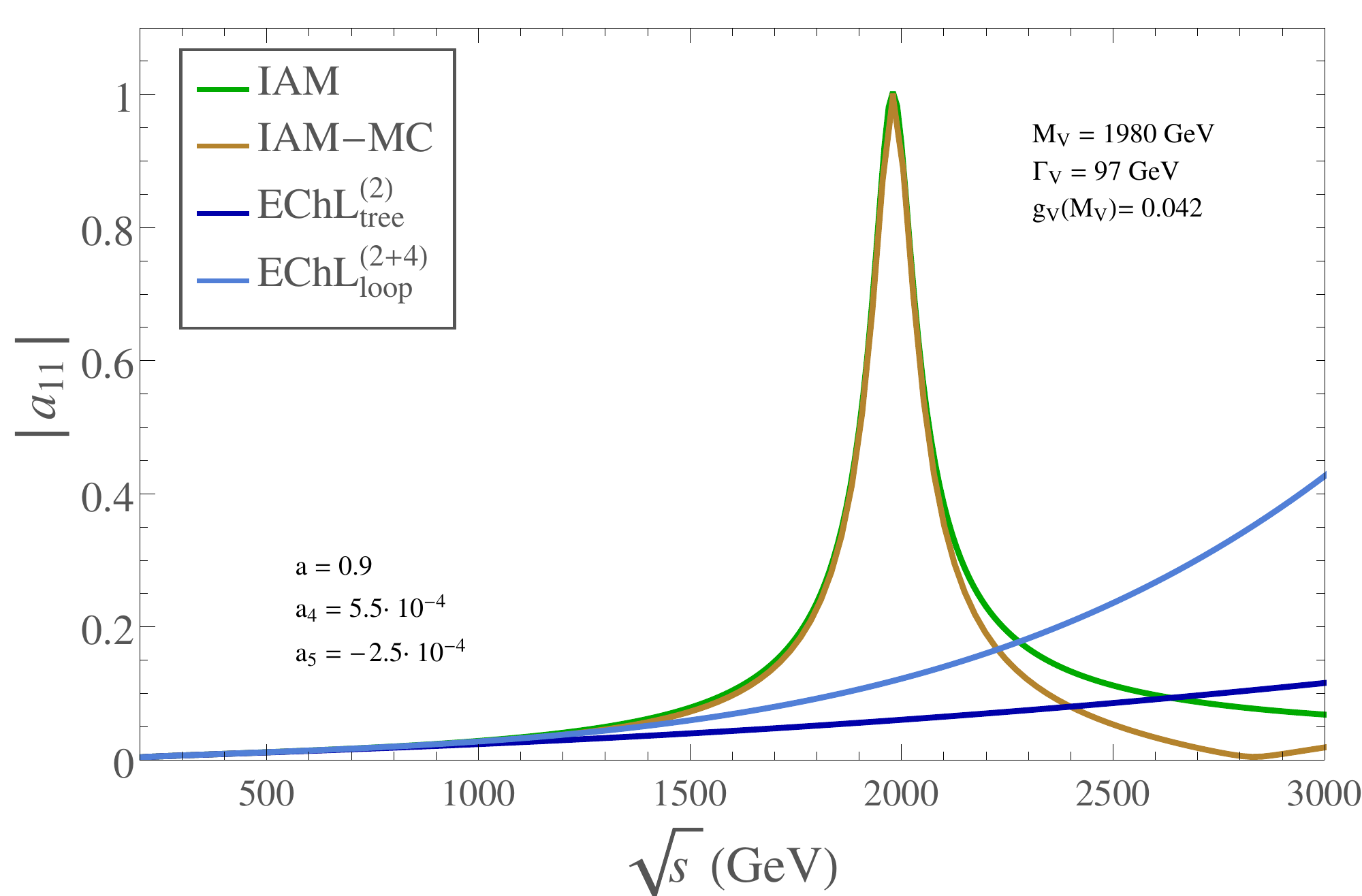}\\
\includegraphics[width=.49\textwidth]{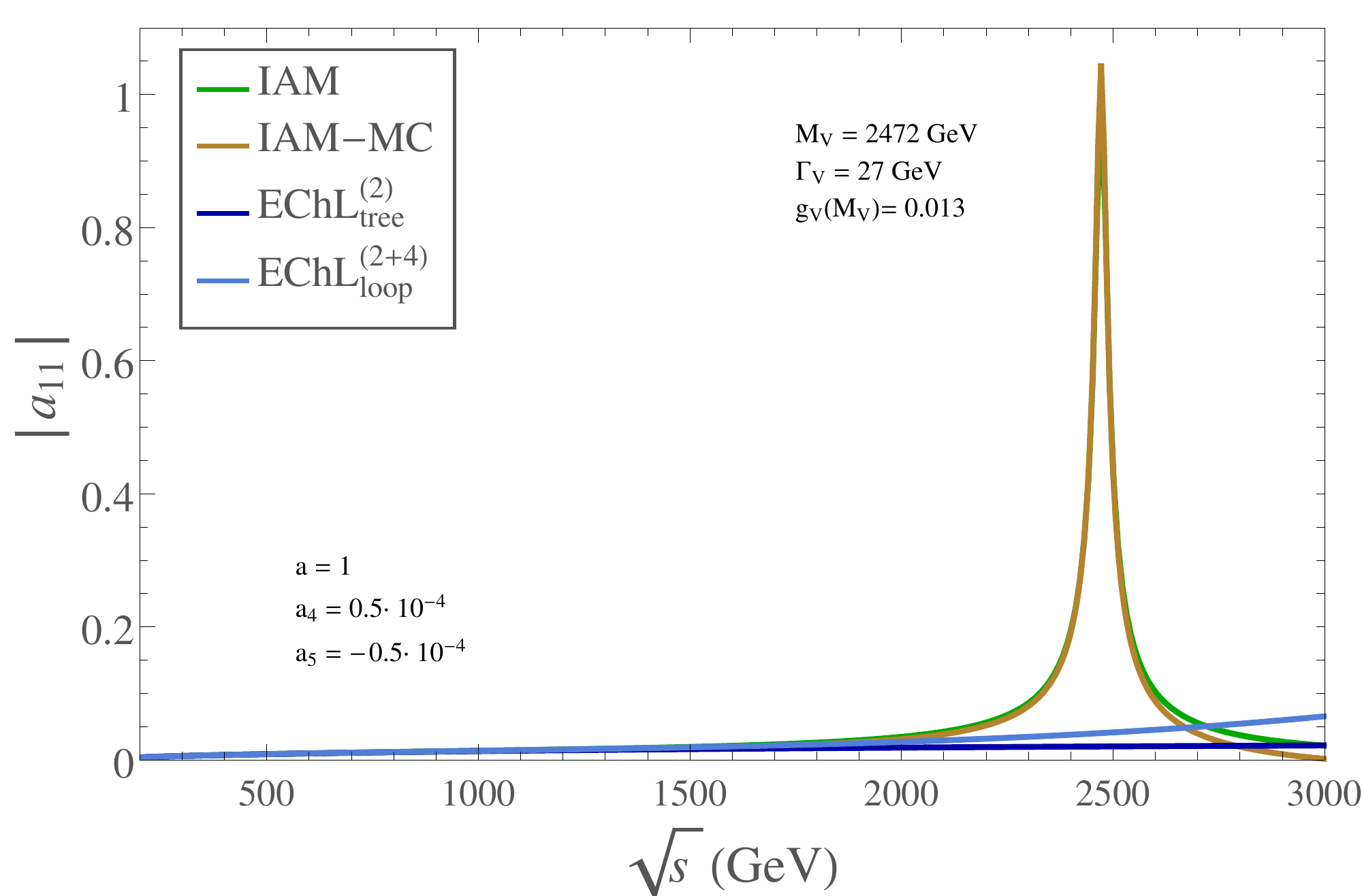}
\includegraphics[width=.49\textwidth]{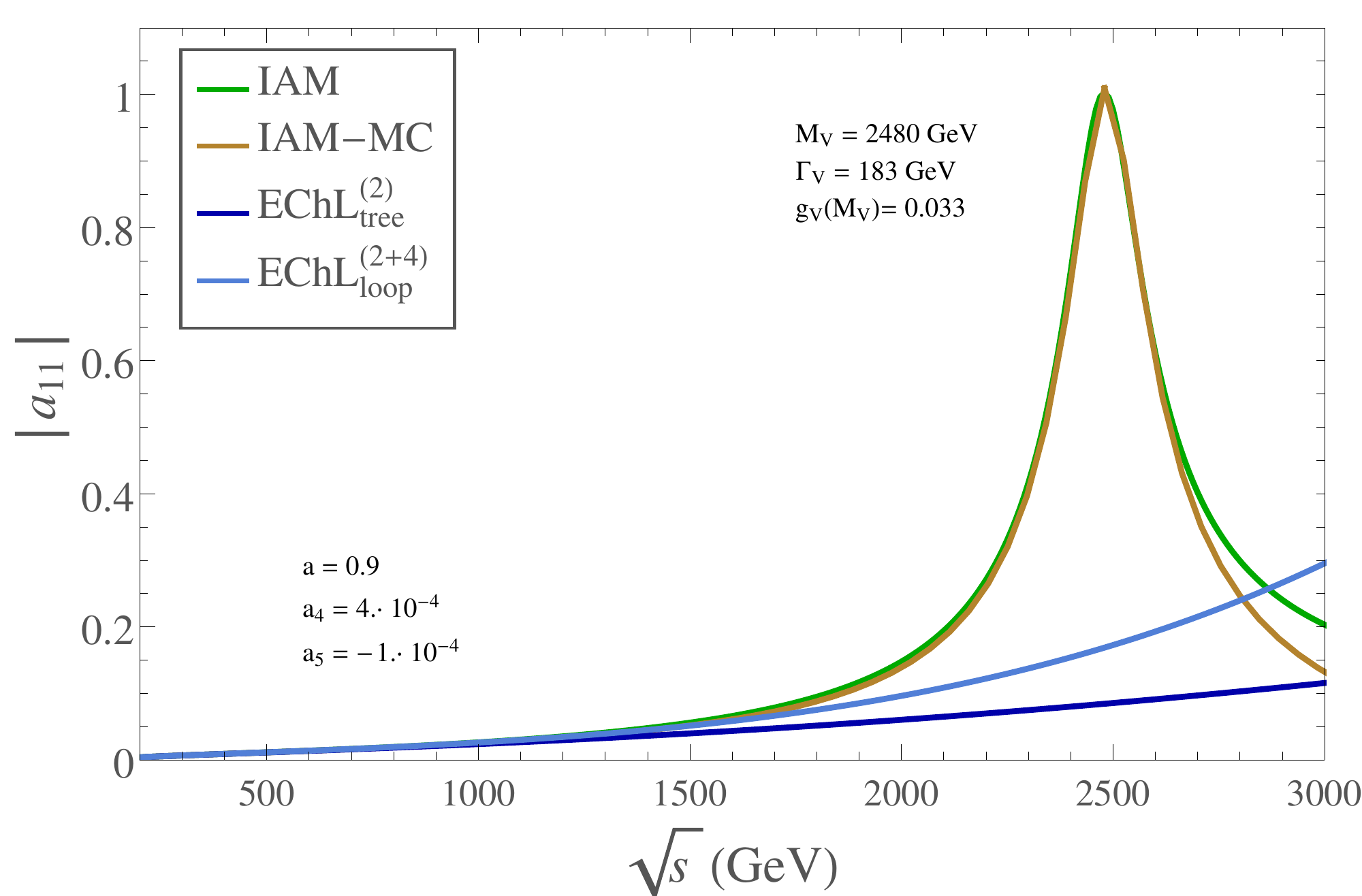}\\
\caption{Predictions of the $|a^{11}|$ partial waves as a function of the center of mass energy $\sqrt{s}$ for all the selected benchmark points in \tabref{tablaBMP}. Different lines correspond to the different models considered in the text:  EChL unitarized with the IAM (green), our IAM-MC model (orange),  non-unitarized EChL up to $\mO(p^2)$ (dark blue) and  non-unitarized EChL up to $\mO(p^4)$ including loop contributions (light blue).}
\label{fig:pw1}
\end{center}
\end{figure}

We present in \figref{fig:pw1} our predictions of the partial waves $a^{11}_{\rm IAM-MC}$ for all the selected benchmark points of \tabref{tablaBMP}. We have also included in these plots the corresponding predictions from the IAM and from the EChL, at both LO and NLO, for comparison. In these plots we clearly see the accuracy of our IAM-MC model in simulating the behaviour of the IAM amplitudes. This happens not only at the region surrounding the resonance, where it is clearly very good, but also below and above the resonance, inside the displayed energy interval of $\sqrt{s} \in (200,3000)$ GeV.

 \begin{figure}[t!]
\begin{center}
\includegraphics[width=.49\textwidth]{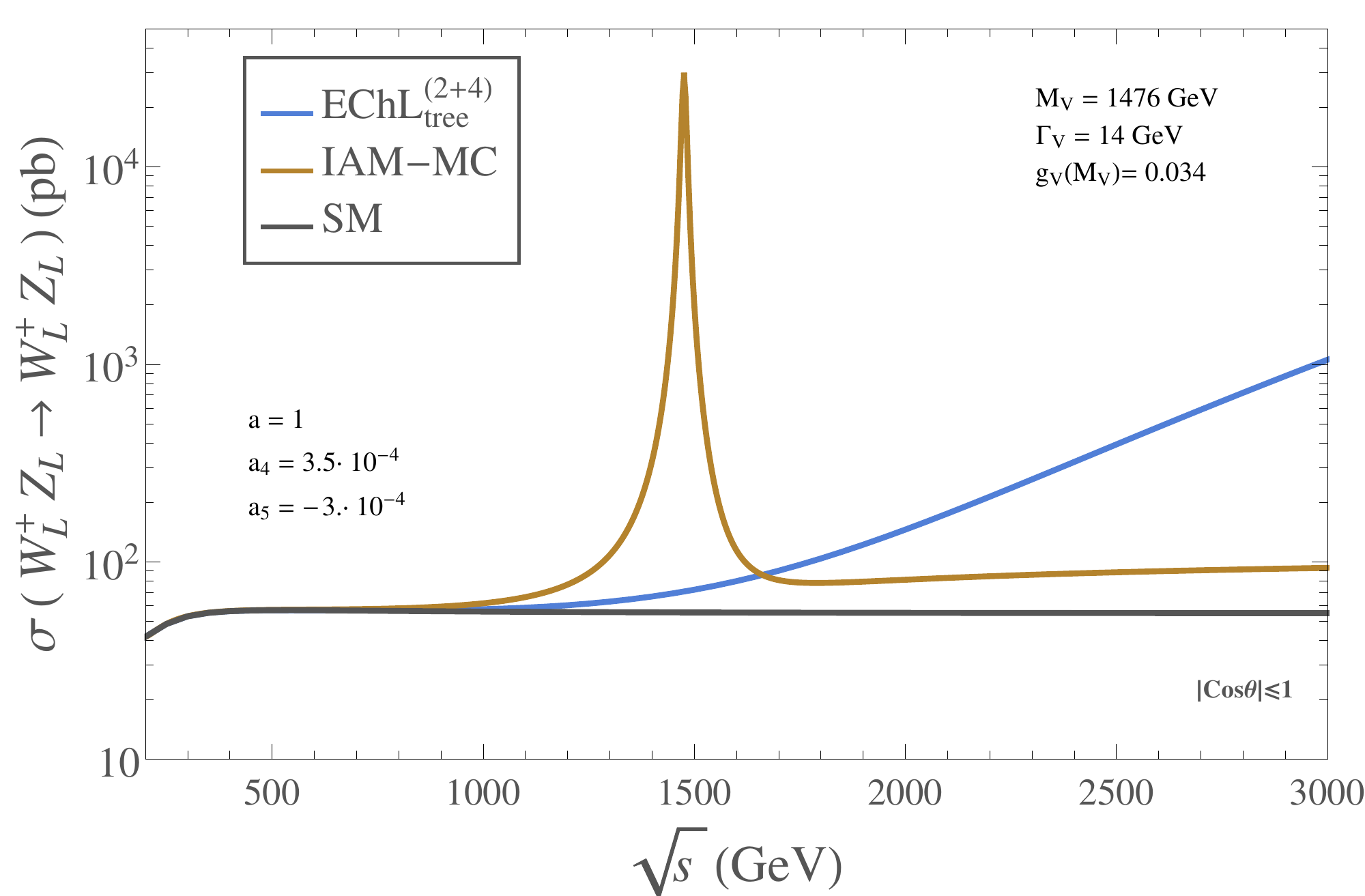}
\includegraphics[width=.49\textwidth]{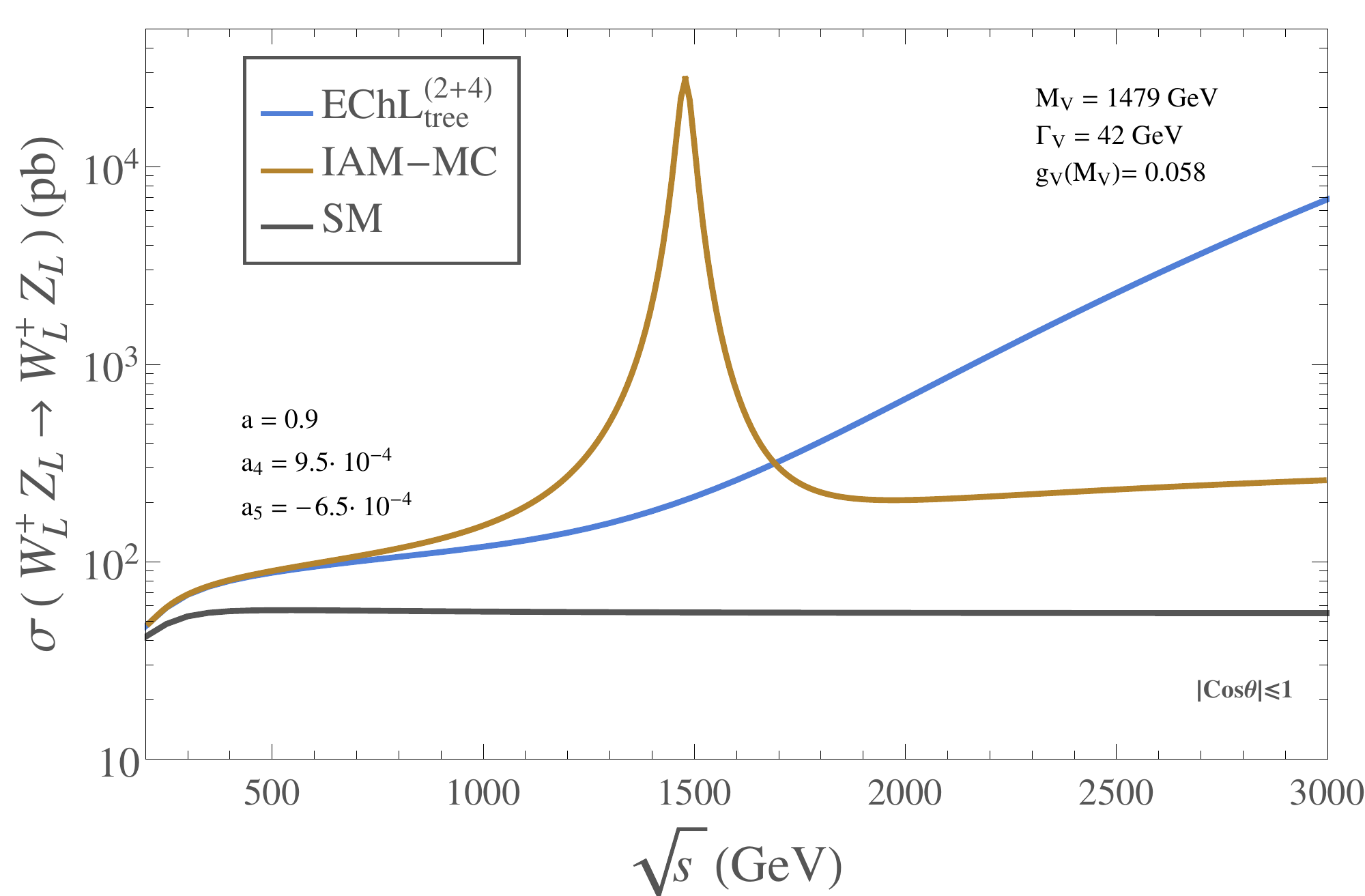}\\
\includegraphics[width=.49\textwidth]{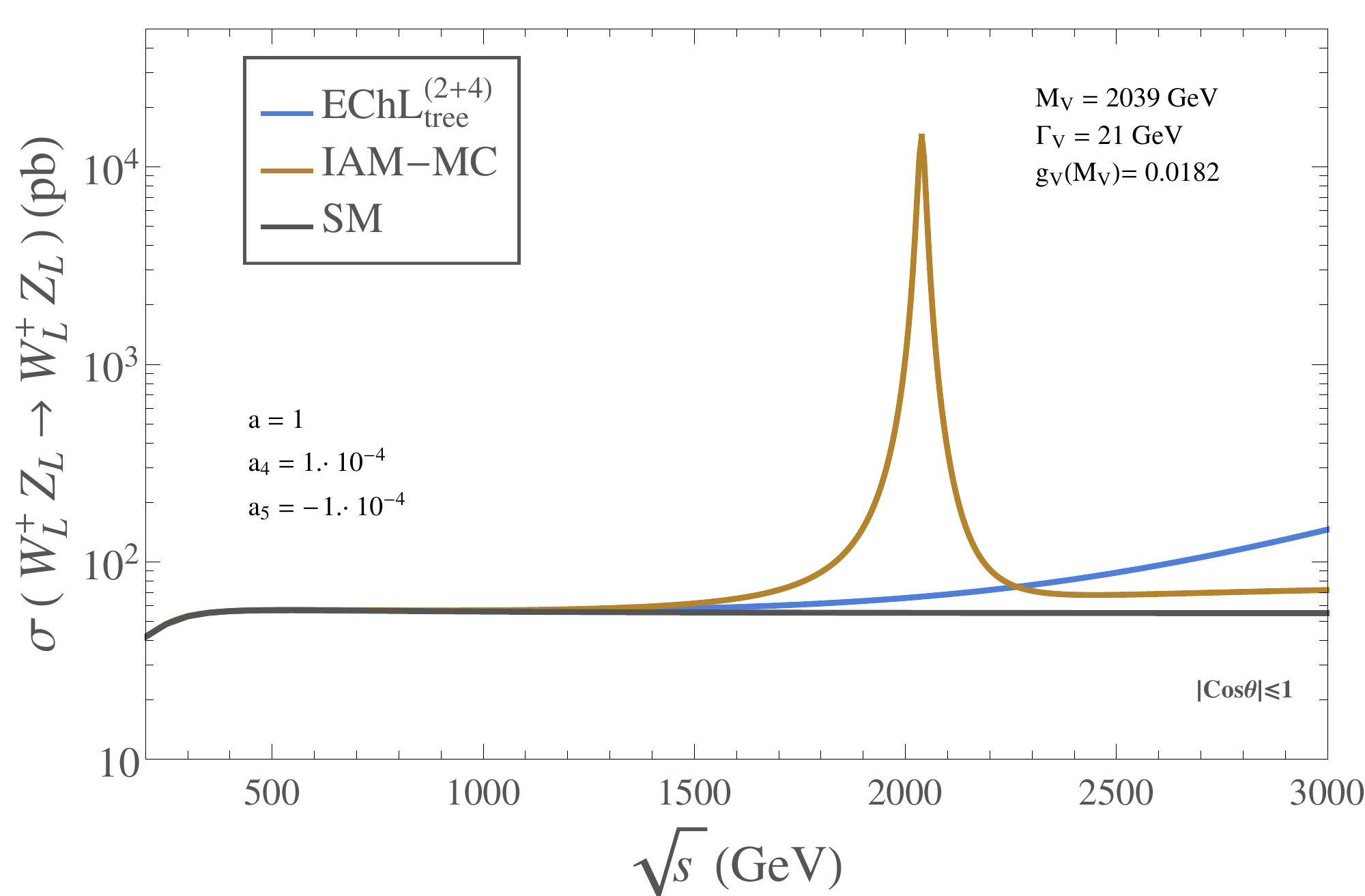}
\includegraphics[width=.49\textwidth]{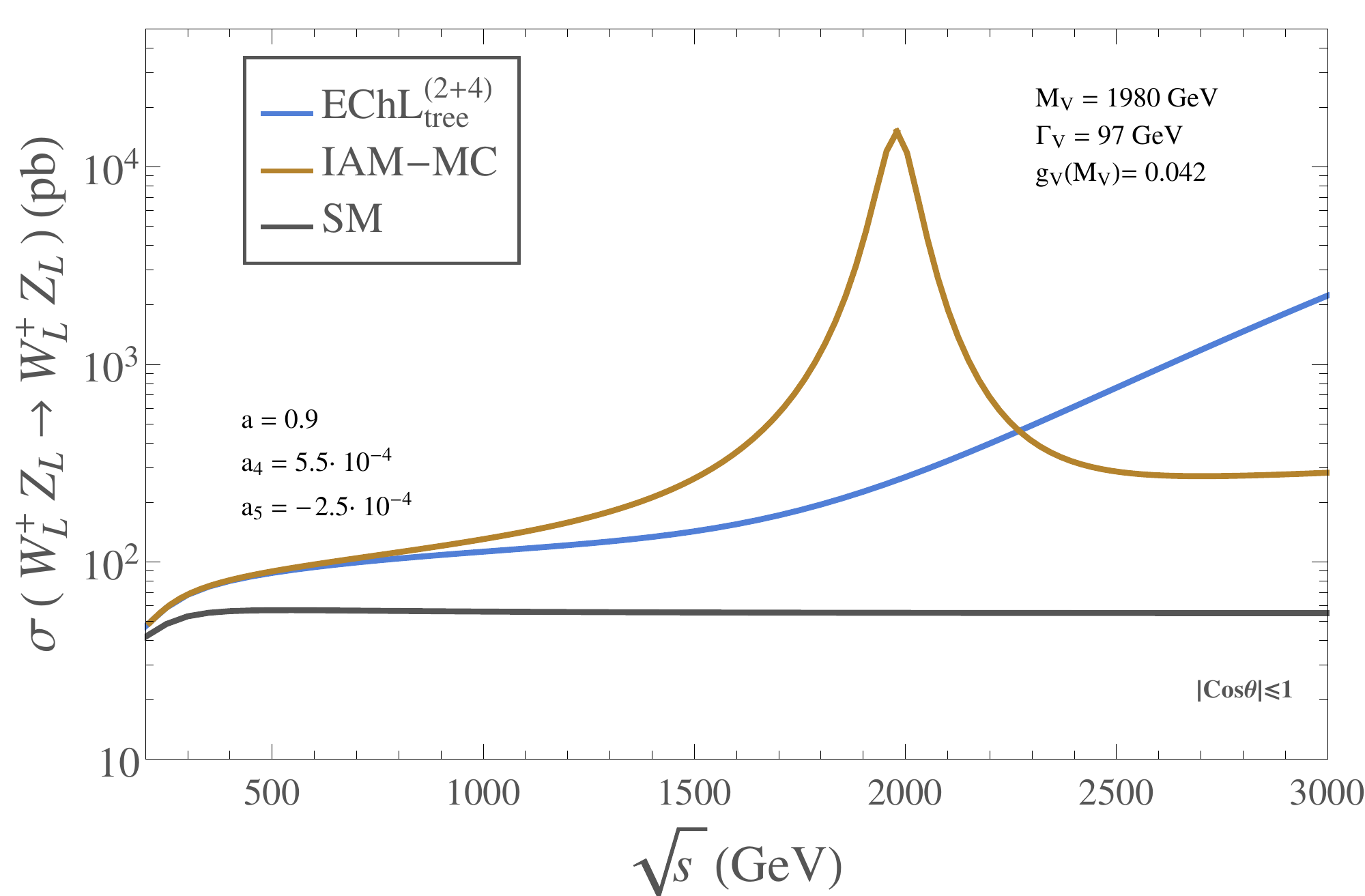}\\
\includegraphics[width=.49\textwidth]{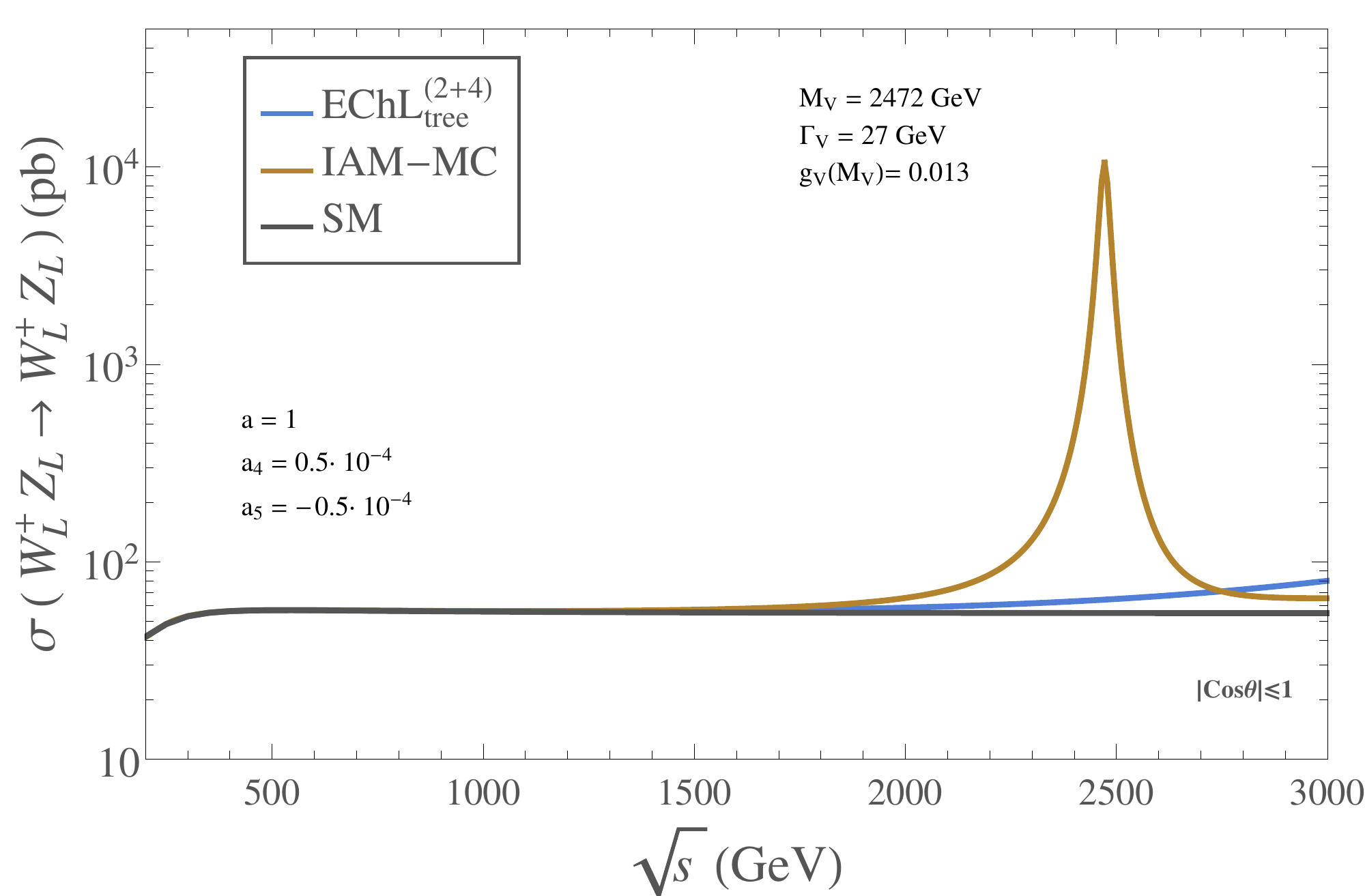}
\includegraphics[width=.49\textwidth]{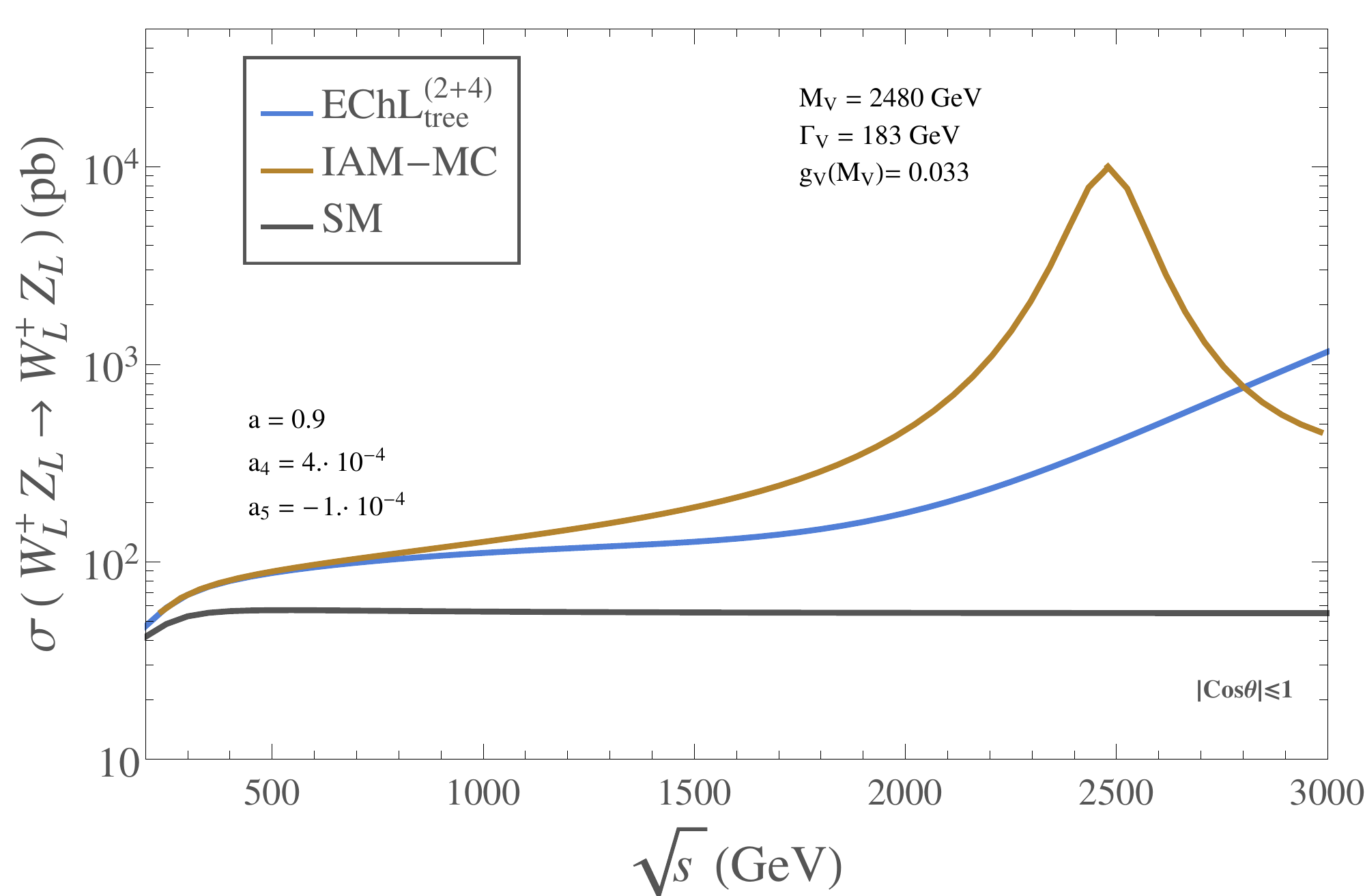}\\
\caption{Predictions of the cross section $\sigma(W^+_LZ_L\to W^+_LZ_L)$ as a function of the center of mass energy $\sqrt{s}$ for all the selected benchmark points in \tabref{tablaBMP} integrated over the whole center of mass scattering angle, $|\cos\theta|\leq1$.
Different lines correspond to the different models considered in the text:
SM (black), our IAM-MC model (orange) and non-unitarized  EChL up to $\mO(p^4)$ (blue).}
\label{fig:xsec1}
\end{center}
\end{figure}

For the numerical computation that is relevant for the forthcoming study of the LHC events we will not use the decomposition in  partial waves, but the complete amplitude instead. This is an important point, since a description of
$\sigma(W_LZ_L \to W_LZ_L)$ in terms of only the lowest partial waves would not give a realistic result for energies away from the resonant region, a we have seen already in Chapter \ref{VBS}. Therefore, before starting the analysis of the LHC events, it is convenient to learn first about the predictions of the cross section at the $WZ \to WZ$ subprocess level.

We present in \figref{fig:xsec1} our numerical results for $\sigma(W_LZ_L \to W_LZ_L)$  within our IAM-MC framework and for the same benchmark points of \tabref{tablaBMP}. In these plots we have also included the predictions from the SM and from the EChL for comparison. What we learn from these figures is immediate: the vector resonances do emerge clearly in the scattering of the longitudinal modes, well above the SM background. We also see that the predictions from the IAM-MC match those from the EChL at low energies, as expected. The main features of the resonances, i.e., the mass, the width and the coupling are obviously manifested in each profile of the resonant IAM-MC lines.  It is also worth mentioning our explicit test that all these cross sections in \figref{fig:xsec1} respect the Froissart unitary bound in \eqref{froisbound}.

So far we have been discussing about the predictions of the scattering amplitudes for the longitudinal gauge boson modes. However, for a realistic study with applications to LHC physics, as we will do in the next section, we must explore also the behavior of the scattering of the transverse modes. In fact, the transverse $W_T$ and $Z_T$ gauge bosons are dominantly radiated from the initial quarks at the LHC, as compared to the longitudinal ones and, consequently, they will be relevant and have to be taken into account in the full computation. Of course we will make our predictions at the LHC taking into account all the polarization channels as it must be.

To compute the various amplitudes $A(W_A Z_B \to W_C W_D)$ with all the polarization possibilities for $A,B,C,D$ being either $L$ or $T$, we proceed as described above for the case of the longitudinal modes. We use the same analytical results for the amplitudes given in the appendices in terms of the generic polarization vectors and substitute there the proper polarization vectors according to the corresponding $L$ or $T$ cases. 

\begin{figure}[t!]
\begin{center}
\includegraphics[width=.49\textwidth]{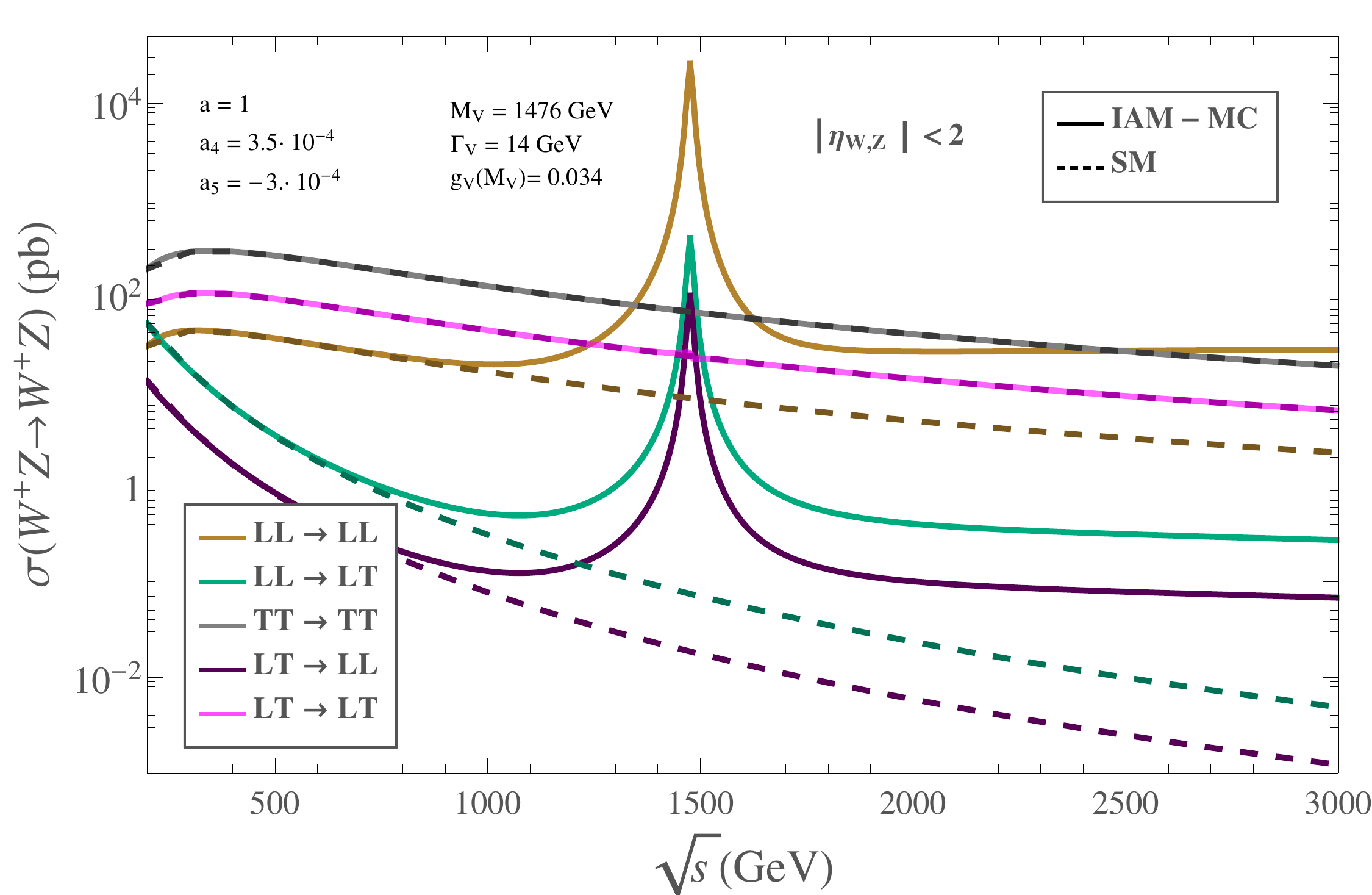}
\includegraphics[width=.49\textwidth]{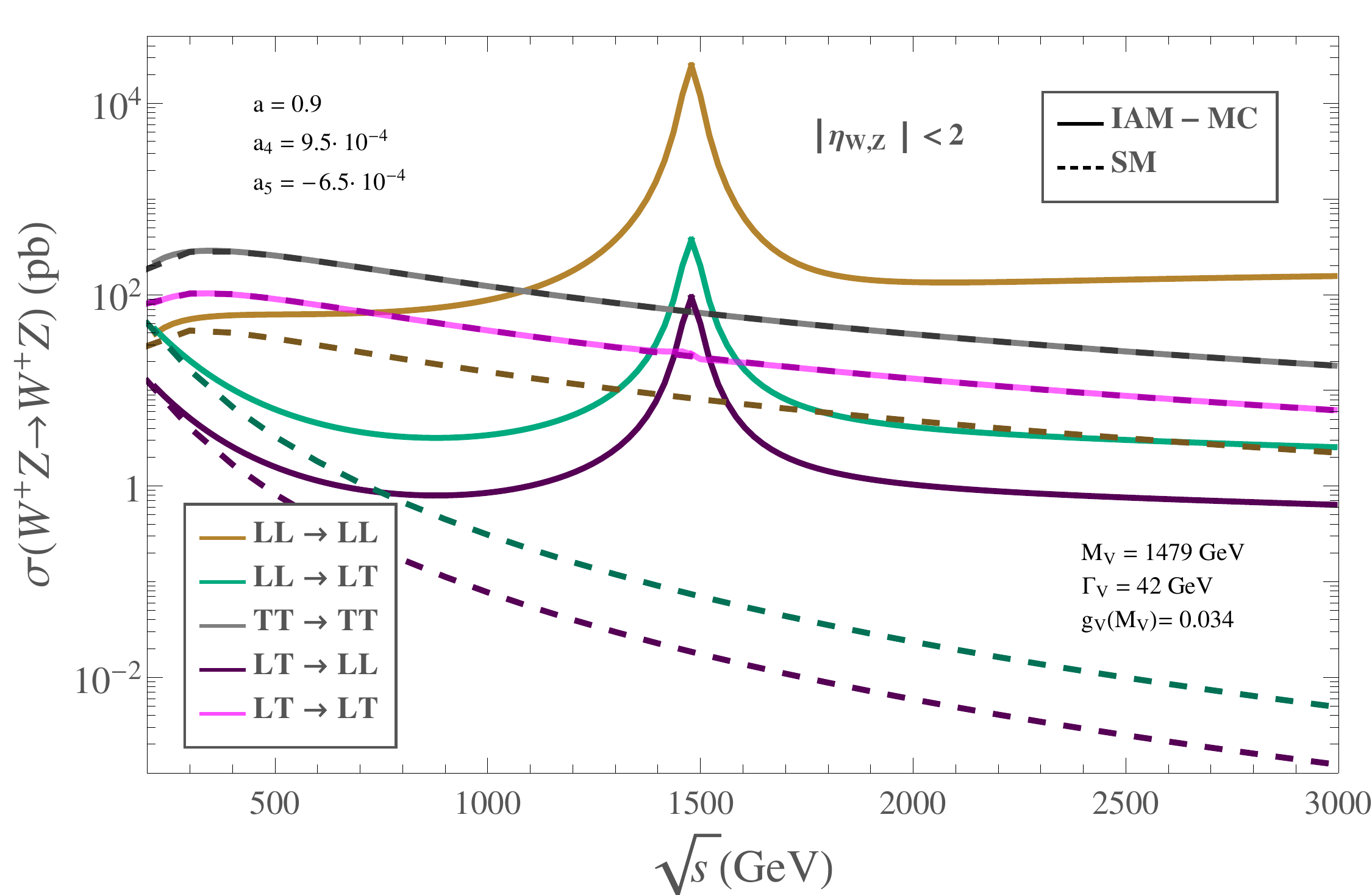}\\
\caption{Cross section $\sigma(W^+Z\to W^+ Z)$ as a function of the center of mass energy $\sqrt{s}$ for  the most relevant polarization channels and for the two selected benchmark points, BP1 (left panel) and BP1' (right panel). Results were obtained imposing a cut on the center of mass scattering angle that corresponds to $|\eta_{W,Z}|<2$. This cut will be used as a detector acceptance cut in the LHC process. Solid lines are the predictions from our IAM-MC model and dashed lines are the predictions from the SM.}
\label{fig:IAM-MC-polarization}
\end{center}
\end{figure}

The numerical results of the cross sections $\sigma(W_A Z_B \to W_C W_D)$ for the most relevant polarizations channels are presented in \figref{fig:IAM-MC-polarization} for the two
benchmark points BP1 and BP1' that we have chosen as illustrative examples. We have also included the corresponding predictions of the cross sections in the SM for comparison. All these results have been computed with FeynArts and FormCalc, and have been checked with MadGraph5.

Regarding this  \figref{fig:IAM-MC-polarization}, one can confirm that at the subprocess level, $WZ \to WZ$,  the scattering of longitudinal modes in our IAM-MC model clearly dominates over the other polarization channels in the region surrounding the resonance. This is  in contrast with the SM case, where the $TT\to TT$ channel dominates by far in the whole energy region studied .  This feature of the IAM-MC was indeed expected since, as already said, the coupling $g_V$ affects mainly to the longitudinal modes. Secondly, the predictions of the resonant peaks in the IAM-MC are clearly above the SM background in all the polarization channels that resonate. Thirdly, we also learn that the $LL\to LL$ channel is not the only one that resonates. In fact, also the $LL\to LT$, $LT\to LL$ and $LT\to LT$ channels  manifest a resonant behavior (barely appreciated in the figure in the $LT\to LT$ case) in the IAM-MC, although with  much lower cross sections at the peak than the dominant $LL\to LL$ channel. In these examples the hierarchy found in the IAM-MC predictions at the peak is the following:
\be
\sigma(LL \to LL) \gg \sigma(LL \to LT) > \sigma(LT \to LL)> \sigma(TT \to TT)
> \sigma(LT \to LT) ,
\ee
where $\sigma(AB\to CD)$ is short-hand notation for $\sigma(W_A Z_B\to W_C Z_D)$, and where $LT$ corresponds to $W_LZ_T+W_TZ_L$. 

Also from \figref{fig:IAM-MC-polarization} one can see that $\sigma(LL\to LT)$ is approximately two orders of magnitude smaller than  $\sigma(LL\to LL)$. Therefore, we conclude that the main features found previously for the $\sigma(W_LZ_L\to W_LZ_L)_{\rm IAM-MC}$, in the region close to the resonance, should emerge in the total cross section, $\sigma(WZ \to WZ)_{\rm{IAM-MC}}$, given the fact that this channel is by far the domminant one. This will be confirmed in the next section.

\section[Sensitivity to vector resonances in  $pp\to WZjj$ at the LHC]{Sensitivity to vector resonances in  $\boldsymbol{pp\to WZjj}$ at the LHC}
\label{LHC}

The process that we wish to explore here is $pp\to WZjj$ at the LHC via the 
 VBS subprocess $WZ \to WZ$ as we did in the previous Chapter for the non-resonant case. Concretely, we select the process  with $W^+$ instead of $W^-$ since the former is  more copiously produced from the initial protons.
We know that thi type of events containing two gauge bosons $W^+$ and $Z$ and two jets in the final state can take place at the LHC in many different ways, not only by means of VBS. Therefore, in order to be able to select efficiently these VBS mediated processes one has to perform the proper optimal cuts in the kinematical variables of the outgoing particles of the collision. These cuts should favor the VBS configuration versus other competing processes. 

In Chapter \ref{VBS} we already saw that the main characteristics of VBS topologies where two very forward/backward high energy jets, with large pseudorapidity separations and with high invariant masses. In this Chapter we will profit from these features in order to reject efficiently the undesired backgrounds that pollute our resonant signal. To this aim, we will study in detail the most relevant backgrounds comparatively towards the signal to obtain the proper selection criteria leading mainly to VBS configurations.

\begin{figure}[t!]
\begin{center}
\includegraphics[width=0.9\textwidth]{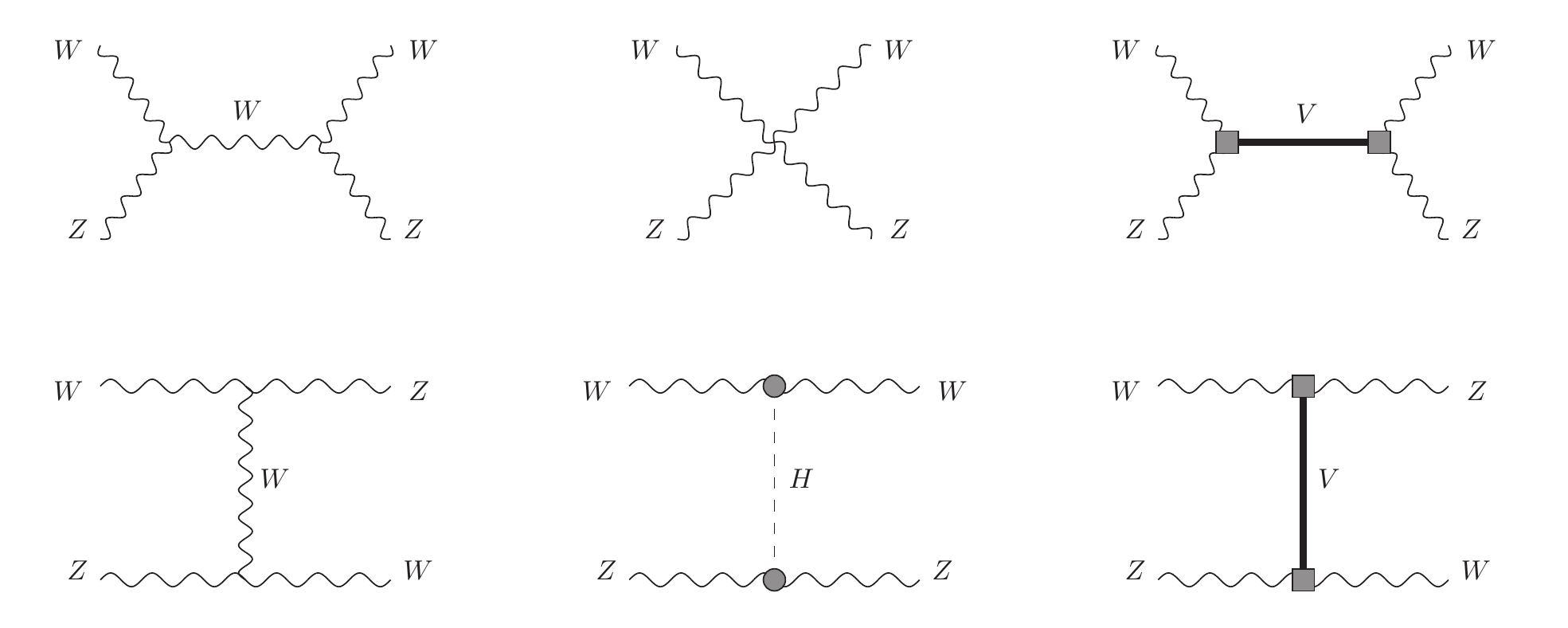}\\
\caption{Feynman diagrams that contribute to the $A(WZ \to WZ)$ tree level amplitude in the IAM-MC and in the unitary gauge. Gray circles represent vertices that are sensitive to the chiral parameter $a$.
Gray squares show vertices with contributions from $\mL_V$. }
\label{fig:DiagramsIAMMC}
\end{center}
\end{figure}

For all the results and plots presented in this section we use MG5 and set the LHC energy to 14 TeV.  For the parton distribution functions we use the NNPDF2.3~\cite{Ball:2013hta} set.  Concretely, the results from our IAM-MC model, which has been described in the previous section, are generated by means of a specific UFO file that contains the model and the  needed four point function $\Gamma_{WZWZ}^{\rm IAM-MC}$.
This $\Gamma_{WZWZ}^{\rm IAM-MC}$ corresponds to the total IAM-MC amplitude  coming from the computation of the diagrams displayed in \figref{fig:DiagramsIAMMC} with the polarization vectors factored out
and is defined  in terms of the IAM-MC model parameters as: 
\begin{align}
-i\,\Gamma_{W^+_\mu Z_\nu W^+_\sigma Z_\lambda}^{\rm IAM-MC} =
-i\,\Gamma_{W^+_\mu Z_\nu W^+_\sigma Z_\lambda}^{\mL_2} 
-i\,\Gamma_{W^+_\mu Z_\nu W^+_\sigma Z_\lambda}^{\mL_V}\,,
\label{fpfUFOMJ}
\end{align}
or, equivalently, extracting the SM amplitude out,
\begin{align}
-i\,\Gamma_{W^+_\mu Z_\nu W^+_\sigma Z_\lambda}^{\rm IAM-MC} =
-i\,\Gamma_{W^+_\mu Z_\nu W^+_\sigma Z_\lambda}^{\rm SM} 
-i\,\Gamma_{W^+_\mu Z_\nu W^+_\sigma Z_\lambda}^{ (a-1)}
-i\,\Gamma_{W^+_\mu Z_\nu W^+_\sigma Z_\lambda}^{\mL_V}\,.
\label{fpfUFO}
\end{align}
The specific computation can be carried out easily with the provided Feynman rules collected in Appendices \ref{FR-SM}, \ref{FR-EChL} and \ref{FR-IAMMC}. Specifically, with the Feynman rules corresponding to the vertices $\VEChL_{W^+_\mu W^-_\nu H}$ and $\VEChL_{Z_\mu Z_\nu H}$ (Appendix \ref{FR-EChL}) and to the vertex $\VIAMMC_{W^+_\mu Z_\nu V^+_\rho}$ (Appendix \ref{FR-IAMMC} with $f_V=0$ as explained previously).

This decomposition turns out to be very convenient to introduce our model in MadGraph, as one can use the SM default model as the basic tool to build the UFO. In this way, we just add up to the SM model files the $\Gamma^{(a-1)}$ and $\Gamma^{\mL_V}$ as four point effective vertices given by:
\begin{align}
-i\,\Gamma_{W^+_\mu Z_\nu W^+_\sigma Z_\lambda}^{(a-1)} &= -\dfrac{g^2}{\cosw^2}\,\dfrac{m_W^2}{t-m_H^2}\,
\big(a^2-1\big)\, g_{\mu\sigma} g_{\nu\lambda}\,,\\
-i\,\Gamma_{W^+_\mu Z_\nu W^+_\sigma Z_\lambda}^{\mL_V} &=\, \dfrac{g^4}{4\, {\rm c}^2_{\rm w}} \Bigg[\dfrac{g_V^2(s)}{s-M_V^2+iM_V\Gamma_V}\Big[h_\nu h_\lambda g_{\mu\sigma}-h_\nu h_\sigma g_{\mu\lambda}-h_\mu h_\lambda g_{\nu\sigma}+h_\mu h_\sigma g_{\nu\lambda}\Big] \nn \\
&\hspace{1.3cm} +\dfrac{g_V^2(u)}{u-M_V^2}\Big[l_\nu l_\lambda g_{\mu\sigma}-l_\lambda h_\sigma g_{\mu\nu} -l_\mu l_\nu g_{\lambda\sigma} + l_\mu l_\sigma g_{\nu\lambda} \Big]\Bigg]\,,\label{fpfUFOparts}
\end{align}
where $h=k_1+k_2$ and $l=k_1-k_4$ following the the total amplitude convention given by $A(W^+(k_1)Z(k_2)\to W^+(k_3)Z(k_4))$. The energy dependent couplings $g_V(s)$ and $g_V(u)$ are the ones defined in \eqrefs{gvenergy}{gvenergytu}.

 This way, with the simplifications assumed in this Chapter, the IAM-MC parameters contained in the UFO file are basically the chiral coefficient $a$ and the vector resonance parameters $M_V$, $\Gamma_V$ and $g_V(M_V)$, which are fixed from the given  input values of $a$, $a_4$ and $a_5$ accordingly to our previous discussion. Concretely, we use the selected points in \figref{fig:contourMW} to make our predictions with MadGraph5 of the signal events at the LHC from the IAM-MC model.
 
\subsection{Study of the most relevant backgrounds}

Regarding the most relevant backgrounds, we only consider here the main irreducible $WZjj$ backgrounds  since we are assuming that the final $W$ and $Z$ gauge bosons can be reasonably identified and disentangled from pure QCD events leading to fake `$WZjj$' configurations. For the same reason, we do not consider either the potential backgrounds from top quarks production and decays. This will be totally justified in the final part of this study where we will focus on the leptonic decays of the final $W$ and $Z$ leading to a very clear signal with three leptons, two jets and missing energy in the final state and with very distinctive kinematics.

We therefore focus here on the two main irreducible SM backgrounds:
\begin{itemize}
\item[1)] The pure SM-EW background, from parton level amplitudes of order
${\cal O}(\alpha^2)$ \newline ${A(q_1 q_2  \to q_3 q_4 WZ)}$.
\item[2)] The mixed SM-QCDEW background, from parton level amplitudes of order
${\cal O}(\alpha \alpha_S)$ $A(q_1 q_2  \to q_3 q_4 WZ)$.
\end{itemize}

\begin{figure}[t!]
\begin{center}
\includegraphics[width=.49\textwidth]{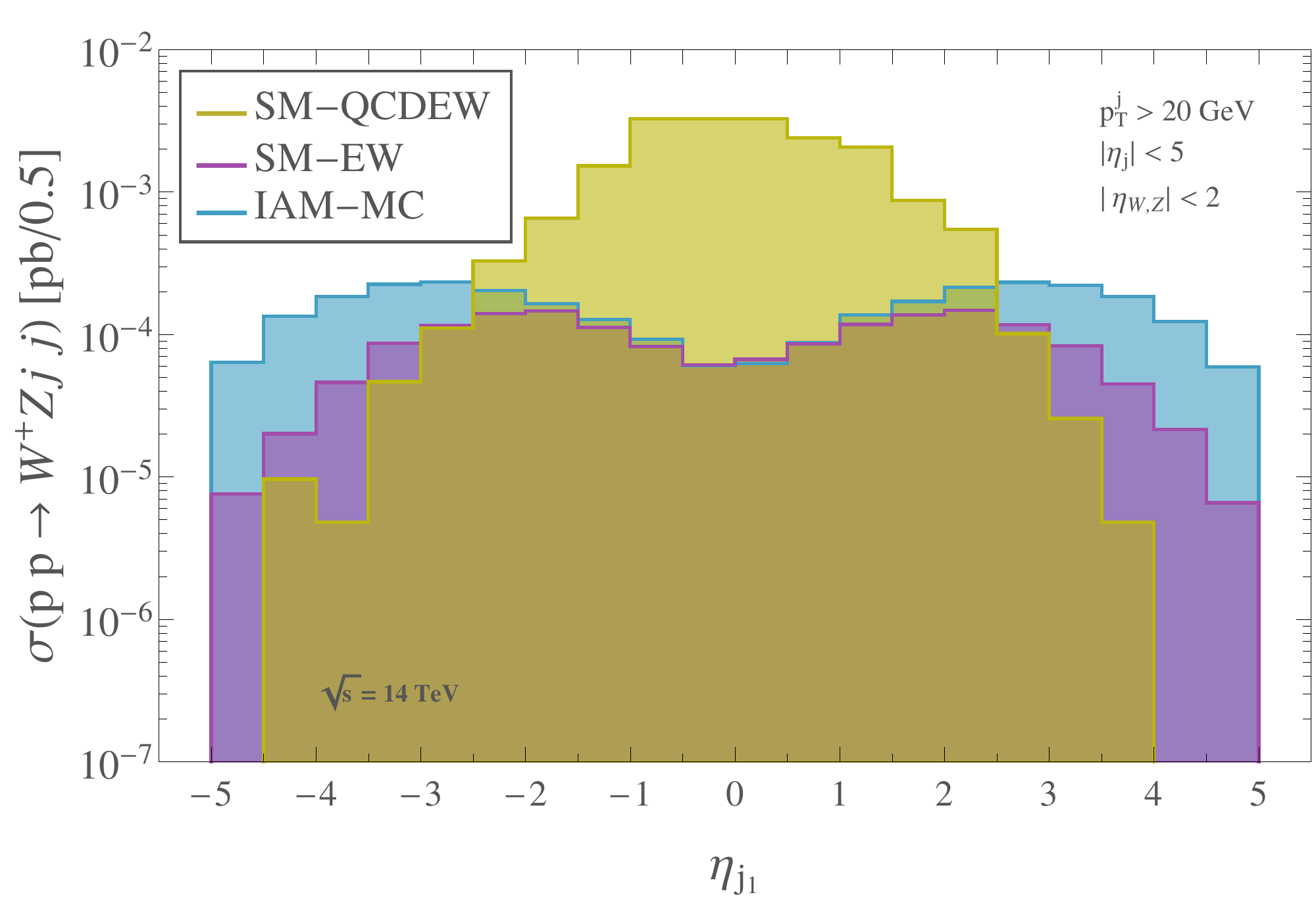}
\includegraphics[width=.49\textwidth]{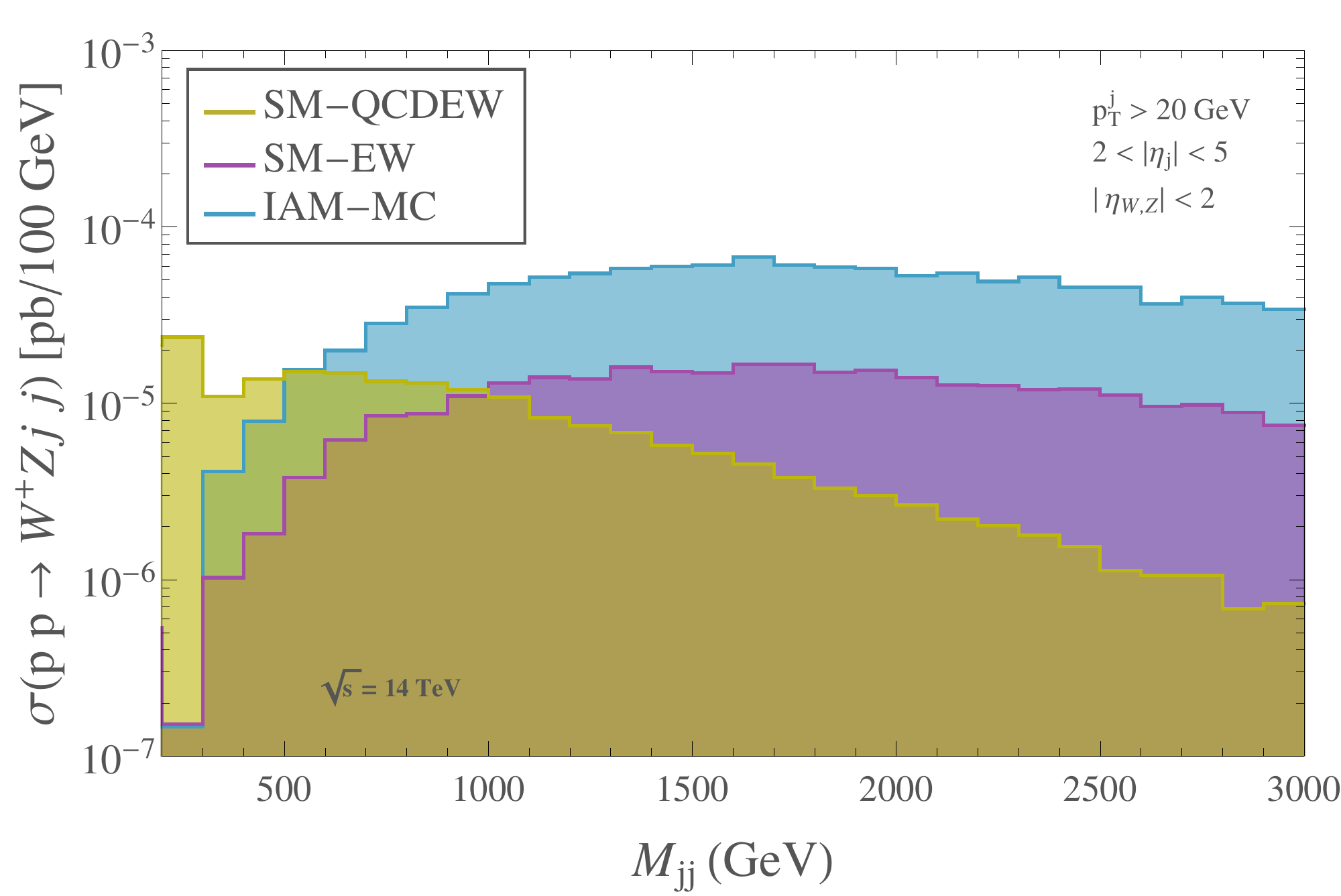}\\
\caption{$\sigma(pp \to W^+Zjj)$ distributions with the pseudorapidity of the outgoing jet $\eta_{j_1}$ (left panel) and with the invariant mass of the final jet pair $M_{jj}$ (right panel).
The predictions for the IAM-MC signal for the selected BP1' scenario (blue) and the two main SM backgrounds, SM-QCDEW (yellow) and SM-EW (purple), are shown separately.}
\label{fig:etajmjj}
\end{center}
\end{figure}

 We show our predictions of the IAM-MC signal for the selected BP1' scenario together with those of the two main irreducible SM-EW and SM-QCDEW backgrounds in   \figref{fig:etajmjj}, for the simple identification cuts specified in the figure. We display the distributions for this signal versus background comparison in the final jet pseudorapidity, $\eta_{j_1}$ (with $j_1$ being the most energetic jet),  and in the invariant mass of the two final jets, $M_{jj}$, since these are the variables that will allow to disentangle the VBS processes from the rest, as we have discussed. 

As we can clearly see in this figure, the signal is mainly produced in the interval $2<|\eta_{j_1}|<5$ and with a rather large jet invariant mass of $M_{jj} > 500$ GeV, whereas the SM-QCDEW background is mainly centrally produced, with $|\eta_{j_1}|  <2$ and at lower invariant masses $M_{jj} <500$ GeV. Therefore, this suggests the following selection of cuts for discriminating the IAM-MC signal from the SM-QCDEW background:
 \begin{align}
&2<|\eta_{j_1,j_2}|<5\,,~~~
\eta_{j_1} \cdot \eta_{j_2} < 0, ~~~
M_{jj}>500\,\,{\rm GeV}\,, \nn \\
 &p_T^{j_1,j_2}>20 ~{\rm GeV} \,, ~~~|\eta_{W,Z}| < 2\,,
\label{optimal-cuts}
\end{align}
where $ \eta_{j_{1,2}}$ are the pseudorapidities of the jets and
$M_{jj}$ is the invariant mass of the jet pair. Notice that the condition $\eta_{j_1} \cdot \eta_{j_2} < 0$ together with the requirement $2<|\eta_{j_1,j_2}|<5$ implies large pseudorapidity difference of the final jets.

Regarding the SM-EW background, as we can see in   \figref{fig:etajmjj}, it has very similar kinematics with respect to our IAM-MC signal in these two jet variables $\eta_{j_1}$ and  $M_{jj}$. This was expected, since, after applying the basic VBS cuts, both receive dominant contributions from the VBS kind of configurations. In order to disentangle our signal from this SM-EW background one has to rely on additional discriminants. As suggested by our previous analysis in section~\ref{sec-model} and by the results presented in Chapter \ref{VBS},  the most powerful of these discriminants would be a devoted study of the final gauge boson polarizations, since the IAM-MC signal produces mainly $W_LZ_Ljj$ events whereas the SM-EW background produces mainly $W_TZ_Tjj$ events.

Nevertheless, as we have argued, sophisticated techniques to distinguish among the polarizations of the final $W$ and $Z$ are not yet well stablished, so we are not going to use a polarization analysis as a discriminant in this work. Thus, we will rely in the following in the most obvious and simple way to discriminate the IAM-MC signal and the SM backgrounds, which is by looking for resonant peaks in the $M_{WZ}$ invariant mass distributions of the unpolarized cross sections.

\subsection{Results for the resonant signal events}
 
 \begin{figure}[t!]
\begin{center}
\includegraphics[width=.49\textwidth]{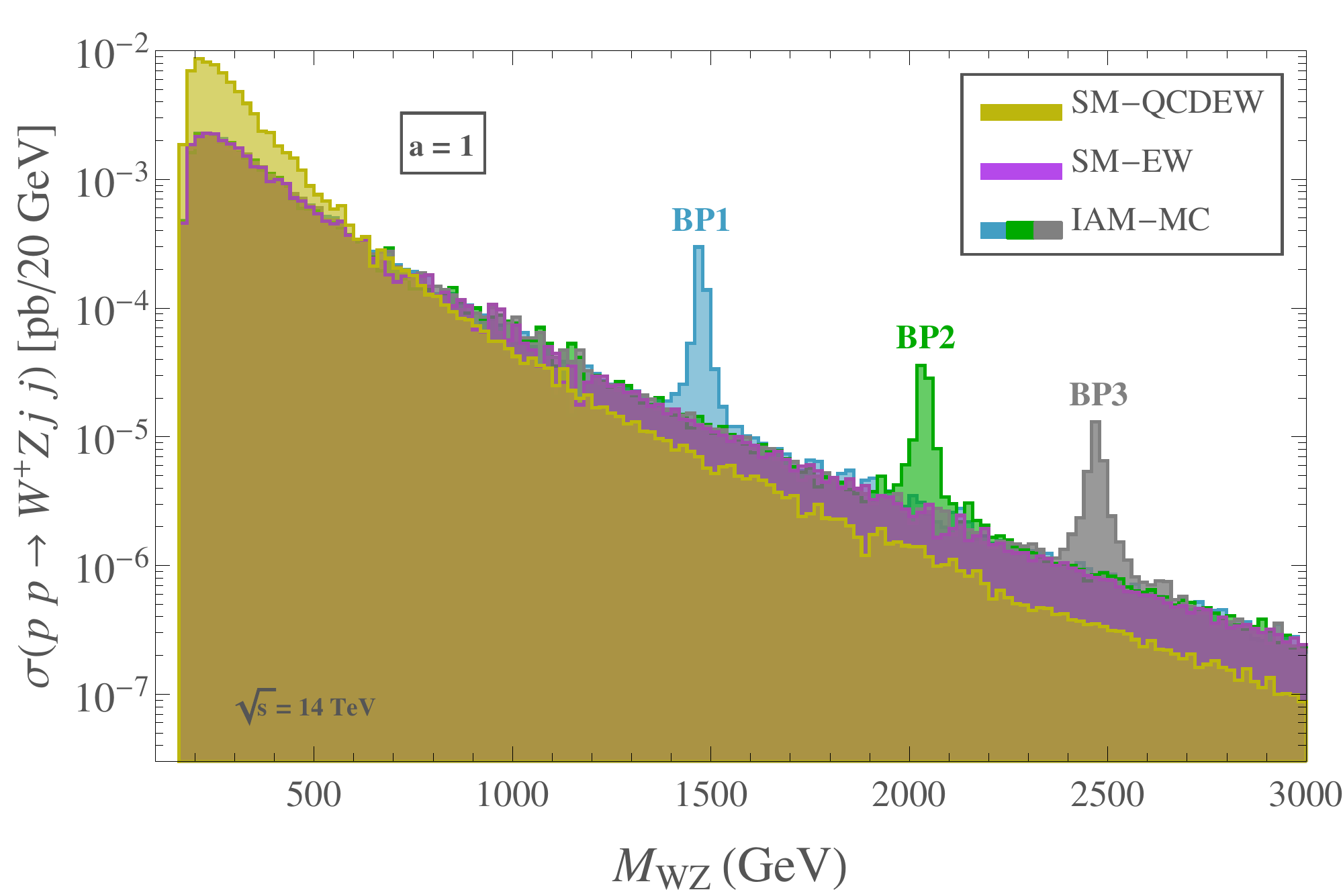}
\includegraphics[width=.49\textwidth]{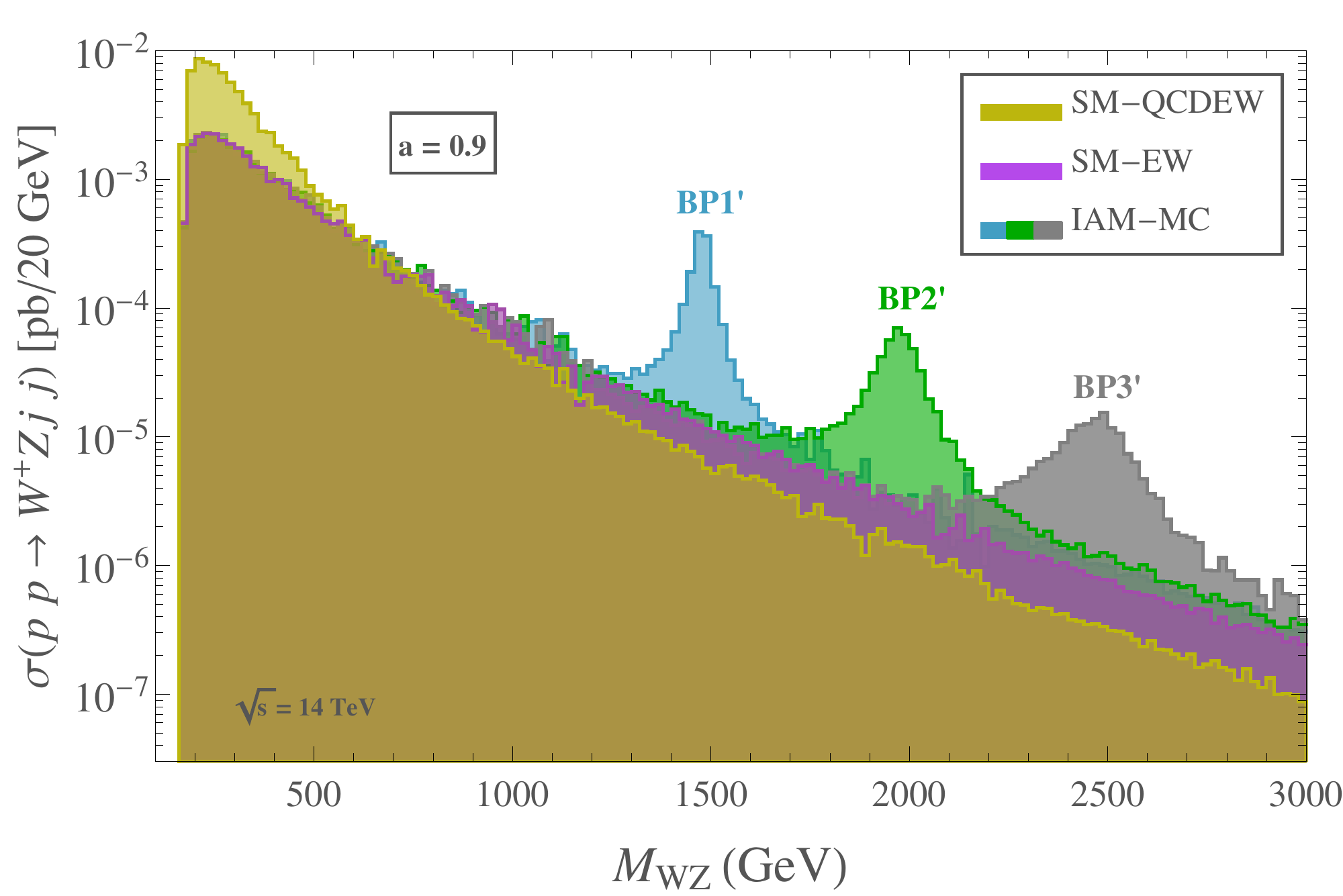}\\
\caption{Predictions of the $\sigma(pp \to W^+Zjj)$ distributions with the invariant mass of the  $WZ$ pair, $M_{WZ}$, for the benchmark points of the IAM-MC model BP1 (blue), PP2 (green), BP3 (gray) in the left panel and BP1' (blue), BP2' (green), BP3'  (gray) in the right panel, and of the two main SM backgrounds, SM-QCDEW (yellow) and SM-EW (purple). The cuts in \eqref{optimal-cuts} have been applied.}
\label{fig:WZdistributions}
\end{center}
\end{figure}

In this subsection we present the main results of our IAM-MC resonant signal events together and compared with the relevant backgrounds explored previously. Our predictions of the  $M_{WZ}$ distributions for the IAM-MC signal and of the two main SM backgrounds, SM-QCDEW and SM-EW, are displayed in \figref{fig:WZdistributions}.   We have summarized in these plots the results for all the selected benchmark points in  \tabref{tablaBMP}, after applying the optimal cuts in \eqref{optimal-cuts}. We see in these figures that the resonant peaks, coming mainly from the interaction of longitudinally polarized gauge bosons, clearly emerge above the SM backgrounds (dominated by the transverse modes) in all these distributions and in all the studied BP scenarios. In order to quantify the statistical significance of these emergent peaks, we define $\significance_{WZ}$ in terms of the predicted events in our IAM-MC model,
${\rm N}(pp\to W^+Zjj)^{\footnotesize \rm IAM-MC}$, and the background events, ${\rm N}(pp\to W^+Zjj)^{\footnotesize \rm SM}$,  as follows:
\be
\significance_{WZ} =\frac{S_{WZ}}{\sqrt{B_{WZ}}}\,,
\label{SS-WZ}
\ee
with,
\begin{align}
S_{WZ}&={\rm N}(pp\to W^+Zjj)^{\footnotesize \rm IAM-MC}- {\rm N}(pp\to W^+Zjj)^{\footnotesize \rm SM}\,, \nn \\
B_{WZ}&={\rm N}(pp\to W^+Zjj)^{\footnotesize \rm SM}\,.
\end{align}
Here the event rates are summed over the interval in $M_{WZ}$
surrounding the corresponding resonance mass. 
In the SM predictions we have summed the purely EW contribution and the QCDEW contributions.
We display in  \tabref{TablasigmasWZ-paper} the results for these $\significance_{WZ}$ of the  $pp\to W^+Z j j$ events, for different LHC luminosities: $\mL=300 ~{\rm fb}^{-1}$, $\mL=1000 ~{\rm fb}^{-1}$ and $\mL=3000 ~{\rm fb}^{-1}$, that are expected for the forthcoming runs~\cite{Barachetti:2120851}. 
We have included the results of two intervals for comparison. 

 \begin{table}[t!]
\begin{center}
\begin{tabular}{ c @{\extracolsep{0.1cm}} c  @{\extracolsep{0.1cm}} c  @{\extracolsep{0.1cm}} c  @{\extracolsep{0.1cm}} c  @{\extracolsep{0.1cm}} c  @{\extracolsep{0.1cm}} c  @{\extracolsep{0.1cm}} c  @{\extracolsep{0.1cm}} c   }
\toprule
\toprule
& & BP1 & BP2 & BP3 & BP1' & BP2' & BP3' \\
\midrule
\multicolumn{1}{c}{\multirow{ 3}{*}{ \begin{sideways}$300\,{\rm fb}^{-1}$\end{sideways} }}& ${\rm N}^{\scalebox{0.7}{\rm IAM-MC}}_{WZ}$ & 89\,(147)& 19\,(25) &4\,(9) & 226\,(412) & 71\,(151) &33\,(59)\\[2pt]
\multicolumn{1}{c}{} &${\rm N}^{\rm SM}_{WZ}$ & 6\,(17)& 2\,(4) &0.3\,(2) & 11\,(45) & 5\,(27) &3\,(14)\\[2pt]
\multicolumn{1}{c}{} &$\significance_{WZ}$ & 34.8\,(31.1)& 10.8\,(9.7) &6\,(5.4) & 64.9\,(54.4) & 28.9\,(23.8) &16.1\,(12)\\[2pt]
\midrule\\[-4ex] \midrule
\multicolumn{1}{c}{\multirow{ 3}{*}{  \begin{sideways} $1000\,{\rm fb}^{-1}$\end{sideways}}} &${\rm N}^{\scalebox{0.7}{\rm IAM-MC}}_{WZ}$ & 298\,(488)& 64\,(82) &13\,(30) & 752\,(1374) & 237\,(504) &110\,(196) \\[2pt]
\multicolumn{1}{c}{}& ${\rm N}^{\rm SM}_{WZ}$ & 19\,(57)& 8\,(15) &1\,(6) & 36\,(151) & 17\,(90) &11\,(46)\\[2pt]
\multicolumn{1}{c}{}& $\significance_{WZ}$ & 63.5\,(56.8)& 19.8\,(17.7) & 11\,(9.9) & 118.5\,(99.4) & 52.7\,(43.5) & 29.3\,(22)\\
\midrule\\[-4ex] \midrule
\multicolumn{1}{c}{\multirow{ 3}{*}{   \begin{sideways}$3000\,{\rm fb}^{-1}$\end{sideways}}} & ${\rm N}^{\scalebox{0.7}{\rm IAM-MC}}_{WZ}$ & 893\,(1465)& 193\,(246) &39\,(89) & 2255\,(4122) &710\,(1511) &331\,(589)\\[2pt]
\multicolumn{1}{c}{}& ${\rm N}^{\rm SM}_{WZ}$ & 58\,(172)& 24\,(44) &3\,(17) & 109\,(454) & 52\,(271) &34\,(139)\\[2pt]
\multicolumn{1}{c}{}& $\significance_{WZ}$ & 110\,(98.5)& 34.3\,(30.6) &19\,(17.1) & 205.3\,(172.2) & 91.3\,(75.3) &50.8\,(38.1)\\
\bottomrule
\bottomrule
\end{tabular}
\vspace{0.4cm}
\caption{Predicted number of $pp\to W^+Z j j$ events of the IAM-MC,  ${\rm N}^{\rm IAM-MC}_{WZ}$, for the selected BP scenarios in \tabref{tablaBMP} and of the SM background (EW+QCDEW), ${\rm N}^{\rm SM}_{WZ}$,  at 14 TeV,  for different LHC luminosities: $\mL=300 ~{\rm fb}^{-1}$, $\mL=1000 ~{\rm fb}^{-1}$ and $\mL=3000 ~{\rm fb}^{-1}$. We also present the corresponding statistical significances, $\significance_{WZ}$, calculated according to \eqref{SS-WZ}. These numbers have been computed summing events in the bins contained in the interval of  $\pm 0.5\,\Gamma_V ~(\pm 2\,\Gamma_V$) around each resonance mass, $M_V$. The cuts in \eqref{optimal-cuts} have been applied.}
 \label{TablasigmasWZ-paper}
\end{center}
\end{table}

First, the events are summed in $M_{WZ}$  over the corresponding narrow
 $(M_V-0.5\,\Gamma_V,$ $ M_V+0.5\,\Gamma_V)$ interval. Second, they are summed over the wider interval around the resonances of $(M_V-2\,\Gamma_V, M_V+2\,\Gamma_V)$. The results differ a bit in the two chosen intervals, as expected, but the conclusions are basically the same: we find very high statistical significances for all the studied BP scenarios in this case of $pp\to W^+Z j j$ events.

\begin{figure}[t!]
\begin{center}
\includegraphics[width=.49\textwidth]{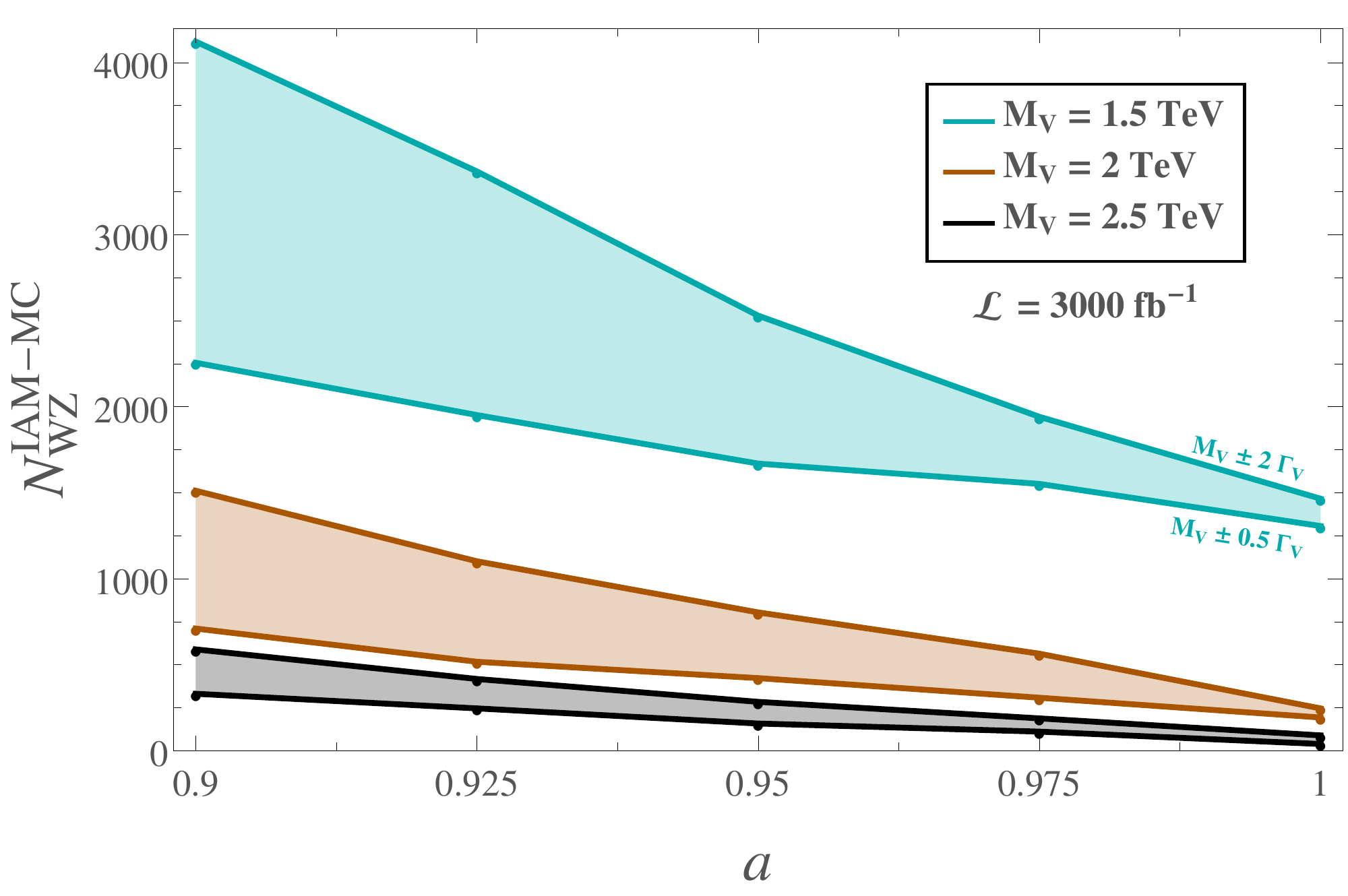}
\includegraphics[width=.48\textwidth]{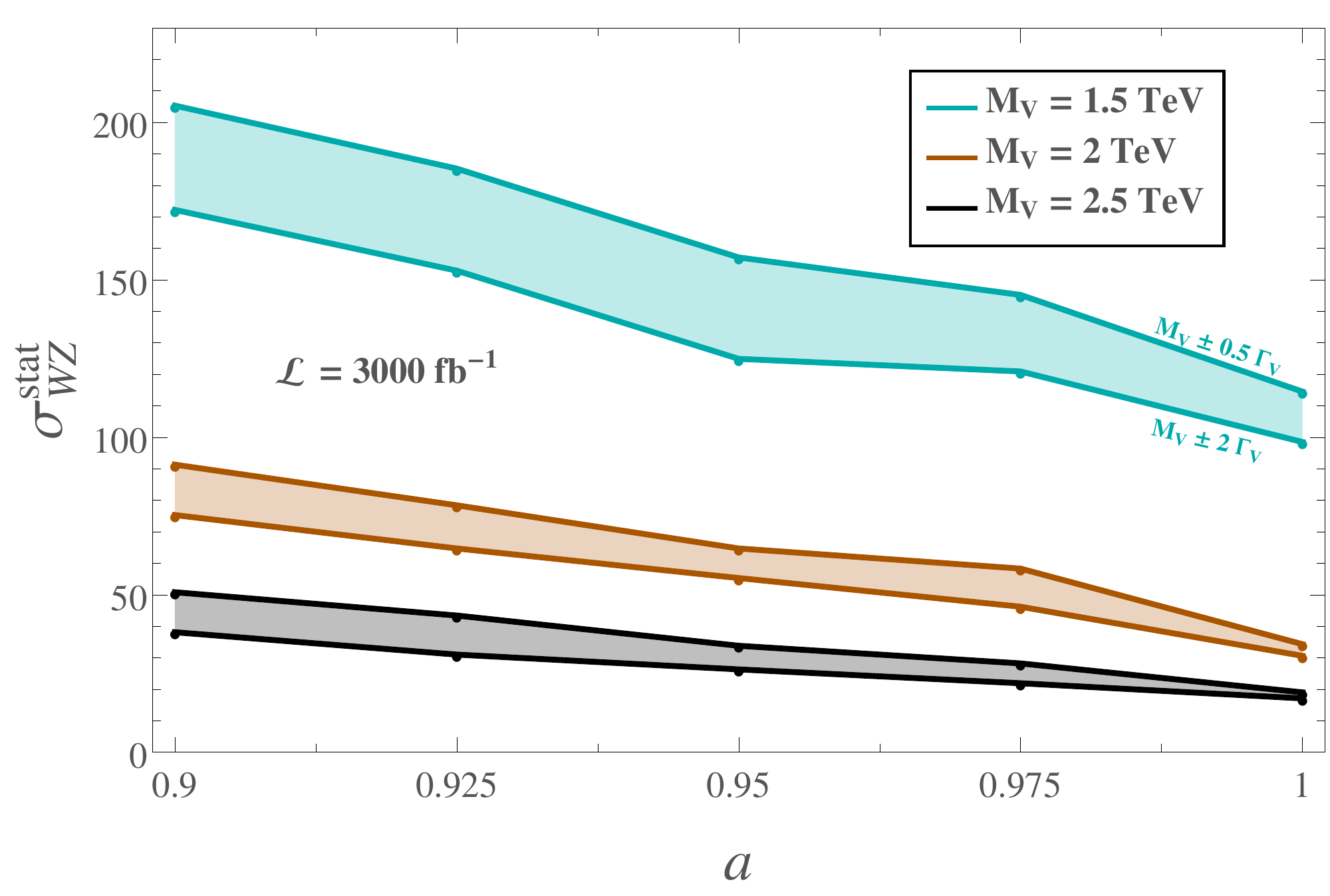}
\caption{Predictions for the number of events, ${\rm N}^{\rm IAM-MC}_{WZ}$ (left panel), and the statistical significance, $\significance_{WZ}$ (right panel), as a function of the parameter $a$ for $\mL=3000~{\rm fb}^{-1}$. Marked points correspond to our  selected benchmark points in \figref{fig:contourMW}. The two lines for each mass are computed by summing events within $\pm 0.5\,\Gamma_V $ and $\pm 2\,\Gamma_V$, respectively.}
\label{fig:asensitivityWZ}
\end{center}
\end{figure}

The predictions in \tabref{TablasigmasWZ-paper} correspond to the selected reference scenarios with the values of the $a$ parameter fixed to the borders of the considered interval $(0.9, 1)$. In order to further study  the sensitivity at the LHC to different values of the $a$ parameter, we have also performed the computation of $W^+Zjj$ events, for the additional benchmark points specified in \figref{fig:contourMW}. The results for these new BP's are collected in \figref{fig:asensitivityWZ}, where we present both the signal event rates, ${\rm N}^{\rm IAM-MC}_{WZ}$, and  the statistical significances, $\significance_{WZ}$, as a function of $a\in(0.9,1)$ for an integrated luminosity of $\mL=3000~{\rm fb}^{-1}$. 

The corresponding rates and significances for the other two luminosities considered here can be easily scaled from these results. The marked points correspond to our selected BP's of \figref{fig:contourMW}. As in \tabref{TablasigmasWZ-paper}, the two lines displayed for each $M_V$ value correspond, respectively,  to summing events in the bins contained in the interval of $\pm 0.5\,\Gamma_V$ and  $\pm 2\,\Gamma_V$ around each resonance mass.

From \figref{fig:asensitivityWZ} it is clear that the high luminosity LHC with  $\mL=3000~{\rm fb}^{-1}$ would be sensitive to all values of $a$ in $(0.9,1)$ through the study of vector resonances with masses of $1.5$, $2$ and $2.5$ TeV. Actually, for this $WZ$ final state, these same conclusions apply to the other two luminosities considered, $\mL=1000~{\rm fb}^{-1}$  and $\mL=300~{\rm fb}^{-1}$. 

The previous results for the statistical significances of $W^+Zjj$ events are really encouraging. The high statistical significances found show that the resonances would be visible if the $W^+$ and $Z$ gauge bosons could be detected as final state particles. However, this is not the real case at colliders, and  one has to reconstruct $W$'s and $Z$'s from their decay products. In particular, the study of the so called `fat jets' in the final state, coming from the hadronic decays of boosted gauge bosons, could lead to a reasonably good reconstruction of the $W^+$ and the $Z$, as we will see in the next Chapter.

The typical signatures of these hadronic events would then consist of four hadronic jets, two thin ones $jj$ triggering the VBS, and two fat ones $JJ$ triggering the final $WZ$.  If this type of signal events were able to be extracted from the QCD backgrounds, the predicted resonances that we show in \figref{fig:WZdistributions} could be very easily discovered. For a fast estimation of the number of signal events that will be obtained by analyzing these kind of hadronic channels with `fat jets' we have performed a naive extrapolation from our results for $WZjj$ events by assuming two hypothetical efficiencies $\epsilon$  for the $W/Z$ reconstruction from `fat jets', which we take from the literature~\cite{Khachatryan:2014hpa, Aad:2015owa,Heinrich:2014kza}, and are usually referred to as `medium' with $\epsilon=0.5$, and `tight' with  $\epsilon=0.25$. The corresponding $JJjj$ signal event rates can be extracted simply by \cite{Heinrich:2014kza}:
\begin{equation} 
{\rm N}^{\rm IAM-MC}_{\rm hadronic}= {\rm N}^{\rm IAM-MC}_{WZ} \times
{\rm BR}(W \to {\rm hadrons}) \times {\rm BR}(Z \to {\rm hadrons})\times 
\epsilon_W \times \epsilon_Z .
\label{fatjetrecfacts}
\end{equation}

\begin{figure}[t!]
\begin{center}
\includegraphics[width=.49\textwidth]{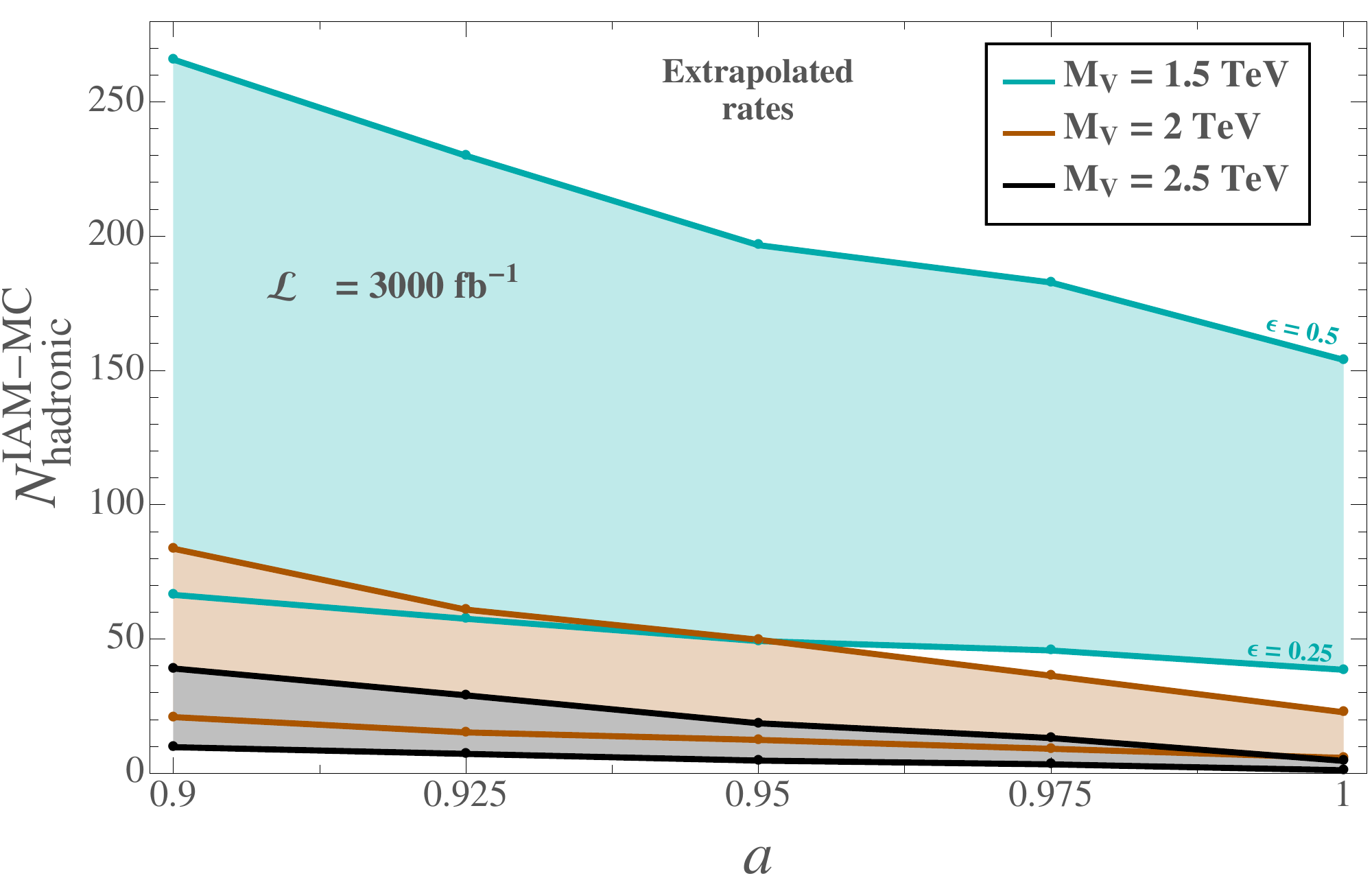}
\includegraphics[width=.48\textwidth]{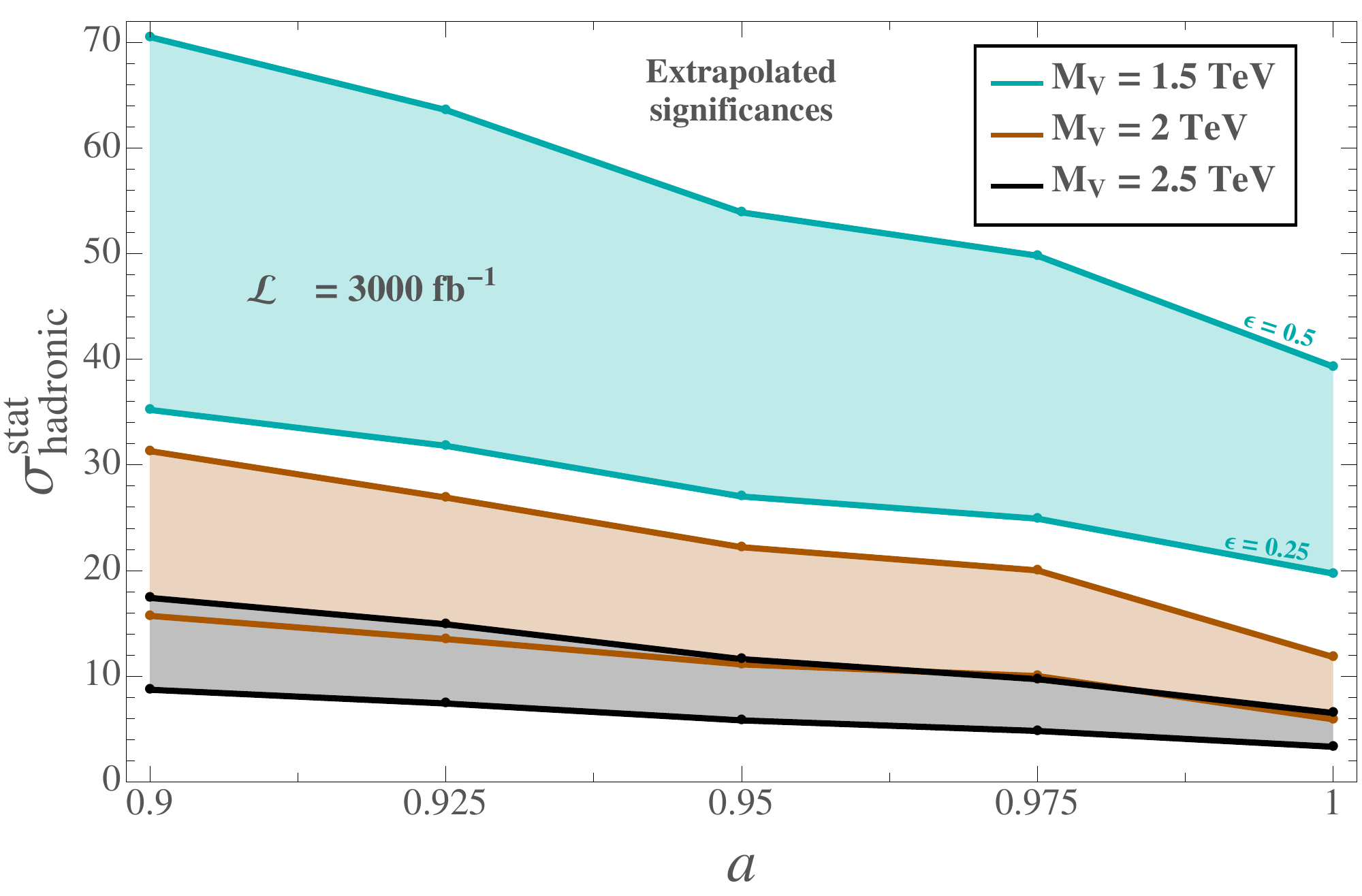}
\caption{Extrapolated $JJjj$ signal event rates from 
\figref{fig:asensitivityWZ} (for $\pm 0.5\,\Gamma_V $), ${\rm N}^{\rm IAM-MC}_{\rm hadronic}$ (left panel), and their corresponding  extrapolated statistical significances (right panel), $\significance_{\rm hadronic}$. The two lines shown for each resonance mass correspond, respectively,  assuming an efficiency in the reconstruction of $W$'s and $Z$'s from the `fat jets' of $\epsilon=0.5$ (upper line) and $\epsilon=0.25$ (lower line).} 
\label{fig:asensitivityhad}
\end{center}
\end{figure}

We show in \figref{fig:asensitivityhad} our predictions for these naively extrapolated number of events (left panel). We show as well a very naive estimate of statistical significances (right panel) computed taking the number of signal events derived from \eqref{fatjetrecfacts}, and considering as backgrounds only the SM-EW and SM-QCDEW ones described above rescaled accordingly to \eqref{fatjetrecfacts} as well. The results are, therefore, very optimistic, since we are neglecting possible new backgrounds that might affect our signal in the purely hadronic case. Nevertheless, it is interesting to shown them here, since in the future these new backgrounds might be efficiently controlled.

The results presented in  \figref{fig:asensitivityhad} are very encouraging and clearly indicate that, if new possible hadronic backgrounds are sufficiently suppressed, a more devoted study of the $W$ and $Z$ hadronic decays leading to `fat jets' the vector resonances of our selected scenarios would all be visible at the high luminosity option of the LHC with $\mL=3000~{\rm fb}^{-1}$. 

Looking at the scaled results for other luminosities, one can see that some of the resonances might be seen already for $\mL=300~{\rm fb}^{-1}$. Concretely, we find that resonances of $M_V\sim 1.5$ TeV could be observed at the LHC with this luminosity with statistical significances larger than 11 (6) for all values of the $a$ parameter if a medium (tight) reconstruction efficiency is assumed. A medium reconstruction efficiency would also allow to find heavier resonances of $M_V\sim$2 (2.5) TeV for values of $a<$0.975 (0.925). The case of $\mL=1000~{\rm fb}^{-1}$, is also very interesting. For this luminosity, the resonances with $M_V$=1.5 TeV and $M_V$=2 TeV could all be seen for any value of the $a$ parameter between 0.9 and 1 and for the two efficiencies considered. The heaviest ones, with masses of $\sim$2.5 TeV, would have significances larger than 3, and therefore could be used to probe values of $a$ in the whole interval studied in this work, if a medium efficiency is assumed. For a tight efficiency, one could still be sensitive to values of the $a$ parameter between 0.9 and 0.95. 

On the other hand, the alternative semileptonic channels  where one final EW gauge boson decays to leptons and the other one to hadrons observed as one fat jet, will also lead to interesting signatures like
$\ell \nu J jj$ and $\ell \ell Jjj$ that are also very promising, with comparable statistics to the previous hadronic channels, as our corresponding naively extrapolated rates (not shown) indicate. For this estimation we have assumed the same naive scenario of having no nwe backgrounds. The potential of these semileptonic channels can also be inferred from the studies in \cite{Aaboud:2016uuk}, where they have been used to notably improve the experimental constraints on $a_4$ and $a_5$ by roughly one order of magnitude, with respect to their previous constraints based on the pure leptonic decays \cite{Aad:2014zda}. 

Nevertheless, our previous estimates of event rates involving `fat jets' although really encouraging are yet too naive, as we have explained, and deserve further studies for a more precise conclusion.  A more realistic and precise computation is needed, but it would require a fully simulated MC analysis of the events with `fat jets' and a good control of the QCD backgrounds and other reducible backgrounds. 

For this reason, in the next Chapter, where we will perform a complete analysis of these fat jet type of events, we will see that the way in which we have estimated the signal events corresponding to the purely hadronic $WZjj$ scenario actually provides a good first approximation to the correct signal event rates. However, since additional backgrounds come into play when studying the purely hadronic case the translation of the statistical significances that can be obtained is not so obvious. We will review all these features in the forthcoming Chapter.

In the present Chapter, however,  we will focus on the cleanest decays of the $W^+$ and $Z$, which are the pure leptonic ones, leading to a final state from the $WZ$ pair with three leptons and one neutrino. Concretely, to unsure a good efficiency in the detection of the final particles we consider just the two first leptonic generations. We propose then to explore at the LHC events of the type $(\ell_1^+\ell_1^-\ell_2^+\slashed{p}_T j_1j_2)$, with ${\ell_{1,2}}$ being either a muon or an electron,  $\slashed{p}_T$ the missing transverse momentum coming from the neutrino, and $j_{1,2}$ the two VBS-tagging jets. The event rates in these leptonic channels suffer from a suppression factor of ${\rm BR}(WZ \to \ell\ell\ell\nu) \simeq 0.014$, but have the advantage of allowing us to reconstruct the invariant mass of the $WZ$ pair in the transverse plane, and also to provide a good reconstruction of the $Z$.

For the present study of the leptonic channels we apply the set of cuts that are partially extracted from Ref.~\cite{Szleper:2014xxa} and optimized as described in the previous background subsection, to make the selection of VBS processes more efficient when having leptons in the final state. These contain all the previous VBS cuts and others, and are summarized by:
\begin{align}
&2<|\eta_{j_{1,2}}|<5\,, ~~~\eta_{j_1} \cdot \eta_{j_2} < 0\,,~~~M_{jj}>500 ~{\rm GeV}\,, \nn \\
&p_T^{j_1,j_2}>20 ~{\rm GeV} \,, \nn \\
&M_Z-10~{\rm GeV} < M_{\ell^+_Z \ell^-_Z} < M_Z+10~{\rm GeV}\,,
\nn \\
&M^T_{WZ}\equiv M^T_{\ell\ell\ell\nu}>500 ~{\rm GeV}\,, \nn \\
&\slashed{p}_T>75 ~{\rm GeV}\,, \nn \\
& p_T^\ell>100 ~{\rm GeV}\,,
\label{lepfin}
\end{align}
where $M_{\ell^+_Z \ell^-_Z}$ is the invariant mass of the lepton pair coming from the Z decay (this means at least one of the two $\ell^+\ell^-$ combinations in the case of $\ell^+\ell^-\ell^+\nu$ with the same lepton flavor),
$\slashed{p}_T$ the transverse missing momentum, $p_T^\ell$ the transverse momentum of the final leptons, and $M^T_{WZ}$ the transverse invariant mass of the $WZ$ pair defined as follows:
\begin{equation}
M^T_{WZ}\equiv M^T_{\ell\ell\ell\nu}=\sqrt{\Big(\sqrt{M^2(\ell\ell\ell)+p_T^2(\ell\ell\ell)} +|\slashed{p}_T|  \Big)^2-\big(\vec{p_T}(\ell\ell\ell)+\vec{\slashed{p}_T}\big)^2}\,,
\label{Mtrans}
\end{equation}
with $M(\ell\ell\ell)$ and $\vec{p_T}(\ell\ell\ell)$ being the invariant mass and the transverse momentum of the three final leptons respectively, and $\vec{\slashed{p}_T}$ the transverse momentum of the neutrino.

\begin{figure}[t!]
\begin{center}
\includegraphics[width=.49\textwidth]{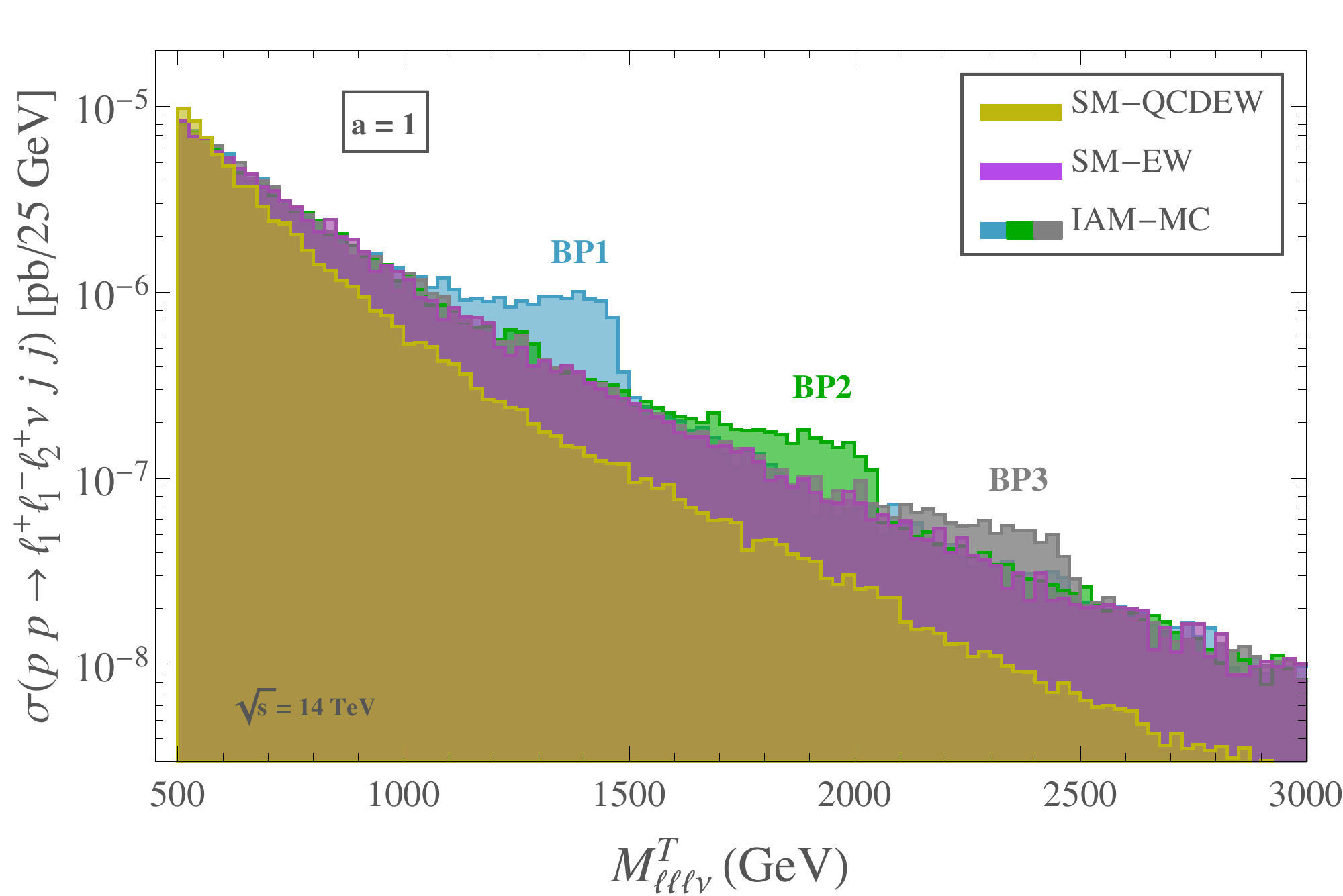}
\includegraphics[width=.49\textwidth]{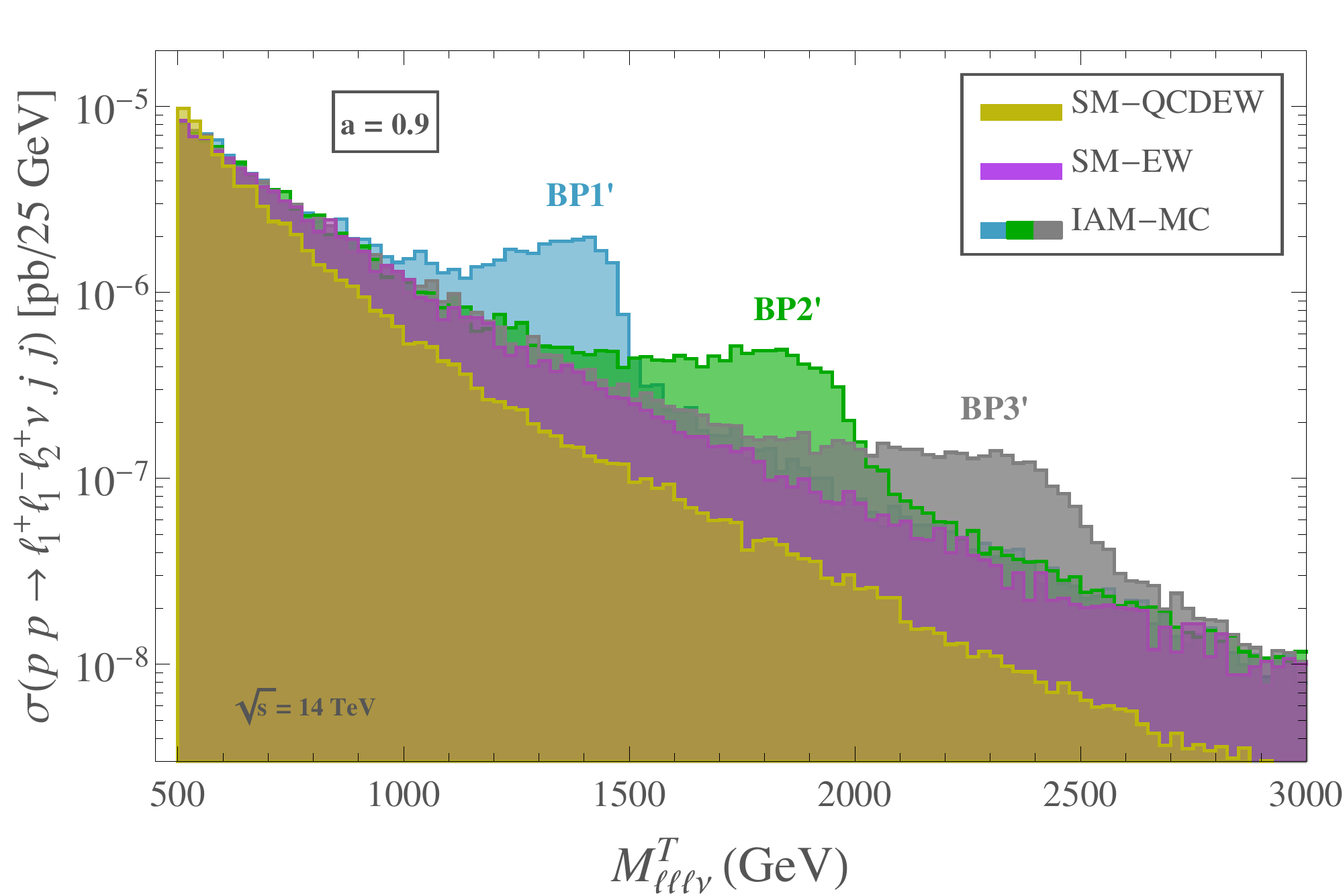}\\
\caption{Predictions of the $\sigma(pp \to \ell_1^+\ell_1^-\ell_2^+ \nu jj)$ distributions with the transverse invariant mass, $M^T_{\ell\ell\ell \nu}$,  for the selected benchmark points of the IAM-MC model BP1 (blue), BP2 (green), BP3 (gray) in the left panel and BP1' (blue), BP2' (green), BP3' (gray) in the right panel, and for the two main SM backgrounds, SM-QCDEW (yellow) and SM-EW (purple). The cuts in \eqref{lepfin} have been applied.}
\label{fig:leptondistributions}
\end{center}
\end{figure}

As before, we generate all the signal, IAM-MC, and background, SM-QCDEW and SM-EW,  events with MG5. The results obtained, after applying the previous cuts in \eqref{lepfin}, are displayed in \figref{fig:leptondistributions}, where the total cross section per bin has been plotted as a function of the transverse invariant mass of the $WZ$ pair as defined in \eqref{Mtrans} .

From this figure we can conclude that the peaks, although smoother, are again clearly seen over the SM backgrounds, specially for the lighter resonances. The shape of the emergent peaks is different than in \figref{fig:WZdistributions}, typically smaller and broader, as corresponding to distributions with the transverse invariant mass, having the maximum at bit lower values, and getting spread in a wider invariant mass range.

Finally, in order to quantify the statistical significance of these emergent peaks, we have computed the quantity $\significance_\ell$, defined in terms of the predicted number of events from the IAM-MC,
${\rm N}(pp\to \ell_1^+\ell_1^-\ell_2^+ \slashed{p}_T jj)^{\footnotesize \rm IAM-MC}$, and the background events,
${\rm N}(pp\to \ell_1^+\ell_1^-\ell_2^+ \slashed{p}_T jj)^{\footnotesize \rm SM}$,  as follows:

\begin{align}
\significance_{\ell} &=\frac{S_{\ell}}{\sqrt{B_{\ell}}}\,,
\label{SS-l}
\end{align}
with,
\begin{align}
S_{\ell}&={\rm N}(pp\to \ell_1^+\ell_1^-\ell_2^+ \slashed{p}_T jj)^{\footnotesize \rm IAM-MC}- {\rm N}(pp\to \ell_1^+\ell_1^-\ell_2^+ \slashed{p} jj)^{\footnotesize \rm SM}\,, \nn \\
B_{\ell}&={\rm N}(pp\to \ell_1^+\ell_1^-\ell_2^+ \slashed{p}_T jj)^{\footnotesize \rm SM}\,.
\end{align}

The final numerical results for $\significance_\ell$ are collected in \tabref{Tablasigmaslep-paper}. Again, we have considered three
 different LHC luminosities: $\mL=300 ~{\rm fb}^{-1}$,
 $\mL=1000 ~{\rm fb}^{-1}$ and $\mL=3000 ~{\rm fb}^{-1}$.
 The numbers of events presented are the results after summing over the intervals in which we have found the largest  statistical significance with at least one IAM-MC event for $\mL=3000 ~{\rm fb}^{-1}$.  
 In particular we consider the following ranges of $M^T_{\ell\ell\ell\nu}$:
\begin{align}
&{\rm BP1:}~1325-1450~{\rm GeV}\,,
&& {\rm BP2:}~1875-2025~{\rm GeV}\,,
&& {\rm BP3:}~2300-2425~{\rm GeV}\,,\nn \\
&{\rm BP1':}~1250-1475~{\rm GeV}\,,
&& {\rm BP2':}~1675-2000~{\rm GeV}\,,
&&{\rm BP3':}~2050-2475~{\rm GeV}\,.
\label{MTintervals}
\end{align} 

\begin{table}[t!]
\begin{center}
\begin{tabular}{cccccccc}
\toprule
\toprule
& & BP1 & BP2 & BP3 & BP1' & BP2' & BP3' \\
\midrule
\multicolumn{1}{c}{\multirow{ 3}{*}{ \begin{sideways}$300\,{\rm fb}^{-1}$\end{sideways} }}& ${\rm N}^{\rm IAM-MC}_{\ell}$ & 2 & 0.5 & 0.1 & 5 & 2 & 0.7 \\[0.5ex]
\multicolumn{1}{c}{} & ${\rm N}^{\rm SM}_{\ell}$ & 1 & 0.4 & 0.1 & 2 &0.6 &0.3 \\[0.5ex]
\multicolumn{1}{c}{} & $\significance_{\ell}$ & 0.9 & - &- &2.8 &1.4 &- \\[0.5ex]
\midrule\\[-4ex]\midrule
\multicolumn{1}{c}{\multirow{ 3}{*}{  \begin{sideways} $1000\,{\rm fb}^{-1}$\end{sideways}}} &${\rm N}^{\rm IAM-MC}_{\ell}$ &7 & 2 & 0.4 &18 &5 &2 \\[0.5ex]
\multicolumn{1}{c}{}& ${\rm N}^{\rm SM}_{\ell}$ &4 & 1 &0.3 &6 &2 &1 \\[0.5ex]
\multicolumn{1}{c}{}& $\significance_{\ell}$&1.6 & 0.3 & -& 5.1&2.5 &1.4 \\[0.5ex]
\midrule\\[-4ex]\midrule
\multicolumn{1}{c}{\multirow{ 3}{*}{   \begin{sideways}$3000\,{\rm fb}^{-1}$\end{sideways}}} & ${\rm N}^{\rm IAM-MC}_{\ell}$ & 22& 5 &1 &53 &16 &7 \\[0.5ex]
\multicolumn{1}{c}{}& ${\rm N}^{\rm SM}_{\ell}$ &12 &4  &1 &17 &6 &3 \\[0.5ex]
\multicolumn{1}{c}{}& $\significance_{\ell}$ & 2.7& 0.6 &0.3 &8.9 &4.4 &2.4 \\[0.5ex]
\bottomrule
\bottomrule
\end{tabular}
\vspace{0.4cm}
\caption{ Predicted number of $pp\to \ell_1^+\ell_1^-\ell_2^+\nu j j$ events of the IAM-MC,  ${\rm N}^{\rm IAM-MC}_{\ell}$, and of the SM background (EW+QCDEW), ${\rm N}^{\rm SM}_{\ell}$,  at 14 TeV,  for different LHC luminosities: $\mL=300 ~{\rm fb}^{-1}$, $\mL=1000 ~{\rm fb}^{-1}$ and $\mL=3000 ~{\rm fb}^{-1}$.  We also present the corresponding statistical significances, $\significance_{\ell}$, calculated according to \eqref{SS-l} after summing events in the intervals collected in \eqref{MTintervals}.
 We only display the value of $\significance_{\ell}$ for the cases in which there is at least one IAM-MC event. The cuts in \eqref{lepfin} have been applied.}
\label{Tablasigmaslep-paper}
\end{center}
\end{table}
 

 As we can see in this \tabref{Tablasigmaslep-paper}, these more realistic statistical significances
 for the leptonic channels, $\significance_\ell$ are considerably smaller than
 the previous $\significance_{WZ}$.  However, we still get scenarios
 with sizable $\significance_\ell$ larger than 3. 
 
Concretely, the scenarios with $a=0.9$ leading to vector resonance masses at and below 2 TeV, could be seen in these leptonic channels at the LHC in its forthcoming high luminosity stages. Particularly, for BP1' with $M_V=1.5$ TeV we get sizeable significances around 3, 5, and 9 for luminosities of 300, 1000 and 3000 ${\rm fb}^{-1}$ respectively, whereas for BP2' with $M_V=2$ TeV the significances are lower, close to 3 for 1000 ${\rm fb}^{-1}$ and slightly above 4 for 3000 ${\rm fb}^{-1}$.

 The scenarios with $a=1$ have comparatively smaller significances, and only the lightest resonances with $M_V=1.5$ TeV , like BP1,   lead to a significance of around 3 for the highest studied luminosity of 3000 ${\rm fb}^{-1}$.  
Notice that there are some cases that we do not consider in our discussion because of the lack of statistics.
The scenarios with heavier resonance masses, at and above 2.5 TeV seem to be very difficult to observe, due to the poor statistics for these masses in the leptonic channels. Only our benchmark point BP3' gets a significance larger that 2 for  3000 ${\rm fb}^{-1}$. Therefore, in order to get more sizable significances in those cases  one would have to perform a more devoted study in other channels like the semileptonic and hadronic ones of the final $WZ$ pair, as we have already commented above.

\begin{figure}[t!]
\begin{center}
\includegraphics[width=.49\textwidth]{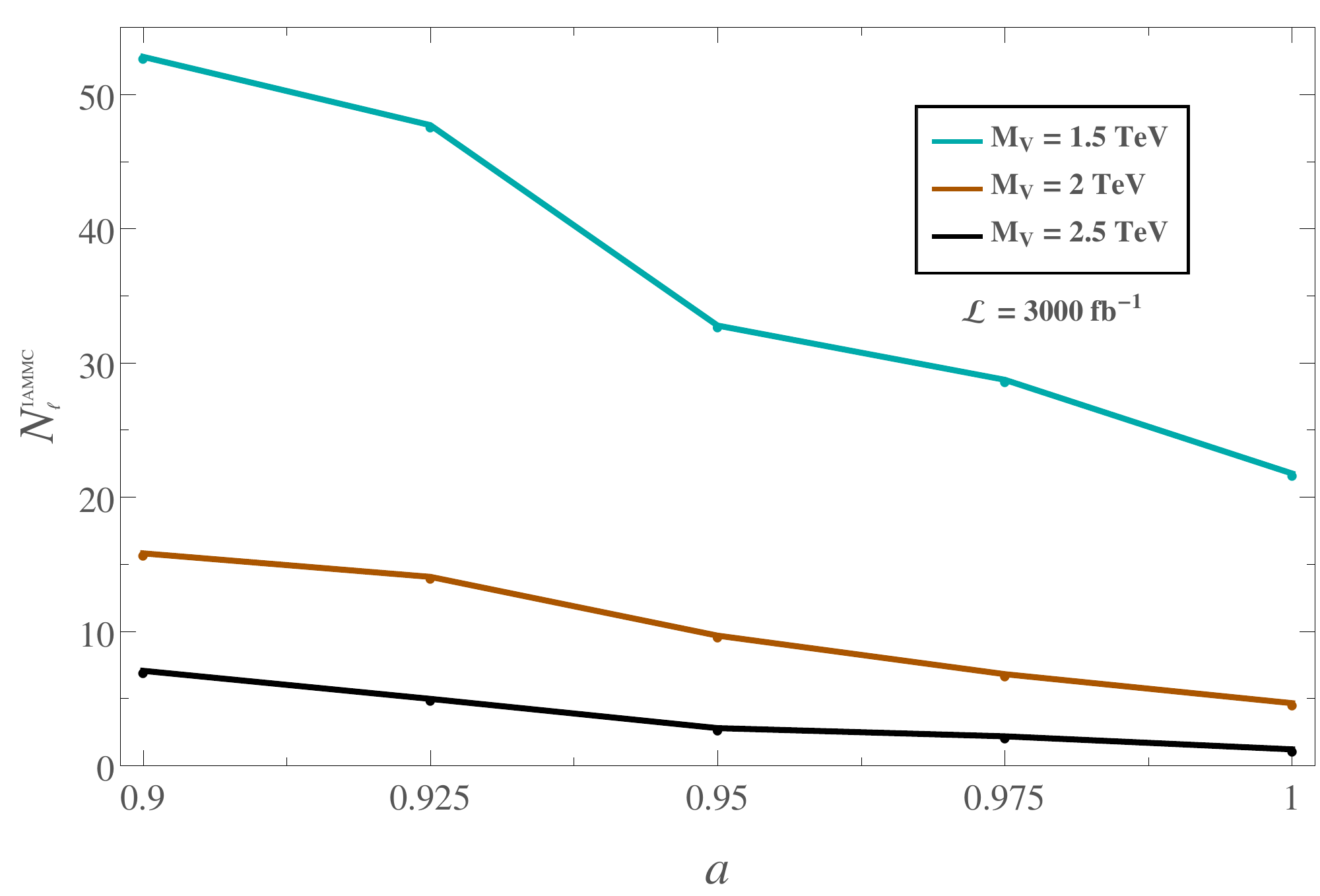}
\includegraphics[width=.48\textwidth]{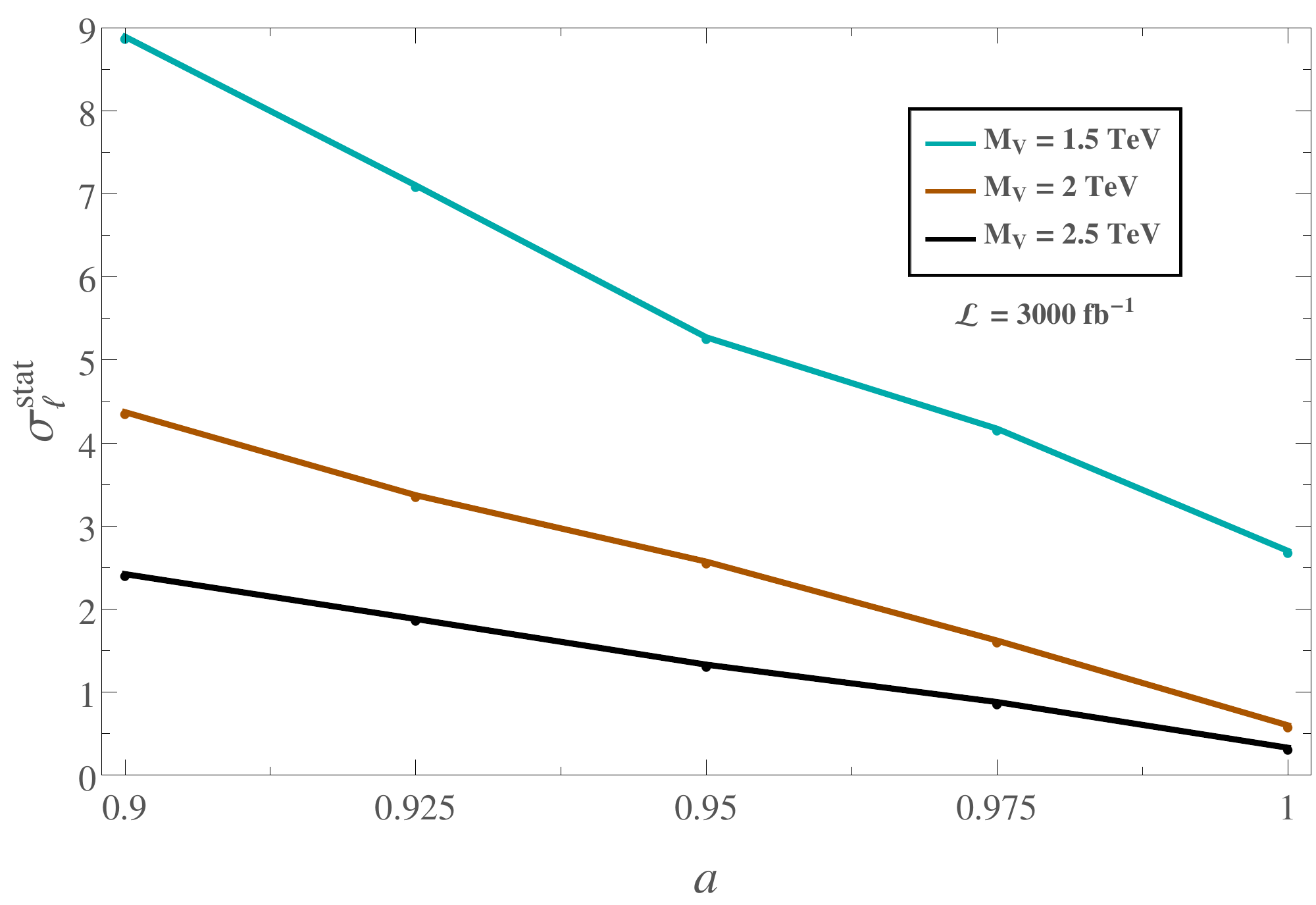}\\
\caption{Predictions for the number of events, ${\rm N}_{\ell}^{\footnotesize \rm IAM-MC}$, (left panel) and the statistical significance, $\significance_{\ell}$, (right panel) as a function of the parameter $a$ for $\mL=3000~{\rm fb}^{-1}$. Marked points correspond to our selected benchmark points in \figref{fig:contourMW}. The cuts in \eqref{lepfin} have been applied.}
\label{fig:asensitivitylep}
\end{center}
\end{figure}

Finally, we have also explored the additional BP points with different values of the $a$ parameter and studied the sensitivities to this parameter in the leptonic channels. The results of the predicted event rates, ${\rm N}^{\rm IAM-MC}_{\ell}$, and statistical significances, $\significance_\ell$, in terms  of the parameter  $a$, within the interval $(0.9,1)$ are displayed in \figref{fig:asensitivitylep}. From this figure we can clearly conclude that, for the highest luminosity  $\mL=3000~{\rm fb}^{-1}$, and for $M_V=1.5$ TeV, there will be good sensitivity to the $a$ parameter, 
with $\significance_\ell$ larger than 3, in the full interval $(0.9,1)$, except for the limiting value of $a=1$ where $\significance_\ell$ is slightly below 3.
For the heavier resonances, we find lower sensitivities, with  $\significance_\ell$ larger than 3 only for $M_V=2$ TeV and $a$ below around 0.94. The case $M_V=2.5$ TeV is not very promising to learn about the parameter $a$ in the fully leptonic channel except, perhaps, for the scenario with the lowest considered value of $a=0.9$ where, as said above for BP3', $\significance_\ell$ gets larger than 2. Nevertheless, this would be strongly improved by exploring other decay channels, as we will see in the next Chapter.

To summarize, in this Chapter we have explored the production and sensitivity to vector resonances at the LHC emerging from the strong interactions of a BSM EWSB sector. We have used the IAM, well known in the context of QCD, to predict the presence of resonances in the spectrum, whose properties are derived form the EChL parameters. 

We have built the IAM-MC model that uses a modified Proca Lagrangian framework to mimic the resonant behaviour of the IAM amplitudes. This IAM-MC framework, where the VBS amplitudes are built from Feynman rules, is very useful  for a Monte Carlo analysis  like the one we have done in the present work with MG5. Our IAM-MC model for the vector resonance production at LHC provides unitary VBS amplitudes (we have checked indeed, that the LHC cross sections respect the Froissart bound given by \eqref{froisbound}), and, therefore, it does not require unphysical {\it ad hoc} cuts to respect unitarity in the study of the signal versus background events.

With this tool, and  focusing on the $pp\to W^+Zjj\to  \ell_1^+\ell_1^-\ell_2^+ \slashed{p}_T jj$ channel which is the most relevant one if one is interested in the study of charged vector resonances in a clean environment, we have computed the statistical significance of selected benchmark points leading to resonance masses in the TeV range. Concretely, we have explored values of $M_V$ of 1.5, 2 and 2.5 TeV as the most interesting cases. 

We have seen that the study of this kind of resonant scenarios at the LHC could lead to observable BSM physics from which we could extract relevant information about the EFT describing the EWSB dynamics. Nevertheless, the results we have obtained for the significances related to these resonances in the purely leptonic channel are modest due to the small rates associated to the leptonic decays of the EW gauge bosons.

In particular, we have seen that with a luminosity of $300~ {\rm fb}^{-1}$ a first hint (with $\significance_\ell$ around 3) of resonances with mass around 1.5 TeV for the case $a=0.9$ could be seen in the leptonic channels. For the first stage of the high luminosity LHC, with $1000~ {\rm fb}^{-1}$, we estimate that these scenarios could  be tested with a high statistical significance larger than 5 and a discovery of these resonances with masses close to 1.5 TeV, like in BP1',  could be done. Interestingly, for the last luminosity considered, $3000~ {\rm fb}^{-1}$,  all the studied scenarios with  resonance masses at and below 2 TeV and with $a=0.9$ could be seen. Concretely, for BP1' and BP2'  we get  $\significance_\ell$ close to 9 and 4 respectively. 

For the heaviest studied resonances, with masses around 2.5 TeV, small hints with $\significance_\ell$  slightly larger than 2  might as well show up in the highest luminosity stage. The sensitivities to other values of $a$ in the interval $(0.9,1)$  have also been explored.  Our numerical results in \figref{fig:asensitivitylep} show that for the highest luminosity  $\mL=3000~{\rm fb}^{-1}$, and for $M_V=1.5$ TeV, there will be good sensitivity to the $a$ parameter in the leptonic channels, with $\significance_\ell$ larger than 3, in the full interval $(0.9,1)$ except for the limiting value of $a=1$ where $\significance_\ell$ is slightly below 3.

 For the heavier resonances, we find lower sensitivities, with  $\significance_\ell$ larger than 3 only for $M_V=2$ TeV and $a$ below around 0.94. The case $M_V=2.5$ TeV does not show appreciable sensitivity to $a$, except for the lowest considered value of $a=0.9$ where,   $\significance_\ell$ gets larger than 2.  

In the light of these results, and even thought some of them are really promising, one expects that the hadronic channels might serve to disentangle these observables more easily, which is why in the next Chapter we will perform a full analysis of the purely hadronic channel of a VBS configuration (in this case the WW one, complementary to the WZ studied here) in the search from dynamically generated EChL resonances.


\chapter[\bfseries Dynamical vector resonances from the EChL in VBS at the LHC: the WW case]{Dynamical vector resonances from the EChL in VBS at the LHC:\\ the WW case}\label{ResonancesWW}
\chaptermark{Dynamical vector resonances from the EChL in VBS at the LHC: the WW case}

As already discussed in the previous Chapter, the emergence of heavy resonances in Vector Boson Scattering would undoubtedly be a remarkable signal of a strongly interacting EWSB sector described by the EChL. Moreover, if these new resonances have masses in the few TeV energy domain, the LHC is then the  proper collider to look for them. In the case in which these resonances couple dominantly to EW gauge bosons, and not to fermions, it is clear that VBS plays the most relevant role in the search for these emergent resonances.

The resonance states emerge as poles in the total EChL amplitude, taking into account the subsequent re-scattering of the EW gauge bosons via VBS type of diagrams, whose effects are indeed very important for the total computation due to the strong character of the interactions involved. This motivates the name ``dynamically generated resonances'' that we also used in Chapter \ref{Resonances}. 

To deal with this resummation process of the re-scattering diagramas and to get unitary predictions at the same time the Inverse Amplitude Method is used, since it  fixes the physical properties of the resonances, like mass, width and couplings to the EW gauge bosons in terms of the EChL parameters. 

In the previous Chapter we performed a devoted study of the production of such resonances at the LHC by studying $WZjj$ events focusing on the leptonic decays of the final gauge bosons. As we argued, the $WZ$ channel was the ideal one to study the charged component, $V^{\pm}$, of the isotriplet of vector resonances, $V^0$, $V^+$ and $V^-$, since it propagated in the $s-$channel. However, in this Chapter we will focus our attention in the neutral component, $V^0$, that may be accessed more efficiently in $W^+W^- \to W^+W^-$ scattering. 

Furthermore,  we will center our analysis on the hadronic decays of the final $W$s, and,  more concretely, in the kinematical regime in which the hadronic decay products of the $W$s are identified as a single, large radius jet. This allows for larger signal rates and for a better reconstruction of the resonance properties than in the leptonic scenario, the large amount of missing energy present in the latter case complicates this task. 

The purely hadronic channel suffers, however, from quite sizable backgrounds, especially regarding the one coming from QCD events. The biggest effort in this Chapter will then be that of optimizing the analysis of the $W$ tagging techniques with fat jets for the observation of the emergent $V^0$ resonance in of $W^+W^-jj$ events. All in all, our analysis aims to explore the sensitivity to the neutral vector resonances $V^0$ with masses between 1.5 and 2.5 TeV at the LHC with $\sqrt{s}=13$ TeV and the planned high luminosity of 3000 ${\rm fb}^{-1}$, paying special attention to the study of efficient cuts to extract the resonant  signal from the QCD background in $pp \to JJjj$ events with VBS configuration, which clearly represents the main challenge of this search.


\section{IAM-MC setup for WW scattering}
\label{model}

The theoretical framework used in the study of the dynamically generated vector resonances in $WW$ scattering is exactly the same as the one introduced in the previous Chapter. We assume a strongly interacting EWSB sector, described by the  EChL, whose scattering amplitudes are unitarized with the IAM. This leads, in general, to an expression of unitary amplitudes that accommodates a pole, whose position depends on the values of the chiral parameters. 

From the characterization of the pole, the resonance properties can be derived in terms of the relevant EChL coefficients. Thus, as explained before, $M_V$, $\Gamma_{V}$ and the resonance couplings to EW gauge bosons will be related to $a$, $a_4$ and  $a_5$, as we set again $b=a^2$. 

Furthermore, since in the study of the $WZ$ scattering performed in Chapter \ref{Resonances} we assumed the isospin limit to compute the IAM amplitudes, and, therefore, to obtain the resonance properties, the three members of the isotriplet of vector resonances $V^0$, $V^+$ and $V^-$ share exactly the same physical characteristics. More specifically, they have the same mass, same width, and same couplings. For this reason, to illustrate our main results in the $WW$ channel, we will use the same benchmark points defined in the previous Chapter that are collected in \tabref{tablaBMP}. These correspond to phenomenologically interesting values of the resonances masses, between 1.5 and 2.5 TeV  that arise directly from IAM amplitudes for specific values of the chiral patameters, as already mentioned.

In order to study the production of the dynamical vector resonances at the LHC by means of a Monte Carlo like MadGraph~\cite{Alwall:2014hca,Frederix:2018nkq}, where the needed input files are not the scattering amplitudes but the interaction vertices themselves, or equivalently the interaction Lagrangian, we use our IAM-MC model described in section \ref{sec-model} of the previous Chapter. 

 \begin{figure}[t!]
\begin{center}
\includegraphics[width=0.9\textwidth]{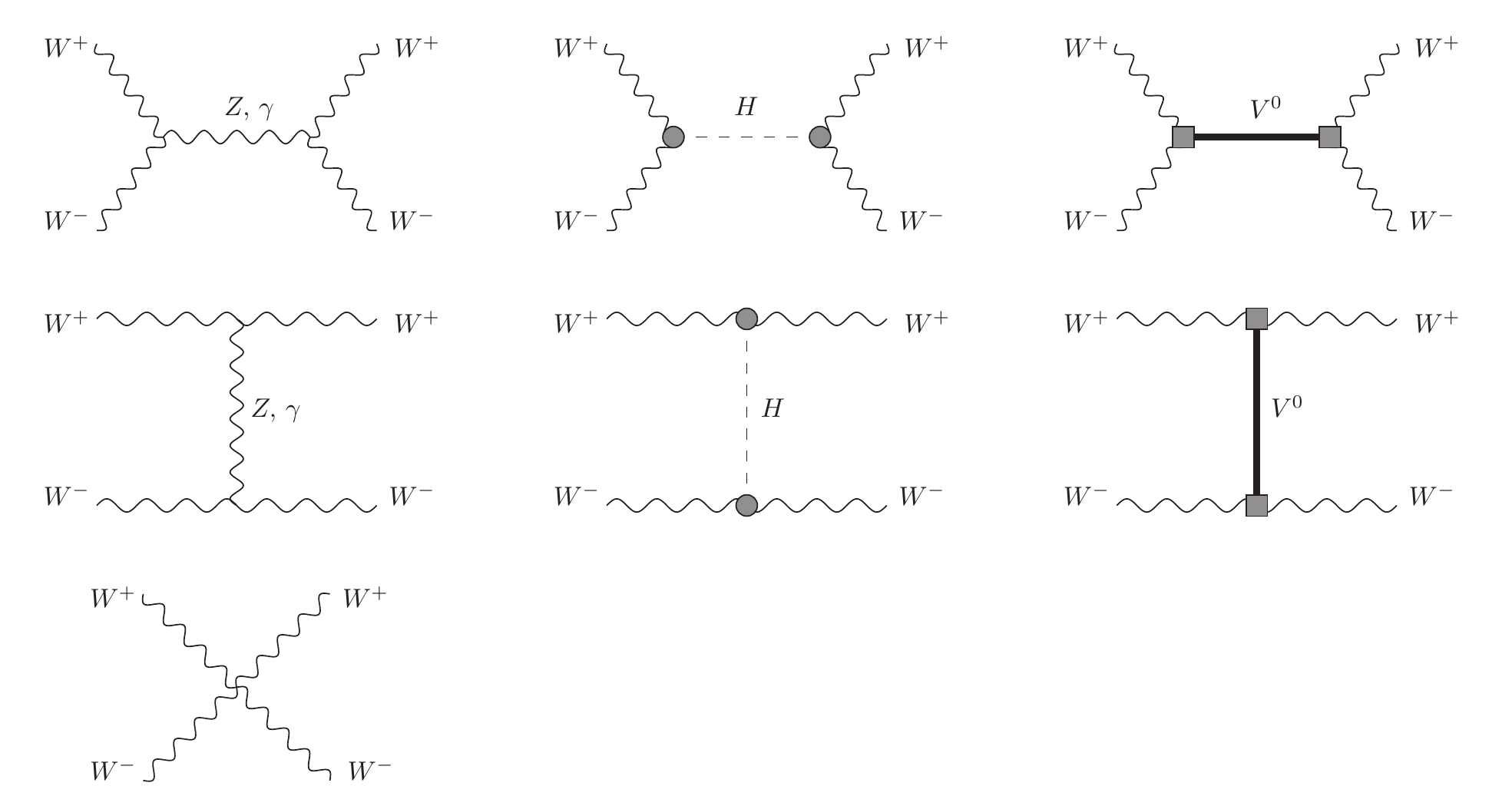}
\caption{Feynman diagrams contributing to the process $W^+W^-\to W^+W^-$ in the Unitary gauge  within our effectiveLagrangian formalism. Gray circles (middle diagrams) represent the $V_{W^+W^-H}$ vertex given in Appendix \ref{FR-EChL} in terms of the EChL parameter $a$ and gray squares (right diagrams) represent the $V_{W^+W^-V^0}$ vertex in Appendix \ref{FR-IAMMC} in terms of the $g_V$ Proca coupling.}
\label{FD}
\end{center}
\end{figure}

Thus, since the vector resonance mass and width are shared by all the states belonging to the isotriplet, the only differences between the $WZ$ and the $WW$ case will be the charge of the resonance that mediates them and the different Feynman diagrams that contribute to each of these channels. The total set of the diagrams contributing to the $W^+W^- \to W^+W^-$ scattering is collected in Figure \ref{FD}, to be compared with that presented in \figref{fig:DiagramsIAMMC} for $W^+Z \to W^+Z$ scattering.

For the practical computation of the full process $pp \to W^+W^-jj$ with MadGraph we have implemented an UFO model with the complete IAM-MC Lagrangian, involving the two relevant pieces $\mL_2$ and $\mL_V$, as before. It is worth mentioning that the full set of diagrams involved in $pp \to W^+W^-jj$, which are generated in MadGraph, include many other diagrams in addition to the subset of diagrams with VBS configuration, so all of them have been taken into account in our numerical computation of the LHC events presented in the next section.

As explained in Chapter \ref{Resonances}, the IAM-MC model contains the  needed four point function $\Gamma_{WWWW}^{\rm IAM-MC}$ that allows to mimic in a Lagrangian or interaction vertex language the IAM resonance properties.
This $\Gamma_{WWWW}^{\rm IAM-MC}$ corresponds to the total IAM-MC amplitude  coming from the computation of the diagrams displayed in \figref{FD} with the polarization vectors factored out and is again defined  in terms of the IAM-MC model parameters as:
\begin{align}
-i\,\Gamma_{W^+_\mu W^-_\nu W^+_\sigma W^-_\lambda}^{\rm IAM-MC} =
-i\,\Gamma_{W^+_\mu W^-_\nu W^+_\sigma W^-_\lambda}^{\rm SM} 
-i\,\Gamma_{W^+_\mu W^-_\nu W^+_\sigma W^-_\lambda}^{(a-1)}
-i\,\Gamma_{W^+_\mu W^-_\nu W^+_\sigma W^-_\lambda}^{\mL_V}\,.
\label{fpfUFOWW}
\end{align}
Here, in the same way as before, $\Gamma^{\rm SM}$ corresponds to the contribution from the SM, $\Gamma^{(a-1)}$ denotes the new effects introduced by  $\mL_2$ with $a\neq1$ with respect to the SM, and $\Gamma^{\mL_V}$ accounts for the new contributions from the dynamically generated resonance.  This computation has been carried out by using the Feynman rules presented in Appendices \ref{FR-SM}, \ref{FR-EChL} and \ref{FR-IAMMC}. Specifically, with those corresponding to the vertex $\VEChL_{W^+_\mu W^-_\nu H}$ (Appendix \ref{FR-EChL}) and to the vertex $\VIAMMC_{W^+_\mu W^-_\nu V^0_\rho}$ (Appendix \ref{FR-IAMMC} with $f_V=0$ as explained in the previous chapter). 

As already discussed, the decomposition defined in \eqref{fpfUFOWW} turns out to be very convenient to introduce our model in MadGraph, as one can use the SM default model as the basic tool to build the UFO. In this way, we just add  to the SM model files the $\Gamma^{(a-1)}$ and $\Gamma^{\mL_V}$ as four point functions given by:
\begin{align}
-i\Gamma_{4W}^{(a-1)} =& 
	-g^2\frac{m_W^2}{t-m_H^2}(a^2-1) g_{\mu\sigma} g_{\nu\lambda} -g^2\frac{m_W^2}{s-m_H^2}(a^2-1) g_{\mu\nu} g_{\sigma\lambda}  \nonumber\\
-i\Gamma_{4W}^{\mL_V} =&  
	\frac{g^4}{4}\bigg[\frac{g_V^2(s)}{s-M_V^2+iM_V\Gamma_V} (h_\nu h_\lambda g_{\mu\sigma}-h_\nu h_\sigma g_{\mu\lambda}-h_\mu h_\lambda g_{\nu\sigma}+h_\mu h_\sigma g_{\nu\lambda}) \nonumber\\
	&+\frac{g_V^2(t)}{t-M_V^2}(l_\nu l_\lambda g_{\mu\sigma}-l_\lambda h_\sigma g_{\mu\nu} -l_\mu l_\nu g_{\lambda\sigma} + l_\mu l_\sigma g_{\nu\lambda})\bigg]
\end{align}
where $h=k_1+k_2$ and $l=k_1-k_3$, following the the total amplitude convention given by $A(W^+(k_1)W^-(k_2)\to W^+(k_3)W^-(k_4))$.

In the above expressions the energy dependent couplings $g_V(s)$ and $g_V(t)$ correspond to those in \eqref{gvenergytu} with $z=s,t$ being the channels in which the resonance $V^0$ is propagating in the present case of  $W^+W^-$ scattering. These non-local interactions are needed in order to ensure unitary predictions, since the Proca Lagrangian itself, i.e., with a constant $g_V$, leads to a violation of unitarity above the resonance mass, as we explicitly shown before. 

With all these considerations in mind we can move on to the analysis of the LHC sensitivity to these neutral vector resonances in WW scattering.


%
\section[Sensitivity to vector resonances in  $pp\to WWjj$ at the LHC]{Sensitivity to vector resonances in  $\boldsymbol{pp\to WWjj}$ at the LHC}
\label{results}
In this section we present the numerical results of the $pp \to W^+W^-jj$ events at the LHC computed within our model for the isotriplet vector resonances  described in the previous sections, selecting exclusively the hadronic decays of the final $W$ gauge bosons. In the present case, as we have said, the relevant vector resonance is the neutral one, $V^0$, in contrast to the study in Chapter \ref{Resonances} of $pp \to WZjj$ events in which the charged component of the isotriplet could be accessed. Furthermore, in this latter study the focus was set mainly on the EW gauge boson leptonic channels, so the present analysis is somehow complementary to that of the previous Chapter.

 In all this section we set the LHC energy to $\sqrt{s}=13$ TeV, and make predictions for all our signal benchmark scenarios defined by the six selected points BP1, BP2, BP3, BP1', BP2', BP3', collected in \tabref{tablaBMP}.  

In order to get a rough estimate of the signal rates at the LHC,  and to learn about the main features of our signal events for the selected  BPs  points, we first analyze them at the naive parton level. We compute the rates for $pp \to W^+W^-jj$ events, with $jj$ denoting quarks, before considering any showering or jet reconstruction algorithm. Then, we apply the suppression factors from the two EW gauge boson hadronic decays given by  $ ({\rm BR}(W \to {\rm hadrons}))^2\sim0.45$.

In this computation of the cross sections for the $pp \to W^+W^-jj$ process we wish to compare the signal rates with the main background rates which, at this parton level, are: 1) SM EW background, with amplitude of ${\cal O}(\alpha^2)$, 2) SM mixed QCDEW background, with amplitude of ${\cal O}(\alpha \alpha_S)$ and 3) top-antitop production from QCD followed by the top (antitop) decay into $b W^+$ $\left({\bar b} W^-\right)$, in which the final bottom and antibottom jets are misidentified as light quark jets $j$. In this latter case we assume a suppression factor due to this misidentification of $bb$ as $jj$ within the range from $(0.2)^ 2$ to $(0.3)^2$ corresponding to the often used b-jet tagging efficiency of 80\% to 70\%.

Since we are interested in events with VBS configuration and within the large invariant mass region of the gauge boson pair, $M_{WW}$, for this parton level computation we have applied in addition to the basic cuts that ensure particle detection, $p_{T_j}>20$ GeV, $\Delta R_{jj} >0.4$, $|\eta_W|<2$, $p_{T_W}>20$ GeV, also the usual VBS cuts given by:
\begin{figure}[t!]
\begin{center}
\includegraphics[width=0.49\textwidth]{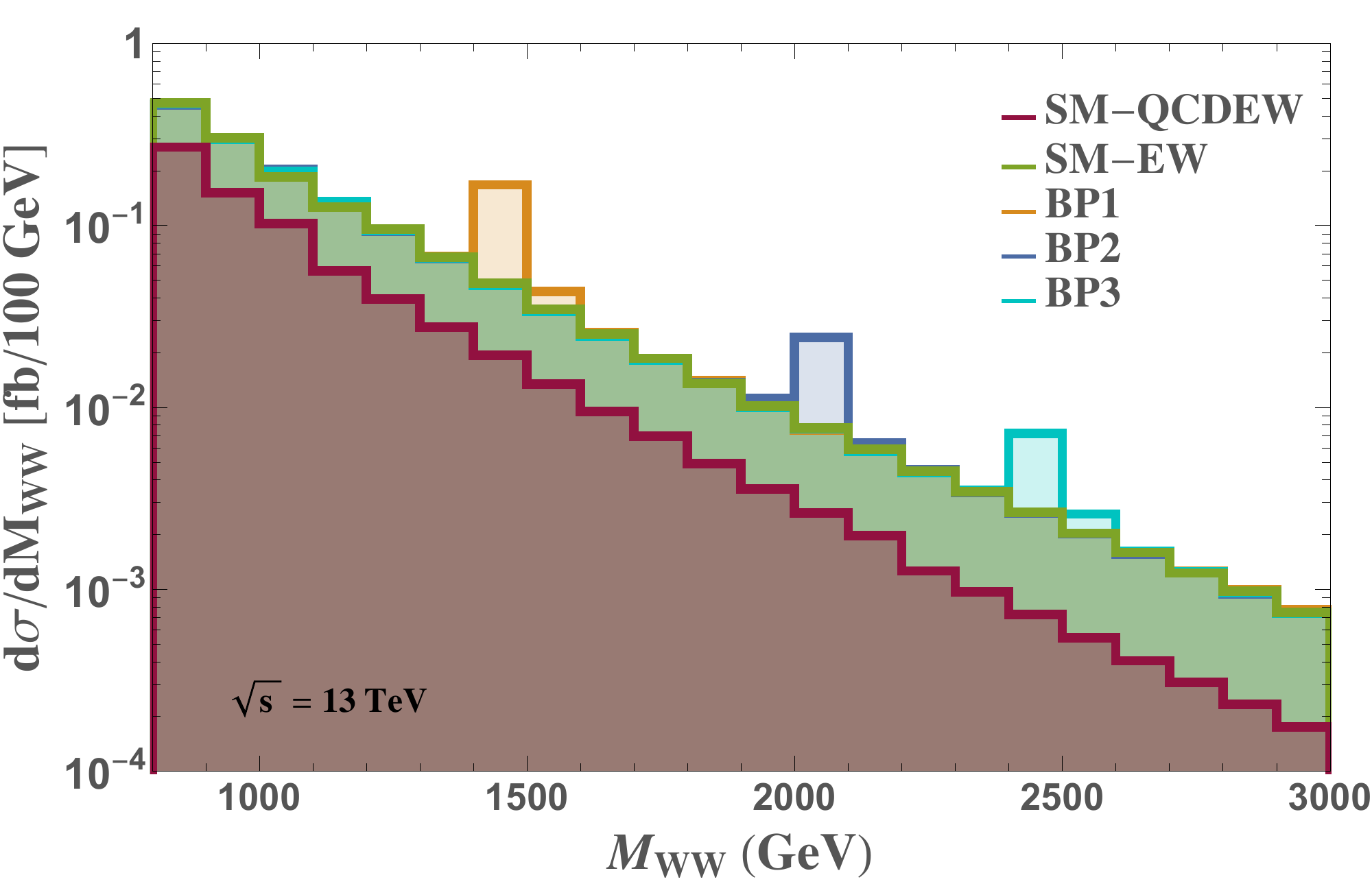}
\includegraphics[width=0.49\textwidth]{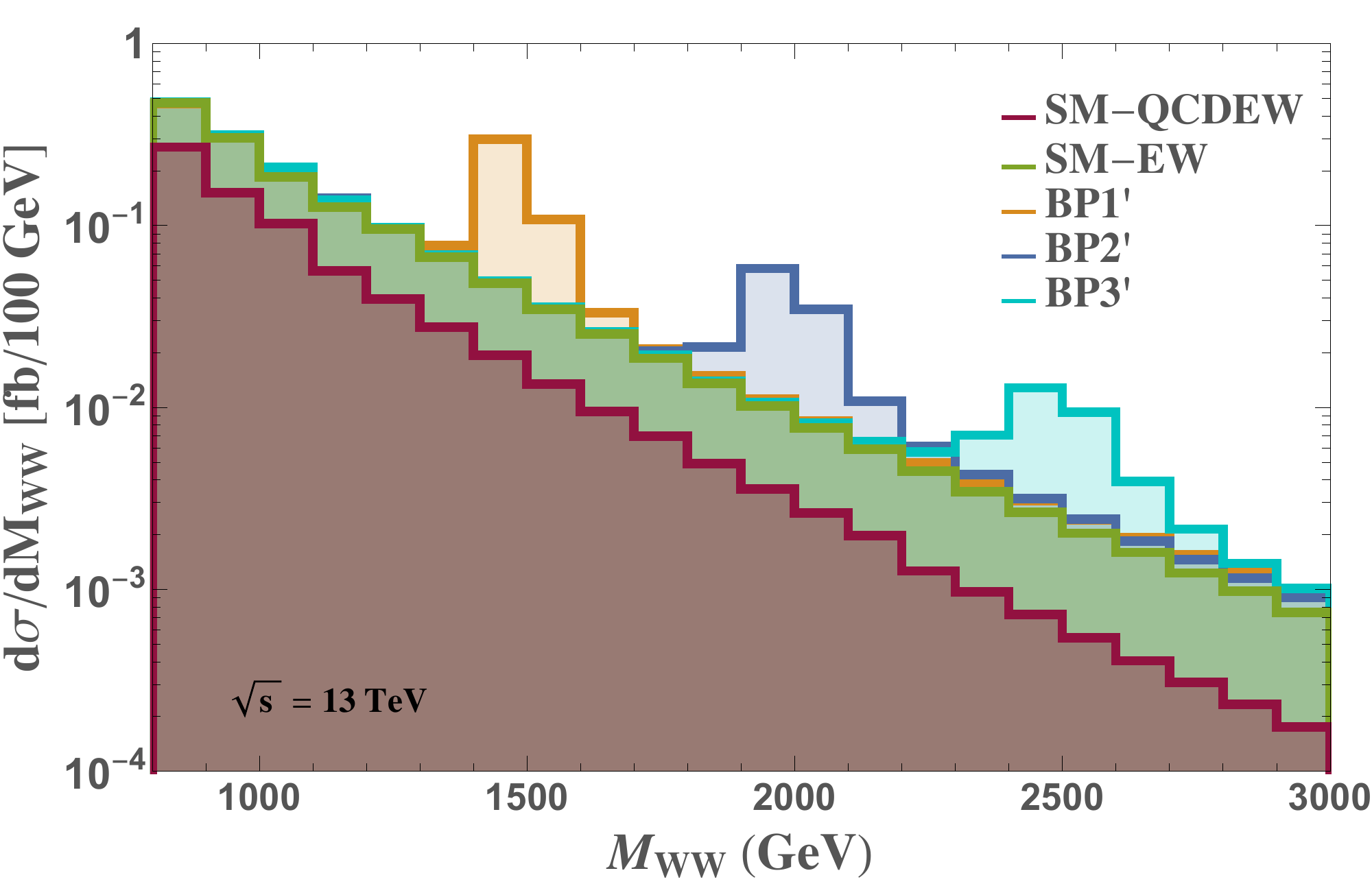}
\caption{Predictions of the  invariant mass, $M_{WW}$, distributions at the parton level. The results presented correspond to the cross sections times the branching ratios of the $W$ bosons to hadrons, $\sigma(pp \to W^+W^-jj)\times({\rm BR}(W\to {\rm hadrons}))^2$. The rates for the EW background (SM-EW), the mixed QCDEW background (SM-QCDEW) and for the selected signal scenarios for the vector resonances given by the BP's defined in \tabref{tablaBMP} are included. The cuts in \eqref{VBScuts} have been applied.}
\label{fig:partonlevel}
\end{center}
\end{figure}
\begin{align}
 &2 < \lvert\eta_{j_1,j_2}\rvert < 5, \nonumber \\
 &\eta_{j_1}\cdot\eta_{j_2} < 0,\nonumber \\\
 & M_{j_1j_2}>500~{\rm GeV},
\label{VBScuts}
\end{align}
where $j_1$ and $j_2$ refer to the two final quarks produced together with the two $W$s.  

The results of these parton level predictions, obtained with MadGraph, are collected in  \figref{fig:partonlevel} and in \tabref{parton}. \figref{fig:partonlevel} shows the distributions in the invariant mass of the $WW$ pair and \tabref{parton} summarizes the total cross sections in the invariant mass region of the $WW$ pair of our interest, i.e., summing events over the interval 800 GeV $<M_{WW}<$ 3000 GeV.

As we can see in \figref{fig:partonlevel} the six studied resonances emerge clearly above the prediction from the SM continuum, which is in turn clearly dominated by the EW background. The other studied backgrounds, the mixed QCDEW and the top-antitop ones are clearly subdominant for these specific configurations. In fact, we have checked explicitly that the main responsible for this strong suppression of the mixed QCDEW and  top-antitop backgrounds are the VBS cuts of \eqref{VBScuts}. In particular, the $t\bar{t}$ background is reduced by a factor of about $10^{-3}$ when applying these VBS cuts in the region of large $M_{WW}$.   

Although these results at the parton level are encouraging, the real challenge is to deal with the difficult task of reconstructing the $W$s from their hadronic decay products. This is the issue that we confront next. 

Concretely, we are going to explore the prospects for the fully hadronic decay channel,
\begin{equation}\label{mc_process}
 pp\to W^+W^-jj,~ W^\pm\to J(jj),
\end{equation}
where the 2 jets ($jj$) coming from the decay of each of the vector bosons are reconstructed as a single fat jet ($J$). See, for instance, ref.~\cite{Aoude:2019cmc} for the semileptonic channel, where only one of the vector bosons decays into 2 jets that are reconstructed as a single fat jet.

We consider three categories of events:
\begin{itemize}
  \item Signal: the prediction from our model of the vector resonances for the process shown in \eqref{mc_process}, with the model parameters set to the corresponding values of one of the BPs of \tabref{tablaBMP}. By construction, this is equal to the SM-EW prediction plus the extra events due to the BSM physics.
  \item SM-EW Background:  SM EW prediction for the process in \eqref{mc_process}.
  \item SM-QCD Background: SM QCD prediction for the process $pp\to jjJJ$  with 2 light jets being on the VBS kinematical region.
  \end{itemize}
Notice that we have not considered other possible backgrounds in the study of this hadronic channel, like those coming from the already mentioned mixed QCDEW and top-antitop backgrounds, since they are well below the SM-EW background. For this reason, we consider only the SM-EW background together with the dominant, and most problematic, QCD background. Notice also that the third category of events from QCD background is quite hard to filter out, as we will describe later.


\begin{table}[t!]
\centering
\vspace{.2cm}
\begin{tabular}{ c |@{\extracolsep{0.45cm}} c  @{\extracolsep{0.45cm}} c @{\extracolsep{0.45cm}} c@{\extracolsep{0.45cm}} c@{\extracolsep{0.45cm}} c@{\extracolsep{0.45cm}} c@{\extracolsep{0.45cm}} c@{\extracolsep{0.45cm}} c@{\extracolsep{0.45cm}} c }
\toprule
\toprule
 &BP1 & BP2  & BP3 & BP1'  & BP2' & BP3' & EW & QCDEW &  $t\bar{t}$ \\[3pt] 
\midrule
$\sigma$ [fb] & 1.57    & 1.46  & 1.44  &  1.8   &  1.55  &  1.51  & 1.43  &  0.71  &   0.11\,(0.24) 
\\[3pt] 
\bottomrule\bottomrule
\end{tabular}
\vspace{0.4cm}

\caption{Parton level predictions for the cross sections times the branching ratios of the $W$ bosons to hadrons, $\sigma(pp \to W^+W^-jj)\times({\rm BR}(W\to {\rm hadrons}))^2$ in fb, corresponding to the signal points, BP1, BP2, BP3, BP1', BP2' and, BP3'. The predictions for the main backgrounds: SM-EW (EW), mixed SM-QCDEW (QCDEW) and SM top-antitop ($t\bar{t}$), are also shown. In the $t\bar{t}$ case the decay chain $t \to W b$ has been considered, and a suppression factor of 0.2 (0.3) for each final $b$-jet being misidentified as a light jet $j$  has been applied. All the results are generated with MadGraph at the parton level, summed over the interval $800 \,{\rm GeV}< M_{WW} < 3000 \, {\rm GeV}$. Cuts in \eqref{VBScuts} have been applied.   
}
\label{parton}
\end{table}

The Monte Carlo chain MadGraph~v5~\cite{Alwall:2014hca,Frederix:2018nkq}, Pythia~8~\cite{Sjostrand:2006za} and Delphes~\cite{deFavereau:2013fsa} is used for this analysis. For the jet reconstruction, we use the FastJet library~\cite{Cacciari:2011ma,Cacciari:2005hq} with the anti-$kT$ algorithm~\cite{Cacciari:2008gp}, both integrated in Delphes. We will also use the boosted objects machinery~\cite{Abdesselam:2010pt,Thaler:2010tr,Thaler:2011gf} integrated in FastJet for $W$-tagging purposes. 
 
For each event, two lists of reconstructed jets are generated with the anti-$k_T$ algorithm, corresponding respectively to the thin (usual) jets, $j$, and to the fat jets, $J$. For the thin-jet one, $R=0.5$ is required, whereas for the fat jet one, $R=0.8$ is required. 
Regarding the cuts on the reconstructed jets, we first apply the following set of initial cuts to the thin jets and to the fat jets, respectively:
\begin{itemize}
\item[1)] Cuts on the thin jets.

We require 2 thin-jets ($j_1$, $j_2$), not b-tagged~\cite{Cacciari:2011ma,Cacciari:2005hq} that
in addition to the detection cuts, $p_{T_{j_1}},p_{T_{j_2}}>20~{\rm GeV}$, $\Delta R_{jj}>0.4$,
verify the VBS cuts,
\begin{align}
 &2 < \lvert\eta_{j_1,j_2}\rvert < 5\,, \nonumber \\
 &\eta_{j_1}\cdot\eta_{j_2} < 0\,, \nonumber \\
 &M_{j_1j_2}>500~{\rm GeV}\,.
\label{VBSthin} 
\end{align}
\item[2)] Cuts on the fat jets

We  require (at least) 2 fat jets, being $J_1$ and $J_2$ the leading (in the sense of largest $p_T$) and sub-leading fat jet respectively. The following basic cuts are set on the transverse momentum, the mass and the rapidity of each fat jet:
\begin{align}
 &p_{T_{J_1}}, p_{T_{J_2}}>200~{\rm GeV}\,, \nonumber \\
 &M_{J_1}, M_{J_2}>20~{\rm GeV}\,, \nonumber \\
 &\lvert\eta_{J_1}\rvert, \lvert\eta_{J_2}\rvert<2\,.
\label{fatjetcuts} 
\end{align}
\end{itemize}
\begin{figure}[t!]
\begin{center}
\includegraphics[width=0.49\textwidth]{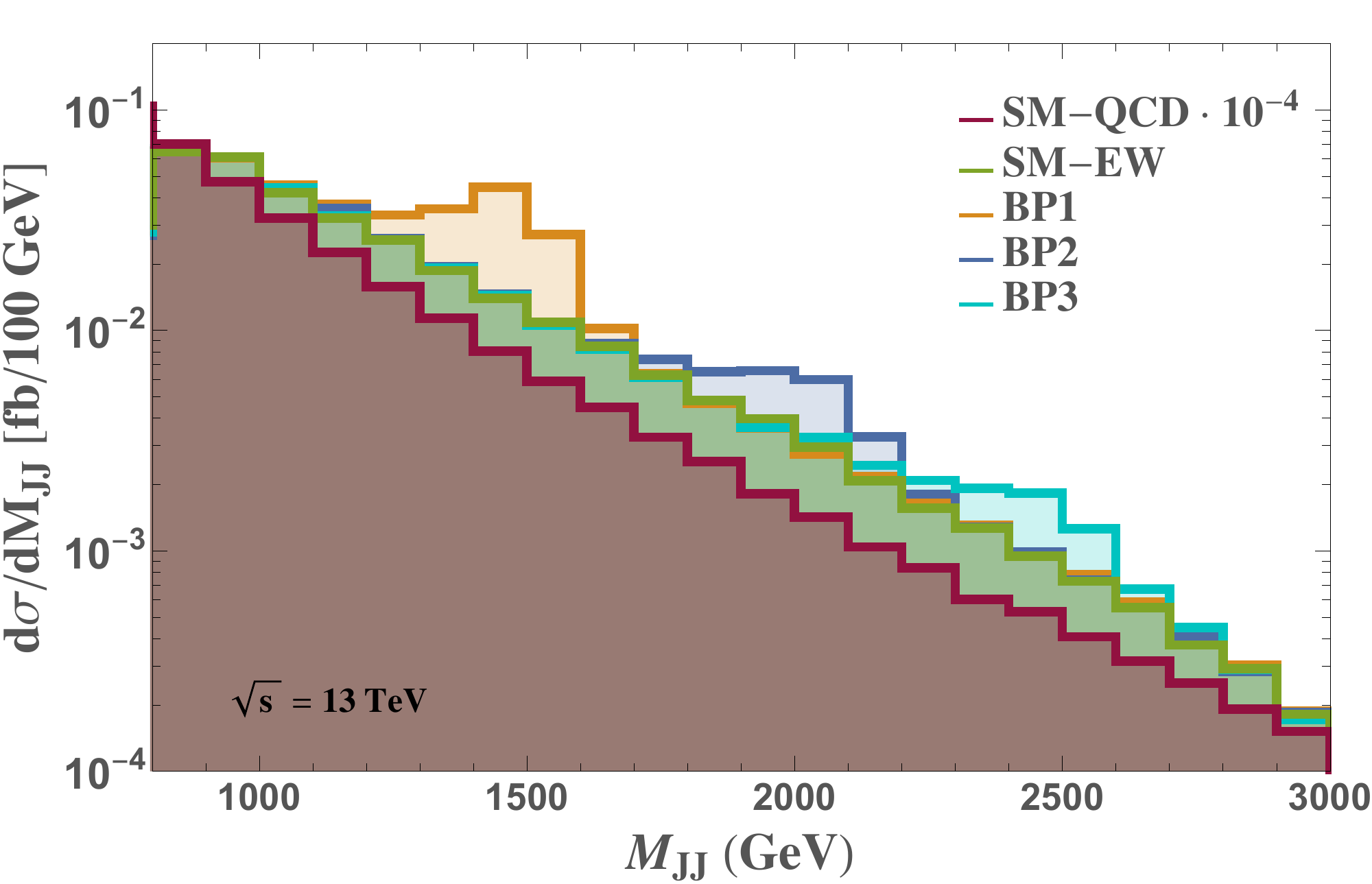}
\includegraphics[width=0.49\textwidth]{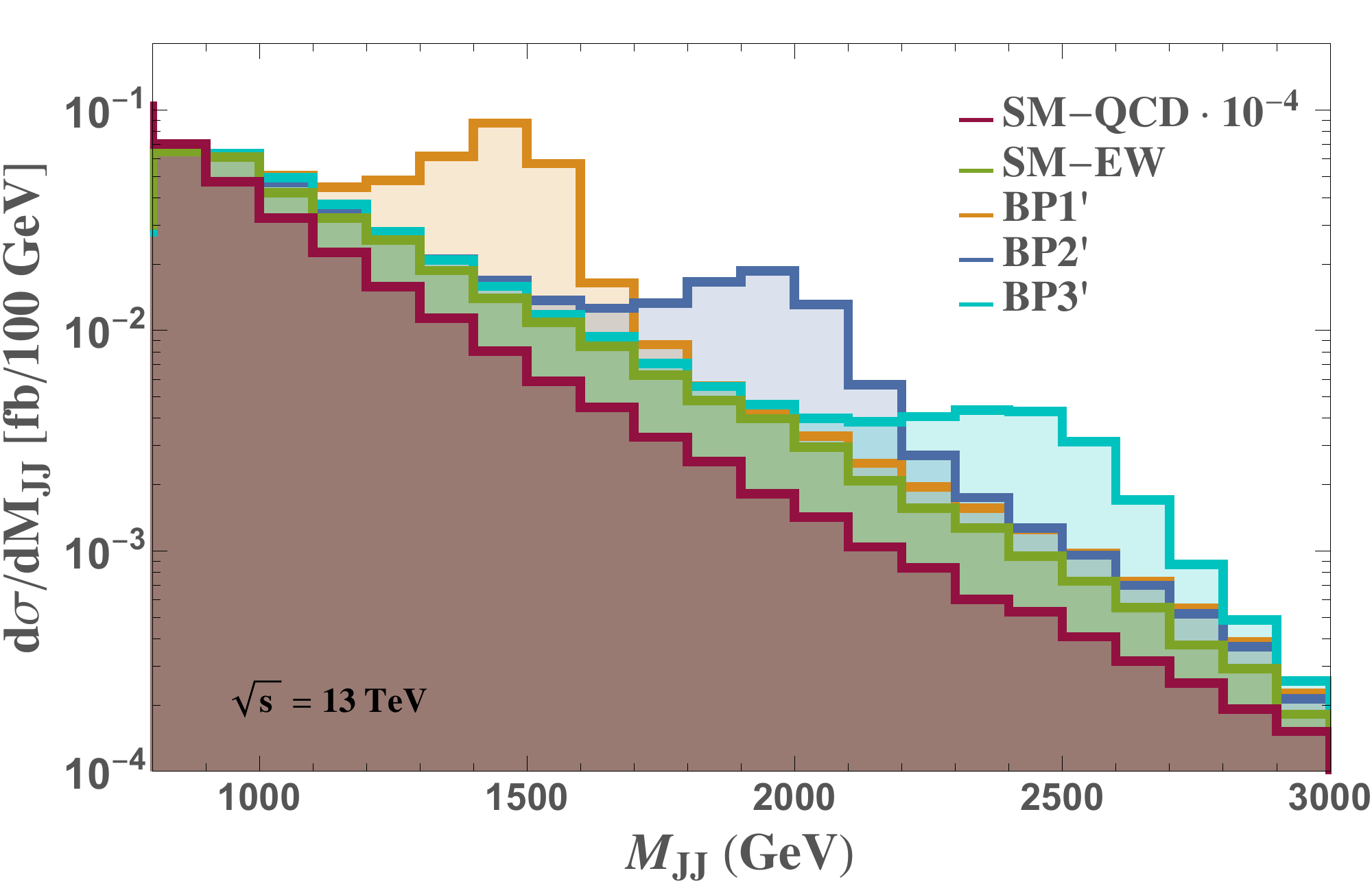}
\caption{Predictions of the cross sections distributions of $JJjj$ events with the invariant mass of the fat jet pair, $M_{JJ}$, after jet reconstruction using  MadGraph+PYTHIA+DELPHES with the anti-$k_T$ algorithm. The rates for the EW background (SM-EW), the pure QCD background (SM-QCD, scaled down by a factor $10^{-4}$) and the selected signal scenarios for the vector resonances given by the BP's defined in \tabref{tablaBMP} are included. The cuts in \eqref{VBSthin} and \eqref{fatjetcuts} have been applied.}
\label{fig:JJlevel}
\end{center}
\end{figure}
If the event is correctly identified, the $j_1$ and $j_2$ VBS jets will be the 2 reconstructed jets coming from the $pp\to W^+W^-jj$ VBS event. The $J_1$ and $J_2$ fat jets will then correspond, directly, to the reconstructed vector bosons $W^\pm\to J(jj)$. By means of 4-momenta conservation, the masses $M_{J_1}$ and $M_{J_2}$ and the total invariant mass $M_{JJ} = (p_{J_1} + p_{J_2} )^2$ of the reconstructed fat jets are identified with those of the original vector bosons coming from the VBS event. Notice that because of the usage of fully hadronic events no information is lost, in contrast to the cases where there are neutrinos in the final state. Hence, the component of momenta parallel to the beamline can be reconstructed. Note also that the requirement $2 < \lvert\eta_{j_1,j_2}\rvert < 5$ for the VBS thin-jets and $\lvert\eta_{J_1,J_2}\rvert<2$ for the reconstructed fat jets coming from the vector bosons means that both objects can be (in principle) easily classified by means of $\eta$ variable since they belong to mutually excluding regions.

It is possible that additional jets collinear with those coming from the hard scattering event are produced by final state radiation. These jets could be reconstructed as additional thin jets, or could lead to a thin jet being also reconstructed as a fat jet. The fat jet constituents would be those coming from the additional radiation process.

With all the above considerations taken into account, we compute our predictions for the three specified event categories obtaining the results summarized in  \figref{fig:JJlevel} and  \tabref{JJ}.  \figref{fig:JJlevel} shows the distributions in the invariant mass of the $JJ$ pair and \tabref{JJ} summarizes the total cross sections in the invariant mass region of the $JJ$ pair of our interest, i.e., summing events over the interval 800 GeV $<M_{JJ}<$ 3000 GeV.

 The main conclusions we learn from these results are the following: first, we see clearly in \figref{fig:JJlevel} that the vector resonances still emerge over the EW background, although with wider peaks than in the previous results at parton level, due to the  typical energy loss in associated to jet reconstruction process. Second, when we compare the $WWjj$ parton level rates in \tabref{parton} with the $JJjj$ rates in  \tabref{JJ} we see that the ratios $JJjj/WWjj$ for the EW processes, i.e., the EW background and the signal BP's, are in the interval $(0.2,0.3)$. This can be interpreted as coming from the rescaling of the parton level results based on an efficiency in each $W$ tagging from each fat jet in the range 
$(0.45,0.55)$. This is indeed in agreement with previous estimates of this efficiency (see, for instance refs.~\cite{Khachatryan:2014hpa,Aad:2015rpa,Aad:2015owa,Heinrich:2014kza}) as we already saw in the $WZ$ case in Chapter \ref{Resonances}.

We also learn from these results of the total $JJjj$ rates that the dangerous QCD background overwhelms both the signal and the EW background by a factor of $10^3-10^4$. Concretely, the total cross section integrated over the interval $800 \, {\rm GeV}< M_{JJ} < 3000 \, {\rm GeV}$ for the QCD background is, according to our result in \tabref{JJ}, 4392 times larger than our largest signal corresponding to the BP1'. This fact is really challenging to deal with. Therefore, in order to improve the signal to background ratios a more refined analysis profiting from the fat jet features is needed. 

\begin{table}[t!]
\centering
\begin{tabular}{ c |@{\extracolsep{0.65cm}} c  @{\extracolsep{0.65cm}} c @{\extracolsep{0.65cm}} c@{\extracolsep{0.65cm}} c@{\extracolsep{0.65cm}} c@{\extracolsep{0.65cm}} c@{\extracolsep{0.65cm}} c@{\extracolsep{0.65cm}} c@{\extracolsep{0.65cm}} c }
\toprule
\toprule
 &BP1 & BP2  & BP3 & BP1'  & BP2' & BP3' & EW & QCD \\[3pt]
\midrule
$\sigma$[fb]&  0.384     & $ 0.322  $ & $ 0.312  $ & $  0.526   $ & $  0.380  $ & $  0.348  $ & $  0.304  $ & $  2310 $ 
\\[3pt]
\bottomrule\bottomrule
\end{tabular}
\vspace{0.4cm}
\caption{ Predictions after jets reconstruction for the cross sections, 
$\sigma(pp \to JJjj) $ in fb, corresponding to the signal points, BP1, BP2, BP3, BP1', BP2' and, BP3', and for the 
main backgrounds: SM-EW (EW) and SM-QCD (QCD).  The results are generated with MadGraph+PYTHIA+DELPHES, summing events over the interval $800 \, {\rm GeV}< M_{JJ} < 3000 \, {\rm GeV}$, and the cuts in \eqref{VBSthin} and \eqref{fatjetcuts} have been applied.   
}
\label{JJ}
\end{table}

We have investigated further into more specific characteristics of the produced fat jets, analyzing in more detail the events for both signal and QCD background in terms of the following  fat jet variables and their optimal cuts: $M_{J_1}$, $M_{J_2}$, $p_{T_{J_1}}$, $p_{T_{J_2}}$, $\Delta\eta_{JJ} = \eta_{J_1} - \eta_{J_2}$, 
$\Delta R_{JJ}=\sqrt{(\Delta (\Phi_{JJ})^ 2+(\Delta\eta_{JJ})^2}$,  and $\tau_{21}=\tau_2/\tau_1$. 
In particular, the latter variable $\tau_{21}$~\cite{Thaler:2010tr} seems to be a very good discriminant for boosted objects studied via fat jets. In fact, it has been already used for $W$-tagging purposes in the case of the semileptonic decay channel~\cite{Aoude:2019cmc} and in the recent ATLAS study~\cite{Aad:2019fbh}.

It is obvious that a cut on the mass variables like $M_J$ and $M_{JJ}$ restricting them to windows around, respectively,  the $M_W$ mass and the corresponding resonance mass $M_{V^0}$ of the studied BP signal,  will improve considerably the signal to background ratio. The only problem imposing these mass windows is that if they are too narrow we may loose too much signal and end up lacking statistics for the analysis. It should be noticed that we are talking about BSM signals over the EW background of, in the best case, $(\sigma_{\rm S}-\sigma_{\rm EW}) \sim 0.222 \, {\rm fb}$, and, therefore,  a total number of events of at most 67 for an integrated luminosity of $L = 300 \, {\rm fb}^{-1}$, and of 670 for $L = 3000\, {\rm fb}^{-1}$. For this reason, we have focused our more refined study on the highest luminosity option.   
 
\begin{figure}[t!]
\begin{center}
\includegraphics[width=0.49\textwidth]{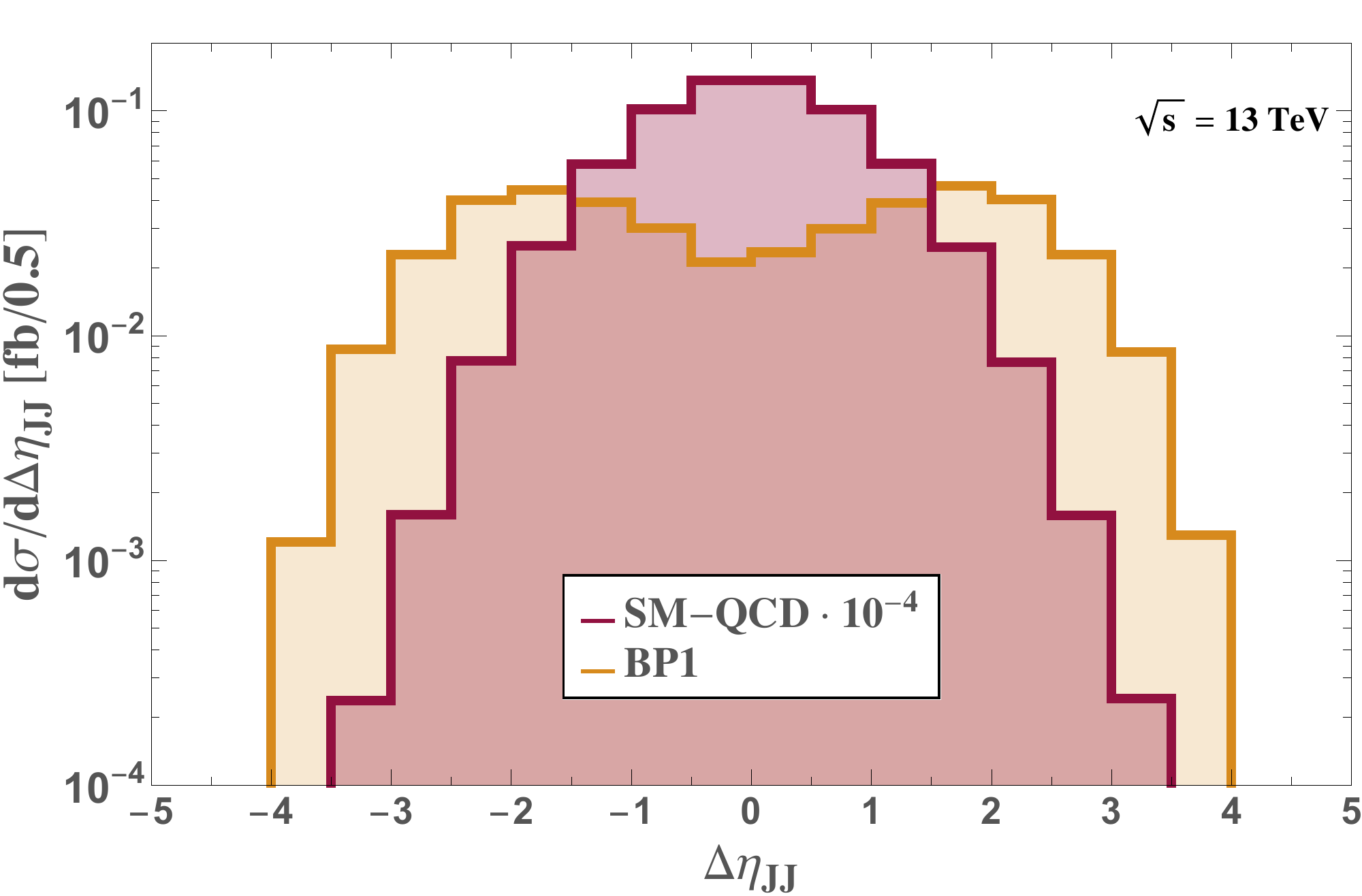}
\includegraphics[width=0.49\textwidth]{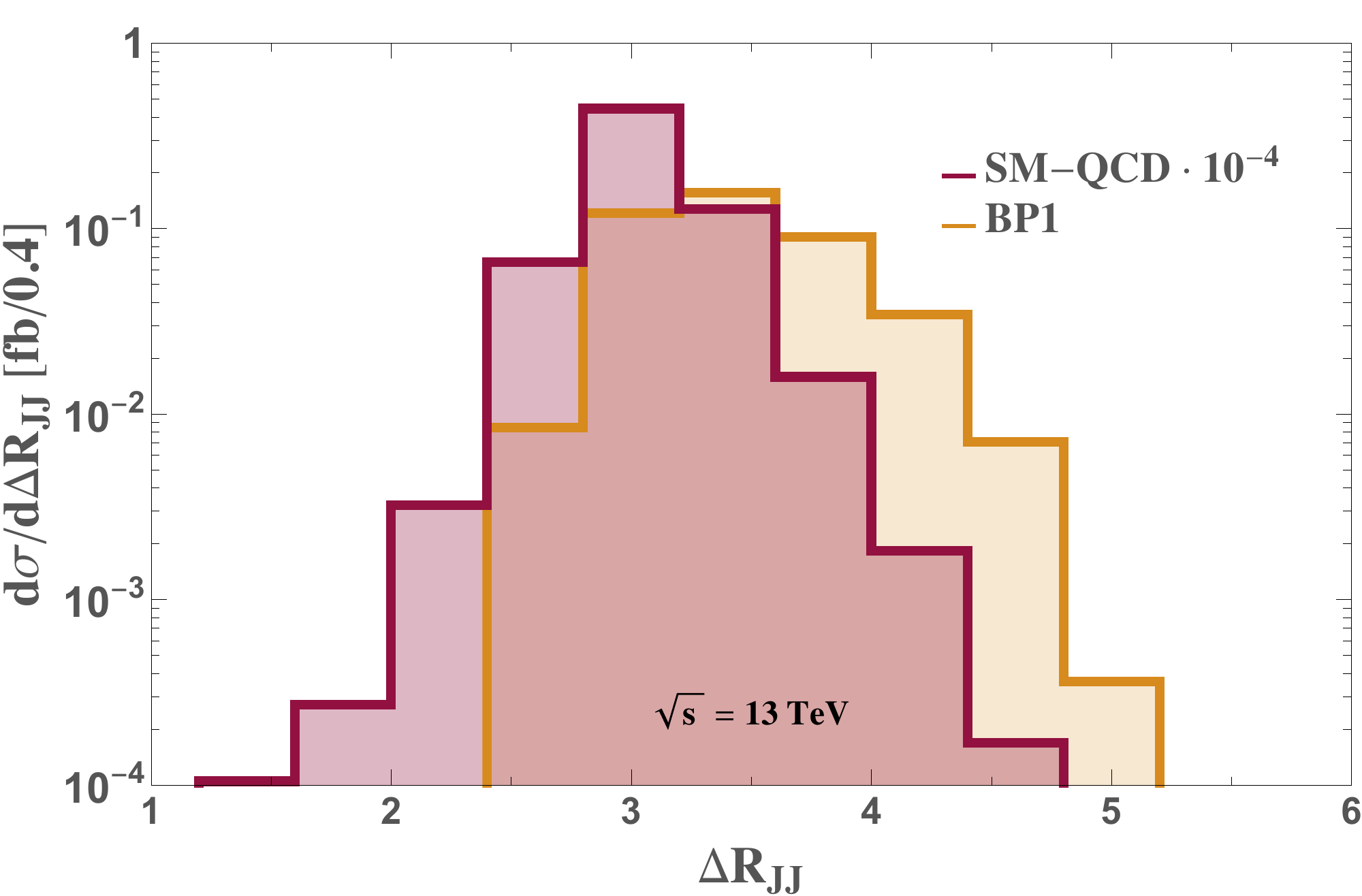}
\includegraphics[width=0.49\textwidth]{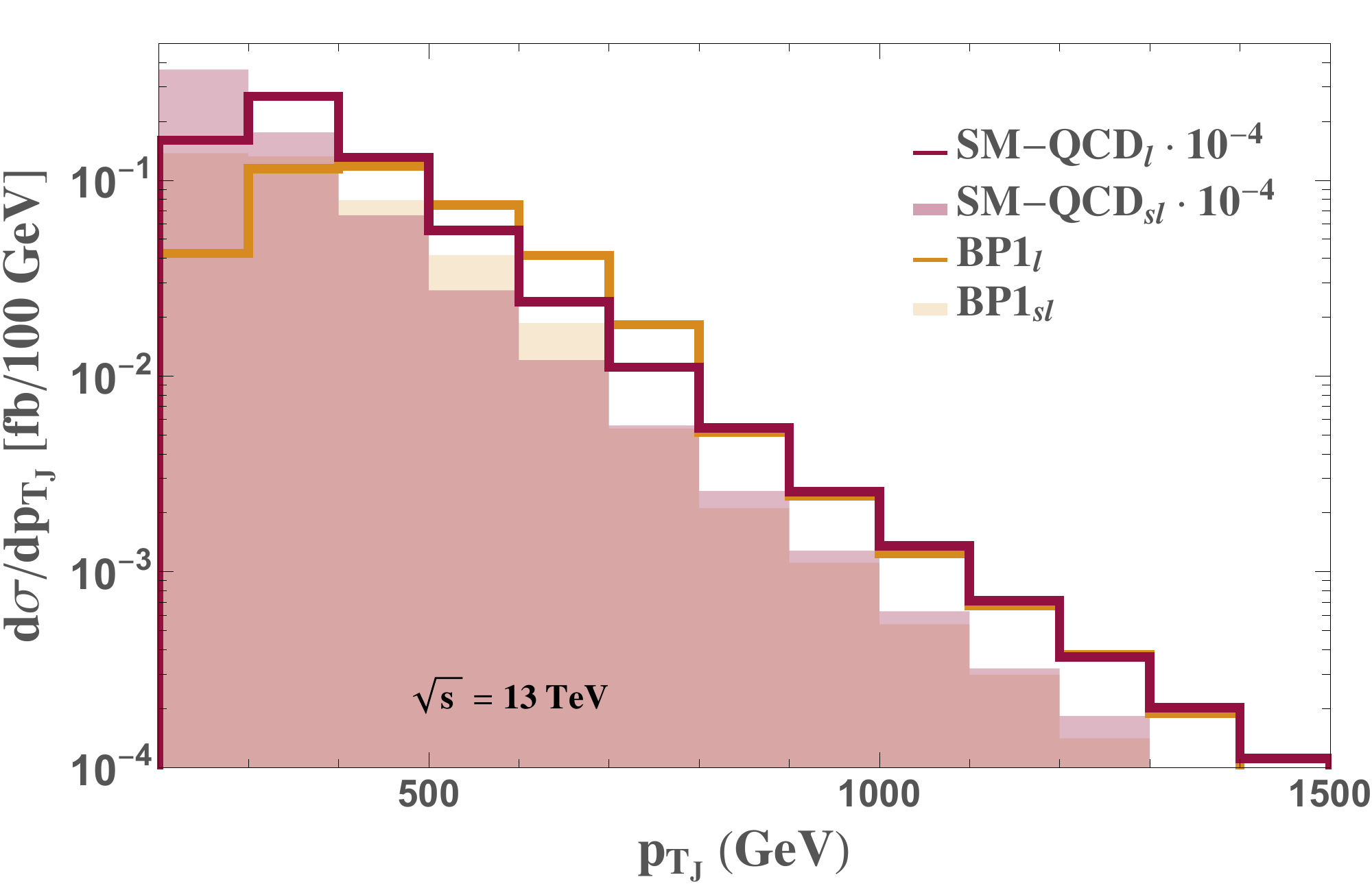}
\includegraphics[width=0.49\textwidth]{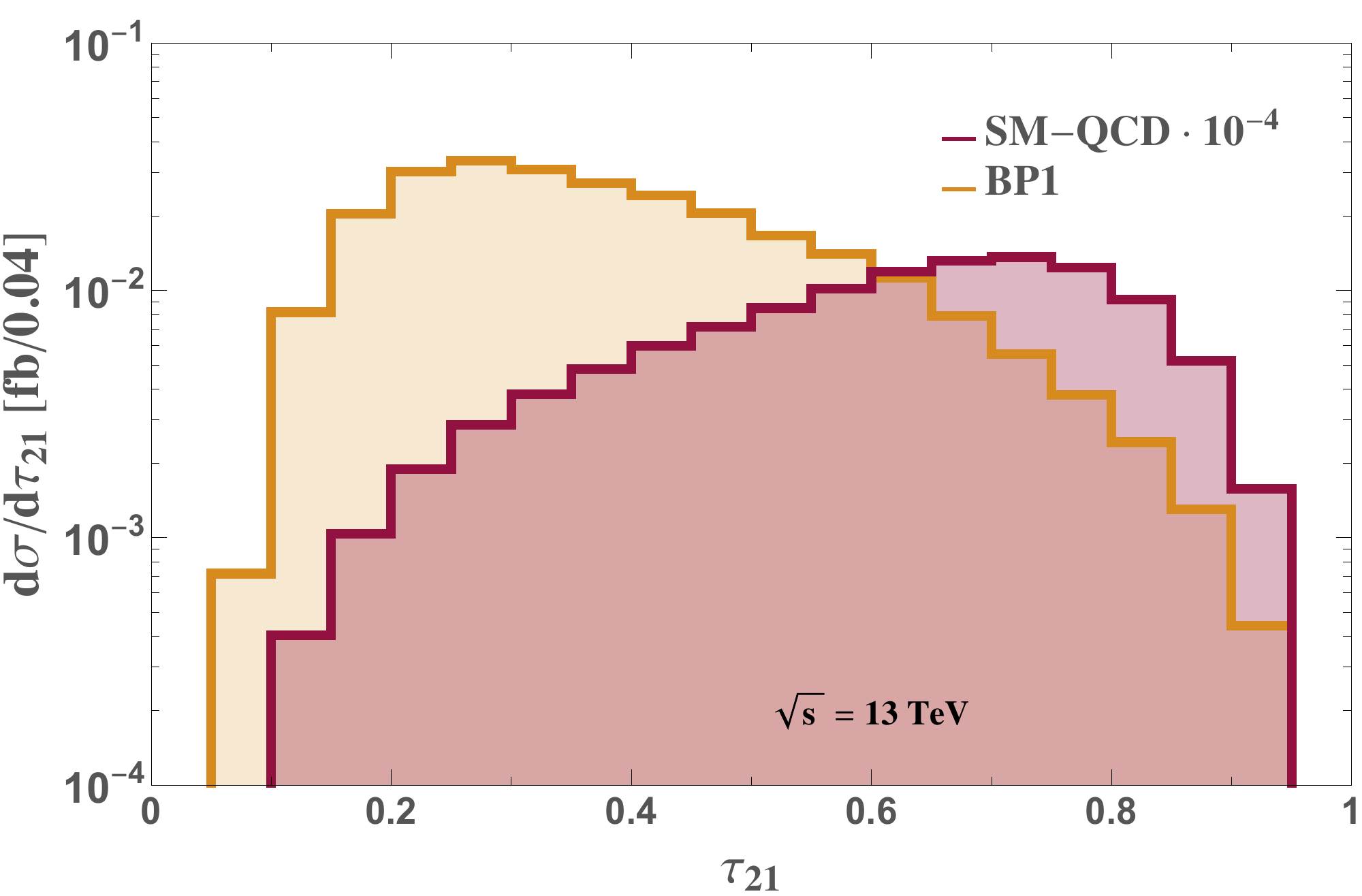}
\caption{Predictions, after jets reconstruction, of the cross sections distributions with the fat jet variables $\Delta\eta_{JJ}$ (upper left plot), $\Delta R_{JJ}$ (upper right plot), $P_{T_J}$ (lower left plot) and $\tau_{21}$ (lower right plot) for the QCD background 
(SM-QCD, scaled down by a factor $10^{-4}$) and for the signal in the  scenario BP1. The cuts in \eqref{VBSthin} and \eqref{fatjetcuts} have been applied. }
\label{fig:Jfeatures}
\end{center}
\end{figure}

Regarding the remaining fat jet variables, we present our results for the distributions of the signal and the QCD background with respect to  $\Delta\eta_{JJ}$, $\Delta R_{JJ}$, $p_{T_{J_i}}$ and $\tau_{21}$ in \figref{fig:Jfeatures}. For the signal we have selected the BP1 case, as an example.

From these figures we learn that the two fat jets from the BP1 signal tend to be more separated, both in $\Delta\eta_{JJ}$ and in $\Delta R_{JJ}$,  than in the QCD case. Also the transverse momenta $P_{T_J}$ of the leading (l) and subleading fat jets (sl) tend to be larger in the signal case than in QCD one. Finally, we find that the $\tau_{21}$ variable, which tests the correctness of the hypothesis of having a fat jet being composed of two light jets (with $\tau_{21}$ close to 1 meaning that this hypothesis is incorrect), is one of the best discriminants in our case. For instance, we have checked explicitly that applying a cut of $\tau_{21}<0.3$  together with 
$60\,{\rm GeV}< M_J < 100\,{\rm GeV}$, in the large invariant mass interval
$1000 \,{\rm GeV}<M_{JJ} <3000  \,{\rm GeV}$, reduces the QCD background by a factor of $2.4 \times 10^{-5}$ whereas the BP1 (BP1') signal is reduced by a milder factor of  $6.3 \times 10^{-2} (7.8 \times 10^{-2})$. Thus, this $\tau_{21}$ variable together with $M_J$ results to be very efficient in reducing the QCD background to a controllable level. Again the only problem is the low statistics of the signal when imposing tight cuts, specially for the heavier resonances.

Finally, in order to make a more systematic exploration looking for the best set of cuts on the fat jet variables, we have performed a full survey considering all the possible combinations of cut choices, including four options for each variable cut. These are set in addition to the basic cuts in \eqref{VBSthin} and \eqref{fatjetcuts}. The following options have been considered:
\begin{itemize}
\item[] $M_J ({\rm GeV})$
 in the interval (50,110), (60,100), (70,95), or no cut.
$J$ refers to both fat jets. 
\item[] $p_T^J ({\rm GeV})$ minimum:
200, 300, 400, 600.
$J$  refers to both fat jets. 
 \item[] $|\Delta \eta_{J_1J_2}|$ minimum:
 0.5, 1.0, 1.5, or no cut.
 \item[] $\Delta R_{J_1J_2}$
 in the interval (2,5), (2.5,4.5), (3,4), or no cut. 
\item[] $\tau_{21}(J)$
 in the interval (0.1,0.4), (0.1,0.35), (0.1,0.3), or no cut.
 $J$  refers to both fat jets. 
\end{itemize}

The considered windows in $M_{JJ}$ are fixed correspondingly for each BP scenario to the interval centered at approximately the resonance mass $m_{V^0}$ $\pm$ 250 GeV. 
In addition, in our search of the optimal cuts, we also allow for events with up to 4 extra thin jets, besides to the two VBS thin jets,  with $\Delta R_{jJ}<0.8$ (angular distance between the non-VBS thin jet $j$ and the closest fat jet $J$).  Regarding the fat jets, we require a minimum of two reconstructed fat jets and a maximum of four, and the variable $M_{JJ}$ is the reconstructed invariant mass of the two leading fat jets. 

In all this study of optimal cuts we have used the highest luminosity option for the LHC of $3000 \, {\rm fb}^{-1}$. Notice that due to the fact that it is not possible to disentangle the $W^+W^-jj$ case from the $W^+W^+jj$ one if one uses $W$ tagging by means of fat jets, we consider in this last analysis both final states contributing in both EW background and signal predictions. In fact, also the $W^-W^-jj$ case would contribute to our final $JJjj$ events but we have checked that it is much smaller than the other two cases. Specifically, for the EW background, we have found the following hierarchy in their corresponding rates at the parton level: $\sigma(W^+W^-jj)=2.4 \, \sigma(W^+W^+jj) =11.2\, \sigma(W^-W^-jj)$. Thus, in the following, we have neglected the contribution from $W^-W^-jj$ and $W^+W^+jj$ in both signal and EW background rates. 


\begin{table}[t!]
\centering
\begin{tabu} to \textwidth {X[c]|X[c]X[c]X[c]X[c]X[c]X[c]}
\toprule
\toprule
& {\footnotesize {\bf BP1'}} & {\footnotesize {\bf BP2'}} &  {\footnotesize {\bf BP3'}} & {\footnotesize {\bf BP1}} & {\footnotesize {\bf BP2}} & {\footnotesize {\bf BP3}} 
\\
\midrule
$\sigma^{\rm stat}$              & 11.7        & 4.1             & 2.3       & 4.6        & 0.77     & 0.52
\\[3pt]                                                                                                             
$\sigma_{\rm EW}^{\rm stat}$               & 18.8         & 8.3             & 5.8       & 7.3        & 1.6      & 1.3
\\[3pt]                                                                                                             
$N_{\rm S}$                           & 103         & 17.2               & 9.7       & 51.9        & 5.7      & 3.6
\\[3pt]                                                                                                             
$N_{\rm EW}$                    & 19.5        & 2.9                & 1.8        & 19.5        & 2.9      & 1.8
\\[3pt]                                                                                                             
$N_{\rm QCD}$                   & 30.8         & 9.5             & 9.8        & 30.8         & 9.5      & 9.8
\\[3pt]                                                                                                             
\midrule                                                                                                            
$p_T^{J_1}$(GeV)                & $>200$           & $>200$      & $>200$      & $>200$      & $>200$    & $>200$     
\\[3pt]                                                                                                             
$p_T^{J_2}$(GeV)                & $>200$           & $>200$      & $>200$      & $>200$      & $>200$    & $>200$  
\\[3pt]                                                                                                             
$\tau_{21}(J_1)$                & $0.1-0.3$    & $0.1-0.3$   & $0.1-0.4$   & $0.1-0.3$   & $0.1-0.3$ & $0.1-0.4$
\\[3pt]
$\tau_{21}(J_2)$                & $0.1-0.4$     & $0.1-0.3$   & $0.1-0.4$   & $0.1-0.4$   & $0.1-0.3$ & $0.1-0.4$ 
\\[3pt]                                                                                                             
$M_{J_1}$(GeV)                   & $70-95$    & $70-95$      & $70-95$    & $70-95$    & $70-95$  & $70-95$  
\\[3pt]   
$M_{J_2}$(GeV)                   & $70-95$    & $50-110$      & $50-110$    & $70-95$    & $50-110$  & $50-110$  
\\[3pt]                                                                                                          
$\lvert\Delta \eta_{J_1J_2}\rvert_{\rm min}$ & no cut         & no cut        & no cut         & no cut       & no cut  & no cut 
\\[3pt]  
$\Delta R_{J_1J_2}$ & no cut          & no cut        & no cut         & no cut         & no cut  & no cut  
\\[3pt]    
$M_{JJ}$\,(TeV)      & $1.5 \pm 0.25$ & $2.0 \pm 0.25$  & $2.5 \pm 0.25$ & $1.5 \pm 0.25$ & $2.0 \pm 0.25$  & $2.5 \pm 0.25$
\\[3pt]
\bottomrule\bottomrule
\end{tabu}
\vspace{0.4cm}
\caption{Results for the optimal cuts and the number of predicted events for the various signal scenarios ($N_{\rm S}$) given by the BPs and for the SM backgrounds, EW ($N_{\rm EW}$) and QCD ($N_{\rm QCD}$), as well as their associated statistical significances defined in \eqref{stat}. Here $J_{1,2}$ are the leading and subleading fat jets respectively. The integrated luminosity is fixed to  $L=3000\,{\rm fb}^{-1}$.}
\label{tablaSIGMAS}
\end{table}

The results of our survey for the optimal cuts are collected in  
\tabref{tablaSIGMAS}. We define our optimal cuts as those that lead to the best statistical signal significance, defined as:
\begin{align}
  \sigma^{\rm stat} &= \frac{N_{\rm S}-N_{\rm EW}}{\sqrt{N_{\rm EW}+N_{\rm QCD}}} \, ,\\
  \sigma_{\rm EW}^{\rm stat}  &= \frac{N_{\rm S}-N_{\rm EW}}{\sqrt{N_{\rm EW}}},
\label{stat}
\end{align}
where $N_{\rm S}$,  $N_{\rm EW}$ and  $N_{\rm QCD}$ refer to the number of events of the signal, EW background and QCD background, respectively, for a  luminosity of $L=3000~{\rm fb}^{-1}$.
 Due to the clear dominance of the QCD background over the EW background and our incapability  of separating both background sources, the most realistic and determining value should be that of the $\sigma^{\rm stat}$ above.

In \tabref{tablaSIGMAS}  we summarize our main findings, where it is manifest that the best significances are found for the resonances with mass close to 1.5 TeV, especially BP1', and next for BP2' with a mass around 2 TeV. The expectations for the resonances with a heavy mass close to 3 YeV are considerably less appealing.  Notice that we have also included $ \sigma_{\rm EW}^{\rm stat}$ in our results in \tabref{tablaSIGMAS} to motivate for a future, more sophisticated analysis that could find out a more efficient way to suppress the difficult QCD background. In that hypothetical case the statistical significances of the signal would improve considerably, as can it be learnt  from the appealing larger values of $ \sigma_{\rm EW}^{\rm stat}$ in this Table.  

In the performed analysis leading to our final results summarised in \tabref{tablaSIGMAS} we have also learnt some interesting features regarding the comparative performance of the various cuts on the studied variables $p_T^{J_{1,2}}$, $\tau_{21}(J_{1,2})$, $M_{J_{1,2}}$, $\lvert\Delta\eta_{J_1J_2}\rvert_{\rm min}$, $\Delta R_{J_1J_2}$, and $M_{JJ}$, which we believe are worth to comment. 

Firstly, as common features to all the studied points, we find that the best cut on $p_T^{J_{1,2}}$ for both fat jets is $p_T^{J_{1,2}}>200$~GeV. Applying a stronger cut on $p_T^{J_{1,2}}$, like the ones considered in our analysis of $300$, $400$ and $600$~GeV, leads generally to a smaller statistical significance of the signal, basically because of the lack of statistics for the signal. It is also common to all points, that $\sigma^{\rm stat}$ is not very sensitive to the considered cuts on $\lvert\Delta \eta_{J_1J_2}\rvert_{\rm min}$ and $\Delta R_{J_1J_2}$, and, therefore, the best option again in order not to loose much signal is the 'no cut' option in both of these variables. For instance, for BP1', varying $\lvert\Delta \eta_{J_1J_2}\rvert_{\rm min}$ from our best option, 'no cut', to 1 changes the prediction of $\sigma^{\rm stat}$ from the largest value in \tabref{tablaSIGMAS} of 11.7 to 10.5. A similar feature is found for the other points. Therefore, the most efficient cuts in reducing the QCD background and leading to the largest $\sigma^{\rm stat}$ for all the benchmark points are those applied on $M_{J_{1,2}}$ and on $\tau_{21}(J_{1,2})$. Since we have done our exploration of cuts for $\tau_{21}(J_{1,2})$ by starting already with quite optimised and similar intervals, we do not find much differences in their corresponding predictions of $\sigma^{\rm stat}$ for all the points. For instance, for BP1', varying $\tau_{21}(J_{1,2})$ from the combination in \tabref{tablaSIGMAS}, respectively,  of ((0.1,0.3), (0.1,0.4)) to ((0.1,0.3),(0.1,0.3)) changes $\sigma^{\rm stat}$ from 11.7 to  11.4. The other combination considered, ((0.1,0.4),(0.1,0.4)) leads to a slightly lower $\sigma^{\rm stat}$ of 8.1. Once again a similar conclusion applies to  the other points. Secondly, the performance of the cuts explored on the variables $M_{J_{1,2}}$ deserves some devoted comments.

These are very relevant variables in the definition of our signal, since they quantify the accuracy in the identification/reconstruction of the two final gauge bosons from the two selected fat jets. Therefore, a priori, the more precision in the determination of $M_{J_{1,2}}$ the better the expected statistical significance of the signal. However, considering too narrow intervals in these variables does not always lead to the best  $\sigma^{\rm stat}$, again because of the lack of statistics for the signal, specially for the heavier resonances. The best option that we have found, as shown in \tabref{tablaSIGMAS}, is: $M_{J_1}$(GeV) in (70,95), $M_{J_2}$(GeV) in (70,95) for BP1' and BP1; and $M_{J_1}$(GeV) in (70,95), $M_{J_2}$(GeV) in (50,110) for BP2', BP2, BP3' and BP3. Changing these to other options leads to lower $\sigma^{\rm stat}$. For instance, for BP1', considering $M_{J_1}$(GeV) in (70,95), $M_{J_2}$(GeV) in (60,100), changes  $\sigma^{\rm stat}$ from our best value of 11.7 to 10.6; and considering $M_{J_1}$(GeV) in (70,95), $M_{J_2}$(GeV) in (50,110) changes it to 9.6. Similarly, for BP2', considering $M_{J_1}$(GeV) in (70,95), $M_{J_2}$(GeV) in (60,100) changes $\sigma^{\rm stat}$ from the best value $4.1$ to $3.9$, and considering  both $M_{J_{1,2}}$(GeV) in (70,95), changes it to 3.0.

 Finally, it is also interesting to comment on what happens if the search is not devoted to the starting optimised windows in $M_{JJ}$, as indicated in \tabref{tablaSIGMAS}, corresponding to the explored intervals centered  at the mass of each resonance. After all, when one compares with data there is not a preferred interval in $M_{JJ}$ where to look at a priori. Thus, it is also interesting to explore the optimal combination of cuts for the overall best performance across all the points in the full explored region of $M_{JJ}$(GeV) in (1000,3000). In this sense, we have checked that the set of cuts specified in \tabref{tablaSIGMAS}, other than $M_{JJ}$,  lead to the best options for all the points. But, as expected, the statistical significance of the signal decreases significantly for all the points compared to the case in which the $M_{JJ}$ window is optimized for each resonance mass.  Specifically, for $M_{JJ}$(GeV) in (1000,3000), we find that $\sigma^{\rm stat}$ for BP1', BP2', BP3', BP1, BP2 and BP3, change our best values in \tabref{tablaSIGMAS} to 4.8, 1.1, 0.5, 1.8, 0.1 and 0.03 respectively. Therefore, in that case, only BP1' could be observed.

To summarize, in this Chapter we have extended our previous UFO model, first applied to the study of the vector resonances emerging in the VBS channel $W^+Z\to W^+Z$, carried out in Chapter \ref{Resonances},  to the different VBS channel $W^+W^-\to W^+W^-$. The charged vector resonance $V^+$ was relevant for $W^+Z\to W^+Z$ scattering since it can propagate in the s-channel of this process, leading to emergent peaks in the cross section. On the other hand, the neutral resonance $V^0$ is the one that can propagate in the s-channel of  the $W^+W^-\to W^+W^-$ process instead. Our study of these resonances at the LHC by means of the process $pp \to WW jj$ is therefore complementary to the previous study. Furthermore, in contrast to Chapter \ref{Resonances} where the focus was set on the fully leptonic channel of the final EW gauge bosons, we devote here our full work to the difficult but very interesting task of disentangling the vector resonance in $pp \to WW jj$ events using  the $W$ gauge boson hadronic decays. 

After performing a detailed study of the final hadronic states in $pp \to JJjj$ events at the LHC with two thin jets having the typical VBS topology and using the two fat jets for the tagging of the two final $W$ bosons, we have concluded that only the lightest and wider resonances might be discovered at the LHC with $L=3000\,{\rm fb}^{-1}$. Other, heavier resonances might reach interesting statistical significances, even above the observation threshold, but would not reach, in principle, the 5$\sigma$ level. These modest results are obtained due to the large amount of QCD events that, even after the selection analysis, overwhelm the signal rates, and that therefore suppose the main channel of this search.

 Despite the fact that the QCD background has been found difficult to deal with, we stay optimistic in what concerns the observation of the dynamically generated vector resonances. It is clear from the results presented in Figs.~\ref{fig:partonlevel} and~\ref{fig:JJlevel}, and in Tables~\ref{parton} and~\ref{JJ}, as well as from those shown in  \tabref{tablaSIGMAS}, that a more devoted study of the background, and, perhaps, of the signal, would lead to very promising prospects for the discovery of these resonances. In this sense, more sophisticated techniques, like deep learning or boost decision trees, might allow to have a better signal-to-background ratio that could lead to a very good sensitivity to the IAM resonances at the LHC. In this sense, we believe that the study of such states is really timely to try to obtain some information about the EChL structure, and, therefore, to the true dynamics of EWSB.


%
%

    \titleformat
{\chapter} 
[display] 
{\fontsize{21pt}{21pt}\scshape\bfseries} 
{} 
{-6.0ex} 
{
    {\color{col1}\rule{\textwidth}{0.8mm}}
    \raggedleft
    {{\fontsize{30pt}{30pt}\selectfont\textcolor{col1}{}}}
} 
[
\vspace{-0.5ex}%
{\color{col1}\rule{\textwidth}{0.8mm}}
] 

%

\chapter*{Conclusions}\label{Conclusions}
\chaptermark{Conclusions}
\addcontentsline{toc}{chapter}{\scshape\bfseries Conclusions}

The Standard Model governs the seas of particle physics and fundamental interactions but high seas exploration is needed if one wants to grasp the treasures of understanding nature at the fundamental level.

It is clear that the SM is one of the most remarkable theories ever postulated, since its predictions match with outstanding agreement the corresponding experimental measurements. Nevertheless, this prodigious framework lacks satisfactory explanations for diverse particle physics phenomena. 

Interestingly, many of the open problems of the SM are or can be related to the spontaneous breaking of the  electroweak symmetry, that allows to generate masses for the SM fermions and EW gauge bosons in a gauge invariant way. Thus, studying the EWSB sector might help us shed some light on many mysteries that nature still has under her sleeve.

In the SM, the EWSB is merely described, but the true nature of its dynamics is in principle unknown. Therefore, beyond the Standard Model physics is required in order to understand its dynamical origin. One of the most efficient ways of parameterizing these BSM contributions is to use effective field theories, since they allow to perform a model independent analysis, considering, so to say, many different ultraviolet scenarios at once.

In Chapter \ref{EChL} we revisited the main features of spontaneous symmetry breaking in the EW theory context, and we introduced different effective descriptions of possible BSM EWSB sectors. We concluded that, since the so-called electroweak chiral Lagrangian with a light Higgs (or Higgs effective field theory) corresponds to the most general setup respecting the relevant symmetries, i.e., the EW chiral and gauge symmetries, it should be the most appropriate tool to study the underlying dynamics of EWSB. Furthermore, apart from being the most general setup, it is also the proper one to describe the low energy interactions among EW gauge bosons in the case of a strongly interacting EWSB sector, which is a well motivated hypothesis nowadays. For this reason, in this Thesis we have relied upon the EChL as our theoretical basis to study deviations in the interactions of the bosonic sector with respect to the SM.

From the phenomenological point of view, vector boson scattering processes have been our key observables in order to look for these deviations. Since the EChL assumes, a priori, strong scalar dynamics, the most clear signals of this kind of new physics should appear in the interactions among EW Goldstone bosons. Nevertheless, they are unphysical particles, so they cannot be observed directly.

There is, however, a tight relation between the EW Goldstone bosons and the longitudinal component of the EW gauge bosons. The equivalence theorem relates the scattering amplitudes of the former particles at high energies with those of the latter, leading to the conclusion that vector boson scattering should be the most sensitive place to look for strongly interacting EWSB sector dynamics. 

Furthermore, in the EChL, the new physics scale results to be at the TeV scale, which is, then well motivated, since it arises from the dimensionful quantity controlling the chiral expansion, which corresponds to $4\pi v\sim 3$ TeV. Because of this, the LHC is the optimal collider to look for new EWSB physics in VBS observables, as it is now exploring the multi-TeV region. 

 For all these reasons, this Thesis represents a phenomenological study of  EChL signals in vector boson scattering observables at the LHC, with the aim of moving towards the comprehension of the true nature of the EWSB sector.
 
To this purpose, in Chapter \ref{VBS}, an exhaustive analysis of the VBS properties has been presented. We have reviewed the main features of these observables at the subprocess level, both in the SM and in the EChL, to then characterize the VBS topologies at the LHC through the kinematical configuration of the two extra jets in the final state. These jets populate the forward and backward regions of the detector and have large pseudorapidity separations as well as large invariant masses, attributes that serve us to reject undesired backgrounds allowing to tag VBS processes quite efficiently at the LHC. 

With these features in mind, and having briefly reviewed, also in Chapter \ref{VBS}, the present status of experimental searches for VBS at the LHC, we have analyzed different examples of how these observables can be used in the pursue of BSM physics signals driven by the EChL.

The first of these examples, corresponding to Chapter  \ref{HH}, concerns the study of the LHC sensitivity to deviations of the Higgs self-coupling $\lambda$ with respect to its SM value in double Higgs productions via VBS. The precise determination of this coupling, especially if it is performed independently from the Higgs mass determination, would allow us to understand the true nature of the Higgs mechanism, and therefore, the EWSB dynamics. 

 After characterizing the subprocesses of our interest, $WW\to HH$ and $ZZ\to HH$, and their direct translation to the LHC case, $pp\to HHjj$, we have focused mainly on the $pp\to HHjj \to b\bar{b}b\bar{b}jj$ process since it benefits from the largest rates. By applying all our selection criteria, based on the VBS characteristic kinematical configuration and in the $HH$ candidates reconstruction, we have given predictions for the sensitivity to BSM $\lambda$ in $pp\to b\bar{b}b\bar{b}jj$ events at the parton level for $\sqrt{s}=14$ TeV and for different future expected luminosities: $L=50,300,1000,3000$ fb${}^{-1}$. Our main results for this channel are summarized in \tabref{Tab:range_4b2j} and in \figref{fig:significances_4b2j}, in which we present the values of $\kappa=\lambda/\lambda_{SM}$ that the LHC would be sensitive to at the $3\sigma$ and at the $5\sigma$ level.
 
We give as well predictions for $pp\to b\bar{b}\gamma\gamma jj$ events, also at the parton level and for $\sqrt{s}=14$ TeV, due to the fact that it provides a cleaner, although with smaller rates, signature. The results of the sensitivities to the Higgs self-coupling in this channel, after applying the proper selection criteria, are collected in \tabref{Tab:range_2b2a2j} and in 
\figref{fig:significances_2b2a2j}. 

Furthermore, we give predictions for the interesting case of $L=1000$ fb${}^{-1}$ of how the sensitivity to BSM $\lambda$ will change from our naive parton level results when taking into account the main distorting effects. We have discussed the impact of $b$-tagging and of $\gamma$ identification efficiencies, of detector effects and of Higgs mass reconstruction resolution, which lead, altogether to a reduction factor of  at most 0.2 in the statistical significance for $pp\to b\bar{b}b\bar{b}jj$ and of at most 0.5 for $pp\to b\bar{b}\gamma\gamma jj$, as it can be seen in \figref{fig:errorband}.

The corresponding changes in the sensitivities to $\kappa$ translate into the fact that, at this luminosity, the LHC will be sensitive to $\kappa > 6.2 \,(7.7)$ at the $3\sigma$ ($5\sigma$) level for the $b\bar{b}b\bar{b} jj$ case and of $\kappa > 7.7 \,(9.4)$ at the $3\sigma$ ($5\sigma$) level for the $b\bar{b}\gamma\gamma jj$ one. In the also interesting case of $L=3000$ fb${}^{-1}$, we get similar reduction factors. The reachable values of $\kappa$ at this last HL-LHC stage via VBS configurations are predicted to be $\kappa > 5.0 \,(6.3)$ at the $3\sigma$ ($5\sigma$) level for the $b\bar{b}b\bar{b} jj$ case and of $\kappa > 6.1 \,(8.0)$ at the $3\sigma$ ($5\sigma$) level for the $b\bar{b}\gamma\gamma jj$ one.

The present study shows that double Higgs production via vector boson scattering at the LHC is a viable and promising window to measure BSM deviations to the Higgs self-coupling and to deeply understand the scalar sector of the SM. Although all simulations are performed at the parton level, without hadronization or detector response simulation, and should be understood as a naive first approximation, we have found that the values of $\lambda$ which are closer to that of the SM, are without a doubt the most challenging ones to observe at this collider. We obtained, however, very competitive results for the sensitivity to BSM $\lambda$ at the LHC. 
Because of this, we believe that the vector boson scattering $HH$ production channel will lead to very interesting (and complementary to those of gluon-gluon fusion) findings about the true nature of the Higgs boson.

 The second example studied in this Thesis concerns the violation of unitarity present in  the scattering of EW gauge bosons. As it was stated in Chapters \ref{EChL} and \ref{VBS}, effective theory predictions typically suffer from unitarity violation problems, due to the energy structure of the operators they contain. These predictions, are, in principle, not compatible with the underlying quantum field theory and cannot be used to interpret experimental data in order to obtain information about the effective theory. To cure this problem unitarization methods are addressed, as explained in Chapter \ref{EChL}. 
 
 There are, nevertheless, many available options to drive non-unitary observables computed with the raw effective theory into unitary ones. This ambiguity supposes a theoretical uncertainty that has to be taken into account when constraining the parameter space of such theories, that, up to now, has been made using one of these prescriptions at a time or no prescription at all. In Chapter \ref{Methods}, we provide a first approximation to quantifying this uncertainty by studying the case of the elastic WZ scattering at the LHC. We first study the values of the NLO coefficients $a_i$ where this violation of unitarity occurs finding that the most relevant operators concerning it correspond to the ones controlled by the chiral coefficients  $a_4$ and $a_5$, which parameterize  the anomalous  interactions among four massive electroweak gauge bosons. Furthermore, at this point, we also analyze the relevance in the unitarization procedure of each of the helicity channels participating in the scattering. Although nowadays most unitarization studies assume that the purely longitudinal scattering is sufficient to understand the unitarity violation problem, we consider of much importance to take into account the whole coupled system of helicity states, since the unitarity conditions relates them.  

We then study the predictions of five unitarization methods: Cut off, Form Factor, Kink, K-matrix and Inverse Amplitude Method, since they are the most used ones in the literature. We do this both at the subprocess level and for the LHC, relying in the latter case upon the Effective W Approximation. We have checked that the predictions of the EWA using the improved functions in \cite{Dawson:1984gx} lead to very good results compared with the full MadGraph predictions both for the SM and the EChL total cross sections, as well as for the corresponding invariant mass distributions. When analyzing each of these methods' predictions compared to those of the raw effective theory and of the SM, one is convinced that they lead to very different results, and that this fact should be taken into account to impose constraints on the parameter space of the effective theory.

For this reason, we construct, based on the ATLAS results for $\sqrt{s}=8$ TeV given in~\cite{Aaboud:2016uuk}, the 95\% exclusion regions in the $[a_4,a_5]$ plane for the various unitarization schemes. The results are contained in \figref{fig:elipse}, from which very interesting features can be extracted. The most important of them is that different methods lead to different parameters constraints and, therefore, it is indeed very clear that using one unitarization method at a time to interpret experimental data does not consider the full effective theory picture. Thus, we infer that since there are many unitarization prescriptions that lead to very different constraints, one should take them all into account in order to provide a reliable bound on the EFT parameters. These different constraints can vary even in an order of magnitude and in the correlation between the $a_4$ and $a_5$ parameters. 

The main conclusion of this Chapter is, therefore, that there is a theoretical uncertainty present in the experimental determination of effective theory parameters due to the unitarization scheme choice. A first approximation to this uncertainty has been quantified analyzing the predictions of $pp\to$ WZ+X events at the LHC from the EChL in terms of $a_4$ and $a_5$ and with different unitarization methods. We believe that it is important to take these uncertainties into account when relying upon experimental values of the constraints of effective theory parameters, in order to consider the full effective theory properties correctly.

These results concern particularly the non-resonant case in which generically smooth deviations from the continuum are expected. Nevertheless, a typical feature of strongly interacting theories, such as the ones described by the EChL, is the appearance of heavy resonances in the spectrum generated dynamically from the strong interactions. Because of this, in Chapters \ref{Resonances} and \ref{ResonancesWW} we study the production and sensitivity to dynamical vector resonances at the LHC in VBS observables. 

In order to study the resonant case we have worked under the framework of the EChL supplemented by the IAM. In addition, we have introduced another effective chiral Lagrangian describing the vector resonances, a modified Proca Lagrangian, that mimics the dynamically generated resonances found by the IAM. The VBS amplitudes computed within the EChL and unitarized with the IAM present poles at $\sqrt{s}=(M_V-i\Gamma_V/2)^2$ for specific values of the chiral parameters. Therefore, this framework provides relation between the physical parameters of the isotriplet of vector resonances, $M_V$ and $\Gamma_V$, and the relevant EChL parameters $a$, $b$, $a_4$ and $a_5$. This approach allows to build a model, that we have called IAM-MC, that is very convenient for a Monte Carlo analysis, since, unlike the IAM, can be directly implemented in terms of four-point functions. This IAM-MC model provides unitary VBS amplitudes that respect the Froissart bound given by \eqref{froisbound}.

With this tool we have studied two different VBS channels in which the IAM isotriplet of vector resonances, $V^0, V^\pm$, could be seen at the LHC. 
In Chapter \ref{Resonances} we have first focused on the $pp\to W^+Zjj$ since only a charged, $V^+$, vector resonance propagates in the $s$-channel, and since it suffers from less severe QCD backgrounds than others.

We have first selected six specific benchmark points in the IAM-MC model parameter space which have vector resonances emerging at mass values that are of phenomenological interest for the searches at the LHC: $M_V=$ 1.5, 2 and 2.5 TeV. These scenarios are used to perform the full study of the Monte Carlo generated events. The remaining nine scenarios that have also been considered have been used to further explore the sensitivity to the $a$ parameter by trying other values in the $[0.9,1]$ interval ($b$ is set to $b=a^2$).      

By studying the  kinematical distributions of the signal, $W^+Zjj$, and the most relevant backgrounds we have  defined the appropriate VBS cuts that allow to differentiate efficiently between the VBS topologies and the rest. With them in mind, we have studied the purely leptonic decays of the WZ channel since, despite the modest resulting event rates, a very clear signal can be achieved. 

Thus, we have fully analyzed the golden leptonic $W$ and $Z$ decay channels leading to a final state with $\ell_1^+\ell_1^-\ell_2^+\nu jj$, $\ell=e,\mu$, and we have presented the results of the appearing resonances in terms of an experimentally measurable variable, the transverse invariant mass of the $\ell_1^+\ell_1^-\ell_2^+\nu$ final leptons. As it is illustrated in \figref{fig:leptondistributions}, the resonant peaks emerge from the SM background, so one might expect them to be observed at the LHC. Our numerical evaluation of the future event rates and sensitivities to these resonant states are summarized in  \tabref{Tablasigmaslep-paper} and in \figref{fig:asensitivitylep}. 

These results in \tabref{Tablasigmaslep-paper} demonstrate that even with a luminosity of $300~ {\rm fb}^{-1}$ a 1.5 TeV resonance corresponding to $a=0.9$ could be seen in the leptonic channels with $\significance_\ell$ around 3.

 For  $L=1000~ {\rm fb}^{-1}$, we estimate that these scenarios could  be tested with a high statistical significance and that a discovery of these resonances with masses close to 1.5 TeV could be done. Interestingly, for the highest luminosity considered, $3000~ {\rm fb}^{-1}$,  all the studied scenarios with  resonance masses at and below 2 TeV and with $a=0.9$ could be seen.  For the heaviest studied resonances  (around 2.5 TeV), small hints with $\significance_\ell$  slightly larger than 2  might as well be observed.
 
 The sensitivities to other values of $a$ in the interval $[0.9,1]$  have also been explored leading to similar conclusions: the study of the lightest resonances will allow to access values of $a$ in the whole interval, while for the heaviest ones only the 0.9 case might be mildly reached. The intermediate case, i.e., masses of around 2 TeV might help to probe values of $a$ between 0.9 and 0.95, closer to the SM expectation.
 
 We have also explored in this Chapter, and in a very naive way, the semileptonic and hadronic decay channels of the $WZ$ scattering. We have concluded that, assuming that no new backgrounds arise in these channels apart form the ones considered at the level of $WZjj$ events, the expectations for the hadronic and semileptonic case are very promising. However, a more definitive conclusion regarding these channels demands a more realistic study including showering and reconstruction of the final jets. This has been performed in the case of $WW$ scattering in Chapter \ref{ResonancesWW}.

In this Chapter \ref{ResonancesWW} we have focused on the study of the vector resonances emerging in the $W^+W^-\to W^+W^-$. In this case, the neutral component of the isotriplet of vector resonances, $V^0$, is the one that resonantes in the s-channel, and therefore, the one that can be accessed efficiently in this VBS process. Our study of these resonances at the LHC by means of the process $pp \to WW jj$ is therefore complementary to the  study in Chapter~\ref{Resonances}. Besides, in contrast to the case in Chapter \ref{Resonances}, we chose to focus on the fully hadronic channel, in which the $W$ boson decays are reconstructed as a single, large radius (fat) jet, $J$, to perform our analysis.

We have studied in detail the final hadronic states in $pp \to JJjj$ events at the LHC with 
two thin jets having the typical VBS topology and using the two fat jets for the tagging of the two final $W$ bosons. After the devoted reconstruction of the final jets we have set all our focus on searching for techniques that can reduce efficiently the most challenging background coming from QCD. We have found that the specific variables $\tau_{21}(J)$, $M_J$ and $M_{JJ}$ are extremely helpful in this concern. 
We have performed a survey on the expected rates for both the signal and the background, as well as for the corresponding statistical significances of the associated BSM signal from the vector resonances, for the six selected benchmark points of \tabref{tablaBMP}. The main results have been summarized in \tabref{tablaSIGMAS} where the optimal cuts that we found are also shown. 
 
With these optimal cuts, only the BP1' largely overcomes the $5\sigma$ of statistical significance at the LHC with a luminosity of $L=3000\,{\rm fb}^{-1}$. The other point with a resonance mass close to 1.5 TeV, BP1, reaches a statistical significance slightly below~5. BP2', with a resonance mass close to 2 TeV, reaches a statistical significance of $4.1\sigma$, whereas BP3', with mass close to 2.5 TeV, a maximum in $\sigma^{\rm stat}$ of~2.3 is obtained. BP2 and BP3, with a statistical significance below~1, are extremely challenging to observe. In the hypothetical situation that the QCD background were efficiently suppressed, all the points, except BP2 and BP3, could be observed, similarly to the conclusions obtained for the charged resonances in the leptonic WZ channel. 

For this reason, the fully hadronic channel via $pp\to W^+W^-jj$ could be studied with $L=3000\,{\rm fb}^{-1}$, with the QCD background being a great challenge. It would be necessary to perform a more detailed study of the SM QCD background, and, of course, to improve the signal vs. background ratio. Some more sophisticated techniques like deep learning, boost decision trees and others could be used to find better optimal cuts in a per-event basis, and to deal with hidden correlations between the different variables, particularly those characterising the fat jets. According to Table 15, BP1', BP2' and BP1 would be clearly detectable at $L=3000\,{\rm fb}^{-1}$.

Summarizing, in this Thesis we have presented a phenomenological analysis of vector boson scattering at the LHC in the search for new physics related to the electroweak symmetry breaking dynamics described by the electroweak chiral Lagrangian. The electroweak symmetry breaking  sector is (or might be) related to a plethora of particle physics phenomena that require explanations which are not contained in the Standard Model and, therefore, the study of this kind of physics might lead us to some of the most important achievements in modern physics. 

Along this Thesis, we have seen that vector boson scattering observables at the LHC are the most sensitive scenarios to look for this type of new interactions. For this reason, the main conclusion of this Thesis is that dedicated searches and analyses of vector boson scattering are the most promising window towards the understanding of the true nature of the electroweak symmetry breaking sector. Our treasure is, for sure, around the corner. Let us keep seeking for it!


\chapter*{Conclusiones}\label{Conclusiones}
\chaptermark{Conclusiones}
\addcontentsline{toc}{chapter}{\scshape\bfseries Conclusiones}

El Modelo Est\'andar gobierna los mares de la f\'isica de part\'iculas, pero la exploraci\'on de los territorios desconocidos de alta mar sigue siendo necesaria si uno quiere alcanzar los tesoros de la comprensi\'on de la naturaleza al nivel m\'as fundamental.

El SM es, sin duda, una de las teor\'ias m\'as excepcionales jam\'as postuladas, ya que sus predicciones coinciden con las medidas experimentales con alt\'isimos grados de precisi\'on. Sin embargo, este prodigioso marco te\'orico no posee la capacidad de explicar algunos fen\'omenos relacionados con la f\'isica de part\'iculas que han sido observados.

Resulta interesante plantearse que muchos de los problemas abiertos del SM est\'an o pueden estar relacionados con la ruptura espont\'anea de la simetr\'ia electrod\'ebil, la cual permite generar masas para los fermiones del SM y para los bosones gauge electrod\'ebiles de una forma en la que la invariancia gauge es manifiesta. Por lo tanto, estudiar el sector de ruptura de la simetr\'ia electrod\'ebil podr\'ia ayudarnos a arrojar algo de luz sobre muchos de los misterios que la naturaleza aun guarda bajo la manga.

A pesar de que en el SM se describe el mecanismo de ruptura de la simetr\'ia electrod\'ebil, la verdadera naturaleza de su din\'amica subyacente aun nos es desconocida, por lo que se requiere f\'isica m\'as all\'a del SM para poder entender este origen din\'amico. Una de las formas m\'as eficientes de parametrizar est\'as contribuciones de nueva f\'isica es utilizar teor\'ias de campo efectivas, ya que permiten realizar un estudio independiente de modelos concretos considerando, de alg\'un modo, diversos posibles escenarios a altas energ\'ias al mismo tiempo.

En el Cap\'itulo \ref{EChL} se present\'o, por tanto, un repaso de las principales caracter\'isticas de la ruptura espont\'anea de la simetr\'ia electrod\'ebil y se introdujeron diferentes descripciones efectivas de posibles sectores asociados a dicha ruptura m\'as all\'a del SM. De entre estas teor\'ias efectivas, se concluy\'o que el llamado Lagrangiano quiral electrod\'ebil con un Higgs ligero (tambi\'en llamado teor\'ia de campos efectiva del Higgs) corresponde al escenario m\'as general que puede construirse respetando las simetr\'ias relevantes: las simetr\'ias electrod\'ebiles quiral y gauge. Adem\'as, el EChL resulta ser la teor\'ia efectiva m\'as apropiada para describir la din\'amica a bajas energ\'ias de las interacciones entre bosones electrod\'ebiles en el contexto de un sector de ruptura de la simetr\'ia electrod\'ebil que interacciona fuertemente, hip\'otesis motivada actualmente. En consecuencia, consideramos que el EChL es la herramienta apropiada para describir la din\'amica subyacente de la ruptura espont\'anea de la simetr\'ia electrod\'ebil, y, por lo tanto, en esta Tesis ha sido empleado como la base te\'orica con la que estudiar desviaciones con respecto del SM en el sector bos\'onico .

Desde el punto de vista fenomenol\'ogico, el {\it scattering} de bosones vectoriales ha sido considerado como el observable clave a la hora de buscar estas desviaciones. Debido a que el EChL asume, a priori, una din\'amica escalar acoplada fuertemente, las se\~nales m\'as claras de este tipo de nueva f\'isica deber\'ian aparecer en las interacciones entre bosones de Goldstone electrod\'ebiles. No obstante, estas part\'iculas no son f\'isicas, y por lo tanto no pueden ser observadas de forma directa.

Existe, sin embargo, una relaci\'on muy estrecha entre los bosones de Goldstone electrod\'ebiles y los componentes longitudinales de los bosones gauge electrod\'ebiles. El teorema de equivalencia relaciona las amplitudes de {\it scattering} de los bosones $W$ y $Z$, a altas energ\'ias, con la de sus bosones de Goldstones correspondientes, lo cual motiva la conclusi\'on de que los procesos de VBS deber\'ian ser los m\'as sensibles a un sector de ruptura de la simetr\'ia electrod\'ebil que interacciona fuertemente.

Adem\'as, en el EChL la escala asociada a la nueva f\'isica est\'a relacionada con la escala del TeV, ya que la cantidad con dimensiones que controla la expansi\'on quiral corresponde a $4\pi v\sim3$ TeV. Por esta raz\'on el LHC es el experimento \'optimo para buscar nueva f\'isica en observables de tipo VBS ya que actualmente se encuentra explorando la regi\'on energ\'etica del TeV.

Por todos estos motivos, esta Tesis respresenta un estudio fenomenol\'ogico de las se\~nales del EChL en observables de tipo VBS en el LHC, con el objetivo de avanzar en la comprensi\'on de la verdadera naturaleza del sector de ruptura de la simetr\'ia electrod\'ebil.

Para ello, en el Cap\'itulo \ref{VBS}, se ha llevado acabo un an\'alisis exhaustivo de las propiedades del VBS a nivel de subproceso tanto en el SM como en el EChL, para, posteriormente, caracterizar las topolog\'ias de tipo VBS en el LHC gracias de la configuraci\'on cinem\'atica de los dos jets del estado final. Estos jets pueblan las regiones del detector {\it forward} y {\it backward} (hacia adelante y hacia atr\'as con respecto al haz de protones) y poseen grandes separaciones en pseudorapidez y altas masas invariantes. Estos atributos resultan  esenciales para rechazar fondos no deseados permitiendo identificar de manera muy eficiente los procesos de VBS en el LHC.

Con estas propiedades en mente, y tras realizar un repaso, tambi\'en en el Cap\'itulo \ref{VBS}, del presente estatus experimental de las b\'usquedas de VBS en el LHC, se han analizado diferentes ejemplos de c\'omo estos observables pueden emplearse en la exploraci\'on de se\~nales de nueva f\'isica descritas por el EChL.

El primero de estos ejemplos, correspondiente al estudio realizado en el Cap\'itulo \ref{HH}, est\'a enfocado a la obtenci\'on de la sensibilidad del LHC a desviaciones del autoacoplamiento del Higgs $\lambda$ con respecto a su valor del SM en la producci\'on de dos bosones de Higgs en VBS. La determinaci\'on precisa de este par\'ametro, especialmente en el caso de que su medida se realizara de forma independiente a la de la masa del Higgs, permitir\'ia comprender la verdadera naturaleza del mecanismo de Higgs, y, por lo tanto, de la ruptura de la simetr\'ia electrod\'ebil.

Tras caracterizar los subprocesos de nuestro inter\'es, $WW\to HH$ y $ZZ\to HH$, y su traducci\'on directa al caso del LHC, $pp\to HHjj$, se ha enfocado el estudio en el proceso completo con cuatro $b-$jets y dos jets de quarks ligeros, $pp\to HHjj \to b\bar{b}b\bar{b}jj$, ya que este posee la mayor secci\'on eficaz. Aplicando todos los criterios de selecci\'on analizados, basados en la cinem\'atica caracter\'istica del VBS y en la reconstrucci\'on de los candidatos a parejas $HH$, se han proporcionado las predicciones para la sensibilidad a valores de $\lambda$ m\'as all\'a del SM en eventos de tipo $pp\to b\bar{b}b\bar{b}jj$ a nivel part\'onico para $\sqrt{s}=14$ TeV y para diferentes luminosidades esperadas: $L=50,300,1000,3000$ fb${}^{-1}$. Los principales resultados quedan recogidos en la Table \ref{Tab:range_4b2j} y en la \figref{fig:significances_4b2j}, en las que se presentan los valores de $\kappa=\lambda/\lambda_{\rm SM}$ a los que el LHC ser\'ia sensible a niveles de $3\sigma$ y $5\sigma$.

Asimismo, tambi\'en se han obtenido las predicciones para el proceso completo con dos fotones, dos $b-$jets y dos jets de quarks ligeros, $pp\to b\bar{b}\gamma\gamma jj$, tambi\'en a nivel part\'onico y para $\sqrt{s}=14$ ya que, \'este proporciona una se\~nal m\'as limpia, aunque tambi\'en m\'as peque\~na. Los resultados de las sensibilidades al autoacoplamiento del Higgs correspondientes a este caso, despu\'es de la aplicaci\'on de los criterios de selecci\'on apropiados, se resumen en la Tabla \ref{Tab:range_2b2a2j}  y en la \figref{fig:significances_2b2a2j}. 

Tambi\'en se muestran las predicciones para el interesante caso de $L=1000$ fb${}^{-1}$ de c\'omo var\'ia la sensibilidad a los valores de $\lambda$ m\'as all\'a del SM cuando se consideran resultados yendo m\'as all\'a del nivel part\'onico na\"if. Se ha discutido el impacto de las eficiencias de identificaci\'on de quarks bottom y de fotones, de los efectos del dectector y de la resoluci\'on de la reconstrucci\'on de la masa del Higgs, dando lugar a un factor de reducci\'on total de, como mucho, 0.2 en la significancia estad\'istica para el caso $pp\to b\bar{b}b\bar{b}jj$ y de 0.5 para el caso $pp\to b\bar{b}\gamma\gamma jj$, como puede verse en la \figref{fig:errorband}.

Los cambios correspondientes a la sensibilidad a $\kappa$ se traducen en el hecho de que, para esta luminosidad, el LHC ser\'ia sensible a $\kappa > 6.2 \,(7.7)$ al nivel de $3\sigma$ ($5\sigma$) para el canal $b\bar{b}b\bar{b} jj$ y a $\kappa > 7.7 \,(9.4)$ para el canal $b\bar{b}\gamma\gamma jj$. Para una luminosidad de $L=3000$ fb${}^{-1}$ se han obtenido factores de reducci\'on similares. De hecho, los valores de $\kappa$ accesibles en esta \'ultima etapa del HL-LHC a trav\'es de configuraciones de tipo VBS corresponden a $\kappa > 5.0 \,(6.3)$ al nivel de $3\sigma$ ($5\sigma$) para el caso $b\bar{b}b\bar{b} jj$ y a  $\kappa > 6.1 \,(8.0)$ para el caso $b\bar{b}\gamma\gamma jj$.

El presente estudio muestra que la producci\'on de dos bosones de Higgs a trav\'es de VBS en el LHC es una opci\'on viable y prometedora para medir desviaciones con respecto al SM del valor del autoacoplamiento del bos\'on de Higgs, as\'i como para comprender de manera profunda el sector escalar del SM. Aunque todas las simulaciones han sido realizadas a nivel part\'onico, sin hadronizaci\'on ni simulaci\'on de respuesta del detector, y, por tanto, deber\'ian ser entendidas como una primera aproximaci\'on na\"if, se ha concluido que los valores de $\lambda$ m\'as cercanos al del SM son, sin duda, los m\'as dif\'iciles de observar en este colisionador. Por otro lado, se han obtenido resultados muy competitivos para la sensibilidad a valores de $\lambda$ m\'as all\'a del SM. Por esta raz\'on creemos firmemente que la producci\'on de dos bosones de Higgs en VBS dar\'a lugar a resultados muy interesantes (adem\'as de complementarios a los correspondientes iniciados en fusi\'on de gluones) sobre la verdadera naturaleza del bos\'on de Higgs.

El segundo ejemplo estudiado en esta Tesis est\'a relacionado con la violaci\'on de unitaridad presente en el scattering de bosones gauge electrod\'ebiles. Tal como se introdujo en los Cap\'itulos \ref{EChL} y \ref{VBS}, las predicciones derivadas de las teor\'ias efectivas sufren problemas de violaci\'on de unitariedad debido a la estructura energ\'etica de los operadores que contienen. Estas predicciones son en principio incompatibles con la te\'orica cu\'antica de campos subyacente y por lo tanto no pueden ser empleadas en la interpretaci\'on de los datos experimentales con el objetivo de obtener informaci\'on sobre la teor\'ia efectiva. Para solucionar este problema se utilizan m\'etodos de unitarizaci\'on, tal como se explic\'o en el Cap\'itulo \ref{EChL}.

Existen muchas opciones para volver unitarios los observables no unitarios calculados directamente desde la teor\'ia efectiva y esta ambig\"uedad supone una incertidumbre te\'orica que ha de tenerse en cuenta al acotar el espacio de par\'ametros de estas teor\'ias. En el Cap\'itulo \ref{Methods} se cuantifica por primera vez esta incertidumbre estudiando el scattering el\'astico $WZ$ en el LHC. Primero se estudian los valores de los coeficientes del Lagrangiano $\mO(p^4)$, $a_i$, para los que tiene lugar la violaci\'on de unitariedad concluyendo que los operadores m\'as relevantes a este respecto son aquellos controlados por $a_4$ y $a_5$, que parametrizan las interacciones an\'omalas de cuatro bosones d\'ebiles. Tambi\'en se analiza la relevancia en el proceso de unitarizaci\'on de los distintos canales de helicidad que participan en el scattering. A pesar de que actualmente en la mayor parte de estudios sobre efectos de unitarizaci\'on se asumen que el canal puramente longitudinal es suficiente para entender el problema de la violaci\'on de unitariedad, en esta Tesis se ha considerado de mucha importancia el hecho de tener en cuenta el sistema completo de canales de helicidad acoplados, ya que la condici\'on de unitaridad los relaciona entre s\'i.

Con todo esto en mente se han estudiado las predicciones de cinco m\'etodos de unitarizaci\'on: Cut off, Form Factor, Kink, K-matrix y M\'etodo de la Amplitud Inversa, ya que son los m\'as com\'unmente utilizados en la literatura, tanto a nivel de subproceso como en el LHC, apoy\'andose en este \'ultimo caso en la aproximaci\'on W efectiva (EWA de sus siglas en ingl\'es). De hecho, se ha comprobado que las predicciones de la EWA utilizando las funciones mejoradas presentadas en \cite{Dawson:1984gx} concuerdan satisfactoriamente con los resultados completos de MadGraph tanto para la secci\'on eficaz del SM como para la del EChL, as\'i como para las distribuciones en masa invariante correspondientes. Al analizar las predicciones de cada uno de estos m\'etodos en comparaci\'on con las de la pura teor\'ia efectiva y las del SM, resulta evidente que cada uno de ellos da lugar a resultados muy diferentes, y que este hecho deber\'ia ser tenido en cuenta a la hora de imponer cotas al espacio de par\'ametros de la teor\'ia efectiva.

Por esta raz\'on, bas\'andose en los resultados de ATLAS para $\sqrt{s}=8$ TeV se ha construido la regi\'on de exclusi\'on en el plano $[a_4,a_5]$ para los distintos m\'etodos de unitarizaci\'on al 95\% de nivel de confianza.  Los resultados correspondientes quedan recogidos en la \figref{fig:elipse}, de la que pueden extraerse conclusiones muy interesantes. La m\'as importante de ellas es la diferencia presente entre las distintas cotas a los par\'ametros que pueden derivarse utilzando un m\'etodo u otro. \'Estas pueden variar incluso en un orden de magnitud y en la correlaci\'on existente entre los par\'ametros $a_4$ y $a_5$, lo cual indica claramente que la utilizaci\'on de un s\'olo m\'etodo a la hora de interpretar los datos experimentales no tiene en cuenta el marco completo de la teor\'ia efectiva. Por lo tanto, se infiere que ya que las diferentes prescripciones de unitarizaci\'on dan lugar a resultados muy diferentes todas deben tenerse en cuenta simult\'aneamente para poder proporcionar una cota fiable a los par\'ametros de la teor\'ia efectiva.

La principal conclusi\'on de este Cap\'itulo \ref{Methods} es, a la vista de los resultados, que sin duda existe una incertidumbre te\'orica asociada a la determinaci\'on experimental de los par\'ametros de la teor\'ia efectiva debido a la elecci\'on del m\'etodo de unitarizaci\'on. Como primera aproximaci\'on, se ha cuantificado dicha incertidumbre analizando las predicciones de eventos $pp\to$ WZ+X en el LHC calculadas en el EChL para diferentes valores de $a_4$ y $a_5$ y unitarizadas de diversas formas. Creemos firmemente que es muy importante tener en cuenta esta incertidumbre a la hora de utilizar las cotas experimentales de los par\'ametros de la teor\'ia efectiva para poder considerar correctamente sus propiedades completas.

Los resultados mostrados conciernen particularmente al caso no resonante, en el que se esperan desviaciones suaves respecto del continuo del SM. Sin embargo, una caracter\'istica t\'ipica de las teor\'ias fuertemente acopladas, como las descritas por el EChL, es la aparici\'on de resonancias pesadas en el espectro generadas din\'amicamente desde las interacciones fuertes. Por este motivo, en los Cap\'itulos \ref{Resonances} y \ref{ResonancesWW}, se ha estudiado la producci\'on y sensibilidad a resonancias din\'amicas vectoriales en observables de VBS en el LHC.

Para poder estudiar el caso resonante, se ha trabajo en el marco del EChL suplementado por el M\'etodo de la Amplitud Inversa (IAM de sus siglas en ingl\'es). Adicionalmente se ha introducido un Lagrangiano quiral efectivo, capaz de describir las resonancias vectoriales, correspondiente a un Lagrangiano de Proca modificado que es capaz de reproducir las propiedades de las resonancias generadas din\'amicamente por la IAM. Las amplitudes del VBS calculadas en el EChL y unitarizadas con la IAM presentan polos en $\sqrt{s}=(M_V-i\Gamma_V/2)^2$ para valores determinados de los par\'ametros quirales. Esto permite construir un modelo, al que se ha bautizado como IAM-MC, que resulta muy \'util para un an\'alisis de eventos de Monte Carlo, ya que, al contrario que la IAM, puede ser implementado directamente en t\'erminos de funciones de cuatro puntos. Adem\'as, el modelo IAM-MC proporciona amplitudes de VBS unitarias que respetan la cota de Froissart dada en la \eqref{froisbound}.

Con esta herramienta se han estudiado dos canales diferentes de VBS en el LHC en los que pueden emerger las resonancias del isotriplete $V^0, V^\pm$ generadas por la IAM. El Cap\'itulo \ref{Resonances} se ha centrado en el estudio del proceso $pp\to W^+Zjj$ debido a que, en este caso, s\'olo la resonancia cargada $V^\pm$ se propaga en el canal $s$ y a que no sufre de fondos asociados tan severos como en otros canales.

Primeramente se han seleccionado seis puntos espec\'ificos en el espacio de par\'ametros del IAM-MC que albergan resonancias vectoriales emergentes con masas interesantes desde el punto de vista fenomenol\'ogico: $M_V=$ 1.5, 2 and 2.5 TeV. Estos escenarios se han empleado para realizar un estudio completo de eventos de Monte Carlo asociados a las resonancias generadas din\'amicamente. Los otros nueve escenarios que tambi\'en se han considerado se han utilizado para explorar en m\'as profundidad la sensibilidad al par\'ametro $a$ en el rango (0.9, 1) ($b$ se ha tomado en todos los casos como $b=a^2$).

Estudiando las distribuciones cinem\'aticas de la se\~nal, $W^+Zjj$, y de los fondos m\'as relevantes, se han definido los cortes de VBS apropiados que permiten diferenciar de manera eficiente los procesos con topolog\'ias de VBS del resto. Con estos cortes en mente se ha estudiado el canal de desintegraci\'on puramente lept\'onico del par $WZ$ ya que, a pesar de sus modestas secciones eficaces, proporciona una se\~nal muy limpia en el LHC.

En este sentido se han presentado los resultados asociados al canal lept\'onico  $\ell_1^+\ell_1^-\ell_2^+\nu jj$, $\ell=e,\mu$  de la aparici\'on de las resonancias en t\'erminos de una variable que puede ser medida experimentalmente: la masa transversa de los leptones finales. Tal como queda ilustrado en la \figref{fig:leptondistributions}, los picos resonantes emergen claramente del fondo del SM, por lo que se esperar\'ia que pudieran ser observados en el LHC. La evaluaci\'on num\'erica de los eventos esperados futuros as\'i como de las sensibilidades a estas resonancias se resume en la Tabla \ref{Tablasigmaslep-paper} y en la \figref{fig:asensitivitylep}.

Los resultados de la Tabla \ref{Tablasigmaslep-paper} demuestran que incluso para una luminosidad de $300~ {\rm fb}^{-1}$, la resonancia de masa 1.5 TeV correspondiente a $a=0.9$ podr\'ia ser detectada en el canal lept\'onico con una significancia de $\significance_\ell\sim$ 3.

Para $L=1000~ {\rm fb}^{-1}$ se ha estimado que los escenarios asociados a este valor ligero de masa, 1.5 TeV, podr\'ian dar lugar al descubrimiento de estos modos resonantes dado que su significancia estad\'istica se aproxima a los 5$\sigma$. Es tambi\'en interesante mencionar que para la luminosidad m\'as alta de las aqu\'i consideradas, $3000~ {\rm fb}^{-1}$, todas las resonancias estudiadas con masas en y por debajo de 2 TeV para el caso $a=0.9$ podr\'ia ser observadas en el LHC. Para las resonancias m\'as pesadas tan s\'olo se alcanzar\'ian significancias ligeramente por encima de 2$\sigma$. 

Las sensibilidades a otros valores de $a$ en el intervalo $(0.9,1)$ tambi\'en se han explorado dando lugar a conclusiones similares: el an\'alisis de las resonancias m\'as ligeras permitir\'ia acceder a todos los valores considerados de $a$, mientras que para las m\'as pesadas \'unicamente el caso $a=0.9$ podr\'ia alcanzarse ligeramente. El escenario intermedio, asociado a masas de 2 TeV, ayudar\'ia a testar valores de $a$ entre 0.9 y 0.95, m\'as pr\'oximos al valor esperado en el SM.

En este Cap\'itulo tambi\'en se han estudiado de una forma muy preliminar los casos hadr\'onico y semilept\'onico del scattering $WZ$. A este respecto se ha concluido que, asumiendo que no existen m\'as fondos en estos canales aparte de los ya considerados al nivel de $WZjj$, las expectativas de observaci\'on de las resonancias son muy prometedoras. No obstante, para poder alcanzar una conclusi\'on m\'as definitiva es necesario llevar a cabo un estudio m\'as realista que tenga en cuenta efectos de hadronizaci\'on y reconstrucci\'on de los jets finales. Este estudio se ha realizado precisamente en el Cap\'itulo \ref{ResonancesWW} para el caso del $WW$ scattering.

Dicho Cap\'itulo \ref{ResonancesWW} se ha centrado en el estudio de las resonancias vectoriales emergentes en el proceso de scattering $W^+W^-\to W^+W^-$. En este caso, el componente neutro del isotriplete de resonancias vectoriales, $V^0$, es el que resuena en el canal $s$, y, por tanto, al que puede accederse en el LHC a trav\'es de este tipo de configuraciones. El an\'alisis realizado para este tipo de resonancias en el proceso $pp \to WW jj$  es, por tanto, complementario al estudio presentado en el Cap\'itulo \ref{Resonances}. Adem\'as, en el Cap\'itulo \ref{ResonancesWW} se ha centrado la exploraci\'on del canal puramente hadr\'onico en el que los productos de desintegraci\'on de cada $W$ se reconstruyen como un s\'olo ({\it fat}) jet $J$, contrariamente al caso lept\'onico analizado en el Cap\'itulo \ref{Resonances}.
 
 Por tanto, en el Cap\'itulo \ref{ResonancesWW} se ha estudiado en detalle el estado final hadr\'onico $JJjj$ en el LHC con dos {\it thin} jets  y dos {\it fat} jets empleados para identificar los bosones $W$. Tras una reconstrucci\'on exhaustiva de los jets finales el an\'alisis se ha centrado en la b\'usqueda de t\'ecnicas con las que reducir de forma eficiente el fondo m\'as problem\'atico, proveniente de eventos de QCD, concluy\'endose que las variables $\tau_{21}(J)$, $M_J$ y $M_{JJ}$ son extremadamente \'utiles al respecto. Con esto en mente, se ha realizado un estudio comparativo de los resultados esperados, tanto para la se\~nal como para el background en los casos presentados en la Tabla \ref{tablaBMP}, de la secci\'on eficaz y de la significancia estad\'istica asociada a las resonancias vectoriales. Los principales resultados obtenidos quedan recogidos en las Figs.~\ref{fig:partonlevel} y \ref{fig:JJlevel}, en las Tablas~\ref{parton} y \ref{JJ}, y en la Tabla \ref{tablaSIGMAS}, donde se muestran, adem\'as, los cortes \'optimos encontrados. 
 
Con los cortes \'optimos presentados en la \tabref{tablaSIGMAS}, tan s\'olo el BP1' alcanza una significancia estad\'istica ampliamente mayor a $5\sigma$ para una luminosidad de $L=3000\,{\rm fb}^{-1}$. Para el otro punto seleccionado con una masa pr\'oxima a 1.5 TeV, el BP1, la significancia estad\'istica resulta ser algo menor a $5\sigma$ para esta misma luminosidad. El punto BP2', con una masa cercana a 2 TeV, llegar\'ia a los 4.1$\sigma$, mientras que el BP3', con una masa pr\'oxima a 2.5 TeV, alcanzar\'ia una significancia m\'axima de 2.3$\sigma$. Los dem\'as puntos, BP2, BP3, con una significancia por debajo de 1 resultar\'ian extremadamente dif\'iciles de observar. S\'olo en el caso hipot\'etico de que el fondo de QCD pudiera ser suprimido eficientemente podr\'ian observarse todas las resonancias escogidas salvo las correspondientes a los casos BP2 y BP3, conclusi\'on similar a la que se lleg\'o en el caso de las resonancias cargadas en desintegraciones lept\'onicas de WZ.
 
 Por estos motivos, se ha concluido que el canal puramente hadr\'onico asociado al proceso $pp\to W^+W^-jj$ podr\'ia ser estudiado en el LHC con una luminosidad de $L=3000\,{\rm fb}^{-1}$ teniendo en cuenta que el control del fondo de QCD representa el mayor reto en dicho estudio. Un an\'alisis m\'as detallado de este fondo, empleando t\'ecnicas m\'as sofisticadas como {\it deep learning} o {\it boost decision trees}, ser\'ia necesario para poder observar las resonancias generadas din\'amicamente por la IAM. Estas t\'ecnicas podr\'ian ser utilizadas para lidiar con correlaciones ocultas entre diferentes variables, en particular aquellas asociadas a los {\it fat} jets. De acuerdo con los resultados de la Tabla \ref{tablaSIGMAS}, los puntos BP1', BP2' y BP1 podr\'ian ser claramente detectados para una luminosidad de $L=3000\,{\rm fb}^{-1}$.
 
Como resumen, podemos decir que en esta Tesis se ha presentado un an\'alisis de la fenomenolog\'ia del scattering de bosones vectoriales en el LHC con el objetivo de buscar nueva f\'isica asociada a la din\'amica de ruptura de la simetr\'ia electrod\'ebil descrita por el Lagrangiano quiral electrod\'ebil. El sector de ruptura espont\'anea de la simetr\'ia electrod\'ebil est\'a (o podr\'ia estar) relacionado con una gran n\'umero de fen\'omenos de la f\'isica de part\'iculas para los que aun no existe una explicaci\'on satisfactoria, y, por tanto, el estudio de su f\'isica asociada podr\'ia conducir a uno de los mayores logros de la f\'isica moderna.

A lo largo de esta Tesis se ha visto que los observables de scattering de bosones vectoriales en el LHC son los m\'as sensibles a este tipo de nueva f\'isica. Por esta raz\'on, la principal conclusi\'on de este trabajo de investigaci\'on es que los estudios dedicados, completos y exhaustivos de este tipo de observables son y ser\'an, sin duda, la ventana m\'as prometedora hacia la comprensi\'on de la verdadera naturaleza del sector de ruptura de la simetr\'ia electrod\'ebil. Nuestro tesoro est\'a a la vuelta de la esquina. !`Sigamos busc\'andolo!


    \titleformat
{\chapter} 
[display] 
{\fontsize{21pt}{21pt}\scshape\bfseries} 
{} 
{-6.0ex} 
{
    {\color{col1}\rule{\textwidth}{0.8mm}}
    \raggedleft
    {{\fontsize{30pt}{30pt}\selectfont\textcolor{col1}{}}}
} 
[
\vspace{-1.0ex}%
{\color{col1}\rule{\textwidth}{0.8mm}}
] 

\begin{appendices}
\chapter*{Appendices}
\chaptermark{Appendices}
\addcontentsline{toc}{chapter}{\scshape\bfseries Appendices}
\renewcommand{\thesection}{\Alph{section}}


\def\VSM{V^{\begin{minipage}{1cm}\scalebox{0.6}{\rm SM}\end{minipage}}}
\def\VEChL{V^{\begin{minipage}{1cm}\scalebox{0.6}{\rm EChL}\end{minipage}}}
\def\VIAMMC{V^{\begin{minipage}{1cm}\scalebox{0.6}{\rm IAM-MC}\end{minipage}}}


\def\FRWWZSM{\begin{align}
\VSM_{W^+_\mu W^-_\rho Z_\nu}=i g \cosw \Big[ g_{\mu\nu}(p_0 - p_+)_\rho + g_{\nu\rho}(p_- - p_0)_\mu +g_{\mu\rho} (p_+ - p_-)_\nu \Big]
\nn\end{align}
}

\def\FRWWASM{
\begin{align}
\VSM_{W^+_\mu W^-_\rho \gamma_\nu}=ie \Big[ g_{\mu\nu}(p_0 - p_+)_\rho + g_{\nu\rho}(p_- - p_0)_\mu +g_{\mu\rho} (p_+ - p_-)_\nu \Big]
\nn\end{align}
}

\def\FRWWHSM{\begin{align}
\VSM_{W^+_\mu W^-_\nu H}=ig\,m_Wg_{\mu\nu}
\nn\end{align}
}

\def\FRZZHSM{
\begin{align}
\VSM_{Z_\mu Z_\nu H}=\dfrac{ig\,m_W}{\cosw^2} g_{\mu\nu}
\nn\end{align}
}

\def\FRWWWWSM{
\begin{align}
\VSM_{W^+_\mu W^-_\nu W^+_\rho W^-_\sigma}=ig^2\Big[2g_{\mu\rho}g_{\nu\sigma}- g_{\mu\nu}g_{\rho\sigma}-g_{\mu\sigma}g_{\nu\rho}\Big]
\nn\end{align}
}

\def\FRWWZZSM{
\begin{align}
\VSM_{W^+_\mu W^-_\rho Z_\nu Z_\sigma}=-ig^2\cosw^2\Big[2g_{\mu\rho}g_{\nu\sigma}- g_{\mu\nu}g_{\rho\sigma}-g_{\mu\sigma}g_{\nu\rho}\Big]
\nn\end{align}
}

\def\FRZZZZSM{
\begin{align}
\VSM_{Z_\mu Z_\rho Z_\nu Z_\sigma}= 0
\nn\end{align}
}

\def\FRWWAASM{
\begin{align}
\VSM_{W^+_\mu \gamma_\nu W^-_\rho \gamma_\sigma}=-ie^2\Big[2g_{\mu\rho}g_{\nu\sigma}- g_{\mu\nu}g_{\rho\sigma}-g_{\mu\sigma}g_{\nu\rho}\Big]
\nn\end{align}
}

\def\FRWWHHSM{
\begin{align}
\VSM_{W^+_\mu W^-_\nu HH}=\frac{ig^2}{2}g_{\mu\nu}
\nn\end{align}
}

\def\FRZZHHSM{
\begin{align}
\VSM_{Z_\mu Z_\nu HH}=\frac{ig^2}{2\cosw^2}g_{\mu\nu}
\nn\end{align}
}


\def\FRWWZEChL{\begin{align}
\VEChL_{W^+_\mu W^-_\rho Z_\nu}=~\VSM_{W^+_\mu W^-_\rho Z_\nu}-\dfrac{ig^3}{\cosw}\Big\{a_3\big[g_{\mu\rho}(p_+-p_-)_\nu
+g_{\mu\nu}(\cosw^2p_0-p_+)_\rho\nn\\+g_{\nu\rho}(p_--\cosw^2p_0)_\mu\big]
+\sinw^2\,(a_1-a_2)\big[p_{0\mu}g_{\nu\rho}-p_{0\rho}g_{\mu\nu}\big]\Big\}
\nn\end{align}
}

\def\FRWWAEChL{
\begin{align}
\VEChL_{W^+_\mu W^-_\rho \gamma_\nu}=~\VSM_{W^+_\mu W^-_\rho \gamma_\nu}+ieg^2(a_1-a_2+a_3)\big[p_{0\mu}g_{\nu\rho}-p_{0\rho}g_{\mu\nu}\big]
\nn\end{align}
}

\def\FRWWHEChL{\begin{align}
\VEChL_{W^+_\mu W^-_\nu H}=\VSM_{W^+_\mu W^-_\nu H}+ ig\,m_W (a-1)\,g_{\mu\nu}
\nn\end{align}
}

\def\FRZZHEChL{
\begin{align}
\VEChL_{Z_\mu Z_\nu H}=\VSM_{Z_\mu Z_\nu H}+ \dfrac{ig\,m_W}{\cosw^2}(a-1)\,g_{\mu\nu}\nn
\nn\end{align}
}

\def\FRWWWWEChL{
\begin{align}
\VEChL_{W^+_\mu W^-_\nu W^+_\rho W^-_\sigma}=~\VSM_{W^+_\mu W^-_\nu W^+_\rho W^-_\sigma}+ig^4\Big\{a_4\big[2g_{\mu\rho}g_{\nu\sigma}+g_{\mu\sigma}g_{\nu\rho}\nn\\
+g_{\mu\nu}g_{\rho\sigma}\big]+2a_5\big[g_{\mu\sigma}g_{\nu\rho}+g_{\mu\nu}g_{\rho\sigma}\big]
~~~~~~~~~~~~\nn\\
-2a_3\big[2g_{\mu\rho}g_{\nu\sigma}-g_{\mu\sigma}g_{\nu\rho}-g_{\mu\nu}g_{\rho\sigma}\big]\Big\}~~~~~~~~~
\nn\end{align}
}

\def\FRWWZZEChL{
\begin{align}
\VEChL_{W^+_\mu W^-_\rho Z_\nu Z_\sigma}=~\VSM_{W^+_\mu W^-_\rho Z_\nu Z_\sigma}+ \dfrac{ig^4}{\cosw^2}\Big\{
a_4\big[g_{\mu\nu}g_{\rho\sigma}+g_{\mu\sigma}g_{\nu\rho}\big]\nn\\
+2a_5\big[g_{\mu\rho}g_{\nu\sigma}\big]
+2\cosw^2a_3\big[2g_{\mu\rho}g_{\nu\sigma}-g_{\mu\nu}g_{\rho\sigma}\nn\\-g_{\mu\sigma}g_{\nu\rho}\big]
\Big\}~~~~~~~~~~~~~~~~~~~~~~~~~~~~~~~~~~~~~~~~~~~~~
\nn\end{align}
}

\def\FRZZZZEChL{
\begin{align}
\VEChL_{Z_\mu Z_\rho Z_\nu Z_\sigma}= \dfrac{2ig^4}{\cosw^4}(a_4+a_5)\big[g_{\mu\nu}g_{\rho\sigma}+g_{\mu\sigma}g_{\nu\rho}+g_{\mu\rho}g_{\nu\sigma}\big]\nn\end{align}
}

\def\FRWWAAEChL{
\begin{align}
\VEChL_{W^+_\mu \gamma_\nu W^-_\rho \gamma_\sigma}=\VSM_{W^+_\mu \gamma_\nu W^-_\rho \gamma_\sigma}
\nn\end{align}
}

\def\FRWWHHEChL{
\begin{align}
\VSM_{W^+_\mu W^-_\nu HH}=~\VSM_{W^+_\mu W^-_\nu HH}+\frac{ig^2}{2}(b-1)g_{\mu\nu}
\nn\end{align}
}

\def\FRZZHHEChL{
\begin{align}
\VEChL_{Z_\mu Z_\nu HH}=\VSM_{Z_\mu Z_\nu HH}+\frac{ig^2}{2\cosw^2}(b-1)g_{\mu\nu}
\nn\end{align}
}


\def\FRWZVIAMMC{
\begin{align}
\VIAMMC_{W^+_\mu Z_\nu V^-_\rho}=-\VIAMMC_{W^-_\mu Z_\nu V^+_\rho}=\dfrac{ig^2}{4\cosw} \Big\{ 2 g_V \big[g_{\rho\mu}p_V {}_{\nu}- g _{\rho\nu}p_V {}_{\mu}\big]\nn\\
+f_V \big[g _{\mu\nu}\left(p_+-p_0 \right)_\rho -g _{\rho\mu}p_+ {}_{\nu} + g_{\rho\nu}p_0 {}_{\mu}\big] \Big\}
\nn\end{align}
}

\def\FRWWVIAMMC{
\begin{align}
\VIAMMC_{W^+_\mu W^-_\nu V^0_\rho}=\dfrac{ig^2}{4} \Big\{ 2 g_V \big[g_{\rho\nu}p_V {}_{\mu}- g _{\rho\mu}p_V {}_{\nu}\big]~~~~~~~~~~~~~~~~~~~~~~~\nn\\
~~~~~~~~~~+f_V \big[g _{\mu\nu}\left(p_--p_+\right)_\rho -g _{\rho\nu}p_- {}_{\mu} + g_{\rho\mu}p_+ {}_{\nu}\big] \Big\}
\nn\end{align}
}

\def\propIAMMC{
\begin{align}
\dfrac{-i}{p^2-M_V^2}\Bigg(  g^{\mu\nu}-\dfrac{p_{V}^{\mu} p_{V}^{\nu}}{M_V^2}\Bigg)
\nn\end{align}
}


\section[Relevant Feynman rules for VBS in the SM]{Relevant Feynman rules for vector boson scattering in the Standard Model} \label{FR-SM}
In this appendix we collect the relevant SM Feynman rules for vector boson scattering processes in the unitary gauge at the tree level. We use here and in the following the short notation $\cosw=\cos\theta_W$. We also label the momenta according to the charge of the associated particle. This way, $p_{\pm,0}$ refers to an incoming $W^\pm$ or a $Z/\gamma$ respectively. 

\begin{center}
\begin{tabular}{ m{2.8cm}m{11cm} }
\includegraphics[scale=0.7]{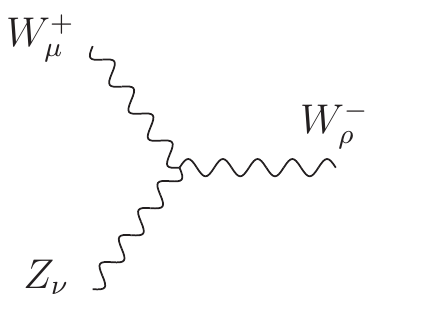} &\vspace{-0.5cm}\begin{minipage}{\textwidth-3.4cm}\FRWWZSM\end{minipage}\\
\includegraphics[scale=0.7]{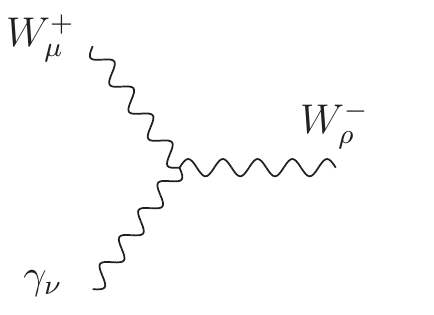} &\vspace{-0.5cm}\begin{minipage}{\textwidth-3.4cm}\FRWWASM\end{minipage}\\
\includegraphics[scale=0.7]{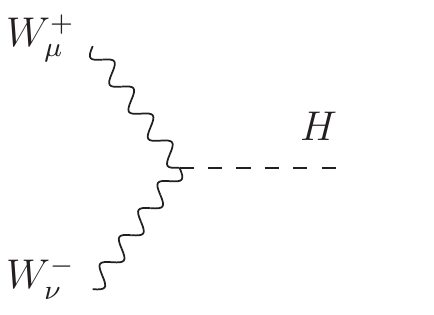} &\vspace{-0.5cm}\begin{minipage}{\textwidth-3.4cm}\FRWWHSM\end{minipage}\\
\includegraphics[scale=0.7]{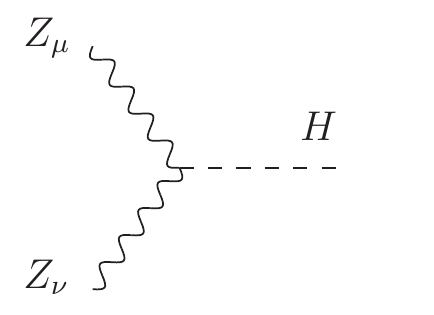} &\vspace{-0.5cm}\begin{minipage}{\textwidth-3.4cm}\FRZZHSM\end{minipage}\\
\includegraphics[scale=0.7]{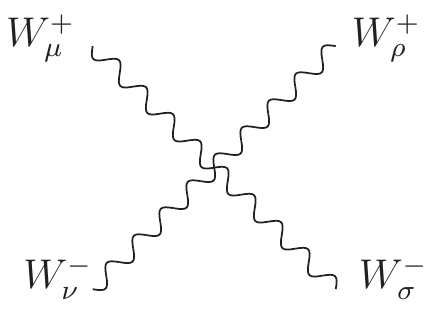} &\vspace{-0.5cm}\begin{minipage}{\textwidth-3.4cm}\FRWWWWSM\end{minipage}\\
\includegraphics[scale=0.7]{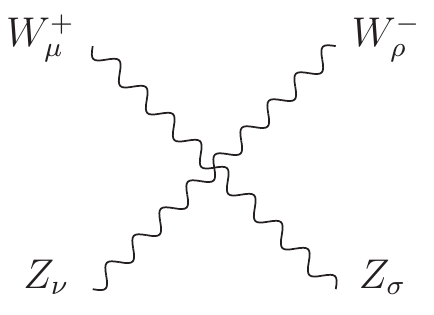} &\vspace{-0.5cm}\begin{minipage}{\textwidth-3.4cm}\FRWWZZSM\end{minipage}\\
\end{tabular}
\end{center}

\begin{center}
\begin{tabular}{ m{2.8cm}m{11cm} }
\includegraphics[scale=0.7]{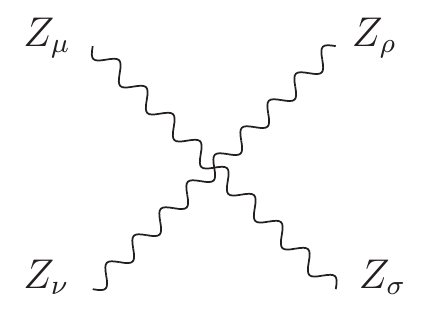} &\vspace{-0.5cm}\begin{minipage}{\textwidth-3.4cm}\FRZZZZSM\end{minipage}\\
\includegraphics[scale=0.7]{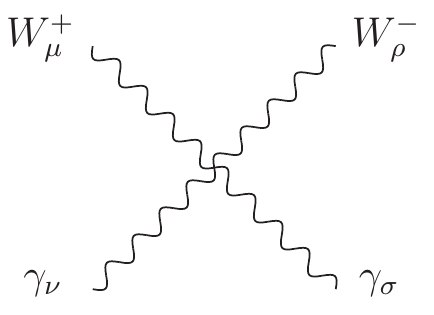} &\vspace{-0.5cm}\begin{minipage}{\textwidth-3.4cm}\FRWWAASM\end{minipage}\\
\includegraphics[scale=0.7]{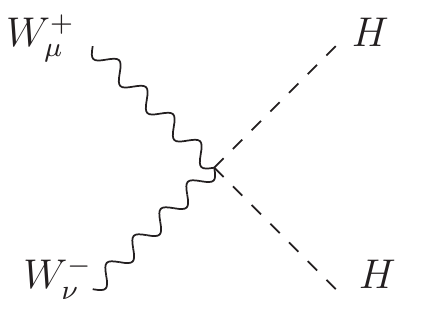} &\vspace{-0.5cm}\begin{minipage}{\textwidth-3.4cm}\FRWWHHSM\end{minipage}\\
\includegraphics[scale=0.7]{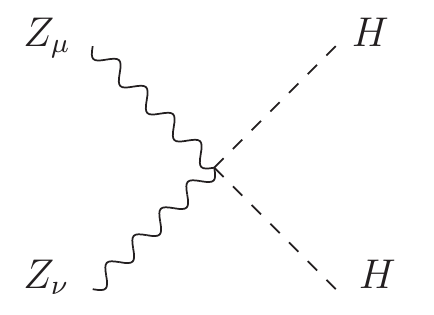} &\vspace{-0.5cm}\begin{minipage}{\textwidth-3.4cm}\FRZZHHSM\end{minipage}\\
\end{tabular}
\end{center}

\section[Relevant Feynman rules for VBS in the EChL]{Relevant Feynman rules for vector boson scattering in the electroweak chiral Lagrangian} \label{FR-EChL}
In this appendix we summarize the relevant EChL Feynman rules for vector boson scattering processes in the unitary gauge at the tree level. These are computed from \eqrefs{eq.L2}{eq.L4} with the same conventions regarding the signs and momenta labelling than in the previous appendix. We present these Feynman rules with the SM part singled out  for an easier comparison.
\begin{center}
\begin{tabular}{ m{2.8cm}m{11cm} }
\includegraphics[scale=0.7]{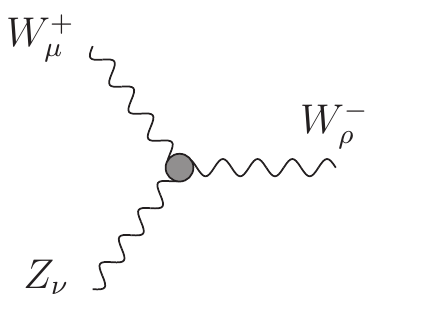} &\vspace{-0.5cm}\begin{minipage}{\textwidth-3.4cm}\FRWWZEChL\end{minipage}\\
\includegraphics[scale=0.7]{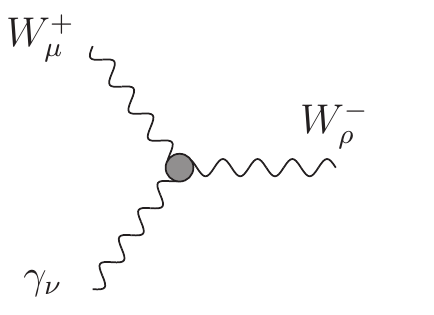} &\vspace{-0.5cm}\begin{minipage}{\textwidth-3.4cm}\FRWWAEChL\end{minipage}\\
\includegraphics[scale=0.7]{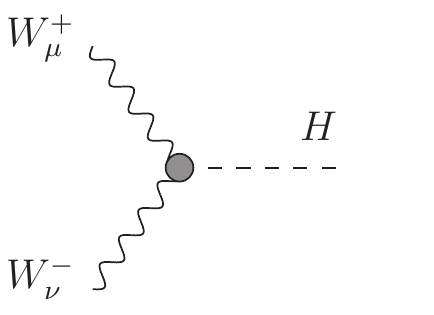} &\vspace{-0.5cm}\begin{minipage}{\textwidth-3.4cm}\FRWWHEChL\end{minipage}\\
\end{tabular}
\end{center}

\begin{center}
\begin{tabular}{ m{2.8cm}m{11cm} }
\includegraphics[scale=0.7]{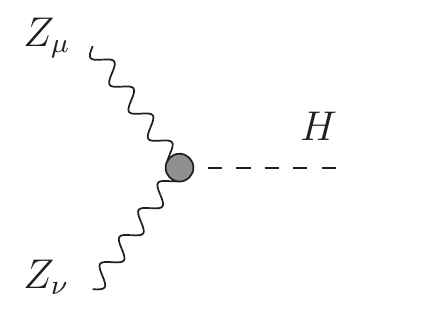} &\vspace{-0.5cm}\begin{minipage}{\textwidth-3.4cm}\FRZZHEChL\end{minipage}\\
\includegraphics[scale=0.7]{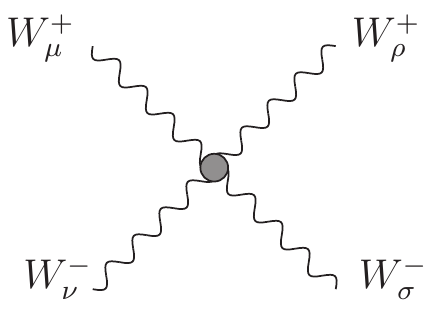} &\vspace{-0.6cm}\begin{minipage}{\textwidth-3.4cm}\FRWWWWEChL\end{minipage}\\
\includegraphics[scale=0.7]{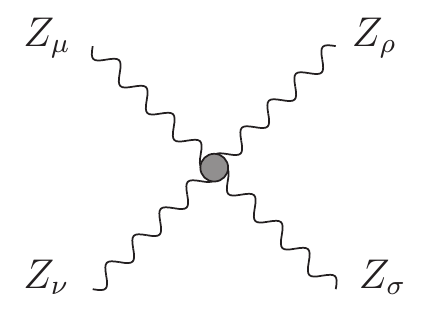} &\vspace{-0.5cm}\begin{minipage}{\textwidth-3.4cm}\FRZZZZEChL\end{minipage}\\
\includegraphics[scale=0.7]{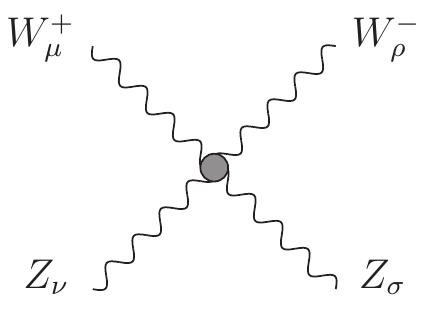} &\vspace{-0.5cm}\begin{minipage}{\textwidth-3.4cm}\FRWWZZEChL\end{minipage}\\
\includegraphics[scale=0.7]{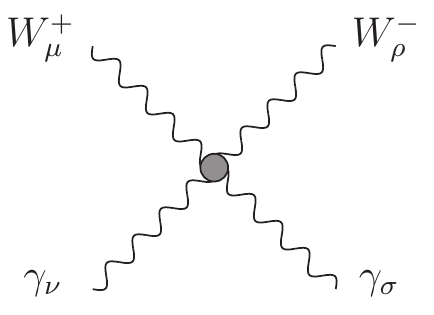} &\vspace{-0.5cm}\begin{minipage}{\textwidth-3.4cm}\FRWWAAEChL\end{minipage}\\
\includegraphics[scale=0.7]{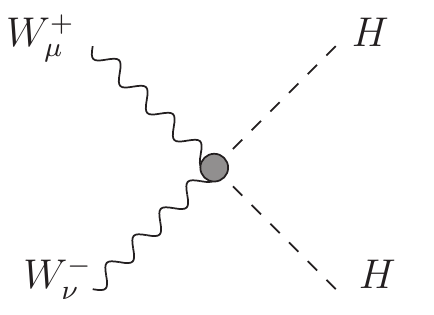} &\vspace{-0.5cm}\begin{minipage}{\textwidth-3.4cm}\FRWWHHEChL\end{minipage}\\
\includegraphics[scale=0.7]{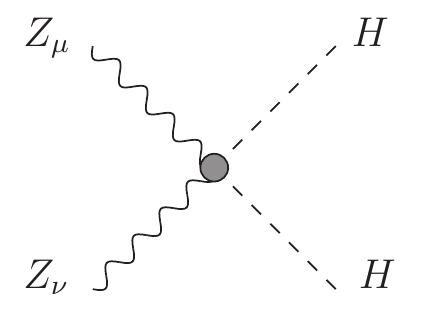} &\vspace{-0.5cm}\begin{minipage}{\textwidth-3.4cm}\FRZZHHEChL\end{minipage}\\
\end{tabular}
\end{center}

\section[Relevant Feynman rules for VBS in the IAM-MC]{Relevant Feynman rules for vector boson scattering in the IAM-MC} \label{FR-IAMMC}

In this appendix we present the relevant IAM-MC Feynman rules for vector boson scattering processes in the unitary gauge at the tree level. These are computed from \eqref{LVugauge} with the same conventions regarding the signs and momenta labelling than in the previous appendices.

\begin{center}
\begin{tabular}{ m{2.8cm}m{11cm} }
\includegraphics[scale=0.7]{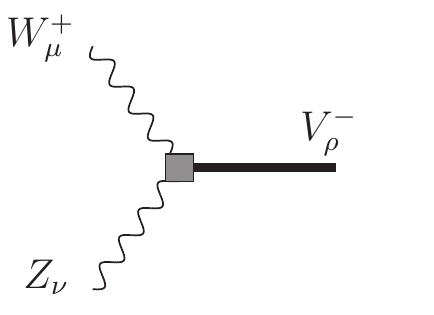} &\vspace{-0.5cm}\begin{minipage}{\textwidth-3.4cm}\FRWZVIAMMC\end{minipage}\\
\end{tabular}
\end{center}

\begin{center}
\begin{tabular}{ m{2.8cm}m{11cm} }
\includegraphics[scale=0.7]{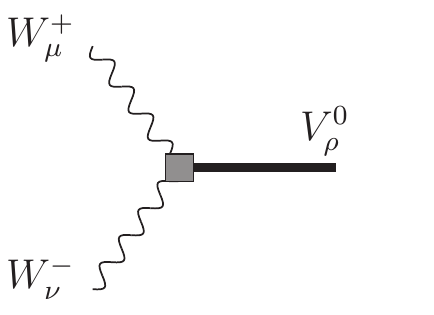} &\vspace{-0.5cm}\begin{minipage}{\textwidth-3.4cm}\FRWWVIAMMC\end{minipage}\\
\end{tabular}
\end{center}

\noindent Furthemore, the resonance propagator reads:

\begin{center}
\begin{tabular}{ m{2.8cm}m{11cm} }
\includegraphics[scale=0.7]{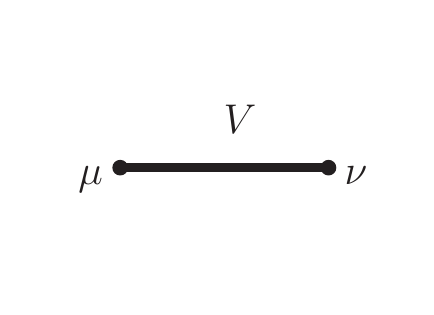} &\vspace{-0.5cm}\begin{minipage}{\textwidth-3.4cm}\propIAMMC\end{minipage}\\
\end{tabular}
\end{center}

\end{appendices}


\chapter*{Agradecimientos}\label{Agradecimientos}
\chaptermark{}

Una Tesis no son s\'olo los a\~nos que pasas matriculado como estudiante de doctorado, es mucho m\'as. Es todo el camino recorrido para llegar hasta aqu\'i, hasta este momento en el que echas la vista atr\'as y te das cuenta de que eres lo que eres y has conseguido lo que has conseguido gracias al apoyo de mucha gente que te ha acompa\~nado en tus peores y en tus mejores momentos.

Si est\'as leyendo esto y me conoces sabr\'as que aunque vaya de dura por la vida en el fondo soy tan blandita que me quiero morir, y sabr\'as que no puedo terminar este inmenso trabajo y esfuerzo sin agradecer a todos y cada uno de los que han contribuido a que se hiciera posible. Entre eso y que me enrollo escribiendo como las persianas, lo siento, pero va a ser largo.

All\'a voy.

En primer lugar querr\'ia dar las gracias a Mar\'ia Jos\'e por el tiempo dedicado a esta Tesis, por la ayuda para conseguir financiaci\'on con la que poder hacer el doctorado y por facilitarme la vida entre Madrid y Par\'is. 

Gracias a Antonio Dobado, Dom\`enec Espriu y Rafael Delgado, miembros honor\'ificos de la Tropa Quiral, por darme la oportunidad de trabajar con vosotros y de aprender de vosotros. Esas reuniones locas con producci\'on de resultados a tiempo real ser\'an dif\'iciles de olvidar. 

Gracias tambi\'en a los miembros del tribunal por dejarse marear con temas burocr\'aticos de la tesis sin queja ninguna (de hecho al rev\'es, siempre con una amabilidad incre\'ible) y por dedicar su valioso tiempo a valorar este trabajo.

En estos momentos me acuerdo tambi\'en de toda la gente del IFT que me ha brindado un lugar en el que trabajar durante cuatro a\~nos y en el que no he podido encontrarme m\'as a gusto. Gracias a las secretarias que nos han salvado la vida una y otra vez y que no podr\'ian ser m\'as encantadoras. Gracias tambi\'en a Susana y a \'Angel por involucrarme en todas las locuras de divulgaci\'on en redes, y gracias a Rebeca, Tiina y Chabely, verdaderas reinas de Invisibles, que se han desvivido por nosotros desde el primer momento. Luca, Enrique, a vosotros tambi\'en quisiera daros las gracias por tratarnos siempre de forma tan cercana y por contestar a todas nuestras preguntas con una brillantez que asusta. No cambi\'eis por favor. Y sobre todo, Luca, nunca pierdas el amor por los perritos, son lo m\'as.

Carlos, para ti me temo que tengo un agradecimiento especial. A ti te debo el haber podido meter la cabeza en la divulgaci\'on a trav\'es de charlas y masterclasses, aunque nunca te perdonar\'e el tener que madrugar para pastorear a los chiquillos. Gracias por estar siempre dispuesto a compartir con nosotros tus historias desternillantes de camisetas chinas que ti\~nen toallas, de robos de coches en Roma o de llamamientos bizarros entre bi\'ologos en festivales folcl\'oricos gallegos. Siempre guardar\'e un recuerdo especial de todas ellas. Gracias tambi\'en por no negarte nunca a contestar a nuestras preguntas aunque termin\'aramos ojipl\'aticos por tu conocimiento apabullante de la f\'isica. Gracias por todo. Espero de verdad que te vaya todo muy bien y que nos veamos en el futuro.

Tampoco pod\'ia faltar el agradecimiento a Juanjo y Ernesto (a los que tambi\'en considero parte del IFT, est\'en donde est\'en), que han sido en cierto modo como mis t\'ios investigadores. Juanjo, eres un torbellino de sabidur\'ia y caos, pero sobre todo, eres una excelente persona. Gracias por acceder a hablar siempre que lo he necesitado, por haberme ense\~nado tanto, por haberme dejado flipando con tus conocimientos de la f\'isica y tu humildad al discutirlos, por haberme hecho reir, por haberme tra\'ido el t\'e chino, por haberme ofrecido las chuches picantes mexicanas y por mil cosas m\'as... se me queda corto este p\'arrafo para agradec\'ertelo todo, amigo. Espero que seas consciente de todo lo que me has aportado y espero que os vaya muy bien tanto a ti como a Ramune como a los peques. Y Ernesto, a ti tengo que agradecerte todas las cosas que he aprendido de ti,  as\'i como toda tu paciencia en las batallas diarias y tu disponibilidad para trabajar siempre. Tambi\'en tu buen rollo constante, tu diplomacia, tu preocupaci\'on por nosotros y tu buena voluntad de ayudarnos siempre que lo necesitamos. Te digo lo mismo: espero que consigas todo lo que te propongas y que te vaya muy bien junto a Mari y a Camilo.

Y en cuanto a otras instituciones aparte del IFT no puedo dejar de agradecer al departamento de f\'isica te\'orica de la UAM. Gracias por darme la oportunidad de dar clases, que es sinceramente una de las cosas con las que m\'as disfruto en la vida. Gracias tambi\'en a Javier y a Anette, porque nos hab\'eis ayudado siempre con una sonrisa.

Merci aussi au LPT d'Orsay, surtout \`a Marie et Jocelyn. Je voudrais \'egalement adresser des remerciements tout particuliers \`a Asmaa, pour sa bont\'e et sa compr\'ehension durant ces deux ans. Asmaa, tu es une personne merveilleuse et tu n'imagines pas combien je te suis reconnaissante de m'avoir accueillie d\`es le premier jour. Cela a \'et\' e une \'epoque difficile pour moi, tu le sais, mais je tiens aussi \`a ce que tu saches que tu as contribu\'e \`a la rendre plus supportable, non seulement parce que tu nous as facilit\' e les connexions Madrid-Paris, mais aussi pour les ``th\'es gourmands'', pour tes histoires hilarantes de ``marabout\'ee'', pour tes commentaires d'encouragement et de soutien... pour tout. Merci du fond du c\oe{}ur, Asmaa, d'\^{e}tre comme tu es. Je me r\'ejouis \'enorm\'ement de t'avoir fr\'equent\'ee pendant toute cette \'epoque et je souhaite sinc\`erement que nous continuerons \`a nous voir \`a l'avenir. 

Y desde luego qu\'e habr\'ia sido de mi sin mis maravillosos profesores. Desde Mar\'ia Jes\'us, que fue la primera y de la que guardo un recuerdo especial, hasta el \'ultimo de vosotros hab\'eis contribuido a que est\'e hoy escribiendo esto, culminando una etapa. Sin embargo, y aunque me acuerdo de muchos (Bego\~na, Marta, Tina, Amador...) tengo que agradecerles en particular a dos de mis profesores no ya el hecho de haberme ense\~nado lengua y literatura por un lado y danza urbana por otro, sino a pensar y a ser. Gracias Fernando y Quique por hacerme cr\'itica, por darme voz, consejos y consuelo. Por ense\~narme que se pueden cambiar las cosas y que hay que querer cambiarlas y por hacerme ver que hay que intentar ser mejor cada d\'ia y seguir adelante. Gracias por darme un lugar en el que escapar de todo siempre que lo necesite, de coraz\'on. Quique, esta tesis no habr\'ia visto la luz de no ser por tus clases en Flow y de las locuras escalando los fines de semana, en las que tambi\'en incluyo a Gema, que no podr\'ia ser mejor Persona. Gracias por todo.

En el terreno de la divulgaci\'on, pero tambi\'en relacionado con la docencia, me gustar\'ia agradecer a todos los profesores de institutos en los que he dado charlas su calurosa acogida y su buen hacer por sus alumnos. Sois un ejemplo. Casildo, Mercedes, Nuria, Beatriz, no os imagin\'ais lo importante que es lo que hac\'eis, de verdad, ojal\'a no dej\'eis de hacerlo nunca. S\'e que probablemente nunca le\'ais estas lineas, pero espero que sep\'ais que aqu\'i ten\'eis a vuestra fan n\'umero uno.

Y de los profesores pasamos a los alumnos, a los que me gustar\'ia dedicar mi m\'as profundo agradecimiento. Han sido tres a\~nos de t\'ecnicas experimentales y dos a\~nos de FAE en los que no podr\'ia haber aprendido m\'as de vosotros. Gracias por hacerme ver que no ten\'ia ni idea de nada hasta que tuve que lidiar con vuestras brillantes preguntas, y gracias por ense\~narme a tener paciencia y a que cuando est\'as en el aula/laboratorio nada importa salvo vosotros. Gracias por prestarme atenci\'on y por haberme motivado a dedicarme a la ense\~nanza. Si de alguien es la culpa, que sep\'ais que es vuestra. 

A mi familia de media y larga distancia tambi\'en tengo que dedicarle un agradecimiento especial y para ello querr\'ia empezar por mis abuelas. Merce, Carmen, sin vosotras no estar\'ia aqu\'i. Siempre, toda mi vida, me he sentido m\'as que apoyada y m\'as que querida por vosotras, y nunca podr\'e agradeceros lo suficiente lo mucho que me hab\'eis cuidado y ense\~nado. Esta tesis es un poquito vuestra tambi\'en. Gracias tambi\'en a mis abuelos por su apoyo constante, a Mari Tere, que es una de las personas m\'as buenas y cari\~nosas que conozco, y a la Chirra, a Juan Carlos y a Juanmi.

De mis abuelos salto a mis primos, a quienes tambi\'en tengo mucho que agradecer. En especial a In\'es, a la que mis amigas aun llaman ``mi prima la guay'' y que, a pesar de no haber mantenido contacto constante (ni falta que ha hecho), me ha ayudado y apoyado cada vez que la he necesitado. Te quiero prima, gracias por todo. Quiero darle las gracias tambi\'en a Nacho (versi\'on peque\~na) por ser tan buen t\'io y por haberme hecho preguntas de todo tipo desde que era un renacuajo. Tambi\'en por tu sonrisa y buen rollo constantes. Gracias primo. 

Y no puedo dejarme a los primos {\it senior}: V\'ictor, con tu humor desternillante, tu capacidad culinaria y tus preguntas de f\'isica; Ana, con tu sonrisa y tu amor por los bichitos que me recuerda mucho al m\'io propio; Antonia, con tu siempre estar pendiente de que estemos todos bien y a gusto; Marta,  dispuesta a colgar unas banderas espirituales en Nepal de tu parte o a tratar a distancia a tu gato... y sobre todo, gracias a Nacho por llevarme en moto desde que era un mico, por ense\~narme (en parte) a bucear, por quitarme muchos miedos, por hacer que me enamorase de los perros desde que ten\'ia seis meses, por todo tu cari\~no, todo tu apoyo y todo lo que me has ense\~nado; y gracias a Lola, simplemente por c\'omo eres y nunca dejar\'as de ser, por querernos tanto a todos y ser tan franca, tan directa, y tan maravillosa, a tu propia manera. Est\'es donde est\'es, esta Tesis tambi\'en es tuya, como del resto de la PRG. 

Tambi\'en de la otra parte de la familia quisiera dar las gracias a Ana, Margarita, y a los Isidros, por formar parte de mi vida desde siempre, por ayudarme cada vez que lo necesito y por saber que siempre puedo contar con ellos. Mil gracias, de verdad. T\'io Jes\'us, tambi\'en a ti quiero agradecerte el haber hecho siempre el esfuerzo de venir a vernos y de disfrutar de nosotros. Gracias de coraz\'on. 

Por \'ultimo en temas familiares, toca la familia con la que no comparto sangre, pero a la que tambi\'en tengo mucho que agradecer. Me gustar\'ia empezar por \'Angel, Cris e Iv\'an, en los que siempre he encontrado un lugar donde estar m\'as que a gusto, donde reirme sin parar, y donde sentirme querida, apoyada y escuchada. Gracias. Gracias tambi\'en a mi t\'io Luis Miguel, que es de las personas m\'as encantadoras, m\'as cari\~nosas y m\'as nobles que he conocido nunca. Gracias por estar absolutamente siempre ah\'i, por quererme tanto desde el primer momento y por hacer el memo conmigo siempre. Te quiero mucho y te agradezco en el alma todo lo que me has apoyado. Un trocito de esta tesis tambi\'en es tuyo. 

Con mi familia eibarresa tampoco comparto sangre, pero s\'i muchas cosas vividas a lo largo de estos (ya unos cuantos) a\~nos. Gracias a Ana y a Juan Luis por acogerme desde el primer d\'ia como si fuera una m\'as de la familia. A Susana y a Pablo por su buen rollo constante y por haberme hecho sentir integrada desde aquel d\'ia en el que yo s\'olo era una chiquilla asustada reci\'en llegada a Deba. Gracias tambi\'en a Niceto y a Carmen por ser siempre tan amables y por preocuparse tanto por nosotros. A Laura por ser la persona con la mayor energ\'ia y las historias m\'as locas que he conocido. Y finalmente gracias a Izaskun por estar siempre tan pendiente de nosotros, por tener siempre tantas ganas de vernos y por cuidarnos tanto. Gracias a todos, de coraz\'on.

Adem\'as de esta, tambi\'en est\'a el resto de la familia que no se elige. Por buena suerte para mi y por mala para ti, que est\'as leyendo esta biblia de agradecimientos, no puedo quejarme de la cantidad ni de la calidad de todos los que caen en esta categor\'ia. 

Las primeras, por antig\"uedad sobre todo (aunque no solo), son Marina, Carmen y Cata, que aunque justo a lo largo de estos a\~nos de tesis no nos hayamos visto tanto, llevamos juntas m\'as de 15 a\~nos compartiendo la vida, tanto la buena como la mala, y s\'e que est\'ais ah\'i siempre, para lo que necesite. Os quiero much\'isimo y tengo mucho por lo que agradeceros. Que sep\'ais que tambi\'en sois inversoras de esta tesis y que en cierta medida tambi\'en os pertenece. Eso s\'i, no os la intent\'eis leer que terminar\'eis queriendo abrirme la cabeza con semejante tocho.

Los siguientes son Clara (+Greg), Laura, Andr\'es, El\'ias y Antonio. Otros a los que veo poco por la ca\'otica mezcla de vidas que representamos, pero a los que tengo que agradecer infinitos momentos de fiesta, risas, charlas, ca\~nas y dem\'as cosas maravillosas de la vida (como jugar a lanzamiento de medusa). Gracias. Gracias por hacer que me sienta igual que si os hubiera visto el d\'ia anterior cada vez que nos reunimos tras meses sin vernos, y gracias por escucharme y por hacer el tonto hasta que casi me ahogue de la risa (creo que no hace falta decir qui\'enes corresponden a esto \'ultimo).

Gracias tambi\'en a los {\it fant\'asticos}, Cynthia, Luis y Cristian, que aunque haya que irse hasta Par\'is para verse sigue mereciendo la pena :) (Cristian a ver si te dejas ver el pelo de una vez). Gracias Cynthia por compartir siempre toda tu vida conmigo, gracias por tu amistad, por tus consejos, por tu forma de escuchar, y, sobre todo, por tu risa. Nunca la pierdas.

Llegamos a los Phmen, que tanto tardaron en descubrir qu\'e significaba el nombre del grupo (ya os vale...). Dequi, Melek, Ram\'on, no puedo agradeceros lo suficiente vuestro buen rollo, vuestro inter\'es en mi y en los dem\'as, vuestras chorradas, vuestra forma autom\'atica de quitarme las rayadas y la mala leche de la cabeza cada vez que os veo, vuestras conversaciones profundas, vuestras bromas sobre cu\'ando acabar\'ia Dequi la carrera o sobre la procedencia almeriense de Melek... formamos una panda muy peculiar que me encanta y por la que os estoy enormemente agradecida. Hab\'eis hecho esta tesis mucho m\'as llevadera, y tambi\'en es en parte vuestra. Gracias.

Simon, Alex, Robert, y Rober tambi\'en tengo un agradecimiento especial para vosotros, aunque solo sea por la de veces que me hab\'eis escuchado daros la lata con mil cosas distintas. Gracias por las ricas comidas de Sim\'on, por el t\'e, por las fotos, gifs y memes de bichitos, por todas las gilipolleces (y horas de estudio) compartidas durante la carrera y el m\'aster y por haberos convertido en grandes amigos.

And these acknowledgments could not finish without a special mention to my {\it Terrier}
friends, Tamiro and Ben. Thank you guys for guiding me all the way through my Bostonian adventure and for helping me survive professor Pi's lectures. Thank you for taking me downtown to eat ``croquetas'' just because you knew I missed them. Thank you for the beers, the laughter and for all your help. Thank you, Ben, for the super long evenings programming ``our pots'' (lol) and thank you, Tamiro, for listening to me every day during five months and for setting up the meeting with Glashow. You became true friends during those crazy days, and this thesis is, in part, yours. Thanks as well to Andrea for calling me Pio and for the chicken dinners. For the guitar lessons, the fabulous road trip along the east coast and for much more. Thank you.

Y ya que hemos pasado por los amigos del colegio, del instituto, y de la universidad (vaya tost\'on madre m\'ia, lo siento), no podr\'ia terminar estos agradecimientos sin hacer menci\'on a todos los IFTeros con los que he tenido el placer de compartir edificio y mucho m\'as a lo largo de estos cuatro a\~nos, aunque hayan venido en varias tandas.

 En la primera etapa, quiero dar las gracias a Diego, Jos\'e Mar\'ia y Santi, mis primeros compis de despacho. Sobre todo a Santi por corregir el aun perfectamente custodiado c\'omic del autoestopista gal\'actico y por haber sido mi moment\'aneo profe de escalada. Gracias a Josu por el kebab de Aranda, por Kentuchy y por mil chorradas m\'as que no caben en estos agradecimientos. Y, sobre todo, por las risas que nos echamos con todas ellas. Gracias a Ana  por las tarde de zumba en el sal\'on, por la guerra conjunta que les ganamos a las cucarachas y por tu brownie. Aun no he probado ninguno mejor. Gracias a Miguel por abrirnos los ojos a las ense\~nanzas de nuestro se\~nor Nima y por ayudarnos desde la carrera incluso con trabajos mandados por profesores sovi\'eticos. Gracias a Reto por su sonrisa constante, por su ayuda cuando se le necesita, por robarme las llaves de la bici cada vez que pasa por el IFT y por portar el estandarte del Eibar, que ah\'i sigue, en tu puerta. Y para terminar esta primera etapa, gracias, de verdad, de coraz\'on a Irene y a V\'ictor. Gracias Irene por recordarme cada vez que te veo lo mucho que me ha llegado a gustar la f\'isica. Gracias por tu inmensa sabidur\'ia, por tu humildad, por tu buen rollo, por todo. Eres y siempre ser\'as una de las personas que m\'as ha marcado (para bien) mi vida acad\'emica, y siempre te lo agradecer\'e en el alma. V\'ictor, a ti no s\'e qu\'e decirte. Que por favor no vuelvas a adue\~narte del altavoz en mi casa para ponerme los Chichos y que te has convertido en un verdadero amigo a lo largo de este tiempo. Eres una de las mejores personas que conozco, uno de los m\'as amigos de sus amigos y adem\'as est\'as completamente majara, no se me ocurre mejor combinaci\'on. No pierdas nunca esa bondad tuya, esa capacidad de escuchar, de interesarte, y de ayudar. No sabes cu\'anto se te agradece, de verdad. !`Ah!, y dale tambi\'en las gracias a Bea por ser como es y por formar parte del team Cersei.
 
De la segunda etapa (aunque a algunos os conozco desde mucho antes de llegar al IFT), querr\'ia agradecer a los principales inversores de la cafeter\'ia del CBM todas las risas, apoyo y buenos momentos durante estos a\~nos. Gracias a \'Alvaro, Raquel, Fer, Salva, Judit, Gallego, Guille y Uga por haberme acompa\~nado a lo largo de este doctorado. Gracias por los pub quizes,  por aquella jam session que hay que repetir, por las fotos de llamas y de gatetes, por escucharme cuando me ha hecho falta, por los ping pongs espor\'adicos y por las cenas, ca\~nas, casas rurales y momentos maravillosos varios :) Gracias en especial a Gallego por ser uno de los mejores compis de despacho que se pueden tener y por su extrema bondad y buen hacer por los dem\'as. Y gracias a Uga, por ser mi hermano durante la carrera, por todo lo que hemos vivido juntos, por convertirte despu\'es en un compa\~nero de piso incre\'ible (aunque te ladre por las ma\~nanas cuando me despierto medio zombie), por toda la ayuda sin la que yo nunca habr\'ia llegado hasta aqu\'i y, sobre todo, por engancharme a la escalada. Gracias de coraz\'on.

Y de la tercera y m\'as reciente etapa, tengo que agradecer de manera muy especial a Rober y a Manu. Este \'ultimo (y m\'as duro) a\~no no habr\'ia sido posible sin vosotros. Gracias por todas las risas que nos hemos echado durante las comidas y en las partidas de pocha de despu\'es (a pesar del nefasto resultado: {\it inyustisia}). Gracias Rober por tu buen car\'acter, por trabajar tant\'isimo y tan bien, por ser el m\'as fair de los players, por tu humor argentino insuperable y por los alfajores. Eres una gran persona y me alegro much\'isimo de haber coincidido contigo este tiempo. Y gracias Manu por ser el mejor hermano peque\~no que pod\'ia pedir. Gracias por darle un poco de vidilla a las clases de FAE, por interesarte luego en hablar con nosotros y por pelear para terminar haciendo el TFM que has hecho aunque haya habido momentos que hayan sido {\it la dolor}. Gracias por tu risa floja y contagiosa, por escucharme cuando lo he necesitado y por no haberme mandado a la mierda con mi T.O.C., especialmente en el tema de los plots. Estoy muy orgullosa de ti, de todo lo que has aprendido, y te agradezco mucho que me hayas dejado formar parte de ello, sobre todo por los momentos desternillante que hemos compartido y porque me has hecho aprender much\'isimo a mi por el camino. Gracias :)

Gracias tambi\'en a Argatxa y a sus girls por haberme acogido y por los buenos momentos que he pasado con vosotros, desde disfrutar los Sanjuanak Eibarren hasta los viajes en coche con Alba, Reto, Eriz y Tania pasando por los puros de Apa y Marta y por las tardes en el local jugando a juegos de mesa, a la play, o viendo los lobos de Arga o Interstellar con preguntas de f\'isica en directo. Gracias. 

\begin{figure}[t!]
\begin{center}
\includegraphics[width=0.3\textwidth]{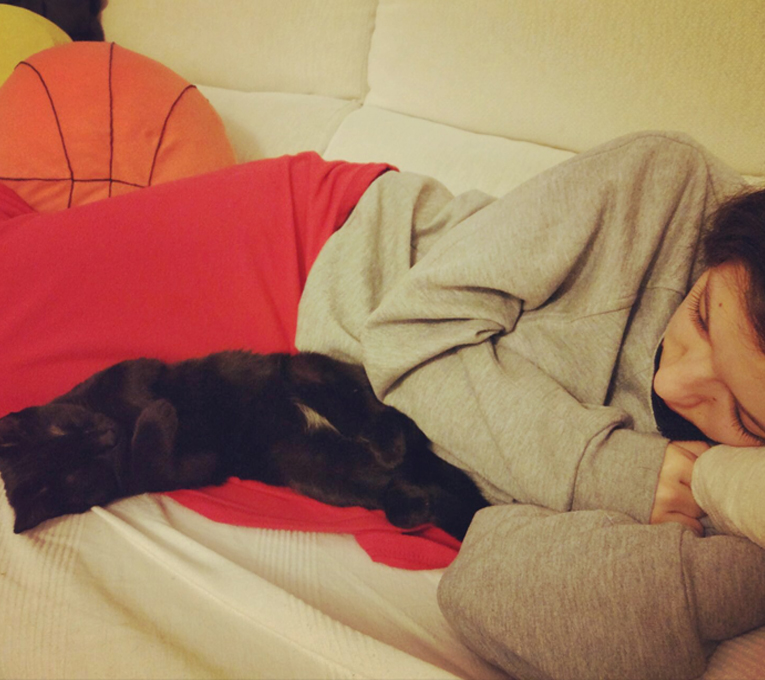}
\includegraphics[width=0.3\textwidth]{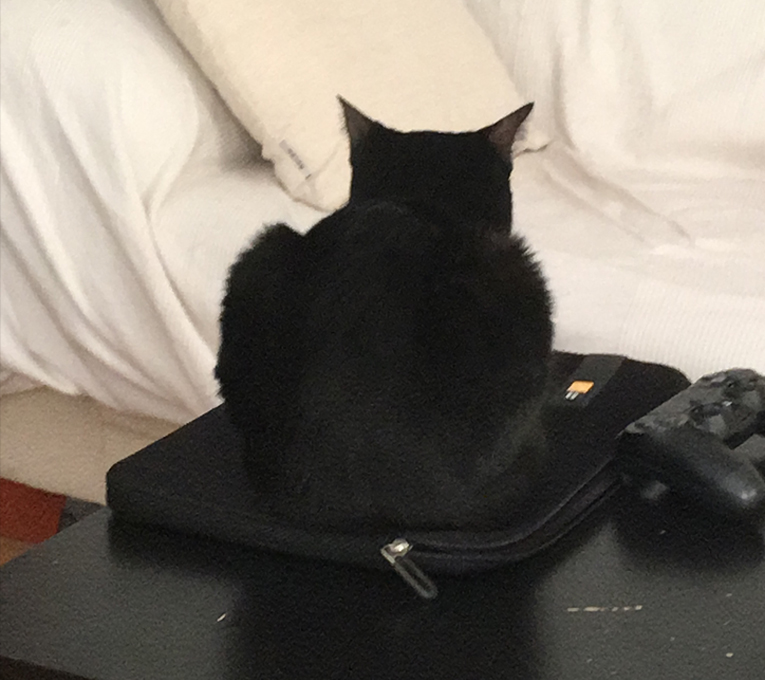}
\includegraphics[width=0.3\textwidth]{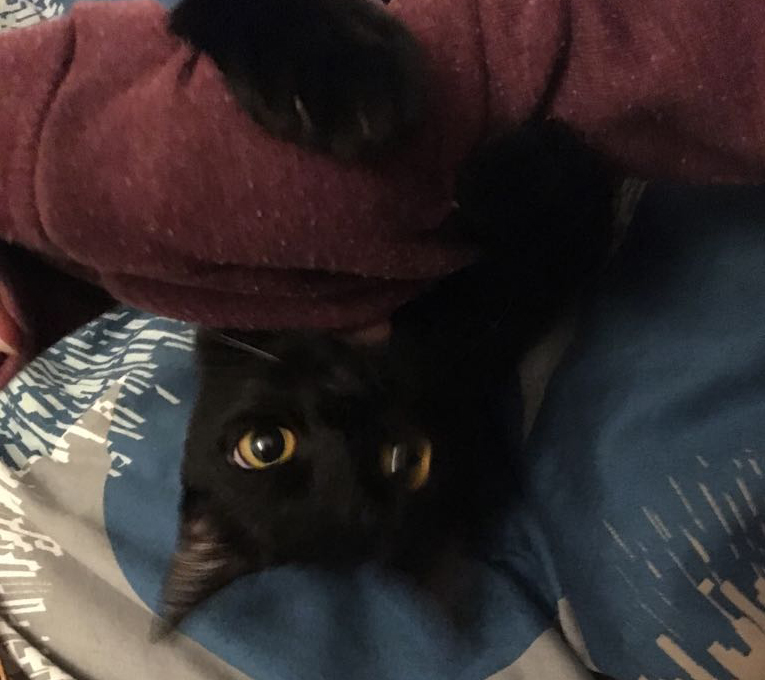}
\caption{Tres estados de mi gata: amorosa, pasota y juquetona. Todos ellos maravillosos.}
\label{wimp}
\end{center}
\end{figure}

Los siguientes a los que tambi\'en tengo mucho que agradecer son mis hermanos peludos. Puede que sea una tonter\'ia, pero esta tesis tambi\'en es de Susto y de Wimpita. 
Llegar a casa y saber que te est\'an esperando con toda la ilusi\'on del mundo es algo que no se paga con dinero y por lo que estoy muy agradecida. Sin esos momentos puede que yo no hubiera llegado hasta aqu\'i. Gracias a Susto por ser literalmente el hermano con el que crec\'i y porque si alguna vez he sentido amor incondicional ha sido el suyo. Y gracias a Wimp por ser la gata m\'as mona de la vida y por querernos a su manera en todas sus versiones (ver \figref{wimp}). Gracias tambi\'en a Monza, por ser mi mami canina cuando era peque\~na, pero tambi\'en a Asen, a Canito, a Blues, a Portia, a Buba, a Sugo, a Kira y, sobre todo, a Drogo. Os quiero.

Pap\'a, Mam\'a, a vosotros ni siquiera s\'e qu\'e deciros. No me da en un p\'arrafo tan corto para expresar lo mucho que os quiero, lo orgullosa que estoy de ser vuestra hija y lo inmensamente agradecida que estoy por todo lo que me hab\'eis dado. El apoyo que he sentido por vuestra parte ha sido constante, durante toda mi vida. Aun con 26 a\~nazos sigo pensando en contaros cualquier 
problema, porque s\'e que puedo contar con vosotros para solucionarlo y s\'e que podr\'e hacerlo siempre. Tambi\'en en las alegr\'ias pienso en vosotros, en lo gratificante que es poder compartirlas juntos, emocionarnos juntos, y meternos juntos con pap\'a porque es un llorica, cosa que me temo que he heredado de \'el, como muchas otras. Soy como soy y he llegado donde he llegado por vosotros, porque hab\'eis estado a mi lado cada momento de cada d\'ia, sin sujetarme, sin tocarme, pero siempre ah\'i para aseguraros de que nunca me cayera, y de que, si lo hac\'ia, supiera siempre levantarme. Me hab\'eis ense\~nado a pensar por mi misma y me hab\'eis inculcado la curiosidad cient\'ifica que llevar\'e conmigo toda la vida. Me hab\'eis ense\~nado a luchar por lo que creo y a que cualquier cosa puede conseguirse y  a sorprenderme y a emocionarme por las cosas que realmente merecen la pena. Sin todo eso esta tesis no se habr\'ia escrito, as\'i que es realmente vuestra. Sois quien me coger\'ia de la mano cerca del agujero de gusano en dos mil ciento y pico, y nunca podr\'e agradeceros eso lo suficiente. 

Javi, para ti tambi\'en se me va a quedar corto el p\'arrafo, me temo. Tengo la sensaci\'on de que nunca te voy a poder agradecer suficiente todo lo que has hecho por mi y todo lo que me has cuidado estos dos a\~nos largos. Has sido el mejor compa\~nero de despacho que podr\'ia haber tenido jam\'as (salvo quiz\'a uno un poco menos sordo, pero no me quejo). Me has hecho los d\'ias soportables, incluso estupendos, me has hecho tener ganas de mirar memes para compartirlos contigo y re\'irnos sin que nadie m\'as sepa que co\~no est\'a pasando, me has llevado y tra\'ido cada d\'ia abordo de {\it badass}, incluso a horas intempestivas, me has regado las plantas, has compartido tu paella (la de verdad, no el arroz con cosas) conmigo... y esto solo como compa\~nero de despacho. Porque no solo has sido mi compa\~nero de despacho y de piso, sino de vida realmente, incluso pel\'andome y cort\'andome un melocot\'on para que no muriera de inanici\'on, u oblig\'andome a ir a bailar contra mi voluntad m\'as firme porque sab\'ias que me sentar\'ia bien, tambi\'en trag\'andote glee (madre m\'ia est\'as fatal)... incluso te has dejado enga\~nar infinitas veces como en lo de llevarte a escalar para que al final te terminara gustando y todo. No tengo sitio para listar todas las cosas que has hecho por mi, y quer\'ia que supieras lo mucho que significan y lo mucho que las valoro, de verdad. Te considero una de las personas m\'as buenas de la vida (no de la m\'ia sino en general) y te agradezco en el alma que hayas estado a mi lado todo este tiempo, porque si no de verdad que creo que me habr\'ia dado una miaja de apechusque y que la hubiera terminado roscando. Por no hablar de que me has permitido compartir a Wimpita contigo, que vale, quiero pensar que algo de culpa de que se haya quedado la tengo yo, pero t\'u tambi\'en la tienes por ser tan blandito y f\'acilmente manipulable, as\'i que gracias Javi, de verdad, de coraz\'on. Esta tesis es en gran medida tuya porque has estado ah\'i cada d\'ia para ayudarme a que se hiciera posible y nunca podr\'e agradec\'ertelo lo suficiente. 

Y Xabi, t\'u est\'as al final de estos agradecimientos pero deber\'ias haber aparecido desde el principio, porque has estado conmigo desde el primer d\'ia de esta tesis (e incluso antes) hasta el final, y s\'e que no ha sido siempre f\'acil. Si de alguien es este trabajo es tuyo, y a ti es sin duda a qui\'en m\'as tengo que agradecerle el que esta tesis haya salido adelante, el que yo haya salido adelante. Fuiste t\'u el que me anim\'o a embarcarme en esto y quiero que sepas que despu\'es de todo no me arrepiento de haberlo hecho. He aprendido much\'isimas cosas, tanto de f\'isica como de la vida, y eso es algo que siempre llevar\'e conmigo. De verdad que llegados a este punto ni siquiera s\'e qu\'e decir, porque s\'e que no voy a ser capaz de transmitirte todo lo que me has dado durante estos a\~nos. Has estado ah\'i siempre, ense\~n\'andome y ayud\'andome paso a paso, incluso cuando te he llamado hist\'erica perdida por haber borrado la tesis entera. Has estado ah\'i en mis momentos malos, ayud\'andome a que fueran buenos, y has estado en los buenos, haciendo que fueran aun mejores. Porque estoy de acuerdo contigo, todo es mejor cuando lo hacemos juntos. Gracias por aguantar mis locuras, mis bajones, mis dudas, mi amor a veces un poco apabullante por los bichitos, mi pasi\'on por las catedrales, mi murga con temas de lo m\'as variado, mis viajes locos paliza, todo... Gracias por recordarme lo mucho que me ha llegado a gustar la f\'isica cuando m\'as lo necesitaba, por darme siempre un lugar al que volver y en el que sentirme a gusto y segura y, sobre todo, gracias por compartir tu vida conmigo. Por favor, no dejes nunca de hacerlo. Gracias por cuidarme tanto y tan bien, incluso desde la distancia. S\'e que no hemos llevado una vida especialmente f\'acil, y lo que nos queda, pero ?`desde cuando nos gustan a nosotros las cosas f\'aciles? A d\'ia de hoy, estoy segura de que merecer\'a todo la pena, y de que un d\'ia nos veremos con Planckito y con una bater\'ia de cocina que ni Masterchef. Gracias por todo. Gracias por ser mi tulip\'an blanco. AMZ Chofi.
\vspace{1cm}

\begin{figure}[H]
\begin{center}
\includegraphics[width=0.3\textwidth]{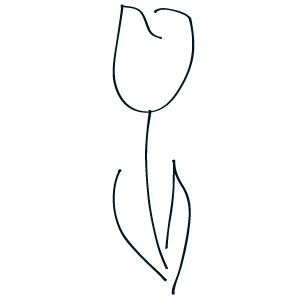}
\end{center}
\end{figure}


\bibliography{bibliography}

%

\end{document}